# Analysis of Chronic Diseases Progression Using Stochastic Modeling

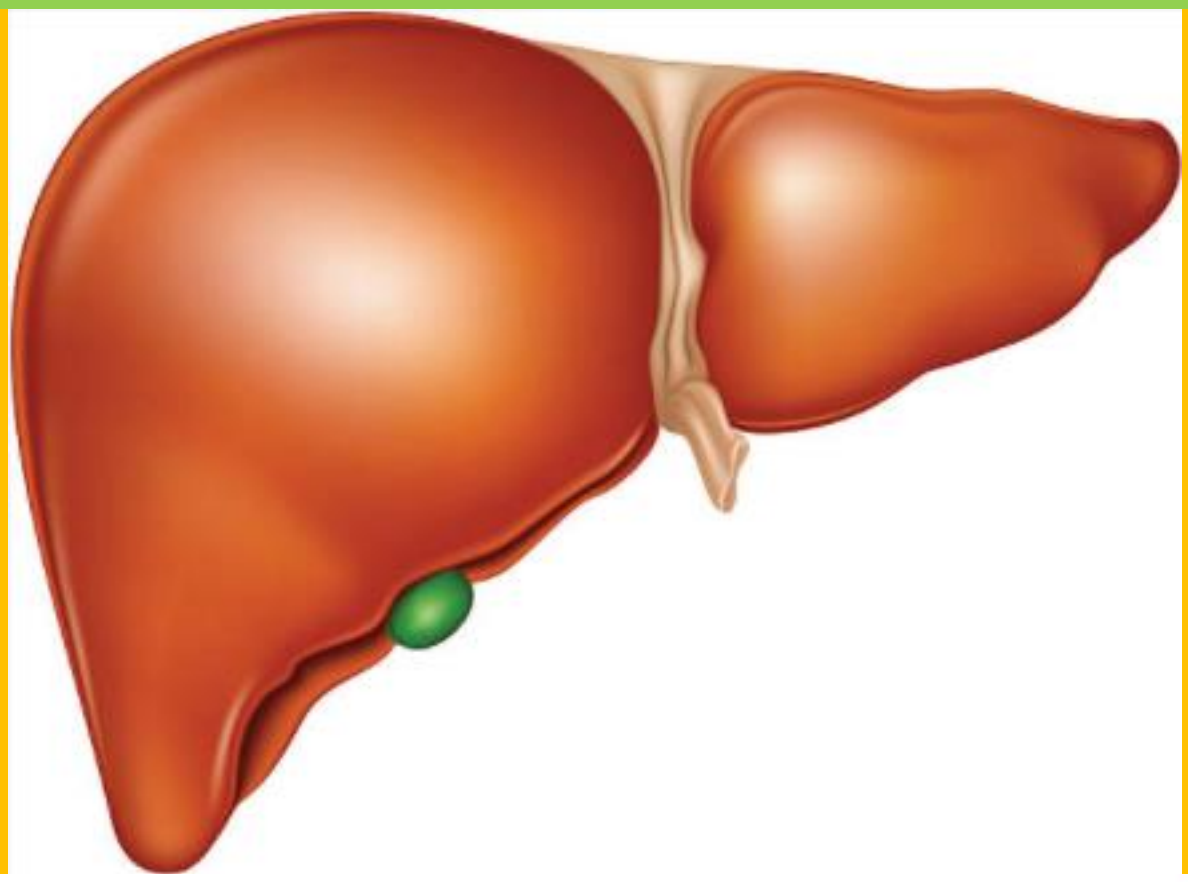

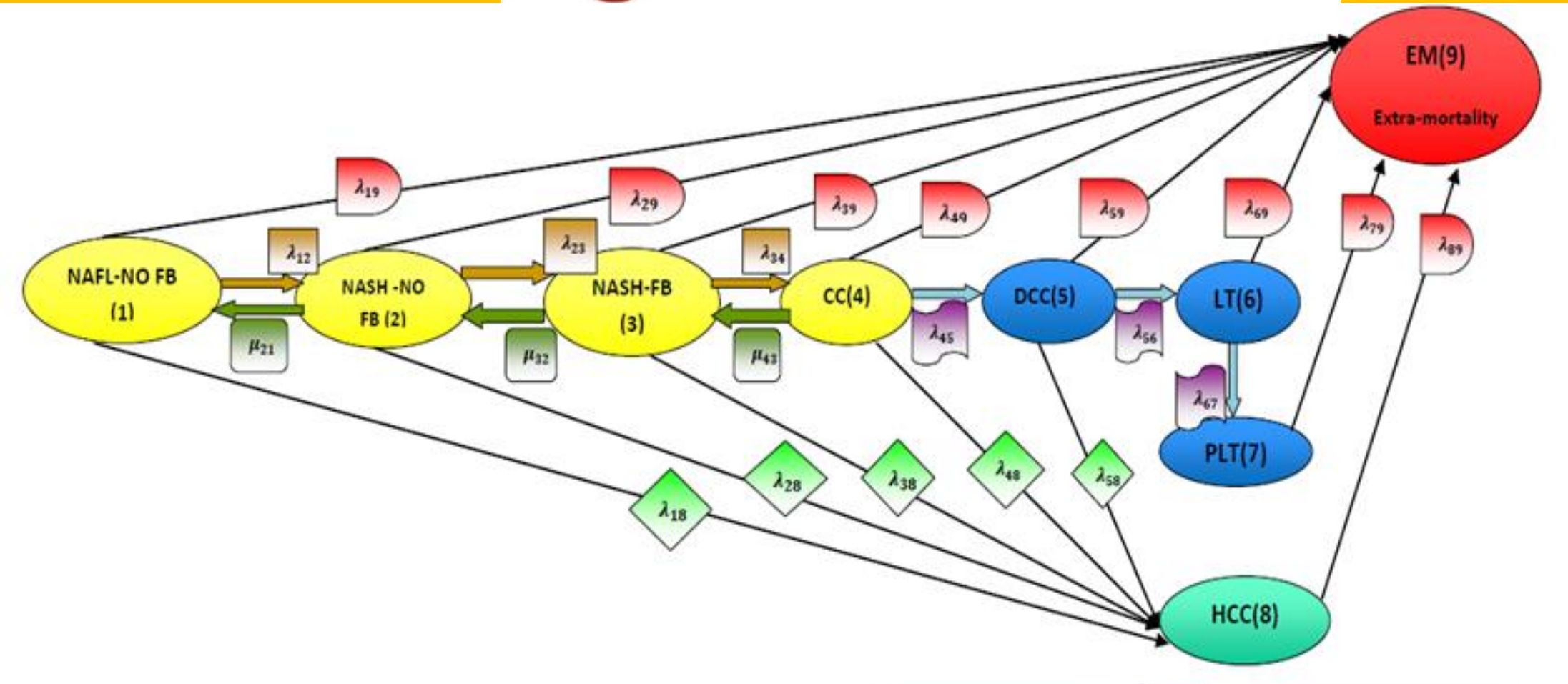

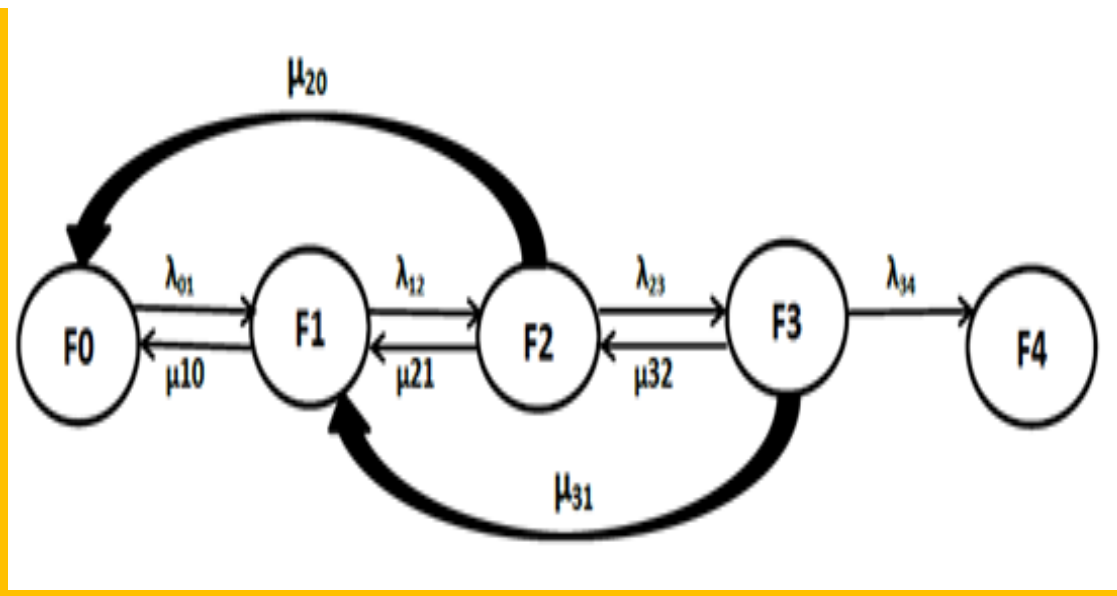

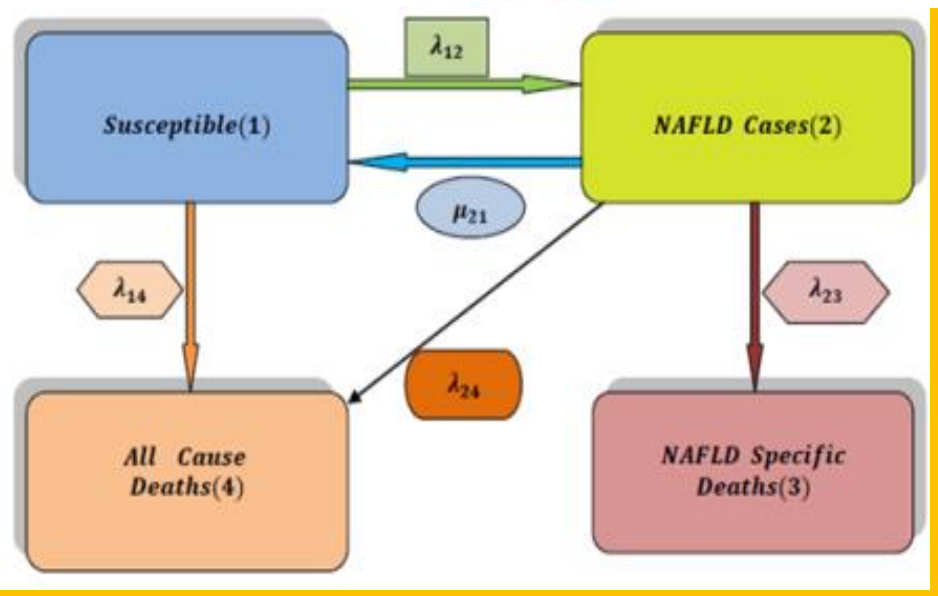

By

**Iman Mohammed Attia Abd-Elkhalik**

**MSc internal medicine, Cairo University.**

**Diploma in Statistics, Cairo University.**

**2021**

# Analysis of Chronic Diseases Progression Using Stochastic Modeling

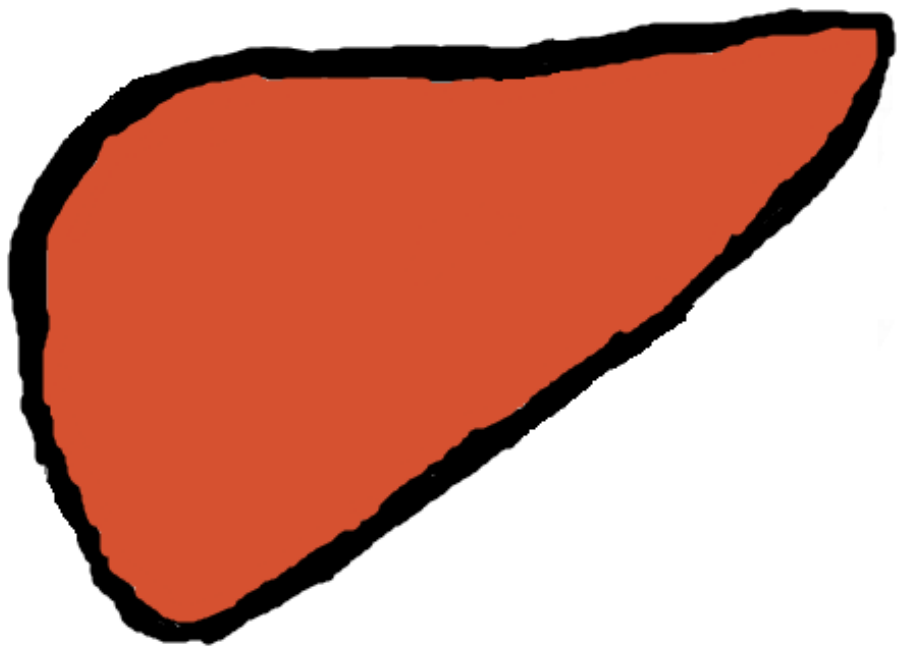

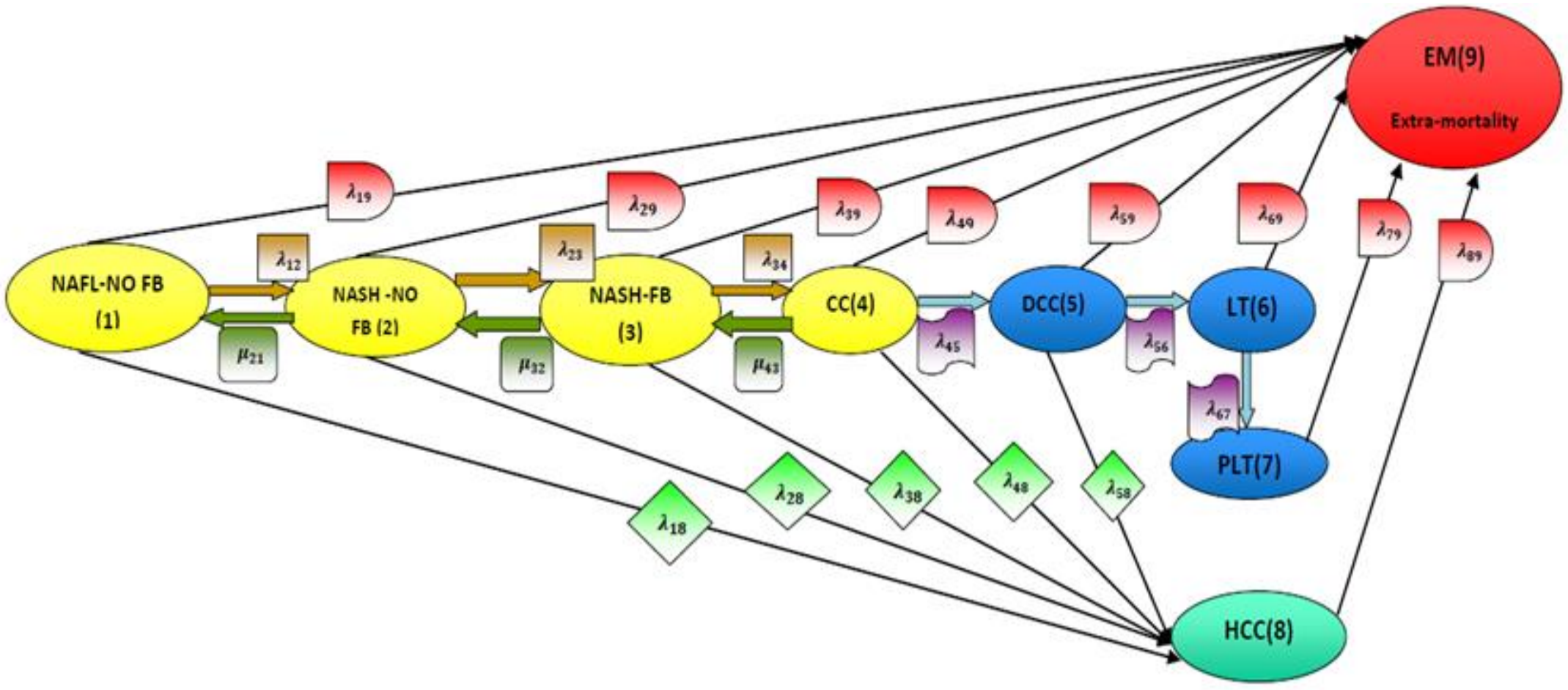

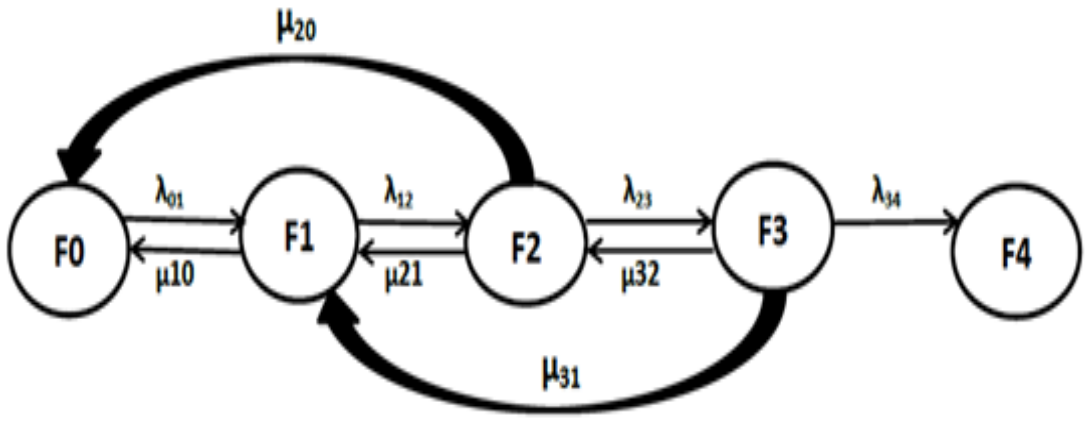

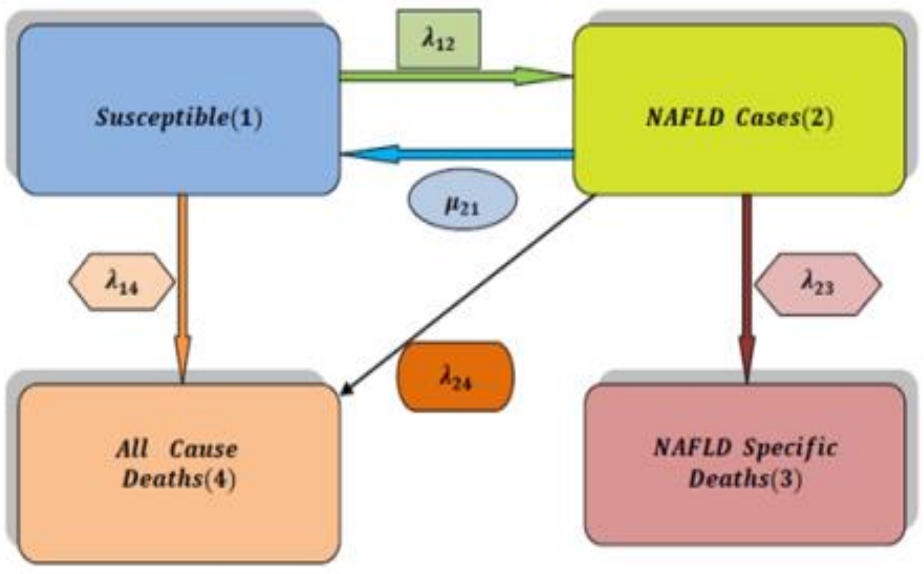

By

**Iman Mohammed Attia Abd-Elkhalik**

**MSc internal medicine, Cairo University.**

**Diploma in Statistics, Cairo University.**

**2021**

## Acknowledgement

I thank God Allah, The All Mighty, The most Generous and The most Glorious for help, easiness and facility making this research accomplished.

I would like to thank all my professors in Faculty of Graduate Studies for Statistical Researches, Cairo University for their effort clarifying lots of facts and ideas since I have been in this faculty in 2015. I would like to appreciate my thank for all who had helped me by all different means and who had clarified ideas and knowledge that aid me in better understanding of the statistical methodologies.

I owe great gratitude to my family especially my mother, my father and my brother for their patience, support, trustworthiness, fidelity and faithfulness in what I am doing, as well as their great graciousness that trigger me for this achievement. I cannot thank them enough for their continuous cooperation, delicacy, kindness and reinforcement. No sufficient thank can be offered to them.

I express great deep grace to my mother, the adversity she had experienced few months before delivering my proposal of this thesis was the inspiring and motivating driver for me to choose and be engaged in this research, as she most probably had non-alcoholic steatohepatitis (NASH) which is a complication of fatal cachexia developed as a result of her multiple sclerosis. The disease had a rapidly progressive course and she did not have a chance to be thoroughly diagnosed before her death. NASH is a subtype of non-alcoholic fatty liver disease (NAFLD), which is a prevalent disease with rapidly increasing epidemic worldwide. A lot of questions have not yet been answered and they need more exploration and research. I hope this work will help to through lights on this disease process from statistical point of view to help physicians and health policy makers have better and more efficient management plans for the challenges they are facing while assessing this worldwide rapidly rising epidemic.

I would like to appreciate informing me with errors, comments, and I may be contacted by electronic email: imanattiafree1972@gmail.com . I would like to refer to the site gettyimages.com/photos/human-liver, as the image of the liver on the cover page of the book had been taken from this site, it is created by AnnaSumska, with DigitalVisionVectors, 165720580.

## Abstract

Epidemiology is the science that studies the occurrence of the disease and numerous mathematical methods are used to study such occurrence. Multistate models are one of these mathematical methods that relay on solving differential equation to get some of the statistical indices. Thus multistate Markov model is a worthy tool to model time to event data such as longitudinal studies conducted on patients on several follow up periods. In medical researches, this technique is used to model disease evolution in which each patient starts in one initial state and eventually ends in absorbing or final state. Continuous Time Markov Chain (CTMC) is one of these multistate models, it is used to estimate transition intensities and probabilities between states, state probabilities distribution at specific time point, mean sojourn time in each state, life expectancy for the patient and expected number of the patients in each state. As the prevalence of obesity and type 2 diabetes have reached an epidemic levels that parallel the rates of widespread distributed non-alcoholic fatty liver disease (NAFLD), CTMC is used to model NAFLD to get better insight about the behavior of such widely prevalent disease worldwide. This helps improve the management of NAFLD as regards early detection and treatment to avoid its morbid complications.

This book handles the analysis of disease evolution over time using continuous time Markov chains and its application on the general abstract form of the disease state with no specification of the states of the disease, then on the unique detailed specification of the states of NAFLD , and lastly on a subset of the early states. This subset of states is used to relate the covariates like: insulin resistance, hyperlipidemia, and hypertension to the rates of transitions among states.

This book throws lights on the work conducted by the author to provide a new approach using maximum likelihood estimation (MLE) and quasi-newton method for prediction of the transition rates among states, and once these rates are obtained the probability transition matrix can be calculated. This approach compensates for the missing values when patients do not commit to the follow up schedule, because it calculates the rate for each interval and then takes weight from each rate corresponding to the proportion of transition counts in each interval from the total transition counts. This approach has been applied to the simplest form of “health, disease, and death” model as well as to the expanded form of the disease constituting the nine states, as each disease has its own characteristic stages in addition to its unique and specific transitions among the states. Also a subset of the “nine states model” that defines the early reversible stages of the disease, pointing to how the fibrosis evolves over time, was utilized to understand the factors that determine its formation, because fibrosis is the ominous predictor for bad outcome and death. The results have yielded that the observed rates approximately equal the estimated rates obtained by MLE used in both the simple and expanded models. Exponentiating this rate matrix as well as analytically solving the differential equations gave the probability transition matrix, but with more stable solution given analytically. Poisson regression was used to relate these rates with the covariate risk factors of the disease like: age, body mass index (BMI), homeostasis measurement assessment-insulin resistance (HOMA-IR2) reflecting insulin resistance, low density lipoprotein cholesterol (LDL-Chol), systolic, and diastolic blood pressure. It has been shown that the most detrimental risk factor for disease progression was insulin resistance, the more resistant to insulin the cells were, the higher the rate of transition to advanced liver fibrosis was. The book contains a hypothetical data for each model to highlight the mathematical statistical concepts used for analysis of such a highly spread disease. In the appendix, a Matlab code is presented to illustrate the calculations used in estimation of rate transition matrix as well as the probability transition matrix. The results obtained from analysis of hypothetical data presented in chapter five are coded by a Matlab code published in site “code ocean” with the following URL: “codeocean.com/capsule/7628018/tree/v1”, in addition the hypothetical data presented in chapter six is coded by Stata and published in the same site with the following URL: “codeocean.com/capsule/4752445/tree/v1”.

# Table of contents

# List of Figures

# List of Tables

# List of abbreviations

| Abbreviation | Meaning |
|---|---|
| AAR | AST to ALT ratio |
| ALT | Alanine Transaminase |
| APRI | AST to Platelet Ratio Index |
| AST | Aspartate Transaminase |
| ATP | Adenosine Triphosphate |
| BMI | Body Mass Index= weight(kg) /height ($m^2$) |
| CAP | Controlled Attenuation Parameter |
| CK-18 | Cytokeratin-18 |
| CT | Computed Tomography |
| CTMC | Continuous time Markov chains |
| CVD | Cardiovascular Disease |
| EASL-SASD-EASO | European Association of Study of Liver Disease-European Association of Study of Diabetes-European Association of Study of Obesity |
| ELF | Enhanced Liver Fibrosis |
| ER | endoplasmic reticulum |
| FIB-4 | Fibrosis Index Score |
| GGT | Gamma Glutamyl Transferase |
| HCC | Hepato-Cellular Carcinoma |
| HCV | Hepatitis C Virus |
| LDL | Low Density Lipoprotein-Cholesterol |
| LSM | Liver Stiffness Measurement |
| MetS | Metabolic Syndrome |
| MOMA-IR | Homeostatic model Assessment of Insulin Resistance |
| MRE | Magnetic Resonance Elastrography |
| MSIR | Maternally Derived Immunity, Susceptible, Infectious, Recovery |
| NAFL | Non-Alcoholic Fatty Liver |
| NAFLD | Non-Alcoholic Fatty Liver Disease |
| NASH | Non-Alcoholic steatohepatitis |
| NFS | NAFLD Fibrosis Score |
| NIVs | Non-Invasive Tests |
| PNPL-3 | Patatin Like Phospholipase Domain Containing 3 Gene |
| SEIR | Susceptible, Exposed, Infectious, Recovery |
| SEIS | Susceptible, Exposed, Infectious, Susceptible |
| SIR | Susceptible, Infectious, Recovery |
| SIRD | Susceptible, Infectious, Recovery,Deceased |
| SIRV | Susceptible, Infectious, Recovery,Vaccinated |
| SIS | Susceptible, Infectious, Susceptible |
| U/S | ultrasonography |
| VCTE | Vibration Controlled Transient Elastography |

# Chapter One: Introduction

Study of disease occurrence is the scope of the epidemiology. It is the science that studies the distribution, forms and factors of the disease as well as the health conditions in a specific population. There are many mathematical models that can describe the disease process according to the stages of the disease. They are described in the form of ordinary differential equations that are deterministic but they also can be formulated with a stochastic frame which is close to reality but with much more complexity to be analyzed. Some of these models used in the infectious diseases is the basic model of SIR described with 3 states (susceptible, infectious, recovered) and it reflects the stages which the patient can pass through when acquiring infection. There are some variations of this model like: SIS model formulated with 2 states ( susceptible and infectious) with no recovery state as in common cold because the patient does not develop immunity, SIRD model described with 4 states( susceptible, infectious, recovered, deceased) and the difference here is that recovery means immunity while the deceased means the chronic complications and sequels of the disease, SIRV model is an extension of SIR and take into consideration the vaccinated stage, MSIR is also an extension to the SIR model but the addition of first stage that reflect the "maternally derived immunity" that the newborns have from their mothers in the first few months after birth, SEIR model that describes the following states (susceptible, exposed, infectious, recovered) with the so called exposed state reflecting the latency or incubation period during which the patient has been infected but the manifestations of the disease is still not apparent and the ability to infect others is still not acquired, SEIS model (susceptible, exposed, infectious, susceptible) is the same as the directly preceded model but with no recovery stage (Bailey, 1975).

There are other models that can describe the chronic disease other than the infectious diseases like the diseases caused by disturbed immunity, cancer, genetic and metabolic diseases. One of these models is the DisMod II model with 4 states (susceptible, diseased, death due to disease, death from any other cause) (Barendregt et al., 2003) .Also multistate models, like the continuous time Markov chains (CTMC), are elaborative to analyze the chronic diseases. CTMC is frequently used to model panel data in various fields of science, including: medicine, sociology, biology, physics and finance. A multistate Markov model is a valuable tool to model time to event data utilized in longitudinal studies conducted on patients on several follow up periods. In medical studies, this technique is used to model illness-death process in which each patient starts in one initial state and eventually ends in absorbing or final state .The patient may experience several events that are related to their original disease or to some other events that are not part of their original disease (Kruijshaar et al., 2002).

There may be several intermediate states in-between the initial and the final state and they can or cannot be visited by the patient. Not all the patients may reach the final state by the end of the study or they may be lost before reaching any intermediate state or final state and so they are called censored patients.

Continuous Time Markov Chain (CTMC) as being one of these multistate models, its main objectives are to identify all the possible movements among the states to estimate the following:

- Transition intensities and probabilities between states
- State transition probabilities at particular time point.
- Mean sojourn time in each state
- Life expectancy for the patient
- Expected number of the patients in each state

Also CTMC is used to incorporate and thus identify covariates that affect the transition intensities aiming to provide opportunity to evaluate the factors that influence various movements among these states.

In this book the light will be thrown on some of the statistical concepts and indices that can be derived using CTMC to analyze the non –alcoholic fatty liver disease (NAFLD). This rapidly increasing metabolic derangement of the liver and its associated complications has huge economic burden on society to allocate resources for investigating this disease process, preventing, and treating it. The study will handle the following chapters:

**Chapter Two:** this chapter will highlight the basic statistical and mathematical definitions as well as the proofs of some theories.

**Chapter Three:** this chapter will simply explain the medical definitions of the NAFLD disease as well as how to investigate it. These concepts will be used to understand the definitions of the states of the NAFLD disease. Also some of the ongoing studies, concerning the treatments, are mentioned.

**Chapter Four:** this chapter will demonstrate the simple model of "health, disease, and death" process and how the statistical indices can be mathematically derived with an applied hypothetical numerical example.

**Chapter Five:** this chapter will clarify the disease process in more extended form as the "disease state" will be expanded to describe the unique states for NAFLD, the transitions among the states, and the ultimate fate of the process. Finally this statistical explanation will be supplemented with an applied hypothetical example.

**Chapter Six:** this chapter will elucidate how the incorporated covariates can be related to the transition rates among the states. A subset of the states taken from the extended model describing the early reversible stages of the disease will be used in the model.

The mathematical approach used in this work can be summarized in the following items:

- Applying quasi-newton method after obtaining the maximum likelihood function will yield the transition rate matrix Q.
- Exponentiation of this Q matrix will give the probability transition matrix, solving the Kolmogorov differential equations to get the probability functions, thereafter substitute Q matrix in these functions to get the final probability functions, this approach yields a more stable transition probability matrix than exponentiation.
- Poisson regression model is used to incorporate the covariates, thus relating these covariates with the transition rates.

**Chapter Seven**: in this chapter the author conveys the conclusions, recommendations and future works. This is followed by appendix A & B that contain the Matlab code to estimate the transition rate matrix and probability rate matrix respectively, while appendix C discusses the hypotheses tests for the models describes in the chapter four, five and six. Appendix 4 contains the hypothetical data of the patients presented in chapter six.

# Chapter Two: Definitions and Notation

## Review of literature

stochastic processes enlarges the concept of the random variable to include time , that is to mean the random variable does not only map an event $s \in \Omega$ ,where $\Omega$ is the sample space ,to some number $X(s)$ but it maps it to different numbers at different times. So this is not only a number $X(s)$ but it is $X(s,t)$ where $t \in T$ and T is called the parameter set of the process and it is usually a mere set of times.

Therefore, stochastic process is defined as a family of random variables $\{X(t,s)|t \in T, s \in \Omega\}$ over a given probability space and is indexed by the parameter $t$. For such a random process is used interchangeably with the stochastic process (Ibe 2013).

Stochastic processes are classified into four types according to nature of the state space and the parameter space: If t is interval of real numbers hence the process is called continuous time stochastic process but if the t is a countable set of positive number the process is called discrete time stochastic process. The state space is either continuous or discrete (Castañeda, Arunachalam, and Dharmaraja 2012).

Examples for such classification:

- Discrete state, discrete time stochastic process:

The number of individuals in a population at the end of the year $t$ is modeled as a stochastic process $\{X(t,s)|t \in T, s \in \Omega\}$ having $T = \{0,1,2,\dots\}$ and state space $\Omega = \{0,1,2,\dots\}$

- Discrete state, continuous time stochastic process:

The number of incoming calls in an interval $[0,t]$. Then the stochastic process $\{X(t,s)|t \in T, s \in \Omega\}$ has $T = \{t: 0 \leq t \leq \infty\}$ and $\Omega = \{0,1,2,\dots\}$

- Continuous state, discrete time stochastic process:

The share price for an asset at the close of trading on day $t$ with $T = \{0,1,2,\dots\}$ & $\Omega = \{t: 0 \leq x \leq \infty\}$

- Continuous state, continuous time stochastic process:

The value of the Dow-Jones index at time $t$ such that $T = \{t: 0 \leq t \leq \infty\}$ and $\Omega = \{t: 0 \leq x \leq \infty\}$

Many chronic diseases have a natural progressive course over time, passing through successive stages comprising the disease process. CTMC is used to model this course of the disease. The “health, disease, and death” model is the simplest disease model. It has three states representing health, disease and death. Transitions are allowed to occur from health to disease to death and health to death. Recovery from disease to health is also permitted as figure (2.1):

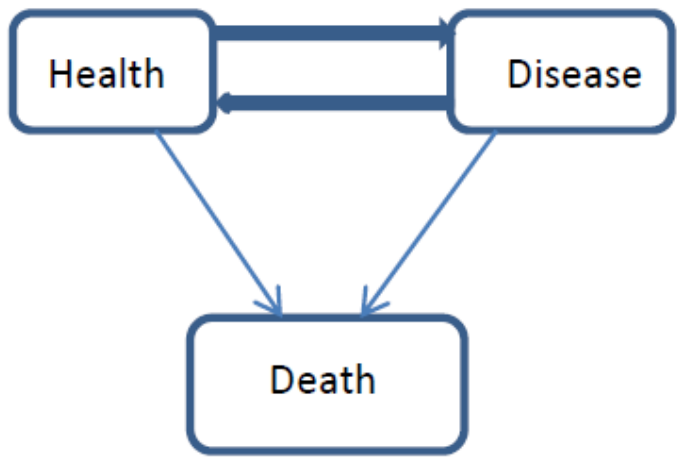

Figure (2. 1): health disease death model

A commonly realistic used model is expressed by a series of successively more sever disease stages, and an absorbing state, often death state. The patient may advance into or recover from adjacent disease stages, or die at any disease stage. Observations of the state $S_i(t)$ are made on a number of individuals $i$ at arbitrary time points $t$, which may vary between individuals, as shown in the figure (2.2) and (2.3):

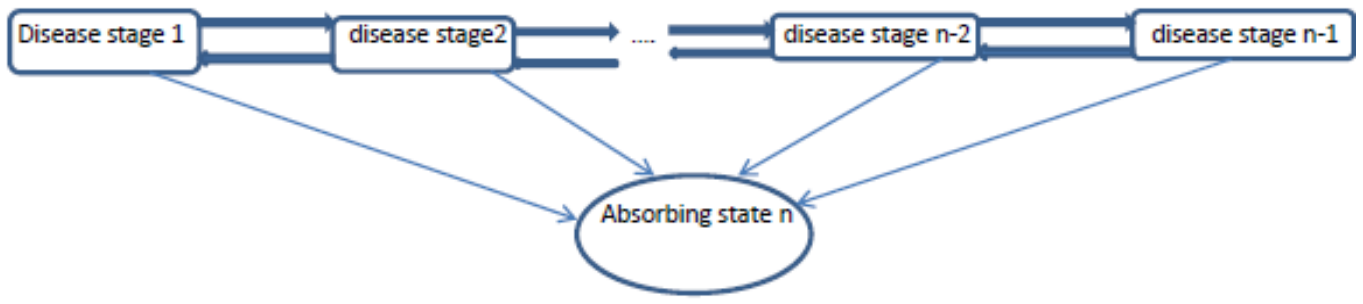

Figure (2. 2): general model for disease progression (reversible progression)

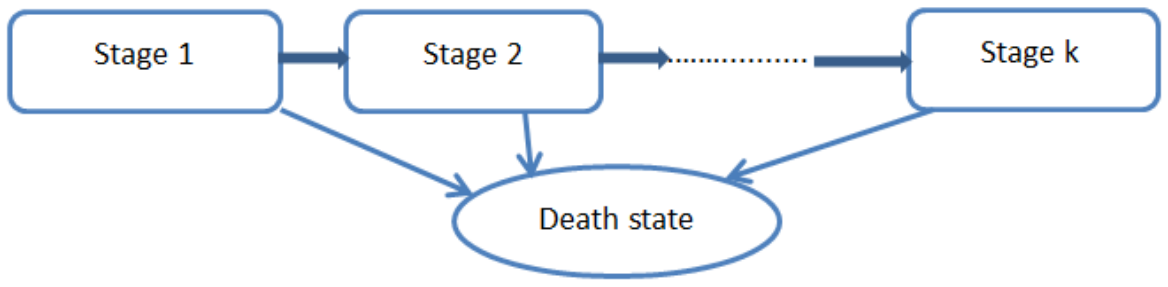

Figure (2. 3): general model for disease progression (irreversible progression)

Estes et al. (2018) used multistate Markov chains to model the epidemic of nonalcoholic fatty liver disease. They forecasted the non-alcoholic fatty liver disease progression to raise 21%, from 83.1 million (2015) to 100.9 million (2030), and the non-alcoholic steato-hepatitis to be elevated 63% from 16.52 million to 27 million cases. The prevalence in adult population aged more than 15 years will be 33.5% in 2030. The incidence of “decompensated liver cirrhosis” will elevate 168% to 105430 cases by 2030, incidence of liver cancer will rise by 137% to 12240 cases and deaths will increase by 178% to 78300 deaths in 2030. Also Younossi et al. (2016) used the multistate Markov chains to construct 5 models in United States and 4 European countries ( Germany, France, Italy, and United Kingdom) for estimating the burden of NAFLD in these countries. The patients move among 9 states with the following results: more than 64 million are estimated to have NAFLD with the annual cost of nearly 103 billion dollars (1613 dollars per patient), while in the 4 European countries there are approximately 52 million persons with NAFLD with annual cost of about 35 billion.

Anwar and Mahmoud (2014) used CTMC to study the progression of the chronic kidney disease and estimate the mean time spent in each stage of disease process as well as estimating life expectancy of a chronic kidney disease patient.

Grover et al. (2014) used time dependent multistate Markov chain to assess progression of liver cirrhosis in patients having hepatitis C virus with various prognostic factors.

Bartolomeo, Trerotoli, and Serio (2011) had employed a hidden Markov model to determine the transition probabilities incorporating various covariates for progression of liver cirrhosis to HCC and to death. Two illness states and one death state was considered and found that male patients have twice the probability of developing HCC as compared to female patients also ample presence of concomitant diseases increases the risk of death in patients with HCC.

Fackrell (2009) demonstrated the usage of the CTMC structure with phase type distribution in modeling the health care system.

Foucher et al. (2005) used semi-Markov model to explicitly define the waiting time distribution based on generalized Weibull distribution giving an extension of the homogenous CTMC based implicitly on exponential waiting time distribution and applying this concept with an example of the evolution of HIV infected patients.

Saint-Pierre et al. (2003) used CTMC with time dependent covariates and Markov model with piece-wise constant intensities to model asthma control evolution.

Jackson and Sharples (2002) measured forced expiratory volume in one second ($FEV_1$) at irregular interval and used hidden Markov models for studying the staged decline in respiratory functions as well as the influencing covariates after developing Brochiolitis Obliterans Syndrome (BOS) following lung transplantation. Jackson 2011 used CTMC to model the outcome of heart long transplant patients.

Pérez-Ocón, Ruiz-Castro, and Gámiz-Pérez (2001) had applied CTMC technique to examine the influence of three post-surgical treatments (chemotherapy, radiotherapy, hormonal therapy) to 300 breast cancer patients on the their life times and relapse times .The survival of the patients in the group where all three treatment combinations are given is more as compared to radiotherapy and chemotherapy group and radiotherapy group alone.

Marshall and Jones (1995) discussed multistate model consisting of 3 transient states representing early stage of diabetic retinopathy and one final absorbing state representing the irreversible stage of retinopathy. They explored the effects of factors influencing the onset, progression and regression of diabetic retinopathy among subjects with insulin-dependent diabetes mellitus under assumption that CTMC determines the transition times between disease stages.

Sharples (1993) modeled the transition rates between grades of coronary occlusive disease from each grade to death following cardiac transplantation. The disease process was graded on a 3 point scale according to amount of narrowing observed in major vessels using serial angiography.

Longini Jr et al. (1989) used a staged Markov model to estimate the distribution and mean length of the incubation period for AIDS from a cohort of 603 HIV infected persons who have been followed through various stages of infection which are modeled into 4 illness stages and one final absorbing death stage.

There are numerous studies over the several past decades addressing these statistical methods for analysis of disease evolution and progression through time.

## 2.1. Some Basic Definitions:

### 2.1.1. Some Basic Statistical and Mathematical Definitions:

**<u>Definition 1:</u> Stochastic Process:**

A real stochastic process is a collection of random variables $\{X_t; t \geq 0\}$ defined on a common probability space $(\Omega, \Im, P)$ with values in $\mathbb{R}$. T is called the index set of the process or parametric space, which is usually a subset of $\mathbb{R}$. The set of the values that the random variable $X_t$ can take is called the state space and is denoted by S . The mapping defined for each fixed $\omega \in \Omega$

$$X(\omega): T \to S$$

$$t \mapsto X_t(\omega)$$

**<u>Definition 2:</u> Markov Process:**

Let $\{X_t; t \geq 0\}$ be a stochastic process defined over a probability space $(\Omega, \Im, P)$ and with a state space $(\mathcal{B})$. $\{X_t; t \geq 0\}$ is a Markov process if for any $0 \leq t_1 \leq t_2 \leq \cdots \leq t_n$ and for any $B \in \mathcal{B}$ ,

$$P\left(X_{t_n} \in B | X_{t_1}, \ldots, X_{t_{n-1}}\right) = P\left(X_{t_n} \in B | X_{t_{n-1}}\right)$$

**Definition 3: Continuous Time Markov Process (CTMC) :**

Let $\{X_t; t \geq 0\}$ be a stochastic process with countable state space S. A process is a continuous time Markov chain if: $P\left(X_{t_n} = j | X_{t_1} = i_1, \ldots, X_{t_{n-1}} = i_{n-1}\right) = P\left(X_{t_n} = j | X_{t_{n-1}} = i_{n-1}\right)$

For all $j, i, \ldots, i_{n-1} \in S$ and for all $0 \leq t_1 \leq t_2 \leq \cdots \leq t_n$

**Definition 4: Homogenous Continuous Time Markov Chain:**

CTMC is homogenous if and only if $P(X_{t+s} = j | X_s = i)$ is independent of $s$ for all t .

**Definition 5: Transition Probability:**

Let $p_{ij}(t)$ be the probability of the transition from state $i$ to state $j$ in an interval of length $t$.

$p_{ij}(t) = (X_{t+s} = j | X_s = i)$ where $s, t \geq 0$. In matrix notation : $P(t) = \left(p_{ij}(t)\right)\ for\ all\ i, j \in s$

It satisfies the following conditions:

$$p_{ij}(0) = \delta_{ij}$$

$$\lim_{t \to 0^+} p_{ij}(t) = \delta_{ij}\ so\ \lim_{t \to 0^+} P(t) = I$$

$$for\ any\ t \geq 0, i, j \in S, 0 \leq p_{ij}(t) \leq 1\ the \sum_{k \in S} p_{ik}(t) = 1$$

$$for\ all\ i, j \in S, for\ any\ s, t \geq 0 :\ p_{ij}(t+s) = \sum_{k \in S} p_{ik}(t) p_{kj}(s)\quad so\quad P(t+s) = P(t)P(s)$$

And this is called chapman-Kolmogorov equation for CTMC.

**Definition 6: Transition Rate Matrix or Intensity Matrix or Infinitesimal Generator**

Let $\{X_t; t \geq 0\}$ be a CTMC , $q_{ij}(t)$ is the rate at which transition occur from state $i$ to state $j$ at time $t$ or $Q(t) = \lim_{\Delta t \to 0} \left\{\frac{P(t, t+\Delta t) - I}{\Delta t}\right\} = P'(0)\ \ Since\ P(0) = I$

$$Q = \begin{bmatrix} -q_0 & q_{01} & q_{02} & \cdots \\ q_{10} & -q_1 & q_{12} & \cdots \\ q_{20} & q_{21} & -q_0 & \cdots \\ \vdots & \vdots & \vdots & \ddots \end{bmatrix}$$

**Definition 7: Stationary Probability Distribution**

Let $\{X_t; t \geq 0\}$ be a CTMC with generator matrix $Q$ and transition probability matrix $P(t)$ . Suppose $\pi = (\pi_0, \pi_1, \ldots)^t$ is nonnegative i.e. $\pi_i \geq 0$ for $i = 0,1,2, \ldots$ So $Q\pi = 0\ and\ \sum_{i=0}^{\infty} \pi_i = 1$ . This $\pi_i$ is called stationary probability distribution .it can also be defined in terms of $P(t)$ such that

$$\pi P(t) = \pi\ , for\ t \geq 0\ , \sum_{i=0}^{\infty} \pi_i = 1\ , \pi_i \geq 0\ for\ i = 0,1,2, \ldots and\ known\ P(t) of\ a\ finite\ process$$

**Definition 8: Embedded Markov Chain (EMC)**

Let $Y_n$ denote a random variable for a CTMC $\{X_t; t \geq 0\}$ at the $n^{th}$jump, $Y_n = X(W_n), n = 0,1,2, ..$

Where $W_i$ is the time at which the $n^{th}$ jump occurs and the $T_i = W_{i+1} - W_i$ is the holding time or the time spent in the state until the next jump occurs at $W_i$. The set of the random variables $\{Y_n\}_{n=0}^{\infty}$ is known as the embedded Markov chain or the jump chain associated with the CTMC $\{X_t; t \geq 0\}$ with a transition matrix $T = t_{ij}$

**Definition 10: Reachability or Accessibility**

State $j$ can be reached from state , $i \to j$ , if $p_{ij}(t) > 0$ for some $t \geq 0$.

**Definition 11: Communicating State**

State $i$ communicate with state $j$ , $(i \leftrightarrow j)$, if $i \to j$ and $j \to i$

The set of states that communicate is called a communication class.

**Definition 12: Irreducibility.**

If every state can be reached from every other state, the Markov chain is irreducible; otherwise, it is said to be reducible.

**Definition 13**: **Closed Class**

A set of states C is closed if it is impossible to reach any state outside of C from a state inside C, $p_{ij}(t) = 0\ ;\ for\ t \geq 0\ and\ \ if\ ;\ \ i \in C\ \ \ and\ \ \ j \notin C$

**Definition 14: First Return Time**

$T_{ii}$ is the first time the chain is in state $i$ after leaving state $i$ , it can occur for $t > 0$

**Definition 15: Recurrent State**

State $i$ is recurrent in a CTMC $\{X_t; t \geq 0\}$ , if the first return time is finite $P\{T_{ii} < \infty | X(0) = i\} = 1$

State $i$ is recurrent in a CTMC $\{X_t; t \geq 0\}$ if and only if state $i$ in the corresponding embedded Markov chain $\{Y_n\}_{n=0}^{\infty}$ is recurrent.

**Definition 16: Transient State**

State $i$ is transient in a CTMC $\{X_t; t \geq 0\}$ , if the first return time is finite $P\{T_{ii} < \infty | X(0) = i\} < 1$

State $i$ is transient in a CTMC $\{X_t; t \geq 0\}$ if and only if state $i$ in the corresponding embedded Markov chain $\{Y_n\}_{n=0}^{\infty}$ is transient.

**2.1.2.Some basic medical definitions:**

**Definition 18**: ATP is the molecule that store energy inside the cell

**Definition 19**: endoplasmic reticulum is a component in cell that acts as a factory responsible for protein synthesis for the cell.

**Definition 20** : Fibro-genesis mechanism through which fibrous tissue is formed i.e the process of fibrous tissue formation.

**Definition 21**: Pathophysiology is the mechanism of the disease by which the disease induces its effects.

## 2.2. Model Specification:

### 2.2.1. Probability Transition Matrix:

The model is specified by a probability transition matrix $P(t)$ whose $(i,j)th$ entry ,$p_{ij}(t)$ is the probability of a transition from state $i$ at time $t$ to some other state $j$ at rate $q_{ij}(t)$ per specified unit of time according to the studied process or system. So continuous time Markov chain is modeled by its matrix of transition rates $Q(t)$ at time $t$ . The probability that a transition occurs from a given source state to a specific destination state depends on both the source as well as the length of the interval of observation. That is to say if the period of observation $\tau = \Delta t$ has a very small duration so the probability of observing a transition from state $i$ at time $t$ to state $j$ at time $t + \Delta t$ during this interval $[t, t + \Delta t)$ i.e. $p_{ij}\ (t, t + \Delta\, \mathrm{t})$ is very small.

So as $\Delta t \to 0,\ p_{ij}\ (t, t + \Delta\, \mathrm{t}) \to 0\ \ \ for\ i \neq j$ .

And from the conservation of probability:

$p_{ii}\ (t, t + \Delta\, \mathrm{t}) \rightarrow 1\ as\ \Delta t \rightarrow 0$ .

On the other hand, as $\Delta t$ enlarges this probability $p_{ij}(t, t + \Delta t)$ increases to the level that the larger the period is, the more probable multiple events will be observed. Nevertheless, the observation periods are adequately selected to be small enough so that the probability of observing multiple events in such small observation period is of order $o(\Delta t)$ , a quantity for which

$$\lim_{\Delta \rightarrow 0} \frac{o(\Delta t)}{\Delta t} = 0$$

Continuous Markov chain exhibits Markov (memoryless) property:

$$P[X(t_{k+1}) = x_{k+1}\ |\ X(t_k) = x_k, X(t_{k-1}) = x_{k-1}, \dots, X(t_0) = x_0\ ] = P[X(t_{k+1}) = x_{k+1} | X(t_k) = x_k]$$

for any $t_0 \leq t_1 \leq \cdots \leq t_k \leq t_{k+1}$ . So if the current state $x_k$ is known, then the value taken by $X(t_{k+1})$ depends only on $x_k$ and not on any past history of the state (no state memory). Also, the amount spent in the current state does not determine the next state (no age memory)(Cassandras and Lafortune 2009).

### 2.2.2. Generator Matrix or the Transition Rate Matrix:

According to (Allen 2010), the transition probabilities $p_{ij}(t)$ are used to obtain transition rates $q_{ij}(t)$. A rate of transition does not depend on the length or duration of observation period, it is an instantaneously defined quantity that indicates the number of transitions that occur per unit time. The $q_{ij}(t)$ is the rate of transition from state $i$ to state $j$ at time t . In non-homogenous Markov chain both $q_{ij}(t)$ and $p_{ij}(t)$ may depend on the time $t$ not the interval $\Delta t$ .

The transition probabilities $p_{ij}(t)$ are continuous and differentiable for $\geq\ 0$ . And at $t = 0$, they equal

$$p_{ij}(0) = 0 \quad and \quad p_{ii}(0) = 1$$

While defining:

$$q_{ij} = \lim_{\Delta t \rightarrow 0} \left\{ \frac{p_{ij}(\Delta t) - p_{ij}(0)}{\Delta t} \right\} = \lim_{\Delta t \rightarrow 0} \left\{ \frac{p_{ij}(\Delta t)}{\Delta t} \right\} \quad for\ i \neq j$$

And

$$q_{ii} = \lim_{\Delta t \rightarrow 0} \left\{ \frac{p_{ii}(\Delta t) - p_{ii}(0)}{\Delta t} \right\} = \lim_{\Delta t \rightarrow 0} \left\{ \frac{p_{ij}(\Delta t) - 1}{\Delta t} \right\} \quad for\ i = j$$

As well as

$$p_{ij}(\Delta t) = q_{ij}(t)\Delta t \quad and \quad \sum_{j \neq i}^{\infty} p_{ij}(\Delta t) = \sum_{j \neq i}^{\infty} q_{ij}(t)\Delta t + o(\Delta t)$$

And from conservation of probability:

$$1 - p_{ii}(\Delta t) = \sum_{j \neq i} p_{ij}(\Delta t) = \sum_{j \neq i} q_{ij}(t)\Delta t + o(\Delta t)$$

$$1 - p_{ii}(\Delta t) = \lim_{\Delta t \rightarrow 0} \sum_{i \neq j}^{\infty} \left\{ \frac{q_{ij}(t)\Delta t + o(\Delta t)}{\Delta t} \right\} = -\sum_{i \neq j}^{\infty} \frac{q_{ij}\ \Delta t}{\Delta t}$$

$$q_{ii} = -\sum_{j=0, j\neq i}^{\infty} q_{ij}(t)$$

In homogenous continuous time Markov chain:

$$q_{ij} = \lim_{\Delta t \to 0} \left\{ \frac{p_{ij}(\Delta t)}{\Delta t} \right\} \quad for \;\; i \neq j \quad ; \;\; q_{ii} = \lim_{\Delta t \to 0} \left\{ \frac{p_{ii}(\Delta t) - 1}{\Delta t} \right\} \; for \; i = j$$

Or in matrix notation:

$$Q = \lim_{\Delta t \to 0} \left\{ \frac{P(\Delta t) - I}{\Delta t} \right\}$$

To sum up

$$Q = \begin{bmatrix} -q_0 & q_{01} & q_{02} & \cdots \\ q_{10} & -q_1 & q_{12} & \cdots \\ q_{20} & q_{21} & -q_0 & \cdots \\ \vdots & \vdots & \vdots & \ddots \end{bmatrix}$$

It is called infinitesimal generator matrix or rate transition matrix of Markov chain

$$Q = P'(0) \quad since \quad P(0) = I$$

If $S$ is a finite or countable state space and $Q = \left(q_{ij}\right)_{i,j \in S}$ then it satisfies the following properties:

$$q_{ij} \geq 0 \;\; for \; i \neq j \;\; and \; q_{ii} \; \leq 0 \;\; for \; all \; i \; , \; \sum_{j \neq i} q_{ij} = -q_{ii} \quad and \; \sum_{j} q_{ij} = 0 \;\; for \; all \; i$$

## 2.3. Chapman-Kolmogorov Equations:

According to (Cassandras and Lafortune 2009), the transition probability is called $p_{ij}(s,t)$ and is defined as $p_{ij}(s,t) = P[X(t) = j | X(s) = i], s \leq t$

To derive the equation; the events of transitions $[X(t) = j | X(s) = i]$ are conditioned on $[X_u = r]$

For some $u$ such that $s \leq u \leq t$ and that the rule of total probability implies

$$p_{ij}(s,t) = \sum_{all\; r} P[X(t) = j | X(u) = r, X(s) = i] \, . \, P[X(u) = r | \; X(s) = i]$$

And in addition the memory-less property ensures the following:

$$P[X(t) = j | X(u) = r, X(s) = i] = P[X(t) = j | X(u) = r] = P_{rj}(u,t)$$

Moreover , $P[X(u) = r | \; X(s) = i] = P_{ir}(s,u)$. Therefore,

$$p_{ij}(s,t) = \sum_{all\; r} P_{ir}(s,u) . P_{rj}(u,t) \qquad , s \leq u \leq t$$

The above Chapman-Kolmogorov equation can be rewritten in matrix notation as:

$H(s,t) = H(s,u)H(u,t) \; , \;\; s \leq u \leq t \;\; : where$

$H(s,t) = \left[p_{ij}(s,t)\right] \;\; and \quad H(s,s) = I \; ( \, identity \; matrix)$

## 2.4. Chapman-Kolmogorov Differential Equation:

### 2.4.1. Forward Chapman-Kolmogorov differential equation:

The Chapman-Kolmogorov equation for time instants $s \leq u \leq t + \Delta t \ \ and \ \ \Delta t > 0$ is

$H(s, t + \Delta \text{t}) = H(s,t)H(t, t + \Delta t\,)$ subtracting $H(s,t)$ from both sides of this equation yields

$H(s, t + \Delta \text{t}) - H(s,t) = H(s,t)[H(t, t + \Delta t\,) - I]$ , then dividing by $\Delta t$ and taking the limit as $\Delta t \to 0$ gives

$$\lim_{\Delta t \to 0} \frac{H(s, t + \Delta t\,) - H(s,t)}{\Delta t} = H(s,t) \ \ \lim_{\Delta t \to 0} \frac{H(t, t + \Delta t\,) - I}{\Delta t}$$

$$\lim_{\Delta t \to 0} \frac{H(s, t + \Delta t\,) - H(s,t)}{\Delta t} = \frac{\partial H(s,t)}{\partial t} \ , is\ the\ partial\ derivative\ \ of\ p_{ij}(s,t)\ with\ respect\ to\ (t)\ \ if\ it\ exits$$

And $Q(t) = \lim_{\Delta t \to 0} \frac{H(t, t + \Delta t\,) - I}{\Delta t}$

$$\lim_{\Delta t \to 0} \frac{H(s, t + \Delta t\,) - H(s,t)}{\Delta t} = H(s,t) \ \ \lim_{\Delta t \to 0} \frac{H(t, t + \Delta t\,) - I}{\Delta t} \ = \frac{\partial H(s,t)}{\partial t} = H(s,t)Q(t) \quad , \ \ s \leq t$$

This is called forward differential equation and in a similar fashion, the backward differential equation can be derived.

According to (Stewart 2009) the forward differential equation can also be derived like this :

The Chapman-Kolmogorov equations for CTMC are derived from Markov property that states

$$p_{ij}(s,t) = \sum_{all\ k} p_{ik}(s)\, p_{kj}(t) \ \ for\ i,j = 0,1,2, \ldots \qquad and\ \ s \leq \ u \ \leq t$$

To transfer from state i at time s to state j at time t , some state k will be visited as an intermediate state between states i and j at an intermediate time u . When the CTMC is homogenous, this is written as following:

$$p_{ij}(t + \Delta t) = \sum_{all\ k} p_{ik}(t)\, p_{kj}(\Delta t) = \sum_{k \neq j} p_{ik}(t)\, p_{kj}(\Delta t) + p_{ij}(t) p_{jj}(\Delta t) \quad for\ t, \Delta t \geq 0$$

Thus :

$$\frac{p_{ij}(t + \Delta t) - p_{ij}(t)}{\Delta t} = \sum_{k \neq j} p_{ik}(t) \frac{p_{kj}(\Delta t)}{\Delta t} + p_{ij}(t) \frac{p_{jj}(\Delta t)}{\Delta t} - \frac{p_{ij}(t)}{\Delta t}$$

$$= \sum_{k \neq j} p_{ik}(t) \frac{p_{kj}(\Delta t)}{\Delta t} + p_{ij}(t) \left( \frac{p_{jj}(\Delta t) - 1}{\Delta t} \right)$$

Taking the limit as $\Delta t \to 0$ and recalling

$$q_{ij} = \ \lim_{\Delta t \to 0} \left\{ \frac{p_{ij}(\Delta t)}{\Delta t} \right\} \ \ for\ \ i \neq j \quad ; \ q_{jj} = \lim_{\Delta t \to 0} \left\{ \frac{p_{jj}(\Delta t) - 1}{\Delta t} \right\} for\ i = j$$

$$\frac{dp_{ij}(t)}{dt} = \sum_{k \neq j} p_{ik}(t)\, q_{kj} + p_{ij}(t) q_{jj}$$

That is to mean the forward differential equation is:

$$\frac{dp_{ij}(t)}{dt} = \sum_{k \neq j} p_{ik}(t)\, q_{kj} \quad for\ i, j = 0,1, \dots; \quad and\ in\ matrix\ notation: \ \frac{dP(t)}{dt} = P(t)Q$$

**2.4.2. The backward differential equation is derived as in the following steps:**

$$p_{ij}(t + \Delta t) = \sum_{all\ k} p_{ik}(\Delta t)\, p_{kj}(t) \qquad for\ t, \Delta t \geq 0$$

$$= \sum_{k \neq j} p_{ik}(\Delta t)\, p_{kj}(t) + p_{ii}(\Delta t) p_{ij}(t)$$

$$\frac{p_{ij}(t + \Delta t) - p_{ij}(t)}{\Delta t} = \sum_{k \neq j} \frac{p_{ik}(\Delta t)}{\Delta t} p_{kj}(t) + \frac{p_{ii}(\Delta t)}{\Delta t} p_{ij}(t) - \frac{p_{ij}(t)}{\Delta t}$$

$$= \sum_{k \neq j} \frac{p_{ik}(\Delta t)}{\Delta t} p_{kj}(t) + \left(\frac{p_{ii}(\Delta t) - 1}{\Delta t}\right) p_{ij}(t)$$

Taking the limit as $\Delta t \to 0$ and recalling

$$q_{ij} = \lim_{\Delta t \to 0} \left\{\frac{p_{ij}(\Delta t)}{\Delta t}\right\} \quad for\ \ i \neq j \quad ; \ q_{ii} = \lim_{\Delta t \to 0} \left\{\frac{p_{ii}(\Delta t) - 1}{\Delta t}\right\} for\ i = j$$

Thus:

$$\frac{dp_{ij}(t)}{dt} = \sum_{k \neq j} q_{ik}\ \ p_{kj}(t) + q_{ii}\ \ p_{ij}(t)$$

That is to mean the forward differential equation is:

$$\frac{dp_{ij}(t)}{dt} = \sum_{k \neq j} q_{ik}\ \ p_{kj}(t) \quad for\ i, j = 0,1, \dots;\ and\ in\ matrix\ notation: \frac{dP(t)}{dt} = QP(t)$$

And the solution for this differential equation is by matrix exponential:

$$P(t) = ce^{Qt} = e^{Qt} = I + \sum_{n=1}^{\infty} \frac{Q^n t^n}{n!}$$

The constant of integration is $c = P(0) = I$ . The matrix exponential can be rather hard and unstable or unsteady to compute.

## 2.5. The sojourn time of the continuous time Markov chain:

Let $\{X(t), t \geq 0\}$ be a homogenous continuous-time Markov chain and it is in a non-absorbing state $i$ at time $t = 0$ . Let $T_i$ be the time until a transition out of state $i$ occurs. Then with $s > 0$ , $t > 0$

$$P\{T_i > s + t | X(0) = i\} = P\{T_i > s + t | X(0) = i, T_i > s\} P\{T_i > s | X(0) = i\}$$

$$= P\{T_i > s + t | X(s) = i\} P\{T_i > s | X(0) = i\}$$

$$= P\{T_i > t | X(0) = i\} P\{T_i > s | X(0) = i\}$$

Therefore,

$$\hat{F}(t + s) = \hat{F}(t) \hat{F}(s)$$

Where

$\hat{F}(t) = P\{T_i > t|X(0) = i\}, t > 0$ . and as long as this is true if and only if $\hat{F}(t) = e^{-u_i t}$ for some positive parameter $u_i > 0$ and $t > 0$ so sojourn time in state $i$ must be exponentially distributed .if markov chain started at time $t = 0$ in state $i$ and has not moved from state $i$ by time $s$ which is equivalent to saying that $P\{T_i > s|X(0) = i\} = 1$, then $P\{T_i > s + t|T_i > s\} = P\{T_i > t\}$ and the continuous random variable time $T_i$ is memoryless. And since the only continuous distribution that has the memoryless property ( the distribution of residual time being equal to the distribution itself) is the exponential distribution, this results in that the duration of time until a transition occurs from state $i$ is exponentially distributed .In a homogenous continuous time Markov chain with a non-absorbing $i$ state which can move to one or more states $j \neq i$ , the mempryless property of the chain compel the duration of time until this transition takes place to be exponentially distributed with a rate of transition $q_{ij}$ . So the time to reach some state $j \neq i$ has an exponential distribution with rate $q_{ij}$ . More to say is that upon exiting state $i$ there are more than one state that can be reached and subsequently a race condition is started to take place and the transition to the winning state occurs, the state which minimizes the sojourn time in state i . and as a fact that the minimum value of a number of exponentially distributed random variables is also an exponentially distributed random variable with rate equal to the sum of the original rates, this drives to the conclusion that time spent in state $i$ of a homogenous continuous Markov chain is exponentially distributed with a parameter of this sojourn time being equal to $u_i = \sum_{i \neq j} q_{ij}$ .Therefore, the probability distribution of the sojourn time in state $i$ is given by:

$$F_i(x) = 1 - e^{-u_i x} \quad , \qquad x \geq 0$$

Where

$$u_i = \sum_{j \neq i} q_{ij} = -q_{ii}$$

Putting all these together leads to realization that sojourn time in any state of a homogenous continuous time Markov chain must be exponentially distributed. This is violated in non-homogenous chain as it is not exponentially distributed (Stewart 2009).

## 2.6. Poisson Process:

The Poisson process is defined as a counting process $\{N(t), t \geq 0\}$ that is stochastic process describing events that occur randomly in a given interval of time length so it is a CTMC $\{X(t): t \geq 0\}$ with state space of non-negative integers $= \{0,1,2, \dots\}$ . It is a non-decreasing function in time, it has stationary increasing independent increment in time, it typically represents the cumulative number of events that have occurred by time $t$ (Shortle et al. 2018) .

Since the exponential distribution is fundamental and essential in modeling CTMC and it is closely and tightly related to the Poisson process that is defined to be a CTMC, the properties of the Poisson process can be derived from the characteristics of the exponential random variable $(T)$. It is a continuous random variable representing a quantity of time with PDF $(t) = \lambda e^{-\lambda t}\, for\, t \geq 0\, and\, \lambda > 0$, while $\lambda$ represents the rate at which the number of events occur per unit time. The cumulative distribution function $(CDF)$, complementary cumulative distribution function $(CCDF)$, mean and variance are as following:

$$F(t) = CDF = P\{T < t\} = 1 - e^{-\lambda t} \quad t \geq 0 \ , \lambda > 0$$

$$\overline{F}(t) = CCDF = P\{T > t\} = e^{-\lambda t} \quad t \geq 0 \ , \lambda > 0$$

$$E(T) = \frac{1}{\lambda} \quad , \quad var = \frac{1}{\lambda^2}$$

The exponential random variable T has the memoryless property that is to say:

$$P\{T > t + s | T > s\} = P\{T > t\} \ , (t, s \geq 0)$$

Proof:

$$P\{T > t + s | T > s\} = \frac{P\{T > t + s, T > s\}}{P\{T > s\}} = \frac{P\{T > t + s\}}{P\{T > s\}} = \frac{e^{-\lambda(t+s)}}{e^{-\lambda s}} = e^{-\lambda t} = P\{T > t\}$$

The exponential distribution is the only continuous distribution that exhibits the memoryless property

Proof :

Continuous function that satisfies $g(t + s) = g(t) + g(s)$ is a function of the form $g(t) = Ct$ , where C is an arbitrary constant. So if the random variable $T$ is memoryless, then $P\{T > t\} = e^{Ct}$ which is the CCDF of an exponential distribution $(with\ C = -\lambda)$ . From laws of conditional probability:

$$P\{T > t\} = P\{T > t + s | T > s\} = \frac{P\{T > t + s, T > s\}}{P\{T > s\}} = \frac{P\{T > t + s\}}{P\{T > s\}}$$

$P\{T > t + s\} = P\{T > t\}.P\{T > s\}$ or $\overline{F}(t + s) = \overline{F}(t)\overline{F}(s)$ and

$\ln \overline{F}(t + s) = ln\overline{F}(t) + ln\overline{F}(s)$ and so $ln\overline{F}(t) = Ct \quad or \quad \overline{F}(t) = e^{Ct}$

There is a close relation between the exponential and Poisson distribution that enables automatically defining one distribution from the other:

Define $P_0(t) = probability\ of\ no\ events\ during\ a\ period\ of\ time\ t$

Given that time till transition occurrence is exponential $(T)$with rate of transition $\lambda$

$P_0(t) = P\{T \geq t\} = 1 - P\{T \leq t\} = 1 - \left(1 - e^{-\lambda t}\right) = e^{-\lambda t}$

For sufficiently small interval $\Delta t \geq 0$

$$P_0(\Delta t) = e^{-\lambda \Delta t} = 1 - \Delta t + \frac{(\lambda \Delta t)^2}{2!} - \cdots = 1 - \lambda \Delta t + o(\Delta t)$$

The exponential distribution assumes that at most during this interval one event can occur. Thus

$$P_1(\Delta t) = P\{T \leq t\} = 1 - e^{-\lambda t} = 1 - e^{-\lambda t} = 1 - \left(1 - \lambda \Delta t + \frac{(\lambda \Delta t)^2}{2!} - \cdots\right) = \lambda \Delta t + o(\Delta t)$$

$P_{\geq 2}(\Delta t) = o(\Delta t)$

Thus a Poisson process has the following properties:
$N(0) = 0$
$P_0(\Delta t) = \{0\ event\ between\ t\ and\ t + \Delta t\} = 1 - \lambda \Delta t + o(\Delta t)$
$P_1(\Delta t) = \{1\ event\ between\ t\ and\ t + \Delta t\} = \lambda \Delta t + o(\Delta t)$
$P_{\geq 2}(\Delta t) = \{2\ or\ more\ events\ between\ t\ and\ t + \Delta t\} = o(\Delta t)$
The numbers of events in non-overlapping intervals are statistically independent; that is, the process has independent increments.

Since the Poisson random variable $(Y)$is a discrete random variable with PMF represented by:

$p_n = e^{-\lambda}\frac{\lambda^n}{n!}$ , $(n = 0,1,2,\dots)$ with mean and variance , $E(Y) = \lambda$ and $var(Y) = \lambda$ respectively; thus the number of events, $N(t)$ being a Poisson process, occurring by time $t$ is a Poisson random variable with mean $\lambda t$ . That is to mean :

$$p_n(t) = e^{-\lambda t}\frac{(\lambda t)^n}{n!} \ ,(n = 0,1,2,\dots) \ where \ p_n(t) = P\{X(t) = n\}$$

proof :

$P_0(t)$ is the probability of no events by time $t$.

$P_0(0, t + \Delta t) = P\{0 \ events \ in \ [0, t] \ and \ 0 \ events \ in \ (t,t + \Delta t]\}$

$= P\{0 \ events \ in \ [0, t] \}. P \{0 \ events \ in \ (t,t + \Delta t]\}$ as $[0, t]$ and $(t,t + \Delta t]$ are disjoint with independent increments

$= P_0(t). \ [1 - \lambda\Delta t + o(\Delta t)]$ rearrange

$P_0(0, t + \Delta t) - P_0(t) = -\lambda\Delta t P_0(t) + o(\Delta t)P_0(t)$ dividing by $\Delta t$ and taking the limit as $\Delta t \to 0$ gives

$$\lim_{\Delta t\to 0}\frac{P_0(0, t + \Delta t) - P_0(t)}{\Delta t} = \lim_{\Delta t\to 0}\frac{-\lambda\Delta t P_0(t)}{\Delta t} + \lim_{\Delta t\to 0}\frac{o(\Delta t)}{\Delta t}P_0(t) = -\lambda P_0(t)$$

$$\frac{dP_0(t)}{dt} = -\lambda P_0(t) \ \ ,P\{N(0) = 0\} = 1 = P_0(0)$$

So the solution to this first order differential equation is

$P_0(t) = e^{-\lambda t}$

For $P_n(t) \ where \ n \geq 1$

$$\begin{aligned}P_n(0, t + \Delta t) = \ & P\{n \ events \ in \ [0, t] \ and \ 0 \ events \ in \ (t,t + \Delta t]\} \\ & +P\{n - 1 \ events \ in \ [0, t] \ and \ 1 \ events \ in \ (t,t + \Delta t]\} \\ & +P\{n - 2 \ events \ in \ [0, t] \ and \ 2 \ events \ in \ (t,t + \Delta t]\} \\ & + \cdots + P\{0 \ events \ in \ [0, t] \ and \ n \ events \ in \ (t,t + \Delta t]\}\end{aligned}$$

$$\begin{aligned}P_n(0, t + \Delta t) = \ & P_n(t)[1 - \lambda\Delta t + o(\Delta t)] + P_{n-1}(t)[\lambda\Delta t + o(\Delta t)] + P_{n-2}(t)[o(\Delta t)] + \cdots \\ & + P_0(t)[o(\Delta t)]\end{aligned}$$

Collecting the term of $o(\Delta t)$ and rearrange:

$P_n(0, t + \Delta t) - P_n(t) = -\lambda\Delta t P_n(t) + \lambda\Delta t P_{n-1}(t) + o(\Delta t)$

Dividing by $\Delta t$ and taking the limit as $\Delta t \to 0$

$$\begin{aligned}\lim_{\Delta t\to 0}\frac{P_n(0, t + \Delta t) - P_n(t)}{\Delta t} &= \lim_{\Delta t\to 0}\frac{-\lambda\Delta t P_n(t)}{\Delta t} + \lim_{\Delta t\to 0}\frac{\lambda\Delta t P_{n-1}(t)}{\Delta t} + \lim_{\Delta t\to 0}\frac{o(\Delta t)}{\Delta t} \\ &= -\lambda P_n(t) + \lambda P_{n-1}(t) \ , \ n \geq 1\end{aligned}$$

$\frac{dP_n(t)}{dt} = -\lambda P_n(t) + \lambda P_{n-1}(t)$ $,P\{N(0) \geq 1\} = 0 = P_{n\geq 1}(0)$ . According to (Allen, 2010), this system of differential equations can be solved iteratively .For $P_1(t)$ which is first order differential equation:

$$\frac{dP_1(t)}{dt} + \lambda P_1(t) = \lambda P_0 = \lambda e^{-\lambda t} \ , \qquad P_1(0) = 0 \ \ ( the \ initial \ condition \ )$$

Multiplying by factor $e^{\lambda t}$ ( integrating factor) yields:

$\frac{d}{dt}\left(e^{\lambda t}P_1(t)\right) = \lambda$ , integrating both side from $(0\ to\ t)$ and applying the initial condition give the solution: $P_1(t) = \lambda t e^{-\lambda t}$ . Proceeding in such a manner for $P_2(t)$ and for all the transition probabilities covering the state space in a process where $\lambda_n = \lambda\ for\ n\ \in\ S$ and $\mu_1 = \mu_2 = \cdots = 0$, will give:

$$p_n(t) = e^{-\lambda t}\frac{(\lambda t)^n}{n!}\ and\ this\ is\ the\ probability\ mass\ function\ of\ Poisson\ distribution.$$

## 2.7. State Probability Distribution:

According to (Cassandras and Lafortune 2009), it is the probability vector $\pi(t)$ which determines the probability that a system will be in a particular state at a specific time point given its initial state probability vector $\pi(0)$, transition rate matrix $Q$ and specified by its state space $X$. This analysis can be conducted in two approaches:

### 2.7.1. Transient Analysis:

Define state probability $\pi_j(t) = P[X(t) = j]$ and condition the event $[X(t) = j]$ on the event $[X(0) = i]$ with a defined $\pi_i(0) = P[X(0) = i]$ . Thereafter the rule of total probability implies that :

$$\pi_j(t) = P[X(t) = j] = \sum_{all\ i} P[X(t) = j|X(0) = i].P[X(0) = i] = \sum_{all\ i} P_{ij}(t)\pi_i(0)$$

This relation can be rewritten in matrix notation as: $\pi(t) = \pi(0)P(t)$

And since $P(t) = e^{Qt}$ thus the state probability vector at time t is given by $\pi(t) = \pi(0)e^{Qt}$

### 2.7.2. Steady State Analysis:

By this approach, steady state behavior of the system is of much interest, great advantage and huge benefit to be calculated. The system is turned on and is working for some time then its performance and accomplishment are tested in the long run to see how all state probabilities have reached some fixed unchangeable values and no longer varying as time elapses. This relay on some basic requisites such as:

Presence and evaluation of the limit: $\pi_j = \sum_{t\to\infty}\pi_j(t)$ . If this limit exists , $\pi_j$ is called steady state, equilibrium or stationary state probability .

If differentiating $\pi(t) = \pi(t)e^{Qt}$with respect to $t$ and substitute $t = 0$ the differential equation is

$\frac{d}{dt}\pi(t) = \pi(t)Q$ and solving such a system to obtain explicit solution even for a simple Markov chain is not a trivial matter. So if the $\pi_j = \sum_{t\to\infty}\pi_j(t)$ exists, this implies that as $t \to \infty$ this quantity $\frac{d}{dt}\pi(t) \to 0$ , since $\pi(t)$ no longer depends on $t$ .Therefore, $\frac{d}{dt}\pi(t) = \pi(0)Q$ reduces to $\pi(t)Q = 0$ .

In an irreducible continuous time Markov chain consisting of positive recurrent states, a unique stationary state probability distribution vector $\pi$ exists such that $\pi_j > 0$ and $\pi_j = \sum_{t\to\infty}\pi_j(t)$ which is independent of the initial state probability vector. Moreover, $\pi$ is determined by solving $\pi Q = 0$ subject to $\sum_{all\ j}\pi_j = 1$ .

## 2.8. The Embedded Markov Chain and State Probabilities:

If the time actually spent in any state of a continuous time Markov chain is ignored or neglected and only the sequence of transitions that actually occurred is considered, a new discrete time Markov chain is obtained. It is known as the embedded Markov chain $(EMC)$ and is also called jump chain (Stewart 2009).

According to (Allen 2010) define $Y_n$ as the $n^{th}$ state visited by the continuous time Markov chain, then $\{Y_n, n = 0,1,2, \dots\}$ is the embedded Markov chain derived from CTMC $\{X(t), t \geq 0\}$.

The embedded Markov chain is a DTMC and it is useful for classifying states in the corresponding CTMC. Define a transition probability matrix $T = (t_{ij})$ for embedded markov chain, $\{Y_n\}_{n=0}^{\infty}$ where $t_{ij} = P\{Y_{n+1} = j | Y_n = i\}$ , this T matrix can be defined using the generator Q matrix .

First, the transition probability $t_{ii} = 0$ , this is an inherited assumption from the definition of the embedded Markov chain that is the state must change , unless state $i$ is absorbing . A state $i$ in the continuous time chain is absorbing if $q_{ii} = 0$ ( rate of change is zero since the state does not change ). Thus ,

$$t_{ii} = \begin{cases} 0, if\ q_{ii} \neq 0 \\ 1, if\ q_{ii} = 0 \end{cases}$$

Second: Since

$$q_{ij} = \lim_{\Delta t \to 0} \left\{ \frac{p_{ij}(\Delta t)}{\Delta t} \right\} \ for\ i \neq j \quad ; -q_{ii} = \lim_{\Delta t \to 0} \left\{ \frac{1 - p_{ii}(\Delta t)}{\Delta t} \right\} \ for\ i = j$$

$$So \quad -\frac{q_{ij}}{q_{ii}} = \lim_{\Delta t \to 0} \left\{ \frac{p_{ij}(\Delta t)}{1 - p_{ii}(\Delta t)} \right\}$$

This is a probability of a transfer from state $i$ to state $j$ , given the process does not remain in state $i$. define $t_{ij} = -\frac{q_{ij}}{q_{ii}}$ , $q_{ii} \neq 0$ . Thus transition probability $t_{ij}$ for $i \neq j$ :

$$t_{ij} = \begin{cases} \frac{q_{ij}}{\sum_{k=0,k\neq i}^{\infty} q_{ki}} = -\frac{q_{ij}}{q_{ii}}, if\ q_{ii} \neq 0 \\ 0 \qquad\qquad\qquad\quad , if\ q_{ii} = 0 \end{cases}$$

$$T = \begin{bmatrix} 0 & -q_{01}/q_{00} & -q_{02}/q_{00} & \cdots \\ -q_{10}/q_{11} & 0 & -q_{12}/q_{11} & \cdots \\ -q_{20}/q_{22} & -q_{21}/q_{22} & 0 & \cdots \\ \vdots & \vdots & \vdots & \ddots \end{bmatrix}$$

Matrix $T$ is a stochastic process, with $0 \leq t_{ij} \leq 1$ & $\sum_{all\ j} t_{ij} = 1\ for\ all\ i$ .The transition probabilities are homogenous (independent of n) .$T^{(n)} = \left(t_{ij}^{(n)}\right)$ , where $t_{ij}^{(n)} = P\{Y_n = j | Y_0 = i\}$ . For a continuous time Markov chain with no absorbing states, matrix $T$ can be expressed in matrix notation as: $T = I - D_Q^{-1} Q$ , where $D_Q = diag\{Q\}$ is a diagonal matrix whose diagonal elements are equal to the diagonal elements of $Q$ . Thus $T$ has the characteristic of a transition probability matrix for a discrete time Markov chain. Many of the properties of the states of a continuous time Markov chain can be deduced from those of its corresponding embedded chain. That is to say; state $i$ can reach state , for any $i \neq j$ , if there is a $t \geq 0$ such that $p_{ij} > 0$ . A CTMC is irreducible if, for any two states $i$ and , there exists real $t_1 \geq 0, t_2 \geq 0$ such that $p_{ij}(t_1) > 0$ and $p_{ji}(t_2) > 0$ . A CTMC is irreducible if and only if its EMC is irreducible, that is to mean for $\neq j$ , $t_{ij} = 0$ if and only if $q_{ij} = 0$ ,and thus, if $Q$ is irreducible , $T$ is irreducible and vice versa. The concepts of communicating class, closed class, absorbing state, recurrence, transience, and irreducibility are all inherited directly from the embedded Markov chain. This is equivalent to say that a state $i$ is transient in CTMC if and only if it is transient in its EMC and the state $j$ is recurrent in CTMC if and only if it is recurrent in its EMC. Unfortunately, the concepts of null recurrence and positive recurrence for CTMC cannot be applied in terms of EMC. Positive recurrence depends on the waiting time $\{W_i\}_{i=0}^{\infty}$ so that EMC alone is not sufficient to define positive recurrence. The basic limit theorem for CTMC is used to determine the mean recurrence time and hence positive or null recurrence. If CTMC $\{X(t): t \geq 0\}$ is non-explosive and irreducible, then for all $i$ and, $\lim_{t\to\infty} P_{ij}(t) = -\frac{1}{q_{ii}\mu_{ii}}$, where $\mu_{ii}$ is the mean recurrence time, $0 \leq \mu_{ii} \leq \infty$ . In particular a finite irreducible CTMC is non-explosive and this limit exists and it's positive. Units of $\mu_{ii}$ are times and the units of $q_{ii}$ are $1/time$ . Since Markov chain is irreducible which means state $i$ cannot be absorbing so the value of the $q_{ii} < 0$ in the limit (if $q_{ii} = 0,$ then $q_{ij} = 0$ which means state $i$ is absorbing).However, if Markov chain is either null recurrent or transient, so this $\mu_{ii}$ can be infinite. And if the chain is null recurrent and irreducible, then the $\lim_{t\to\infty} P_{ij}(t) = 0$ . On the other hand, if the chain is positive recurrent and irreducible, then the limit is strictly positive. For the existence of a positive limit for the finite Markov chain, all that is needed is to show that the generator matrix Q is irreducible. That is to mean, a finite irreducible CTMC is positive recurrent. (Allen, 2010), (Stewart 2009).

As long as there are no concepts of time steps at which transitions either occur or do not occur in CTMC, so the concepts of periodicity is meaningless. Thus, to have an ergodic state in CTMC, the chain should be positive recurrent. Therefore, Markov chain is ergodic if all its states are ergodic. To sum up, a finite irreducible Markov chain is ergodic (Allen, 2010).

# Chapter Three: Non-Alcoholic Fatty Liver Disease

## 3.1. Prevalence

NAFLD is one of the most common chronic liver disease .Changing life style behavior during the last few decades regarding the eating habits and sedentary life, such that the consumption of high fat and fructose diet resembling the western diets accompanied by lack of exercise, have increased dramatically all through the world . This globally increases the prevalence of obesity and type 2 diabetes worldwide. NAFLD, with its various phenotypes, is most probably discovered accidentally during routine ultrasonography follow up or community surveys using magnetic resonance spectroscopy, while tests of liver enzymes underrate the true prevalence(Younossi et al. 2017).Figure (3.1) illustrates the prevalence of the NAFLD worldwide and the distribution of the PNPLA3 genotype:

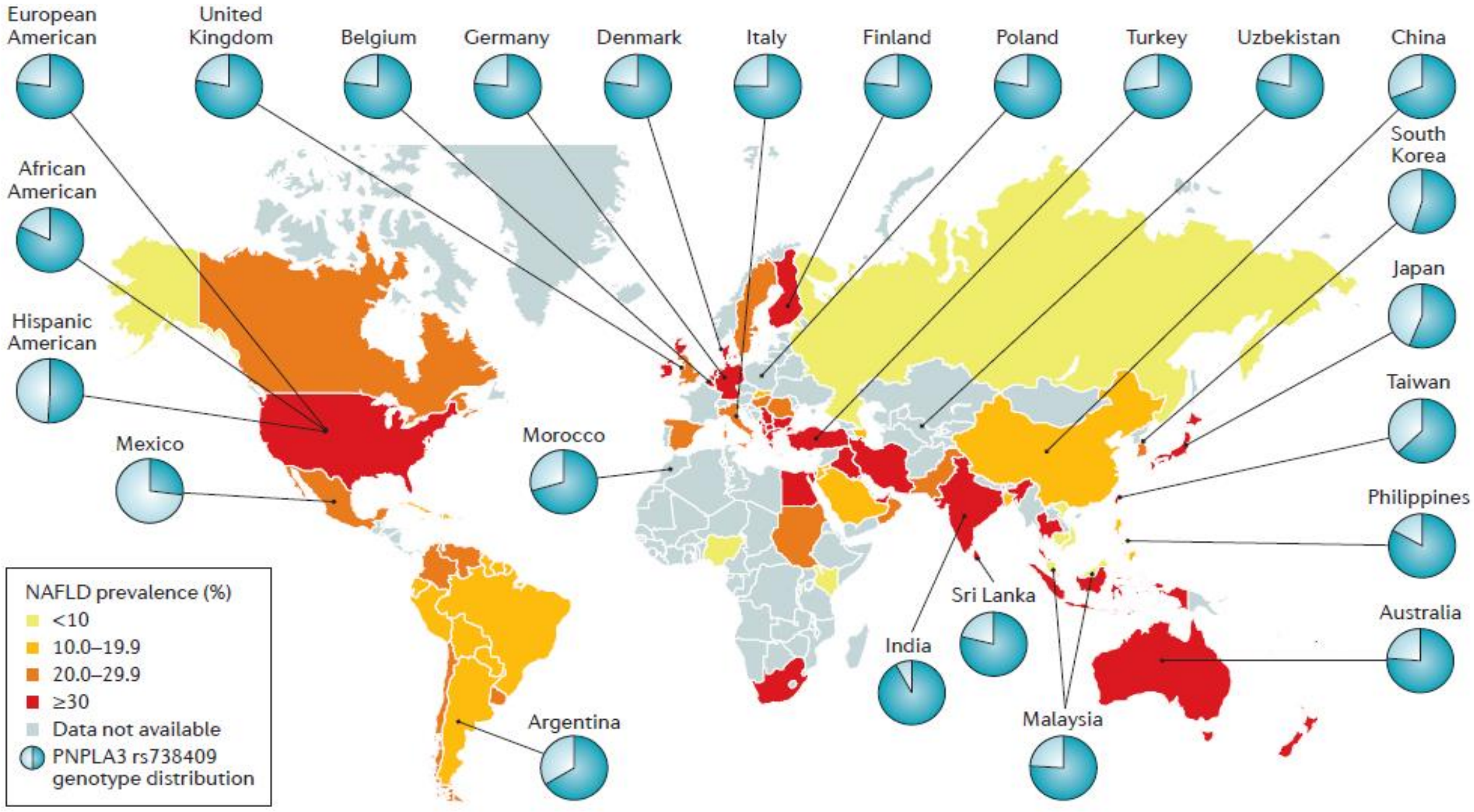

Figure (3. 1): worldwide estimated prevalence of NAFLD and the distribution of PNPLA3 genotype

### 3.1.1. NAFLD in USA:

In USA the prevalence is estimated to be 24% using U/S while it is 21% if noninvasive methods like, Fatty Liver Index, are used. It varies according to ethnicity and so Hispanic Americans are first to come preceding American of European origin and lastly the African Americans come, although this last group has the highest levels of hypertension and obesity. Some studies reveal that the Latino patients having NASH were less aged, with physical exercise being less and much more carbohydrate ingestion than the white patients of non-Latino descent; however these patients with NASH were more susceptible and liable to insulin resistance than the white Latino patients. Within the ethnicity group the prevalence exhibits variations according to the parent country: it is higher at 33% in American of Mexican descent, at 16% in those of Dominican descent while it has a prevalence of 18% in people of Puerto Rican descent, even after controlling for other risk factors (sex, age, waist circumference, BMI, serum HOMA, serum HDL, hypertension, serum C-reactive protein and triglycerides as well as level of education) using multivariate analysis (Younossi et al. 2017).

### 3.1.2. NAFLD in South America:

Using U/S, the prevalence was considered to be 30% such as in Brazil, though also using the U/S, it was 23% in Chile. In Columbia it was 26.6% in men using U/S .However, countries like: Paraguay, Peru, Uruguay, Ecuador, and Argentina had reported prevalence rates to be as minimum as 13% (Peru) up to as maximum as 24 %( Uruguay) (Younossi et al. 2017).

### 3.1.3. NAFLD in Europe:

In spite of the fact that prevalence has wide variations according to the method used to reveal NAFLD, nearly one-quarter of the European population suffers from NAFLD. Meta-analysis study released in 2016 documented a prevalence of 23.71% in Europe, ranging from 5% to 44% in various countries. The data from Germany estimated a 30% prevalence rate using U/S. NAFLD accounted for deranged liver tests is 26.4% in England. In France simple steatosis was observed in 26.8% of liver biopsies due to questionable deranged liver tests, with 32.7 % of those biopsies had NASH. In northern Italy, NAFLD evaluated by U/S was similar in those having and those free of liver disease (25% in contrast with 20%, p=0.203), diagnosed by disturbed levels of liver enzymes or hepatitis B surface antigen and /or anti-HCV positivity. Epidemiological statistical data from Spain reported resembling rate of 25.8%. From Romania, a study conducted on 3005 in-patients population not suffering from any liver disease, revealed that NAFLD evaluated by sonography was present in 20% of those hospitalized patients, whereas a Hungarian study reported 22.6% prevalence rate of fatty liver detected by U/S. (Younossi et al. 2017).

### 3.1.4. NAFLD in Asia-Pacific and Africa:

There are enormous variations between countries comprising this region due to huge differences between its countries in levels of economic, educational and political aspects as well as health-care systems that all impact the individuals' lifestyle, nutritional culture and sedentary attitudes. Data are not available in a comprehensive form, due to lack of epidemiological statistical surveys extending throughout the country for assessment of fatty liver, and subsequently there are marked dissimilarities in the prevalence of NAFLD over time and between different regions within the same nation. Prevalence in Chengdu (Southwest China) was 12.5%. It was 15% in Shanghai (east china) and 17% in Guangdong (south China) while in central China; it was 24.5%. Furthermore, a study of 7152 employees in shanghai utilizing U/S for detecting fatty liver released in 2012 estimated NAFLD prevalence to be as great as 38.17%. In a study from Hong Kong, "proton magnetic resonance spectroscopy" determined the quantity of fat in the liver and estimated its prevalence rate as 28.8%; although it was 19.3% in non-obese individuals while it was 60.5% among the obese, while the study emanating from Taiwan reported that the prevalence was 11.4% in the general population, however; it was 50% in elderly and 66.4% in persons with sedentary lifestyle like in taxi drivers. It was 25% in Japan in 2005 as diagnosed by U/S. Using the same modality, the prevalence in 141610 individuals from South Korea was 27.3% in 2013 (Younossi et al. 2017). Table (3.1) lists risk factors of metabolic syndrome.

South Asia and Indian regions are towards quick and speedy urbanized changes in the social and economic aspects. In the rural India preserving their traditional cultural diet and lifestyles, the prevalence rate is low (9%), whereas in urban areas the prevalence rate fluctuates between 16 and 32% as the lifestyle mimics the trends in western countries. Similar variations in rates between rural and urban areas (5-30%) were obtained from smaller surveys in Indonesia, Malaysia, Singapore, and Sri-Lanka. Data from Africa are scant. In Nigeria, the rate was 9.5-16.6% in diabetic individuals and 1.2-4.5% in non-diabetic individuals. Comparably, the prevalence in overweight or obese South African individuals was 45-50%, while in 2014; it was 20% in the population studied from Sudan (Younossi et al. 2017).

Table (3. 1): Risk factors for NAFLD

| Risk factor | Effect |
|---|---|
| **Age** | High risk for NAFLD and advanced fibrosis if age>45 years |
| **Metabolic syndrome** | The more criteria there are , the higher probability of NASH and advanced fibrosis is . |
| **sex** | Male > famale |
| **ethnicity** | High in Hispanics, intermediate in white , low in black |
| **Dietary factors** | Diets with high contents of cholesterol, saturated fats and fructose but low in carbohydrates increase the risk. caffeine is beneficial |
| **Obstructive sleep apnea** | It increases risk for NASH and advanced fibrosis |
| **Genetic factors** | PNPLA 3 and TM6SF2 increase risk of NASH and advanced fibrosis, PNPLA3 increase risk for HCC. |

Metabolic syndrome is defined as having abdominal obesity identified by waist circumference > 94 cm for males and > 80 for females in eastern countries while it is >102 for males and > 88 for females in western countries. Plus 2 or more of the following:

1. Blood glucose ≥100 mg/dL or drug treate diabetes.
2. Arterial pressure ≥ 130/85 mmgh or drug treated for hypertension.
3. Triglyceride levels ≥150 mg/dL or drug treated for high level in the blood.
4. HDL cholesterol levels < 40mg/dL for males and <50 mg/dL for females or drug treated.

## 3.2. Lean NAFLD:

It is NAFLD in which obesity is not present. It was first described in Asian population and it can occur in 10-20% of Europeans and Americans. Elevated visceral adiposity as well as increased dietary intake of fructose and fat, along with high risk genetic background affecting the metabolism, may be correlated with lean NAFLD. Large portion of these individuals belong to the “metabolically obese, normal weight phenotype”, reported in at least 5% of the western population. These individuals are characterized by being non-obese, more frequently lacking physical activity with defective insulin sensitivity, high cardiovascular risk and fatty liver resulting from low capacity to store fat in adipose tissue, dysfunctional mitochondrial adaptation as well as increased hepatic triglyceride synthesis “de novo lipogenesis”(Conus et al. 2007)

### 3.2.1. Prevalence of non-obese NAFLD:

Table (3.2) summarizes the prevalence of non-obese NAFLD according to the definition that non-obese NAFLD is in accordance to BMI $< 30\ kg/m^2$ for people in western countries and $< 25kg/m^2$ for people in eastern countries, while lean NAFLD is recognized as the existence of NAFLD in population with BMI $< 25kg/m^2$ of western countries descent and $< 23kg/m^2$ for people of eastern countries descent (Kim and Kim 2017).

As regard the histology of non-obese NAFLD, there are limited scarce data on it but both non-obese NAFLD and obese-NAFLD are supposed to have structural features that are indiscernible from each other. Table (3.3) summarizes the available data (Kim and Kim 2017).

Younossi et al. (2017) mentioned causes of NAFLD in lean individuals:

1. Environmental reasons: High cholesterol and/or high fructose diet.
2. Acquired lipodystrophy as a result of ingestion of anti-retroviral drugs for HIV
3. Genetic causes: “PNPLA3 variants” and congenial metabolic defects; like: “familial hypo-beta-lipo-protinemia”, “congenital lipodystrophy”, and “lysosomal acid lipase deficiency”.
4. Endocrine factors: “polycystic ovary syndrome”, “growth hormone or hypothyroidism deficiency”.
5. Drug-related causes: tamoxifen, methotrexate or amiodarone.
6. Other reasons: “jejunoileal bypass”, “total parenteral nutrition”, or starvation.

Ruderman et al. (1981) (Ruderman and Schneider 1981) were the first to suggest that the "metabolically obese normal weight" (MONW) subjects , although having BMI < 25kg/m2 they have some features that make them a "high risk" group to develop atherosclerosis and its subsequent cardiovascular events, thus they had proposed a scoring system of 22 criteria in 1998 to recognize those MONW persons. This scoring system shown in table (3.4) has not been confirmed, validated or widely accepted in medical societies (Ruderman et al. 1998). Some authors used different definitions or criteria to identify individuals with MONW syndrome as illustrated in table (3.5) (Conus et al. 2007).

NAFLD is clearly obvious in patients with metabolic derangements associated with obesity, such as dyslipidemia, type II diabetes, and metabolic syndrome. However, it is clear without doubt that NAFLD is not present in all obese persons and, more seriously, NAFLD can also be found in non-obese individuals. Although NAFLD in non-obese subjects has been recorded to occur in all ethnicities covering all age groups, it seems to be realized in a more frequent rate in Asian, even when criteria for specific BMI, respecting the ethnicity restrictions to define obesity, are used. Liver biopsies reveal that the prevalence of NASH and fibrosis does not differ significantly between non-obese and obese patients once they have developed NAFLD. Visceral adiposity against general adiposity, high cholesterol and fructose ingestion, and genetic background like "patatin-like phospholipase domain-containing 3" may accompany non-obese NAFLD. Generally speaking, NASH carries high mortality rates, mainly from cardiovascular events, independent of other metabolic risk factors. Although data concerning the mortality burden in non-obese NAFLD patients are not well established, it may be substantial to characterize the "high risk" group of them and control their "metabolic milieu". For the present time, lifestyle changes to decrease visceral obesity, in the form of good dietary habits and exercise, continue to be the standard care in non-obese NAFLD patients. Table (3.6) summarizes clinical metabolic factors accompanying non-obese NAFLD in various studies (Kim and Kim 2017).

Lipo-dystrophy phenotype is a disease causing loss or absence of subcutaneous fat (fat under the skin), ectopic accumulation of the triglyceride in liver and skeletal muscle, and severe "insulin resistance". The most common cause of this disease is anti-HIV drugs (HAART therapy) given to HIV patients, but some genetic types are present, although relatively rare. Sarcopenia is a condition of generalized progressive loss of skeletal muscle mass, function and strength with raised plasma amino acids concentrations, with or without obesity, leading to NAFLD. Mutations in the gene coding for cholesteryl ester transfer protein (CETP) responsible for exchange of triglycerides between lipoproteins is associated with NAFLD in young Caucasian female with BMI < 25, and reported prevalence in lean homozygotes type is over 30 %. Also mutation in the gene encoding for interferon lambda 4, which has impact on regulation of "innate immunity" is associated with necro-inflammation, severe fibrosis, and NASH phenotype of the disease, especially in non-obese subjects. Deficiency of the enzyme phosphatidylethanolamine N-methyltransferase (PEMT) involved in phophatidylcholines synthesis in liver cells is associated with NASH development in subjects on high sucrose high fat diet, although they have normal triglyceride and cholesterol levels, and they are with low BMI and normal insulin sensitivity, highlighting deficiency of PEMT as an etiological factor for lean NASH. Subjects with history of fetal growth retardation are prone to development of adipocyte insulin resistance, which is maintained through neonatal period and during adulthood, which put them on risk of acquiring severe form of pediatric NAFLD presented by increased activity of the disease at histology, with normal BMI and less common prevalence of family history of type 2 diabetes, pointing at less genetic effects for NAFLD in these patients. Some lean patients with NASH have very low levels of "gut microbiome" like: Ruminococcus and lactobacilli in comparison to overweight and obese NASH patients, as NAFLD has been associated with decreased rate of bacteroidetes and increased rate of Prevotella, Porphyromonas, and ethanol-producing bacteria (Younes and Bugianesi 2019).

Table (3. 2): Prevalence of lean NAFLD and non-obese NAFLD

| | | | | | Prevalence | |
|---|---|---|---|---|---|---|
| **Study** | **population** | **N** | **Detection** | **BMI cut-off value, kg/m$^2$** | **Non-obese NAFLD** | **Over-all NAFLD** |
| **Western** | | | | | | |
| ***Lean NAFLD*** | | | | | | |
| (Bellentani et al. 2000) | Italy, community-based (non-obese) | 257 | U/S | <25 | 16.4% | 58.3% |
| ***Non-obese NAFLD*** | | | | | | |
| (Browning et al. 2004) | US, population-based (Dallas heart study) | 2287 | MRS | <30 | white =20% Hispanic=26% | 31% |
| (Foster et al. 2013) | US, population-based (MESA) | 3056 | CT | <30 | 11.3% in African Am. | |
| **Eastern** | | | | | | |
| ***Lean NAFLD*** | | | | | | |
| (Fan et al. 2005) | China, population-based | 3175 | U/S | <23 | 3.3% | 20.8% |
| (Das et al. 2010) | India, community-based | 1911 | U/S, liver biopsy | 75% of NAFLD had BMI<25,54% had BMI <23 | | 8.7% |
| (Sinn et al. 2012) | Korea,community-based (non-obese, non-diabetic) | 5878 | U/S | ≥18.5,<23<br>≥18.5,<25 | 16%(lean)<br>45%(non-obese) | 27.4% |
| ***Nonobese NAFLD*** | | | | | | |
| (Omagari et al. 2002) | Japan,community-based (nonobese, nondiabetic) | 3432 | U/S | 11.2% of NAFLD are non-alcoholic with BMI < 25 | | 21.8% |
| (Hae et al. 2004) | Korea,community-based (nonobese, nondiabetic) | 768 | U/S | ≥18.5,<25<br>≥25,<30 | 16.1%<br>34.4% | 23.4% |
| (Chen et al. 2006) | Taiwan, community-based | 3245 | U/S | <25 | 4.2% | 11.5% |
| (Park et al. 2006) | Korea, community-based | 6648 | U/S | <25<br>25-30(ob.) | 9.8%<br>33.4% | 18.7% |
| (Dassanayake et al. 2009) | Sri-Lanke, community-based (Ragama health study) | 2985 | U/S | <25 | 16.7% | 32.6% |
| (Fu et al. 2009) | Taiwan, community- based (adolescents) | 220 | U/S | <85$^{th}$ percentile | 16% in Normal BMI | 39.8% |
| (Kwon et al. 2012) | Korea, community-based | 29994 | U/S | <25 | 12.6% | 20.1% |
| (Xu et al. 2013) | China, community-based | 6905 | U/S | <25 | 7.3% (non-obese) | 8.88% |
| (Bagheri Lankarani et al. 2013) | Iran, population-based | 819 | U/S | <25 | 16.4% with Normal BMI | 21.5% |
| (Wei et al. 2015) | Hong Kong, community-based | 911 | MRS | <25 | 19.3% (non-obese) | 28.8% |

Table (3. 3): Prevalence of lean NAFLD and non-obese NAFLD according to histology

| | | | | **Prevalence** | | |
|---|---|---|---|---|---|---|
| **Study** | **Region** | **Biopsy-Proven NAFLD** | **BMI Cut-Off Value,Kg/M$^2$** | **Non-Obese NAFLD** | **NASH** | **Fibrosis** |
| **Lean NAFLD** | | | | | | |
| (Marchesini et al. 2003) | Italy | 163(biopsied) out of 304 NAFLD patients | <25(normal)<br>≥25 to <30(over-weight) | n=112(without metabolic syndrome)<br>n=51( with MetS | 65%(<25)<br>73%(<25-30)<br>84%(≥25)<br>120 cases out of 163= NASH | 4 cases cirrhosis, 3 of them with MetS and 1 case without MetS |
| (Manco et al. 2011) | Italy (children) | N=66 | 85≤BMI<95 Percentile For age And sex (over-weight) | 21patient= 31.8% (overweight) | n=24 (NAS≥5 ) | n=8 (F3) |
| (Akyuz, Yesil, and Yilmaz 2014) | Turkey | 483 | <25<br>Lean(n=37)<br>Over-weight (n=446) | 7.6% Prevalence of Lean NAFLD | NAS<br>5 patients with NAS (4-6)<br>(BMI ≥ 25)<br>5patients with NAS (2-7)<br>(BMI < 25) | Fibrosis stage<br>1 patients with fibrosis (0-2)<br>(BMI ≥ 25)<br>0 patients with fibrosis (0-1)<br>(BMI < 25) |
| (Cruz et al. 2014) | International | 1090 | <25 | 11.5% | | |
| (Leung et al. 2017) | Hong Kong | 307<br>lean<br>(n=29),<br>Over-weight<br>(n=43) | <23<br>Non-obese<br>(n=72) | 29/307<br>=9.4% | NASH<br>11/29<br>=41%(<23)<br>19/43<br>=45%(23-24.9) | ≥F3<br>7/29=<br>26%(<25)<br>11/43=<br>26%(23-24.9) |
| **Non-Obese** | | | | | | |
| (Das et al. 2010) | India. (n=1911) | 36(biopsied),median BMI=25.6(18.7-37.3) | | NAFL by US CT(n=167), 75%(not over-weight) | 11/36=31% (NASH) (NAS≥5) | 4/167=2.4% (cirrhosis). Prev. in entire popualtion= 4/1911=.2% |
| (Vos et al. 2011) | Belgium | 1777 had underwent Liver Biopsy.<br>lean NAFLD(n=31)<br>Obese-NAFLD(n=48)<br>Total=79<br>Non-diabetic | BMI < 30<br>Persons with missing values for variables were removed . | 50 Out of 1777(2.8%)<br>And the same 50 out of 130 cryptogenic liver disease ( 38%) are lean NAFLD | NASH(61%) with BMI <30(out of 31 persons)<br>NASH(85%) with BMI >30(out of 48 persons) | ≥F2<br>19%(BMI<30) (out of n=31)<br>47%( BMI ≥30)<br>( out of n=48) |
| (Alam et al. 2014) | Bangladesh | n=465 by US (nonobese=119, obese=346)<br>n=220(biopsied), (non-obese=56), (obese=164) | Non-obese has BMI <25 | Prevalence of NAFLD<br>=119/465<br>=25.6% | NASH<br>=30/56<br>=53.6%(<25)<br>77/164<br>=47%(≥25) | ≥F2<br>=11/56<br>=20%(<25)<br>31/164<br>=19%(≥25) |
| (Leung et al. 2017) | Hong Kong | 307<br>Overwight<br>(n=43),<br>Lean<br>(n=29) | <25<br>Non-obese<br>(n=72) | 72/307<br>=23.5% | NASH<br>30/72<br>=44%(<25)<br>121/235<br>=52%(≥25) | ≥F3<br>18/72<br>=26%(<25)<br>64/235<br>=28%(≥25) |

Table (3. 4): Proposed scoring method for identifying MONW individuals

| Presence of associated disease or biochemical abnormalities | points |
|---|---|
| **Hyper-gylcemia:** | |
| Type 2 diabetes | 4 |
| Impaired glucose tolerance test | 4 |
| Gestational diabetes | 3 |
| Impaired fasting glucose (110-125 mg/dL) | 2 |
| **Hyper-triglyceridemia:** | |
| Triglyceride>150mg/dL; HDL-cholestrol<35 | 3 |
| Triglyceride>150mg/dL | 2 |
| Triglyceride>100-150mg/dL | 1 |
| **Essential Hypertension:** | |
| Blood pressure >140/90 mmHg | 2 |
| Blood pressure 125-140/85-90 mmHg | 1 |
| Polycystic ovaries | 4 |
| Premature coronary heart disease ( under age 60 y) | 3 |
| Uric acid (>8 mg/dL) | 2 |
| **Family History ( First-Degree Relative ):** | |
| Type 2 diabetes or impaired glucose tolerance | 3 |
| Essential hypertension ( under age 60 y) | 2 |
| Hypertriglyceridemia | 3 |
| Premature coronary heart disease ( under age 60) | 2 |
| **Presence Of Predisposing Factors:** | |
| Low birth weight (<2.5 kg) | 2 |
| Inactivity(<90 min aerobic exercise/week) | 2 |
| **Evidence Of Mild Obesity Or Central Adiposity ( Maximum 4 Points):** | |
| Weight gain: >4,8,12 kg after age 18 years , after age 18 years (W), 21 y (M) | 1-3 |
| BMI: 23-25 , 25-27 kg/m$^2$ | 1,2 |
| **Waist ( Inches):** | |
| 28-30 , >30(W) | 1,2 |
| 34-36 , >36 ( M) | 1,2 |
| Ethnic group at high risk | 1-3 |

Note: MONW individuals score 7 points or greater

Meigs et al 2006 reported the study (community-based) conducted by them on 2902 subjects(55 % women) without diabetes or cardiovascular diseases (CVD) to evaluate the risks for both events stratified by BMI and the presence of MetS or IR. The results were that out of 2092 subjects, 2.6% were normal weight with MetS (resembling MONW), 8.1 % were obese without MetS, 33.8% were normal weight without MetS, 13.9% were obese with MetS. Of the normal-weight subjects, 7.1% had MetS while among the obese people 37% did not have MetS. As regards IR, of the normal-weight subjects, 7.7% were insulin resistant while among the obese people 44.3% were insulin sensitive. Their data had revealed that in white community (the study population), the prevalence of MONW was about 3% overall and 7% among normal-weight subjects (BMI < 25) while the metabolically healthy obese (MHO) persons had prevalence of about 8% overall of the Framingham community and 37% of obese persons without MetS. They found that risk factors for type 2 diabetes and CVD that cluster with central obesity( increased waist circumference beyond the cutoff values) like: low level of HDL-chol, elevated levels of triglycerides, impaired fasting glucose, and hypertension, were present with relative high frequency in normal-weight subjects with MetS than in obese subjects without MetS , over 7-11 years of follow up. They suggested that MHO subjects may have subclinical vascular diseases and thus this warrant longer follow up studies to evaluate type 2 diabetes and CVD when these events develop, also these patients are younger than obese patients with MetS, but as they get older.

Table (3. 5): Criteria used to identify MONW subjects

| Reference | Criteria |
|---|---|
| (Dvorak et al. 1999) | BMI < 26.3kg/$m^2$. Glucose disposal <8mg.$min^{-1}$.$kg^{-1}$ of fat free mass(FFM) in euglycemic hyperinsulinemic clamp. 71 young women. MONW(n=13)with impaired insulin sensitivity (Glucose disposal < 8mg.$min^{-1}$.$kg^{-1}$ ), normal(n=58) with normal insulin sensitivity (Glucose disposal > 8mg.$min^{-1}$.$kg^{-1}$ ). |
| (Katsuki et al. 2003) | BMI <25 kg/$m^2$. 40 person .MONW(n=20) with peripheral insulin resistance, normal(n=20) with normal peripheral resistance determined by euglycemic hyperinsulinemic clamp ( glucose infusion rate GIR ).significant correlation between visceral fat area (VFA) or serum TG and GIF in MONW but not subcutaneous or total fat area. Multiple regression showed (VFA) & serum TG were significantly associated with GIF in all MONW and normal weight persons. |
| (Molero-Conejo et al. 2003) | n=167 boys and girls (14-17 yr), HOMA-IR and HOMA β-cell index (markers of insulin resistance), lean (BMI <25 kg/$m^2$) , obese otherwise, central fat distribution (subscapular/triceps ratio), total chol., LDL-chol, HDL-chol, TG, systolic and diastolic blood pressure, Fasting serum insulin > 84 pmol/L |
| (Conus et al. 2004) | BMI <30 kg/$m^2$, 96 young women, MONW(n=12),non-MONW(n=84), same BMI for both groups, MONW(HOMA >1.69), non-MONW (HOMA ≤1.69). |
| (St-Onge, Janssen, and Heymsfield 2004) | n=7602(male and female of Hispanic, Black and White ethnicities, age >20yr) BMI=18.5-26.9 kg/$m^2$ , divided into 4 groups ( $g_1$=18.5-20.9, $g_2$=21-22.9, $g_3$=23-24.9, $g_4$=25-26.9), prevalence of each of the metabolic syndrome components, as well as overall MetS highly statistically increases with increases in BMI, after controlling for other confounders such as non-modifiers: age, ethnicity, education, income, and modifiers: alcohol, physical activity, carbohydrate, fat and fiber intakes. |
| (Meigs et al. 2006) | BMI <25 hg/$m^2$<br>Metabolic syndrome ( ATP III) or insulin resistance ( top quartile of HOMA) |

Table (3. 6): Risk factors for NAFLD in non-obese with associated odd ratio

| Study | Risk Factors For Non-Obese NAFLD |
|---|---|
| (Omagari et al. 2002) | NAFLD non-overweight men (% fat 'OR=1.166 ', FBG 'OR=1.024', TG 'OR=1.004').<br>NAFLD non-overweight women (% fat 'OR=1.182', FBG 'OR=1.038', TG 'OR=1.013'). |
| (Hae et al. 2004) | In normal weight (Male 'OR=3.1', triglyceride 'OR=1.004',waist circumference 'OR=1.14', $logHOMA_{IR}$ 'OR=5.7'),in over weight (age 'OR=1.04', BMI 'OR=1.8', $logHOMA_{IR}$ 'OR=17.5'). |
| (Chen et al. 2006) | In non-obese persons with BMI <25 (age from 40 to 64 'OR=2.35', triglyceride (TG) 'OR=2.48'). Elevated ALT is association not causal relation 'OR=15.45' |
| (Das et al. 2010) | Persons with BMI<25 (BMI 'OR=1.2', biceps skin fold thickness 'OR=1.2') . |
| (Sinn et al. 2012) | Whole persons (Age 'OR=1.01', HOMA2-IR 'OR=3.34', triglyceride 'OR=1.84', HDL 'OR=1.61', FBG 'OR=1.4', waist circum. 'OR=1.57', (BMI>23) 'OR=2.46',MetS 'OR=2.42' ). |
| (Wei et al. 2015) | In non-obese persons (BMI 'OR=1.33', waist circumference 'OR=1.11', $HemoglobinAI_C$ 'OR=1.83', HOMA-IR 'OR=1.24', ferritin 'OR=1.001', PNPLA3 polymorphism 'OR=4.37'). |

## 3.3. Macroscopic and Microscopic Picture of the Liver:

The liver is the biggest internal organ in the human body; it comprises 2% of the weight of the adult and more than 500 functions are allocated to the liver. Releasing urea and albumin into the blood sets it as an endocrine gland. Releasing bile into the intestine defines it as an exocrine gland. While synthesizing and storing glycogen and triglyceride determines it as a depot or repository organ. It is also a regulator for the metabolism of lipid, carbohydrate,and amino acid (Ehrlich et al. 2019). In figure (3.2) the liver location in human body is shown, while figure (3.3) shows the histological microscopic picture of the liver.

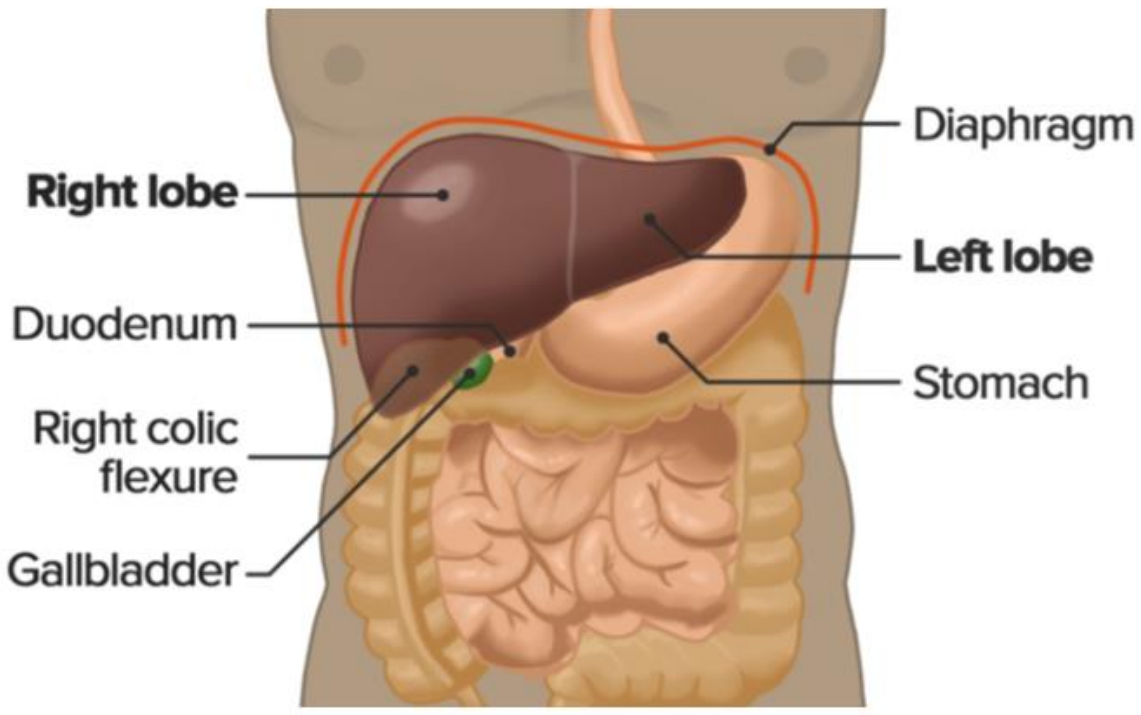

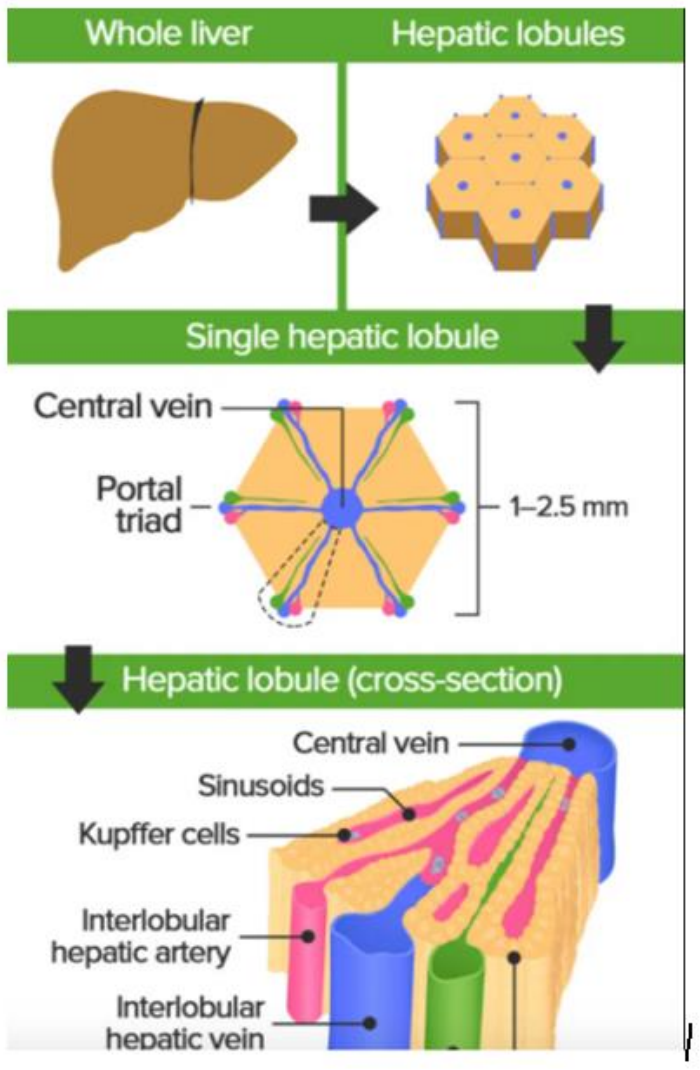

Figure (3. 2): microscopic and macroscopic picture of the liver

*On the left: location of the liver in the abdomen of the human with right and left lobes. On the right: the histological unit of the liver, hepatic lobule with its hexagonal shape like honey comb and how the central vein, portal vein as well as the hepatic artery are arranged in relation to the bile ducts tributaries. source: lecturio.com/concepts/liver*

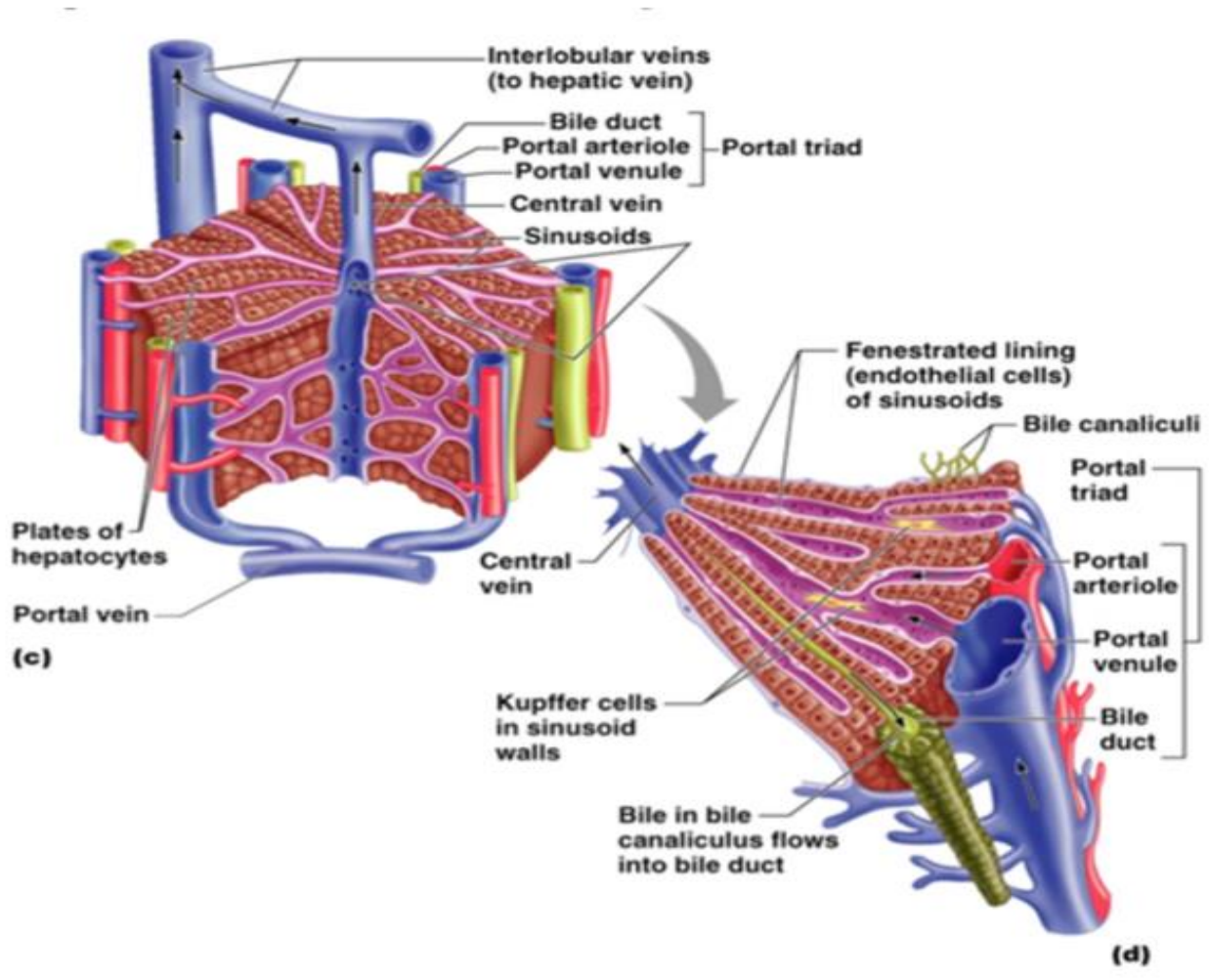

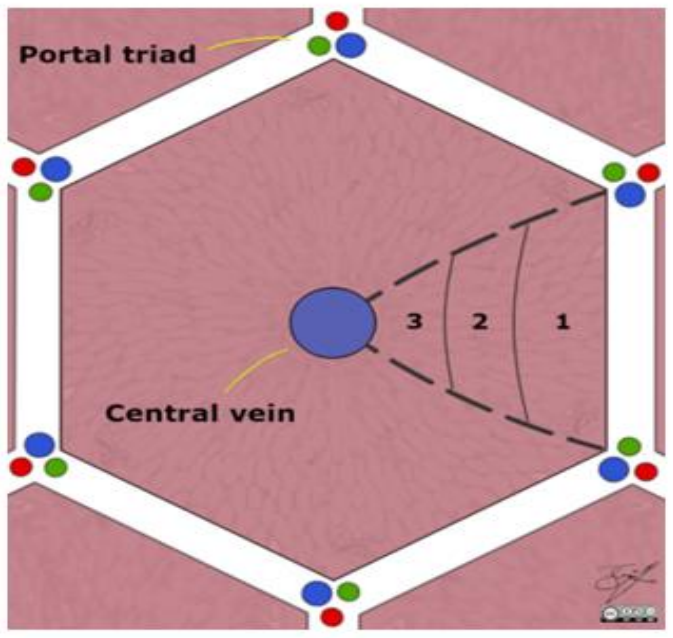

Figure (3. 3): the hepatic acinus or lobule (hepatic functional unit)

*The hepatic functional unit with its hexagonal shape, at the vertices are the portal triads composed of branch of hepatic artery(red), portal vein(blue) and bile duct( green)and in the middle of the lobule is the central vein. Blood flows from the periphery of the lobule (portal triad) to the central vein in the following order: zone1 (peri-portal), zone2(intermediate), zone3 (centri-lobular, peri-venular). Source on the left: quizlet.com/150763356/liver-physiology-flash-cards/ .on the right radiopaedia.org/cases/hepatic-functional-unit?lang=gb*

## 3.4. Pathogenesis of NAFLD:

NAFLD is characterized by adipose tissue inability to manage the chronic state of energy excess that will manifest as a systemic disease. This adipose tissue elicits inflammatory milieu and state of insulin resistance; both of which hinder the capacity of adipose tissue to accumulate fat and enhances ectopic deposition of fat mainly in liver, skeletal muscle and heart. The excessive accumulation of lipid inside the hepatocytes delineates the presence of steatosis (NAFL). Moreover, some of this fat is lipotoxic to the liver especially when present in excessive amount. This fatty hepatic milieu in the presence of some other insults like: microbiota derived substances, the released hepatokines from the liver as well as the released adipokines from adipocytes promotes inflammation. This brings about hepatocytes injury and death, a state that is called (NASH). It is a state of inflammation accompanied by hepatocyte ballooning and death. This state once it has developed; it accelerates the mitochondrial dysfunction, endoplasmic reticulum stress, uncontrollable generation of reactive oxygen species, excess autophagy and enhanced activity of inflammasomes. When lipotoxicity is chronic; the accumulated dead cells encourage wound healing process that is mainly composed of fibro-genesis. If this reparative process is kept going either if the insult has been resolved or if the injurious agent cannot be eliminated, the net result will be a state of progressive fibrosis resulting ultimately in cirrhosis and other comorbidities like hepatocellular carcinoma (HCC). The regenerative and repair responses to the different specific causes of “chronic liver diseases” are monotonous that is to mean; the causative agent that induces liver disease is specific while the regenerative process is the same regardless the nature of this agent, and it is not hindered unless this agent is removed (Boyer and Lindor 2016). However; in the presence of individual, genetic and heritable differences once the fibrosis process has started in response to this agent, it is maintained and may account for why only 10% to 20% of patients with NASH develop progressive fibrosis and need more aggressive treatment, and why that frequency is relatively preserved across numerous dissimilar classes of liver diseases (Loomba and Sanyal 2013). Figure (3.4) shows the pathophysiology of NAFLD and interactions between various factors involved in this disease (Akshintala et al. 2021)

## 3.5. Definition and Terminology:

NAFLD covers a spectrum of hepatic diseases containing two essential disease phenotypes. The first one is nonalcoholic fatty liver (NAFL; simple steatosis), where triglycerides accumulate in more than 5% of the hepatocytes without histologic evidence of inflammation, cellular injury, or fibrosis. The second one is non-alcoholic steato-hepatitis (NASH), where steatosis comes with histologic features of necro-inflammation and hepatocyte ballooning degeneration (with or without evidence of fibrosis). These phenotypes may evolve to cirrhosis, liver-related mortality and in some cases to hepatocellular carcinoma (HCC). NAFLD, by definition, occurs when excessively consuming alcohol is absent; a cutoff point of not more than 20g/day for women and not more than 30g/day for men is used to distinguish and separate it from alcohol-related liver disease. Primary NAFLD is the disease related to obesity, insulin resistance, hypertension, and criteria of metabolic syndrome, most of NAFLD patients reside in this category. The secondary NAFLD points toward a category of patients with causes other than metabolic syndrome associated conditions; such as drug or toxin induced fatty liver as well as rare inherited genetic metabolic diseases (Boyer and Lindor 2016). Table (3.7) lists the causes of NAFLD.

NAFLD is a disease characterized by wide variation in the disease seriousness and disease consequences reflecting the interactions between extrinsic and intrinsic factors. Although it is initiated by high fat and carbohydrate diet associated with sedentary lifestyle composing some of the extrinsic factors, the contributors of intrinsic factors presented in genetic background crucially determine how the patient responds to the excess caloric intake and metabolic stressors. These factors are known as modifiers or adaptors (Ribeiro et al. 2004)(Brand and Esteves 2005).

### 3.5.1. Environmental Modifiers (Extrinsic Factors):

Diets high in fructose and fat while deficient with antioxidants, intestinal micro-biome (imbalance between protective and harmful intestinal bacteria), alcohol consumption in obese patients, presence of other conditions or diseases such as hemochromatosis, obstructive sleep apnea syndrome, and hepatitis C are some of the identifiable factors. These factors are incriminated for increasing rate of progression of fibrosis in NAFLD patients resulting in bad outcome (Park et al. 2007)(Baffy 2005).

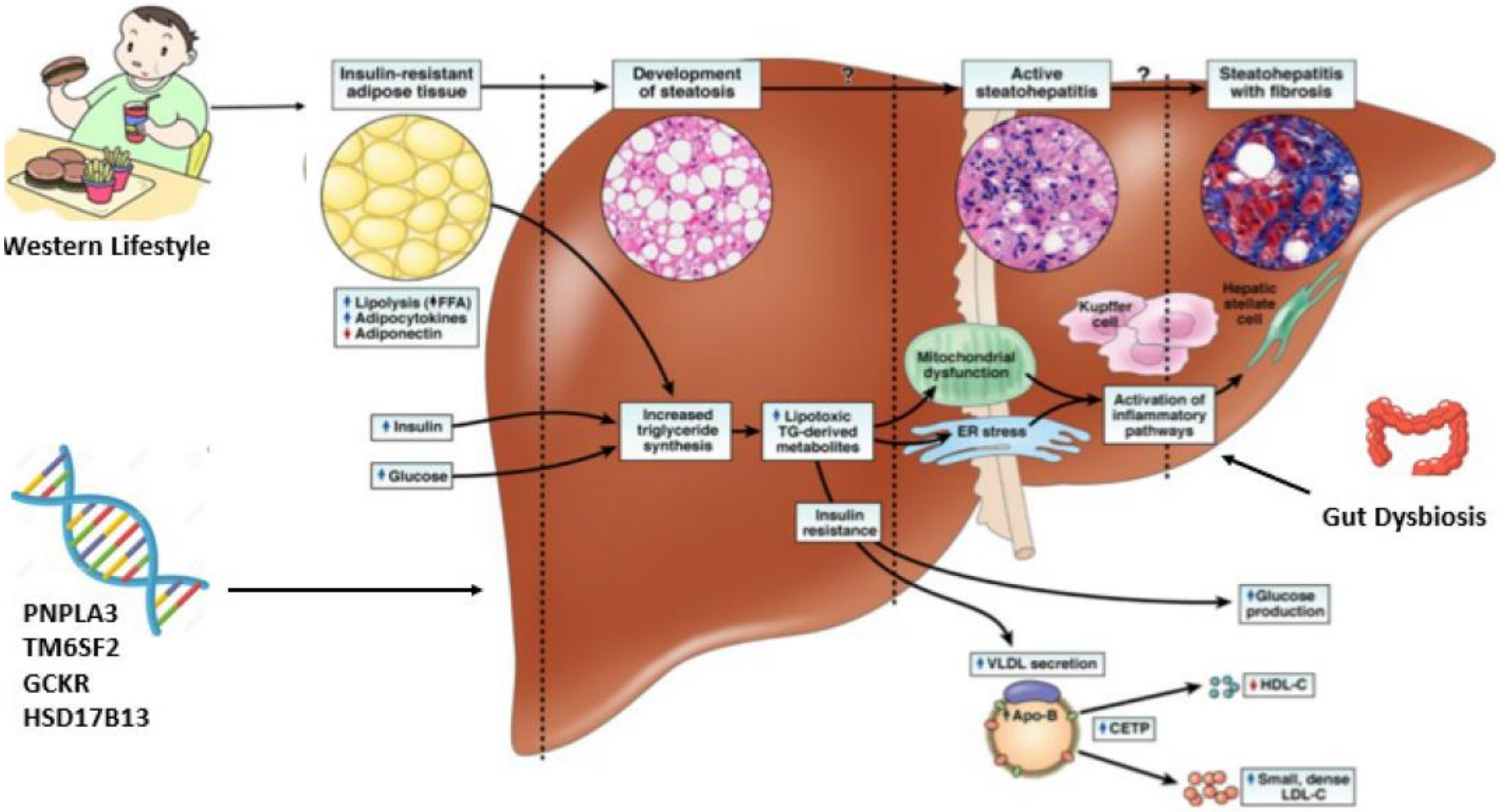

Figure (3. 4): pathophysiology of NAFLD

*With high-fat diet, adipocyte (fat cell) acclimates to chronic energy surplus by hypertrophy (increase its size) and hyperplasia (increase its number by successive divisions) to accommodate fat deposition (triglyceride, TG) and thus protects tissues other than fat cells to accumulate fat. Tissues such as liver, skeletal muscles, pancreas, heart and other tissues have limited capacity to exhibit hypertrophy and hyperplasia to store excess energy in the form of TG. When adaptive adipocytic protective mechanisms of storage are exceeded, hypertrophic cells exhibit insulin resistance (IR) as well as become like macrophages due to the gene expression pattern that is similar between them and so can produce adipocytokines that induce IR, attract macrophages and elicit systemic subclinical inflammation. Adipose tissue undergoes excessive lipolysis as a consequence of IR and adipocytokines, releasing large amount of free fatty acids into the blood. Increased hepatic FFA uptake and de novo lipogenesis in the presence of hepatic IR reflected by impaired B oxidation of these FFA, impaired ketogenesis, mitochondrial maladaptation with overloaded mitochondrial electron chain and impaired synthesis of ATPs will lead to accumulation of toxic lipids like ceramides, diacyleglycerols and lysophosphatidylcholines. Reactive oxygen species (ROS), Lipid peroxides, resulting from mitochondrial maladaptation as well as the toxic lipids promote more IR milieu and inflammatory environment. This environment causes apoptosis of liver cells (programmed cell death) which is a potent stimulant to kupffer cells to be active and thus activating hepatic stellate cells to transform into myo-fibrobalsts that deposit extracellular matrix and collagen. Also this environment causes endoplasmic reticulum (ER) stress and impaired autophagy (unsuccessful elimination of dysfunctional organelles and protein inside the cell) leading to DNA damage and apoptosis (cell death) and leading to infammasome activation (digestion of dead liver cell by immune systems). Gut micro-biome altered by excess carbohydrate and fat diets stimulate an inflammatory milieu inside the heaptocytes and thus augmenting the effect of toxic lipids, lipid peroxides, misfolded proteins resulting from ER stress and damaged DNA. NAFLD is a heritable disease; presence of some genotypes accelerates progression to NASH and fibrosis.*

### 3.5.2. Genetic Modifiers (Intrinsic Factors):

The genes implicated in NAFLD severity and progression are classified into four functional categories: adaptors of insulin sensitivity, adaptors of lipid metabolism, adaptors of progression to NASH (adaptors of oxidative stress, endotoxin outcome, or cytokine activity), and lastly adaptors of fibrosis development and progression (Yang et al. 2000)(Fromenty et al. 2004).Table (3.8) summarizes some of these genes. Of these genes, the only validated genes that can be considered of proven significant to be used for screening and stratifying the high risk patient for NASH is patatin-like phospholipase domain-containing 3 gene “PNPLA3” and the trans-membrane6 superfamily2 gene “TM6SF2” (Petta et al. 2009) .

Familial aggregation studies, epidemiological, and twin studies have demonstrated consistent possible explainable function of heritability in pathogenesis of NAFLD, clarifying how susceptibility to NAFLD and aggressive course of NASH exhibits markedly observed interpersonal variability. Moreover, these individuals share environmental risk factors, but genetic risk factors can explicate almost up to more than 50% of variability in the course of NAFLD among patients. Heritability can explain about 60% of variability in ALT levels reflecting hepatic fat content. Also it can approximately interpret about 50% of GGT level.

The first step to identify specific gene is to establish heritability of a trait, then identify this specific gene by linkage and association analysis. Nutritional factors and activity levels must be accounted for, and adjusted in the analysis, because the shared environmental confounders affect the heritability assessment.

Minor G allele of PNPLA3 deranges the hydrolysis of triglyceride, leading to increased hepatic fat accumulation, thus this gene increases risk of NASH progression and advanced fibrosis by 1.5 fold, as it has direct role on stellate cell function in NASH which is responsible for increased extracellular matrix deposition and enhanced fibro-genesis. It causes increased susceptibility to hepatocellular carcinoma by fivefold in homozygous carriers, independent of confounders like: age, gender, BMI or diabetes. The protein encoded by PNPLA3 is called “adiponutrin”. The malfunction gene is “PNPLA3-148 MM”.

Exercise and weight loss decrease hepatic fat content in homozygous carriers of this allele (GG), along with, low carbohydrate and caloric diet reduce the fat accumulation in the liver. The reduction is significant in homozygous carriers than the heterozygous carriers and this may be due to elevated baseline level. This emphasizes the complexity of the NAFLD and its various phenotypes, stressing on the profound interaction between the genetic background and the environmental factors. However, mere reduction in caloric intake and exercise can oppose and negate this intense heritable effect for progression of NASH (Dongiovanni, Anstee, and Valenti 2013).

Table (3. 7 ) Causes of steatosis and steatohepatitis

| |
|---|
| ***Acquired Metabolic And /Or Nutritional Disorders:*** |
| Metabolic Syndrome( Obesity, Insulin Resistance, Type2 Diabetes0<br>Starvation And Cachexia<br>Protein Malnutrition( Kwashiorker, Anorexia Nervosa)<br>Dietary Choline Deficiency<br>Total Parenteral Nutrition<br>Acute Fatty Liver Of Pregnancy<br>HELLP |
| ***Drugs*** |
| Amiodarone, Aspirin, Chloroquine, Corticosteroids, Methotrexate, Nsaids, Treatment Of HIV Infection, Estrogens, Tamoxifen, Tetracycline, Valproic Acid. |
| ***Toxins*** |
| Ethanol, Petrochemicals, Heavy Metals |
| ***Rare Monogenic Disease:*** |
| Abetalipoproteinemia, Cholesterol Ester Storage Disease<br>Familial Combined Hyperlipidemia, Familial Hypobetalipoproteinemia<br>Glycogen Storage Disease, Inherited Defects In Fatty Acid B-Oxidation<br>Lecithin-Cholesterol Acyl-Transferase Deficiency, Lipodystrophy<br>Lysosomal Acid Lipase Deficiency, Ornithine Trans-Carbamylase Deficiency, Wilson Disease. |
| ***Infections And Immunological Conditions:*** |
| Chronic Hepatitis C ( Genotype 3)<br>Bacterial Overgrowth Following Jejunoileal Bypass<br>Celiac Disease<br>Reye Syndrome |

Some other proteins (glucokinase regulator =GCKR) are associated with PNPLA3 protein product that regulate both glucose breakdown and triglyceride breakdown respectively, so malfunction of both genes “PNPLA3 I148M and CGKR P446L polymorphisms” leads to unrestricted glycolysis (glucose breakdown is unchecked) and subsequently this favors and facilitates hepatic lipogenesis (increase liver fat synthesis). This heritability effect expounds at most one third of variability in hepatic fat content among obese children coming from European descent (Santoro et al. 2012).

"Trans-membrane 6 superfamily 2 gene" (TM6SF2) encodes for a protein that act as lipid transporter out of the liver. Carriage of both copies of malfunction form, E167K variant, increases the risk of progressive NASH and advanced fibrosis by 2 folds independent of other confounders as well as PNPLA3. Carriage of minor allele is associated with NAFLD, whether carriage of major allele is accompanied with dyslipidemia and high risk of myocardial infarction and cardiovascular disease. This is due to the fact that the price paid to protect the liver is the increased risk of CVD, as these patients tends to export the fat out of the liver at the expense of being up-taken by the blood vessels, and hence atherosclerosis will ensue, causing myocardial infarctions and strokes (Boyer and Lindor 2016) .

Table (3. 8): Some of the genes possibly involved in modifying the NAFLD course

| **Group Of Genes** | **Example** |
|---|---|
| **Glucose Metabolism And Insulin Resistance** | Insulin receptor genes, PPARϒ, RBP-4, TNF, adiponectin IL-6, SOCS, PPA2 |
| **Genes For Lipid Metabolism** | |
| Liver fat synthesis | RXR, LXR, SREBP, ChREBP, fatty acid synthase, Acetyl CoA carboxylase, leptin, adiponectin |
| Liver Fat export | Apolipoprorein B, apolipoprotein C-III, MTP |
| Fatty acid oxidation | Adiponectin, PPARα, carnitine palmitoyl transferase |
| **Genes Of Statohepatitis Risk** | |
| Genes for oxidative stress And endotoxin response | Adiponectin,TNFα, SOD,GST, GSH, TLR, MnSOD, MAO |
| Genes for adiponectin and their receptors | Adiponectin, leptin, resistin, RBP4, AFABP, adiponectin receptors, leptin receptors. |
| Genes for cytokines and their receptors | TNFα, IL-6, IL-10, MCP1, TNFR |
| **Genes For Fibrogenesis** | Adiponectin, leptin, angiotensinogen, steroids, KFL6, TGFβ, CTGF, MMP, TIMP |

**PPARϒ** : *peroxisome proliferator-activated receptorϒ*, **RBP-4**:*retinol binding protein 4*, **TNFα**: *tumor necrosis factorα*, **IL-6** *: interleukin 6*, **SOCS**: *suppressor of cytokine signaling*, **PPA2**: *protein phosphastase A2*, **RXR**: *retinoid X receptors*, **LXR**: *liver X receptor*, **SREBP**: *sterol responsive element binding protein*, **ChREBP**: *carbohydrate responsive element binding protein*, **MTP**: *microsomal triglyceride transfer protein*, **PPARα**: *peroxisome proliferator-activated receptorα*, **SOD**: *superoxide dismutase*, **GST**: *glutathione transferase*, **GSH**: *glutathione perioxidase*, **TLR**: *Toll-like receptor*, **MnSOD**: *manganese superoxide dismutase*, **MAO**: *monoamine oxidase*, **AFABP**: *adipocyte fatty acid binding protein*, **IL-10**: *interleukin 10*, **MCP 1**: *monocyte chemo-attractant protein-1*, **TNFR:** *tumor necrosis factor receptor,* **KFL6** : *Kupper-like factor 6*, **TGFβ**: *tissue growth factorβ*, **CTGF**: *connective tissue growth factor*, **MMP**: *matrix metalloproteinase*, **TIMP**: *tissue inhibitor of matrix metalloproteinase*

## 3.6. Dynamic model of NAFLD:

As discussed above some NAFL patients can evolve to NASH and advanced fibrosis, on the other hand some of those NASH patients can regress to NAFL as time elapses. Thus the mere presence of NAFLD or NASH on baseline liver biopsy examination has no universal prognostic impact on the course of the disease, although NAFL patients have lowest risk for evolution to fibrosis as compared to NASH patients. So it is hypothesized that in early stages of NAFLD, the patients cycle between NAFL and NASH. Regardless the biopsy results being NAFL or NASH, about 80% of them are slow progressors, that is to mean, they are unlikely to progress further beyond mild fibrosis ( F0 to F2), while approximately 20% manifest rapid fibrosis progression and develop advanced fibrosis and cirrhosis(F3 to F4) within a few years (De and Duseja 2020).The majority of NAFLD patients are slow progressors to evolve from F0 to F1, however; subset of these patients are rapid progressors to evolve from F0 to F3 or F4. Study accomplished by (McPherson et al. 2015) on 108 patients having sequential liver biopsy with median interval of 6.6 years, 42% were progressors, 40% had stable fibrosis, while 18% were regressors. Figure (3.5) shows the dynamic model of NAFLD.

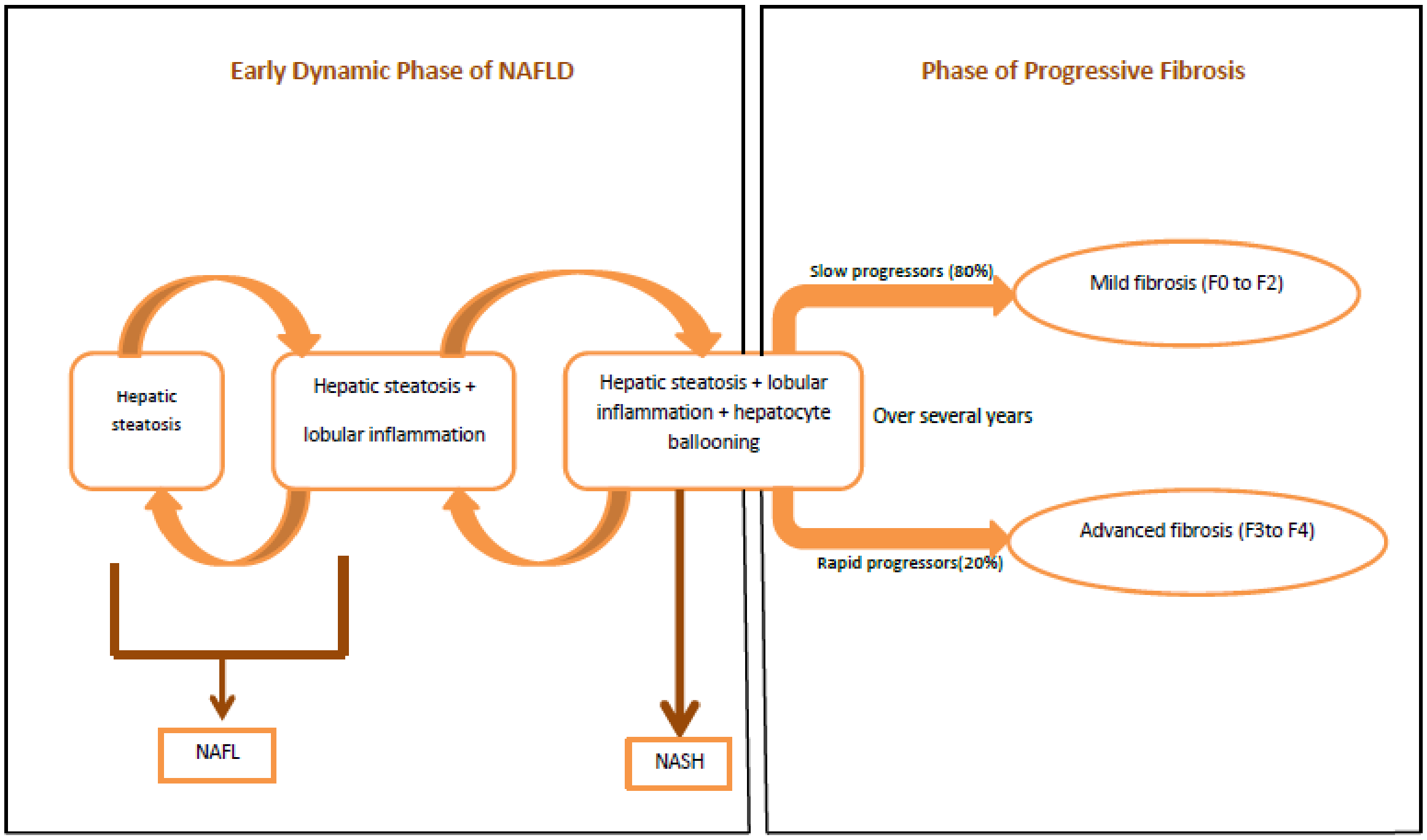

Figure (3. 5): dynamic model of NAFLD

## 3.7. Diagnosis of NAFLD:

The practice guidelines of the American Association for the study of Liver Disease define NAFLD by the presence of the following features (Chalasani et al. 2018):

- The existance of hepatic steatosis established by histology or imaging.
- Exclusion of other secondary causes of hepatic steatosis, especially significant alcohol consumption.
- Exclusion of any other causes of chronic liver disease like drug, infection, autoimmune diseases, hereditary, genetic, or metabolic causes like: hepatitis B and C infection, HIV, autoimmune hepatitis, primary sclerosing cholangitis, primary biliary cirrhosis, Celiac disease, Wilson's disease, hemochromatosis, α-1 antitrypsin deficiency, cystic fibrosis and porphoryia .

Assessment of the patient in clinical practice should evaluate the disease activity that is to mean, whether the disease is NAFL or NASH, the stage of fibrosis and the severity of the risk factors such as insulin resistance (IR) and metabolic syndrome (MetS) components.

### 3.7.1. Liver biopsy:

NAFLD is characterized by three main histopathologic features: steatosis, liver injury "steatohepatitis" and fibrosis. These features differ between adults and children. The following discussion will concern the features of the disease in adults. Liver biopsy permits direct examination of liver tissue to evaluate inflammatory disease activity (grade of disease) and evolution of fibrosis to cirrhosis (stage of fibrosis). However, one of the major limitations that face liver biopsy is sampling error because < 1/50000 of the total hepatic volume is sampled at a specific single time point with the heterogeneity of NAFLD distribution throughout the hepatic parenchyma. Also accurate diagnosis is observer-dependent and influenced by pathologist experience. To overcome these limitations, core length biopsy of 15 mm or more taken by at least a 16 gauge needle is the required biopsy minimum enough to provide an adequate enough sample. And to standardize the biopsy evaluation, a validated well-defined scoring system should be used. This system should have easy applicability, marked reproducibility, and a valid intra-observer and inter-observer consistency. NAFLD activity score (NAS) proposed by U.S. national institutes of health-sponsored NASH CRN gathers the assessment of steatosis, inflammation and ballooning to create NAS ranging from 0 to 8 points and a distinct fibrosis score ranging from 0 to 4 (Kleiner et al. 2005). It is a useful and beneficial research tool for use in clinical trials but it is not a suitable prognostic tool to use in clinical practice. As a result and to avoid these limitations in the routine clinical practice, an improvement so called the "steatosis-activity-fibrosis" (SAF) has been developed (Bedossa et al. 2012). Using SAF; steatosis, activity and fibrosis are assessed apart from each other and then an algorithm is implemented to categorize biopsies into one of the three diagnostic groups: normal, NAFL, or NASH. Distinguishing steatosis from markers of cellular injury (ballooning and lobular inflammation) better differentiate NASH from more indolent course of liver disease (NAFL).Table (3.9) illustrates the comparison between NAS and SAF score systems. Figure (3.6) illustrates the SAF algorithm. Figure (3.7) demonstrates the microscopic picture encountered in NAFLD. Figure (3.8) shows the stages of fibrosis in NAFLD.

Although, liver biopsy is the gold standard method for diagnosis of NAFLD and its phenotypes as well as for staging fibrosis, some disadvantages hamper its use as a routine and follow up investigation in clinical practice like: hospital admission, invasiveness, elevated cost, reported mortality although minimally low, sample error and it is a poorly suited and suitable diagnostic test for such a highly prevalent condition. Therefore, development of noninvasive techniques has become mandatory to assess the disease activity, fibrosis stage as well as to detect hepatocellular carcinoma (HCC). They are also used to guide therapy plan in routine clinical practice because they are more suitable than liver biopsy (Castera et al. 2019).

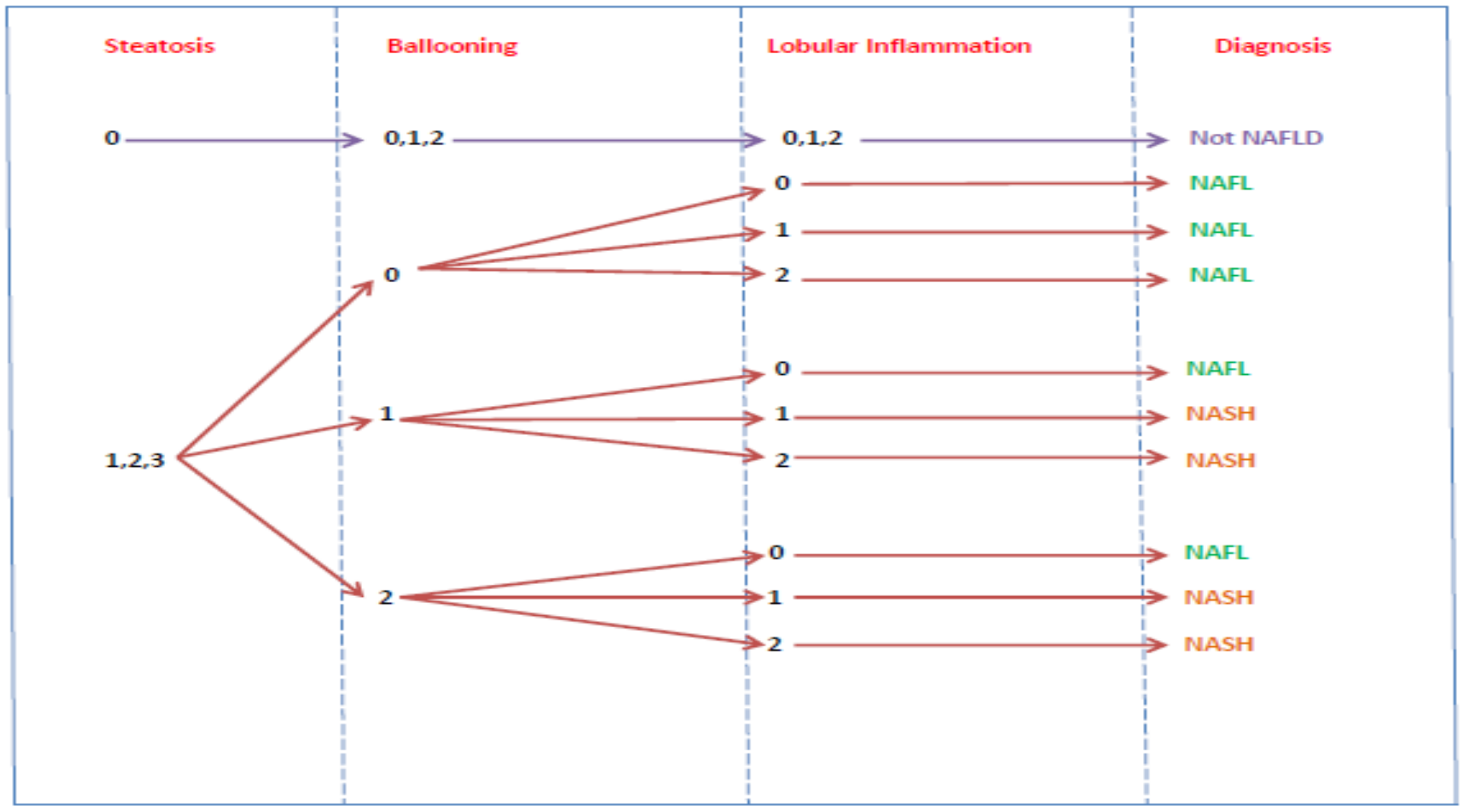

Figure (3. 6): steatosis-activity-fibrosis score algorithm (SAF)

*Presence of Stage 0 steatosis regardless the grade of ballooning or lobular inflammation excludes NAFLD, while more than 5% steatosis with the presence of both ballooning and lobular inflammation of any grade establishes the NASH diagnosis. On the other hand, more than 5% of steatosis in the presence of ballooning or inflammation of any degree point to NAFLD diagnosis.*

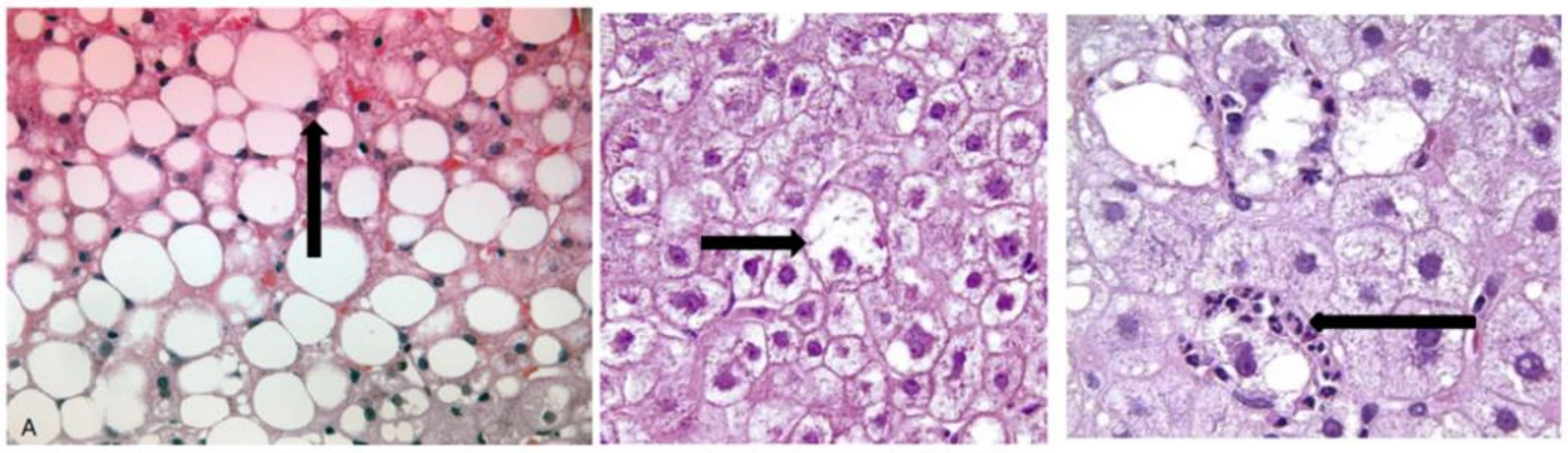

Figure (3. 7): microscopic picture of pathological changes encountered in NAFLD

*Picture (A) on the left shows the steatosis where lipid droplets accumulate in hepatocytes and displace the nucleus to the periphery against the cell wall giving a picture of a ring. The picture in the middle illustrates ballooned hepatocytes with striated cytoplasm. The picture on the right shows the inflammatory cells (polymorph-nuclear leucocytes). source:abdominalkey.com/fatty-liver-disease-2/.*

Table (3. 9): Comparisons between NAS and SAF score systems

| **NAFLD activity score(NAS)** | | **SAF** | |
|---|---|---|---|
| **Histologic features** | **Category definition** | **Histologic features** | **Category definition** |
| **Steatosis** | 0=<5% | **Steatosis** | 0=<5% |
| | 1=5-33% | | 1=5-33% |
| | 2=34-66% | | 2=34-66% |
| | 3=>66% | | 3=>66% |
| | + | | =steatosis score(S0-3) |
| **Hepatocyte Ballooning** | 0= none | **Hepatocyte Ballooning** | 0=none |
| | 1=few | | 1=clusters of round hepatocytes with pale cytoplasm |
| | 2=many | | 2=same as grade 1 with enlarged hepatocytes(> 2times normal size) |
| | + | | + |
| **Inflammation** | 0=none | **Inflammation** | 0=none |
| | 1= 1-2 foci per 20 times field | | 1= <2 foci per 20 times field |
| | 2= 2-4 foci per 20 times field | | 2= >2 foci per 20 times field |
| | 3= >4 foci per 20 times field | | |
| | =NAFLD activity score (0-8) | | =activity score (A 0-4) |
| **Fibrosis** | 0=no fibrosis | **Fibrosis** | 0=no fibrosis |
| | 1a=zone 3 mild peri-sinusoidal fibrosis | | 1a=zone 3 mild peri-sinusoidal fibrosis |
| | 1b=zone 3 moderate peri-sinusoidal fibrosis | | 1b=zone 3 moderate peri-sinusoidal fibrosis |
| | 1c=periportal/portal fibrosis only | | 1c=periportal/portal fibrosis only |
| | 2=zone 3 plus portal/periportal fibrosis | | 2=zone 3 plus portal/periportal fibrosis |
| | 3= bridging fibrosis | | 3= bridging fibrosis |
| | 4=cirrhosis | | 4=cirrhosis |
| | =fibrosis score (F0-4) | | =fibrosis score (F0-4) |

The following 4 figures illustrate the stages of fibrosis histologically.

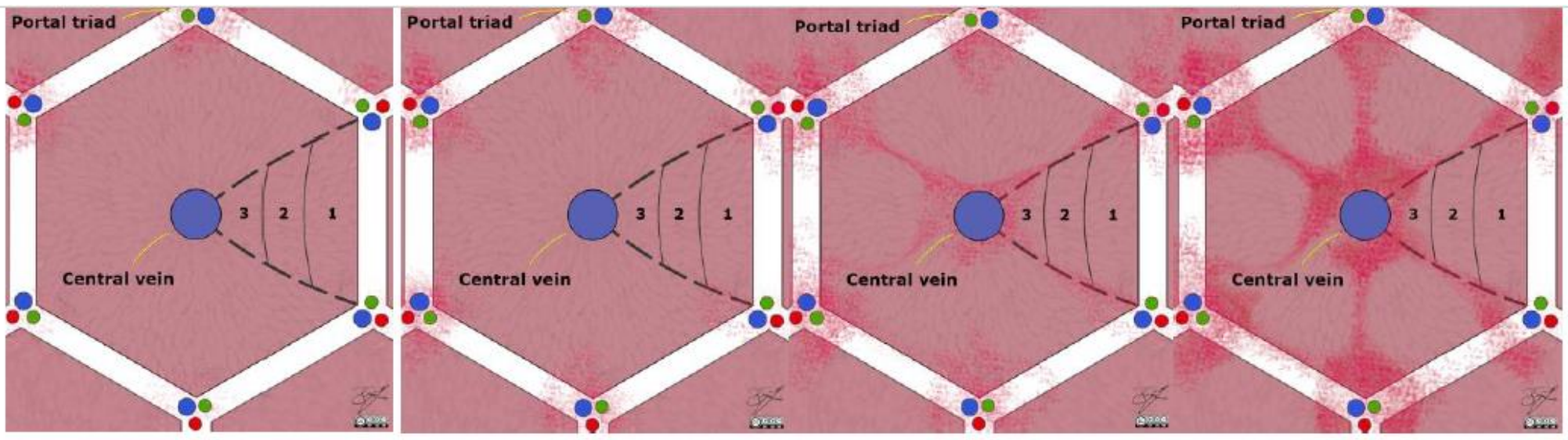

F1: Fibrous expansion of portal triad

F2: Fibrous septa extending from portal triad

F3: Bridging fibrous septa extending from portal triad to central vein

F4: Clear bands of fibrosis forming nodules

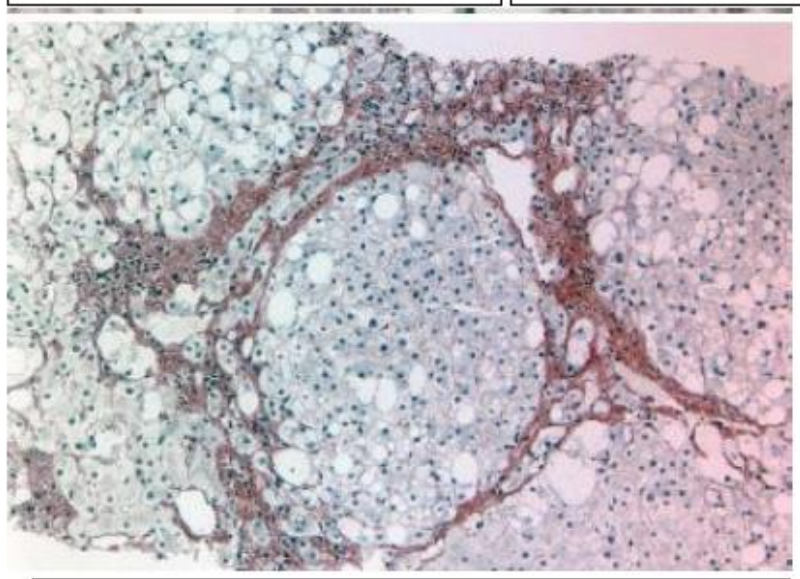

Hepatic nodules (disorganized hepatocytes surrounded or entangled by fibrous tissue)

Figure (3. 8): stages of liver fibrosis
*Source: radiopaedia.org/cases/hepatic-functional-unit?lang=gb ..*

### 3.7.2. Noninvasive Tests for Diagnosis of NAFLD:

The noninvasive tests can be subdivided into two main major groups that can be used to diagnose steatosis, activity, and fibrosis. Noninvasive techniques include serological biological tests as well as imaging physical techniques.

The “two-hit theory” of steato-hepatitis assumes that the evolution of steatosis is a mandatory precedent to establishment of steatohepatitis. Imaging tests (ultrasound, CT, and MRI) often record the severity of steatosis. This significant correlation between the quantified severity of steatosis demonstrated by imaging modalities, and the histological grading of steatosis is proposed by (Chalasani et al. 2008) to be used as an additive variable, towards increasing risks of NASH and zone 3 fibrosis in NAFLD patients. But the study conducted by them is a cross sectional study that could not address whether the degree of steatosis severity affects the rate of progression to steatohepatitis, in other word, whether patients with severe steatosis exhibit faster rate of evolution to full-blown picture of NASH than those with lesser degree of steatosis. Moreover, they demonstrated that how the variability of severity of inflammation, steatosis and hepatocyte ballooning between individuals affecting fibrosis progression rate urgently need large-scale longitudinal studies to be conducted. Furthermore, they pointed out that if the correlation between imaging reporting steatosis severity and histology grading this severity, as well as the correlation between this histology and evolution of steatohepatitis are strongly asserted, then quantifying the degree of hepatic steatosis by imaging could be used as a representative therapeutic end-point in pilot studies or clinical trials treating NASH patients and developing new drugs, as numerous pharmaceutical corporations inventing drugs for treating metabolic derangements like obesity, hyperlipidemia, and diabetes are routinely conducting “MR spectroscopy” to check up on “hepatic steatosis”.

#### *3.7.2.1. Imaging Tests:*

The existence of NAFLD can be detected radio-logically using the ultrasonography, computed tomography (CT) and magnetic resonance imaging (MRI). Sensitivity and specificity widely vary between modalities. Ultrasonography can detect steatosis if hepatic fat content is ≥20% with mean sensitivity ranges from 73.3 to 90.5%; however, this sensitivity is limited if less than 30% of hepatocytes are steatotic and this is observed with progression to advanced fibrosis as the degree of steatosis decreases. In addition, it cannot precisely discriminate between various stages of fibrosis or detect coexistent NASH. It is also highly operator dependent, but it has advantages of being widely available, low cost, and no radiation exposure. Ultrasonography-based vibration controlled transient elastrography (VCTE) accomplished by Fibro-scan uses controlled attenuation parameter (CAP) to grade steatosis. It measures “US attenuation” by liver fat content. It measures grade >0 steatosis with optimal cutoff 248 dB/m and area under the receiver operating characteristic curve of 0.825, and grade >1 steatosis with optimal cutoff 268 dB/m and area under the receiver operating characteristic curve of 0.865, thus CAP is useful as liver-specific, easy to be carried out, has valuable “cost-effective” measurement, and reproducible within patients as well as it can be repeated to supervise changes in steatosis. Because Fibro-scan has several advantages like: being painless to the patients, rapid to perform, prompt availability of results, increased intra- and inter-operator reproducibility and good validated diagnostic accuracy in many studies, thus it is a worldwide used-modality to assess hepatic fibrosis. As it is good for diagnosis of stage 3 fibrosis (84 % sensitivity and 83% specificity) with area under ROC being 0.93% and cutoff value of 8.7 kPa, and it is excellent for stage 4 fibrosis (92% sensitivity and 88% specificity) with area under ROC being 0.95 and cutoff value of 10.3 Kpa, however it has somewhat minimal accuracy for the diagnosis of stage 2 fibrosis (79% sensitivity and 76% specificity) with area under the ROC being 0.84 and cutoff value of 7 kPa. These results were obtained by Wong et al. (2010) utilizing M probe. It has some inconveniences with raised failure rate in morbidly obese patients and due to: dense chest wall, ascites, cramped intercostal spaces, and experience immaturity of the operator. And to overcome these obstacles, VCTE device has three distinct probes for stiffness evaluation in various situations: S probe for children, M probe for adults, and XL probe for overweight patients. “Magnetic resonance elastography” (MRE) is an MRI-based methodology to quantitatively image liver stiffness. The liver stiffness cut-off values for diagnosing mild stage of liver fibrosis >1 ranges from 2.5 to 3.02 kPa with sensitivity of 44-75% and specificity of 77-91%; while in advanced fibrosis >3, the cutoff value ranges from 2.99 to 4.8 kPa with sensitivity of 33-91% and specificity of 80-94%. Area under the ROC using MRE that corresponds to diagnosis of stage >1 liver fibrosis ranges from 0.772 to 0.86, of stage >2 liver fibrosis ranges from 0.856 to 0.89, of stage >3 liver fibrosis ranges from 0.87 to 0.981, and of stage >4 liver fibrosis ranges from 0.882 to 0.993. MRE is better than other biomarkers, scoring systems and US-based elastography for diagnosis of liver fibrosis (Yoneda et al. 2018)(Boyer and Lindor 2016)(Boyer and Lindor 2016).

#### *3.7.2.2. Serological Tests:*

The serological markers for recognizing stages of liver fibrosis are classified into indirect markers that mirror perturbations in hepatic functions but not that in collagen turnover (AST/ALT and platelet levels) and direct markers that are concerned with fibro-genesis and extracellular matrix cycle. Although serum aminotransferase is frequently used in clinical practice as marker for NASH, it should not be used to identify NASH because it has poor predictive value for detection of NASH. Serum ALT value of more than twice the upper normal limit i.e. more than 70 U/L has 50% sensitivity and 61% specificity for NASH diagnosis. Normal ranges of ALT levels have been shown in 80% of patients with hepatic fat deposits. More studies should be conducted to find out cutoff values for early and better detection of NASH.

Serum cytokeratin (CK)-18 is an indicator or sign for "hepatocyte apoptosis" and it has been extensively investigated to discriminate steatosis from NASH. Cut-off value of CK-18 of more than 240 U/L has 76.7% sensitivity and 95% specificity for detecting NASH. Another study showed a value of 270 U/L with 64 % sensitivity and 76 % specificity. Further studies are compulsory required for a validated consistent cutoff value (Paul 2020).

There are many scoring systems to detect fibrosis. One of the most commonly used systems for detection of fibrosis is the AST to platelet ratio index (APRI). Tapper et al. (2014) reported that APRI greater than 1 had 30% sensitivity and 92.8% specificity and it was most significant predictor of advanced fibrosis. It is calculated using this formula:

$$APRI = \frac{\frac{AST(IU/L)}{AST\ ,ULN(IU/L)}}{platelet\ count(10^9/L)}$$

where ULN is the upper limit of normal of the AST enzyme measured in IU/L
Another scoring system is the FIB-4; it is calculated using this formula:

$$FIB4 = \frac{age(years) \times AST(IU/L)}{platelet(10^9/L) \times \sqrt{ALT(IU/L)}}$$

values less than 1.3 have a negative predictive value of 90% for stage F3 to stage F4 fibrosis, whereas values more than 2.67 have a positive predictive value of 80 % (Shah et al. 2009)

NAFLD fibrosis score (NFS) is another widely used score calculated with routinely measured parameters using this formula: (Angulo, Jason M Hui, et al. 2007)

$$NFS = -1.675 + .037 * age(\ years\ ) + .094 * BMI\ (kg/m^2) + 1.13 * diabetis(yes = 1, no = 0) + .99 * AAR - .013 * platelet(10^9/L) - .66 * albumin(g/dL)$$

where $AAR$ is aspartate transaminase (AST) to alanine transaminase (ALT) ratio.

Enhanced liver fibrosis (ELF) test is a commercial panel of markers concentrating on matrix turnover. It has accuracy identical to or a little bit better than that of the "NFS". Both of them have been shown to have a predictive effect on the outcome and mortality of the disease on the long run (Lichtinghagen et al. 2013). ELF can be calculated using this formula:

$$ELF\ score = -7.412 + \ 0.681 * ln(HA) + 0.775 * ln(PIIINP) + \ 0.494 * ln(TIMP - 1)$$

where $HA$ is "hyaluronic acid", $PIIINP$ is "amino-terminal pro-peptide of type III pro-collagen" and $TIMP$ is "tissue inhibitor of metalloproteinase-1".

Fibro-test can be used to exclude advanced fibrosis (F3-F4). It is a commercial marker panel with a patent algorithm composed of total bilirubin, $\alpha_2$-macroglobulin, gamma glutamyl transferase (GGT), hapto-globin, and apo-lipoprotein A1, corrected for gender, age, and BMI. A fibro-test score less than 0.3 is considered low risk for fibrosis with 77% sensitivity, 77%specificity, and 9% NPV (Imbert-Bismut et al. 2001)(Boyer and Lindor 2016). Table (3.10) shows some of the serological tests.

Fatty liver index (FLI) is easy to calculate. It can help physician to initially choose individual for imaging study like Ultrasongrahy or MR spectroscopy, to further assess hepatic steatosis, to sequentially assess them non-invasively for advanced fibrosis using other prediction system scores and blood biomarkers, to follow them up during treatment, and to enroll patients in epidemiological studies.

$$FLI = \frac{e^{(0.953*ln(TG)+0.139*BMI+0.718*ln(GGT)+0.053*waist\ circumference-15.745)}}{1+e^{(0.953*ln(TG)+0.139*BMI+0.718*ln(GGT)+0.053*waist\ circumference-15.745)}} \times 100$$

where TG is plasma triglyceride level measured in mg/dL and GGT is gamma glutamyl transferase. Values < 30 can be used to rule out hepatic steatosis with 87% sensitivity, while; values ≥ 60 are used to rule in hepatic steatosis with 86% specificity. According to the "standardized regression coefficients", the largest achievement of FLI prediction ability is attributed to waist circumference, followed by BMI, triglycerides, and GGT; as depicted by values of these coefficients: 0.356, 0.353, 0.308, and 0.278 respectively (Bedogni et al. 2006).

NAFLD liver fat score and liver fat equation supply the general practitioner with a simple inexpensive and non-invasive tool to predict liver fat content in susceptible individuals, which can be quantitatively measured by histology and proton magnetic resonance spectroscopy on contrary to NAFLD liver fat content, which estimates it qualitatively and thus has limited sensitivity.

$$NAFLD\ liver\ fat\ score = -2.89 + 1.18 * metabolic\ syndrome(yes = 1, no = 0)$$
$$+0.45 * diabetes(yes = 2, no = 0) + 0.15 * insulin(mU/L) + 0.04 * AST(U/L) - 0.94 * AST/ALT$$

where insulin and AST are measured after 8 hours fasting and diabetes is of type 2. Values > -0.64 is optimal cutoff values is with 86% sensitivity and 71% specificity. Other values are used to increase sensitivity and specificity: cutoff ≥ -1.413 gives 95% sensitivity and 52% specificity while cutoff value ≥ 1.257 gives 95% specificity and 51% sensitivity (Kotronen et al. 2009).

Table (3. 10): Some of the selected non-invasive tests

| Test | Cut-Off Value | Sensitivity | Specificity | PPV | NPV |
|---|---|---|---|---|---|
| **APRI** | ***High risk :*** >1 | 27% | 89% | 37% | 84% |
| **NFS predict advanced fibrosis(F3-F4)** | ***High risk :*** NFS > 0.676 | 51% | 98% | 90% | 85% |
| | ***In-determinant:*** NFS = -1.455 to 0.676 | | | | |
| | ***Low risk:*** NFS <-1.455 | 82% | 77% | 56% | 93% |
| **FIB-4 Predict advanced fibrosis(F3-F4)** | ***High risk :*** FIB-4 >2.67 | 33% | 98% | 80% | 83% |
| | ***In-determinant:*** FIB-4 = 1.3 to 2.67 | | | | |
| | ***Low risk:*** FIB-4 <1.3 | 74% | 71% | 43% | 90% |
| **ELF predict advanced fibrosis (F3-F4)** | ***High risk :*** ELF >0.3576 | 80 % | 90% | 71% | 94% |
| | ***Low risk :*** ELF < .3576 | | | | |

### 3.7.3. Diagnostic Algorithms of NAFLD:

Liver biopsy as previously mentioned is the gold standard method to diagnose the degree of steatosis, inflammation and fibrosis as well as to determine the presence or absence of liver-related morbidities like hepatocellular carcinoma (HCC). It has limitations like invasiveness, high cost, 1% to 3% risk of major complications, estimated .01% mortality, and infeasibility to be done to individuals with highly prevalent biological process like NAFLD. This drives noninvasive procedures to be used to stratify persons with NAFLD for future management regarding all possible modalities of treatment and prognostic follow-up. A lot of algorithms have been proposed by many societies all over the world to illustrate the suggested plan for physicians to follow aiming to achieve early diagnosis, treatment and follow up of persons with NAFLD, thereby reducing liver-related morbidities and mortalities. Here are some of these algorithms; almost all of them start with risk stratification of individuals to identify patients with advanced fibrosis who are liable for progression to severely morbid complications. The extent of liver fibrosis is the most prognostic factor for such progression. Table (3.10) lists the common risk factors for NAFLD.

* Figure (3.9) points out suggested algorithm to use for risk stratification in NAFLD patients utilizing the noninvasive tests (Castera et al. 2019).
* Figure (3.10) demonstrates the EASL-EASD-EASO clinical practice guidelines for management of NAFLD (Marchesini et al. 2016).
* Figure (3.11) demonstrates the EASL-EASD-EASO clinical practice guidelines for management of NAFLD in Type 2 Diabetes (Sberna et al. 2018).
* Figure (3.12) illustrates the German guidelines (DGVS) for identifying patients at risk for NAFLD(Blank et al. 2020)
* Figure (3.13) shows clinical strategy of diagnosing and following patients with NAFLD based on invasive and noninvasive methods on behave of the Japanese society of gastroenterology (Yoneda et al. 2018).
* Figure (3.14) demonstrates when to do a liver biopsy (Machado and Cortez-Pinto 2013).
* Figure (3.15) depicts the approach to risk stratification that serially implement tests with high negative predictive value (NPV) that liver biopsy is selected for a highly at risk patient of advanced liver fibrosis/cirrhosis. FIB-4 score or a validated commercial fibrosis panel may be substituted for NAFLD fibrosis score or one-stage screen for fibrosis developed with a different modality (Boyer and Lindor 2016).

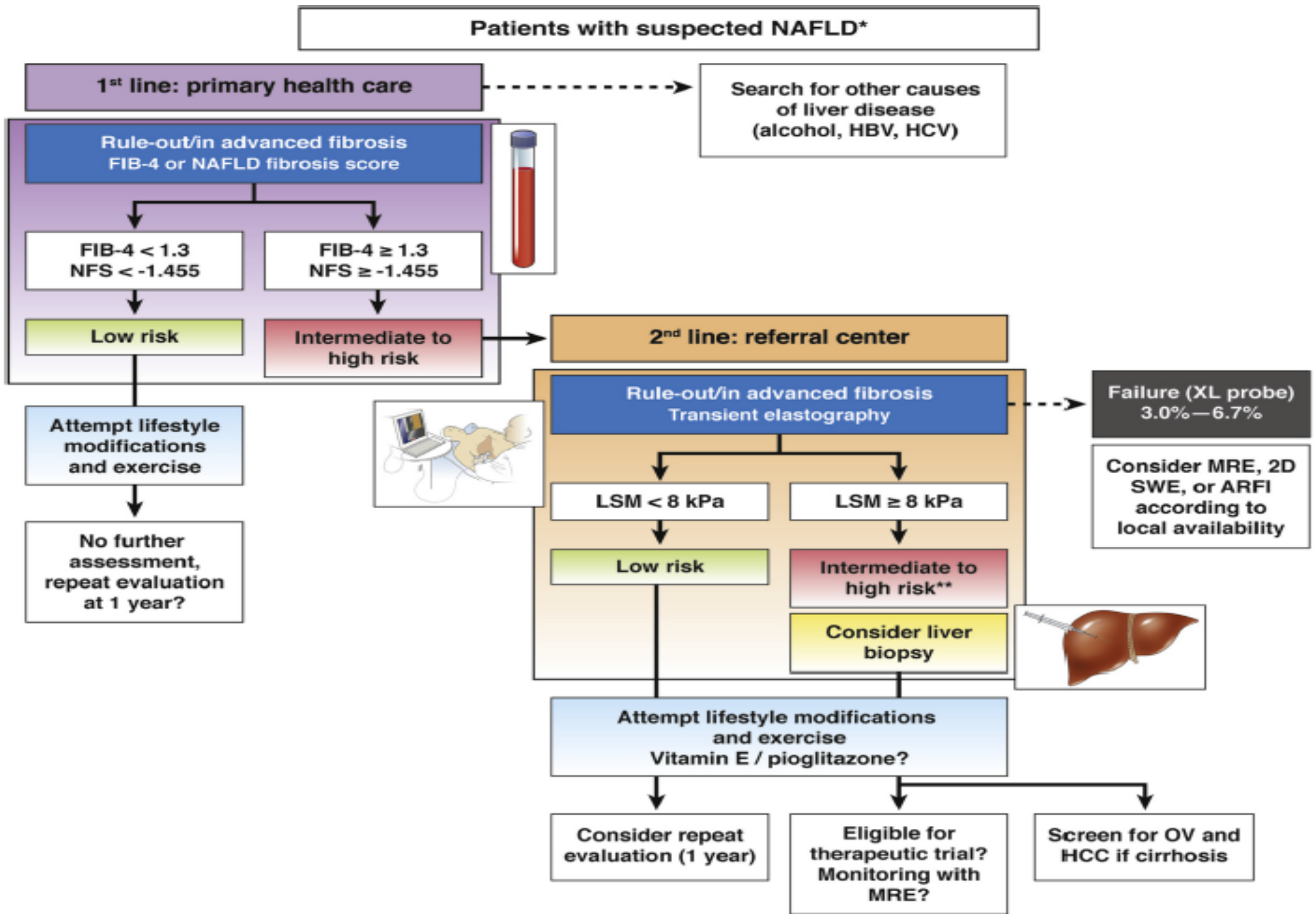

Figure (3. 9): risk stratification and management of NAFLD patients

*NALF is suspected if steatosis is detected by U/S or abnormal elevation of liver enzymes in high risk patients ( type 2 diabetes, obesity, or metabolic syndrome) with exclusion of other causes of chronic liver disease. Sequential tests are performed according to local availability and the situation of use: in primary health care units, the first line tools that are inexpensive, simple, noninvasive, and widely available are the serum biomarkers like FIB-4 or NFS with high negative predictive value (88%-95%) to negate advanced fibrosis. Low risk patients of fibrosis (55-58% of cases, with FIB-4<1.3 or NFS<-1.455) are given no further assessments other than lifestyle modifications and exercises. Intermediate risk patients (30% of cases with FIB-4 =1.3 to 3.25 or NFS=-1.455 to 0.672) as well as high risk patients of advanced fibrosis (12%-15% of cases with FIB-4>3.25 or NFS >0.672 and positive predictive value 75%-90%) are referred to specialized center for LSM( liver stiffness measurements) using transient elastography (TE) in fasting state with M probe for patients with "skin-liver capsule distance" <25 mm otherwise XL probe is used for more obese patients. low risk patients for advanced fibrosis( LSM <8 kPa; NPV= 94%-100%) should repeat the assessment within 1 year . patient with intermediate risk ( LSM =8-10 kPa) or high risk ( LSM ≥10 kPa; PPV=47%-70%) of having advanced fibrosis should undergo liver biopsy. According to availability there are alternatives to the scoring system like the patent commercial tests such as Fibro-test or Fibro-meter. In case of failure of XL-probe of TE, there are alternatives such as MRE especially in patients with BMI >35 kg/m$^2$. All patients should be offered lifestyle modifications and exercise.*

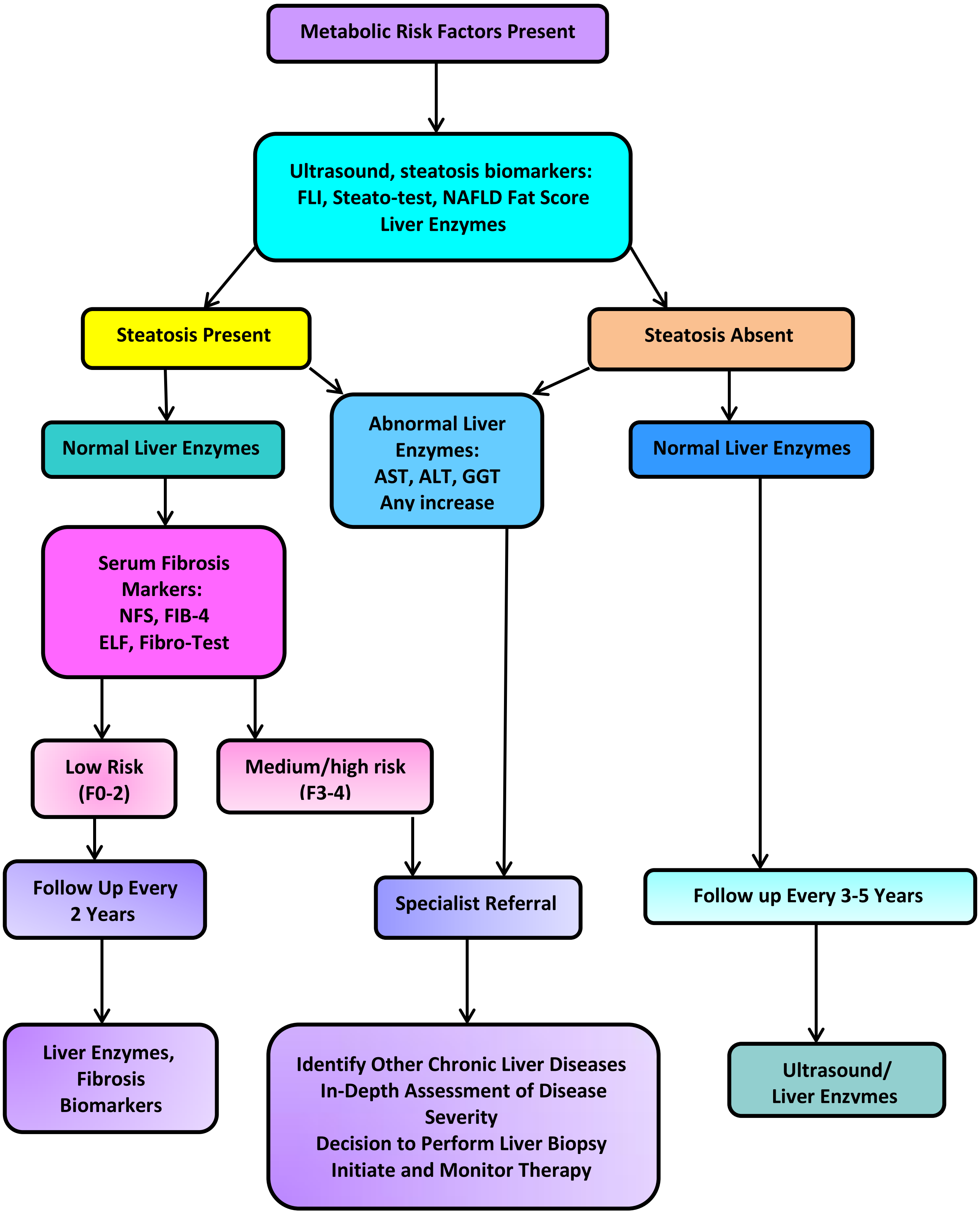

Figure (3. 10): EASL-EASD-EASO clinical practice guidelines for identification of NAFLD patients

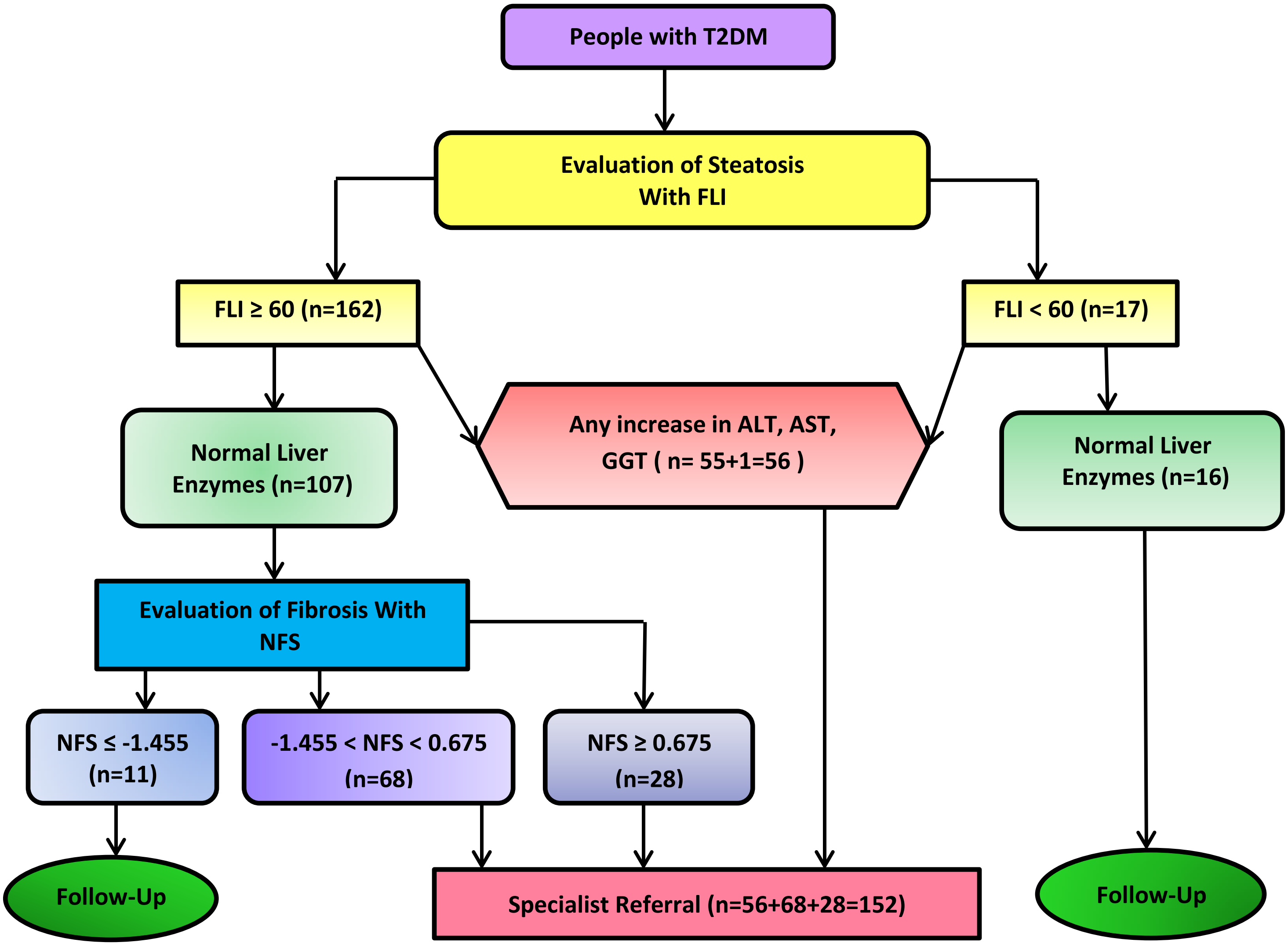

Figure (3. 11): EASL-EASD-EASO clinical practice guidelines for management of NAFLD in type 2 diabetes

*Sberna et al. 2017 conducted a single-center retrospective observational study to evaluate the application of the EASL-EASD-EASO in type 2 diabetic patients; it showed up that there was an excessive referral rate to hepatologists in liver clinic. As illustrated in the above figure, 179 diabetic patients were screened for the presence of steatosis with FLI. Any patient with increased liver enzymes, whether with (steatosis absent i.e.) FLI < 60 (n=1), or with (steatosis present i.e.) FLI ≥ 60 (n=55) was referred to a liver clinic (a total of 56). Sixteen patients with FLI <60 and with normal liver enzymes were assigned to follow up every 3-5 years. The remaining 107 patients with FLI ≥ 60 and with normal liver enzymes were evaluated for liver fibrosis with NFS. Of those, 11 patients with NFS ≤ -1.455 were decided to be followed up every 2 years, 68 patients with NFS levels between -1.455 and 0.675, and 28 patients with NFS ≥ 0.675 were referred to a liver clinic. Thus the total referral rate was approximately 85%. Further assessment of fibrosis score of those 56 patients with elevated liver enzyme could reduce the number of the patients to be referred to a liver clinic, which will be illustrated in the study conducted by Blank et al 2020, as shown in the German guidelines.*

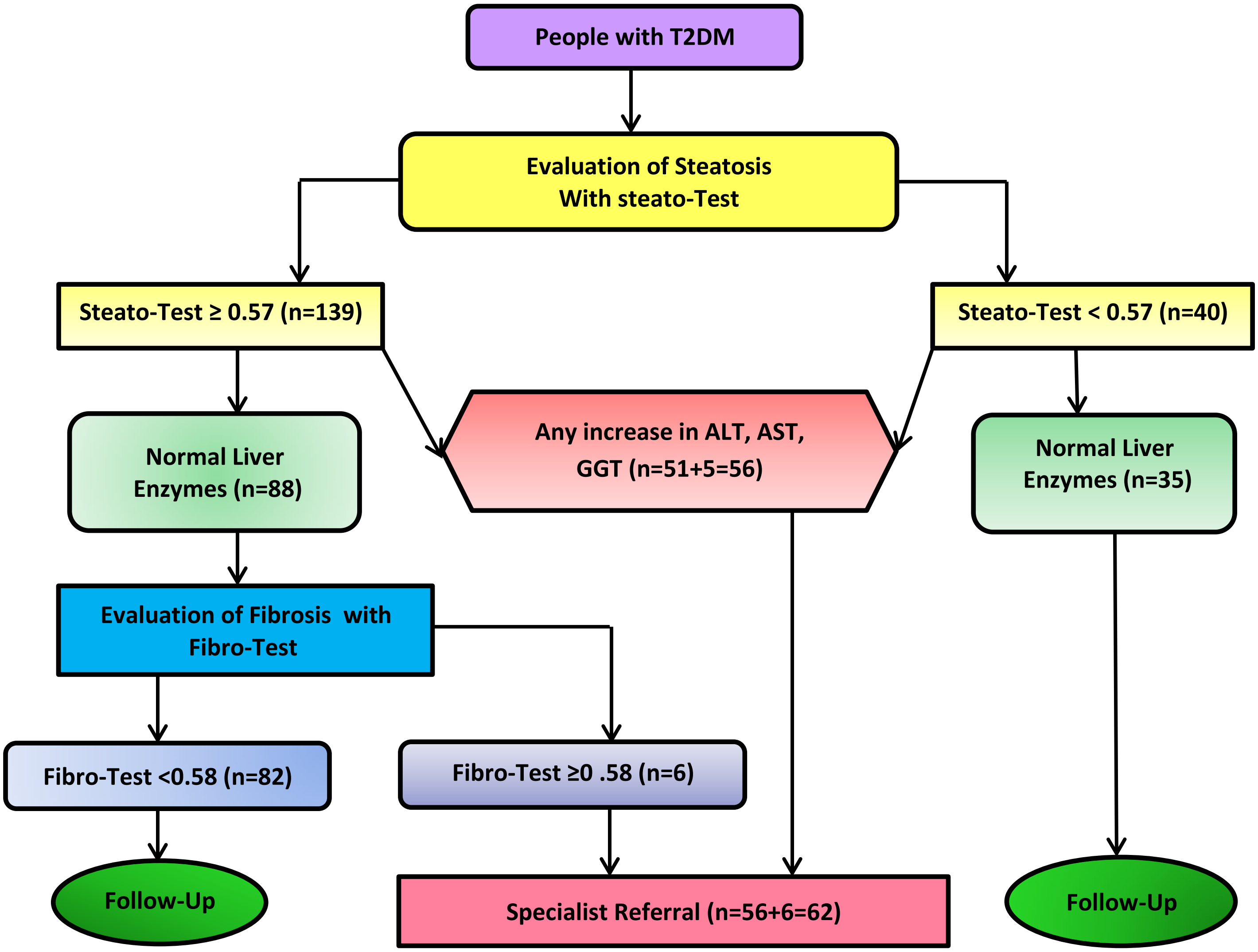

*Figure (3.11.1):EASL-EASD-EASO clinical practice guidelines for management od NAFLD in type 2 diabetic using steatotest and fibrotest*

*But when Sberna et al. 2017 used another system to evaluate the application of the EASL-EASD-EASO on type 2 diabetic patients, it showed up that the referral rate to liver clinic decreased but still high. As illustrated in the above figure, 179 diabetic patients were screened for the presence of steatosis with steatoTest. Any patient with increased liver enzymes, whether with (steatosis absent i.e.) steatoTest < 0.57 (n=5), or with (steatosis present i.e.) steatoTest ≥ 0.57 (n=51) was referred to a liver clinic (a total of 56). Thirty five patients with steatoTest < 0.57 and with normal liver enzymes were assigned to follow up every 3-5 years. The remaining 88 patients with steatoTest ≥ 0.57 and with normal liver enzymes were evaluated for liver fibrosis with FibroTest. Of those, 82 patients with FibroTest < 0.58 were decided to be followed up every 2 years while 6 patients with FibroTest ≥ 0.58 were referred to a liver clinic. Thus the total referral rate was approximately 34.6% (62 patients). Sberna et al. reported that: ''it would not be possible to refer such a high proportion of people with Type 2 diabetes to a liver clinic. The application of EASL-EASD-EASO guidelines cannot be used in clinical practice in people with Type 2 diabetes. It is essential to develop specific steatosis scores and fibrosis scores for people with Type 2 diabetes in order to improve the selection of patients to be referred to a liver clinic ''.*

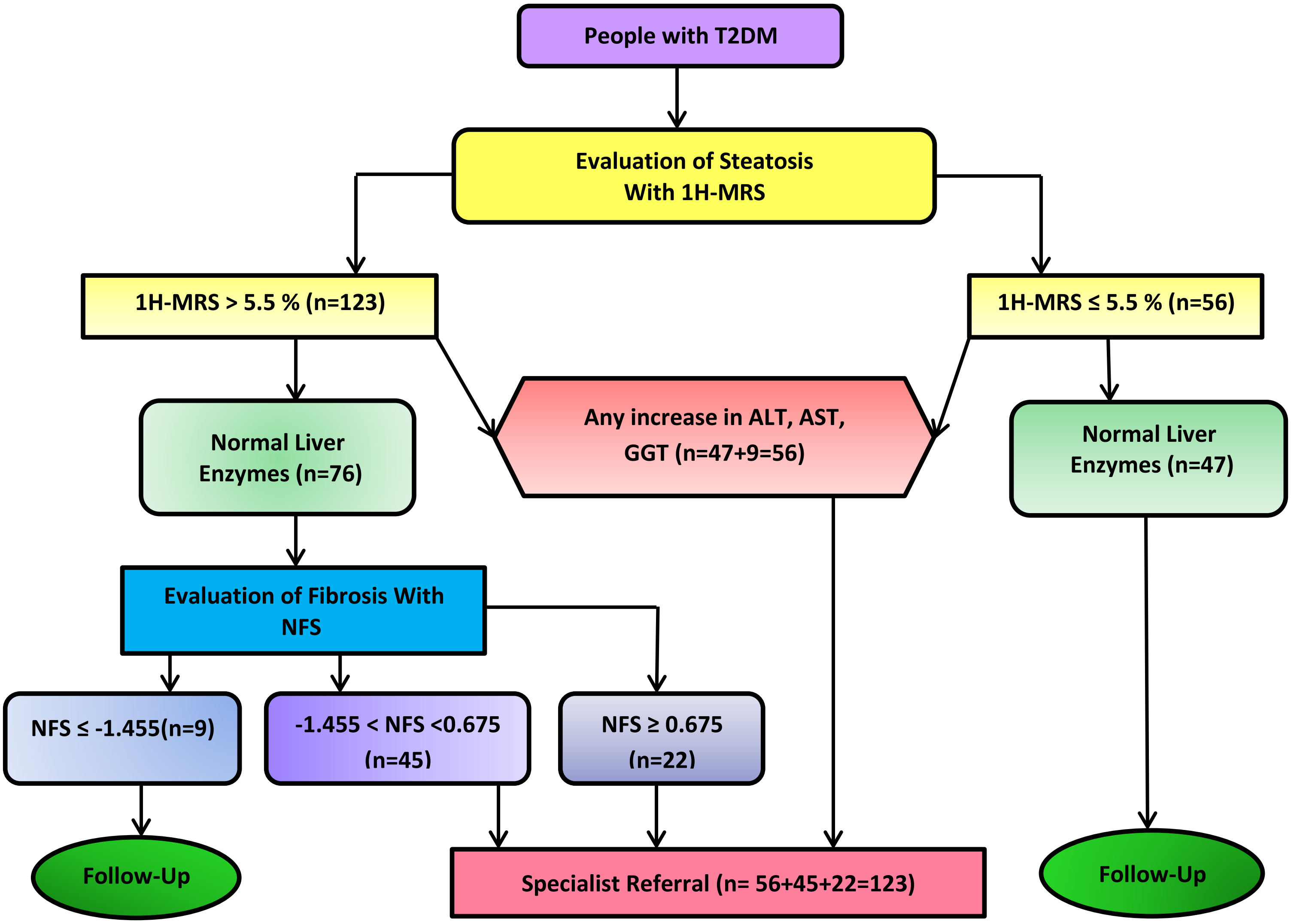

*Figure (3.11.2):EASL-EASD-EASO clinical practice guidelines for management of NAFLD in type 2 diabetic using 1H-MRS and NFS*

*Sberna et al. 2017 used another system. As illustrated in the above figure, 179 diabetic patients were screened for the presence of steatosis with 1H-MRS (proton magnetic resonance spectroscopy); it is very sensitive in detecting steatosis and can reproducibly quantify fat content in the liver. Any patient with increased liver enzymes, whether with (steatosis absent i.e.) hepatic fat content ≤ 5.5% (n=9), or with (steatosis present i.e.) hepatic fat content > 5.5% (n=47) was referred to a liver clinic (a total of 56). Forty seven patients with hepatic fat content ≤ 5.5% and with normal liver enzymes were assigned to follow up every 3-5 years. The remaining 76 patients with hepatic fat content > 5.5% and with normal liver enzymes were evaluated for liver fibrosis with NFS. Of those, 9 patients with NFS ≤ -1.455 were decided to be followed up every 2 years, 45 patients with NFS levels between -1.455 and 0.675 and 22 patients with NFS ≥ 0.675 were referred to a liver clinic. So the total referral rate was 68.7%, less than 85% obtained when FLI was initially used to detect steatosis, as; FLI was mainly designed to detect steatosis in general population not in diabetic patients and spectroscopy is more sensitive than FLI to detect the hepatic fat content.*

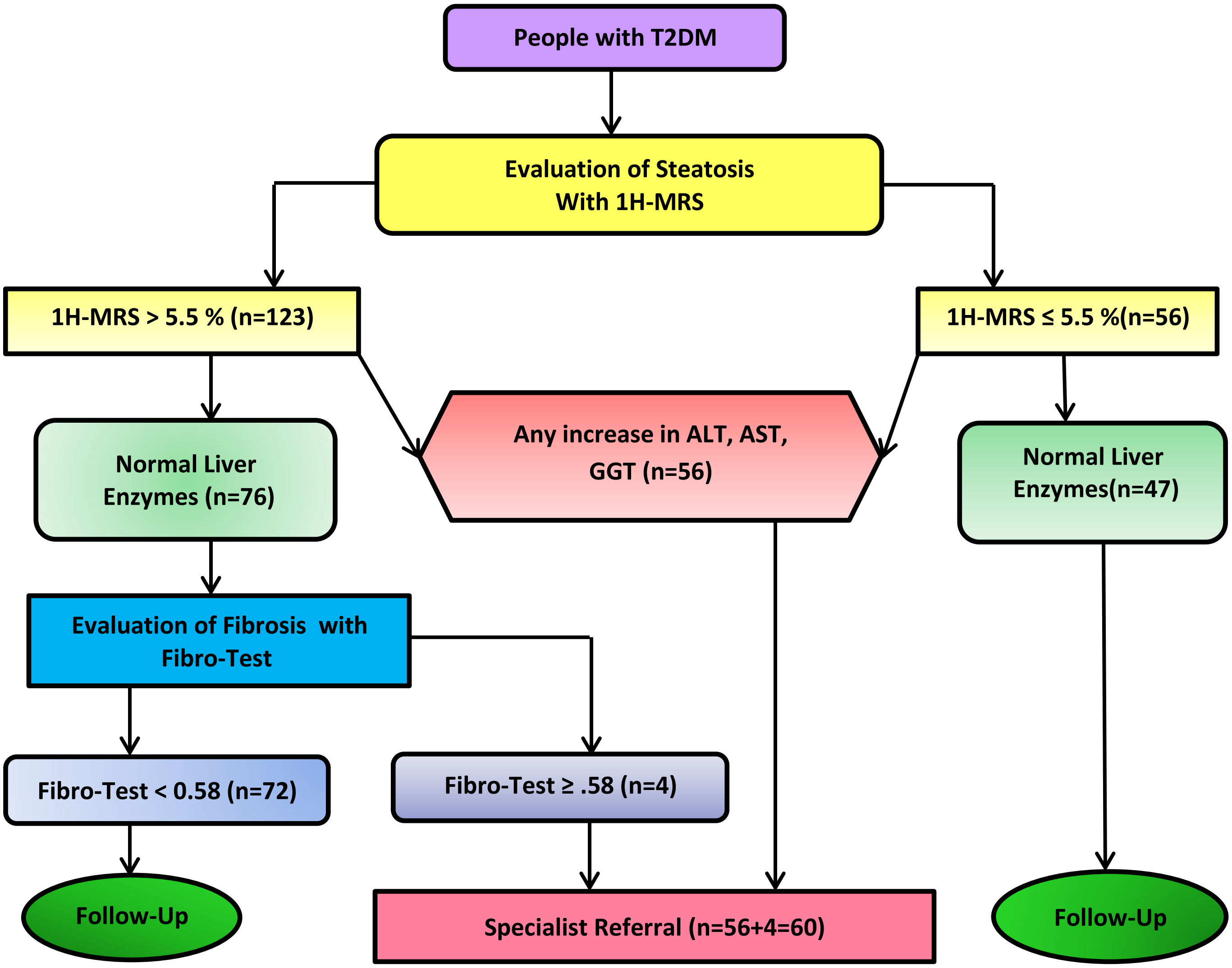

*Figure (3.11.3):EASL-EASD-EASO clinical practice guidelines for management od NAFLD in type 2 diabetic using 1H-MRS and fibrotest*

*Sberna et al. 2017 used another system. As illustrated in the above figure, 179 diabetic patients were screened for the presence of steatosis with 1H-MRS, as in the previous figure. Any patient with increased liver enzymes, whether with (steatosis absent i.e.) hepatic fat content ≤ 5.5% (n=9), or with (steatosis present i.e.) hepatic fat content > 5.5% (n=47) was referred to a liver clinic (a total of 56). Forty seven patients with hepatic fat content ≤ 5.5% and with normal liver enzymes were assigned to follow up every 3-5 years. The remaining 76 patients with hepatic fat content > 5.5% and with normal liver enzymes were evaluated for liver fibrosis with FibroTest. Of those, 72 patients with FibroTest < 0.58 were decided to be followed up every 2 years while 4 patients with FibroTest ≥ 0.58 were referred to a liver clinic. Thus the total referral rate was 33.5% (60 patients). This is the same rate when SteatoTest was initially used to detect steatosis followed by FibroTest to detect fibrosis, which was still high.*

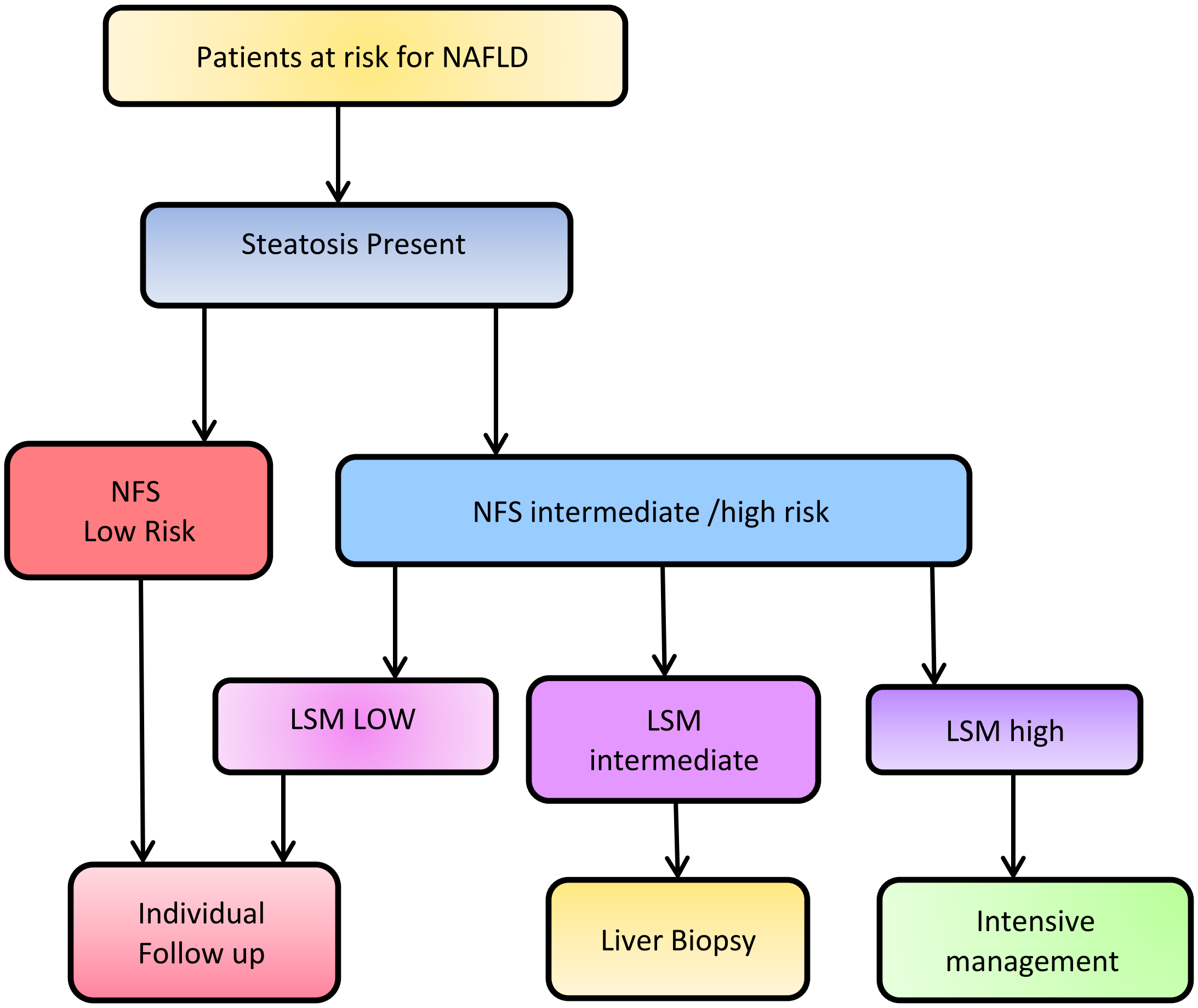

Figure (3. 12 ): German guidelines (DGVS) for identifying patients at risk for NAFLD

*Blank et al. 2020 assessed the guidelines performance of the EASL-EASD-EASO and the DGVS, which take into account the all possible variant of LSM that is lacking in the EASL guidelines, however; age adapted cutoffs should be considered to increase the performance. The study evaluated risk stratification and different management policies, to decrease the referral rates to special center, avoid cost intensive biomarkers and elastography, thus optimizing management plans. Univariate and multivariate analysis propose AST as single marker for risk stratification because it is highly elevated in patients with elevated LSM ( marker for fibrosis) and thus this would classify 75% for long term follow up and 25% are referred to the specialist with 46% sensitivity and 88% specificity. Also in this study, NFS and LSM were applied as stated in the German Guideline, with sensitivity 47-75%, while; the EASL-EASD-EASO guidelines have 30-52% specificity. When Blank et al. applied German algorithm, LSM was used to refer around 76% to referral center and when EASL-EASD-EASO was applied, the referral rate was 77%, thus the algorithms are sub-optimal. Also LSM are not used in the primary care unit. Duo to high referral rate to specialized center, Blank et al. recommended that a simpler algorithm is urgently needed.*

The next figure is:

Figure (3. 13): Japanese society of gastroenterology guidelines for diagnosis and management of NAFLD based on non-invasive tests

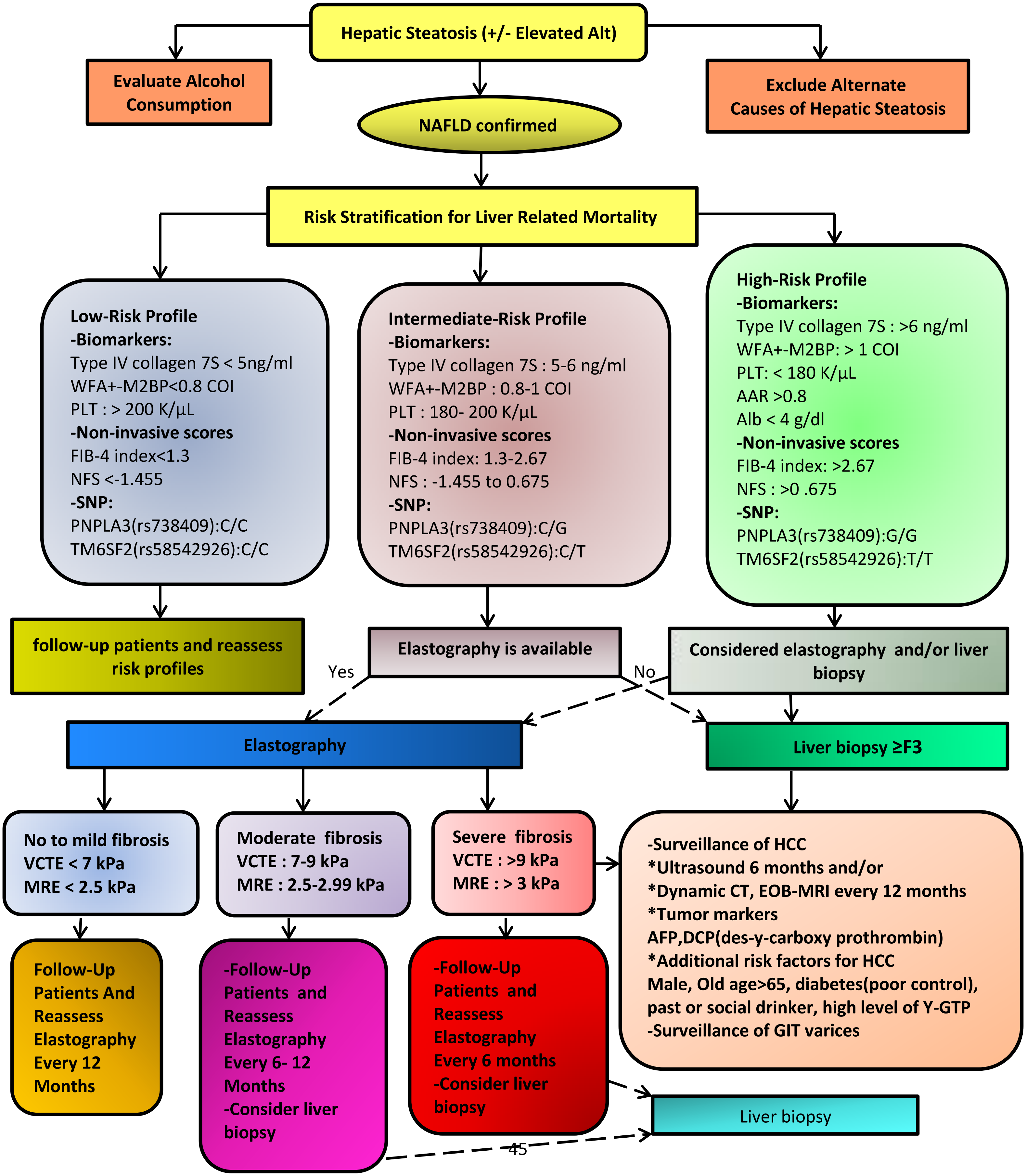
Hepatic Steatosis (+/- Elevated Alt)
Evaluate Alcohol Consumption
Exclude Alternate Causes of Hepatic Steatosis
NAFLD confirmed
Risk Stratification for Liver Related Mortality
Low-Risk Profile
-Biomarkers:
Type IV collagen 7S < 5ng/ml
WFA+-M2BP<0.8 COI
PLT : > 200 K/µL
-Non-invasive scores
FIB-4 index<1.3
NFS <-1.455
-SNP:
PNPLA3(rs738409):C/C
TM6SF2(rs58542926):C/C
Intermediate-Risk Profile
-Biomarkers:
Type IV collagen 7S : 5-6 ng/ml
WFA+-M2BP : 0.8-1 COI
PLT : 180- 200 K/µL
-Non-invasive scores
FIB-4 index: 1.3-2.67
NFS : -1.455 to 0.675
-SNP:
PNPLA3(rs738409):C/G
TM6SF2(rs58542926):C/T
High-Risk Profile
-Biomarkers:
Type IV collagen 7S : >6 ng/ml
WFA+-M2BP: > 1 COI
PLT: < 180 K/µL
AAR >0.8
Alb < 4 g/dl
-Non-invasive scores
FIB-4 index: >2.67
NFS : >0 .675
-SNP:
PNPLA3(rs738409):G/G
TM6SF2(rs58542926):T/T
follow-up patients and reassess risk profiles
Elastography is available
Yes
No
Considered elastography and/or liver biopsy
Elastography
Liver biopsy ≥F3
No to mild fibrosis
VCTE < 7 kPa
MRE < 2.5 kPa
Moderate fibrosis
VCTE : 7-9 kPa
MRE : 2.5-2.99 kPa
Severe fibrosis
VCTE : >9 kPa
MRE : > 3 kPa
-Surveillance of HCC
*Ultrasound 6 months and/or
*Dynamic CT, EOB-MRI every 12 months
*Tumor markers
AFP,DCP(des-y-carboxy prothrombin)
*Additional risk factors for HCC
Male, Old age>65, diabetes(poor control),
past or social drinker, high level of Y-GTP
-Surveillance of GIT varices
Follow-Up Patients And Reassess Elastography Every 12 Months
-Follow-Up Patients and Reassess Elastography Every 6- 12 Months
-Consider liver biopsy
-Follow-Up Patients and Reassess Elastography Every 6 months
-Consider liver biopsy
Liver biopsy

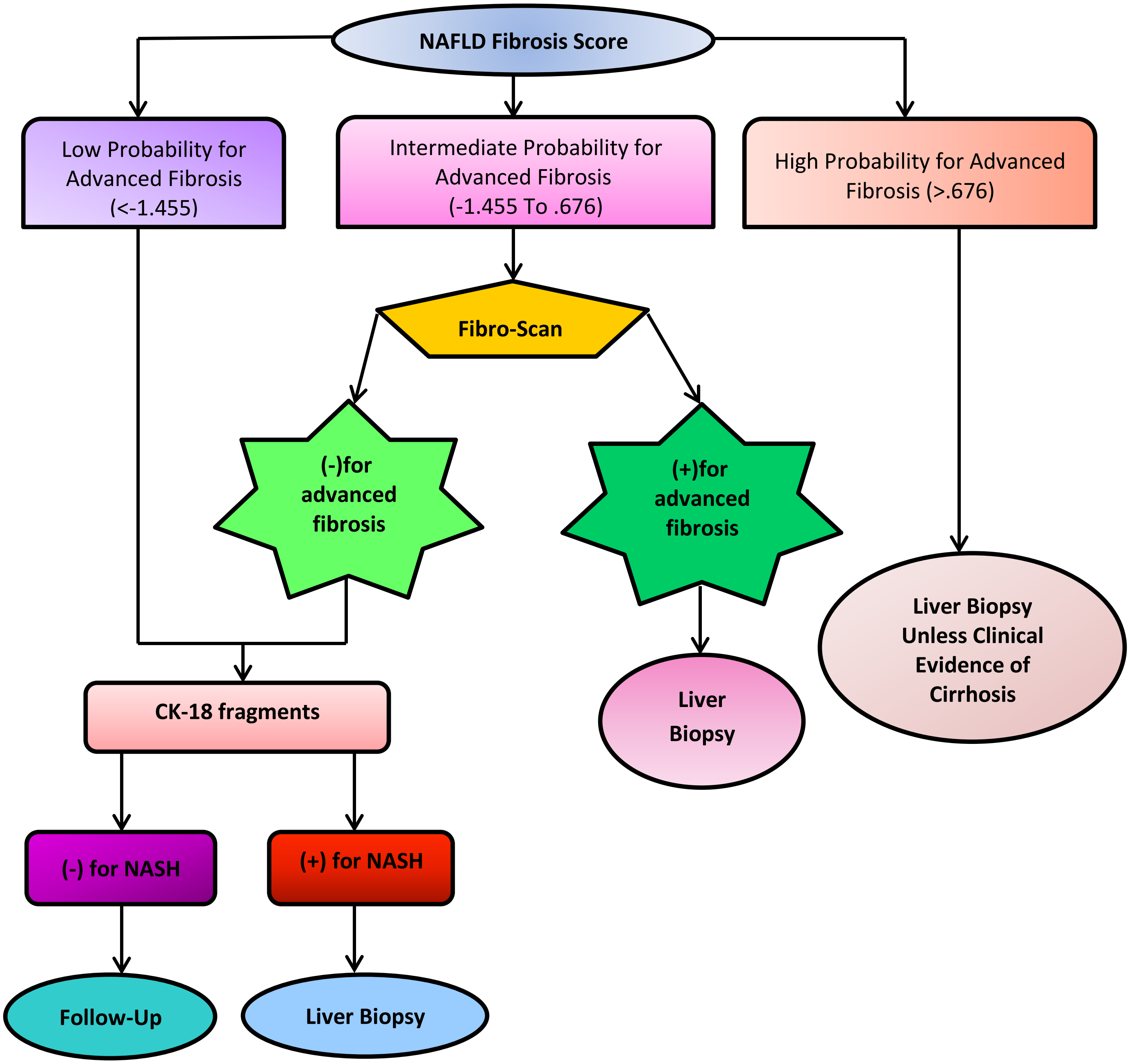

Figure (3. 14): when to do a liver biopsy?

*NFS and fibro-scan are used initially to evaluate the severity of fibrosis as it is the most significant prognostic value. Clinically if the patient shows picture of cirrhosis and no doubt for the etiology of liver disease; no further assessment is needed. On the other hand if the NFS suggests advanced fibrosis so 90% of patients will have fibrosis but cirrhosis must be excluded and a liver biopsy is highly recommended as its presence change greatly the management. But in case NFS reveals low probability for advanced fibrosis; due to the 93% NPV of the test, advanced fibrosis can be ruled out with high confidence. However NASH should be excluded and this can be attained by CK-18 fragments measurement. If it does not suggest NASH, patient can be put on the follow up schedule every 2 years. But if its value suggests NASH, liver biopsy is mandatory for confirmation as it is false positive in 14% of the patients. Unclassified patients with NFS, fibro-scan can add information for their diagnosis with the aid of liver biopsy to confirm cirrhosis in case fibro-scan suggests advanced fibrosis.*

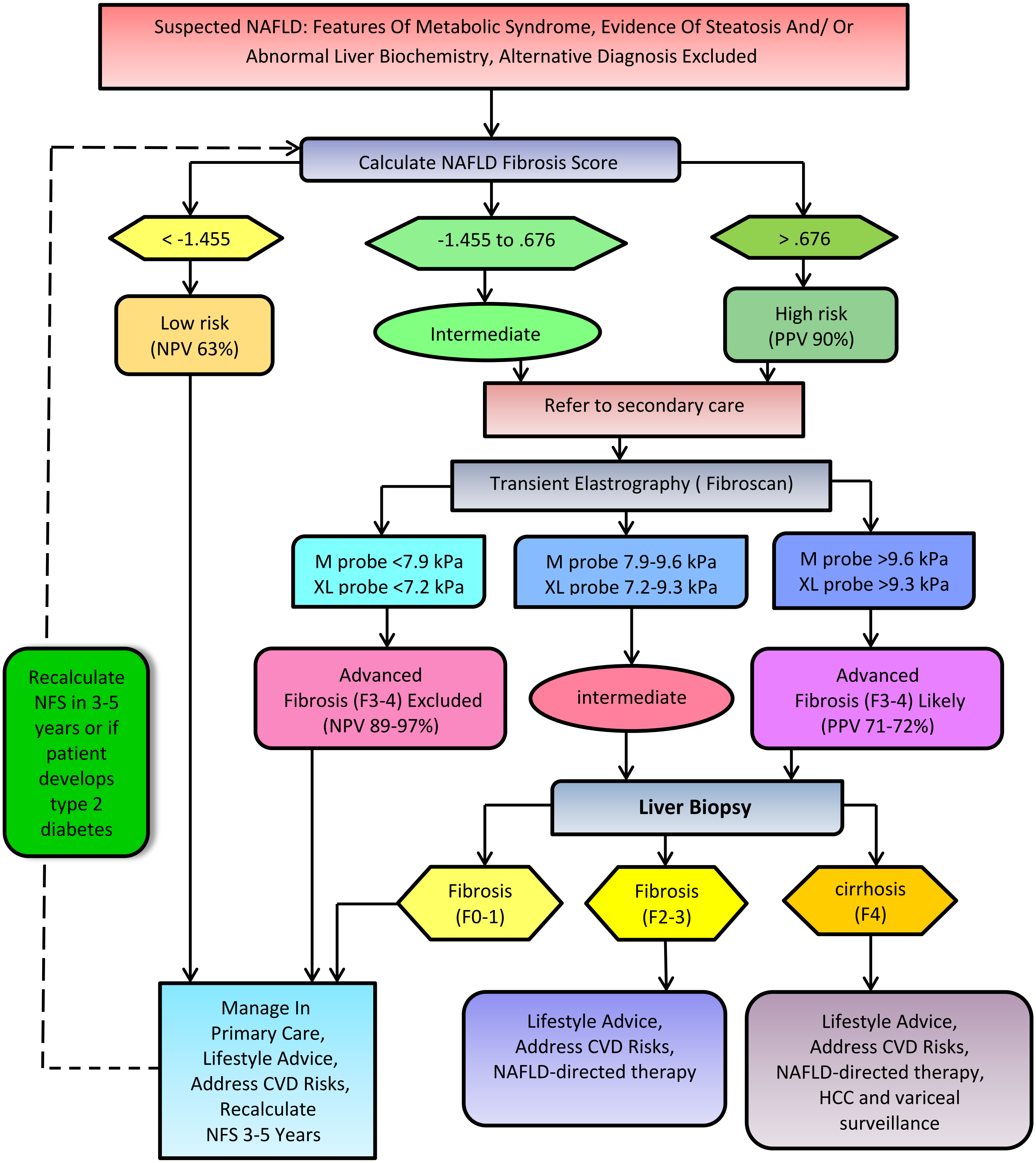

Figure (3. 15): the approach to risk stratification and management of NAFLD patients

## 3.8. Treatment.

The keystone in treatment of NAFLD must start with treatment of risk factors like obesity, dyslipidemia, diabetes and hypertension. Initially the patients must continue to change lifestyle as diet habits and avoidance of sedentary life. This can be achieved by reducing the caloric intake presented in high carbohydrate and fat diets to achieve weight loss of 5% to 10 % of total body weight which will be reflected in improvement of NAFLD histo-pathologic picture. This must be accompanied by exercise, although the best exercise program has not yet been decided; therefore large randomized control trials (RCTs) are needed to determine the type and duration of beneficial exercise. Caffeinated coffee reduces both hepatic fibrosis in NASH as well as incidence of hepatocellular carcinoma (HCC) in NAFLD. Pharmacological treatment is not yet well established. Despite the enormous progress that have been achieved to develop drugs that target different mechanisms involved in pathogenesis of NAFLD, most of them are still under trials; either phase II or phase III trials, and for the present meanwhile no drugs have been approved by the U.S. Food and Drug Administration for treatment of NAFLD (Boyer and Lindor 2016). Figure (3.16) highlights the various classes of the drugs. Figure (3.17) shows the treatment of patients according to histo-pathological findings

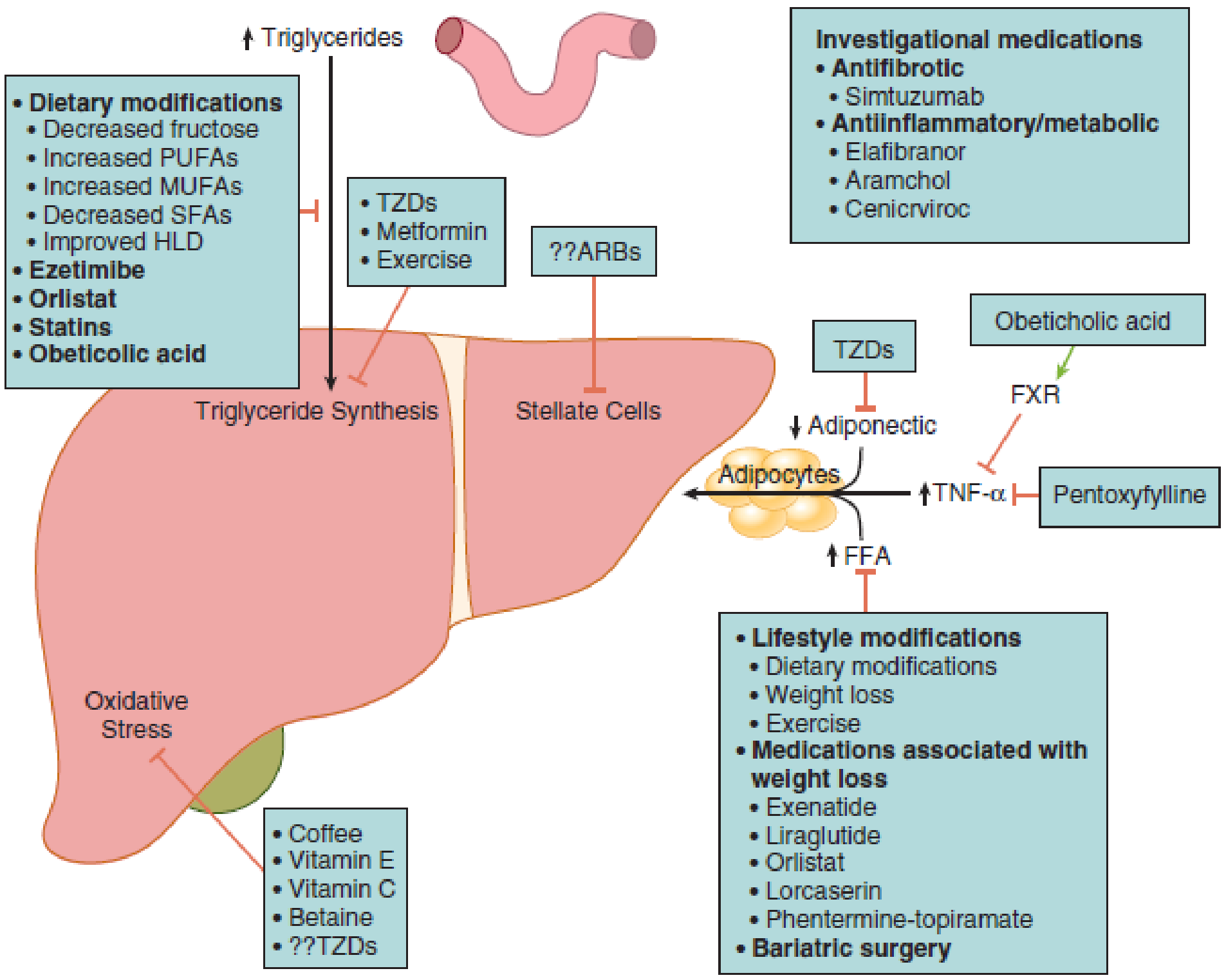

Figure (3. 16): classes of drugs according to mechanisms of action that targets a specific pathophysiological step in the disease process

**Next figure:**

Figure (3. 17): treatment according to histological lesions in liver

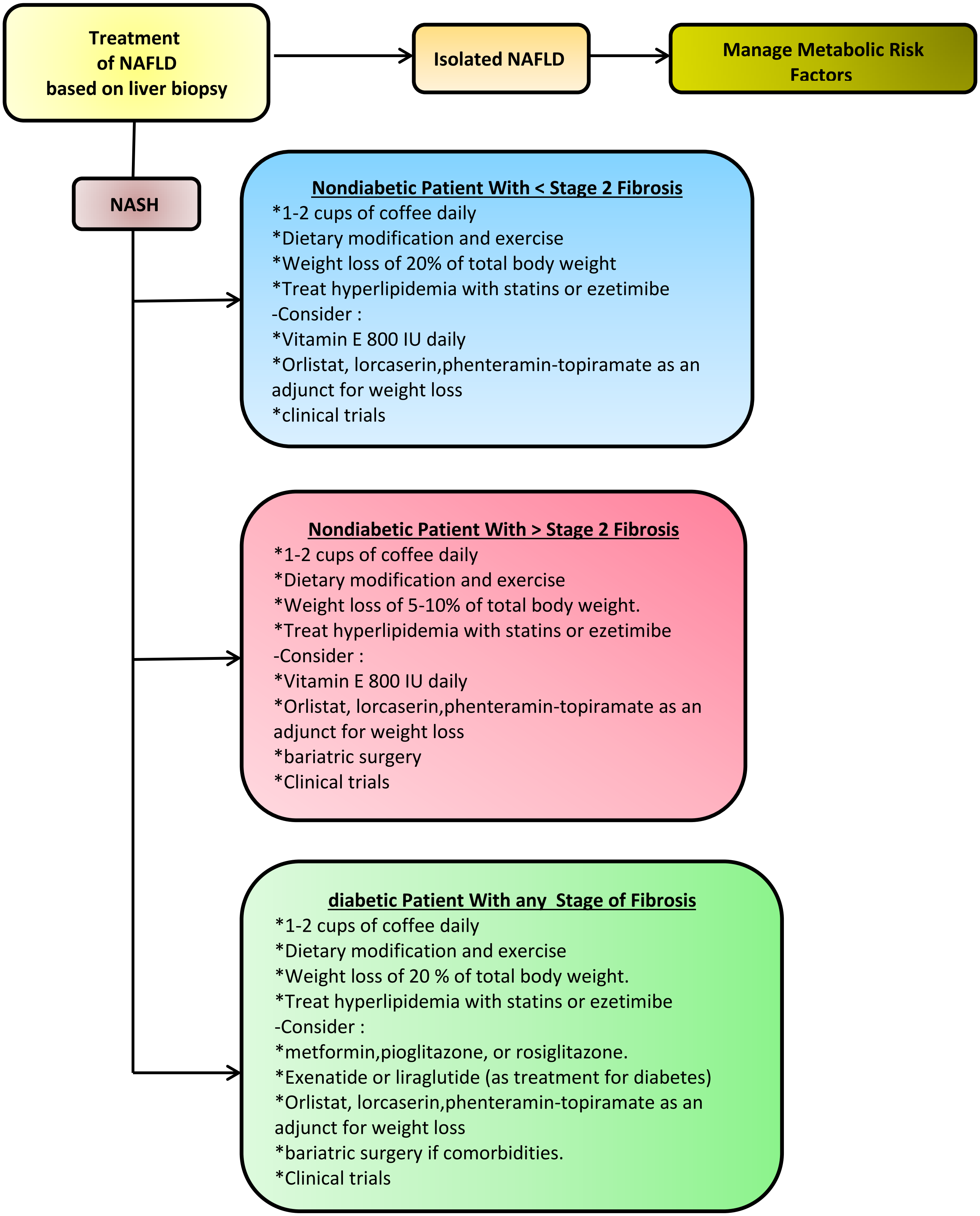
Treatment of NAFLD based on liver biopsy
Isolated NAFLD
Manage Metabolic Risk Factors
NASH
Nondiabetic Patient With < Stage 2 Fibrosis
*1-2 cups of coffee daily
*Dietary modification and exercise
*Weight loss of 20% of total body weight
*Treat hyperlipidemia with statins or ezetimibe
-Consider :
*Vitamin E 800 IU daily
*Orlistat, lorcaserin,phenteramin-topiramate as an adjunct for weight loss
*clinical trials
Nondiabetic Patient With > Stage 2 Fibrosis
*1-2 cups of coffee daily
*Dietary modification and exercise
*Weight loss of 5-10% of total body weight.
*Treat hyperlipidemia with statins or ezetimibe
-Consider :
*Vitamin E 800 IU daily
*Orlistat, lorcaserin,phenteramin-topiramate as an adjunct for weight loss
*bariatric surgery
*Clinical trials
diabetic Patient With any Stage of Fibrosis
*1-2 cups of coffee daily
*Dietary modification and exercise
*Weight loss of 20 % of total body weight.
*Treat hyperlipidemia with statins or ezetimibe
-Consider :
*metformin,pioglitazone, or rosiglitazone.
*Exenatide or liraglutide (as treatment for diabetes)
*Orlistat, lorcaserin,phenteramin-topiramate as an adjunct for weight loss
*bariatric surgery if comorbidities.
*Clinical trials

Here are some of the classes of the drug that are still under investigations (Neuschwander-Tetri 2017).

1. Improve insulin resistance: Thiazolidinediones; like pioglitazone.
2. Reduce de novo lipogenesis: inhibitors of ACC1 and SCD as well as FXR ligands (obeticholic acid)
3. Increase the metabolism: PPARα ligands, mitochondrial uncouplers, thyromemitics.
4. Reduce the injury: antioxidant, reduce ER stress
5. Reduce cell death: anti-apoptotic
6. Reduce inflammation: pentoxifyllin, CCR2/CCR5 inhibitors, TLR-4 inhibitors.
7. Reduce fibrosis

Bariatric surgery is reserved to morbid obesity and once the patient develops liver cirrhosis, so he is candidate for liver transplantation.

For the meantime large numbers of medications are under trials, but once they are released in the markets, several challenges will be met. Some of them are: what are the drugs that should be initially chosen to start treatment with, how to switch between drugs, which drugs to be added or to be removed, what is the duration of treatment, what are the tools to assess response to treatment, what is the target end point in the treatment plan and how to maintain this target. All these questions need to be answered, and they are the challenges facing the physicians in the near future.

## 3.9. Future Challenges:

The future challenges can be grouped into five categories:

### 3.9.1. First Challenge: large number of patients with advanced fibrosis due to NASH is underestimated.

This is due to imperfect plan for whom to screen for NASH and how to apply this in high-risk individuals due to scant cost-effectiveness studies. Most guidelines relies on measurements of ALT and ultrasound to detect NAFLD and NASH, but the definition of raised liver enzymes especially ALT must emphasize that any rise in ALT is considered abnormal (Prati et al. 2002) , its value does not have to be 2 or 3 times above the normal upper limit to consider it significant for the presence of NASH (Mofrad et al. 2003). The limitation of ultrasound is its low sensitivity to detect the presence of NAFLD if the percentage of hepatocytes that is loaded with fat is less than 20% (Dasarathy et al. 2009). Also it is not specific; as it mainly relies on raised echogenicity of the liver to diagnose NAFLD, which may be caused by fibrosis or other infiltrative processes. The combination of both liver enzymes and ultrasound as screening methodology may be beneficial to detect NAFLD in high risk population (Portillo-Sanchez et al. 2015). Screening these individuals for the presence of advanced fibrosis as a prognostic indictor of disease progression to cirrhosis with the subsequent liver related mortality must be of great concern to primary care physician. This can also be attained with the use of "Vibration-Controlled Transient Elastography" (VCTE) or "Magnetic Resonance Elastography" (MRE), but the high expense and limited availability in primary care settings make their use infeasible (Chalasani et al. 2018). Instead NFS and FIB-4 scoring systems make it easier to identify such patients and subsequently referring them to a specialist (Angulo et al. 2007). New noninvasive scoring systems are emerged but the lack of external validation on different ethnicities warrant urgent studies to be conducted, furthermore; cost-effectiveness researches are mandatory (Alkhouri, Lawitz, and Noureddin 2019).

### 3.9.2. Second Challenge: Who should have liver biopsy before treatment?

Due to emergence of new drugs such as the anti-fibrotic drugs, the need to determine the presence of advanced fibrosis is urgent. Liver biopsy is invasive procedure with complications of bleeding, pneumothorax and even death. Thus, new noninvasive scoring systems are continuously developed to assess presence of F3 and F4 fibrosis before starting treatment. Longitudinal studies, that determine the level of combined biological biomarkers like ELF and radiological marker such as LSM before approval and release of some drugs in the market, are needed to appraise whether such biomarkers and algorithms reflect the stage of fibrosis prior to initiating the treatment (Wong et al. 2018). And so the biopsy is only reserved for patients with discordant or intermediate levels (Newsome et al. 2020).

#### 3.9.3. Third Challenge: how to determine the treatment response noninvasively?

Anti-fibrotic drugs such as obeticholic acid and selonsertib are in their final stages in the phase trials and the target of their usage is the amelioration of liver fibrosis by one stage seen on liver histology obtained by liver biopsy after one year of treatment to decide whether to continue with the drug or not. According to data obtained from phase II trials; approximately 45% of patients exhibited a response rate evaluated by one stage improvement per year and so they will need repeated liver biopsy every year to assess this response rate which is unworkable. Tests like FIB-4 and NFS are useful for case detection but are less likely to be fruitful for treatment response follow up. Tests such as ELF and VCTE have acceptable accuracy in detecting presence of NASH and staging the fibrosis, but further longitudinal studies are needed to consider the noninvasive evaluation of treatment response if the drug confirms to be beneficial. If for example combination of biomarker and imaging tests such as ELF and VCTE indicate improvement after one year so the drug should be continued otherwise it could be discontinued. If the results of the tests are discordant then liver biopsy might be helpful (Alkhouri et al. 2019). Some already existing noninvasive tests need further external validation and to be evaluated on different ethnicities.

#### 3.9.4. Fourth Challenge: How to modify the treatment according to the initial response and the follow up investigations?

After one year improvement in stage of fibrosis, would it be preferable to get further improvement or continuation of the achieved improvement is the target goal of treatment? What should be the target goal of treatment and if it is reachable, the up-coming question is how to maintain it? When to say there is no response to treatment and so the drug is fruitless and when to say it shows partial response and so the drug is partially effective? If the initial drug has not shown the desired improvement, should another drug be added or switch to another drug, and in either cases what is this drug? After reaching the desired level of improvement; should the patient continue for life on this drug or should he stop? All these questions are not answered up to this point. Another question that should be asked is whether the anti-fibrotic treatment is reserved only for patients with advanced fibrosis (F3-F4) excluding the patient with mild and moderate fibrosis (F1-F2). Some suggestions were proposed; as the fibrosis is a score system on a 5 point scale (F0-F4) then 20% reduction of LSM from the baseline recorded by VCTE or reduction to less than the cut-off for F3 could be considered an appropriate response after one year of treatment. As an example for biological marker; a 2-point reduction in ELF score can be delineated as a good response for treatment. Evaluating treatment response with noninvasive tests (such as VCTE and ELT) should be included within phase III trials and issued prior to a drug's approval (Alkhouri et al. 2019).

#### 3.9.5. Fifth Challenge: Should the first drug to start with be individualized according to baseline predictors?

Evaluating baseline predictors like baseline serum bile acids and C4 (complement 4) as well as quantifying baseline cytokeratin 18 fragments (a marker of hepatocyte apoptosis) that may reflect the underlying pathophysiologic mechanism for the disease progression will help the physician to determine which is the most probable beneficial anti-fibrotic drug to start with. Also the side effects and the associated comorbidities will have a lot of impact to choose such a drug. Further longitudinal prospective studies are needed to validate these concepts (Konerman, Jones, and Harrison 2018).

Figure (3.18) demonstrates a proposed algorithm for primary care physicians to screen and identify patients with NAFLD and advanced fibrosis, while figure (3.19) depicts the proposed algorithm deciding treatment response to fibrotic NASH medications.

Alkhouri, Lawitz, and Noureddin( 2019)proposed a NASH management policy or plan:

1. Garbling for NAFLD in high-risk populations is highly advisable.
2. Patients at high risk for advanced fibrosis are identified by combined implementation of biological biomarkers and physical imaging tests while the liver biopsy is preserved for discordant results.
3. If advanced fibrosis is detected, the first drug to be initially used is likely to pose anti-fibrotic properties like obeticholic acid (OCA) or selonsertib (SEL).
4. Reassess the treatment response after 1 year and then on a yearly basis. If noninvasive tests (NIVTs) demonstrate sufficient response, then continue on this single agent.
5. If the NIVTs show no adequate response, so stop and shift to another drug.
6. If the response is partial, so adding another drug should be taken into account.

For current meantime no approved drug treatment for NAFLD/NASH is available; however, abundant drugs are being scrutinized in phase II and phase III clinical trials. The outcome from these trails has been favorable and hopeful as regard amelioration in histological features of the disease like: steatosis, inflammation and fibrosis. As NAFLD is a heterogeneous complex disease process with a chronic prolonged course evolving over time, examining the effectiveness of these drugs on the long run as well as their safety and effectiveness is crucial, especially when taken into account that some of these drugs have serious and possible metabolic side effects. Heading the substantial and remarkable impediments for registration (by increasing the awareness and alertness of the seriousness of the disease by both physicians and patients) into clinical trials is of paramount, as well as improving the design of these trials will play a tremendous role for achieving success in these trials, and hence approval of such therapeutic drugs. This can be accomplished by utilizing the non-invasive markers of steatosis, inflammation, and fibrosis; in addition to, the clinical trial designs. In FLINT study, conducted on 283 non-cirrhotic NASH patients taking 25 mg daily obeticholic acid; the improvement in histology detected by NAS was 2 points or more with no deterioration of fibrosis and 35% of patients taking OCA had a decrement in fibrosis score by at least one stage in comparison with 19 % in the placebo arm.

REGENERATE study (still in progress with estimated primary completion date on September 2025 as shown in clinicaltrials.gov official site) will evaluate safety and efficacy of Obeticholic Acid(OCA) in NASH patients with fibrosis whom are randomizes to a daily dose of 25 mg, 10 mg, and placebo, with end points like: amelioration of fibrosis by at least one stage and decaying of NASH with no deterioration of fibrosis. At 18 month of randomization, liver biopsy revealed statistically significant histological amelioration of fibrosis and decaying of NASH with no deterioration in fibrosis for both 10 mg and 25 mg doses. In GOLDEN study, conducted on 274 NASH patients, 120 mg Elafibranor taken daily for 52 weeks induced decaying of moderate to severe NASH in meaningfully higher percentage of patients in comparison to placebo, moreover; these patients also showed lowering in fibrosis stage compared to non-resolving NASH patients. RESOLVE-IT trial (last update was on November 30, 2020, as shown in clinicalrials.gov official site, but the study is still in progress according to(Guirguis et al. 2020)) emerged in May 2020 had shown that 19.2% of patients, on 120 mg daily Elafibranor, had NASH decay without deterioration of fibrosis compared to 14.7% in the placebo group, which was not statistically significant. Furthermore, 24.5% of patients had shown fibrosis amelioration of more than one stage compared to 22.4% in the placebo group, which was also not statistically significant.

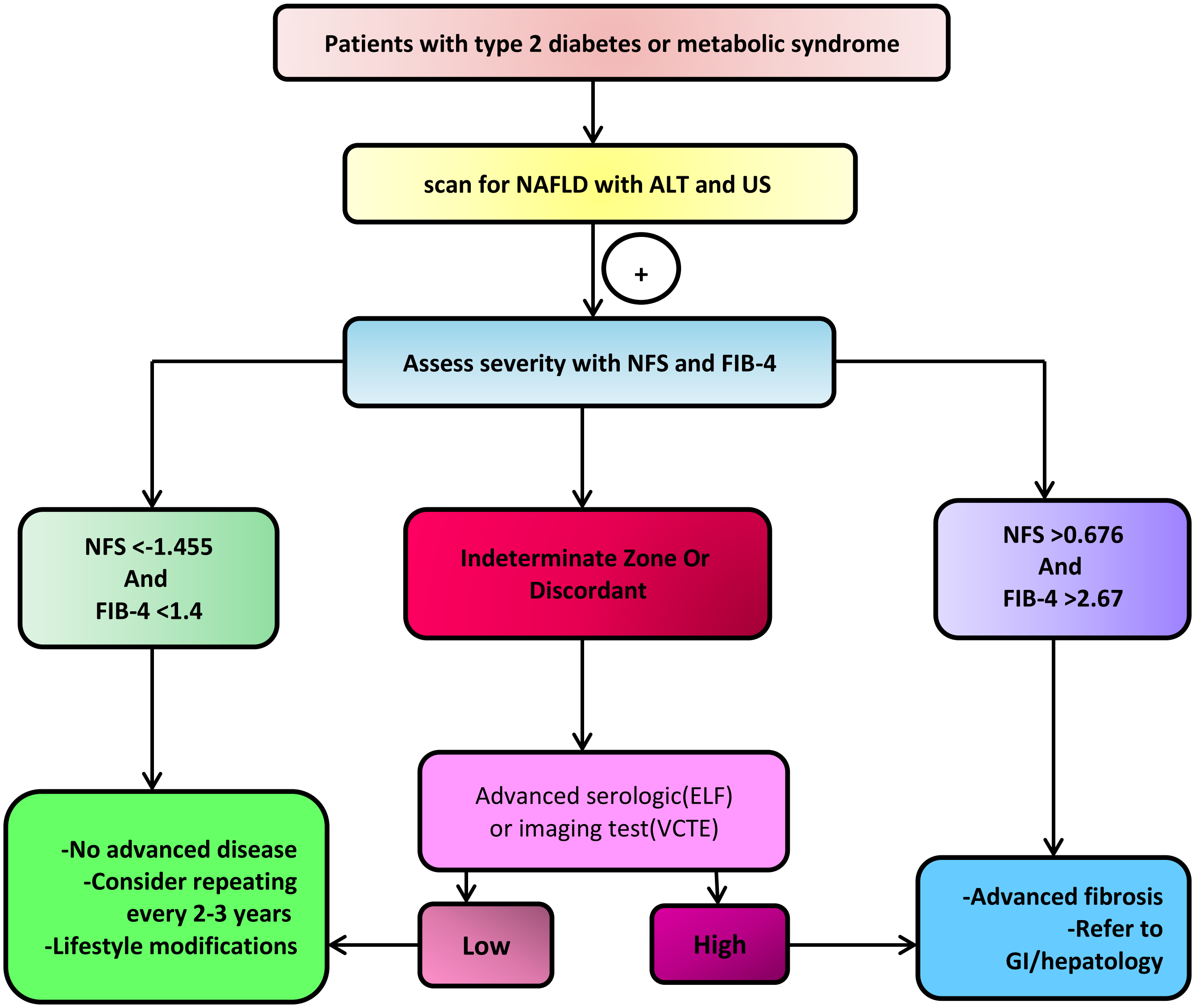

Figure (3. 18): algorithm for primary care physician to detect and identify patients with NAFLD and advanced fibrosis

In CENTAUR trial, conducted over 289 patients taking cenicriviroc (CVC), 150 mg daily and placebo for 52 weeks, no comparative betterment in NAS between NASH group and placebo was seen, however; there was one stage or more amelioration of fibrosis with no deterioration of NASH in the group taking the CVC compared with placebo group. The AURORA trial (primary completion dates are October 2021 according to clinicaltrails.gov site and October 2028 according to (Guirguis et al. 2020)) will evaluate long-term safety and efficacy of 150 mg daily CVC for the treatment of fibrosis in NASH adult patients at 2 phases, the first has endpoint of at least one stage amelioration of fibrosis without deterioration of NASH at month 12, and phase 2 has end point that is cirrhosis, liver-related outcome as HCC, and all causes of mortality. In a small, open-label, randomized phase II trial including 72 biopsy-proven NASH patients (NAS ≥ 5 and stage 2-3 liver fibrosis) receiving 18 mg daily Selonsertib for 24 weeks ,there was significant improvement in liver disease activity, fibrosis, stiffness, liver fat content, and progression to cirrhosis (Alkhouri, Poordad, and Lawitz 2018).

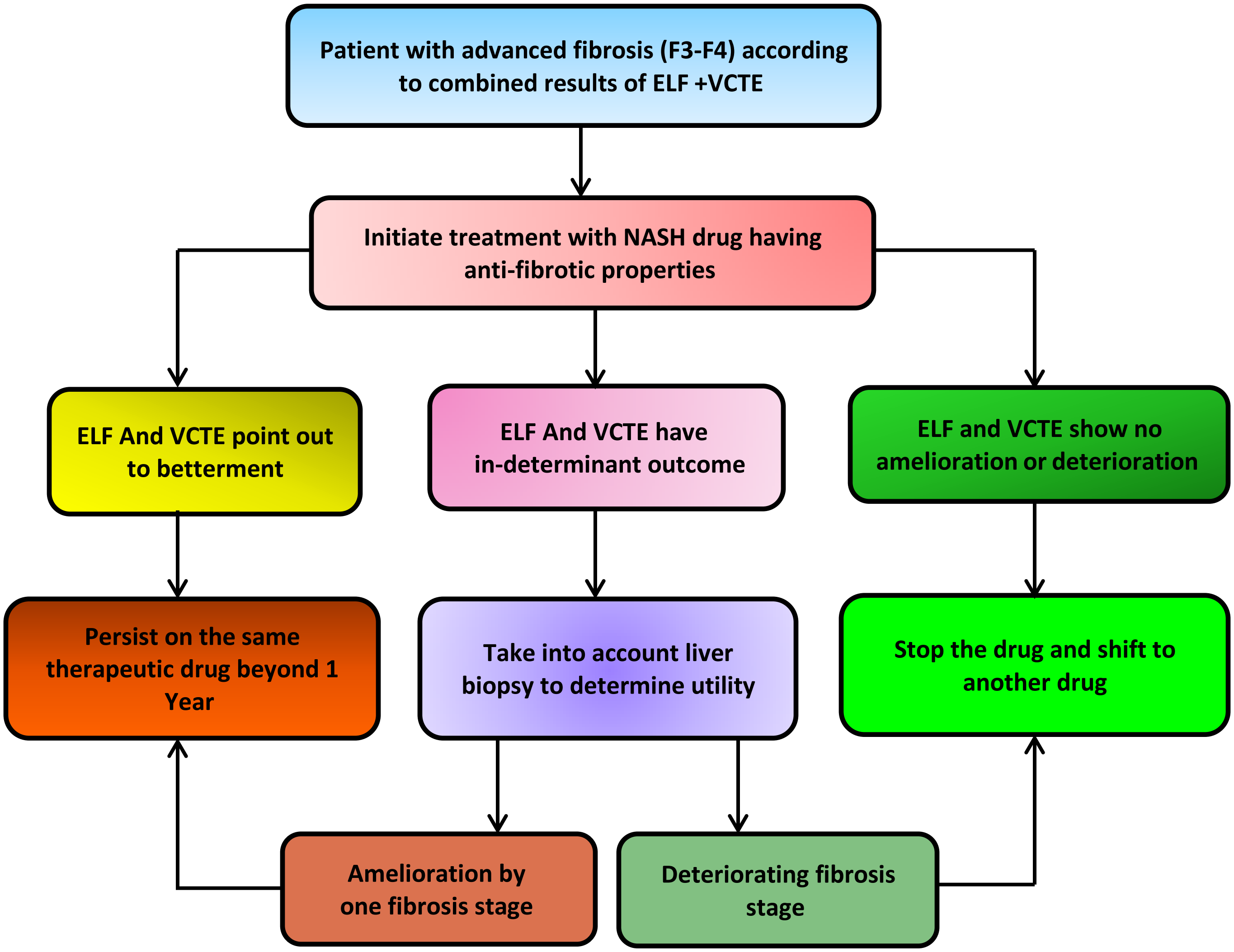

Figure (3. 19 ): plan of treatment according to response to fibrotic NASH medications

FLINT, GOLDEN, and CENTAUR are phase IIb placebo-controlled RCT( randomized control trial), while REGENERATE, RESOLVE-IT, and AURORA are randomized, placebo-controlled, double-blinded, multicenter phase III trials

# Chapter Four: CTMC Analyzing NAFLD Progression (Small Model)

Studying natural history of disease during which individuals start at one initial state then as time passes the patients move from one state to another, can be investigated by using multistate Markov chains. Evolution of the disease over different phases can be monitored by taking repeated observations of the disease stage at pre-specified time points following entry into the study. Disease stage is recorded at time of observation while the exact time of state change is unobserved. NAFLD is a multistage disease process; in its simplest form has a general structure model as depicted in Figure (4.1).

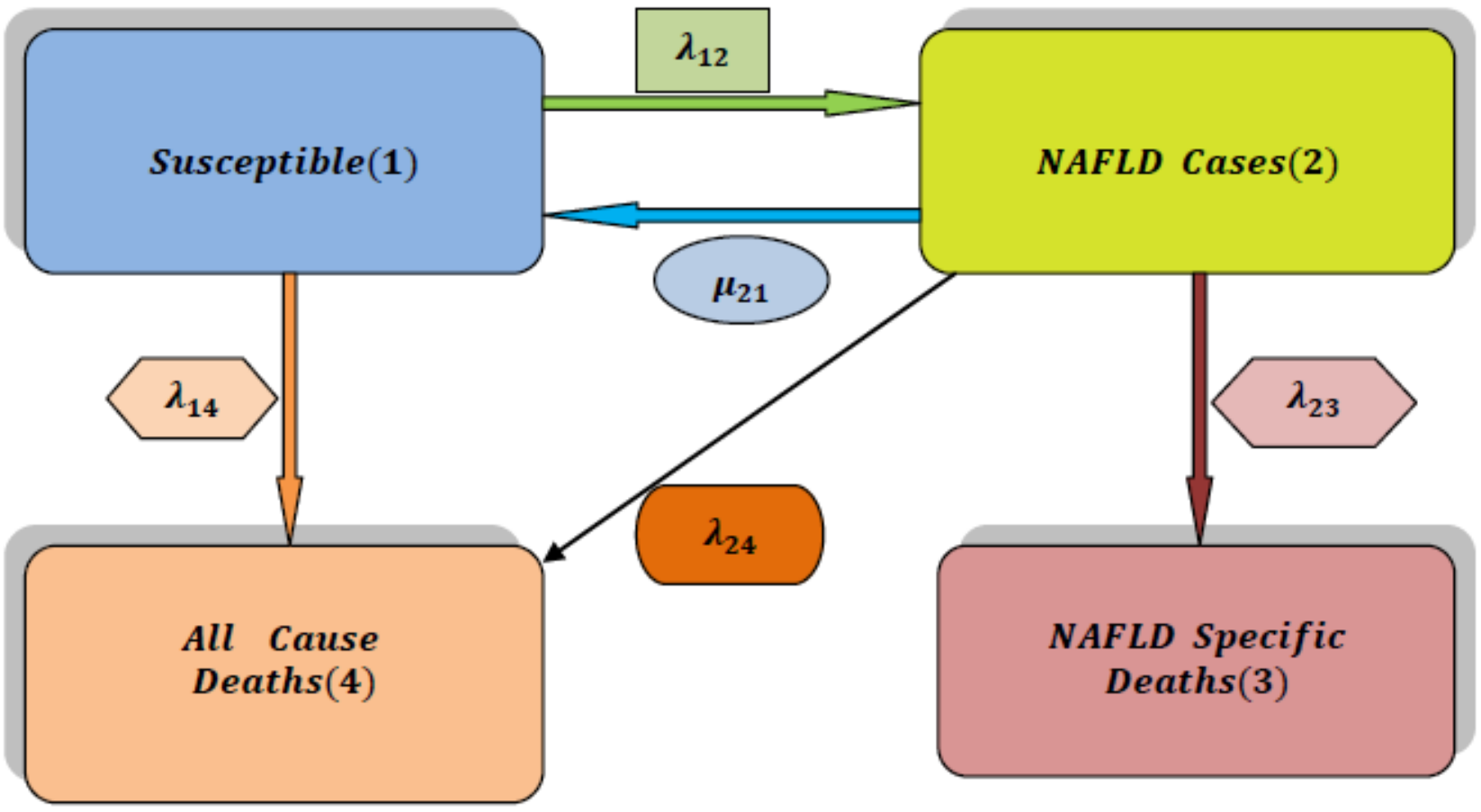

Figure (4. 1): general model structure .(Younossi et al. 2016)

NAFLD stages are modeled as time homogenous CTMC , that is to mean $P_{ij}(\Delta t)$ depends on $\Delta t$ and not on $t$ ,with constant transition intensities $\lambda_{ij}$ over time, exponentially distributed time spent within each state and patients' events follow Poisson distribution. The states are: one for the susceptible cases (state 1) and one for NAFLD cases (state 2) and two absorbing states ; one for the death due to NAFLD (state 3) and one for death due to any other cause (state 4). The transition rate $\lambda_{12}$ is the rate of progression from state 1 to state 2, while the transition rate $\mu_{12}$ is the regression rate from state 2 to state 1. The transition rate $\lambda_{23}$ is the progression rate from state 2 to state 3 and $\lambda_{24}$ is the rate of progression from state 2 to state 4. For simplicity, all individuals are assumed to enter the disease process at stage one and they are all followed up with the same length of time interval between measurements.

In this chapter the transition probabilities and transition rates are thoroughly discussed. Also mean sojourn time and its variance are reviewed as well as state probability distribution and its covariance matrix. This is followed by exploration of the life expectancy of the patients and the expected numbers of patients in each state. Lastly a hypothetical numerical example is used to illustrate these concepts.

### 4.1. Transition Probability Matrix

The transitions can occur at any point in time. The rates at which these transitions occur are constant over time and thus are independent of t that is to say the transition of patient from $state\ i\ at\ time = t\ \ to\ state\ j\ \ at\ \ t = t\ + s\ where\ s = \Delta t$ depends on difference between two consecutive time points. And it's defined as $\theta_{ij}\,(t) = \lim_{\Delta t\to 0}\frac{P_{ij}(\Delta t)-I}{\Delta t}$ or the Q matrix.

For the above multistate Markov model demonstrating the NAFLD disease process; the forward Kolomogrov differential equations are the following:

$$\frac{d}{dt}P_{ij}(t) = \begin{bmatrix} P_{11} & P_{12} & P_{13} & P_{14} \\ P_{21} & P_{22} & P_{23} & P_{24} \\ 0 & 0 & P_{33} & 0 \\ 0 & 0 & 0 & P_{44} \end{bmatrix} \begin{bmatrix} -(\lambda_{12}+\lambda_{14}) & \lambda_{12} & 0 & \lambda_{14} \\ \mu_{21} & -(\mu_{21}+\lambda_{23}+\lambda_{24}) & \lambda_{23} & \lambda_{24} \\ 0 & 0 & 0 & 0 \\ 0 & 0 & 0 & 0 \end{bmatrix}$$

The Kolmogrove differential equations:

$$\frac{dP_{11}}{dt} = -P_{11}(\lambda_{12}+\lambda_{14}) + P_{12}\mu_{21}\ , \frac{dP_{12}}{dt} = P_{11}\lambda_{12} - P_{12}(\mu_{21}+\lambda_{23}+\lambda_{24}), \frac{dP_{13}}{dt} = P_{12}\lambda_{23}, \frac{dP_{14}}{dt} = P_{11}\lambda_{14} + P_{12}\lambda_{24}$$

$$\frac{dP_{21}}{dt} = -P_{21}(\lambda_{12}+\lambda_{14}) + P_{22}\mu_{21}\ , \frac{dP_{22}}{dt} = P_{21}\lambda_{12} - P_{22}(\mu_{21}+\lambda_{23}+\lambda_{24}), \frac{dP_{23}}{dt} = P_{22}\lambda_{23}, \frac{dP_{24}}{dt} = P_{21}\lambda_{14} + P_{22}\lambda_{24}$$

$P_{33} = 1\ , P_{44} = 1$

The solution of this system of equations will give the $P_{ij}(t)$

To solve the set of probabilities in the first row:

The first 2 equations are:

$$\frac{dP_{11}(t)}{dt} = -P_{11}(\lambda_{12}+\lambda_{14}) + P_{12}\mu_{21} \qquad , \qquad \frac{dP_{12}(t)}{dt} = P_{11}\lambda_{12} - P_{12}(\mu_{21}+\lambda_{23}+\lambda_{24})$$

$let \quad \lambda_{12}+\lambda_{14} = \gamma_1\ , \mu_{21}+\lambda_{23}+\lambda_{24} = \gamma_2$

*To get* $P_{11}$

$DP_{11} + \gamma_1 P_{11} - \mu_{21}P_{12} = 0 \qquad (1) \rightarrow \ (D+\gamma_1)P_{11} - \mu_{21}P_{12} = 0 \qquad (3)$

$DP_{12} + \gamma_2 P_{12} - \lambda_{12}P_{11} = 0 \qquad (2) \rightarrow \ -\lambda_{12}P_{11} + (D+\gamma_2)P_{12} = 0 \qquad (4)$

*Multiply equation* $(3)$ *by* $(D+\gamma_2)$ *and multiply equation* $(4)$ *by* $\mu_{21}$

$(D+\gamma_1)(D+\gamma_2)P_{11} - (D+\gamma_2)\mu_{21}P_{12} = 0 \quad (5)$

$-\lambda_{12}\,\mu_{21}\ P_{11} + (D+\gamma_2)\,\mu_{21}\,P_{12} = 0 \qquad (6)$

Add the above equations:

$[(D+\gamma_1)(D+\gamma_2) - \lambda_{12}\mu_{21}]\ P_{11} = 0 \qquad \& \qquad [D^2 + (\gamma_1+\gamma_2)D + \gamma_1\gamma_2 - \lambda_{12}\mu_{21}]\,P_{11} = 0$

$$\therefore\ w_1 = \frac{-(\gamma_1+\gamma_2) - \sqrt{(\gamma_1+\gamma_2)^2 - 4\,\gamma_1\gamma_2 + 4\lambda_{12}\mu_{21}}}{2} \quad \& \quad w_2 = \frac{-(\gamma_1+\gamma_2) + \sqrt{(\gamma_1+\gamma_2)^2 - 4\,\gamma_1\gamma_2 + 4\lambda_{12}\mu_{21}}}{2}$$

$(\gamma_1+\gamma_2)^2 - 4\,\gamma_1\gamma_2 + 4\lambda_{12}\mu_{21} > 0$

$P_{11} = c_1 e^{w_1 t} + c_2 e^{w_2 t}$

*To get* $P_{12}$

$(D+\gamma_1)P_{11} - \mu_{21}P_{12} = 0 \qquad (3)$

$-\lambda_{12}P_{11} + (D+\gamma_2)P_{12} = 0 \qquad (4)$

*Multiply equation* $(3)$ *by* $\lambda_{12}$ *and multiply equation* $(4)$ *by* $(D+\gamma_1)$

$(D+\gamma_1)\,\lambda_{12}\,P_{11} - \lambda_{12}\,\mu_{21}P_{12} = 0 \quad (5)$

$-(D+\gamma_1)\ \lambda_{12}\,P_{11} + (D+\gamma_1)(D+\gamma_2)\ P_{12} = 0 \qquad (6)$

Add the above equations :

$[(D+\gamma_1)(D+\gamma_2) - \lambda_{12}\mu_{21}]\ P_{12} = 0 \ \ \& \ \ [D^2 + (\gamma_1+\gamma_2)D + \gamma_1\gamma_2 - \lambda_{12}\mu_{21}]\,P_{12} = 0$

$$w_1 = \frac{-(\gamma_1+\gamma_2) - \sqrt{(\gamma_1+\gamma_2)^2 - 4\,\gamma_1\gamma_2 + 4\lambda_{12}\mu_{21}}}{2} \quad \& \quad w_2 = \frac{-(\gamma_1+\gamma_2) + \sqrt{(\gamma_1+\gamma_2)^2 - 4\,\gamma_1\gamma_2 + 4\lambda_{12}\mu_{21}}}{2}$$

$(\gamma_1+\gamma_2)^2 - 4\,\gamma_1\gamma_2 + 4\lambda_{12}\mu_{21} > 0$

$P_{12} = c_3 e^{w_1 t} + c_4 e^{w_2 t}$

*Substitute in*: $DP_{11} + \gamma_1 P_{11} - \mu_{21}P_{12} = 0$ *to solve for constants*

$c_1\,w_1 e^{w_1 t} + c_2\,w_2 e^{w_2 t} + c_1\gamma_1\ e^{w_1 t} + c_2\ \gamma_1 e^{w_2 t} - \mu_{21}c_3 e^{w_1 t} - \mu_{21}\,c_4 e^{w_2 t} = 0$

$(c_1\,w_1 + c_1\gamma_1 - \mu_{21}c_3)\ e^{w_1 t} + (c_2\,w_2 + c_2\ \gamma_1 - \mu_{21}\,c_4)e^{w_2 t} = 0$

| | |
|---|---|
| $c_1\,\mathrm{w}_1 + c_1\gamma_1 - \mu_{21}c_3 = 0$<br>$\mu_{21}c_3 = c_1\,\mathrm{w}_1 + c_1\gamma_1$<br>$c_3 = \frac{c_1}{\mu_{21}}(\mathrm{w}_1 + \gamma_1)$ | $\mathrm{c}_2\,\mathrm{w}_2 + \mathrm{c}_2\,\gamma_1 - \mu_{21}\,\mathrm{c}_4 = 0$<br>$\mu_{21}\,\mathrm{c}_4 = \mathrm{c}_2\,\mathrm{w}_2 + \mathrm{c}_2\,\gamma_1$<br>$\mathrm{c}_4 = \frac{\mathrm{c}_2}{\mu_{21}}(\mathrm{w}_2 + \gamma_1)$ |

$P_{11} = c_1 e^{w_1 t} + c_2 e^{w_2 t}$, $P_{12} = c_3 e^{w_1 t} + c_4 e^{w_2 t} = \frac{c_1}{\mu_{21}}(\mathrm{w}_1 + \gamma_1)\, e^{w_1 t} + \frac{\mathrm{c}_2}{\mu_{21}}(\mathrm{w}_2 + \gamma_1)\, e^{w_2 t}$

Using initial values at :

$P_{11}(0) = 1 \rightarrow c_1 + \mathrm{c}_2 = 1 \rightarrow c_1 = 1 - \mathrm{c}_2$

$P_{12}(0) = 0 \rightarrow \frac{c_1}{\mu_{21}}(\mathrm{w}_1 + \gamma_1) + \frac{\mathrm{c}_2}{\mu_{21}}(\mathrm{w}_2 + \gamma_1) = 0$

$\therefore\; c_2 = \left(\frac{\mathrm{w}_1 + \gamma_1}{\mathrm{w}_1 - \mathrm{w}_2}\right)$ and $c_1 = (1 - c_2) = 1 - \frac{(\gamma_1 + \mathrm{w}_1)}{(\mathrm{w}_1 - \mathrm{w}_2)} = \left(\frac{\mathrm{w}_2 + \gamma_1}{\mathrm{w}_2 - \mathrm{w}_1}\right)$

$P_{11} = \left(\frac{\mathrm{w}_2 + \gamma_1}{\mathrm{w}_2 - \mathrm{w}_1}\right) e^{w_1 t} + \left(\frac{\mathrm{w}_1 + \gamma_1}{\mathrm{w}_1 - \mathrm{w}_2}\right) e^{w_2 t}$

$P_{12} = \frac{c_1}{\mu_{21}}(\mathrm{w}_1 + \gamma_1)\, e^{w_1 t} + \frac{\mathrm{c}_2}{\mu_{21}}(\mathrm{w}_2 + \gamma_1)\, e^{w_2 t}$

$P_{12} = \left(\frac{\mathrm{w}_2 + \gamma_1}{\mathrm{w}_2 - \mathrm{w}_1}\right)\left(\frac{\mathrm{w}_1 + \gamma_1}{\mu_{21}}\right) e^{w_1 t} + \left(\frac{\mathrm{w}_1 + \gamma_1}{\mathrm{w}_1 - \mathrm{w}_2}\right)\left(\frac{\mathrm{w}_2 + \gamma_1}{\mu_{21}}\right) e^{w_2 t}$

$let: \left(\frac{\mathrm{w}_2 + \gamma_1}{\mathrm{w}_2 - \mathrm{w}_1}\right) = A_1$ , $\left(\frac{\mathrm{w}_1 + \gamma_1}{\mathrm{w}_1 - \mathrm{w}_2}\right) = \mathrm{A}_2$

$\left(\frac{\mathrm{w}_2 + \gamma_1}{\mathrm{w}_2 - \mathrm{w}_1}\right)\left(\frac{\mathrm{w}_1 + \gamma_1}{\mu_{21}}\right) = \mathrm{A}_3$ , $\left(\frac{\mathrm{w}_1 + \gamma_1}{\mathrm{w}_1 - \mathrm{w}_2}\right)\left(\frac{\mathrm{w}_2 + \gamma_1}{\mu_{21}}\right) = \mathrm{A}_4$

$\therefore\; P_{11} = A_1 e^{w_1 t} + \mathrm{A}_2 e^{w_2 t}$ *and* $P_{12} = \mathrm{A}_3\, e^{w_1 t} + \mathrm{A}_4\, e^{w_2 t}$

*To get* $P_{13}$ *where* $\frac{dP_{13}}{dt} = P_{12}\lambda_{23}$

$\frac{dP_{13}}{dt} = \lambda_{23}(\mathrm{A}_3\, e^{w_1 t} + \mathrm{A}_4\, e^{w_2 t}) = \lambda_{23}\mathrm{A}_3\, e^{w_1 t} + \lambda_{23}\,\mathrm{A}_4\, e^{w_2 t}$

$P_{13} = \left[\lambda_{23}\mathrm{A}_3 \frac{e^{w_1 t}}{\mathrm{w}_1} - \frac{\lambda_{23}\mathrm{A}_3}{\mathrm{w}_1}\right] + \left[\lambda_{23}\,\mathrm{A}_4 \frac{e^{w_2 t}}{\mathrm{w}_2} - \frac{\lambda_{23}\,\mathrm{A}_4}{\mathrm{w}_2}\right]$

$P_{13} = \frac{\lambda_{23}\mathrm{A}_3}{\mathrm{w}_1}(e^{w_1 t} - 1) + \frac{\lambda_{23}\,\mathrm{A}_4}{\mathrm{w}_2}(e^{w_2 t} - 1)$

$let\; \frac{\lambda_{23}\mathrm{A}_3}{\mathrm{w}_1} = \mathrm{A}_5$ , $\frac{\lambda_{23}\,\mathrm{A}_4}{\mathrm{w}_2} = A_6$

$\therefore\; P_{13} = A_5\,(e^{w_1 t} - 1) + \mathrm{A}_6\,(e^{w_2 t} - 1)$

*To get* $P_{14}$ *where* $\frac{dP_{14}}{dt} = \lambda_{14}(A_1 e^{w_1 t} + \mathrm{A}_2 e^{w_2 t}) + \lambda_{24}(\mathrm{A}_3\, e^{w_1 t} + \mathrm{A}_4\, e^{w_2 t})$

$\frac{dP_{14}}{dt} = (\lambda_{14} A_1 + \lambda_{24}\,\mathrm{A}_3) e^{w_1 t} + (\lambda_{14}\,\mathrm{A}_2 + \lambda_{24}\,\mathrm{A}_4) e^{w_2 t}$ , $let: (\lambda_{14} A_1 + \lambda_{24}\,\mathrm{A}_3) = G_1$ , $(\lambda_{14}\,\mathrm{A}_2 + \lambda_{24}\,\mathrm{A}_4) = \mathrm{G}_2$

$\frac{dP_{14}}{dt} = G_1 e^{w_1 t} + \mathrm{G}_2\, e^{w_2 t}$

$P_{14} = \left[G_1 \frac{e^{w_1 t}}{\mathrm{w}_1} - \frac{G_1}{\mathrm{w}_1}\right] + \left[\mathrm{G}_2 \frac{e^{w_2 t}}{\mathrm{w}_2} - \frac{\mathrm{G}_2}{\mathrm{w}_2}\right] = \frac{G_1}{\mathrm{w}_1}(e^{w_1 t} - 1) + \frac{\mathrm{G}_2}{\mathrm{w}_2}(e^{w_2 t} - 1)$

$let: \frac{G_1}{\mathrm{w}_1} = A_7$ , $\frac{\mathrm{G}_2}{\mathrm{w}_2} = A_8 \rightarrow \therefore\; P_{14} = A_7(e^{w_1 t} - 1) + A_8(e^{w_2 t} - 1)$

To solve the set of probabilities in the second row:

$\frac{dP_{21}}{dt} = -P_{21}(\lambda_{12} + \lambda_{14}) + P_{22}\mu_{21}$ , $\frac{dP_{22}}{dt} = P_{21}\lambda_{12} - P_{22}(\mu_{21} + \lambda_{23} + \lambda_{24})$

*previously* $\lambda_{12} + \lambda_{14} = \gamma_1$ *and* $\mu_{21} + \lambda_{23} + \lambda_{24} = \gamma_2$

*Using similar steps as for $p_{11}$, $p_{12}$*

$P_{21} = c_5 e^{w_1 t} + \mathrm{c}_6 e^{w_2 t}$ ; where $\mathrm{w}_1$ and $\mathrm{w}_2$ as previously defined

*To get* $P_{22}$ *Using similar steps as for $p_{11}$, $p_{12}$*

$\therefore\ P_{22} = c_7 e^{w_1 t} + c_8 e^{w_2 t}$

$Substitute\ in: DP_{21} + \gamma_1 P_{21} - \mu_{21} P_{22} = 0\ \ to\ get\ constant$

$c_5\, w_1 e^{w_1 t} + c_6\, w_2 e^{w_2 t} + c_5 \gamma_1\, e^{w_1 t} + c_6\ \gamma_1 e^{w_2 t} - \mu_{21} c_7 e^{w_1 t} - \mu_{21}\, c_8 e^{w_2 t} = 0$

$(c_5\, w_1 + c_5 \gamma_1 - \mu_{21} c_7)\, e^{w_1 t} + (c_6\, w_2 + c_6 \gamma_1 - \mu_{21}\, c_8) e^{w_2 t} = 0$

| | |
|---|---|
| $c_5\, w_1 + c_5 \gamma_1 - \mu_{21} c_7 = 0$<br>$\mu_{21} c_7 = c_5\, w_1 + c_5 \gamma_1$<br>$c_7 = \dfrac{c_5}{\mu_{21}} (w_1 + \gamma_1)$ | $c_6\, w_2 + c_6\, \gamma_1 - \mu_{21}\, c_8 = 0$<br>$\mu_{21}\, c_8 = c_6\, w_2 + c_6\, \gamma_1$<br>$c_8 = \dfrac{c_6}{\mu_{21}} (w_2 + \gamma_1)$ |

$P_{21} = c_5 e^{w_1 t} + c_6 e^{w_2 t}, P_{22} = c_7 e^{w_1 t} + c_8 e^{w_2 t} = \dfrac{c_5}{\mu_{21}} (w_1 + \gamma_1)\, e^{w_1 t} + \dfrac{c_6}{\mu_{21}} (w_2 + \gamma_1)\, e^{w_2 t}$

Using initial values at :

$P_{21}(0) = 0 \rightarrow c_5 + c_6 = 0 \rightarrow c_5 = -c_6$

$P_{22}(0) = 1 \rightarrow \dfrac{c_5}{\mu_{21}} (w_1 + \gamma_1) + \dfrac{c_6}{\mu_{21}} (w_2 + \gamma_1) = 1$

$P_{21} = \left(\dfrac{\mu_{21}}{w_1 - w_2}\right) (e^{w_1 t} - e^{w_2 t})\ ,\ P_{22} = \left(\dfrac{w_1 + \gamma_1}{w_1 - w_2}\right) e^{w_1 t} + \left(\dfrac{w_2 + \gamma_1}{w_2 - w_1}\right) e^{w_2 t}$

$let\ \left(\dfrac{\mu_{21}}{w_1 - w_2}\right) = A_9\ ,\ \left(\dfrac{w_1 + \gamma_1}{w_1 - w_2}\right) = A_{10}\ ,\ \left(\dfrac{w_2 + \gamma_1}{w_2 - w_1}\right) = A_{11}$

$\therefore\ P_{21} = A_9\, (e^{w_1 t} - e^{w_2 t})\ \ and\ \ P_{22} = A_{10}\ e^{w_1 t} + A_{11}\, e^{w_2 t}$

*To get* $P_{23}$ *where* $\frac{dP_{23}}{dt} = P_{22} \lambda_{23}$

$\dfrac{dP_{23}}{dt} = \lambda_{23} (A_{10}\, e^{w_1 t} + A_{11}\, e^{w_2 t}) = \lambda_{23} A_{10}\, e^{w_1 t} + \lambda_{23}\, A_{11}\, e^{w_2 t}$

$P_{23} = \left[\lambda_{23} A_{10}\ \dfrac{e^{w_1 t}}{w_1} - \dfrac{\lambda_{23} A_{10}}{w_1}\right] + \left[\lambda_{23}\, A_{11}\ \dfrac{e^{w_2 t}}{w_2} - \dfrac{\lambda_{23}\, A_{11}}{w_2}\right]$

$P_{23} = \dfrac{\lambda_{23} A_{10}}{w_1}\ (e^{w_1 t} - 1) + \dfrac{\lambda_{23}\, A_{11}}{w_2}\ (e^{w_2 t} - 1)$

$let\ \dfrac{\lambda_{23} A_{10}}{w_1} = A_{12}\ ,\ \dfrac{\lambda_{23}\, A_{11}}{w_2} = A_{13}$

$\therefore\ P_{23} = A_{12}\ (e^{w_1 t} - 1) + A_{13}\, (e^{w_2 t} - 1)$

*To get* $P_{24}$

$\dfrac{dP_{24}}{dt} = P_{21} \lambda_{14} + P_{22} \lambda_{24} = (\lambda_{14} A_9 + \lambda_{24}\, A_{10}) e^{w_1 t} + (\lambda_{24}\ A_{11} - \lambda_{14}\, A_9) e^{w_2 t}$

$let\ :\ (\lambda_{14} A_9 + \lambda_{24}\, A_{10}) = G_3\ ,\ (\lambda_{24}\ A_{11} - \lambda_{14}\, A_9) = G_4$

$\dfrac{dP_{24}}{dt} = G_3 e^{w_1 t} + G_4\, e^{w_2 t}$

$P_{24} = \left[G_3 \dfrac{e^{w_1 t}}{w_1} - \dfrac{G_3}{w_1}\right] + \left[G_4\ \dfrac{e^{w_2 t}}{w_2} - \dfrac{G_4}{w_2}\right] = \dfrac{G_3}{w_1} (e^{w_1 t} - 1) + \dfrac{G_4}{w_2} (e^{w_2 t} - 1)$

$let\ :\ \dfrac{G_3}{w_1} = A_{14}\ ,\ \dfrac{G_4}{w_2} = A_{15}\ ,\ \therefore\ P_{24} = A_{14} (e^{w_1 t} - 1) + A_{15} (e^{w_2 t} - 1)$

$P_{33} = 1\ \ and\ \ P_{44} = 1$

## 4.2. Maximum Likelihood Estimation of the Q Matrix

Let $n_{ijr}$ be the number of individuals in state $i$ at $t_{r-1}$ and in state $j$ at time $t_r$ . Conditioning on the distribution of individuals among states at $t_0$ , then the likelihood function for $\theta$ is

$L(\theta) = \prod_{r=1}^{w} \left\{ \prod_{i,j=1}^{k} \left[P_{ij}(t_{r-1}, t_r)\right]^{n_{ijr}} \right\}, where\ k\ is\ the\ index\ of\ the\ \ number\ of\ states$

$$\log L(\theta) = \sum_{r=1}^{\tau}\sum_{i.j=1}^{k} n_{ijr}\, log\, P_{ij}(t_{r-1}, t_r)\ ,\ where\ \ \tau = (t_r - t_{r-1})$$

According to Kalbfleisch & Lawless (1985), applying Quasi-Newton method to estimate the rates mandates calculating the score function which is a vector –valued function for the required rates and it's the first derivative of the probability transition function with respect to $\theta$ .The second derivative is assumed to be zero .

$$S(\theta) = \frac{\partial}{\partial\theta_h}\log L = \sum_{r=1}^{\tau}\sum_{i,j=1}^{k} n_{ijr}\frac{\partial P_{ij}(\tau)/\partial\theta_h}{P_{ij}(\tau)}\ \ ,h = 1,2,3,4,5\ \ while\ P_{ij}(\tau) = \frac{n_{ijr}}{n_{i+}}$$

$$where:\ \ \theta_1 = \lambda_{12}\ ,\quad \theta_2 = \lambda_{14}\ ,\quad \theta_3 = \mu_{21}\ ,\quad \theta_4 = \lambda_{23}\ ,\quad \theta_5 = \lambda_{24}$$

$$\frac{n_{ijr}}{P_{ij}(\tau)} = n_{i+}\ \ ,\qquad such\ that\ \ n_{i+} = \sum_{j=1}^{k} n_{ijr}$$

$$S(\theta) = \tau\, e^{\Lambda\tau}d\Lambda\ and\ it's\ scaled\ 4\ times\ by\ n_{1+}\ and\ another\ 4\ times\ by\ n_{2+}\ for\ each\ \tau$$

$\Lambda$ is the eigenvalues for each Q matrix in each $\tau$

$$\frac{\partial^2}{\partial\theta_g\partial\theta_h}\log L = \sum_{r=1}^{\tau}\sum_{i,j=1}^{k} n_{ijr}\left\{\frac{\partial^2 P_{ij}(\tau)/\partial\theta_g\partial\theta_h}{P_{ij}(\tau)} - \frac{\partial P_{ij}(\tau)/\partial\theta_g \partial P_{ij}(\tau)/\partial\theta_h}{P_{ij}^2(\tau)}\right\}$$

Assuming the second derivative is zero and $\frac{n_{ijr}}{P_{ij}(\tau)} = n_{i+}$ then

$$M_{ij}(\theta) = \frac{\partial^2}{\partial\theta_g\partial\theta_h}\log L = -\sum_{r=1}^{\tau}\sum_{i,j=1}^{k} n_{i+}\frac{\partial P_{ij}(\tau)/\partial\theta_g \partial P_{ij}(\tau)/\partial\theta_h}{P_{ij}(\tau)}$$

The Quasi-Newton formula is

$$\theta_1 = \theta_0 + [M(\theta_0)]^{-1}S(\theta_0)$$

According to Klotz & Sharples ( 1994) the initial $\theta_0 = \frac{n_{ijr}}{n_{i+}}\ for\ \Delta t = 1$

Applying this to NALFD process:

$$Q = \begin{bmatrix} -(\lambda_{12}+\lambda_{14}) & \lambda_{12} & 0 & \lambda_{14} \\ \mu_{21} & -(\mu_{21}+\lambda_{23}+\lambda_{24}) & \lambda_{23} & \lambda_{24} \\ 0 & 0 & 0 & 0 \\ 0 & 0 & 0 & 0 \end{bmatrix}$$

$$let\ \ (\lambda_{12}+\lambda_{14}) = \gamma_1\ \ ,\ (\mu_{21}+\lambda_{23}+\lambda_{24}) = \gamma_2$$

$$\frac{\partial}{\partial\theta}P_{ij}(t) = t\, e^{\Lambda t}d\Lambda, where\ \Lambda\ are\ the\ eiqenvalues\ of\ the\ Q\ matrix, to\ get\ d\Lambda\ with\ respect\ to\ each\ rate:$$

$$first : solve\ for\ eigenvalues\ which\ are\ the\ solution\ of\ the\ |Q - \rho I| = 0$$

$$second :\ differentaite\ each\ value\ with\ respect\ to\ all\ rates, as\ shown\ below$$
$$|Q - \rho I|\ gives\ the\ characteristic\ polynomial\ of\ this\ matrix :$$
$$[-\lambda_{12}\mu_{21} + (\gamma_1+\rho)(\gamma_2+\rho)](-\rho)(-\rho) = 0$$
$$[-\lambda_{12}\mu_{21} + (\gamma_1+\rho)(\gamma_2+\rho)](-\rho)(-\rho) = \rho^2[\rho^2 + (\gamma_1+\gamma_2)\rho + \gamma_1\gamma_2 - \lambda_{12}\mu_{21}] = 0$$

$$\rho = \{0,0,\rho_3,\rho_4\}$$

$$\rho_3 = \frac{-(\gamma_1+\gamma_2) - \sqrt{[(\gamma_1+\gamma_2)]^2 - 4\gamma_1\gamma_2 + 4\mu_{21}\lambda_{12}}}{2} = \frac{-(\gamma_1+\gamma_2) - \sqrt{.}}{2}$$

$$\rho_4 = \frac{-(\gamma_1+\gamma_2) + \sqrt{[(\gamma_1+\gamma_2)]^2 - 4\gamma_1\gamma_2 + 4\mu_{21}\lambda_{12}}}{2} = \frac{-(\gamma_1+\gamma_2) + \sqrt{.}}{2}$$

$$[(\gamma_1+\gamma_2)]^2 - 4\gamma_1\gamma_2 + 4\mu_{21}\lambda_{12} = \sqrt{.}$$

$$\sqrt{.} = (\lambda_{12}^2 + \lambda_{14}^2 + \mu_{21}^2 + \lambda_{23}^2 + \lambda_{24}^2 + 2\lambda_{12}\lambda_{14} + 2\lambda_{12}\mu_{21} + 2\lambda_{23}\mu_{21} + 2\lambda_{24}\mu_{21} + 2\lambda_{23}\lambda_{24} - 2\lambda_{12}\lambda_{23} - 2\lambda_{12}\lambda_{24} - 2\lambda_{14}\mu_{21} - 2\lambda_{14}\lambda_{23} - 2\,\lambda_{14}\lambda_{24})^{.5}$$

$$\rho_3 = \frac{1}{2}\{-\lambda_{12} - \lambda_{14} - \mu_{21} - \lambda_{23} - \lambda_{24} - (\lambda_{12}^2 + \lambda_{14}^2 + \mu_{21}^2 + \lambda_{23}^2 + \lambda_{24}^2 + 2\lambda_{12}\lambda_{14} + 2\lambda_{12}\mu_{21} + 2\lambda_{23}\mu_{21} + 2\lambda_{24}\mu_{21} + 2\lambda_{23}\lambda_{24} - 2\lambda_{12}\lambda_{23} - 2\lambda_{12}\lambda_{24} - 2\lambda_{14}\mu_{21} - 2\lambda_{14}\lambda_{23} - 2\,\lambda_{14}\lambda_{24})^{.5}\}$$

$$\rho_4 = \frac{1}{2}\{-\lambda_{12} - \lambda_{14} - \mu_{21} - \lambda_{23} - \lambda_{24} + (\lambda_{12}^2 + \lambda_{14}^2 + \mu_{21}^2 + \lambda_{23}^2 + \lambda_{24}^2 + 2\lambda_{12}\lambda_{14} + 2\lambda_{12}\mu_{21} + 2\lambda_{23}\mu_{21} + 2\lambda_{24}\mu_{21} + 2\lambda_{23}\lambda_{24} - 2\lambda_{12}\lambda_{23} - 2\lambda_{12}\lambda_{24} - 2\lambda_{14}\mu_{21} - 2\lambda_{14}\lambda_{23} - 2\,\lambda_{14}\lambda_{24})^{.5}\}$$

$\theta_1 = \lambda_{12}$ , $\theta_2 = \lambda_{14}$ , $\theta_3 = \mu_{21}$ , $\theta_4 = \lambda_{23}$ , $\theta_5 = \lambda_{24}$

$$\frac{\partial}{\partial\lambda_{12}}\rho_3 = -\frac{1}{2} - \frac{1}{2}\left(\frac{1}{2}\right)(.)^{-.5}(2\lambda_{12} + 2\lambda_{14} + 2\mu_{21} - 2\lambda_{23} - 2\lambda_{24}) = -\frac{1}{2} - \frac{1}{2}(.)^{-.5}(\lambda_{12} + \lambda_{14} + \mu_{21} - \lambda_{23} - \lambda_{24})$$

$$\frac{\partial}{\partial\lambda_{14}}\rho_3 = -\frac{1}{2} - \frac{1}{2}\left(\frac{1}{2}\right)(.)^{-.5}(2\lambda_{14} + 2\lambda_{12} - 2\mu_{21} - 2\lambda_{23} - 2\,\lambda_{24}) = -\frac{1}{2} - \frac{1}{2}(.)^{-.5}(\lambda_{12} + \lambda_{14} - \mu_{21} - \lambda_{23} - \lambda_{24})$$

$$\frac{\partial}{\partial\mu_{21}}\rho_3 = -\frac{1}{2} - \frac{1}{2}\left(\frac{1}{2}\right)(.)^{-.5}(2\mu_{21} + 2\lambda_{12} + 2\lambda_{23} + 2\lambda_{24} - 2\lambda_{14}) = -\frac{1}{2} - \frac{1}{2}(.)^{-.5}(\mu_{21} + \lambda_{12} + \lambda_{23} + \lambda_{24} - \lambda_{14})$$

$$\frac{\partial}{\partial\lambda_{23}}\rho_3 = -\frac{1}{2} - \frac{1}{2}\left(\frac{1}{2}\right)(.)^{-.5}(2\lambda_{23} + 2\mu_{21} + 2\lambda_{24} - 2\lambda_{12} - 2\lambda_{14}) = -\frac{1}{2} - \frac{1}{2}(.)^{-.5}(\lambda_{23} + \mu_{21} + \lambda_{24} - \lambda_{12} - \lambda_{14})$$

$$\frac{\partial}{\partial\lambda_{24}}\rho_3 = -\frac{1}{2} - \frac{1}{2}\left(\frac{1}{2}\right)(.)^{-.5}(2\lambda_{24} + 2\mu_{21} + 2\lambda_{23} - 2\lambda_{12} - 2\lambda_{14}) = -\frac{1}{2} - \frac{1}{2}(.)^{-.5}(\lambda_{24} + \mu_{21} + \lambda_{23} - \lambda_{12} - \lambda_{14})$$

$$\frac{\partial}{\partial\lambda_{12}}\rho_4 = -\frac{1}{2} + \frac{1}{2}\left(\frac{1}{2}\right)(.)^{-.5}(2\lambda_{12} + 2\lambda_{14} + 2\mu_{21} - 2\lambda_{23} - 2\lambda_{24}) = -\frac{1}{2} + \frac{1}{2}(.)^{-.5}(\lambda_{12} + \lambda_{14} + \mu_{21} - \lambda_{23} - \lambda_{24})$$

$$\frac{\partial}{\partial\lambda_{14}}\rho_4 = -\frac{1}{2} + \frac{1}{2}\left(\frac{1}{2}\right)(.)^{-.5}(2\lambda_{14} + 2\lambda_{12} - 2\mu_{21} - 2\lambda_{23} - 2\,\lambda_{24}) = -\frac{1}{2} + \frac{1}{2}(.)^{-.5}(\lambda_{12} + \lambda_{14} - \mu_{21} - \lambda_{23} - \lambda_{24})$$

$$\frac{\partial}{\partial\mu_{21}}\rho_4 = -\frac{1}{2} + \frac{1}{2}\left(\frac{1}{2}\right)(.)^{-.5}(2\mu_{21} + 2\lambda_{12} + 2\lambda_{23} + 2\lambda_{24} - 2\lambda_{14}) = -\frac{1}{2} + \frac{1}{2}(.)^{-.5}(\mu_{21} + \lambda_{12} + \lambda_{23} + \lambda_{24} - \lambda_{14})$$

$$\frac{\partial}{\partial\lambda_{23}}\rho_4 = -\frac{1}{2} + \frac{1}{2}\left(\frac{1}{2}\right)(.)^{-.5}(2\lambda_{23} + 2\mu_{21} + 2\lambda_{24} - 2\lambda_{12} - 2\lambda_{14}) = -\frac{1}{2} + \frac{1}{2}(.)^{-.5}(\lambda_{23} + \mu_{21} + \lambda_{24} - \lambda_{12} - \lambda_{14})$$

$$\frac{\partial}{\partial\lambda_{24}}\rho_4 = -\frac{1}{2} + \frac{1}{2}\left(\frac{1}{2}\right)(.)^{-.5}(2\lambda_{24} + 2\mu_{21} + 2\lambda_{23} - 2\lambda_{12} - 2\lambda_{14}) = -\frac{1}{2} + \frac{1}{2}(.)^{-.5}(\lambda_{24} + \mu_{21} + \lambda_{23} - \lambda_{12} - \lambda_{14})$$

To get maximum likelihood differentiate each $P_{ij}$ with respect to each rate

$$\frac{\partial}{\partial\theta_h}P_{ij}(t) = t\,e^{\Lambda t}\,d\,\Lambda$$

$$t\,e^{\Lambda t}\,d\,\Lambda = t\,e^{\rho_3 t}\begin{bmatrix}\frac{\partial\rho_3}{\partial\lambda_{12}}\\ \frac{\partial\rho_3}{\partial\lambda_{14}}\\ \frac{\partial\rho_3}{\partial\mu_{21}}\\ \frac{\partial\rho_3}{\partial\lambda_{23}}\\ \frac{\partial\rho_3}{\partial\lambda_{24}}\end{bmatrix} + t\,e^{\rho_4 t}\begin{bmatrix}\frac{\partial\rho_4}{\partial\lambda_{12}}\\ \frac{\partial\rho_4}{\partial\lambda_{14}}\\ \frac{\partial\rho_4}{\partial\mu_{21}}\\ \frac{\partial\rho_4}{\partial\lambda_{23}}\\ \frac{\partial\rho_4}{\partial\lambda_{24}}\end{bmatrix}, substitute\ t = 1$$

$t\,e^{\Lambda t}\,d\,\Lambda\ \ at\ t = 1$

$$= e^{\rho_3}\begin{bmatrix} -\frac{1}{2}-\frac{1}{2}(.)^{-.5}(\lambda_{12}+\lambda_{14}+\mu_{21}-\lambda_{23}-\lambda_{24}) \\ -\frac{1}{2}-\frac{1}{2}(.)^{-.5}(\lambda_{14}+\lambda_{12}-\mu_{21}-\lambda_{23}-\lambda_{24}) \\ -\frac{1}{2}-\frac{1}{2}(.)^{-.5}(\mu_{21}+\lambda_{12}+\lambda_{23}+\lambda_{24}-\lambda_{14}) \\ -\frac{1}{2}-\frac{1}{2}(.)^{-.5}(\lambda_{23}+\mu_{21}+\lambda_{24}-\lambda_{12}-\lambda_{14}) \\ -\frac{1}{2}-\frac{1}{2}(.)^{-.5}(\lambda_{24}+\mu_{21}+\lambda_{23}-\lambda_{12}-\lambda_{14}) \end{bmatrix} + e^{\rho_4}\begin{bmatrix} \frac{-1}{2}+\frac{1}{2}(.)^{-.5}(\lambda_{12}+\lambda_{14}+\mu_{21}-\lambda_{23}-\lambda_{24}) \\ \frac{-1}{2}+\frac{1}{2}(.)^{-.5}(\lambda_{12}+\lambda_{14}-\mu_{21}-\lambda_{23}-\lambda_{24}) \\ \frac{-1}{2}+\frac{1}{2}(.)^{-.5}(\mu_{21}+\lambda_{12}+\lambda_{23}+\lambda_{24}-\lambda_{14}) \\ \frac{-1}{2}+\frac{1}{2}(.)^{-.5}(\lambda_{23}+\mu_{21}+\lambda_{24}-\lambda_{12}-\lambda_{14}) \\ \frac{-1}{2}+\frac{1}{2}(.)^{-.5}(\lambda_{24}+\mu_{21}+\lambda_{23}-\lambda_{12}-\lambda_{14}) \end{bmatrix}$$

$$= \begin{bmatrix} -\frac{e^{\rho_3}}{2}-\frac{e^{\rho_4}}{2}+(.)^{-.5}(\lambda_{12}+\lambda_{14}+\mu_{21}-\lambda_{23}-\lambda_{24})\left[\frac{e^{\rho_4}}{2}-\frac{e^{\rho_3}}{2}\right] \\ -\frac{e^{\rho_3}}{2}-\frac{e^{\rho_4}}{2}+(.)^{-.5}(\lambda_{12}+\lambda_{14}-\mu_{21}-\lambda_{23}-\lambda_{24})\left[\frac{e^{\rho_4}}{2}-\frac{e^{\rho_3}}{2}\right] \\ -\frac{e^{\rho_3}}{2}-\frac{e^{\rho_4}}{2}+(.)^{-.5}(\mu_{21}+\lambda_{12}+\lambda_{23}+\lambda_{24}-\lambda_{14})\left[\frac{e^{\rho_4}}{2}-\frac{e^{\rho_3}}{2}\right] \\ -\frac{e^{\rho_3}}{2}-\frac{e^{\rho_4}}{2}+(.)^{-.5}(\lambda_{23}+\mu_{21}+\lambda_{24}-\lambda_{12}-\lambda_{14})\left[\frac{e^{\rho_4}}{2}-\frac{e^{\rho_3}}{2}\right] \\ -\frac{e^{\rho_3}}{2}-\frac{e^{\rho_4}}{2}+(.)^{-.5}(\lambda_{24}+\mu_{21}+\lambda_{23}-\lambda_{12}-\lambda_{14})\left[\frac{e^{\rho_4}}{2}-\frac{e^{\rho_3}}{2}\right] \end{bmatrix}$$

$$v_1 = -\frac{e^{\rho_3}}{2}-\frac{e^{\rho_4}}{2}+(.)^{-.5}(\lambda_{12}+\lambda_{14}+\mu_{21}-\lambda_{23}-\lambda_{24})\left[\frac{e^{\rho_4}}{2}-\frac{e^{\rho_3}}{2}\right]$$
$$v_2 = -\frac{e^{\rho_3}}{2}-\frac{e^{\rho_4}}{2}+(.)^{-.5}(\lambda_{12}+\lambda_{14}-\mu_{21}-\lambda_{23}-\lambda_{24})\left[\frac{e^{\rho_4}}{2}-\frac{e^{\rho_3}}{2}\right]$$
$$v_3 = -\frac{e^{\rho_3}}{2}-\frac{e^{\rho_4}}{2}+(.)^{-.5}(\mu_{21}+\lambda_{12}+\lambda_{23}+\lambda_{24}-\lambda_{14})\left[\frac{e^{\rho_4}}{2}-\frac{e^{\rho_3}}{2}\right]$$
$$v_4 = -\frac{e^{\rho_3}}{2}-\frac{e^{\rho_4}}{2}+(.)^{-.5}(\lambda_{23}+\mu_{21}+\lambda_{24}-\lambda_{12}-\lambda_{14})\left[\frac{e^{\rho_4}}{2}-\frac{e^{\rho_3}}{2}\right]$$
$$v_5 = -\frac{e^{\rho_3}}{2}-\frac{e^{\rho_4}}{2}+(.)^{-.5}(\lambda_{24}+\mu_{21}+\lambda_{23}-\lambda_{12}-\lambda_{14})\left[\frac{e^{\rho_4}}{2}-\frac{e^{\rho_3}}{2}\right]$$

$$M(\theta) = \begin{bmatrix} v_1 \\ v_2 \\ v_3 \\ v_4 \\ v_5 \end{bmatrix}\begin{bmatrix} v_1 & v_2 & v_3 & v_4 & v_5 \end{bmatrix} = \begin{bmatrix} v_1^2 & v_1v_2 & v_1v_3 & v_1v_4 & v_1v_5 \\ v_2v_1 & v_2^2 & v_2v_3 & v_2v_4 & v_2v_5 \\ v_3v_1 & v_3v_2 & v_3^2 & v_3v_4 & v_3v_5 \\ v_4v_1 & v_4v_2 & v_4v_3 & v_4^2 & v_4v_5 \\ v_5v_1 & v_5v_2 & v_5v_3 & v_5v_4 & v_5^2 \end{bmatrix}$$

According to Klotz & Sharples ( 1994) $\tau = t_r - t_{r-1} = \Delta t \;\; and \;\; P_{ij} = \frac{n_{ij}}{n_{i+}}$

$$S(\theta) = \frac{\partial Log\ L}{\partial \theta_h} = \sum_{\Delta t \geq 1}^{3} \sum_{i,j=1}^{k} n_{ij} \frac{\partial P_{ij}(\Delta t)/\partial \theta_h}{P_{ij}(\Delta t)} = \sum_{\Delta t \geq 1}^{3} \sum_{i,j=1}^{k} n_{ij} \frac{\partial P_{ij}(\Delta t)/\partial \theta_h}{n_{ij}/n_{i+}} = \sum_{\Delta t = 1}^{3} \sum_{i,j=1}^{k} n_{i+} \frac{\partial P_{ij}(\Delta t)}{\partial \theta_h}$$
$$= \sum_{\Delta t=1}^{3} \sum_{i,j=1}^{k} n_{i+}\, t\, e^{\Lambda t}\, d\, \Lambda$$

$the\ resulting\ vectort\ (te^{\Lambda t}\ d\ \Lambda)\ is\ scaled\ by\ a\ factor\ = \left(\frac{n_{ij}(\Delta t)}{P_{ij}(\Delta t)}\right)$ 8 $times\ ;\ one\ for\ each\ pdf$

$the\ followings\ \ are\ the\ scalers$ :

$i.e\ \frac{n_{11}}{p_{11}} = n_{1+}, \frac{n_{12}}{p_{12}} = n_{1+}, \frac{n_{13}}{p_{13}} = n_{1+}, \frac{n_{14}}{p_{14}} = n_{1+}, \frac{n_{21}}{p_{21}} = n_{2+}, \frac{n_{22}}{p_{22}} = n_{2+}, \frac{n_{23}}{p_{23}} = n_{2+}, \frac{n_{24}}{p_{24}} = n_{2+},$

$all\ at(\Delta t)\ then\ the\ scaled\ vectors\ are\ summed\ up\ \ to\ get\ the\ score\ function\ .$

$This\ score\ function\ is\ used\ in\ quasi-Newton\ Raphson\ method$ :

According to Kalbfleisch & Lawless (1985)the second derivative is assumed to be zero , the score function is crossed product with itself and scaled for each pdf with the scalers :
$i.e\ \frac{n_{1+}}{p_{11}}, \frac{n_{1+}}{p_{12}}, \frac{n_{1+}}{p_{13}}, \frac{n_{1+}}{p_{14}}, \frac{n_{2+}}{p_{21}}, \frac{n_{2+}}{p_{22}}, \frac{n_{2+}}{p_{23}}, \frac{n_{2+}}{p_{24}}$ the scaled matrices are summed up to get the hessian matrix $M(\theta_0)$

$$\frac{\partial^2 Log\ L}{\partial\theta_g\partial\theta_h} = \sum_{\Delta t=1}^{3}\sum_{i,j}^{k} n_{ij}\left[\frac{\partial^2 P_{ij}(\Delta t)/\partial\theta_g\partial\theta_h}{P_{ij}(\Delta t)} - \frac{\partial P_{ij}(\Delta t)/\partial\theta_g \partial P_{ij}(\Delta t)/\partial\theta_h}{P_{ij}^2(\Delta t)}\right] \quad , \quad where\ P_{ij} = \frac{n_{ij}}{n_{i+}}$$

$$\frac{\partial^2 Log\ L}{\partial\theta_g\partial\theta_h} = \sum_{\Delta t=1}^{3}\sum_{i,j}^{k}\left(P_{ij}\ \ n_{i+}\right)\left[\frac{\partial^2 P_{ij}(\Delta t)/\partial\theta_g\partial\theta_h}{P_{ij}(\Delta t)} - \frac{\partial P_{ij}(\Delta t)/\partial\theta_g \partial P_{ij}(\Delta t)/\partial\theta_h}{P_{ij}^2(\Delta t)}\right]$$

$$\frac{\partial^2 Log\ L}{\partial\theta_g\partial\theta_h} = \sum_{\Delta t=1}^{3}\sum_{i,j}^{k}\left(P_{ij}\ \ n_{i+}\right)\left[\frac{0}{P_{ij}(\Delta t)} - \frac{\partial P_{ij}(\Delta t)/\partial\theta_g \partial P_{ij}(\Delta t)/\partial\theta_h}{P_{ij}^2(\Delta t)}\right]$$

$$\frac{\partial^2 Log\ L}{\partial\theta_g\partial\theta_h} = -\sum_{\Delta t=1}^{3}\sum_{i,j}^{k} n_{i+}\frac{\partial P_{ij}(\Delta t)/\partial\theta_g \partial P_{ij}(\Delta t)/\partial\theta_h}{P_{ij}(\Delta t)}$$

Quasi-Newton Raphson method formula:
$\theta_1 = \theta_0 + M(\theta_0)^{-1}S(\theta_0)$

Substituting in Quasi-Newton formula by the initial value, then the score and inverse of the hessian matrix are calculated to give the estimated rates.

$$S(\theta) = (4n_{1+})\begin{bmatrix} v_1 \\ v_2 \\ v_3 \\ v_4 \\ v_5 \end{bmatrix} + (4n_{2+})\begin{bmatrix} v_1 \\ v_2 \\ v_3 \\ v_4 \\ v_5 \end{bmatrix} = (4n_{1+} + 4n_{2+})\begin{bmatrix} v_1 \\ v_2 \\ v_3 \\ v_4 \\ v_5 \end{bmatrix} \quad , \ for\ each\ \Delta t$$

$$M(\theta) = (4n_{1+} + 4n_{2+})^2\begin{bmatrix} v_1 \\ v_2 \\ v_3 \\ v_4 \\ v_5 \end{bmatrix}\begin{bmatrix} v_1 & v_2 & v_3 & v_4 & v_5 \end{bmatrix} = (4n_{1+} + 4n_{2+})^2\begin{bmatrix} v_1^2 & v_1v_2 & v_1v_3 & v_1v_4 & v_1v_5 \\ v_2v_1 & v_2^2 & v_2v_3 & v_2v_4 & v_2v_5 \\ v_3v_1 & v_3v_2 & v_3^2 & v_3v_4 & v_3v_5 \\ v_4v_1 & v_4v_2 & v_4v_3 & v_4^2 & v_4v_5 \\ v_5v_1 & v_5v_2 & v_5v_3 & v_5v_4 & v_5^2 \end{bmatrix}$$

$$M(\theta) = (4n_{1+} + 4n_{2+})^2\begin{bmatrix} v_1^2 & v_1v_2 & v_1v_3 & v_1v_4 & v_1v_5 \\ v_2v_1 & v_2^2 & v_2v_3 & v_2v_4 & v_2v_5 \\ v_3v_1 & v_3v_2 & v_3^2 & v_3v_4 & v_3v_5 \\ v_4v_1 & v_4v_2 & v_4v_3 & v_4^2 & v_4v_5 \\ v_5v_1 & v_5v_2 & v_5v_3 & v_5v_4 & v_5^2 \end{bmatrix} = \begin{bmatrix} O & R \\ X & Y \end{bmatrix} then\ this\ matrix\ will\ be\ scaled$$

$M(\theta)\ is\ a\ singular\ matrix$

$$where\ O = \begin{bmatrix} v_1^2 & v_1v_2 & v_1v_3 \\ v_2v_1 & v_2^2 & v_2v_3 \\ v_3v_1 & v_3v_2 & v_3^2 \end{bmatrix}\ and\ it\ has\ O^{-1}, \ \text{so}\ \ [M(\theta)]^{-1} = \begin{bmatrix} O^{-1} & 0 \\ 0 & 0 \end{bmatrix}$$

## 4.3. Mean Sojourn Time

It is the mean time spent by a patient in a given state i of the process. It is calculated in relations to transition rates $\hat{\theta}$. These times are independent and exponentially distributed random variables with mean $\frac{1}{\lambda_i}$ where $\lambda_i = -\lambda_{ii}$ ; $i = 1,2,3,4$ .Denoting mean sojourn time by $s_i$ for state i at visits 1,2,…

$$s_1 = \frac{1}{(\lambda_{12} + \lambda_{14})} \quad , \qquad and \quad s_2 = \frac{1}{(\mu_{21} + \lambda_{23} + \lambda_{24})}$$

According to Kalbfleisch & Lawless (1985) the asymptotic variance of this time is calculated by applying multivariate delta method:

$$var(s_i) = \left[\left(q_{ii}(\hat{\theta})\right)^{-2}\right]^2 \sum_{h=1}^{5}\sum_{g=1}^{5} \frac{\partial q_{ii}}{\partial \theta_g}\frac{\partial q_{ii}}{\partial \theta_h} [\mathrm{M}(\theta)]^{-1}|_{\theta=\hat{\theta}}$$

These times are independent so covariance between them is zero

$$var(s_i) = \left[\left(q_{ii}(\hat{\theta})\right)^{-2}\right]^2 \sum_{h=1}^{5}\sum_{g=1}^{5} \frac{\partial q_{ii}}{\partial \theta_g}\frac{\partial q_{ii}}{\partial \theta_h} [\mathrm{M}(\theta)]^{-1}|_{\theta=\hat{\theta}}$$

$$var(s_i) = \left[\left(q_{ii}(\hat{\theta})\right)^{-2}\right]^2 \sum_{h=1}^{5}\sum_{g=1}^{5} \left[\frac{\partial q_{ii}}{\partial \theta_h}\right]^T [\mathrm{M}(\theta)]^{-1}|_{\theta=\hat{\theta}} \frac{\partial q_{ii}}{\partial \theta_g}$$

$$\frac{\partial q_{ii}}{\partial \theta_h} = \begin{bmatrix} \frac{\partial q_{11}}{\partial \lambda_{12}} \\ \frac{\partial q_{11}}{\partial \lambda_{14}} \\ \frac{\partial q_{22}}{\partial \mu_{21}} \\ \frac{\partial q_{22}}{\partial \lambda_{23}} \\ \frac{\partial q_{33}}{\partial \lambda_{24}} \end{bmatrix} = \begin{bmatrix} -1 \\ -1 \\ -1 \\ -1 \\ -1 \end{bmatrix} \quad , \quad \frac{\partial q_{ii}}{\partial \theta_g} = \begin{bmatrix} \frac{\partial q_{11}}{\partial \lambda_{12}} \\ \frac{\partial q_{11}}{\partial \lambda_{14}} \\ \frac{\partial q_{22}}{\partial \mu_{21}} \\ \frac{\partial q_{22}}{\partial \lambda_{23}} \\ \frac{\partial q_{33}}{\partial \lambda_{24}} \end{bmatrix} = \begin{bmatrix} -1 \\ -1 \\ -1 \\ -1 \\ -1 \end{bmatrix}$$

$$var(s_i) = \left[\left(q_{ii}(\hat{\theta})\right)^{-2}\right]^2 \sum_{h=1}^{5}\sum_{g=1}^{5} \left[\frac{\partial q_{ii}}{\partial \theta_h}\right]^T [\mathrm{M}(\theta)]^{-1}|_{\theta=\hat{\theta}} \frac{\partial q_{ii}}{\partial \theta_g}$$

$$var(s_1) = \frac{1}{(\lambda_{12} + \lambda_{14})^4} [-1 \quad -1 \quad -1 \quad -1 \quad -1][\mathrm{M}(\theta)]^{-1}|_{\theta=\hat{\theta}} \begin{bmatrix} -1 \\ -1 \\ -1 \\ -1 \\ -1 \end{bmatrix}$$

$$var(s_2) = \frac{1}{(\mu_{21} + \lambda_{23} + \lambda_{24})^4} [-1 \quad -1 \quad -1 \quad -1 \quad -1][\mathrm{M}(\theta)]^{-1}|_{\theta=\hat{\theta}} \begin{bmatrix} -1 \\ -1 \\ -1 \\ -1 \\ -1 \end{bmatrix}$$

$[M(\theta)]^{-1}|_{\theta=\hat{\theta}}$ *as calculated by MLE of rate matrix*

## 4.4. State Probability Distribution:

It is the probability distribution for each state at a specific time point given the initial probability distribution. Thus using the rule of total probability; a solution describing the transient behavior of a chain characterized by Q and an initial condition $\pi(0)$ is obtained by direct substitution to solve:

$\pi(t) = \pi(0)P(t)$ , that is to mean to get the probability of distribution NAFLD system after a certain period of time; the following equation must be solved:

$$[\pi_1 \quad \pi_2 \quad \pi_3 \quad \pi_4] = [\pi_{1(0)} \quad \pi_{2(0)} \quad \pi_{3(0)} \quad \pi_{4(0)}] \begin{bmatrix} P_{11} & P_{12} & P_{13} & P_{14} \\ P_{21} & P_{22} & P_{23} & P_{24} \\ 0 & 0 & P_{33} & 0 \\ 0 & 0 & 0 & P_{44} \end{bmatrix},$$

$$where\ \pi_{3(0)} = \pi_{4(0)} = 0\ , as\ both\ are\ death\ states$$

$$\pi_1 = \pi_{1(0)}P_{11} + \pi_{2(0)}P_{21} \quad , \quad \pi_2 = \pi_{1(0)}P_{12} + \pi_{2(0)}P_{22}$$
$$\pi_3 = \pi_{1(0)}P_{13} + \pi_{2(0)}P_{23} \quad , \quad \pi_4 = \pi_{1(0)}P_{14} + \pi_{2(0)}P_{24}$$

Thus to obtain stationary probability distribution when $t$ goes to infinity or in other words when the process does not depend on time; call $\lambda_{12} + \lambda_{14} = \gamma_1$ *and* $\mu_{21} + \lambda_{23} + \lambda_{24} = \gamma_2$

$$\pi(t) = \pi(0)P(t) = \pi(0)e^{Qt} \ , \quad differentiating\ both\ sides$$

$$\frac{d}{dt}\pi(t)\bigg|_{t=0} = \pi(0)Q$$

$$\frac{d}{dt}\pi(t)\bigg|_{t=0} = [\pi_{0(1)} \quad \pi_{0(2)} \quad \pi_{0(3)} \quad \pi_{0(4)}] \begin{bmatrix} -\gamma_1 & \lambda_{12} & 0 & \lambda_{14} \\ \mu_{21} & -\gamma_2 & \lambda_{23} & \lambda_{24} \\ 0 & 0 & 0 & 0 \\ 0 & 0 & 0 & 0 \end{bmatrix}$$

$$\frac{d}{dt}\pi_1(t)\bigg|_{t=0} = -\pi_{0(1)}\gamma_1 + \pi_{0(2)}\mu_{21} \quad , \quad \frac{d}{dt}\pi_2(t)\bigg|_{t=0} = \pi_{0(1)}\lambda_{12} - \pi_{0(2)}\gamma_2$$
$$\frac{d}{dt}\pi_3(t)\bigg|_{t=0} = \pi_{0(2)}\lambda_{23} \quad , \quad \frac{d}{dt}\pi_4(t)\bigg|_{t=0} = \pi_{0(1)}\lambda_{14} + \pi_{0(2)}\lambda_{24}$$

*By solving these equations, the vector* $[\pi_1 \quad \pi_2 \quad \pi_3 \quad \pi_4]$ *at specific time point is obtained*
Solving these differential equations even for simple chains is not a trivial matter.

Thus as $t \to \infty$ , the $\frac{d}{dt}\pi_j(t) = 0$ , since $\pi_z(t)$does not depend on time.

$$\frac{d}{dt}\pi(t) = \pi(t)Q \qquad will\ reduce\ to\ \ \pi(t)Q = 0$$

$$By\ solving\ \ \pi Q = 0,\ \ subject\ to\ \sum_{all\ z} \pi_z\ = 1\ , the\ state\ probability\ distribution\ is\ obtained$$

$$[\pi_{(1)} \quad \pi_{(2)} \quad \pi_{(3)} \quad \pi_{(4)}] \begin{bmatrix} -\gamma_1 & \lambda_{12} & 0 & \lambda_{14} \\ \mu_{21} & -\gamma_2 & \lambda_{23} & \lambda_{24} \\ 0 & 0 & 0 & 0 \\ 0 & 0 & 0 & 0 \end{bmatrix} = \begin{bmatrix} 0 \\ 0 \\ 0 \\ 0 \end{bmatrix}$$

$$-\pi_{(1)}\gamma_1 + \pi_{(2)}\mu_{21} = 0\ , \quad \pi_{(1)}\lambda_{12} - \pi_{(2)}\gamma_2 = 0\ , \quad \pi_{(2)}\lambda_{23} = 0\ , \quad \pi_{(1)}\lambda_{14} + \pi_{(2)}\lambda_{24} = 0$$
$$1 = \pi_{(1)} + \pi_{(2)} + \pi_{(3)} + \pi_{(4)}$$

The above equations are expressed in matrix notation as:

$$\begin{bmatrix} -\gamma_1 & \mu_{21} & 0 & 0 \\ \lambda_{12} & -\gamma_2 & 0 & 0 \\ 0 & \lambda_{23} & 0 & 0 \\ \lambda_{14} & \lambda_{24} & 0 & 0 \\ 1 & 1 & 1 & 1 \end{bmatrix} \begin{bmatrix} \pi_1 \\ \pi_2 \\ \pi_3 \\ \pi_4 \end{bmatrix} = \begin{bmatrix} 0 \\ 0 \\ 0 \\ 0 \\ 1 \end{bmatrix} \quad \rightarrow \quad \begin{bmatrix} \pi_1 \\ \pi_2 \\ \pi_3 \\ \pi_4 \end{bmatrix} = \begin{bmatrix} 0 \\ 0 \\ \pi_3 \\ 1 - \pi_3 \end{bmatrix}$$

### 4.4.1. Asymptotic Covariance of the Stationary Distribution

To obtain this, multivariate delta method is used as well as the following function of the

$Q'\pi = 0$ ,  $as\ \pi\ is\ not\ a\ simple\ function\ of\ \theta$

$$\frac{\partial}{\partial\theta}F(\theta_h,\pi_i) = \frac{\partial}{\partial\theta_h}(Q'\pi_i) = 0 \quad , \qquad with\ \ implicit\ differentiation \quad ,$$

$$\frac{\partial}{\partial\theta_h}F(\theta_h,\pi_i) = \frac{\partial}{\partial\theta_h}(Q'\pi_i) = [Q']\left[\frac{\partial}{\partial\theta_h}\pi_i\right] + \pi_i\left[\frac{\partial}{\partial\theta_h}Q'\right]^T \quad ,let's\ call \quad \pi_i\left[\frac{\partial}{\partial\theta_h}Q'\right]^T = C(\theta)\ \ is\ a\ matrix$$

$\left[\frac{\partial}{\partial\theta_h}\pi\right]$ $this\ is\ a\ matrix\ that\ gives\ all\ derivatives\ of\ \pi_1, \pi_2, \pi_3, \pi_4\ with\ respect\ to\ each\ \lambda_{12}, \lambda_{14}, \mu_{21}, \lambda_{23}, \lambda_{24}$

$$\left[\frac{\partial}{\partial\theta_h}\pi\right] = -[Q']^{-1}C(\theta)\ \ ,\pi(\theta) = \begin{bmatrix}\pi_1\\ \pi_2\\ \pi_3\\ \pi_4\end{bmatrix}\ is\ a\ column\ vector$$

$$C(\theta) = \pi(\theta)\left[\frac{\partial}{\partial\theta_h}Q'\right]^T = \begin{bmatrix}0\\ 0\\ \pi_3\\ 1-\pi_3\end{bmatrix}\begin{bmatrix}1 & 1 & 1 & 1 & 1\end{bmatrix} = \begin{bmatrix}0 & 0 & 0 & 0 & 0\\ 0 & 0 & 0 & 0 & 0\\ \pi_3 & \pi_3 & \pi_3 & \pi_3 & \pi_3\\ 1-\pi_3 & 1-\pi_3 & 1-\pi_3 & 1-\pi_3 & 1-\pi_3\end{bmatrix}$$

$$\pi(\theta)\left[\frac{\partial}{\partial\theta_h}Q'\right]^T = C(\theta)\ \ is\ a\ matrix\ of\ \ size\ \ (4\ \times\ 5)$$

$$Q' = \begin{bmatrix}-\gamma_1 & \mu_{21} & 0 & 0\\ \lambda_{12} & -\gamma_2 & 0 & 0\\ 0 & \lambda_{23} & 0 & 0\\ \lambda_{14} & \lambda_{24} & 0 & 0\end{bmatrix}\ is\ a\ singular\ matrix\ , thus$$

$[Q']^{-1}$ $obtained\ by\ \ pseudoinverse\ \ using\ \ SVD\ \ and\ resulting\ in\ \ 4\ by\ 4\ matrix$

$$let:\quad A(\theta) = \left[\frac{\partial}{\partial\theta_h}\pi_i\right] = -[Q']^{-1}C(\theta) = \begin{bmatrix}\frac{\partial\pi_1}{\partial\lambda_{12}} & \frac{\partial\pi_1}{\partial\lambda_{14}} & \frac{\partial\pi_1}{\partial\mu_{21}} & \frac{\partial\pi_1}{\partial\lambda_{23}} & \frac{\partial\pi_1}{\partial\lambda_{24}}\\ \frac{\partial\pi_2}{\partial\lambda_{12}} & \frac{\partial\pi_2}{\partial\lambda_{14}} & \frac{\partial\pi_2}{\partial\mu_{21}} & \frac{\partial\pi_2}{\partial\lambda_{23}} & \frac{\partial\pi_2}{\partial\lambda_{24}}\\ \frac{\partial\pi_3}{\partial\lambda_{12}} & \frac{\partial\pi_3}{\partial\lambda_{14}} & \frac{\partial\pi_3}{\partial\mu_{21}} & \frac{\partial\pi_3}{\partial\lambda_{23}} & \frac{\partial\pi_3}{\partial\lambda_{24}}\\ \frac{\partial\pi_4}{\partial\lambda_{12}} & \frac{\partial\pi_4}{\partial\lambda_{14}} & \frac{\partial\pi_4}{\partial\mu_{21}} & \frac{\partial\pi_4}{\partial\lambda_{23}} & \frac{\partial\pi_4}{\partial\lambda_{24}}\end{bmatrix}\quad is\ \ 4\ by\ 5\ matrix$$

Using multivariate delta method

$var(\pi_i) = A(\theta)var(\theta)A(\theta)'$ ,  $where\ \ var(\theta) = [M(\theta)]^{-1}\ \ and\ \ \ i = 1,2,3,4$

### 4.5. Life Expectancy of Patient in NAFLD Disease Process:

The disease process is composed of state 1 and state 2 which are transient states, while state 3 and state 4 both are absorbing states. So partitioning the Q matrix into 4 sets

$$Q = \begin{bmatrix}-(\lambda_{12}+\lambda_{14}) & \lambda_{12} & 0 & \lambda_{14}\\ \mu_{21} & -(\mu_{21}+\lambda_{23}+\lambda_{24}) & \lambda_{23} & \lambda_{24}\\ 0 & 0 & 0 & 0\\ 0 & 0 & 0 & 0\end{bmatrix} = \begin{bmatrix}B & A\\ 0 & 0\end{bmatrix}$$

$$Where\ \ B = \begin{bmatrix}-(\lambda_{12}+\lambda_{14}) & \lambda_{12}\\ \mu_{21} & -(\mu_{21}+\lambda_{23}+\lambda_{24})\end{bmatrix}\ ,\qquad A = \begin{bmatrix}0 & \lambda_{14}\\ \lambda_{23} & \lambda_{24}\end{bmatrix}\quad \&\ \ A = BZ$$

$$let\quad (\lambda_{12}+\lambda_{14}) = \gamma_1\quad ,\quad (\mu_{21}+\lambda_{23}+\lambda_{24}) = \gamma_2$$

$$so\ \ Z = \begin{bmatrix} \frac{\lambda_{12}\lambda_{23}}{\mu_{21}\lambda_{12} - \gamma_1\gamma_2} & \frac{\lambda_{12}(\mu_{21}\lambda_{14} + \lambda_{24}\gamma_1) - \lambda_{14}(\mu_{21}\lambda_{12} - \gamma_1\gamma_2)}{\gamma_1(\mu_{21}\lambda_{12} - \gamma_1\gamma_2)} \\ \frac{\gamma_1\lambda_{23}}{\mu_{21}\lambda_{12} - \gamma_1\gamma_2} & \frac{(\mu_{21}\lambda_{14} + \lambda_{24}\gamma_1)}{\mu_{21}\lambda_{12} - \gamma_1\gamma_2} \end{bmatrix}$$

$[\acute{P}(t) \quad \acute{P}_k(t)] = [P(t) \quad P_k(t)]\begin{bmatrix} B & A \\ 0 & 0 \end{bmatrix}$ *can be written as*

$\acute{P}(t) = P(t)B$

$\acute{P}_k(t) = P(t)A$

*The solution to* $\acute{P}(t) = P(t)B$ *is* $P(t) = P(0)e^{BT}$ *then* $\acute{P}_k(t) = P(0)e^{BT}A$

$$P_k(t) = A\frac{e^{Bt}}{B}\Big|_{t=0}^{t=t} = A\left[\frac{e^{Bt}}{B} - \frac{1}{B}\right] = \frac{A}{B}[e^{Bt} - 1] = \frac{BZ}{B}[e^{Bt} - 1] = Z[e^{Bt} - 1]$$

$$and \quad e^{BT} = 1 + Bt + \frac{(Bt)^2}{2!} + \frac{(Bt)^3}{3!} + \frac{(Bt)^4}{4!} + \cdots = \sum_{j=0}^{\infty}\frac{(Bt)^j}{j!}$$

*If* $\tau_k$ *is the time taken from state i to reach the absorbing death state from the initil time*

$F_k(t) = pr[\tau_k \le t] = pr[X(t) = k] = P_k(t) = Z[e^{Bt} - 1]$

The moment theory for Laplace transform can be used to obtain the mean of the time which has the above cumulative distribution function.

CTMC can be written in a Laplace transform such that:

$[sP^*(s) - P(0) \quad sP^*{}_k(s)] = [P^*(s) \quad P^*{}_k(s)]\begin{bmatrix} B & A \\ 0 & 0 \end{bmatrix}$

$\therefore\ sP^*(s) - P(0) = P^*(s)B$

$\therefore\ sP^*{}_k(s) = P^*(s)A$

Rearrange :

$\therefore\ sP^*(s) - P^*(s)B = P(0)$

$P^*(s)\ [sI - B] = P(0) \rightarrow \quad P^*(s) \ = P(0)[sI - B]^{-1}$

$\therefore\ sP^*{}_k(s) = P^*(s)A \ \rightarrow\ P^*{}_k(s) = \frac{1}{s}\ P^*(s)A = \frac{1}{s}\ P(0)[sI - B]^{-1}A$

$F^*{}_k(s) = \frac{1}{s}\ P(0)[sI - B]^{-1}A$

$f^*{}_k(s) = s\,F^*{}_k(s) = P(0)[sI - B]^{-1}A \quad ; \ where \ \ A = BZ$

Mean time to absorption:

$$E(\tau_k) = (-1)\frac{df^*{}_k(s)}{ds}\Big|_{s=0} = P(0)[sI - B]^{-2}A|_{s=0} = P(0)[B]^{-1}\,Z = [B]^{-1}\,Z$$

$E(\tau_k) = [B]^{-1}\,Z$ solving this equation:

$$E(\tau_{ik}) = \begin{bmatrix} \frac{-\gamma_2}{\gamma_1\gamma_2 - \mu_{21}\lambda_{12}} & \frac{-\lambda_{12}}{\gamma_1\gamma_2 - \mu_{21}\lambda_{12}} \\ \frac{-\mu_{21}}{\gamma_1\gamma_2 - \mu_{21}\lambda_{12}} & \frac{-\gamma_1}{\gamma_1\gamma_2 - \mu_{21}\lambda_{12}} \end{bmatrix}\begin{bmatrix} \frac{\lambda_{12}\lambda_{23}}{\mu_{21}\lambda_{12} - \gamma_1\gamma_2} & \frac{\lambda_{12}(\mu_{21}\lambda_{14} + \lambda_{24}\gamma_1) - \lambda_{14}(\mu_{21}\lambda_{12} - \gamma_1\gamma_2)}{\gamma_1(\mu_{21}\lambda_{12} - \gamma_1\gamma_2)} \\ \frac{\gamma_1\lambda_{23}}{\mu_{21}\lambda_{12} - \gamma_1\gamma_2} & \frac{(\mu_{21}\lambda_{14} + \lambda_{24}\gamma_1)}{\mu_{21}\lambda_{12} - \gamma_1\gamma_2} \end{bmatrix}$$

$$E(\tau_{13}) = \frac{\lambda_{12}\lambda_{23}(\gamma_1 + \gamma_2)}{(\gamma_1\gamma_2 - \mu_{21}\lambda_{12})^2} \quad , \quad E(\tau_{14}) = \frac{\lambda_{12}(\mu_{21}\lambda_{14} + \lambda_{24}\gamma_1)(\gamma_1 + \gamma_2)}{\gamma_1\ (\gamma_1\gamma_2 - \mu_{21}\lambda_{12})^2} + \frac{\lambda_{14}\gamma_2}{\gamma_1(\gamma_1\gamma_2 - \mu_{21}\lambda_{12})}$$

$$E(\tau_{23}) = \frac{\lambda_{23}(\mu_{21}\lambda_{12} + \gamma_1^2)}{(\gamma_1\gamma_2 - \mu_{21}\lambda_{12})^2} \quad , \quad E(\tau_{24}) = \frac{(\mu_{21}\lambda_{14} + \lambda_{24}\gamma_1)(\mu_{21}\lambda_{12} + \gamma_1^2)}{\gamma_1\,(\gamma_1\gamma_2 - \mu_{21}\lambda_{12})^2} + \frac{\mu_{21}\lambda_{14}}{\gamma_1(\gamma_1\gamma_2 - \mu_{21}\lambda_{12})}$$

### 4.6. Expected Number of Patients in Each State

Let $u(0)$ be the size of patients in a specific state at specific time $t = 0$ .The initial size of patients $u(0) = \sum_{j=1}^{2} u_j\,(0)$ , as there are 2 transient states and 2 absorbing states, where $u_j(0)$ is the initial size or number of patients in state $j$ at time $t = 0$ given that $u_3(0) = 0$ and $u_4(0) = 0$ i.e initial size of patients in state 3 and state 4 ( both are absorbing death state) are zero at initial time point $t = 0$ . As the transition or the movement of the patients among states are independent so at the end of the whole time interval $(0, t)$ and according to Chiang( 1968), there will be $u_j(t)$ patients in state 1 and in state 2 at time $t$ , also there will be $u_3(t)$patients in state 3 (death state) at time $t$ and $u_4(t)$ patients in state 4 (death state) at time $t$ .

$$E\left[u_j(t)|u_j(0)\right] = \sum_{j=1,i=1}^{2} u_i(0)P_{ij}(t) \quad , \quad i\&\, j = 1,2$$

$$E\left[u_j(t)|u_i(0)\right] = \sum_{j=3,i=1}^{j=4,i=2} u_i(0)P_{ij}(t) \quad , \quad i = 1,2 \text{ and } j = 3,4$$

$$E\left[u_j(t)|u_i(0)\right] = [u_1(0) \quad u_2(0) \quad 0 \quad 0]\begin{bmatrix} P_{11} & P_{12} & P_{13} & P_{14} \\ P_{21} & P_{22} & P_{23} & P_{24} \\ 0 & 0 & P_{33} & 0 \\ 0 & 0 & 0 & P_{44} \end{bmatrix} = [u_1(t) \quad u_2(t) \quad u_3(t) \quad u_4(t)]$$

$u_1(t) = u_1(0)P_{11} + u_2(0)P_{21}$ , $u_2(t) = u_1(0)P_{12} + u_2(0)P_{22}$ , $u_3(t) = u_1(0)P_{13} + u_2(0)P_{23}$ , $u_4(t) = u_1(0)P_{14} + u_2(0)P_{24}$

### 4.7. Hypothetical Numerical Example

To illustrate the above concepts and discussion, a hypothetical numerical example is introduced. It does not represent real data but it is for demonstrative purposes

A study was conducted over 8 years on 310 patients with risk factors for developing NAFLD such as type 2 diabetes mellitus, obesity, and hypertension acting alone or together as a metabolic syndrome. The patients were decided to be followed up every year by a liver biopsy to identify the NAFLD cases, but the actual observations were recorded as shown in the table (4.6). The following 5 successive tables (4.1), (4.2), (4.3), (4.5) illustrate a summary for counts of transitions in in the whole period of study and in various lengths of time intervals:

**Table (4.1) demonstrates Numbers of observed transitions among states of NAFLD process during different time intervals $\Delta t = 1, 2, 3$ years**

| $\Delta t$ | Transitions among states | | | | | | | |
|---|---|---|---|---|---|---|---|---|
| | (1,1) | (1,2) | (1,3) | (1,4) | (2,1) | (2,2) | (2,3) | (2,4) |
| 1 | 330 | 163 | 45 | 12 | 5 | 185 | 45 | 15 |
| 2 | 70 | 30 | 10 | 1 | 2 | 20 | 13 | 4 |
| 3 | 21 | 8 | 7 | 3 | 1 | 6 | 3 | 1 |
| | 421 | 201 | 62 | 16 | 8 | 211 | 61 | 20 |

**Table (4.2) demonstrates total counts of transitions throughout whole period of the study (8 years)**

| | State 1 | State 2 | State 3 | State 4 | Total counts |
|---|---|---|---|---|---|
| State 1 | 421 | 201 | 62 | 16 | 700 |
| State 2 | 8 | 211 | 61 | 20 | 300 |
| State 3 | 0 | 0 | 0 | 0 | 0 |
| State 4 | 0 | 0 | 0 | 0 | 0 |
| Total | 429 | 412 | 123 | 36 | 1000 |

**Table (4.3) demonstrates the observed counts of transitions during time interval $\Delta t = 1\ year$**

| | State 1 | State 2 | State 3 | State 4 | Total counts |
|---|---|---|---|---|---|
| State 1 | 330 | 163 | 45 | 12 | 550 |
| State 2 | 5 | 185 | 45 | 15 | 250 |
| State 3 | 0 | 0 | 0 | 0 | 0 |
| State 4 | 0 | 0 | 0 | 0 | 0 |
| Total | 335 | 348 | 90 | 27 | 800 |

**Table (4.4) demonstrates the observed counts of transitions during time interval $\Delta t = 2\ years$**

| | State 1 | State 2 | State 3 | State 4 | Total counts |
|---|---|---|---|---|---|
| State 1 | 70 | 30 | 10 | 1 | 111 |
| State 2 | 2 | 20 | 13 | 4 | 39 |
| State 3 | 0 | 0 | 0 | 0 | 0 |
| State 4 | 0 | 0 | 0 | 0 | 0 |
| Total | 72 | 50 | 23 | 5 | 150 |

**Table (4.5) demonstrates the observed counts of transitions during time interval $\Delta t = 3\ years$**

| | State 1 | State 2 | State 3 | State 4 | Total counts |
|---|---|---|---|---|---|
| State 1 | 21 | 8 | 7 | 3 | 39 |
| State 2 | 1 | 6 | 3 | 1 | 11 |
| State 3 | 0 | 0 | 0 | 0 | 0 |
| State 4 | 0 | 0 | 0 | 0 | 0 |
| Total | 22 | 14 | 10 | 4 | 50 |

**The next table (4.6) demonstrates the layout of the data for the persons studied over the 8 years:**

| Patient ID | time | | | | | | | | | Patient ID | | | | | | | | | |
|---|---|---|---|---|---|---|---|---|---|---|---|---|---|---|---|---|---|---|---|
| | 0 | 1 | 2 | 3 | 4 | 5 | 6 | 7 | 8 | | 0 | 1 | 2 | 3 | 4 | 5 | 6 | 7 | 8 |
| 1 | 1 | 1 | 2 | 2 | | | | | | 53 | 1 | 1 | 1 | 2 | 2 | 4 | | | |
| 2 | 1 | 1 | 2 | 1 | | 2 | | 1 | | 54 | 1 | | 2 | | 2 | | | | |
| 3 | 1 | 3 | | | | | | | | 55 | 1 | | 2 | | 2 | | | | |
| 4 | 1 | 1 | 1 | 1 | | | | | | 56 | 1 | 1 | 1 | 1 | 2 | 2 | 4 | | |
| 5 | 1 | 2 | 1 | | 2 | | 2 | | | 57 | 1 | | 2 | | 2 | | | | |
| 6 | 1 | 2 | 1 | | 2 | | 2 | | | 58 | 1 | 1 | 1 | 1 | 1 | 1 | 2 | 2 | 4 |
| 7 | 1 | 2 | 1 | | 2 | | 2 | | | 59 | 1 | 2 | 2 | | 2 | | | | |
| 8 | 1 | 3 | | | | | | | | 60 | 1 | 1 | 1 | 1 | 3 | | | | |
| 9 | 1 | 1 | 1 | 2 | 2 | 4 | | | | 61 | 1 | 3 | | | | | | | |
| 10 | 1 | 3 | | | | | | | | 62 | 1 | 1 | 1 | 2 | 2 | 4 | | | |
| 11 | 1 | 1 | 1 | 2 | 1 | | | | | 63 | 1 | 1 | 1 | 1 | 1 | 2 | 2 | | 2 |
| 12 | 1 | 3 | | | | | | | | 64 | 1 | 1 | 2 | 2 | | 2 | | 2 | |
| 13 | 1 | 1 | 1 | 3 | | | | | | 65 | 1 | 1 | 1 | 1 | 2 | 2 | | 2 | |
| 14 | 1 | 3 | | | | | | | | 66 | 1 | 1 | 2 | 2 | | 2 | | | |
| 15 | 1 | 1 | 1 | 1 | 1 | 3 | | | | 67 | 1 | 1 | 1 | 1 | 2 | 2 | | 2 | |
| 16 | 1 | 1 | 1 | 1 | 1 | 1 | 1 | 1 | 2 | 68 | 1 | 2 | 2 | | 2 | | | | |
| 17 | 1 | 1 | 2 | 2 | 4 | | | | | 69 | 1 | 2 | 2 | 4 | | | | | |
| 18 | 1 | 1 | 1 | 1 | 2 | | | | | 70 | 1 | 1 | | 1 | | 1 | | 1 | |
| 19 | 1 | 1 | 1 | 1 | 1 | | 2 | | 1 | 71 | 1 | 1 | 1 | 1 | | 1 | | 1 | |
| 20 | 1 | 1 | 1 | 1 | 1 | 1 | 1 | | | 72 | 1 | 1 | 1 | 1 | 1 | 1 | 2 | 2 | 2 |
| 21 | 1 | 1 | 3 | | | | | | | 73 | 1 | 2 | 2 | | 2 | | | | |
| 22 | 1 | 1 | 2 | 2 | 4 | | | | | 74 | 1 | 2 | 2 | | 2 | | | | |
| 23 | 1 | 3 | | | | | | | | 75 | 1 | 1 | 1 | 3 | | | | | |
| 24 | 1 | 3 | | | | | | | | 76 | 1 | 1 | 1 | 3 | | | | | |
| 25 | 1 | 3 | | | | | | | | 77 | 1 | 2 | 2 | 2 | 2 | 2 | 2 | | 2 |
| 26 | 1 | 3 | | | | | | | | 78 | 1 | 1 | 1 | | 1 | | 1 | | 1 |
| 27 | 1 | 3 | | | | | | | | 79 | 1 | 1 | 1 | 1 | 2 | 2 | 2 | 2 | |
| 28 | 1 | 1 | 2 | 2 | 4 | | | | | 80 | 1 | 3 | | | | | | | |
| 29 | 1 | 4 | | | | | | | | 81 | 1 | 3 | | | | | | | |
| 30 | 1 | 4 | | | | | | | | 82 | 1 | 1 | 1 | 2 | 2 | | | | |
| 31 | 1 | 4 | | | | | | | | 83 | 1 | 1 | 1 | 1 | 1 | 2 | 2 | | 2 |
| 32 | 1 | 1 | 2 | 2 | 4 | | | | | 84 | 1 | 1 | 1 | 1 | 1 | 2 | 2 | | |
| 33 | 1 | 2 | 2 | 4 | | | | | | 85 | 1 | 2 | 2 | 2 | 2 | 2 | 2 | | 3 |
| 34 | 1 | 2 | 2 | | 2 | | | | | 86 | 1 | 2 | 2 | 4 | | | | | |
| 35 | 1 | 2 | 2 | | 2 | | | | | 87 | 1 | | 1 | | 1 | | 1 | | 1 |
| 36 | 1 | 3 | | | | | | | | 88 | 1 | | 1 | | 1 | | 1 | | 1 |
| 37 | 1 | 3 | | | | | | | | 89 | 1 | | 1 | | 1 | | 1 | | 1 |
| 38 | 1 | 1 | 2 | 2 | 4 | | | | | 90 | 1 | 1 | 1 | 1 | 2 | 2 | | 3 | |
| 39 | 1 | 4 | | | | | | | | 91 | 1 | 1 | 1 | 1 | 1 | 2 | 2 | | 3 |
| 40 | 1 | 4 | | | | | | | | 92 | 1 | | 1 | | 1 | | 1 | | 1 |
| 41 | 1 | 4 | | | | | | | | 93 | 1 | | 1 | | 1 | | 1 | | 1 |
| 42 | 1 | 1 | 1 | 2 | 2 | 4 | | | | 94 | 1 | 1 | 1 | 3 | | | | | |
| 43 | 1 | 4 | | | | | | | | 95 | 1 | 1 | 1 | 1 | 2 | 2 | | 3 | |
| 44 | 1 | 4 | | | | | | | | 96 | 1 | 1 | 1 | 1 | 3 | | | | |
| 45 | 1 | 4 | | | | | | | | 97 | 1 | 1 | 1 | 1 | 1 | | 1 | | 1 |
| 46 | 1 | 1 | 3 | | | | | | | 98 | 1 | 2 | 2 | | 3 | | | | |
| 47 | 1 | 4 | | | | | | | | 99 | 1 | 1 | 1 | 2 | 2 | | 3 | | |
| 48 | 1 | 4 | | | | | | | | 100 | 1 | 1 | 1 | 2 | 2 | | 3 | | |
| 49 | 1 | 1 | 1 | 1 | 1 | 2 | 2 | 4 | | 101 | 1 | 1 | 1 | 1 | 2 | 2 | | 3 | |
| 50 | 1 | 1 | 1 | 1 | 1 | 1 | 2 | 2 | | 102 | 1 | 2 | 2 | | 3 | | | | |
| 51 | 1 | 3 | | | | | | | | 103 | 1 | 1 | 1 | 3 | | | | | |
| 52 | 1 | 3 | | | | | | | | 104 | 1 | 1 | 1 | 3 | | | | | |

| Patient ID | time | | | | | | | | | Patient ID | time | | | | | | | | |
|---|---|---|---|---|---|---|---|---|---|---|---|---|---|---|---|---|---|---|---|
| | 0 | 1 | 2 | 3 | 4 | 5 | 6 | 7 | 8 | | 0 | 1 | 2 | 3 | 4 | 5 | 6 | 7 | 8 |
| 105 | 1 | 1 | 1 | 3 | | | | | | 157 | 1 | 1 | 1 | 1 | 1 | 2 | 2 | 3 | |
| 106 | 1 | 3 | | | | | | | | 158 | 1 | 2 | 2 | 3 | | | | | |
| 107 | 1 | 1 | 1 | 1 | 1 | 2 | 2 | | 3 | 159 | 1 | 2 | 2 | | | | | | |
| 108 | 1 | | 1 | | 1 | | 1 | | 1 | 160 | 1 | 1 | 1 | 2 | 2 | 3 | | | |
| 109 | 1 | 1 | 1 | 1 | 2 | 2 | | 3 | | 161 | 1 | 1 | 1 | 2 | 2 | 3 | | | |
| 110 | 1 | 1 | 1 | 1 | 1 | 3 | | | | 162 | 1 | 1 | 1 | 1 | 1 | 2 | 2 | 3 | |
| 111 | 1 | 2 | 2 | | 3 | | | | | 163 | 1 | 2 | 2 | | | | | | |
| 112 | 1 | 1 | 1 | 1 | 3 | | | | | 164 | 1 | 1 | 1 | 2 | 2 | | | | |
| 113 | 1 | 2 | 2 | | 3 | | | | | 165 | 1 | 2 | 2 | | | | | | |
| 114 | 1 | 1 | 1 | 3 | | | | | | 166 | 1 | 1 | 1 | 2 | 2 | 3 | | | |
| 115 | 1 | 1 | 1 | 3 | | | | | | 167 | 1 | 2 | 2 | | | | | | |
| 116 | 1 | 2 | 2 | | 4 | | | | | 168 | 1 | 1 | 1 | 1 | 1 | 1 | | | |
| 117 | 1 | 1 | 1 | 3 | | | | | | 169 | 1 | 2 | 2 | 2 | 3 | | | | |
| 118 | 1 | 1 | 1 | 1 | | 1 | | 1 | | 170 | 1 | 2 | 2 | 2 | | | | | |
| 119 | 1 | 1 | 1 | 1 | 1 | 2 | 2 | | | 171 | 1 | 1 | 2 | 2 | 3 | | | | |
| 120 | 1 | | 1 | | 1 | | 1 | | 1 | 172 | 1 | 2 | 2 | 2 | 2 | 2 | | | |
| 121 | 1 | 1 | 1 | | 1 | | 1 | | 1 | 173 | 1 | 2 | 2 | 3 | | | | | |
| 122 | 1 | 1 | 1 | 1 | 3 | | | | | 174 | 1 | 2 | 2 | | | | | | |
| 123 | 1 | 1 | 1 | 1 | 1 | 1 | | 2 | | 175 | 1 | 1 | 2 | 2 | 2 | 2 | | | |
| 124 | 1 | 1 | 1 | 1 | 1 | 1 | 2 | 2 | | 176 | 1 | 2 | 2 | | | | | | |
| 125 | 1 | 2 | 2 | | 4 | | | | | 177 | 1 | 2 | 2 | 2 | 2 | 2 | | | |
| 126 | 1 | 2 | 2 | | 4 | | | | | 178 | 1 | 2 | 2 | 3 | | | | | |
| 127 | 1 | 1 | 1 | 1 | 1 | 2 | 2 | | 4 | 179 | 1 | 1 | 1 | 2 | 2 | | | | |
| 128 | 1 | 1 | 1 | 1 | 1 | 2 | 2 | | | 180 | 1 | 2 | 2 | 2 | 2 | 3 | | | |
| 129 | 1 | 1 | 2 | 2 | 3 | | | | | 181 | 1 | 2 | 2 | 3 | | | | | |
| 130 | 1 | 2 | 2 | | | | | | | 182 | 1 | 3 | | | | | | | |
| 131 | 1 | 1 | 1 | 3 | | | | | | 183 | 1 | 3 | | | | | | | |
| 132 | 1 | 2 | 2 | | | | | | | 184 | 1 | 1 | 1 | 1 | 1 | 2 | 2 | 2 | |
| 133 | 1 | 1 | 1 | 1 | | 1 | | 1 | | 185 | 1 | 2 | 2 | 2 | 2 | 3 | | | |
| 134 | 1 | 2 | 2 | | | | | | | 186 | 1 | 2 | 2 | | | | | | |
| 135 | 1 | 2 | 2 | | | | | | | 187 | 1 | 1 | 1 | 1 | 2 | 2 | 2 | 2 | |
| 136 | 1 | 1 | 1 | 2 | 2 | 3 | | | | 188 | 1 | 2 | 2 | | | | | | |
| 137 | 1 | 2 | 2 | | | | | | | 189 | 1 | 1 | 1 | 2 | 2 | 3 | | | |
| 138 | 1 | 2 | 2 | 2 | 2 | 2 | | | | 190 | 1 | 1 | 1 | 1 | 2 | 2 | 3 | | |
| 139 | 1 | 1 | 1 | 1 | 1 | | | | | 191 | 1 | 2 | 2 | 2 | 3 | | | | |
| 140 | 1 | 2 | 2 | 2 | 2 | 2 | | | | 192 | 1 | 2 | 2 | | | | | | |
| 141 | 1 | 1 | 1 | 1 | 1 | 1 | 1 | 2 | 2 | 193 | 1 | 1 | 2 | 2 | | | | | |
| 142 | 1 | | 1 | | 1 | | 1 | | 1 | 194 | 1 | 2 | 2 | | | | | | |
| 143 | 1 | 1 | 1 | 1 | 1 | 3 | | | | 195 | 1 | 1 | 1 | 1 | 2 | 2 | 3 | | |
| 144 | 1 | | 1 | | 1 | | 1 | | 1 | 196 | 1 | 2 | 2 | | | | | | |
| 145 | 1 | 1 | 1 | 2 | 2 | 2 | 2 | 3 | | 197 | 1 | 2 | 2 | | | | | | |
| 146 | 1 | | 1 | | | | | | | 198 | 1 | 1 | 1 | 2 | 2 | 3 | | | |
| 147 | 1 | 1 | 1 | 2 | 2 | 2 | 2 | 3 | | 199 | 1 | 2 | 2 | | | | | | |
| 148 | 1 | | 1 | | 1 | | 1 | | 1 | 200 | 1 | 1 | 1 | 2 | 2 | 3 | | | |
| 149 | 1 | 1 | 1 | 1 | 1 | 2 | 2 | | | 201 | 1 | 2 | 2 | | | | | | |
| 150 | 1 | | 1 | | 1 | | 1 | | 1 | 202 | 1 | 1 | 1 | 1 | 3 | | | | |
| 151 | 1 | | 1 | | 1 | | 1 | | 1 | 203 | 1 | 2 | 2 | | | | | | |
| 152 | 1 | 1 | 1 | 1 | 2 | 2 | 3 | | | 204 | 1 | 1 | 1 | 1 | 1 | 1 | 3 | | |
| 153 | 1 | 3 | | | | | | | | 205 | 1 | 2 | 2 | | | | | | |
| 154 | 1 | 2 | 2 | 3 | | | | | | 206 | 1 | 1 | 1 | 1 | 2 | 2 | 3 | | |
| 155 | 1 | 2 | 2 | 3 | | | | | | 207 | 1 | 2 | 2 | | | | | | |
| 156 | 1 | | 1 | | 1 | | 1 | | 1 | 208 | 1 | 1 | 1 | 1 | 1 | 1 | 1 | | |

| Patient ID | time | | | | | | | | | Patient ID | time | | | | | | | | |
|---|---|---|---|---|---|---|---|---|---|---|---|---|---|---|---|---|---|---|---|
| | 0 | 1 | 2 | 3 | 4 | 5 | 6 | 7 | 8 | | 0 | 1 | 2 | 3 | 4 | 5 | 6 | 7 | 8 |
| 209 | 1 | 2 | 2 | 3 | | | | | | 261 | 1 | | 2 | | | 2 | | | |
| 210 | 1 | 2 | 2 | | | | | | | 262 | 1 | | 2 | | | | | | |
| 211 | 1 | 1 | 1 | 1 | 2 | 2 | | | | 263 | 1 | | 2 | | | | | | |
| 212 | 1 | 2 | 2 | 3 | | | | | | 264 | 1 | | 2 | | | 2 | | | |
| 213 | 1 | 1 | 1 | 2 | 2 | 3 | | | | 265 | 1 | | 2 | | | | | | |
| 214 | 1 | 2 | 2 | | | | | | | 266 | 1 | | 2 | | | | | | |
| 215 | 1 | 1 | 1 | 1 | 2 | 2 | 3 | | | 267 | 1 | | 2 | | | | | | |
| 216 | 1 | 2 | 2 | 3 | | | | | | 268 | 1 | | 2 | | | 2 | | | |
| 217 | 1 | 2 | 2 | 3 | | | | | | 269 | 1 | | 2 | | | | | | |
| 218 | 1 | 2 | 2 | | | | | | | 270 | 1 | | 2 | | | | | | |
| 219 | 1 | 2 | 2 | | | | | | | 271 | 1 | | 2 | | | | | | |
| 220 | 1 | 1 | 1 | 2 | 2 | 3 | | | | 272 | 1 | | 2 | | | | | | |
| 221 | 1 | 1 | 1 | 1 | 1 | 2 | 2 | 3 | | 273 | 1 | | 2 | | | | | | |
| 222 | 1 | 2 | 2 | 3 | | | | | | 274 | 1 | | 2 | | | | | | |
| 223 | 1 | 2 | 2 | 3 | | | | | | 275 | 1 | | 2 | | | | | | |
| 224 | 1 | 1 | 2 | 2 | 3 | | | | | 276 | 1 | | 3 | | | | | | |
| 225 | 1 | 3 | | | | | | | | 277 | 1 | | 3 | | | | | | |
| 226 | 1 | 1 | 1 | 2 | 2 | 3 | | | | 278 | 1 | | 3 | | | | | | |
| 227 | 1 | 3 | | | | | | | | 279 | 1 | | 3 | | | | | | |
| 228 | 1 | 1 | 1 | 1 | 1 | 2 | 2 | 3 | | 280 | 1 | | 3 | | | | | | |
| 229 | 1 | 2 | 2 | 3 | | | | | | 281 | 1 | | 3 | | | | | | |
| 230 | 1 | 3 | | | | | | | | 282 | 1 | | 3 | | | | | | |
| 231 | 1 | 1 | 1 | 1 | 1 | 2 | 2 | | | 283 | 1 | | 3 | | | | | | |
| 232 | 1 | 2 | 2 | | | | | | | 284 | 1 | | 3 | | | | | | |
| 233 | 1 | 1 | 1 | 1 | 1 | 1 | 1 | 3 | | 285 | 1 | | 3 | | | | | | |
| 234 | 1 | 2 | 2 | 3 | | | | | | 286 | 1 | | 4 | | | | | | |
| 235 | 1 | 2 | 2 | | | | | | | 287 | 1 | | | 1 | | | | | |
| 236 | 1 | 1 | 1 | 1 | 3 | | | | | 288 | 1 | | | 1 | | | | | |
| 237 | 1 | 2 | 2 | | | | | | | 289 | 1 | | | 1 | | | 1 | | |
| 238 | 1 | 1 | 1 | 2 | 2 | 3 | | | | 290 | 1 | | | 1 | | | 2 | | |
| 239 | 1 | 2 | 2 | | | | | | | 291 | 1 | | | 1 | | | 2 | | |
| 240 | 1 | 1 | 1 | 1 | 1 | 2 | 2 | | | 292 | 1 | | | 1 | | | 3 | | |
| 241 | 1 | 2 | 2 | | | | | | | 293 | 1 | | | 1 | | | 1 | | |
| 242 | 1 | 1 | 2 | 2 | 3 | | | | | 294 | 1 | | | 1 | | | 3 | | |
| 243 | 1 | 2 | 2 | | | | | | | 295 | 1 | | | 1 | | | 2 | | |
| 244 | 1 | 2 | 2 | | | | | | | 296 | 1 | | | 1 | | | 3 | | |
| 245 | 1 | 1 | 2 | | | 1 | 1 | | | 297 | 1 | | | 1 | | | 1 | | |
| 246 | 1 | 2 | | | | | | | | 298 | 1 | | | 1 | | | 2 | | |
| 247 | 1 | 1 | 2 | 3 | | | | | | 299 | 1 | | | 1 | | | 3 | | |
| 248 | 1 | 2 | | | | | | | | 300 | 1 | | | 1 | | | 1 | | |
| 249 | 1 | 1 | 2 | | | 4 | | | | 301 | 1 | | | 1 | | | 2 | | |
| 250 | 1 | 2 | | | 3 | | | | | 302 | 1 | | | 1 | | | 3 | | |
| 251 | 1 | 2 | | | | | | | | 303 | 1 | | | 1 | | | 3 | | |
| 252 | 1 | 2 | | | 3 | | | | | 304 | 1 | | | 2 | | | | | |
| 253 | 1 | 2 | | | 3 | | | | | 305 | 1 | | | 2 | | | | | |
| 254 | 1 | 4 | | | | | | | | 306 | 1 | | | 2 | | | | | |
| 255 | 1 | | 2 | | | 2 | | | | 307 | 1 | | | 3 | | | | | |
| 256 | 1 | | 2 | | | 2 | | | | 308 | 1 | | | 4 | | | | | |
| 257 | 1 | | 2 | | | 2 | | | | 309 | 1 | | | 4 | | | | | |
| 258 | 1 | | 2 | | | | | | | 310 | 1 | | | 4 | | | | | |
| 259 | 1 | | 2 | | | | | | | | | | | | | | | | |
| 260 | 1 | | 2 | | | | | | | | | | | | | | | | |

### 4.7.1. Steps to obtain the statistical indices:

**1- Step one:** Use the count table in each interval to get estimated rates and variance-covariance matrix in each interval.

Analyzing the rates in first interval $\Delta t = 1$ and the same procedures are repeated for $\Delta t = 2,3$

$\rho_3 = -.37443$ , $\rho_4 = -.20757$

$$\begin{bmatrix} v_1 \\ v_2 \\ v_3 \\ v_4 \\ v_5 \end{bmatrix} = \begin{bmatrix} -.71195 \\ -.72692 \\ -.5488 \\ -.77332 \\ -.77332 \end{bmatrix}, \rightarrow 4(550) + 4(250) \begin{bmatrix} -.71195 \\ -.72692 \\ -.5488 \\ -.77332 \\ -.77332 \end{bmatrix} = \begin{bmatrix} -2278.24439 \\ -2326.14092 \\ -1756.17225 \\ -2474.62015 \\ -2474.62015 \end{bmatrix} = scaled\ S(\theta)$$

$$scaled\ S(\theta)[scaled\ S(\theta)]^T = M(\theta)$$

$$M(\theta) = \begin{bmatrix} 5190397 & 5299517 & 4000989.6 & 5637789 & 5637789 \\ 5299517 & 5410932 & 4085104 & 5756315 & 5756315 \\ 4000989.6 & 4085104 & 3084141 & 4345859 & 4345859 \\ 5637789 & 5756315 & 4345859 & 6123745 & 6123745 \\ 5637789 & 5756315 & 4345859 & 6123745 & 6123745 \end{bmatrix}$$

$$M(\theta)\ is\ multiplied\ by\ \left(\frac{550^2}{330} + \frac{550^2}{163} + \frac{550^2}{45} + \frac{550^2}{12} + \frac{250^2}{5} + \frac{250^2}{185} + \frac{250^2}{45} + \frac{250^2}{15}\right) = 53096.45 \cong 53096$$

$$M(\theta) = \begin{bmatrix} 2.76E+11 & 2.81E+11 & 2.12E+11 & 2.99E+11 & 2.99E+11 \\ 2.81E+11 & 2.87E+11 & 2.17E+11 & 3.06E+11 & 3.06E+11 \\ 2.12E+11 & 2.17E+11 & 1.64E+11 & 2.31E+11 & 2.31E+11 \\ 2.99E+11 & 3.06E+11 & 2.31E+11 & 3.25E+11 & 3.25E+11 \\ 2.99E+11 & 3.06E+11 & 2.31E+11 & 3.25E+11 & 3.25E+11 \end{bmatrix}$$

$$[scaled\ M(\theta)]^{-1} = \begin{bmatrix} 1.34E-09 & -1.6E-09 & 3.78E-10 & 0 & 0 \\ 6.14E-10 & -3.1E-09 & 3.27E-09 & 0 & 0 \\ -2.6E-09 & 6.14E-09 & -4.82E-09 & 0 & 0 \\ 0 & 0 & 0 & 0 & 0 \\ 0 & 0 & 0 & 0 & 0 \end{bmatrix}$$

$$[scaled\ M(\theta)]^{-1} \times scaled\ S(\theta) = [scaled\ M(\theta)]^{-1} \times scaled\ S(\theta)$$

$$\theta_1 = \theta_0 + [scaled\ M(\theta)]^{-1} scaled\ S(\theta) = \begin{bmatrix} .30000001 \\ .02200007 \\ .0200001 \\ .18 \\ .06 \end{bmatrix}$$

Repeating this procedure for $\Delta t = 2\ and\ \Delta t = 3$ will give the following vectors respectively (substitute for t=2 and t=3 in their intervals):

$$for\ \Delta t = 2\ \rightarrow \hat{\vec{\theta}} = \begin{bmatrix} .27 \\ .009 \\ .05 \\ .333 \\ .103 \end{bmatrix} obtained\ in\ the\ first\ iteration\ with\ a\ difference\ from\ the\ initial\ values\ = \begin{bmatrix} 1.42E-09 \\ -7.1E-08 \\ -9.6E-08 \\ 0 \\ 0 \end{bmatrix}$$

$$for\ \Delta t = 3 \rightarrow \hat{\vec{\theta}} = \begin{bmatrix} .206172 \\ .077985 \\ .091339 \\ .273 \\ .091 \end{bmatrix} obtained\ in\ second\ iteration\ with\ a\ difference\ from\ the\ initial\ values = \begin{bmatrix} 1.17E-03 \\ 9.87E-04 \\ 3.40E-04 \\ 0 \\ 0 \end{bmatrix}$$

As noted from this procedure in all time intervals, the initial values are almost the estimated values regardless of the interval.

observed counts of transitions during time interval Δt=1

| | state 1 | state 2 | state 3 | state 4 | total |
|---|---|---|---|---|---|
| state 1 | 330 | 163 | 45 | 12 | 550 |
| state 2 | 5 | 185 | 45 | 15 | 250 |
| state 3 | 0 | 0 | 0 | 0 | 0 |
| state 4 | 0 | 0 | 0 | 0 | 0 |

initial rates

| | |
|---|---|
| lambda12 | 0.3 |
| lambda14 | 0.022 |
| mu21 | 0.02 |
| lambda23 | 0.18 |
| lambda24 | 0.06 |

initial Q matrix

| | | | |
|---|---|---|---|
| -0.322 | 0.3 | 0 | 0.022 |
| 0.02 | -0.26 | 0.18 | 0.06 |
| 0 | 0 | 0 | 0 |
| 0 | 0 | 0 | 0 |

step 1: calculate eigenvalues for the above initial Q matrix Δt=1 :

| |
|---|
| -0.37443 |
| -0.20757 |

step 2 : calculate t*e^(eigenvalue*t)*partial differentiation of eigenvalue with respect to specific theta or rate

| |
|---|
| -0.71195 |
| -0.72692 |
| -0.5488 |
| -0.77332 |
| -0.77332 |

step 3 : scale the above score function by a factor equals 4(550)+4(250)=3200

| |
|---|
| -2278.244 |
| -2326.141 |
| -1756.172 |
| -2474.62 |
| -2474.62 |

step 4 : multiply the scaled score function with the transposed scaled score function to get the hessian matrix

| | | | | |
|---|---|---|---|---|
| 5190397 | 5299517 | 4000990 | 5637789 | 5637789 |
| 5299517 | 5410932 | 4085104 | 5756315 | 5756315 |
| 4000990 | 4085104 | 3084141 | 4345859 | 4345859 |
| 5637789 | 5756315 | 4345859 | 6123745 | 6123745 |
| 5637789 | 5756315 | 4345859 | 6123745 | 6123745 |

step 5 : scale the above hessian matrix by a factor equals 53096.45 ≈53096

| | | | | |
|---|---|---|---|---|
| 2.76E+11 | 2.81E+11 | 2.12E+11 | 2.99E+11 | 2.99E+11 |
| 2.81E+11 | 2.87E+11 | 2.17E+11 | 3.06E+11 | 3.06E+11 |
| 2.12E+11 | 2.17E+11 | 1.64E+11 | 2.31E+11 | 2.31E+11 |
| 2.99E+11 | 3.06E+11 | 2.31E+11 | 3.25E+11 | 3.25E+11 |
| 2.99E+11 | 3.06E+11 | 2.31E+11 | 3.25E+11 | 3.25E+11 |

step 6 : invert the scaled hessian matrix

| | | | | |
|---|---|---|---|---|
| 1.34E-09 | -1.60E-09 | 3.78E-10 | 0.00E+00 | 0.00E+00 |
| 6.14E-10 | -3.10E-09 | 3.27E-09 | 0.00E+00 | 0.00E+00 |
| -2.60E-09 | 6.14E-09 | -4.82E-09 | 0.00E+00 | 0.00E+00 |
| 0.00E+00 | 0.00E+00 | 0.00E+00 | 0.00E+00 | 0.00E+00 |
| 0.00E+00 | 0.00E+00 | 0.00E+00 | 0.00E+00 | 0.00E+00 |

step 7 : multiply the inverted scaled hessian matrix with the scaled score function

| |
|---|
| 5.53E-09 |
| 6.55E-08 |
| 1.04E-07 |
| 0.00E+00 |
| 0.00E+00 |

step 8 : apply the Quasi-Newton formula( add the vector of initial rate values with the above calculated vector )

| |
|---|
| 3.00E-01 |
| 2.20E-02 |
| 2.00E-02 |
| 1.80E-01 |
| 6.00E-02 |

observed counts of transitions during time interval Δt=2

| | state 1 | state 2 | state 3 | state 4 | total |
|---|---|---|---|---|---|
| state 1 | 70 | 30 | 10 | 1 | 111 |
| state 2 | 2 | 20 | 13 | 4 | 39 |
| state 3 | 0 | 0 | 0 | 0 | 0 |
| state 4 | 0 | 0 | 0 | 0 | 0 |

initial rates

| | |
|---|---|
| lambda12 | 0.27 |
| lambda14 | 0.009 |
| mu21 | 0.05 |
| lambda23 | 0.333 |
| lambda24 | 0.103 |

initial Q matrix

| | | | |
|---|---|---|---|
| -0.279 | 0.27 | 0 | 0.009 |
| 0.05 | -0.486 | 0.333 | 0.103 |
| 0 | 0 | 0 | 0 |
| 0 | 0 | 0 | 0 |

step 1: calculate eigenvalues for the above initial Q matrix Δt=2 :

| |
|---|
| -0.5381 |
| -0.2269 |

step 2 : calculate t*e^(eigenvalue*t)*partial differentiation of eigenvalue with respect to specific theta or rate

| |
|---|
| -1.0773 |
| -1.17187 |
| -0.26961 |
| -0.78033 |
| -0.78033 |

step 3 : scale the above score function by a factor equals 4(111)+4(39)=600

| |
|---|
| -646.378 |
| -703.124 |
| -161.769 |
| -468.196 |
| -468.196 |

step 4 : multiply the scaled score function with the transposed scaled score function to get the hessian matrix

| | | | | |
|---|---|---|---|---|
| 417804.4 | 454483.7 | 104564 | 302631.7 | 302631.7 |
| 454483.7 | 494383 | 113744 | 329199.8 | 329199.8 |
| 104563.8 | 113743.5 | 26169.2 | 75739.57 | 75739.57 |
| 302631.7 | 329199.8 | 75739.6 | 219207.6 | 219207.6 |
| 302631.7 | 329199.8 | 75739.6 | 219207.6 | 219207.6 |

step 5 : scale the above hessian matrix by a factor equals 15476.614

| | | | | |
|---|---|---|---|---|
| 6.47E+09 | 7.03E+09 | 1.60E+09 | 4.68E+09 | 4.68E+09 |
| 7.03E+09 | 7.65E+09 | 1.76E+09 | 5.09E+09 | 5.09E+09 |
| 1.60E+09 | 1.76E+09 | 4.10E+08 | 1.20E+09 | 1.20E+09 |
| 4.68E+09 | 5.09E+09 | 1.20E+09 | 3.39E+09 | 3.39E+09 |
| 4.68E+09 | 5.09E+09 | 1.20E+09 | 3.39E+09 | 3.39E+09 |

step 6 : invert the scaled hessian matrix

| | | | | |
|---|---|---|---|---|
| -1.38E-08 | -8.62E-08 | 4.30E-07 | 0.00E+00 | 0.00E+00 |
| -8.62E-08 | 8.62E-08 | -2.98E-08 | 0.00E+00 | 0.00E+00 |
| 4.30E-07 | -2.98E-08 | -1.59E-06 | 0.00E+00 | 0.00E+00 |
| 0.00E+00 | 0.00E+00 | 0.00E+00 | 0.00E+00 | 0 |
| 0.00E+00 | 0.00E+00 | 0.00E+00 | 0.00E+00 | 0 |

step 7 : multiply the inverted scaled hessian matrix with the scaled score function

| |
|---|
| 1.42E-09 |
| -7.10E-08 |
| -9.60E-08 |
| 0.00E+00 |
| 0.00E+00 |

step 8 : apply the Quasi-Newton formula( add the vector of initial rate values with the above calculated vector )

| |
|---|
| 2.70E-01 |
| 9.00E-03 |
| 5.00E-02 |
| 3.33E-01 |
| 1.03E-01 |

observed counts of transitions during time interval Δt=3

| | state 1 | state 2 | state 3 | state 4 | total |
|---|---|---|---|---|---|
| state 1 | 21 | 8 | 7 | 3 | 39 |
| state 2 | 1 | 6 | 3 | 1 | 11 |
| state 3 | 0 | 0 | 0 | 0 | 0 |
| state 4 | 0 | 0 | 0 | 0 | 0 |

first iteration :

initial rates

| | |
|---|---|
| lambda12 | 0.205 |
| lambda14 | 0.077 |
| mu21 | 0.091 |
| lambda23 | 0.273 |
| lambda24 | 0.091 |

initial Q matrix

| | | | |
|---|---|---|---|
| -0.282 | 0.205 | 0 | 0.077 |
| 0.091 | -0.455 | 0.273 | 0.091 |
| 0 | 0 | 0 | 0 |
| 0 | 0 | 0 | 0 |

step 1: calculate eigenvalues for the above initial Q matrix Δt=3 :

| |
|---|
| -0.53017 |
| -0.20683 |

step 2 : calculate t*e^(eigenvalue*t)*partial differentiation of eigenvalue with respect to specific theta or rate

| |
|---|
| -1.09831 |
| -1.3802 |
| -0.2093 |
| -0.84431 |
| -0.84431 |

step 3 : scale the above score function by a factor equals 4(39)+4(11)=200

| |
|---|
| -219.663 |
| -276.039 |
| -41.8608 |
| -168.862 |
| -168.862 |

step 4 : multiply the scaled score function with the transposed scaled score function to get the hessian matrix

| | | | | |
|---|---|---|---|---|
| 48251.83 | 60635.61 | 9195.258 | 37092.81 | 37092.81 |
| 60635.61 | 76197.69 | 11555.21 | 46612.64 | 46612.64 |
| 9195.258 | 11555.21 | 1752.323 | 7068.705 | 7068.705 |
| 37092.81 | 46612.64 | 7068.705 | 28514.49 | 28514.49 |
| 37092.81 | 46612.64 | 7068.705 | 28514.49 | 28514.49 |

step 5 : scale the above hessian matrix by a factor equals 1289.338

| | | | | |
|---|---|---|---|---|
| 6.22E+07 | 7.82E+07 | 1.19E+07 | 4.78E+07 | 4.78E+07 |
| 7.82E+07 | 9.82E+07 | 1.49E+07 | 6.01E+07 | 6.01E+07 |
| 1.19E+07 | 1.49E+07 | 2.26E+06 | 9.11E+06 | 9.11E+06 |
| 4.78E+07 | 6.01E+07 | 9.11E+06 | 3.68E+07 | 3.68E+07 |
| 4.78E+07 | 6.01E+07 | 9.11E+06 | 3.68E+06 | 3.68E+07 |

step 6 : invert the scaled hessian matrix

| | | | | |
|---|---|---|---|---|
| 1.15E+00 | -6.20E-01 | -1.97E+00 | 0.00E+00 | 0.00E+00 |
| -6.20E-01 | 2.10E-01 | 1.87E+00 | 0.00E+00 | 0.00E+00 |
| -1.97E+00 | 1.87E+00 | -1.98E+00 | 0.00E+00 | 0.00E+00 |
| 0.00E+00 | 0.00E+00 | 0.00E+00 | 0.00E+00 | 0 |
| 0.00E+00 | 0.00E+00 | 0.00E+00 | 0.00E+00 | 0 |

step 7 : multiply the inverted scaled hessian matrix with the scaled score function

| |
|---|
| 1.17E-03 |
| 9.87E-04 |
| 3.40E-04 |
| 0.00E+00 |
| 0.00E+00 |

step 8 : apply the Quasi-Newton formula( add the vector of initial rate values with the above calculated vector )

| |
|---|
| 2.06E-01 |
| 7.79E-03 |
| 9.13E-02 |
| 2.73E-01 |
| 9.10E-02 |

second iteration :

initial theta from previous iteration

| |
|---|
| 0.206174 |
| 0.077987 |
| 0.09134 |
| 0.273 |
| 0.091 |

final inverted scaled hessian matrix :

| | | | | |
|---|---|---|---|---|
| 1.229517 | -0.92897 | -0.317 | 0 | 0 |
| -0.92897 | 0.7567 | -0.12256 | 0 | 0 |
| -0.317 | -0.12256 | 2.47316 | 0 | 0 |
| 0 | 0 | 0 | 0 | 0 |
| 0 | 0 | 0 | 0 | 0 |

step 1:

| |
|---|
| -0.53148 |
| -0.20802 |

step 2:

| |
|---|
| -1.09044 |
| -1.37232 |
| -0.20777 |
| -0.84405 |
| -0.84405 |

step 3:

| |
|---|
| -218.087 |
| -274.464 |
| -41.5547 |
| -168.809 |
| -168.809 |

step 5:

| | | | | |
|---|---|---|---|---|
| 61323578 | 77176056 | 11684702 | 47467151 | 47467151 |
| 77176056 | 97126487 | 14705261 | 59737668 | 59737668 |
| 11684702 | 14705261 | 2226424 | 9044474 | 9044474 |
| 47467151 | 59737668 | 9044474 | 36741666 | 36741666 |
| 47467151 | 59737668 | 9044474 | 36741666 | 36741666 |

step 6;

| | | | | |
|---|---|---|---|---|
| 1.2295 | -0.92897 | -0.317 | 0 | 0 |
| -0.92897 | 0.7567 | -0.12256 | 0 | 0 |
| -0.317 | -0.12256 | 2.47316 | 0 | 0 |
| 0 | 0 | 0 | 0 | 0 |
| 0 | 0 | 0 | 0 | 0 |

step 7:

| |
|---|
| -0.00032 |
| 0.002558 |
| 0.000447 |
| 0 |
| 0 |

step 8 :

| |
|---|
| 0.206 |
| 0.078 |
| 0.091 |
| 0.273 |
| 0.091 |

**2-step two**: get final rate and variance-covariance matrix

If the scaled score function in each iteration is weighted according to the contribution of the counts of transitions in this interval to the whole number of transitions (1000 transitions) and summed up, this will give

$$(.8)\begin{bmatrix} .3 \\ .022 \\ .02 \\ .18 \\ .06 \end{bmatrix} + (.15)\begin{bmatrix} .3 \\ .022 \\ .02 \\ .18 \\ .06 \end{bmatrix} + (.05)\begin{bmatrix} .206 \\ .078 \\ .091 \\ .273 \\ .091 \end{bmatrix} = \begin{bmatrix} .2908 \\ .02285 \\ .02805 \\ .2076 \\ .068 \end{bmatrix}$$

Also the weighted sum of the inversed scaled hessian matrix should be used as the variance -covariance matrix of parameter $\theta$

$$[scaled\ M(\theta)]^{-1} = \begin{bmatrix} .061475 & -.04645 & -.01585 & 0 & 0 \\ -.04645 & .037836 & -.00613 & 0 & 0 \\ -.01585 & -.00613 & .123658 & 0 & 0 \\ 0 & 0 & 0 & 0 & 0 \\ 0 & 0 & 0 & 0 & 0 \end{bmatrix}$$

| final estimated rates from first interval | weighted estimated rates from first interval |
|---|---|
| 3.00E-01 | 0.24 |
| 2.20E-02 | 0.0176 |
| 2.00E-02 | 0.016 |
| 1.80E-01 | 0.144 |
| 6.00E-02 | 0.048 |
| | |
| **final estimated rates from second interval** | **weighted estimated rates from second interval** |
| 2.70E-01 | 0.0405 |
| 9.00E-03 | 0.00135 |
| 5.00E-02 | 0.0075 |
| 3.33E-01 | 0.04995 |
| 1.03E-01 | 0.01545 |
| | |
| **final estimated rates from third interval** | **weighted estimated rates from third interval** |
| 2.06E-01 | 0.0103 |
| 7.80E-02 | 0.0039 |
| 9.10E-02 | 0.00455 |
| 2.73E-01 | 0.01365 |
| 9.10E-02 | 0.00455 |
| | |
| | **weighted sum of estimated rates over the 3 intervals** |
| | 0.2908 |
| | 0.02285 |
| | 0.02805 |
| | 0.2076 |
| | 0.068 |

inverted scaled hessian matrix from first interval calc.

| | | | | |
|---|---|---|---|---|
| 1.34E-09 | -1.60E-09 | 3.78E-10 | 0.00E+00 | 0.00E+00 |
| 6.14E-10 | -3.10E-09 | 3.27E-09 | 0.00E+00 | 0.00E+00 |
| -2.60E-09 | 6.14E-09 | -4.82E-09 | 0.00E+00 | 0.00E+00 |
| 0.00E+00 | 0.00E+00 | 0.00E+00 | 0.00E+00 | 0.00E+00 |
| 0.00E+00 | 0.00E+00 | 0.00E+00 | 0.00E+00 | 0.00E+00 |

weightd inverted scaled hessian matrix from first interval

| | | | | |
|---|---|---|---|---|
| 1.07E-09 | -1.28E-09 | 3.02E-10 | 0.00E+00 | 0.00E+00 |
| 4.91E-10 | -2.48E-09 | 2.62E-09 | 0.00E+00 | 0.00E+00 |
| -2.08E-09 | 4.91E-09 | -3.86E-09 | 0.00E+00 | 0.00E+00 |
| 0.00E+00 | 0.00E+00 | 0.00E+00 | 0.00E+00 | 0.00E+00 |
| 0.00E+00 | 0.00E+00 | 0.00E+00 | 0.00E+00 | 0.00E+00 |

inverted scaled hessian matrix from second interval calc.

| | | | | |
|---|---|---|---|---|
| -1.38E-08 | -8.62E-08 | 4.30E-07 | 0.00E+00 | 0.00E+00 |
| -8.62E-08 | 8.62E-08 | -2.98E-08 | 0.00E+00 | 0.00E+00 |
| 4.30E-07 | -2.98E-08 | -1.59E-06 | 0.00E+00 | 0.00E+00 |
| 0.00E+00 | 0.00E+00 | 0.00E+00 | 0.00E+00 | 0 |
| 0.00E+00 | 0.00E+00 | 0.00E+00 | 0.00E+00 | 0 |

weighted inverted scaled hessian matrix from second interval

| | | | | |
|---|---|---|---|---|
| -2.07E-09 | -1.29E-08 | 6.45E-08 | 0.00E+00 | 0.00E+00 |
| -1.29E-08 | 1.29E-08 | -4.47E-09 | 0.00E+00 | 0.00E+00 |
| 6.45E-08 | -4.47E-09 | -2.38E-07 | 0.00E+00 | 0.00E+00 |
| 0.00E+00 | 0.00E+00 | 0.00E+00 | 0.00E+00 | 0.00E+00 |
| 0.00E+00 | 0.00E+00 | 0.00E+00 | 0.00E+00 | 0.00E+00 |

inverted scaled hessian matrix from third inteval calc.

| | | | | |
|---|---|---|---|---|
| 1.23E+00 | -0.92897 | -0.317 | 0 | 0 |
| -9.29E-01 | 7.57E-01 | -1.23E-02 | 0 | 0 |
| -3.17E-01 | -0.12256 | 2.47316 | 0 | 0 |
| 0 | 0 | 0 | 0 | 0 |
| 0 | 0 | 0 | 0 | 0 |

weighted inverted scaled hessian matrix from third interval

| | | | | |
|---|---|---|---|---|
| 0.061476 | -0.04645 | -0.01585 | 0 | 0 |
| -0.04645 | 0.037836 | -0.00061 | 0 | 0 |
| -0.01585 | -0.00613 | 0.123658 | 0 | 0 |
| 0 | 0 | 0 | 0 | 0 |
| 0 | 0 | 0 | 0 | 0 |

weighted sum of inverted scaled hessian matricesin 3 interval

| | | | | |
|---|---|---|---|---|
| 0.061476 | -0.04645 | -0.01585 | 0 | 0 |
| -0.04645 | 0.037836 | -0.00613 | 0 | 0 |
| -0.01585 | -0.00613 | 0.12658 | 0 | 0 |
| 0 | 0 | 0 | 0 | 0 |
| 0 | 0 | 0 | 0 | 0 |

**3-step three: Calculating the Mean Sojourn Time and its Variance:**

It is the average amount of time spent by a patient in the state:

$$E(s_1) = \frac{1}{\lambda_{12} + \lambda_{14}} = \frac{1}{.2908 + .02285} = 3.19 year$$

$$E(s_2) = \frac{1}{\mu_{21} + \lambda_{23} + \lambda_{24}} = \frac{1}{.02805 + .2076 + .068} = 3.29\ year\ ,$$

$$var(s_i) = \left[\left(q_{ii}(\hat{\theta})\right)^{-2}\right]^2 \sum_{h=1}^{5}\sum_{g=1}^{5} \left[\frac{\partial q_{ii}}{\partial \theta_h}\right]^T \ [\mathrm{M}(\theta)]^{-1}|_{\theta=\hat{\theta}} \frac{\partial q_{ii}}{\partial \theta_g}$$

$$where\ [\mathrm{M}(\theta)]^{-1}\ is\ the\ weighted\ sum\ of\ inverted\ scaled\ hessian\ matrix$$

$$var(s_1) = \frac{1}{(.2908 + .02285)^4}[-1 \quad -1 \quad -1 \quad -1 \quad -1][\mathrm{M}(\theta)]^{-1}|_{\theta=\hat{\theta}} \begin{bmatrix} -1 \\ -1 \\ -1 \\ -1 \\ -1 \end{bmatrix} = 8.898$$

$$var(s_2) = \frac{1}{(.02805 + .2076 + .068)^4}[-1 \quad -1 \quad -1 \quad -1 \quad -1][\mathrm{M}(\theta)]^{-1}|_{\theta=\hat{\theta}} \begin{bmatrix} -1 \\ -1 \\ -1 \\ -1 \\ -1 \end{bmatrix} = 10.129$$

**4-step four:** State Probability Distribution and expected number of patients at specific time point

Once the rate matrix is obtained, these estimated rates are substituted into the calculated Pdfs from the solved differential equations to get the state probability distribution at any point in time as well as the expected number of patients.

Studying a cohort of 3000 patients with the initial distribution $[.7 \quad .3 \quad 0 \quad 0]$, and initial numbers of patients in each state are $[2100 \quad 900 \quad 0 \quad 0]$.

At 1 year the state probability distribution is approximately:

$$P(1) = [.7 \quad .3 \quad 0 \quad 0]\begin{bmatrix} .734 & .214 & .025 & .027 \\ .021 & .741 & .179 & .059 \\ 0 & 0 & 1 & 0 \\ 0 & 0 & 0 & 1 \end{bmatrix} = [.52 \quad .372 \quad .071 \quad .037]$$

And the expected numbers of patients in each state is:

$$[2100 \quad 900 \quad 0 \quad 0]\begin{bmatrix} .734 & .214 & .025 & .027 \\ .021 & .741 & .179 & .059 \\ 0 & 0 & 1 & 0 \\ 0 & 0 & 0 & 1 \end{bmatrix} = [1559 \quad 1117 \quad 214 \quad 110]$$

At 20 years the state probability distribution is approximately:

$$P(20) = [.7 \quad .3 \quad 0 \quad 0]\begin{bmatrix} .006 & .019 & .675 & .3 \\ .002 & .006 & .742 & .25 \\ 0 & 0 & 1 & 0 \\ 0 & 0 & 0 & 1 \end{bmatrix} = [.0048 \quad .0151 \quad .6951 \quad .285]$$

And the expected numbers of patients in each state is:

$$[2100 \quad 900 \quad 0 \quad 0]\begin{bmatrix} .006 & .019 & .675 & .3 \\ .002 & .006 & .742 & .25 \\ 0 & 0 & 1 & 0 \\ 0 & 0 & 0 & 1 \end{bmatrix} = [15 \quad 45 \quad 2085 \quad 855]$$

At 60 years the state probability distribution is approximately:

$$P(60) = [.7 \quad .3 \quad 0 \quad 0]\begin{bmatrix} 0 & 0 & .7 & .3 \\ 0 & 0 & .75 & .25 \\ 0 & 0 & 1 & 0 \\ 0 & 0 & 0 & 1 \end{bmatrix} = [0 \quad 0 \quad .715 \quad .285]$$

And the expected numbers of patients in each state is:

$$[2100 \quad 900 \quad 0 \quad 0]\begin{bmatrix} 0 & 0 & .7 & .3 \\ 0 & 0 & .75 & .25 \\ 0 & 0 & 1 & 0 \\ 0 & 0 & 0 & 1 \end{bmatrix} = [0 \quad 0 \quad 2145 \quad 855]$$

**Asymptotic Covariance of the Stationary Distribution :**At 60 years the state probability distribution is $[0 \quad 0 \quad .715 \quad .285]$, so to calculate the asymptotic covariance of the stationary distribution follow the upcoming steps:

step 1 : calculate the state distribution at the prespecified time point :

for this model calculate the probability transition matrix at 60 years:it is approximately

| | | | |
|---|---|---|---|
| 0 | 0 | 0.7 | 0.3 |
| 0 | 0 | 0.75 | 0.25 |
| 0 | 0 | 1 | 0 |
| 0 | 0 | 0 | 1 |

for a cohort of 3000 patients with 2100 susceptible and 900 affected patients

the initial probability distribution vector is ( .7 , .3 , 0 , 0 )

multiplying this initial vector with the pdf at this specific time point

the result is this vector ( 0, 0 , .7, .3 ) which is the state probability distribution at 60 years

| |
|---|
| 0 |
| 0 |
| 0.7 |
| 0.3 |

step 2 : partialy differentiate the transposed rate matrix with respect to each lambda

the result is a vector of five ones (1 , 1 , 1 ,  1 , 1 )

| | | | | |
|---|---|---|---|---|
| 1 | 1 | 1 | 1 | 1 |

step 3 : multiply the column vector of state probability distribution with the row vector of ones

the result is a matrix, let it be C ( Ɵ)  :

| | | | | |
|---|---|---|---|---|
| 0 | 0 | 0 | 0 | 0 |
| 0 | 0 | 0 | 0 | 0 |
| 0.7 | 0.7 | 0.7 | 0.7 | 0.7 |
| 0.3 | 0.3 | 0.3 | 0.3 | 0.3 |

step 4 : calculate the pseudoinverse of transposed rate matrix ( final estimated rate matrix)

using the singular value decomposition by multiplying the following 3 matrices in this order  :

| | |
|---|---|
| 0.62517014 | 0.7804885 |
| 0.7804885 | -0.62517 |
| 0 | 0 |
| 0 | 0 |
| 3.947297 | 0 |
| 0 | 1.9611806 |

| | | | |
|---|---|---|---|
| -0.68758724 | -0.2178745 | 0.63958 | 0.265883 |
| -0.51448875 | 0.81741806 | -0.25453 | -0.0484 |

the result of the above multiplication is the following matrix :

| | | | |
|---|---|---|---|
| -2.484298 | 0.71354873 | 1.18870046 | 0.58205 |
| -1.487533 | -1.67344506 | 2.28250053 | 0.87848 |
| 0 | 0 | 0 | 0 |
| 0 | 0 | 0 | 0 |

then multiply this matrix by -1 to obtain

| | | | |
|---|---|---|---|
| 2.484298 | -0.71354873 | -1.1887005 | -0.58205 |
| 1.487533 | 1.67344506 | -2.2825005 | -0.87848 |
| 0 | 0 | 0 | 0 |
| 0 | 0 | 0 | 0 |

step 5 : multiply the matrices from step 3 and step 4 to obtain A(Ɵ) matrix :

| | | | |
|---|---|---|---|
| 2.484298 | -0.71354873 | -1.1887005 | -0.58205 |
| 1.487533 | 1.67344506 | -2.2825005 | -0.87848 |
| 0 | 0 | 0 | 0 |
| 0 | 0 | 0 | 0 |

| | | | | |
|---|---|---|---|---|
| 0 | 0 | 0 | 0 | 0 |
| 0 | 0 | 0 | 0 | 0 |
| 0.7 | 0.7 | 0.7 | 0.7 | 0.7 |
| 0.3 | 0.3 | 0.3 | 0.3 | 0.3 |

the result is A(Ɵ) matrix

| | | | | |
|---|---|---|---|---|
| -1.0142211 | -1.0142211 | -1.0142211 | -1.0142211 | -1.0142211 |
| -1.877929 | -1.877929 | -1.877929 | -1.877929 | -1.877929 |
| 0 | 0 | 0 | 0 | 0 |
| 0 | 0 | 0 | 0 | 0 |

step 6 : multiply ( A ) matrix with inversed hessian matrix ( M ) with the transposed of (A) matrix :

| | | | | |
|---|---|---|---|---|
| -1.0142211 | -1.0142211 | -1.0142211 | -1.0142211 | -1.0142211 |
| -1.877929 | -1.877929 | -1.877929 | -1.877929 | -1.877929 |
| 0 | 0 | 0 | 0 | 0 |
| 0 | 0 | 0 | 0 | 0 |
| 0.061476 | -0.04645 | -0.01585 | 0 | 0 |
| -0.04645 | 0.037836 | -0.00613 | 0 | 0 |
| -0.01585 | -0.00613 | 0.12658 | 0 | 0 |
| 0 | 0 | 0 | 0 | 0 |
| 0 | 0 | 0 | 0 | 0 |

| | | | |
|---|---|---|---|
| -1.0142211 | -1.877929 | 0 | 0 |
| -1.0142211 | -1.877929 | 0 | 0 |
| -1.0142211 | -1.877929 | 0 | 0 |
| -1.0142211 | -1.877929 | 0 | 0 |
| -1.0142211 | -1.877929 | 0 | 0 |

to obtain the asymptotic covariance matrix  :

| | | | |
|---|---|---|---|
| 0.0885764 | 0.16400785 | 0 | 0 |
| 0.1640078 | 0.30367644 | 0 | 0 |
| 0 | 0 | 0 | 0 |
| 0 | 0 | 0 | 0 |

$C(\theta)$ matrix is calculated as in the following steps:

$$C(\theta) = \pi(\theta)\left[\frac{\partial}{\partial\theta_h}Q'\right]^T = \begin{bmatrix} 0 \\ 0 \\ .7 \\ .3 \end{bmatrix} [1 \quad 1 \quad 1 \quad 1 \quad 1] = \begin{bmatrix} 0 & 0 & 0 & 0 & 0 \\ 0 & 0 & 0 & 0 & 0 \\ .7 & .7 & .7 & .7 & .7 \\ .3 & .3 & .3 & .3 & .3 \end{bmatrix}$$

Then $\left[\frac{\partial}{\partial\theta_h}\pi\right] = -[Q']^{-1}C(\theta)$, $-[Q']^{-1}$ is calculated taking into account that $Q'$is a singular matrix and its inverse (the pseudoinverse) is obtained via singular value decomposition (SVD).

$$Q' = \begin{bmatrix} -.31365 & .02805 & 0 & 0 \\ .2908 & -.30365 & 0 & 0 \\ 0 & .2076 & 0 & 0 \\ .02285 & .068 & 0 & 0 \end{bmatrix}$$ , by SVD $Q' = U\Sigma V^T$

$$Q' = \begin{bmatrix} -.68758 & -.51449 & 0 & 0 \\ -.21787 & .81742 & 0 & 0 \\ .63957 & -.25453 & 1 & 0 \\ .26588 & -.04839 & 0 & 1 \end{bmatrix} \begin{bmatrix} .25334 & 0 & 0 & 0 \\ 0 & .50989 & 0 & 0 \\ 0 & 0 & 0 & 0 \\ 0 & 0 & 0 & 0 \end{bmatrix} \begin{bmatrix} .62517 & .78049 & 0 & 0 \\ .78049 & -.62517 & 0 & 0 \\ 0 & 0 & 1 & 0 \\ 0 & 0 & 0 & 1 \end{bmatrix}$$

$$[Q']^+ = \begin{bmatrix} .62517 & .78049 \\ .78049 & -.62517 \\ 0 & 0 \\ 0 & 0 \end{bmatrix} \begin{bmatrix} 3.9473 & 0 \\ 0 & 1.96118 \end{bmatrix} \begin{bmatrix} -.68758 & -.2178 & .63958 & .26588 \\ -.51449 & .81742 & -.25453 & -.0484 \end{bmatrix}$$

$$[Q']^+ = \begin{bmatrix} -2.48429 & .71355 & 1.1887 & .58205 \\ -1.48753 & -1.6734 & 2.2825 & .87848 \\ 0 & 0 & 0 & 0 \\ 0 & 0 & 0 & 0 \end{bmatrix}$$

$$A(\theta) = -[Q']^+C(\theta) = \begin{bmatrix} -1.01422 & -1.01422 & -1.01422 & -1.01422 & -1.01422 \\ -1.8779 & -1.8779 & -1.8779 & -1.8779 & -1.8779 \\ 0 & 0 & 0 & 0 & 0 \\ 0 & 0 & 0 & 0 & 0 \end{bmatrix}$$

$$A(\theta)[scaled\ M(\theta)]^{-1}[A(\theta)]^T = \begin{bmatrix} .088589 & .16401 & 0 & 0 \\ .16401 & .30368 & 0 & 0 \\ 0 & 0 & 0 & 0 \\ 0 & 0 & 0 & 0 \end{bmatrix}$$

**5-step five: Life Expectancy of the Patient (mean time to absorption):**

$$E(\tau_{ik}) = [B]^{-1}Z = \begin{bmatrix} -3.48691 & -3.33935 \\ -.32212 & -3.60174 \end{bmatrix} \begin{bmatrix} -.69325 & -.30675 \\ -.74772 & -.25228 \end{bmatrix} = \begin{bmatrix} 4.9142 & 1.9121 \\ 2.9164 & 1.0074 \end{bmatrix}$$

$E(\tau_{13}) = 4.9142\ years, E(\tau_{14}) = 1.9121\ years, E(\tau_{23}) = 2.9164\ years, E(\tau_{24}) = 1.0074\ years$

Mean time spent by the susceptible individuals in state 1 is approximately 3 years and 2 months, and in state 2 is approximately 3 years and 3.5 months. According to American Association for the study of Liver Disease (Chalasani et al. 2018),the most common cause of death in patients with NAFLD is cardiovascular disease (CVD) independent of other metabolic comorbidities , whether the liver-related mortality is the second or third cause of death among patients with NAFLD. Cancer-related mortality is among the top three causes of death in subjects with NAFLD. As shown from the calculations; mean time to absorption can be classified into : mean time from state 1( susceptible individuals with risk factors) to state 3 ( liver-related mortality)is approximately 5 years, while the mean time from state 1 to state 4 ( for example CVD as an example for causes of death other than liver-related mortality causes) is approximately 2 years .The mean time from state 2( NAFLD) to state 3 ( liver-related mortality ) is approximately 3 years while it decreases to approximately 1 year from state 2 ( NAFLD) to state 4 ( other causes than liver-related mortality).

# Chapter Five: CTMC Analyzing NAFLD Progression (Big Model)

NAFLD is a multistage disease process consisting of 9 stages as depicted in figure 1. As shown in the figure; the patient can move across the stages of the disease process. While the remission rates are allowed from stage 4 (compensated liver cirrhosis) to the earlier stages, patient progresses to HCC and liver transplantation once he arrives to stage 5(decompensated liver cirrhosis) and remission rates are not allowed. Death state can be reached from any state. The patient can move from the first 5 stages to stage 8 (HCC) with higher rate of progression from stage 4(CC) or stage 5 (DCC) to stage 8( HCC) compared to first 3 stages. A brief definition of each stage is illustrated below the figure. The following indices are calculated :

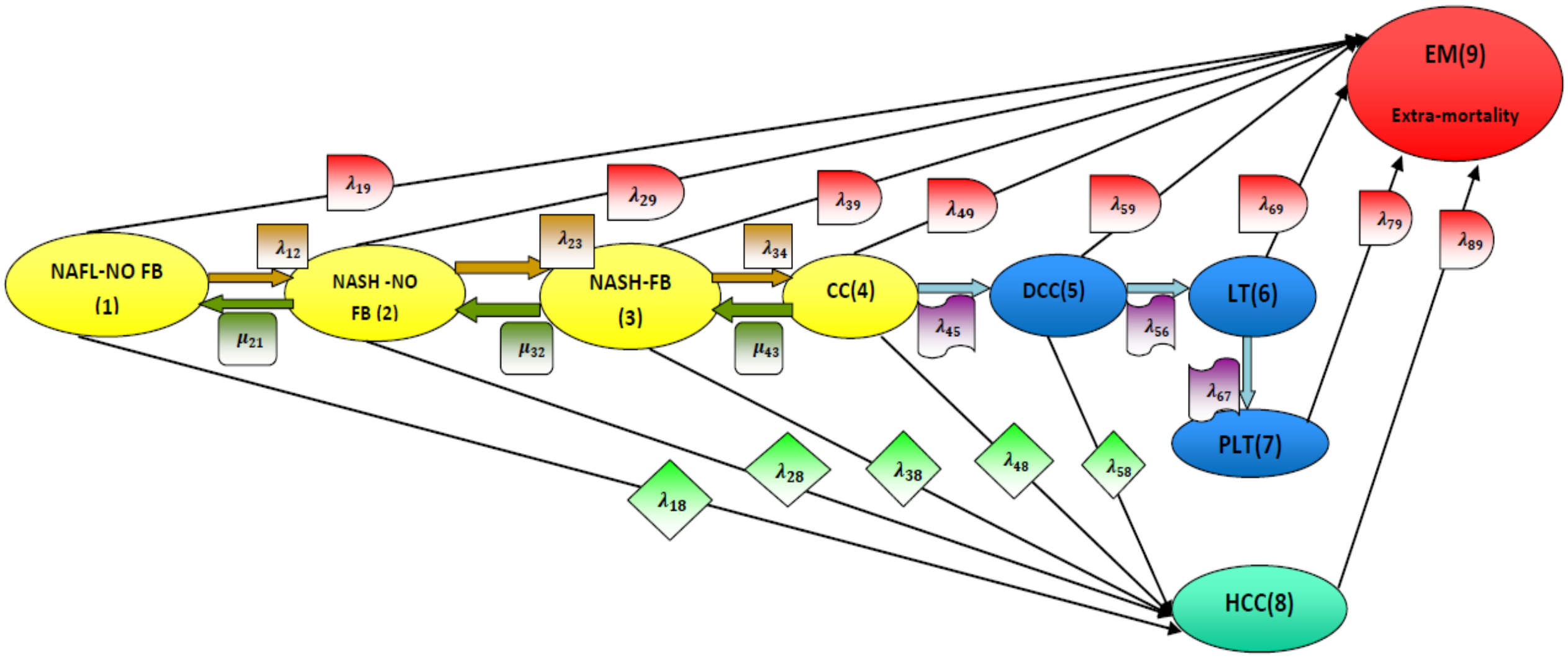

Figure (5. 1): expanded form of the disease model structure . (Younossi et al. 2016)

*NAFLD-NO FB = nonalcoholic fatty liver disease with no fibrosis (stage 1). NASH-NO FB = nonalcoholic steato-hepatitis with no fibrosis (stage 2). NASH-FB = nonalcoholic steato-hepatitis with fibrosis (stage 3). CC= compensated cirrhosis (stage 4). DCC= de-compensated cirrhosis ( stage 5). LT= liver transplant( stage 6). PLT =post liver transplant ( stage 7). HCC =hepato-cellular carcinoma ( stage 8). EM= extra-mortality ( stage 9).*

## 5.1.Probability transition matrix:

For the multistate Markov model demonstrating the NAFLD disease process; the forward Kolomogrov differential equations are the following:

$$\frac{d}{dt}P_{ij}(t) =$$

$$\begin{bmatrix} P_{11} & P_{12} & P_{13} & P_{14} & P_{15} & P_{16} & P_{17} & P_{18} & P_{19} \\ P_{21} & P_{22} & P_{23} & P_{24} & P_{25} & P_{26} & P_{27} & P_{28} & P_{29} \\ P_{31} & P_{32} & P_{33} & P_{34} & P_{35} & P_{36} & P_{37} & P_{38} & P_{39} \\ P_{41} & P_{42} & P_{43} & P_{44} & P_{45} & P_{46} & P_{47} & P_{48} & P_{49} \\ 0 & 0 & 0 & 0 & P_{55} & P_{56} & P_{57} & P_{58} & P_{59} \\ 0 & 0 & 0 & 0 & 0 & P_{66} & P_{67} & 0 & P_{69} \\ 0 & 0 & 0 & 0 & 0 & 0 & P_{77} & 0 & P_{79} \\ 0 & 0 & 0 & 0 & 0 & 0 & 0 & P_{88} & P_{89} \\ 0 & 0 & 0 & 0 & 0 & 0 & 0 & 0 & P_{99} \end{bmatrix} \begin{bmatrix} -\gamma_1 & \lambda_{12} & 0 & 0 & 0 & 0 & 0 & \lambda_{18} & \lambda_{19} \\ \mu_{21} & -\gamma_2 & \lambda_{23} & 0 & 0 & 0 & 0 & \lambda_{28} & \lambda_{29} \\ 0 & \mu_{32} & -\gamma_3 & \lambda_{34} & 0 & 0 & 0 & \lambda_{38} & \lambda_{39} \\ 0 & 0 & \mu_{43} & -\gamma_4 & \lambda_{45} & 0 & 0 & \lambda_{48} & \lambda_{49} \\ 0 & 0 & 0 & 0 & -\gamma_5 & \lambda_{56} & 0 & \lambda_{58} & \lambda_{59} \\ 0 & 0 & 0 & 0 & 0 & -\gamma_6 & \lambda_{67} & 0 & \lambda_{69} \\ 0 & 0 & 0 & 0 & 0 & 0 & -\lambda_{79} & 0 & \lambda_{79} \\ 0 & 0 & 0 & 0 & 0 & 0 & 0 & -\lambda_{89} & \lambda_{89} \\ 0 & 0 & 0 & 0 & 0 & 0 & 0 & 0 & 0 \end{bmatrix}$$

*The Kolmogrov differential equations are:*

*The set of equations of the first row:*

$$\frac{d\,P_{11}(t)}{dt} = -(\lambda_{12}+\lambda_{18}+\lambda_{19})P_{11}(t)+\mu_{21}P_{12}(t), \frac{d\,P_{12}(t)}{dt} = \lambda_{12}\,P_{11}(t)-(\lambda_{23}+\lambda_{28}+\lambda_{29}+\ \mu_{21})P_{12}(t)+\mu_{32}P_{13}(t)$$

$$\frac{d\,P_{13}(t)}{dt} = \lambda_{23}\,P_{12}(t)-(\lambda_{34}+\lambda_{38}+\lambda_{39}+\ \mu_{32})P_{13}(t)+\mu_{43}P_{14}(t), \frac{d\,P_{14}(t)}{dt} = \lambda_{34}\,P_{13}(t)-(\lambda_{45}+\lambda_{48}+\lambda_{49}+\ \mu_{43})P_{14}(t)$$

$$\frac{d\,P_{15}(t)}{dt} = \lambda_{45}\,P_{14}(t)-(\lambda_{56}+\lambda_{58}+\lambda_{59})P_{15}(t)\,, \frac{d\,P_{16}(t)}{dt} = \lambda_{56}\,P_{15}(t)-(\lambda_{67}+\lambda_{69})P_{16}(t), \frac{d\,P_{17}(t)}{dt} = \lambda_{67}\,P_{16}(t)-\lambda_{79}\,P_{17}(t)$$

$$\frac{d\,P_{18}(t)}{dt} = \lambda_{18}\,P_{11}(t)+\lambda_{28}\,P_{12}(t)+\lambda_{38}\,P_{13}(t)+\lambda_{48}\,P_{14}(t)+\lambda_{58}\,P_{15}(t)-\lambda_{89}\,P_{18}(t)$$

$$\frac{d\,P_{19}(t)}{dt} = \lambda_{19}\,P_{11}(t)+\lambda_{29}\,P_{12}(t)+\lambda_{39}\,P_{13}(t)+\lambda_{49}\,P_{14}(t)+\lambda_{59}\,P_{15}(t)+\lambda_{69}\,P_{16}(t)+\lambda_{79}\,P_{17}(t)+\lambda_{89}\,P_{18}(t)$$

*The set of equations of the second row:*

$$\frac{d\,P_{21}(t)}{dt} = -(\lambda_{12}+\lambda_{18}+\lambda_{19})P_{21}(t)+\mu_{21}P_{22}(t), \frac{d\,P_{22}(t)}{dt} = \lambda_{12}\,P_{21}(t)-(\lambda_{23}+\lambda_{28}+\lambda_{29}+\ \mu_{21})P_{22}(t)+\mu_{32}P_{23}(t)$$

$$\frac{d\,P_{23}(t)}{dt} = \lambda_{23}\,P_{22}(t)-(\lambda_{34}+\lambda_{38}+\lambda_{39}+\ \mu_{32})P_{23}(t)+\mu_{43}P_{24}(t), \frac{d\,P_{24}(t)}{dt} = \lambda_{34}\,P_{23}(t)-(\lambda_{45}+\lambda_{48}+\lambda_{49}+\ \mu_{43})P_{24}(t)$$

$$\frac{d\,P_{25}(t)}{dt} = \lambda_{45}\,P_{24}(t)-(\lambda_{56}+\lambda_{58}+\lambda_{59})P_{25}(t), \frac{d\,P_{26}(t)}{dt} = \lambda_{56}\,P_{25}(t)-(\lambda_{67}+\lambda_{69})P_{26}(t), \frac{d\,P_{27}(t)}{dt} = \lambda_{67}\,P_{26}(t)-\lambda_{79}\,P_{27}(t)$$

$$\frac{d\,P_{28}(t)}{dt} = \lambda_{18}\,P_{21}(t)+\lambda_{28}\,P_{22}(t)+\lambda_{38}\,P_{23}(t)+\lambda_{48}\,P_{24}(t)+\lambda_{58}\,P_{25}(t)-\lambda_{89}\,P_{28}(t)$$

$$\frac{d\,P_{29}(t)}{dt} = \lambda_{19}\,P_{21}(t)+\lambda_{29}\,P_{22}(t)+\lambda_{39}\,P_{23}(t)+\lambda_{49}\,P_{24}(t)+\lambda_{59}\,P_{25}(t)+\lambda_{69}\,P_{26}(t)+\lambda_{79}\,P_{27}(t)+\lambda_{89}\,P_{28}(t)$$

*The set of equations of the third row:*

$$\frac{d\,P_{31}(t)}{dt} = -(\lambda_{12}+\lambda_{18}+\lambda_{19})P_{31}(t)+\mu_{21}P_{32}(t), \frac{d\,P_{32}(t)}{dt} = \lambda_{12}\,P_{31}(t)-(\lambda_{23}+\lambda_{28}+\lambda_{29}+\ \mu_{21})P_{32}(t)+\mu_{32}P_{33}(t)$$

$$\frac{d\,P_{33}(t)}{dt} = \lambda_{23}\,P_{32}(t)-(\lambda_{34}+\lambda_{38}+\lambda_{39}+\ \mu_{32})P_{33}(t)+\mu_{43}P_{34}(t), \frac{d\,P_{34}(t)}{dt} = \lambda_{34}\,P_{33}(t)-(\lambda_{45}+\lambda_{48}+\lambda_{49}+\ \mu_{43})P_{34}(t)$$

$$\frac{d\,P_{35}(t)}{dt} = \lambda_{45}\,P_{34}(t)-(\lambda_{56}+\lambda_{58}+\lambda_{59})P_{35}(t), \frac{d\,P_{36}(t)}{dt} = \lambda_{56}\,P_{35}(t)-(\lambda_{67}+\lambda_{69})P_{36}(t), \frac{d\,P_{37}(t)}{dt} = \lambda_{67}\,P_{36}(t)-\lambda_{79}\,P_{37}(t)$$

$$\frac{d\,P_{38}(t)}{dt} = \lambda_{18}\,P_{31}(t)+\lambda_{28}\,P_{32}(t)+\lambda_{38}\,P_{33}(t)+\lambda_{48}\,P_{34}(t)+\lambda_{58}\,P_{35}(t)-\lambda_{89}\,P_{38}(t)$$

$$\frac{d\,P_{39}(t)}{dt} = \lambda_{19}\,P_{31}(t)+\lambda_{29}\,P_{32}(t)+\lambda_{39}\,P_{33}(t)+\lambda_{49}\,P_{34}(t)+\lambda_{59}\,P_{35}(t)+\lambda_{69}\,P_{36}(t)+\lambda_{79}\,P_{37}(t)+\lambda_{89}\,P_{38}(t)$$

*The set of equations of the fourth row:*

$$\frac{d\,P_{41}(t)}{dt} = -(\lambda_{12}+\lambda_{18}+\lambda_{19})P_{41}(t)+\mu_{21}P_{42}(t), \frac{d\,P_{42}(t)}{dt} = \lambda_{12}\,P_{41}(t)-(\lambda_{23}+\lambda_{28}+\lambda_{29}+\ \mu_{21})P_{42}(t)+\mu_{32}P_{43}(t)$$

$$\frac{d\,P_{43}(t)}{dt} = \lambda_{23}\,P_{42}(t)-(\lambda_{34}+\lambda_{38}+\lambda_{39}+\ \mu_{32})P_{43}(t)+\mu_{43}P_{44}(t), \frac{d\,P_{44}(t)}{dt} = \lambda_{34}\,P_{43}(t)-(\lambda_{45}+\lambda_{48}+\lambda_{49}+\ \mu_{43})P_{44}(t)$$

$$\frac{d\,P_{45}(t)}{dt} = \lambda_{45}\,P_{44}(t)-(\lambda_{56}+\lambda_{58}+\lambda_{59})P_{45}(t), \frac{d\,P_{46}(t)}{dt} = \lambda_{56}\,P_{45}(t)-(\lambda_{67}+\lambda_{69})P_{46}(t), \frac{d\,P_{47}(t)}{dt} = \lambda_{67}\,P_{46}(t)-\lambda_{79}\,P_{47}(t)$$

$$\frac{d\,P_{48}(t)}{dt} = \lambda_{18}\,P_{41}(t)+\lambda_{28}\,P_{42}(t)+\lambda_{38}\,P_{43}(t)+\lambda_{48}\,P_{44}(t)+\lambda_{58}\,P_{45}(t)-\lambda_{89}\,P_{48}(t)$$

$$\frac{d\,P_{49}(t)}{dt} = \lambda_{19}\,P_{41}(t) + \lambda_{29}\,P_{42}(t) + \lambda_{39}\,P_{43}(t) + \lambda_{49}\,P_{44}(t) + \lambda_{59}\,P_{45}(t) + \lambda_{69}\,P_{46}(t) + \lambda_{79}\,P_{47}(t) + \lambda_{89}\,P_{48}(t)$$

Last 13 equations :

$$\frac{d\,P_{55}(t)}{dt} = -(\lambda_{56} + \lambda_{58} + \lambda_{59})P_{55}(t), \frac{d\,P_{56}(t)}{dt} = \lambda_{56}\,P_{55}(t) - (\lambda_{67} + \lambda_{69})P_{56}(t), \frac{d\,P_{57}(t)}{dt} = \lambda_{67}\,P_{56}(t) - \lambda_{79}\,P_{57}(t)$$

$$\frac{d\,P_{58}(t)}{dt} = \lambda_{58}\,P_{55}(t) - \lambda_{89}\,P_{58}(t), \frac{d\,P_{59}(t)}{dt} = \lambda_{59}\,P_{55}(t) + \lambda_{69}\,P_{56}(t) + \lambda_{79}\,P_{57}(t) + \lambda_{89}\,P_{58}(t), \frac{d\,P_{66}(t)}{dt} = -(\lambda_{67} + \lambda_{69})P_{66}(t)$$

$$\frac{d\,P_{67}(t)}{dt} = \lambda_{67}\,P_{66}(t) - \lambda_{79}P_{67}(t), \frac{d\,P_{69}(t)}{dt} = \lambda_{69}\,P_{66}(t) + \lambda_{79}P_{67}(t), \frac{d\,P_{77}(t)}{dt} = -\lambda_{79}P_{77}(t), \frac{d\,P_{79}(t)}{dt} = \lambda_{79}P_{77}(t), \frac{d\,P_{88}(t)}{dt} = -\,\lambda_{89}P_{88}(t)$$

$$\frac{d\,P_{89}(t)}{dt} = \lambda_{89}P_{88}(t), P_{99}(t) = 1$$

**<u>The Kolmogrov Differential Equations For The First 4 Probabilities In The First 4 Rows Are:</u>**

$$\frac{d\,P_{ij}(t)}{dt} = P(t)Q(t) = \begin{bmatrix} P_{11} & P_{12} & P_{13} & P_{14} \\ P_{21} & P_{22} & P_{23} & P_{24} \\ P_{31} & P_{32} & P_{33} & P_{34} \\ P_{41} & P_{42} & P_{43} & P_{44} \end{bmatrix} \begin{bmatrix} -\gamma_1 & \lambda_{12} & 0 & 0 \\ \mu_{21} & -\gamma_2 & \lambda_{23} & 0 \\ 0 & \mu_{32} & -\gamma_3 & \lambda_{34} \\ 0 & 0 & \mu_{43} & -\gamma_4 \end{bmatrix}$$

$let's\ \ call\quad (\lambda_{12} + \lambda_{18} + \lambda_{19}) = \gamma_1 \quad , \qquad (\lambda_{23} + \lambda_{28} + \lambda_{29} + \mu_{21}) = \gamma_2$

$(\lambda_{34} + \lambda_{38} + \lambda_{39} + \mu_{32}) = \gamma_3 \quad , \qquad (\lambda_{45} + \lambda_{48} + \lambda_{49} + \mu_{43}) = \gamma_4$

***<u>The set of equations of the first row: ( first 4 probabilities )</u>***

The differential equations for the first 4 PDFs' as stated previously solved using Laplace method :

$\mathcal{L}\left\{P_{ij}(t)\right\}\ \ is\,presented\,as\ \ P^*_{ij}(s)\ \ , \mathcal{L}\left\{P'_{ij}(t)\right\} = s\mathcal{L}\left\{P_{ij}(t)\right\} - P_{ij}(0) = s\,P^*_{ij}(s) - P_{ij}(0)$

$$\frac{d\,P_{11}(t)}{dt} = -\gamma_1 P_{11}(t) + \mu_{21}P_{12}(t)$$

$s\,P^*_{11}(s) - P_{11}(0) = -\gamma_1\ P^*_{11}(s) + \mu_{21}\ P^*_{12}(s) \quad \rightarrow \qquad s\,P^*_{11}(s) - 1 = -\gamma_1\ P^*_{11}(s) + \mu_{21}\ P^*_{12}(s)$

$(s\ +\ \gamma_1)\ P^*_{11}(s) - \mu_{21}\ P^*_{12}(s) = 1\ \ldots\ldots\ldots\ldots\ldots\ldots\ldots\ldots\ldots\ldots\ldots\ldots\ldots\ldots\ \ (1)$

$$\frac{d\,P_{12}(t)}{dt} = \lambda_{12}\,P_{11}(t) - (\lambda_{23} + \lambda_{28} + \lambda_{29} + \mu_{21})P_{12}(t) + \mu_{32}P_{13}(t)$$

$s\,P^*_{12}(s) - P_{12}(0) = \lambda_{12}\ P^*_{11}(s) - \gamma_2\ P^*_{12}(s) + \mu_{32}\ P^*_{13}(s) \rightarrow s\,P^*_{12}(s) - 0\ -\ \lambda_{12}\ P^*_{11}(s) + \gamma_2\ P^*_{12}(s) - \mu_{32}\ P^*_{13}(s) = 0$

$(s + \gamma_2)\ P^*_{12}(s)\ -\ \lambda_{12}\ P^*_{11}(s) - \mu_{32}\ P^*_{13}(s) = 0 \ldots\ldots\ldots\ldots\ldots\ldots.. \qquad (2)$

$$\frac{d\,P_{13}(t)}{dt} = \lambda_{23}\,P_{12}(t) - (\lambda_{34} + \lambda_{38} + \lambda_{39} + \mu_{32})P_{13}(t) + \mu_{43}P_{14}(t)$$

$s\,P^*_{13}(s) - P_{13}(0) = \lambda_{23}\ P^*_{12}(s) - \gamma_3\ P^*_{13}(s) + \mu_{43}\ P^*_{14}(s) \rightarrow\ s\,P^*_{13}(s) - 0 = \lambda_{23}\ P^*_{12}(s) - \gamma_3\ P^*_{13}(s) + \mu_{43}\ P^*_{14}(s)$

$(s + \gamma_3)\ P^*_{13}(s)\quad -\ \lambda_{23}\ P^*_{12}(s) - \mu_{43}\ P^*_{14}(s) = 0 \ldots\ldots\ldots\ldots\ldots\ldots\ldots\ldots\ldots.\ (3)$

$$\frac{d\,P_{14}(t)}{dt} = \lambda_{34}\,P_{13}(t) - (\lambda_{45} + \lambda_{48} + \lambda_{49} + \mu_{43})P_{14}(t)$$

$s\,P^*_{14}(s) - P_{14}(0) = \lambda_{34}\ P^*_{13}(s) - \gamma_4\ P^*_{14}(s) \quad \rightarrow \quad s\,P^*_{14}(s) - 0 = \lambda_{34}\ P^*_{13}(s) - \gamma_4\ P^*_{14}(s)$

$(s+\gamma_4)\ P_{14}^*(s) - \lambda_{34}\ P_{13}^*(s) = 0$ ……………………………. (4)

Putting these equations (1,2,3,4) in matrix notation : $MP_{ij}^*(s) = Z$

$$\begin{bmatrix} (s+\gamma_1) & -\mu_{21} & 0 & 0 \\ -\lambda_{12} & (s+\gamma_2) & -\mu_{32} & 0 \\ 0 & -\lambda_{23} & (s+\gamma_3) & -\mu_{43} \\ 0 & 0 & -\lambda_{34} & (s+\gamma_4) \end{bmatrix} \begin{bmatrix} P_{11}^*(s) \\ P_{12}^*(s) \\ P_{13}^*(s) \\ P_{14}^*(s) \end{bmatrix} = \begin{bmatrix} 1 \\ 0 \\ 0 \\ 0 \end{bmatrix} \ .\ Then\ apply\ Cramer\ rule\ to\ solve\ for\ P_{ij}^*(s) = \frac{D_{P_{ij}^*(s)}}{D}$$

***<u>The determinant in the denominator :</u>***

$To\ solve\ for\ \ P_{ij}^*(s)\ using\ cramer\ rule\ ,the\ determinanat\ \ is\ calculated\ for : D\ \ ,\ D_{P_{11}^*(s)}\ ,\ D_{P_{12}^*(s)}\ ,\ D_{P_{13}^*(s)}\ ,\ D_{P_{14}^*(s)}$

$Put\ M\ in\ the\ form \begin{bmatrix} A & B \\ 0 & C \end{bmatrix}\ to\ calculate\ the\ determinant\ in\ a\ blocked\ partition\ matrix$

$$\left\| \begin{bmatrix} (s+\gamma_1) & -\mu_{21} & 0 & 0 \\ -\lambda_{12} & (s+\gamma_2) & -\mu_{32} & 0 \\ 0 & -\lambda_{23} & (s+\gamma_3) & -\mu_{43} \\ 0 & 0 & -\lambda_{34} & (s+\gamma_4) \end{bmatrix} \right\|$$

$$D = [(s+\gamma_1)(s+\gamma_2) - \lambda_{12}\,\mu_{21}] \times \left[ \left( \frac{-\lambda_{23}\,\mu_{32}(s+\gamma_1) - \lambda_{12}\,\mu_{21}(s+\gamma_3) + (s+\gamma_1)(s+\gamma_2)(s+\gamma_3)}{(s+\gamma_1)(s+\gamma_2) - \lambda_{12}\,\mu_{21}} \right)(s+\gamma_4) - \lambda_{34}\,\mu_{43} \right]$$

$$D = [(s+\gamma_1)(s+\gamma_2) - \lambda_{12}\,\mu_{21}][(s+\gamma_3)(s+\gamma_4) - \lambda_{34}\,\mu_{43}] - \lambda_{23}\,\mu_{32}(s+\gamma_1)(s+\gamma_4)$$

$$D = s^4 + (w_1+w_3)s^3 + (w_2 + w_4 + \ w_1 w_3 - \lambda_{23}\,\mu_{32}\ )\ s^2 + (w_2 w_3 + w_1 w_4 - \ w_5)s + \ (w_2 w_4 - \ w_6)\ \ ,\qquad where:$$

$w_1 = \gamma_1 + \gamma_2 = (\lambda_{12} + \lambda_{18} + \lambda_{19}) + (\lambda_{23} + \lambda_{28} + \lambda_{29} + \mu_{21})\ ,\qquad w_2 = \gamma_1(\lambda_{23} + \lambda_{28} + \lambda_{29}) + \mu_{21}(\lambda_{18} + \lambda_{19})$

$w_3 = \gamma_3 + \gamma_4 = (\lambda_{34} + \lambda_{38} + \lambda_{39} + \mu_{32}) + \ (\lambda_{45} + \lambda_{48} + \lambda_{49} + \mu_{43})\ ,\qquad w_4 = \gamma_3(\ \lambda_{45} + \lambda_{48} + \lambda_{49}) + \mu_{43}(\lambda_{38} + \lambda_{39} + \mu_{32})$

$w_5 = \lambda_{23}\,\mu_{32}\ (\gamma_1 + \ \gamma_4\ )\ \ ,\qquad w_6 = \lambda_{23}\,\mu_{32}\,\gamma_1\,\gamma_4 = \lambda_{23}\,\mu_{32}\ [(\lambda_{12} + \lambda_{18} + \lambda_{19})(\lambda_{45} + \lambda_{48} + \lambda_{49} + \mu_{43})]$

Determinant in the denominator is a polynomial of the 4 th degree with the following roots: $r_1$, $r_2$, $r_3$, $r_4$

$D = \ (s - r_1)(s - r_2)(s - r_3)(s - r_4)$, Then

- ***<u>expand through first column to get</u>***:

$$D_{P_{11}^*(s)} = \left\| \begin{bmatrix} 1 & -\mu_{21} & 0 & 0 \\ 0 & (s+\gamma_2) & -\mu_{32} & 0 \\ 0 & -\lambda_{23} & (s+\gamma_3) & -\mu_{43} \\ 0 & 0 & -\lambda_{34} & (s+\gamma_4) \end{bmatrix} \right\| = ([s+\gamma_2][(s+\gamma_3)(s+\gamma_4) - \lambda_{34}\,\mu_{43}]) - \lambda_{23}\,\mu_{32}\ (s+\gamma_4)$$

$$D_{P_{11}^*(s)} = s^3 + (\gamma_2 + \gamma_3 + \gamma_4)\ s^2 + (\gamma_2\gamma_3 + \gamma_2\gamma_4 + \gamma_3\gamma_4 - \lambda_{34}\,\mu_{43}\ - \lambda_{23}\,\mu_{32})\ s - \lambda_{23}\,\mu_{32}\,\gamma_4 - \lambda_{34}\mu_{43}\,\gamma_2 + \gamma_2\gamma_3\gamma_4$$

- ***<u>expand through second column to get</u>*** :

$$D_{P_{12}^*(s)} = \left\| \begin{bmatrix} (s+\gamma_1) & 1 & 0 & 0 \\ -\lambda_{12} & 0 & -\mu_{32} & 0 \\ 0 & 0 & (s+\gamma_3) & -\mu_{43} \\ 0 & 0 & -\lambda_{34} & (s+\gamma_4) \end{bmatrix} \right\| = \lambda_{12}[(s+\gamma_3)(s+\gamma_4) - \lambda_{34}\,\mu_{43}]$$

$$D_{P_{12}^*(s)} = \lambda_{12}s^2 + \lambda_{12}\ (\gamma_3 + \gamma_4)s + \lambda_{12}\,\gamma_3\gamma_4\ - \ \lambda_{12}\,\lambda_{34}\,\mu_{43}$$

- ***<u>expand through the third column to get</u>*** :

$$D_{P^*_{13}(s)} = \left|\begin{bmatrix} (s+\gamma_1) & -\mu_{21} & 1 & 0 \\ -\lambda_{12} & (s+\gamma_2) & 0 & 0 \\ 0 & -\lambda_{23} & 0 & -\mu_{43} \\ 0 & 0 & 0 & (s+\gamma_4) \end{bmatrix}\right| = \lambda_{12}\lambda_{23}\ s + \lambda_{12}\lambda_{23}\ \gamma_4$$

- ***expand through fourth column to get :***

$$D_{P^*_{14}(s)} = \left|\begin{bmatrix} (s+\gamma_1) & -\mu_{21} & 0 & 1 \\ -\lambda_{12} & (s+\gamma_2) & -\mu_{32} & 0 \\ 0 & -\lambda_{23} & (s+\gamma_3) & 0 \\ 0 & 0 & -\lambda_{34} & 0 \end{bmatrix}\right| = \lambda_{12}\ \lambda_{23}\lambda_{34}$$

Using the above technique for the set of equations of the second row ( first 4 probabilities) gives the following matrix:

$$\begin{bmatrix} (s+\gamma_1) & -\mu_{21} & 0 & 0 \\ -\lambda_{12} & (s+\gamma_2) & -\mu_{32} & 0 \\ 0 & -\lambda_{23} & (s+\gamma_3) & -\mu_{43} \\ 0 & 0 & -\lambda_{34} & (s+\gamma_4) \end{bmatrix}\begin{bmatrix} P^*_{21}(s) \\ P^*_{22}(s) \\ P^*_{23}(s) \\ P^*_{24}(s) \end{bmatrix} = \begin{bmatrix} 0 \\ 1 \\ 0 \\ 0 \end{bmatrix}, \textit{then solve for} P^*_{ij}(s) \textit{ by calculating } \ D_{P^*_{21}(s)},\ D_{P^*_{22}(s)},\ D_{P^*_{23}(s)},\ D_{P^*_{24}(s)}$$

- ***expand through first column to get***:

$$D_{P^*_{21}(s)} = \left|\begin{bmatrix} 0 & -\mu_{21} & 0 & 0 \\ 1 & (s+\gamma_2) & -\mu_{32} & 0 \\ 0 & -\lambda_{23} & (s+\gamma_3) & -\mu_{43} \\ 0 & 0 & -\lambda_{34} & (s+\gamma_4) \end{bmatrix}\right| = \mu_{21}[(s+\gamma_3)(s+\gamma_4) - \lambda_{34}\,\mu_{43}]$$

$$D_{P^*_{21}(s)} = \mu_{21}s^2 + (\mu_{21}\,\gamma_3 + \mu_{21}\,\gamma_4)\,s + \mu_{21}\,\gamma_3\,\gamma_4 - \mu_{21}\lambda_{34}\,\mu_{43}$$

- ***expand through second column to get :***

$$D_{P^*_{22}(s)} = \left|\begin{bmatrix} (s+\gamma_1) & 0 & 0 & 0 \\ -\lambda_{12} & 1 & -\mu_{32} & 0 \\ 0 & 0 & (s+\gamma_3) & -\mu_{43} \\ 0 & 0 & -\lambda_{34} & (s+\gamma_4) \end{bmatrix}\right| = (s+\gamma_1)[(s+\gamma_3)(s+\gamma_4) - \lambda_{34}\,\mu_{43}]$$

$$D_{P^*_{22}(s)} = s^3 + (\gamma_1+\gamma_3+\gamma_4)s^2 + (\gamma_1\gamma_3 + \gamma_1\gamma_4 + \gamma_3\gamma_4 - \lambda_{34}\,\mu_{43})s - \lambda_{34}\,\mu_{43}\gamma_1 + \gamma_1\gamma_3\gamma_4$$

- ***expand through third column to get :***

$$D_{P^*_{23}(s)} = \left|\begin{bmatrix} (s+\gamma_1) & -\mu_{21} & 0 & 0 \\ -\lambda_{12} & (s+\gamma_2) & 1 & 0 \\ 0 & -\lambda_{23} & 0 & -\mu_{43} \\ 0 & 0 & 0 & (s+\gamma_4) \end{bmatrix}\right| = \lambda_{23}(s+\gamma_4)(s+\gamma_1) = \lambda_{23}s^2 + (\lambda_{23}\gamma_1 + \lambda_{23}\gamma_4)s + \lambda_{23}\gamma_1\gamma_4$$

- ***expand through fourth column to get :***

$$D_{P^*_{24}(s)} = \left|\begin{bmatrix} (s+\gamma_1) & -\mu_{21} & 0 & 0 \\ -\lambda_{12} & (s+\gamma_2) & -\mu_{32} & 1 \\ 0 & -\lambda_{23} & (s+\gamma_3) & 0 \\ 0 & 0 & -\lambda_{34} & 0 \end{bmatrix}\right| = \lambda_{23}\lambda_{34}\,(s+\gamma_1) = \lambda_{23}\lambda_{34}\,s + \lambda_{23}\lambda_{34}\gamma_1$$

Using the above technique for the set of equations of the third row ( first 4 probabilities) gives the following matrix:

$$\begin{bmatrix} (s+\gamma_1) & -\mu_{21} & 0 & 0 \\ -\lambda_{12} & (s+\gamma_2) & -\mu_{32} & 0 \\ 0 & -\lambda_{23} & (s+\gamma_3) & -\mu_{43} \\ 0 & 0 & -\lambda_{34} & (s+\gamma_4) \end{bmatrix}\begin{bmatrix} P^*_{31}(s) \\ P^*_{32}(s) \\ P^*_{33}(s) \\ P^*_{34}(s) \end{bmatrix} = \begin{bmatrix} 0 \\ 0 \\ 1 \\ 0 \end{bmatrix}, \textit{then solve for } P^*_{ij}(s) \textit{ by calculating } \ D_{P^*_{31}(s)},\ D_{P^*_{32}(s)},\ D_{P^*_{33}(s)},\ D_{P^*_{34}(s)}$$

- ***expand through first column to get***:

$$D_{P^*_{31}(s)} = \left|\begin{bmatrix} 0 & -\mu_{21} & 0 & 0 \\ 0 & (s+\gamma_2) & -\mu_{32} & 0 \\ 1 & -\lambda_{23} & (s+\gamma_3) & -\mu_{43} \\ 0 & 0 & -\lambda_{34} & (s+\gamma_4) \end{bmatrix}\right| = \mu_{21}\,\mu_{32}\ (s+\gamma_4) = \mu_{21}\,\mu_{32}\,s + \mu_{21}\,\mu_{32}\gamma_4$$

- ***expand through second column to get*** :

$$D_{P^*_{32}(s)} = \left|\begin{bmatrix}(s+\gamma_1) & 0 & 0 & 0\\ -\lambda_{12} & 0 & -\mu_{32} & 0\\ 0 & 1 & (s+\gamma_3) & -\mu_{43}\\ 0 & 0 & -\lambda_{34} & (s+\gamma_4)\end{bmatrix}\right| = \mu_{32}(s+\gamma_1)(s+\gamma_4) = \mu_{32}\,s^2 + (\gamma_1\,\mu_{32} + \gamma_4\,\mu_{32})s + \mu_{32}\gamma_1\gamma_4$$

- ***expand through third column to get*** :

$$D_{P^*_{33}(s)} = \left|\begin{bmatrix}(s+\gamma_1) & -\mu_{21} & 0 & 0\\ -\lambda_{12} & (s+\gamma_2) & 0 & 0\\ 0 & -\lambda_{23} & 1 & -\mu_{43}\\ 0 & 0 & 0 & (s+\gamma_4)\end{bmatrix}\right| = (s+\gamma_4)[(s+\gamma_1)(s+\gamma_2) - \mu_{21}\lambda_{12}]$$

$$D_{P^*_{33}(s)} = s^3 + (\gamma_1+\gamma_2+\gamma_4)s^2 + (\gamma_1\gamma_2 + \gamma_1\gamma_4 + \gamma_2\gamma_4 - \mu_{21}\lambda_{12})s - \mu_{21}\lambda_{12}\gamma_4 + \gamma_1\gamma_2\gamma_4$$

- ***expand through fourth column to get*** :

$$D_{P^*_{34}(s)} = \left|\begin{bmatrix}(s+\gamma_1) & -\mu_{21} & 0 & 0\\ -\lambda_{12} & (s+\gamma_2) & -\mu_{32} & 0\\ 0 & -\lambda_{23} & (s+\gamma_3) & 1\\ 0 & 0 & -\lambda_{34} & 0\end{bmatrix}\right|$$

$$D_{P^*_{34}(s)} = \lambda_{34}\,[(s+\gamma_1)(s+\gamma_2) - \mu_{21}\lambda_{12}] = \lambda_{34}\,s^2 + (\lambda_{34}\gamma_1 + \lambda_{34}\gamma_2)s + \lambda_{34}\gamma_1\gamma_2 - \lambda_{12}\lambda_{34}\mu_{21}$$

Using the above technique for the set of equations of the third row ( first 4 probabilities) gives the following matrix:

$$\begin{bmatrix}(s+\gamma_1) & -\mu_{21} & 0 & 0\\ -\lambda_{12} & (s+\gamma_2) & -\mu_{32} & 0\\ 0 & -\lambda_{23} & (s+\gamma_3) & -\mu_{43}\\ 0 & 0 & -\lambda_{34} & (s+\gamma_4)\end{bmatrix}\begin{bmatrix}P^*_{41}(s)\\ P^*_{42}(s)\\ P^*_{43}(s)\\ P^*_{44}(s)\end{bmatrix} = \begin{bmatrix}0\\0\\0\\1\end{bmatrix}$$

*, then solve for* $P^*_{ij}(s)$ *by calculating* $D_{P^*_{41}(s)}$ , $D_{P^*_{42}(s)}$ , $D_{P^*_{43}(s)}$ , $D_{P^*_{44}(s)}$

- ***expand through first column to get***:

$$D_{P^*_{41}(s)} = \left|\begin{bmatrix}0 & -\mu_{21} & 0 & 0\\ 0 & (s+\gamma_2) & -\mu_{32} & 0\\ 0 & -\lambda_{23} & (s+\gamma_3) & -\mu_{43}\\ 1 & 0 & -\lambda_{34} & (s+\gamma_4)\end{bmatrix}\right| = \mu_{21}\,\mu_{32}\,\mu_{43}$$

- ***expand through second column to get*** :

$$D_{P^*_{42}(s)} = \left|\begin{bmatrix}(s+\gamma_1) & 0 & 0 & 0\\ -\lambda_{12} & 0 & -\mu_{32} & 0\\ 0 & 0 & (s+\gamma_3) & -\mu_{43}\\ 0 & 1 & -\lambda_{34} & (s+\gamma_4)\end{bmatrix}\right| = \mu_{32}\,\mu_{43}\,(s+\gamma_1) = \mu_{32}\,\mu_{43}\,s + \mu_{32}\,\mu_{43}\,\gamma_1$$

- ***expand through third column to get*** :

$$D_{P^*_{43}(s)} = \left|\begin{bmatrix}(s+\gamma_1) & -\mu_{21} & 0 & 0\\ -\lambda_{12} & (s+\gamma_2) & 0 & 0\\ 0 & -\lambda_{23} & 0 & -\mu_{43}\\ 0 & 0 & 1 & (s+\gamma_4)\end{bmatrix}\right| = \mu_{43}\,[(s+\gamma_1)(s+\gamma_2) - \mu_{21}\lambda_{12}]$$

$$D_{P^*_{43}(s)} = \mu_{43}s^2 + (\gamma_1\mu_{43} + \gamma_2\mu_{43})s + \gamma_2\gamma_1\mu_{43} - \lambda_{12}\mu_{21}\mu_{43}$$

- ***expand through fourth column to get*** :

$$D_{P^*_{44}(s)} = \left|\begin{bmatrix}(s+\gamma_1) & -\mu_{21} & 0 & 0\\ -\lambda_{12} & (s+\gamma_2) & -\mu_{32} & 0\\ 0 & -\lambda_{23} & (s+\gamma_3) & 0\\ 0 & 0 & -\lambda_{34} & 1\end{bmatrix}\right| = -\lambda_{23}\,\mu_{32}\,(s+\gamma_1) + (s+\gamma_3)\,[(s+\gamma_1)(s+\gamma_2) - \mu_{21}\lambda_{12}]$$

$$D_{P^*_{44}(s)} = s^3 + (\gamma_1+\gamma_2+\gamma_3)s^2 + (\gamma_1\gamma_2 + \gamma_1\gamma_3 + \gamma_2\gamma_3 - \mu_{21}\lambda_{12} - \lambda_{23}\,\mu_{32})s - \lambda_{23}\,\mu_{32}\,\gamma_1 - \mu_{21}\lambda_{12}\gamma_3 + \gamma_1\gamma_2\gamma_3$$

***Then Using partial fraction to get inverse laplace:***

$$D_{P_{11}^*(s)} = s^3 + (\gamma_2 + \gamma_3 + \gamma_4)\, s^2 + (\gamma_2\gamma_3 + \gamma_2\gamma_4 + \gamma_3\gamma_4 - \lambda_{34}\,\mu_{43} - \lambda_{23}\,\mu_{32})\, s - \lambda_{23}\,\mu_{32}\,\gamma_4 - \lambda_{34}\mu_{43}\,\gamma_2 + \gamma_2\gamma_3\gamma_4$$

$$P_{11}^*(s) = \frac{D_{P_{11}^*(s)}}{D} = \frac{A_{11}}{(s-r_1)} + \frac{B_{11}}{(s-r_2)} + \frac{C_{11}}{(s-r_3)} + \frac{D_{11}}{(s-r_4)}$$

$$D_{P_{11}^*(s)} = A_{11}(s-r_2)(s-r_3)(s-r_4) + B_{11}(s-r_1)(s-r_3)(s-r_4) + C_{11}(s-r_1)(s-r_2)(s-r_4) + D_{11}(s-r_1)(s-r_2)(s-r_3)$$

$$R.H.S. = [A_{11}(s-r_2)(s^2-(r_3+r_4)s+r_3\,r_4)] + [B_{11}(s-r_1)(s^2-(r_3+r_4)s+r_3\,r_4)] + [C_{11}(s-r_1)(s^2-(r_2+r_4)s+r_2\,r_4)] + [D_{11}(s-r_1)(s^2-(r_2+r_3)s+r_2\,r_3)]$$

$$R.H.S. = A_{11}\{s^3 - (r_2 + r_3 + r_4)\, s^2 + (r_2\,r_3 + r_2\,r_4 + r_3\,r_4)\, s - r_2\,r_3\,r_4\} +$$

$$B_{11}\{s^3 - (r_1 + r_3 + r_4)\, s^2 + (r_1\,r_3 + r_1\,r_4 + r_3\,r_4)\, s - r_1\,r_3\,r_4\} +$$

$$C_{11}\{s^3 - (r_1 + r_2 + r_4)\, s^2 + (r_1\,r_2 + r_1\,r_4 + r_2\,r_4)\, s - r_1\,r_2\,r_4\} +$$

$$D_{11}\{s^3 - (r_1 + r_2 + r_3)\, s^2 + (r_1\,r_2 + r_1\,r_3 + r_2\,r_3)\, s - r_1\,r_2\,r_3\}$$

$$R.H.S. = [A_{11} + B_{11} + C_{11} + D_{11}]s^3 +$$

$$[-(r_2 + r_3 + r_4)A_{11} - (r_1 + r_3 + r_4)B_{11} - (r_1 + r_2 + r_4)C_{11} - (r_1 + r_2 + r_3)\, D_{11}]\, s^2 +$$

$$[(r_2\,r_3 + r_2\,r_4 + r_3\,r_4)A_{11} + (r_1\,r_3 + r_1\,r_4 + r_3\,r_4)B_{11} + (r_1\,r_2 + r_1\,r_4 + r_2\,r_4)C_{11} + (r_1\,r_2 + r_1\,r_3 + r_2\,r_3)D_{11}]s +$$

$$[-r_2\,r_3\,r_4\,A_{11} - r_1\,r_3\,r_4\,B_{11} - r_1\,r_2\,r_4\,C_{11} - r_1\,r_2\,r_3\,D_{11}]$$

$$\because\ D_{P_{11}^*(s)} = s^3 + (\gamma_2 + \gamma_3 + \gamma_4)\, s^2 + (\gamma_2\gamma_3 + \gamma_2\gamma_4 + \gamma_3\gamma_4 - \lambda_{34}\,\mu_{43} - \lambda_{23}\,\mu_{32})\, s - \lambda_{23}\,\mu_{32}\,\gamma_4 - \lambda_{34}\mu_{43}\,\gamma_2 + \gamma_2\gamma_3\gamma_4$$

$$\therefore\ D_{P_{11}^*(s)} = R.H.S.$$

***Equating R.H.S.with L.H.S.:***

$$[A_{11} + B_{11} + C_{11} + D_{11}]s^3 = s^3$$

$$[-(r_2 + r_3 + r_4)A_{11} - (r_1 + r_3 + r_4)B_{11} - (r_1 + r_2 + r_4)C_{11} - (r_1 + r_2 + r_3)\, D_{11}]\, s^2 = (\gamma_2 + \gamma_3 + \gamma_4)\, s^2$$

$$[(r_2\,r_3 + r_2\,r_4 + r_3\,r_4)A_{11} + (r_1\,r_3 + r_1\,r_4 + r_3\,r_4)B_{11} + (r_1\,r_2 + r_1\,r_4 + r_2\,r_4)C_{11} + (r_1\,r_2 + r_1\,r_3 + r_2\,r_3)D_{11}]s$$

$$= (\gamma_2\gamma_3 + \gamma_2\gamma_4 + \gamma_3\gamma_4 - \lambda_{34}\,\mu_{43} - \lambda_{23}\,\mu_{32})\, s$$

$$[-r_2\,r_3\,r_4\,A_{11} - r_1\,r_3\,r_4\,B_{11} - r_1\,r_2\,r_4\,C_{11} - r_1\,r_2\,r_3\,D_{11}] = -\lambda_{23}\,\mu_{32}\,\gamma_4 - \lambda_{34}\mu_{43}\,\gamma_2 + \gamma_2\gamma_3\gamma_4$$

$$A_{11} + B_{11} + C_{11} + D_{11} = 1$$

$$-(r_2 + r_3 + r_4)A_{11} - (r_1 + r_3 + r_4)B_{11} - (r_1 + r_2 + r_4)C_{11} - (r_1 + r_2 + r_3)\, D_{11} = \gamma_2 + \gamma_3 + \gamma_4$$

$$(r_2\,r_3 + r_2\,r_4 + r_3\,r_4)A_{11} + (r_1\,r_3 + r_1\,r_4 + r_3\,r_4)B_{11} + (r_1\,r_2 + r_1\,r_4 + r_2\,r_4)C_{11} + (r_1\,r_2 + r_1\,r_3 + r_2\,r_3)D_{11} =$$

$$\gamma_2\gamma_3 + \gamma_2\gamma_4 + \gamma_3\gamma_4 - \lambda_{34}\,\mu_{43} - \lambda_{23}\,\mu_{32}$$

$$-r_2\,r_3\,r_4\,A_{11} - r_1\,r_3\,r_4\,B_{11} - r_1\,r_2\,r_4\,C_{11} - r_1\,r_2\,r_3\,D_{11} = -\lambda_{23}\,\mu_{32}\,\gamma_4 - \lambda_{34}\mu_{43}\,\gamma_2 + \gamma_2\gamma_3\gamma_4$$

This set of the four equations will be used repeatedly to calculate the inverse Laplace transform for the first four probabilities in the first four rows, the difference will be in the resultant vector for each $D_{P^*_{ij}(s)}$ : in matrix notation

$$\begin{bmatrix} 1 & 1 & 1 & 1 \\ -(r_2+r_3+r_4) & -(r_1+r_3+r_4) & -(r_1+r_2+r_4) & -(r_1+r_2+r_3) \\ (r_2 r_3 + r_2 r_4 + r_3 r_4) & (r_1 r_3 + r_1 r_4 + r_3 r_4) & (r_1 r_2 + r_1 r_4 + r_2 r_4) & (r_1 r_2 + r_1 r_3 + r_2 r_3) \\ -r_2 r_3 r_4 & -r_1 r_3 r_4 & -r_1 r_2 r_4 & -r_1 r_2 r_3 \end{bmatrix} \begin{bmatrix} A_{11} \\ B_{11} \\ C_{11} \\ D_{11} \end{bmatrix} = \begin{bmatrix} 1 \\ \gamma_2+\gamma_3+\gamma_4 \\ \gamma_2\gamma_3+\gamma_2\gamma_4+\gamma_3\gamma_4-\lambda_{34}\mu_{43}-\lambda_{23}\mu_{32} \\ -\lambda_{23}\mu_{32}\gamma_4-\lambda_{34}\mu_{43}\gamma_2+\gamma_2\gamma_3\gamma_4 \end{bmatrix}$$

$$let \begin{bmatrix} 1 & 1 & 1 & 1 \\ -(r_2+r_3+r_4) & -(r_1+r_3+r_4) & -(r_1+r_2+r_4) & -(r_1+r_2+r_3) \\ (r_2 r_3 + r_2 r_4 + r_3 r_4) & (r_1 r_3 + r_1 r_4 + r_3 r_4) & (r_1 r_2 + r_1 r_4 + r_2 r_4) & (r_1 r_2 + r_1 r_3 + r_2 r_3) \\ -r_2 r_3 r_4 & -r_1 r_3 r_4 & -r_1 r_2 r_4 & -r_1 r_2 r_3 \end{bmatrix} = K(r)$$

$$\begin{bmatrix} A_{11} \\ B_{11} \\ C_{11} \\ D_{11} \end{bmatrix} = \begin{bmatrix} 1 & 1 & 1 & 1 \\ -(r_2+r_3+r_4) & -(r_1+r_3+r_4) & -(r_1+r_2+r_4) & -(r_1+r_2+r_3) \\ (r_2 r_3 + r_2 r_4 + r_3 r_4) & (r_1 r_3 + r_1 r_4 + r_3 r_4) & (r_1 r_2 + r_1 r_4 + r_2 r_4) & (r_1 r_2 + r_1 r_3 + r_2 r_3) \\ -r_2 r_3 r_4 & -r_1 r_3 r_4 & -r_1 r_2 r_4 & -r_1 r_2 r_3 \end{bmatrix}^{-1} \begin{bmatrix} 1 \\ \gamma_2+\gamma_3+\gamma_4 \\ \gamma_2\gamma_3+\gamma_2\gamma_4+\gamma_3\gamma_4-\lambda_{34}\mu_{43}-\lambda_{23}\mu_{32} \\ -\lambda_{23}\mu_{32}\gamma_4-\lambda_{34}\mu_{43}\gamma_2+\gamma_2\gamma_3\gamma_4 \end{bmatrix}$$

$$\begin{bmatrix} A_{11} \\ B_{11} \\ C_{11} \\ D_{11} \end{bmatrix} = [K(r)]^{-1} \begin{bmatrix} 1 \\ \gamma_2+\gamma_3+\gamma_4 \\ \gamma_2\gamma_3+\gamma_2\gamma_4+\gamma_3\gamma_4-\lambda_{34}\mu_{43}-\lambda_{23}\mu_{32} \\ -\lambda_{23}\mu_{32}\gamma_4-\lambda_{34}\mu_{43}\gamma_2+\gamma_2\gamma_3\gamma_4 \end{bmatrix}$$

$[K(r)]^{-1}$ *is used repeatedly in the calculations of the first 4 PDF's in the first 4 rows of the probability matrix*

$$P^*_{11}(s) = \frac{D_{P^*_{11}(s)}}{D} = \frac{A_{11}}{(s-r_1)} + \frac{B_{11}}{(s-r_2)} + \frac{C_{11}}{(s-r_3)} + \frac{D_{11}}{(s-r_4)}, use\ inverse\ laplace\ of\ both\ sides$$

$$\therefore P_{11}(t) = A_{11}\, e^{r_1 t} + B_{11}\, e^{r_2 t} + C_{11}\, e^{r_3 t} + D_{11} e^{r_4 t}$$

$$For\ \ P^*_{12}(s)\ : \frac{\lambda_{12} s^2 + \lambda_{12}\,(\gamma_3+\gamma_4)s + \lambda_{12}\,\gamma_3\gamma_4 - \lambda_{12}\,\lambda_{34}\,\mu_{43}}{D} = \frac{A_{12}}{(s-r_1)} + \frac{B_{12}}{(s-r_2)} + \frac{C_{12}}{(s-r_3)} + \frac{D_{12}}{(s-r_4)}$$

The same procedure is used to solve the following PDFs' ; what differs is the numerator of both sides, but the denominators on both sides are the same for all upcoming equations. In matrix notation :

$$\begin{bmatrix} A_{12} \\ B_{12} \\ C_{12} \\ D_{12} \end{bmatrix} = [K(r)]^{-1} \begin{bmatrix} 0 \\ \lambda_{12} \\ \lambda_{12}\,(\gamma_3+\gamma_4) \\ \lambda_{12}\,\gamma_3\gamma_4 - \lambda_{12}\,\lambda_{34}\,\mu_{43} \end{bmatrix}$$

$$\therefore P^*_{12}(s) = \frac{D_{P^*_{12}(s)}}{D} = \frac{A_{12}}{(s-r_1)} + \frac{B_{12}}{(s-r_2)} + \frac{C_{12}}{(s-r_3)} + \frac{D_{12}}{(s-r_4)}, \qquad so\ \ P_{12}(t) = A_{12}\, e^{r_1 t} + B_{12}\, e^{r_2 t} + C_{12}\, e^{r_3 t} + D_{12} e^{r_4 t}$$

*For* $P^*_{13}(s)$:

$$\frac{\lambda_{12}\lambda_{23}\ s + \lambda_{12}\lambda_{23}\ \gamma_4}{D} = \frac{A_{13}}{(s-r_1)} + \frac{B_{13}}{(s-r_2)} + \frac{C_{13}}{(s-r_3)} + \frac{D_{13}}{(s-r_4)}\ \ ,\ In\ matrix\ notation:$$

$$\begin{bmatrix} A_{13} \\ B_{13} \\ C_{13} \\ D_{13} \end{bmatrix} = [K(r)]^{-1} \begin{bmatrix} 0 \\ 0 \\ \lambda_{12}\lambda_{23} \\ \lambda_{12}\lambda_{23}\ \gamma_4 \end{bmatrix}$$

$$\therefore P^*_{13}(s) = \frac{D_{P^*_{13}(s)}}{D} = \frac{A_{13}}{(s-r_1)} + \frac{B_{13}}{(s-r_2)} + \frac{C_{13}}{(s-r_3)} + \frac{D_{13}}{(s-r_4)}, so\ P_{13}(t) = A_{13}\, e^{r_1 t} + B_{13}\, e^{r_2 t} + C_{13}\, e^{r_3 t} + D_{13} e^{r_4 t}$$

$$For\ \ P^*_{14}(s)\ : \frac{\lambda_{12}\lambda_{23}\,\lambda_{34}}{D} = \frac{A_{14}}{(s-r_1)} + \frac{B_{14}}{(s-r_2)} + \frac{C_{14}}{(s-r_3)} + \frac{D_{14}}{(s-r_4)}\quad ,\ In\ matrix\ notation:$$

$$\begin{bmatrix} A_{14} \\ B_{14} \\ C_{14} \\ D_{14} \end{bmatrix} = [K(r)]^{-1} \begin{bmatrix} 0 \\ 0 \\ 0 \\ \lambda_{12}\lambda_{23}\lambda_{34} \end{bmatrix}$$

$$\therefore P_{14}^*(s) = \frac{D_{P_{14}^*(s)}}{D} = \frac{A_{14}}{(s-r_1)} + \frac{B_{14}}{(s-r_2)} + \frac{C_{14}}{(s-r_3)} + \frac{D_{14}}{(s-r_4)}, \qquad so \;\; P_{14}(t) = A_{14}\, e^{r_1 t} + B_{14}\, e^{r_2 t} + C_{14}\, e^{r_3 t} + D_{14} e^{r_4 t}$$

$$For \;\; P_{21}^*(s): \rightarrow \begin{bmatrix} A_{21} \\ B_{21} \\ C_{21} \\ D_{21} \end{bmatrix} = [K(r)]^{-1} \begin{bmatrix} 0 \\ \mu_{21} \\ \mu_{21}\,\gamma_3 + \mu_{21}\,\gamma_4 \\ \mu_{21}\,\gamma_3\,\gamma_4 - \mu_{21}\lambda_{34}\,\mu_{43} \end{bmatrix}$$

$$\therefore P_{21}^*(s) = \frac{D_{P_{21}^*(s)}}{D} = \frac{A_{21}}{(s-r_1)} + \frac{B_{21}}{(s-r_2)} + \frac{C_{21}}{(s-r_3)} + \frac{D_{21}}{(s-r_4)}, \qquad so \; P_{21}(t) = A_{21}\, e^{r_1 t} + B_{21}\, e^{r_2 t} + C_{21}\, e^{r_3 t} + D_{21} e^{r_4 t}$$

$$For \;\; P_{22}^*(s): \rightarrow \begin{bmatrix} A_{22} \\ B_{22} \\ C_{22} \\ D_{22} \end{bmatrix} = [K(r)]^{-1} \begin{bmatrix} 1 \\ \gamma_1 + \gamma_3 + \gamma_4 \\ \gamma_1\gamma_3 + \gamma_1\gamma_4 + \gamma_3\gamma_4 - \lambda_{34}\,\mu_{43} \\ -\lambda_{34}\,\mu_{43}\gamma_1 + \gamma_1\gamma_3\gamma_4 \end{bmatrix}$$

$$\therefore \;\; P_{22}^*(s) = \frac{D_{P_{22}^*(s)}}{D} = \frac{A_{22}}{(s-r_1)} + \frac{B_{22}}{(s-r_2)} + \frac{C_{22}}{(s-r_3)} + \frac{D_{22}}{(s-r_4)}, so \; P_{22}(t) = A_{22}\, e^{r_1 t} + B_{22}\, e^{r_2 t} + C_{22}\, e^{r_3 t} + D_{22} e^{r_4 t}$$

$$For \;\; P_{23}^*(s): \rightarrow \begin{bmatrix} A_{23} \\ B_{23} \\ C_{23} \\ D_{23} \end{bmatrix} = [K(r)]^{-1} \begin{bmatrix} 0 \\ \lambda_{23} \\ \lambda_{23}(\gamma_1 + \gamma_4) \\ \lambda_{23}\gamma_1\gamma_4 \end{bmatrix}$$

$$\therefore P_{23}^*(s) = \frac{D_{P_{23}^*(s)}}{D} = \frac{A_{23}}{(s-r_1)} + \frac{B_{23}}{(s-r_2)} + \frac{C_{23}}{(s-r_3)} + \frac{D_{23}}{(s-r_4)}, so \; P_{23}(t) = A_{23}\, e^{r_1 t} + B_{23}\, e^{r_2 t} + C_{23}\, e^{r_3 t} + D_{23} e^{r_4 t}$$

$$For \;\; P_{24}^*(s): \rightarrow \begin{bmatrix} A_{24} \\ B_{24} \\ C_{24} \\ D_{24} \end{bmatrix} = [K(r)]^{-1} \begin{bmatrix} 0 \\ 0 \\ \lambda_{23}\lambda_{34} \\ \lambda_{23}\lambda_{34}\gamma_1 \end{bmatrix}$$

$$\therefore \;\; P_{24}^*(s) = \frac{D_{P_{24}^*(s)}}{D} = \frac{A_{24}}{(s-r_1)} + \frac{B_{24}}{(s-r_2)} + \frac{C_{24}}{(s-r_3)} + \frac{D_{24}}{(s-r_4)}, so \; P_{24}(t) = A_{24}\, e^{r_1 t} + B_{24}\, e^{r_2 t} + C_{24}\, e^{r_3 t} + D_{24} e^{r_4 t}$$

$$For \;\; P_{31}^*(s): \rightarrow \begin{bmatrix} A_{31} \\ B_{31} \\ C_{31} \\ D_{31} \end{bmatrix} = [K(r)]^{-1} \begin{bmatrix} 0 \\ 0 \\ \mu_{21}\,\mu_{32} \\ \mu_{21}\,\mu_{32}\gamma_4 \end{bmatrix}$$

$$\therefore P_{31}^*(s) = \frac{D_{P_{31}^*(s)}}{D} = \frac{A_{31}}{(s-r_1)} + \frac{B_{31}}{(s-r_2)} + \frac{C_{31}}{(s-r_3)} + \frac{D_{31}}{(s-r_4)}, P_{31}(t) = A_{31}\, e^{r_1 t} + B_{31}\, e^{r_2 t} + C_{31}\, e^{r_3 t} + D_{31} e^{r_4 t}$$

$$For \;\; P_{32}^*(s): \rightarrow \begin{bmatrix} A_{32} \\ B_{32} \\ C_{32} \\ D_{32} \end{bmatrix} = [K(r)]^{-1} \begin{bmatrix} 0 \\ \mu_{32} \\ \mu_{32}(\gamma_1 + \gamma_4) \\ \mu_{32}\gamma_1\gamma_4 \end{bmatrix}$$

$$\therefore P_{32}^*(s) = \frac{D_{P_{32}^*(s)}}{D} = \frac{A_{32}}{(s-r_1)} + \frac{B_{32}}{(s-r_2)} + \frac{C_{32}}{(s-r_3)} + \frac{D_{32}}{(s-r_4)}, so \; P_{32}(t) = A_{32}\, e^{r_1 t} + B_{32}\, e^{r_2 t} + C_{32}\, e^{r_3 t} + D_{32} e^{r_4 t}$$

$$For\ \ P_{33}^*(s):\rightarrow \begin{bmatrix} A_{33} \\ B_{33} \\ C_{33} \\ D_{33} \end{bmatrix} = [K(r)]^{-1} \begin{bmatrix} 1 \\ \gamma_1+\gamma_2+\gamma_4 \\ \gamma_1\gamma_2+\gamma_1\gamma_4+\gamma_2\gamma_4-\ \mu_{21}\lambda_{12} \\ \gamma_1\gamma_2\gamma_4-\ \mu_{21}\lambda_{12}\gamma_4 \end{bmatrix}$$

$$\therefore P_{33}^*(s) = \frac{D_{P_{33}^*(s)}}{D} = \frac{A_{33}}{(s-r_1)} + \frac{B_{33}}{(s-r_2)} + \frac{C_{33}}{(s-r_3)} + \frac{D_{33}}{(s-r_4)}\ \ , \quad so\ P_{33}(t) = A_{33}\ e^{r_1 t} + B_{33}\ e^{r_2 t} + C_{33}\ e^{r_3 t} + D_{33} e^{r_4 t}$$

$$For\ \ P_{34}^*(s):\rightarrow \begin{bmatrix} A_{34} \\ B_{34} \\ C_{34} \\ D_{34} \end{bmatrix} = [K(r)]^{-1} \begin{bmatrix} 0 \\ \lambda_{34} \\ \lambda_{34}(\gamma_1+\gamma_2) \\ \lambda_{34}\gamma_1\gamma_2-\lambda_{12}\lambda_{34}\mu_{21} \end{bmatrix}$$

$$\therefore P_{34}^*(s) = \frac{D_{P_{34}^*(s)}}{D} = \frac{A_{34}}{(s-r_1)} + \frac{B_{34}}{(s-r_2)} + \frac{C_{34}}{(s-r_3)} + \frac{D_{34}}{(s-r_4)}\ \ , so\ P_{34}(t) = A_{34}\ e^{r_1 t} + B_{34}\ e^{r_2 t} + C_{34}\ e^{r_3 t} + D_{34} e^{r_4 t}$$

$$For\ \ P_{41}^*(s):\rightarrow \begin{bmatrix} A_{41} \\ B_{41} \\ C_{41} \\ D_{41} \end{bmatrix} = [K(r)]^{-1} \begin{bmatrix} 0 \\ 0 \\ 0 \\ \mu_{21}\,\mu_{32}\ \ \mu_{43} \end{bmatrix}$$

$$\therefore\ P_{41}^*(s) = \frac{D_{P_{41}^*(s)}}{D} = \frac{A_{41}}{(s-r_1)} + \frac{B_{41}}{(s-r_2)} + \frac{C_{41}}{(s-r_3)} + \frac{D_{41}}{(s-r_4)}\ \ , so\ P_{41}(t) = A_{41}\ e^{r_1 t} + B_{41}\ e^{r_2 t} + C_{41}\ e^{r_3 t} + D_{41} e^{r_4 t}$$

$$For\ \ P_{42}^*(s):\rightarrow \begin{bmatrix} A_{42} \\ B_{42} \\ C_{42} \\ D_{42} \end{bmatrix} = [K(r)]^{-1} \begin{bmatrix} 0 \\ 0 \\ \mu_{32}\ \ \mu_{43} \\ \mu_{32}\,\mu_{43}\,\gamma_1 \end{bmatrix}$$

$$\therefore P_{42}^*(s) = \frac{D_{P_{42}^*(s)}}{D} = \frac{A_{42}}{(s-r_1)} + \frac{B_{42}}{(s-r_2)} + \frac{C_{42}}{(s-r_3)} + \frac{D_{42}}{(s-r_4)}\ \ , so\ \ P_{42}(t) = A_{42}\ e^{r_1 t} + B_{42}\ e^{r_2 t} + C_{42}\ e^{r_3 t} + D_{42} e^{r_4 t}$$

$$For\ \ P_{43}^*(s):\rightarrow \begin{bmatrix} A_{43} \\ B_{43} \\ C_{43} \\ D_{43} \end{bmatrix} = [K(r)]^{-1} \begin{bmatrix} 0 \\ \mu_{43} \\ \mu_{43}(\gamma_1+\gamma_2) \\ \gamma_2\gamma_1\mu_{43}-\lambda_{12}\mu_{21}\mu_{43} \end{bmatrix}$$

$$\therefore P_{43}^*(s) = \frac{D_{P_{43}^*(s)}}{D} = \frac{A_{43}}{(s-r_1)} + \frac{B_{43}}{(s-r_2)} + \frac{C_{43}}{(s-r_3)} + \frac{D_{43}}{(s-r_4)}\ , so\ P_{43}(t) = A_{43}\ e^{r_1 t} + B_{43}\ e^{r_2 t} + C_{43}\ e^{r_3 t} + D_{43} e^{r_4 t}$$

$$For\ \ P_{44}^*(s):\rightarrow \begin{bmatrix} A_{44} \\ B_{44} \\ C_{44} \\ D_{44} \end{bmatrix} = [K(r)]^{-1} \begin{bmatrix} 1 \\ \gamma_1+\gamma_2+\gamma_3 \\ \gamma_1\gamma_2+\gamma_1\gamma_3+\ \gamma_2\gamma_3-\mu_{21}\lambda_{12}-\lambda_{23}\ \ \mu_{32} \\ -\lambda_{23}\ \ \mu_{32}\,\gamma_1-\ \mu_{21}\lambda_{12}\gamma_3+\gamma_1\gamma_2\gamma_3 \end{bmatrix}$$

$$\therefore\ P_{44}^*(s) = \frac{D_{P_{44}^*(s)}}{D} = \frac{A_{44}}{(s-r_1)} + \frac{B_{44}}{(s-r_2)} + \frac{C_{44}}{(s-r_3)} + \frac{D_{44}}{(s-r_4)}\ , so\ P_{44}(t) = A_{44}\ e^{r_1 t} + B_{44}\ e^{r_2 t} + C_{44}\ e^{r_3 t} + D_{44} e^{r_4 t}$$

**Solving the Last 5 Probabilities in the First Row by Using the Method of Integrating Factor**

$$\frac{d\,P_{15}(t)}{dt} = \lambda_{45}\,P_{14}(t) - (\lambda_{56}+\lambda_{58}+\lambda_{59})P_{15}(t)\,,\ \ let\ \ (\lambda_{56}+\lambda_{58}+\lambda_{59}) = w$$

$$\frac{d\,P_{15}(t)}{dt} = \lambda_{45}\,P_{14}(t) - wP_{15}(t)\ \ , so\ \frac{d\,P_{15}(t)}{dt} + wP_{15}(t) = \lambda_{45}\,P_{14}(t)\,, multiply\ both\ sides\ by\ integrating\ factor\ e^{wt}$$

$$e^{wt}\ \ \frac{d\,P_{15}(t)}{dt} + w\,e^{wt}\ \ P_{15}(t) = \lambda_{45}\ \ e^{wt}\ \ \ P_{14}(t)\ \ \ , \qquad substitute\ for\ \ P_{14}(t)$$

$$e^{wt} \frac{d\,P_{15}(t)}{dt} + w\,e^{wt}\,P_{15}(t) = \lambda_{45}\,e^{wt}\,[A_{14}\,e^{r_1 t} + B_{14}\,e^{r_2 t} + C_{14}\,e^{r_3 t} + D_{14}e^{r_4 t}]$$

$$\frac{d}{dt}\,e^{wt}\,P_{15}(t) = \lambda_{45}\,A_{14}\,e^{(w+r_1)t} + \lambda_{45}\,B_{14}\,e^{(w+r_2)t} + \lambda_{45}\,C_{14}\,e^{(w+r_3)t} + \lambda_{45}\,D_{14}e^{(w+r_4)t}$$

$let\ \ \lambda_{45}\,A_{14} = G_1\ ,\ \lambda_{45}\,B_{14} = G_2\ ,\ \lambda_{45}\,C_{14} = G_3\ ,\ \lambda_{45}\,D_{14} = G_4\ and\ integrate\ both\ sides\ from\ (0\ to\ t)$

$$e^{wt}\,P_{15}(t) = G_1\left[\frac{e^{(w+r_1)t}}{w+r_1} - \frac{1}{w+r_1}\right] + G_2\left[\frac{e^{(w+r_2)t}}{w+r_2} - \frac{1}{w+r_2}\right] + G_3\left[\frac{e^{(w+r_3)t}}{w+r_3} - \frac{1}{w+r_3}\right] + G_4\left[\frac{e^{(w+r_4)t}}{w+r_4} - \frac{1}{w+r_4}\right]$$

$multiply\ both\ sides\ by\ e^{-wt}$

$$P_{15}(t) = \frac{G_1}{w+r_1}\,[e^{r_1 t} - e^{-wt}] + \frac{G_2}{w+r_2}\,[e^{r_2 t} - e^{-wt}] + \frac{G_3}{w+r_3}\,[e^{r_3 t} - e^{-wt}] + \frac{G_4}{w+r_4}\,[e^{r_4 t} - e^{-wt}]$$

$$P_{15}(t) = \frac{G_1}{w+r_1}\,e^{r_1 t} + \frac{G_2}{w+r_2}e^{r_2 t} + \frac{G_3}{w+r_3}\,e^{r_3 t} + \frac{G_4}{w+r_4}e^{r_4 t} - \left[\frac{G_1}{w+r_1} + \frac{G_2}{w+r_2} + \frac{G_3}{w+r_3} + \frac{G_4}{w+r_4}\right]e^{-wt}$$

$$let\ \frac{G_1}{w+r_1} = F_1\ ,\quad \frac{G_2}{w+r_2} = F_2\ ,\quad \frac{G_3}{w+r_3} = F_3\ ,\frac{G_4}{w+r_4} = F_4\ ,\quad -\left[\frac{G_1}{w+r_1} + \frac{G_2}{w+r_2} + \frac{G_3}{w+r_3} + \frac{G_4}{w+r_4}\right] = F_5$$

$$P_{15}(t) = F_1\,e^{r_1 t} + F_2 e^{r_2 t} + F_3\,e^{r_3 t} + F_4 e^{r_4 t} + F_5 e^{-wt}$$

$$\frac{d\,P_{16}(t)}{dt} = \lambda_{56}\,P_{15}(t) - (\lambda_{67} + \lambda_{69})P_{16}(t)\ ,\qquad let\ \ (\lambda_{67} + \lambda_{69}) = u$$

$$\frac{d\,P_{16}(t)}{dt} = \lambda_{56}\,P_{15}(t) - uP_{16}(t)\ ,so\ \ \frac{d\,P_{16}(t)}{dt} + uP_{16}(t) = \lambda_{56}\,P_{15}(t)\ ,multiply\ both\ sides\ by\ integrating\ factor\ e^{ut}$$

$$e^{ut}\,\frac{d\,P_{16}(t)}{dt} + u\,e^{ut}\,P_{16}(t) = \lambda_{56}\,e^{ut}\,P_{15}(t)\ ,\qquad substitute\ for\ P_{15}(t)$$

$$e^{ut}\,\frac{d\,P_{16}(t)}{dt} + u\,e^{ut}\,P_{16}(t) = \lambda_{56}\,e^{ut}\,[F_1\,e^{r_1 t} + F_2 e^{r_2 t} + F_3\,e^{r_3 t} + F_4 e^{r_4 t} + F_5 e^{-wt}]$$

$$\frac{d}{dt}\,e^{ut}\,P_{16}(t) = \lambda_{56}\,F_1\,e^{(u+r_1)t} + \lambda_{56}\,F_2\,e^{(u+r_2)t} + \lambda_{56}\,F_3\,e^{(u+r_3)t} + \lambda_{56}\,F_4 e^{(u+r_4)t} + \lambda_{56}\,F_5 e^{(u-w)t}$$

$let\ \ \lambda_{56}\,F_1 = G_5\ ,\ \lambda_{56}\,F_2 = G_6\ ,\ \lambda_{56}\,F_3 = G_7\ ,\ \lambda_{56}\,F_4 = G_8\ ,\lambda_{56}\,F_5 = G_9\ and\ integrate\ both\ sides\ from\ (0\ to\ t)$

$$\frac{d}{dt}e^{ut}\,P_{16}(t) = G_5\,e^{(u+r_1)t} + G_6\,e^{(u+r_2)t} + G_7\,e^{(u+r_3)t} + G_8 e^{(u+r_4)t} + G_9\,e^{(u-w)t}$$

$$e^{ut}\,P_{16}(t) = G_5\left[\frac{e^{(u+r_1)t}}{u+r_1} - \frac{1}{u+r_1}\right] + G_6\left[\frac{e^{(u+r_2)t}}{u+r_2} - \frac{1}{u+r_2}\right] + G_7\left[\frac{e^{(u+r_3)t}}{u+r_3} - \frac{1}{u+r_3}\right] + G_8\left[\frac{e^{(u+r_4)t}}{u+r_4} - \frac{1}{u+r_4}\right] + G_9\left[\frac{e^{(u-w)t}}{u-w} - \frac{1}{u-w}\right]$$

$multiply\ both\ sides\ by\ e^{-ut}$

$$P_{16}(t) = \frac{G_5}{u+r_1}\,[e^{r_1 t} - e^{-ut}] + \frac{G_6}{u+r_2}\,[e^{r_2 t} - e^{-ut}] + \frac{G_7}{u+r_3}\,[e^{r_3 t} - e^{-ut}] + \frac{G_8}{u+r_4}\,[e^{r_4 t} - e^{-ut}] + \frac{G_9}{u-w}\,[e^{-wt} - e^{-ut}]$$

$$P_{16}(t) = \frac{G_5}{u+r_1}\,e^{r_1 t} + \frac{G_6}{u+r_2}e^{r_2 t} + \frac{G_7}{u+r_3}\,e^{r_3 t} + \frac{G_8}{u+r_4}e^{r_4 t} + \frac{G_9}{u-w}e^{-wt} - \left[\frac{G_5}{u+r_1} + \frac{G_6}{u+r_2} + \frac{G_7}{u+r_3} + \frac{G_8}{u+r_4} + \frac{G_9}{u-w}\right]e^{-ut}$$

$$let\ \frac{G_5}{u+r_1} = F_6\ ,\ \frac{G_6}{u+r_2} = F_7\ ,\frac{G_7}{u+r_3} = F_8\ ,\frac{G_8}{u+r_4} = F_9,\frac{G_9}{u-w} = F_{10}\ ,-\left[\frac{G_5}{u+r_1} + \frac{G_6}{u+r_2} + \frac{G_7}{u+r_3} + \frac{G_8}{u+r_4} + \frac{G_9}{u-w}\right] = F_{11}$$

$$P_{16}(t) = F_6\,e^{r_1 t} + F_7 e^{r_2 t} + F_8\,e^{r_3 t} + F_9 e^{r_4 t} + F_{10}e^{-wt} + F_{11}e^{-ut}$$

$$\frac{d\,P_{17}(t)}{dt} = \lambda_{67}\,P_{16}(t) - \lambda_{79}\,P_{17}(t)\,, \qquad so\ \ \frac{d\,P_{17}(t)}{dt} + \lambda_{79}\,P_{17}(t) = \lambda_{67}\,P_{16}(t)\quad,\ multiply\ both\ sides\ by\ e^{\lambda_{79}t}$$

$$e^{\lambda_{79}t}\ \frac{d\,P_{17}(t)}{dt} + \lambda_{79}\ \ e^{\lambda_{79}t}\ P_{17}(t) = \lambda_{67}\ \ e^{\lambda_{79}t}\ P_{16}(t)\quad,\ \ substitute\ \ for\ P_{16}(t)$$

$$\frac{d}{dt}\ \ e^{\lambda_{79}t}\ P_{17}(t)\ = \lambda_{67}\ \ e^{\lambda_{79}t}[\ F_6\,e^{r_1t} + F_7e^{r_2t} + F_8\,e^{r_3t} + F_9e^{r_4t} + \ F_{10}e^{-wt} + \ F_{11}e^{-ut}]$$

$$\frac{d}{dt}\ \ e^{\lambda_{79}t}\ P_{17}(t)\ = \lambda_{67}\,F_6\,e^{(\lambda_{79}+r_1)t} + \lambda_{67}\,F_7e^{(\lambda_{79}+r_2)t} + \lambda_{67}\,F_8\,e^{(\lambda_{79}+r_3)t} + \lambda_{67}\,F_9e^{(\lambda_{79}+r_4)t} + \lambda_{67}\,F_{10}e^{(\lambda_{79}-w)t} + \lambda_{67}\,F_{11}e^{(\lambda_{79}-u)t}$$

$let\quad \lambda_{67}\,F_6 = G_{10}\,,\qquad \lambda_{67}\,F_7 = \ G_{11},\qquad \lambda_{67}\,F_8 = G_{12}\ ,\qquad \lambda_{67}\,F_9 = G_{13},\qquad \lambda_{67}\,F_{10} = G_{14}\,,\qquad \lambda_{67}\,F_{11} = G_{15}$

$$e^{\lambda_{79}t}\ \ P_{17}(t) = G_{10}\left[\frac{e^{(\lambda_{79}+r_1)t}}{\lambda_{79}+r_1} - \frac{1}{\lambda_{79}+r_1}\right] + G_{11}\left[\frac{e^{(\lambda_{79}+r_2)t}}{\lambda_{79}+r_2} - \frac{1}{\lambda_{79}+r_2}\right] + G_{12}\left[\frac{e^{(\lambda_{79}+r_3)t}}{\lambda_{79}+r_3} - \frac{1}{\lambda_{79}+r_3}\right] + G_{13}\left[\frac{e^{(\lambda_{79}+r_4)t}}{\lambda_{79}+r_4} - \frac{1}{\lambda_{79}+r_4}\right] + G_{14}\left[\frac{e^{(\lambda_{79}-w)t}}{\lambda_{79}-w} - \frac{1}{\lambda_{79}-w}\right] + G_{15}\left[\frac{e^{(\lambda_{79}-u)t}}{\lambda_{79}-u} - \frac{1}{\lambda_{79}-u}\right]$$

*multiply both sides by* $e^{-\lambda_{79}t}$

$$P_{17}(t)\ = \frac{G_{10}}{\lambda_{79}+r_1}\left[e^{r_1t} - e^{-\lambda_{79}t}\right] + \frac{G_{11}}{\lambda_{79}+r_2}\left[e^{r_2t} - e^{-\lambda_{79}t}\right] + \frac{G_{12}}{\lambda_{79}+r_3}\left[e^{r_3t} - e^{-\lambda_{79}t}\right] + \frac{G_{13}}{\lambda_{79}+r_4}\left[e^{r_4t} - e^{-\lambda_{79}t}\right] + \frac{G_{14}}{\lambda_{79}-w}\left[e^{-wt} - e^{-\lambda_{79}t}\right] + \frac{G_{15}}{\lambda_{79}-u}\left[e^{-ut} - e^{-\lambda_{79}t}\right]$$

$$P_{17}(t)\ = \frac{G_{10}}{\lambda_{79}+r_1}\ e^{r_1t} + \frac{G_{11}}{\lambda_{79}+r_2}\ e^{r_2t} + \frac{G_{12}}{\lambda_{79}+r_3}e^{r_3t} + \frac{G_{13}}{\lambda_{79}+r_4}\ e^{r_4t} + \frac{G_{14}}{\lambda_{79}-w}\ e^{-wt} + \frac{G_{15}}{\lambda_{79}-u}\ e^{-ut}$$

$$-\left[\frac{G_{10}}{\lambda_{79}+r_1} + \frac{G_{11}}{\lambda_{79}+r_2} + \frac{G_{12}}{\lambda_{79}+r_3} + \frac{G_{13}}{\lambda_{79}+r_4} + \frac{G_{14}}{\lambda_{79}-w} + \frac{G_{15}}{\lambda_{79}-u}\right]e^{-\lambda_{79}t}$$

$$let\quad \frac{G_{10}}{\lambda_{79}+r_1} = F_{12}\,, \frac{G_{11}}{\lambda_{79}+r_2} = F_{13}\,, \frac{G_{12}}{\lambda_{79}+r_3} = \ F_{14}\ , \frac{G_{13}}{\lambda_{79}+r_4} = F_{15}\ , \frac{G_{14}}{\lambda_{79}-w} = \ F_{16}\ , \frac{G_{15}}{\lambda_{79}-u} = \ F_{17},$$

$$-\left[\frac{G_{10}}{\lambda_{79}+r_1} + \frac{G_{11}}{\lambda_{79}+r_2} + \frac{G_{12}}{\lambda_{79}+r_3} + \frac{G_{13}}{\lambda_{79}+r_4} + \frac{G_{14}}{\lambda_{79}-w} + \frac{G_{15}}{\lambda_{79}-u}\right] = F_{18}$$

$$P_{17}(t)\ = F_{12}\,e^{r_1t} + F_{13}\,e^{r_2t} + F_{14}e^{r_3t} + F_{15}\,e^{r_4t} + F_{16}\,e^{-wt} + F_{17}\,e^{-ut} + F_{18}\,e^{-\lambda_{79}t}$$

$$\frac{d\,P_{18}(t)}{dt} = \lambda_{18}\,P_{11}(t) + \lambda_{28}\,P_{12}(t) + \lambda_{38}\,P_{13}(t) + \lambda_{48}\,P_{14}(t) + \lambda_{58}\,P_{15}(t) - \lambda_{89}\,P_{18}(t)\,, rearrange:$$

$$\frac{d\,P_{18}(t)}{dt} + \lambda_{89}\,P_{18}(t) = \lambda_{18}\,P_{11}(t) + \lambda_{28}\,P_{12}(t) + \lambda_{38}\,P_{13}(t) + \lambda_{48}\,P_{14}(t) + \lambda_{58}\,P_{15}(t)$$

*multiply both sides by* $e^{\lambda_{89}t}$ *and substitute for* $P_{11}(t), P_{12}(t), P_{13}(t)\,, P_{14}(t), P_{15}(t)$

$$e^{\lambda_{89}t}\ \frac{d\,P_{18}(t)}{dt} + \lambda_{89}e^{\lambda_{89}t}\,P_{18}(t) = \lambda_{18}e^{\lambda_{89}t}\ P_{11}(t) + \lambda_{28}e^{\lambda_{89}t}\,P_{12}(t) + \lambda_{38}e^{\lambda_{89}t}\,P_{13}(t) + \lambda_{48}e^{\lambda_{89}t}\,P_{14}(t) + \lambda_{58}e^{\lambda_{89}t}\,P_{15}(t)$$

$$\frac{d}{dt}\ e^{\lambda_{89}t}P_{18}(t) = \lambda_{18}e^{\lambda_{89}t}\ P_{11}(t) + \lambda_{28}e^{\lambda_{89}t}\,P_{12}(t) + \lambda_{38}e^{\lambda_{89}t}\,P_{13}(t) + \lambda_{48}e^{\lambda_{89}t}\,P_{14}(t) + \lambda_{58}e^{\lambda_{89}t}\,P_{15}(t)$$

$$\because \lambda_{18}e^{\lambda_{89}t}\ P_{11}(t) = \lambda_{18}e^{\lambda_{89}t}\ [A_{11}e^{r_1t} + \ B_{11}e^{r_2t} + C_{11}e^{r_3t} + \ D_{11}e^{r_4t}] = \lambda_{18}A_{11}e^{(\lambda_{89}+r_1)t} + \lambda_{18}B_{11}e^{(\lambda_{89}+r_2)t} + \lambda_{18}C_{11}e^{(\lambda_{89}+r_3)t} + \lambda_{18}D_{11}e^{(\lambda_{89}+r_4)t}$$

$$\because \lambda_{28}e^{\lambda_{89}t}\,P_{12}(t) = \lambda_{28}e^{\lambda_{89}t}\,[A_{12}e^{r_1t} + B_{12}e^{r_2t} + C_{12}e^{r_3t} + D_{12}e^{r_4t}]$$
$$= \lambda_{28}\,A_{12}e^{(\lambda_{89}+r_1)t} + \lambda_{28}\,B_{12}e^{(\lambda_{89}+r_2)t} + \lambda_{28}C_{12}e^{(\lambda_{89}+r_3)t} + \lambda_{28}\,D_{12}e^{(\lambda_{89}+r_4)t}$$

$$\because \lambda_{38}e^{\lambda_{89}t}\,P_{13}(t) = \lambda_{38}e^{\lambda_{89}t}[A_{13}e^{r_1t} + B_{13}e^{r_2t} + C_{13}e^{r_3t} + D_{13}e^{r_4t}]$$
$$= \lambda_{38}\,A_{13}e^{(\lambda_{89}+r_1)t} + \lambda_{38}\,B_{13}e^{(\lambda_{89}+r_2)t} + \lambda_{38}C_{13}e^{(\lambda_{89}+r_3)t} + \lambda_{38}\,D_{13}e^{(\lambda_{89}+r_4)t}$$

$$\because \lambda_{48}e^{\lambda_{89}t}\,P_{14}(t) = \lambda_{48}e^{\lambda_{89}t}[A_{14}e^{r_1t} + B_{14}e^{r_2t} + C_{14}e^{r_3t} + D_{14}e^{r_4t}]$$
$$= \lambda_{48}A_{14}e^{(\lambda_{89}+r_1)t} + \lambda_{48}B_{14}e^{(\lambda_{89}+r_2)t} + \lambda_{48}C_{14}e^{(\lambda_{89}+r_3)t} + \lambda_{48}\,D_{14}e^{(\lambda_{89}+r_4)t}$$

$$\because \lambda_{58}e^{\lambda_{89}t}\,P_{15}(t) = \lambda_{58}e^{\lambda_{89}t}[\,F_1\,e^{r_1t} + F_2e^{r_2t} + F_3\,e^{r_3t} + F_4e^{r_4t} + F_5e^{-wt}]$$
$$= \lambda_{58}F_1\,e^{(\lambda_{89}+r_1)t} + \lambda_{58}F_2e^{(\lambda_{89}+r_2)t} + \lambda_{58}F_3\,e^{(\lambda_{89}+r_3)t} + \lambda_{58}F_4e^{(\lambda_{89}+r_4)t} + \lambda_{58}F_5e^{(\lambda_{89}-w)t}$$

$$\frac{d}{dt}\,e^{\lambda_{89}t}P_{18}(t) = [\lambda_{18}A_{11} + \lambda_{28}\,A_{12} + \lambda_{38}\,A_{13} + \lambda_{48}A_{14} + \lambda_{58}F_1]e^{(\lambda_{89}+r_1)t} + [\lambda_{18}B_{11} + \lambda_{28}\,B_{12} + \lambda_{38}\,B_{13} + \lambda_{48}B_{14} + \lambda_{58}F_2]e^{(\lambda_{89}+r_2)t}$$

$$+[\lambda_{18}C_{11} + \lambda_{28}\,C_{12} + \lambda_{38}\,C_{13} + \lambda_{48}C_{14} + \lambda_{58}F_3]e^{(\lambda_{89}+r_3)t} + [\lambda_{18}D_{11} + \lambda_{28}\,D_{12} + \lambda_{38}\,D_{13} + \lambda_{48}D_{14} + \lambda_{58}F_4]e^{(\lambda_{89}+r_4)t} + \lambda_{58}F_5e^{(\lambda_{89}-w)t}$$

*let* $[\lambda_{18}A_{11} + \lambda_{28}\,A_{12} + \lambda_{38}\,A_{13} + \lambda_{48}A_{14} + \lambda_{58}F_1] = G_{16}$ , $[\lambda_{18}B_{11} + \lambda_{28}\,B_{12} + \lambda_{38}\,B_{13} + \lambda_{48}B_{14} + \lambda_{58}F_2] = G_{17}$

$[\lambda_{18}C_{11} + \lambda_{28}\,C_{12} + \lambda_{38}\,C_{13} + \lambda_{48}C_{14} + \lambda_{58}F_3] = G_{18}$ , $[\lambda_{18}D_{11} + \lambda_{28}\,D_{12} + \lambda_{38}\,D_{13} + \lambda_{48}D_{14} + \lambda_{58}F_4] = G_{19}$ , $\lambda_{58}F_5 = G_{20}$

$$\frac{d}{dt}\,e^{\lambda_{89}t}P_{18}(t) = G_{16}e^{(\lambda_{89}+r_1)t} + G_{17}e^{(\lambda_{89}+r_2)t} + G_{18}e^{(\lambda_{89}+r_3)t} + G_{19}e^{(\lambda_{89}+r_4)t} + G_{20}e^{(\lambda_{89}-w)t}$$ *, integrate both sides :*

$$e^{\lambda_{89}t}P_{18}(t) = G_{16}\left[\frac{e^{(\lambda_{89}+r_1)t}}{\lambda_{89}+r_1} - \frac{1}{\lambda_{89}+r_1}\right] + G_{17}\left[\frac{e^{(\lambda_{89}+r_2)t}}{\lambda_{89}+r_2} - \frac{1}{\lambda_{89}+r_2}\right] + G_{18}\left[\frac{e^{(\lambda_{89}+r_3)t}}{\lambda_{89}+r_3} - \frac{1}{\lambda_{89}+r_3}\right] + G_{19}\left[\frac{e^{(\lambda_{89}+r_4)t}}{\lambda_{89}+r_4} - \frac{1}{\lambda_{89}+r_4}\right]$$
$$+ G_{20}\left[\frac{e^{(\lambda_{89}-w)t}}{\lambda_{89}-w} - \frac{1}{\lambda_{89}-w}\right]$$

*multiply both sides by* $e^{-\lambda_{89}t}$

$$P_{18}(t) = \frac{G_{16}}{\lambda_{89}+r_1}\left[e^{r_1t} - e^{-\lambda_{89}t}\right] + \frac{G_{17}}{\lambda_{89}+r_2}\left[e^{r_2t} - e^{-\lambda_{89}t}\right] + \frac{G_{18}}{\lambda_{89}+r_3}\left[e^{r_3t} - e^{-\lambda_{89}t}\right] + \frac{G_{19}}{\lambda_{89}+r_4}\left[e^{r_4t} - e^{-\lambda_{89}t}\right] + \frac{G_{20}}{\lambda_{89}-w}\left[e^{-wt} - e^{-\lambda_{89}t}\right]$$

$$P_{18}(t) = \frac{G_{16}}{\lambda_{89}+r_1}e^{r_1t} + \frac{G_{17}}{\lambda_{89}+r_2}e^{r_2t} + \frac{G_{18}}{\lambda_{89}+r_3}e^{r_3t} + \frac{G_{19}}{\lambda_{89}+r_4}e^{r_4t} + \frac{G_{20}}{\lambda_{89}-w}e^{-wt} - \left[\frac{G_{16}}{\lambda_{89}+r_1} + \frac{G_{17}}{\lambda_{89}+r_2} + \frac{G_{18}}{\lambda_{89}+r_3} + \frac{G_{19}}{\lambda_{89}+r_4} + \frac{G_{20}}{\lambda_{89}-w}\right]e^{-\lambda_{89}t}$$

*let* $\frac{G_{16}}{\lambda_{89}+r_1} = F_{19}$ , $\frac{G_{17}}{\lambda_{89}+r_2} = F_{20}$ , $\frac{G_{18}}{\lambda_{89}+r_3} = F_{21}$ , $\frac{G_{19}}{\lambda_{89}+r_4} = F_{22}$ , $\frac{G_{20}}{\lambda_{89}-w} = F_{23}$ $-\left[\frac{G_{16}}{\lambda_{89}+r_1} + \frac{G_{17}}{\lambda_{89}+r_2} + \frac{G_{18}}{\lambda_{89}+r_3} + \frac{G_{19}}{\lambda_{89}+r_4} + \frac{G_{20}}{\lambda_{89}-w}\right] = F_{24}$

$$P_{18}(t) = F_{19}e^{r_1t} + F_{20}e^{r_2t} + F_{21}e^{r_3t} + F_{22}e^{r_4t} + F_{23}e^{-wt} + F_{24}e^{-\lambda_{89}t}$$

$$\frac{d\,P_{19}(t)}{dt} = \lambda_{19}\,P_{11}(t) + \lambda_{29}\,P_{12}(t) + \lambda_{39}\,P_{13}(t) + \lambda_{49}\,P_{14}(t) + \lambda_{59}\,P_{15}(t) + \lambda_{69}\,P_{16}(t) + \lambda_{79}\,P_{17}(t) + \lambda_{89}\,P_{18}(t)$$

Substitute for each $P_{ij}(t)$ then integrate both sides:

$$\frac{d\,P_{19}(t)}{dt} = \lambda_{19}\,P_{11}(t) + \lambda_{29}\,P_{12}(t) + \lambda_{39}\,P_{13}(t) + \lambda_{49}\,P_{14}(t) + \lambda_{59}\,P_{15}(t) + \lambda_{69}\,P_{16}(t) + \lambda_{79}\,P_{17}(t) + \lambda_{89}\,P_{18}(t)$$

$$\because \lambda_{19}\,P_{11}(t) = \lambda_{19}[A_{11}e^{r_1t} + B_{11}e^{r_2t} + C_{11}e^{r_3t} + D_{11}e^{r_4t}] = \lambda_{19}A_{11}e^{r_1t} + \lambda_{19}\,B_{11}e^{r_2t} + \lambda_{19}C_{11}e^{r_3t} + \lambda_{19}\,D_{11}e^{r_4t}$$

$$\because \lambda_{29}\,P_{12}(t) = \lambda_{29}[A_{12}e^{r_1t} + B_{12}e^{r_2t} + C_{12}e^{r_3t} + D_{12}e^{r_4t}] = \lambda_{29}\,A_{12}e^{r_1t} + \lambda_{29}\,B_{12}e^{r_2t} + \lambda_{29}C_{12}e^{r_3t} + \lambda_{29}\,D_{12}e^{r_4t}$$

$$\because \lambda_{39}\,P_{13}(t) = \lambda_{39}[A_{13}e^{r_1t} + B_{13}e^{r_2t} + C_{13}e^{r_3t} + D_{13}e^{r_4t}] = \lambda_{39}\,A_{13}e^{r_1t} + \lambda_{39}\,B_{13}e^{r_2t} + \lambda_{39}C_{13}e^{r_3t} + \lambda_{39}\,D_{13}e^{r_4t}$$

$$\because \lambda_{49}\,P_{14}(t) = \lambda_{49}[A_{14}e^{r_1t} + B_{14}e^{r_2t} + C_{14}e^{r_3t} + D_{14}e^{r_4t}] = \lambda_{49}\,A_{14}e^{r_1t} + \lambda_{49}B_{14}e^{r_2t} + \lambda_{49}C_{14}e^{r_3t} + \lambda_{49}\,D_{14}e^{r_4t}$$

$$\because \lambda_{59}\,P_{15}(t) = \lambda_{59}[\,F_1\,e^{r_1t} + F_2e^{r_2t} + F_3\,e^{r_3t} + F_4e^{r_4t} + F_5e^{-wt}] = \lambda_{59}F_1\,e^{r_1t} + \lambda_{59}F_2e^{r_2t} + \lambda_{59}F_3\,e^{r_3t} + \lambda_{59}F_4e^{r_4t} + \lambda_{59}F_5e^{-wt}$$

$$\because \lambda_{69}\,P_{16}(t) = \lambda_{69}[F_6\,e^{r_1t} + F_7e^{r_2t} + F_8\,e^{r_3t} + F_9e^{r_4t} + F_{10}e^{-wt} + F_{11}e^{-ut}]$$
$$= \lambda_{69}\,F_6\,e^{r_1t} + \lambda_{69}F_7e^{r_2t} + \lambda_{69}F_8\,e^{r_3t} + \lambda_{69}\,F_9e^{r_4t} + \lambda_{69}\,F_{10}e^{-wt} + \lambda_{69}\,F_{11}e^{-ut}$$

$\because \lambda_{79}\, P_{17}(t) = \lambda_{79}\big[F_{12}\, e^{r_1 t} + F_{13}\, e^{r_2 t} + F_{14} e^{r_3 t} + F_{15}\, e^{r_4 t} + F_{16}\, e^{-wt} + F_{17}\, e^{-ut} + F_{18}\, e^{-\lambda_{79} t}\big]$

$$= \lambda_{79}\, F_{12}\, e^{r_1 t} + \lambda_{79} F_{13}\, e^{r_2 t} + \lambda_{79} F_{14} e^{r_3 t} + \lambda_{79} F_{15}\, e^{r_4 t} + \lambda_{79} F_{16}\, e^{-wt} + \lambda_{79} F_{17}\, e^{-ut} + \lambda_{79} F_{18}\, e^{-\lambda_{79} t}$$

$\because \lambda_{89}\, P_{18}(t) = \lambda_{89}\big[F_{19} e^{r_1 t} + F_{20} e^{r_2 t} + F_{21} e^{r_3 t} + F_{22} e^{r_4 t} + F_{23} e^{-wt} + F_{24} e^{-\lambda_{89} t}\big]$

$$= \lambda_{89}\, F_{19} e^{r_1 t} + \lambda_{89}\, F_{20} e^{r_2 t} + \lambda_{89} F_{21} e^{r_3 t} + \lambda_{89} F_{22} e^{r_4 t} + \lambda_{89} F_{23} e^{-wt} + \lambda_{89} F_{24} e^{-\lambda_{89} t}$$

$$\frac{d\, P_{19}(t)}{dt} =$$

$= (\lambda_{19} A_{11} + \lambda_{29}\, A_{12} + \lambda_{39}\, A_{13} + \lambda_{49}\, A_{14} + \lambda_{59}\, F_1 + \lambda_{69}\, F_6 + \lambda_{79}\, F_{12} + \lambda_{89}\, F_{19}) e^{r_1 t} +$

$(\lambda_{19} B_{11} + \lambda_{29}\, B_{12} + \lambda_{39}\, B_{13} + \lambda_{49}\, B_{14} + \lambda_{59}\, F_2 + \lambda_{69}\, F_7 + \lambda_{79}\, F_{13} + \lambda_{89}\, F_{20}) e^{r_2 t} +$

$(\lambda_{19} C_{11} + \lambda_{29}\, C_{12} + \lambda_{39}\, C_{13} + \lambda_{49}\, C_{14} + \lambda_{59}\, F_3 + \lambda_{69}\, F_8 + \lambda_{79}\, F_{14} + \lambda_{89}\, F_{21}) e^{r_3 t} +$

$(\lambda_{19} D_{11} + \lambda_{29}\, D_{12} + \lambda_{39}\, D_{13} + \lambda_{49}\, D_{14} + \lambda_{59}\, F_4 + \lambda_{69}\, F_9 + \lambda_{79}\, F_{15} + \lambda_{89}\, F_{22}) e^{r_4 t} +$

$(\lambda_{59}\, F_5 + \lambda_{69}\, F_{10} + \lambda_{79} F_{16} + \lambda_{89} F_{23}) e^{-wt} + (\lambda_{69}\, F_{11} + \lambda_{79} F_{17})\, e^{-ut} + \lambda_{79} F_{18}\, e^{-\lambda_{79} t} + \lambda_{89} F_{24} e^{-\lambda_{89} t}$

$Let \quad (\lambda_{19} A_{11} + \lambda_{29}\, A_{12} + \lambda_{39}\, A_{13} + \lambda_{49}\, A_{14} + \lambda_{59}\, F_1 + \lambda_{69}\, F_6 + \lambda_{79}\, F_{12} + \lambda_{89}\, F_{19}) = G_{21}$

$(\lambda_{19} B_{11} + \lambda_{29}\, B_{12} + \lambda_{39}\, B_{13} + \lambda_{49}\, B_{14} + \lambda_{59}\, F_2 + \lambda_{69}\, F_7 + \lambda_{79}\, F_{13} + \lambda_{89}\, F_{20}) = G_{22}$

$(\lambda_{19} C_{11} + \lambda_{29}\, C_{12} + \lambda_{39}\, C_{13} + \lambda_{49}\, C_{14} + \lambda_{59}\, F_3 + \lambda_{69}\, F_8 + \lambda_{79}\, F_{14} + \lambda_{89}\, F_{21}) = G_{23}$

$(\lambda_{19} D_{11} + \lambda_{29}\, D_{12} + \lambda_{39}\, D_{13} + \lambda_{49}\, D_{14} + \lambda_{59}\, F_4 + \lambda_{69}\, F_9 + \lambda_{79}\, F_{15} + \lambda_{89}\, F_{22}) = G_{24}\, , (\lambda_{59}\, F_5 + \lambda_{69}\, F_{10} + \lambda_{79} F_{16} + \lambda_{89} F_{23}) = G_{25}$

$, (\lambda_{69}\, F_{11} + \lambda_{79} F_{17}) = G_{26}\;, \;\; \lambda_{79} F_{18} = G_{27}\;, \qquad \lambda_{89} F_{24} = G_{28}$

$$\frac{d\, P_{19}(t)}{dt} = G_{21} e^{r_1 t} + G_{22} e^{r_2 t} + G_{23} e^{r_3 t} + G_{24} e^{r_4 t} + G_{25} e^{-wt} + G_{26} e^{-ut} + G_{27} e^{-\lambda_{79} t} + G_{28} e^{-\lambda_{89} t}$$

$$P_{19}(t) = G_{21}\left[\frac{e^{r_1 t}}{r_1} - \frac{1}{r_1}\right] + G_{22}\left[\frac{e^{r_2 t}}{r_2} - \frac{1}{r_2}\right] + G_{23}\left[\frac{e^{r_3 t}}{r_3} - \frac{1}{r_3}\right] + G_{24}\left[\frac{e^{r_4 t}}{r_4} - \frac{1}{r_4}\right] + G_{25}\left[\frac{e^{-wt}}{-w} - \frac{1}{-w}\right] + G_{26}\left[\frac{e^{-ut}}{-u} - \frac{1}{-u}\right] + G_{27}\left[\frac{e^{-\lambda_{79} t}}{-\lambda_{79}} - \frac{1}{-\lambda_{79}}\right] + G_{28}\left[\frac{e^{-\lambda_{89} t}}{-\lambda_{89}} - \frac{1}{-\lambda_{89}}\right]$$

$$P_{19}(t) = \frac{G_{21}}{r_1}[e^{r_1 t} - 1] + \frac{G_{22}}{r_2}[e^{r_2 t} - 1] + \frac{G_{23}}{r_3}[e^{r_3 t} - 1] + \frac{G_{24}}{r_4}[e^{r_4 t} - 1] + \frac{G_{25}}{w}[1 - e^{-wt}] + \frac{G_{26}}{u}[1 - e^{-ut}] + \frac{G_{27}}{\lambda_{79}}[1 - e^{-\lambda_{79} t}] + \frac{G_{28}}{\lambda_{89}}[1 - e^{-\lambda_{89} t}]$$

$$P_{19}(t) = \frac{G_{21}}{r_1} e^{r_1 t} + \frac{G_{22}}{r_2} e^{r_2 t} + \frac{G_{23}}{r_3} e^{r_3 t} + \frac{G_{24}}{r_4} e^{r_4 t} - \frac{G_{25}}{w} e^{-wt} - \frac{G_{26}}{u} e^{-ut} - \frac{G_{27}}{\lambda_{79}} e^{-\lambda_{79} t} - \frac{G_{28}}{\lambda_{89}} e^{-\lambda_{89} t} + \left[-\frac{G_{21}}{r_1} - \frac{G_{22}}{r_2} - \frac{G_{23}}{r_3} - \frac{G_{24}}{r_4} + \frac{G_{25}}{w} + \frac{G_{26}}{u} + \frac{G_{27}}{\lambda_{79}} + \frac{G_{28}}{\lambda_{89}}\right]$$

$let \;\; \frac{G_{21}}{r_1} = F_{25}\, , \frac{G_{22}}{r_2} = F_{26}\, , \frac{G_{23}}{r_3} = F_{27}\, , \frac{G_{24}}{r_4} = F_{28}\, , -\frac{G_{25}}{w} = F_{29}\, , -\frac{G_{26}}{u} = F_{30}\, , -\frac{G_{27}}{\lambda_{79}} = F_{31}\, , -\frac{G_{28}}{\lambda_{89}} = F_{32},$

$$\left[-\frac{G_{21}}{r_1} - \frac{G_{22}}{r_2} - \frac{G_{23}}{r_3} - \frac{G_{24}}{r_4} + \frac{G_{25}}{w} + \frac{G_{26}}{u} + \frac{G_{27}}{\lambda_{79}} + \frac{G_{28}}{\lambda_{89}}\right] = F_{33}$$

$$P_{19}(t) = F_{25} e^{r_1 t} + F_{26} e^{r_2 t} + F_{27} e^{r_3 t} + F_{28} e^{r_4 t} + F_{29} e^{-wt} + F_{30} e^{-ut} + F_{31} e^{-\lambda_{79} t} + F_{32} e^{-\lambda_{89} t} + F_{33}$$

**Solving The Last 5 Probabilities In The second Row By Using The Method Of Integrating Factor**

Using the same procedure as in the last 5 probabilities in the first row, but what differ are the coefficients for each PDFs':

$for\;\; P_{25}(t)\!: \;\; \lambda_{45}\;\, A_{24} = H_1 \;\;, \;\; \lambda_{45}\, B_{24} = H_2 \;\;\;, \;\; \lambda_{45}\, C_{24} = H_3 \;\;, \;\; \lambda_{45}\, D_{24} = H_4$

$$and \quad \frac{H_1}{w+r_1} = K_1 \;, \quad \frac{H_2}{w+r_2} = K_2 \;, \quad \frac{H_3}{w+r_3} = K_3 \;, \frac{H_4}{w+r_4} = K_4, \quad -\left[\frac{H_1}{w+r_1} + \frac{H_2}{w+r_2} + \frac{H_3}{w+r_3} + \frac{H_4}{w+r_4}\right] = K_5$$

$$P_{25}(t) = K_1\, e^{r_1 t} + K_2 e^{r_2 t} + K_3\, e^{r_3 t} + K_4 e^{r_4 t} + K_5 e^{-wt}$$

$$for\ P_{26}(t):\ \lambda_{56}\, K_1 = H_5 \;,\ \lambda_{56}\, K_2 = H_6 \;,\ \lambda_{56}\, K_3 = H_7 \;,\ \lambda_{56}\, K_4 = H_8 \;, \lambda_{56}\, K_5 = H_9$$

$$and \quad \frac{H_5}{u+r_1} = K_6 \,, \frac{H_6}{u+r_2} = K_7 \,, \frac{H_7}{u+r_3} = K_8 \,, \frac{H_8}{u+r_4} = K_9, \frac{H_9}{u-w} = K_{10} \,, -\left[\frac{H_5}{u+r_1} + \frac{H_6}{u+r_2} + \frac{H_7}{u+r_3} + \frac{H_8}{u+r_4} + \frac{H_9}{u-w}\right] = K_{11}$$

$$P_{26}(t) = K_6\, e^{r_1 t} + K_7 e^{r_2 t} + K_8\, e^{r_3 t} + K_9 e^{r_4 t} + K_{10} e^{-wt} + K_{11} e^{-ut}$$

$$for\ P_{27}(t):\ \lambda_{67}\, K_6 = H_{10} \,, \quad \lambda_{67}\, K_7 = H_{11}, \quad \lambda_{67}\, K_8 = H_{12} \;, \quad \lambda_{67}\, K_9 = H_{13}, \quad \lambda_{67}\, K_{10} = H_{14} \,, \quad \lambda_{67}\, K_{11} = H_{15}$$

$$and \quad \frac{H_{10}}{\lambda_{79}+r_1} = K_{12} \,, \quad \frac{H_{11}}{\lambda_{79}+r_2} = K_{13} \,, \quad \frac{H_{12}}{\lambda_{79}+r_3} = K_{14} \;, \quad \frac{H_{13}}{\lambda_{79}+r_4} = K_{15} \;, \quad \frac{H_{14}}{\lambda_{79}-w} = K_{16} \;, \quad \frac{H_{15}}{\lambda_{79}-u} = K_{17}$$

$$-\left[\frac{H_{10}}{\lambda_{79}+r_1} + \frac{H_{11}}{\lambda_{79}+r_2} + \frac{H_{12}}{\lambda_{79}+r_3} + \frac{H_{13}}{\lambda_{79}+r_4} + \frac{H_{14}}{\lambda_{79}-w} + \frac{H_{15}}{\lambda_{79}-u}\right] = K_{18}$$

$$P_{27}(t) = K_{12}\, e^{r_1 t} + K_{13}\, e^{r_2 t} + K_{14} e^{r_3 t} + K_{15}\, e^{r_4 t} + K_{16}\, e^{-wt} + K_{17}\, e^{-ut} + K_{18}\, e^{-\lambda_{79} t}$$

$$for\ P_{28}(t):\ [\lambda_{18} A_{21} + \lambda_{28}\, A_{22} + \lambda_{38}\, A_{23} + \lambda_{48} A_{24} + \lambda_{58} K_1] = H_{16} \,, [\lambda_{18} B_{21} + \lambda_{28}\, B_{22} + \lambda_{38}\, B_{23} + \lambda_{48} B_{24} + \lambda_{58} K_2] = H_{17}$$

$$[\lambda_{18} C_{21} + \lambda_{28}\, C_{22} + \lambda_{38}\, C_{23} + \lambda_{48} C_{24} + \lambda_{58} K_3] = H_{18} \,, [\lambda_{18} D_{21} + \lambda_{28}\, D_{22} + \lambda_{38}\, D_{23} + \lambda_{48} D_{24} + \lambda_{58} K_4] = H_{19} \,, \lambda_{58} K_5 = H_{20}$$

$$and \quad \frac{H_{16}}{\lambda_{89}+r_1} = K_{19} \,, \frac{H_{17}}{\lambda_{89}+r_2} = K_{20} \,, \frac{H_{18}}{\lambda_{89}+r_3} = K_{21} \,, \frac{H_{19}}{\lambda_{89}+r_4} = K_{22} \,, \frac{H_{20}}{\lambda_{89}-w} = K_{23}$$

$$, -\left[\frac{H_{16}}{\lambda_{89}+r_1} + \frac{H_{17}}{\lambda_{89}+r_2} + \frac{H_{18}}{\lambda_{89}+r_3} + \frac{H_{19}}{\lambda_{89}+r_4} + \frac{H_{20}}{\lambda_{89}-w}\right] = K_{24}$$

$$P_{28}(t) = K_{19} e^{r_1 t} + K_{20} e^{r_2 t} + K_{21} e^{r_3 t} + K_{22} e^{r_4 t} + K_{23} e^{-wt} + K_{24} e^{-\lambda_{89} t}$$

$$for\ P_{29}(t):\ (\lambda_{19} A_{21} + \lambda_{29}\, A_{22} + \lambda_{39}\, A_{23} + \lambda_{49}\, A_{24} + \lambda_{59}\, K_1 + \lambda_{69}\, K_6 + \lambda_{79}\, K_{12} + \lambda_{89}\, K_{19}) = H_{21}$$

$$(\lambda_{19} B_{21} + \lambda_{29}\, B_{22} + \lambda_{39}\, B_{23} + \lambda_{49}\, B_{24} + \lambda_{59}\, K_2 + \lambda_{69}\, K_7 + \lambda_{79}\, K_{13} + \lambda_{89}\, K_{20}) = H_{22}$$

$$(\lambda_{19} C_{21} + \lambda_{29}\, C_{22} + \lambda_{39}\, C_{23} + \lambda_{49}\, C_{24} + \lambda_{59}\, K_3 + \lambda_{69}\, K_8 + \lambda_{79}\, K_{14} + \lambda_{89}\, K_{21}) = H_{23}$$

$$(\lambda_{19} D_{21} + \lambda_{29}\, D_{22} + \lambda_{39}\, D_{23} + \lambda_{49}\, D_{24} + \lambda_{59}\, K_4 + \lambda_{69}\, K_9 + \lambda_{79}\, K_{15} + \lambda_{89}\, K_{22}) = H_{24} \,, (\lambda_{59}\, K_5 + \lambda_{69}\, K_{10} + \lambda_{79} K_{16} + \lambda_{89} K_{23}) = H_{25} \,, (\lambda_{69}\, K_{11} + \lambda_{79} K_{17}) = H_{26}$$

$$\lambda_{79} K_{18} = H_{27} \;, \quad \lambda_{89} K_{24} = H_{28}$$

$$and \quad \frac{H_{21}}{r_1} = K_{25} \;, \frac{H_{22}}{r_2} = K_{26} \,, \frac{H_{23}}{r_3} = K_{27} \;, \frac{H_{24}}{r_4} = K_{28} \;, \; -\frac{H_{25}}{w} = K_{29} \;, -\frac{H_{26}}{u} = K_{30} \;, \; -\frac{H_{27}}{\lambda_{79}} = K_{31} \,, \quad -\frac{H_{28}}{\lambda_{89}} = K_{32},$$

$$+\left[-\frac{H_{21}}{r_1} - \frac{H_{22}}{r_2} - \frac{H_{23}}{r_3} - \frac{H_{24}}{r_4} + \frac{H_{25}}{w} + \frac{H_{26}}{u} + \frac{H_{27}}{\lambda_{79}} + \frac{H_{28}}{\lambda_{89}}\right] = K_{33}$$

$$P_{29}(t) = K_{25} e^{r_1 t} + K_{26} e^{r_2 t} + K_{27} e^{r_3 t} + K_{28} e^{r_4 t} + K_{29} e^{-wt} + K_{30} e^{-ut} + K_{31} e^{-\lambda_{79} t} + K_{32} e^{-\lambda_{89} t} + K_{33}$$

**Solving The Last 5 Probabilities In The third Row By Using The Method Of Integrating Factor**

Using the same procedure as in the last 5 probabilities in the first row, but what differ are the coefficients for each PDFs':

$$for\ P_{35}(t):\ \lambda_{45}\ A_{34} = L_1 \;,\ \lambda_{45}\, B_{34} = L_2 \;,\ \lambda_{45}\, C_{34} = L_3 \;,\ \lambda_{45}\, D_{34} = L_4 \;,$$

$$and \quad \frac{L_1}{w+r_1} = M_1 \;, \quad \frac{L_2}{w+r_2} = M_2 \;, \quad \frac{L_3}{w+r_3} = M_3 \;, \frac{L_4}{w+r_4} = M_4, \quad -\left[\frac{L_1}{w+r_1} + \frac{L_2}{w+r_2} + \frac{L_3}{w+r_3} + \frac{L_4}{w+r_4}\right] = M_5$$

$P_{35}(t) = M_1 e^{r_1 t} + M_2 e^{r_2 t} + M_3 e^{r_3 t} + M_4 e^{r_4 t} + M_5 e^{-wt}$

$for\ P_{36}(t){:}\ \lambda_{56}\ M_1 = L_5\ ,\ \lambda_{56} M_2 = L_6\ ,\ \lambda_{56} M_3 = L_7\ ,\ \lambda_{56} M_4 = L_8\ ,\lambda_{56} M_5 = L_9\ ,$

$$and\ \frac{L_5}{u+r_1} = M_6\ ,\ \frac{L_6}{u+r_2} = M_7\ ,\frac{L_7}{u+r_3} = M_8\ ,\frac{L_8}{u+r_4} = M_9, \frac{L_9}{u-w} = M_{10}\ ,-\left[\frac{L_5}{u+r_1} + \frac{L_6}{u+r_2} + \frac{L_7}{u+r_3} + \frac{L_8}{u+r_4} + \frac{L_9}{u-w}\right] = M_{11}$$

$P_{36}(t) = M_6 e^{r_1 t} + M_7 e^{r_2 t} + M_8 e^{r_3 t} + M_9 e^{r_4 t} + M_{10} e^{-wt} + M_{11} e^{-ut}$

$for\ P_{37}(t){:}\ \lambda_{67} M_6 = L_{10}\ ,\quad \lambda_{67} M_7 = L_{11},\quad \lambda_{67} M_8 = L_{12}\ ,\quad \lambda_{67} M_9 = L_{13},\quad \lambda_{67} M_{10} = L_{14}\ ,\quad \lambda_{67} M_{11} = L_{15}$

$$and\quad \frac{L_{10}}{\lambda_{79}+r_1} = M_{12}\ ,\quad \frac{L_{11}}{\lambda_{79}+r_2} = M_{13}\ ,\quad \frac{L_{12}}{\lambda_{79}+r_3} = M_{14}\ ,\quad \frac{L_{13}}{\lambda_{79}+r_4} = M_{15}\ ,\quad \frac{L_{14}}{\lambda_{79}-w} = M_{16}\ ,\quad \frac{L_{15}}{\lambda_{79}-u} = M_{17},$$

$$-\left[\frac{L_{10}}{\lambda_{79}+r_1} + \frac{L_{11}}{\lambda_{79}+r_2} + \frac{L_{12}}{\lambda_{79}+r_3} + \frac{L_{13}}{\lambda_{79}+r_4} + \frac{L_{14}}{\lambda_{79}-w} + \frac{L_{15}}{\lambda_{79}-u}\right] = M_{18}$$

$P_{37}(t) = M_{12} e^{r_1 t} + M_{13} e^{r_2 t} + M_{14} e^{r_3 t} + M_{15} e^{r_4 t} + M_{16} e^{-wt} + M_{17} e^{-ut} + M_{18} e^{-\lambda_{79} t}$

$for\ P_{38}(t){:}\ [\lambda_{18} A_{31} + \lambda_{28} A_{32} + \lambda_{38} A_{33} + \lambda_{48} A_{34} + \lambda_{58} M_1] = L_{16}\quad ,\quad [\lambda_{18} B_{31} + \lambda_{28} B_{32} + \lambda_{38} B_{33} + \lambda_{48} B_{34} + \lambda_{58} M_2] = L_{17}$

$[\lambda_{18} C_{31} + \lambda_{28} C_{32} + \lambda_{38} C_{33} + \lambda_{48} C_{34} + \lambda_{58} M_3] = L_{18}\ ,\ [\lambda_{18} D_{31} + \lambda_{28} D_{32} + \lambda_{38} D_{33} + \lambda_{48} D_{34} + \lambda_{58} M_4] = L_{19}\ ,\ \lambda_{58} M_5 = L_{20}$

$$and\ \frac{L_{16}}{\lambda_{89}+r_1} = M_{19}\ ,\frac{L_{17}}{\lambda_{89}+r_2} = M_{20}\ ,\frac{L_{18}}{\lambda_{89}+r_3} = M_{21}\ ,\frac{L_{19}}{\lambda_{89}+r_4} = M_{22}\ ,\quad \frac{L_{20}}{\lambda_{89}-w} = M_{23},$$

$$-\left[\frac{L_{16}}{\lambda_{89}+r_1} + \frac{L_{17}}{\lambda_{89}+r_2} + \frac{L_{18}}{\lambda_{89}+r_3} + \frac{L_{19}}{\lambda_{89}+r_4} + \frac{L_{20}}{\lambda_{89}-w}\right] = M_{24}$$

$P_{38}(t) = M_{19} e^{r_1 t} + M_{20} e^{r_2 t} + M_{21} e^{r_3 t} + M_{22} e^{r_4 t} + M_{23} e^{-wt} + M_{24} e^{-\lambda_{89} t}$

$for\ P_{39}(t){:}\ (\lambda_{19} A_{31} + \lambda_{29} A_{32} + \lambda_{39} A_{33} + \lambda_{49} A_{34} + \lambda_{59} M_1 + \lambda_{69} M_6 + \lambda_{79} M_{12} + \lambda_{89} M_{19}) = L_{21}$

$(\lambda_{19} B_{31} + \lambda_{29} B_{32} + \lambda_{39} B_{33} + \lambda_{49} B_{34} + \lambda_{59} M_2 + \lambda_{69} M_7 + \lambda_{79} M_{13} + \lambda_{89} M_{20}) = L_{22}$

$(\lambda_{19} C_{31} + \lambda_{29} C_{32} + \lambda_{39} C_{33} + \lambda_{49} C_{34} + \lambda_{59} M_3 + \lambda_{69} M_8 + \lambda_{79} M_{14} + \lambda_{89} M_{21}) = L_{23}$

$(\lambda_{19} D_{31} + \lambda_{29} D_{32} + \lambda_{39} D_{33} + \lambda_{49} D_{34} + \lambda_{59} M_4 + \lambda_{69} M_9 + \lambda_{79} M_{15} + \lambda_{89} M_{22}) = L_{24}$

$(\lambda_{59} M_5 + \lambda_{69} M_{10} + \lambda_{79} M_{16} + \lambda_{89} M_{23}) = L_{25}\ ,\quad (\lambda_{69} M_{11} + \lambda_{79} M_{17}) = L_{26}\quad ,\quad \lambda_{79} M_{18} = L_{27},\quad \lambda_{89} M_{24} = L_{28}$

$$and\ \frac{L_{21}}{r_1} = M_{25}\ ,\quad \frac{L_{22}}{r_2} = M_{26}\ ,\frac{L_{23}}{r_3} = M_{27}\ ,\frac{L_{24}}{r_4} = M_{28}\ ,\ -\frac{L_{25}}{w} = M_{29}\ ,-\frac{L_{26}}{u} = M_{30}\ ,\ -\frac{L_{27}}{\lambda_{79}} = M_{31}\ ,-\frac{L_{28}}{\lambda_{89}} = M_{32}$$

$$,+\left[-\frac{L_{21}}{r_1} - \frac{L_{22}}{r_2} - \frac{L_{23}}{r_3} - \frac{L_{24}}{r_4} + \frac{L_{25}}{w} + \frac{L_{26}}{u} + \frac{L_{27}}{\lambda_{79}} + \frac{L_{28}}{\lambda_{89}}\right] = M_{33}$$

$P_{39}(t) = M_{25} e^{r_1 t} + M_{26} e^{r_2 t} + M_{27} e^{r_3 t} + M_{28} e^{r_4 t} + M_{29} e^{-wt} + M_{30} e^{-ut} + M_{31} e^{-\lambda_{79} t} + M_{32} e^{-\lambda_{89} t} + M_{33}$

**Solving The Last 5 Probabilities In The fourth Row By Using The Method Of Integrating Factor**

Using the same procedure as in the last 5 probabilities in the first row, but what differ are the coefficients for each PDFs':

$for\ P_{45}(t)\ \ \lambda_{45}\ A_{44} = Y_1\ ,\ \lambda_{45} B_{44} = Y_2\ ,\ \lambda_{45} C_{44} = Y_3\ ,\ \lambda_{45} D_{44} = Y_4\ ,$

$$and\ \frac{Y_1}{w+r_1} = R_1\ ,\quad \frac{Y_2}{w+r_2} = R_2\ ,\quad \frac{Y_3}{w+r_3} = R_3\ ,\frac{Y_4}{w+r_4} = R_4,\quad -\left[\frac{Y_1}{w+r_1} + \frac{Y_2}{w+r_2} + \frac{Y_3}{w+r_3} + \frac{Y_4}{w+r_4}\right] = R_5$$

$P_{45}(t) = R_1 e^{r_1 t} + R_2 e^{r_2 t} + R_3 e^{r_3 t} + R_4 e^{r_4 t} + R_5 e^{-wt}$

$for\ P_{46}(t){:}\ \lambda_{56}\ R_1 = Y_5\ ,\ \lambda_{56} R_2 = Y_6\ ,\ \lambda_{56} R_3 = Y_7\ ,\ \lambda_{56} R_4 = Y_8\ ,\quad \lambda_{56} R_5 = Y_9\ , integrate\ both\ sides\ from\ (0\ to\ t)$

$$and \quad \frac{Y_5}{u+r_1} = R_6 \ , \ \frac{Y_6}{u+r_2} = R_7 \ , \frac{Y_7}{u+r_2} = R_8 \ , \frac{Y_8}{u+r_2} = R_9 , \frac{Y_9}{u-w} = R_{10} \ , -\left[\frac{Y_5}{u+r_1} + \frac{Y_6}{u+r_1} + \frac{Y_7}{u+r_1} + \frac{Y_8}{u+r_1} + \frac{Y_9}{u-w}\right] = R_{11}$$

$$P_{46}(t) = R_6\, e^{r_1 t} + R_7 e^{r_2 t} + R_8\, e^{r_3 t} + R_9 e^{r_4 t} + R_{10} e^{-wt} + R_{11} e^{-ut}$$

$$for \ P_{47}(t): \ \lambda_{67} R_6 = Y_{10} \ , \quad \lambda_{67} R_7 = Y_{11}, \quad \lambda_{67} R_8 = Y_{12} \ , \quad \lambda_{67} R_9 = Y_{13}, \quad \lambda_{67} R_{10} = Y_{14} \ , \quad \lambda_{67} R_{11} = Y_{15}$$

$$and \quad \frac{Y_{10}}{\lambda_{79}+r_1} = R_{12} \ , \quad \frac{Y_{11}}{\lambda_{79}+r_2} = R_{13} \ , \quad \frac{Y_{12}}{\lambda_{79}+r_3} = R_{14} \ , \quad \frac{Y_{13}}{\lambda_{79}+r_4} = R_{15} \ , \quad \frac{Y_{14}}{\lambda_{79}-w} = R_{16} \ , \quad \frac{Y_{15}}{\lambda_{79}-u} = R_{17},$$

$$-\left[\frac{Y_{10}}{\lambda_{79}+r_1} + \frac{Y_{11}}{\lambda_{79}+r_2} + \frac{Y_{12}}{\lambda_{79}+r_3} + \frac{Y_{13}}{\lambda_{79}+r_4} + \frac{Y_{14}}{\lambda_{79}-w} + \frac{Y_{15}}{\lambda_{79}-u}\right] = R_{18}$$

$$P_{47}(t) = R_{12}\, e^{r_1 t} + R_{13}\, e^{r_2 t} + R_{14} e^{r_3 t} + R_{15}\, e^{r_4 t} + R_{16}\, e^{-wt} + R_{17}\, e^{-ut} + R_{18}\, e^{-\lambda_{79} t}$$

$$for \ \ P_{48}(t): \ \ [\lambda_{18} A_{41} + \lambda_{28} A_{42} + \lambda_{38} A_{43} + \lambda_{48} A_{44} + \lambda_{58} R_1] = Y_{16} \ , [\lambda_{18} B_{41} + \lambda_{28} B_{42} + \lambda_{38} B_{43} + \lambda_{48} B_{44} + \lambda_{58} R_2] = Y_{17}$$

$$[\lambda_{18} C_{41} + \lambda_{28} C_{42} + \lambda_{38} C_{43} + \lambda_{48} C_{44} + \lambda_{58} R_3] = Y_{18} \ , [\lambda_{18} D_{41} + \lambda_{28} D_{42} + \lambda_{38} D_{43} + \lambda_{48} D_{44} + \lambda_{58} R_4] = Y_{19} \ , \quad \lambda_{58} R_5 = Y_{20}$$

$$and \quad \frac{Y_{16}}{\lambda_{89}+r_1} = R_{19} \ , \frac{Y_{17}}{\lambda_{89}+r_2} = R_{20} \ , \frac{Y_{18}}{\lambda_{89}+r_3} = R_{21} \ , \frac{Y_{19}}{\lambda_{89}+r_4} = R_{22} \ , \quad \frac{Y_{20}}{\lambda_{89}-w} = R_{23}$$

$$-\left[\frac{Y_{16}}{\lambda_{89}+r_1} + \frac{Y_{17}}{\lambda_{89}+r_2} + \frac{Y_{18}}{\lambda_{89}+r_3} + \frac{Y_{19}}{\lambda_{89}+r_4} + \frac{Y_{20}}{\lambda_{89}-w}\right] = R_{24}$$

$$P_{48}(t) = R_{19} e^{r_1 t} + R_{20} e^{r_2 t} + R_{21} e^{r_3 t} + R_{22} e^{r_4 t} + R_{23} e^{-wt} + R_{24} e^{-\lambda_{89} t}$$

$$for\, P_{49}(t): \ \ (\lambda_{19} A_{41} + \lambda_{29} A_{42} + \lambda_{39} A_{43} + \lambda_{49} A_{44} + \lambda_{59} R_1 + \lambda_{69} R_6 + \lambda_{79} R_{12} + \lambda_{89} R_{19}) = Y_{21}$$

$$(\lambda_{19} B_{41} + \lambda_{29} B_{42} + \lambda_{39} B_{43} + \lambda_{49} B_{44} + \lambda_{59} R_2 + \lambda_{69} R_7 + \lambda_{79} R_{13} + \lambda_{89} R_{20}) = Y_{22}$$

$$(\lambda_{19} C_{41} + \lambda_{29} C_{42} + \lambda_{39} C_{43} + \lambda_{49} C_{44} + \lambda_{59} R_3 + \lambda_{69} R_8 + \lambda_{79} R_{14} + \lambda_{89} R_{21}) = Y_{23}$$

$$(\lambda_{19} D_{41} + \lambda_{29} D_{42} + \lambda_{39} D_{43} + \lambda_{49} D_{44} + \lambda_{59} R_4 + \lambda_{69} R_9 + \lambda_{79} R_{15} + \lambda_{89} R_{22}) = Y_{24} \ , (\lambda_{59} R_5 + \lambda_{69} R_{10} + \lambda_{79} R_{16} + \lambda_{89} R_{23}) = Y_{25}$$

$$(\lambda_{69} R_{11} + \lambda_{79} R_{17}) = Y_{26} \quad , \ \lambda_{79} R_{18} = Y_{27} \quad , \ \lambda_{89} R_{24} = Y_{28}$$

$$and \ \ \frac{Y_{21}}{r_1} = R_{25} \quad , \quad \frac{Y_{22}}{r_2} = R_{26} \ , \frac{Y_{23}}{r_3} = R_{27} \ \ , \frac{Y_{24}}{r_4} = R_{28} \ \ , \ \ -\frac{Y_{25}}{w} = R_{29} \ \ , -\frac{Y_{26}}{u} = R_{30} \ , \ \ -\frac{Y_{27}}{\lambda_{79}} = R_{31} \ , -\frac{Y_{28}}{\lambda_{89}} = R_{32}$$

$$, + \left[-\frac{Y_{21}}{r_1} - \frac{Y_{22}}{r_2} - \frac{Y_{23}}{r_3} - \frac{Y_{24}}{r_4} + \frac{Y_{25}}{w} + \frac{Y_{26}}{u} + \frac{Y_{27}}{\lambda_{79}} + \frac{Y_{28}}{\lambda_{89}}\right] = R_{33}$$

$$P_{49}(t) = R_{25} e^{r_1 t} + R_{26} e^{r_2 t} + R_{27} e^{r_3 t} + R_{28} e^{r_4 t} + R_{29} e^{-wt} + R_{30} e^{-ut} + R_{31} e^{-\lambda_{79} t} + R_{32} e^{-\lambda_{89} t} + R_{33}$$

Solving the last 12 equations in the system by integrating factor:

$$\frac{d\, P_{55}(t)}{dt} = -(\lambda_{56} + \lambda_{58} + \lambda_{59}) P_{55}(t) \ \ , \quad let \ \ (\lambda_{56} + \lambda_{58} + \lambda_{59}) = w \quad \to so \quad P_{55}(t) = e^{-wt} \ \ ,$$

$$\frac{d\, P_{56}(t)}{dt} = \lambda_{56}\, P_{55}(t) - (\lambda_{67} + \lambda_{69}) P_{56}(t) \ \ , \qquad let \ ( \ \lambda_{67} + \lambda_{69}) = u$$

$$\therefore \ P_{56}(t) = = \frac{\lambda_{56}}{u-w}[e^{-wt} - e^{-ut}] \ \ , \ and \quad let \ \ \frac{\lambda_{56}}{u-w} = \ v_1 \quad so \ \ P_{56}(t) = v_1 \ \ (e^{-wt} - e^{-ut})$$

$$\frac{d\, P_{57}(t)}{dt} = \lambda_{67}\, P_{56}(t) - \lambda_{79}\, P_{57}(t) \quad \to \quad \frac{d}{dt}\, e^{\lambda_{79} t} \ \ P_{57}(t) = \quad \lambda_{67}\, v_1 \ \ e^{(\lambda_{79}-w)t} - \ \lambda_{67}\, v_1 e^{(\lambda_{79}-u)t}$$

$$e^{\lambda_{79} t} \ \ P_{57}(t) = \quad \lambda_{67}\, v_1 \left[\frac{e^{(\lambda_{79}-w)t}}{\lambda_{79}-w} - \frac{1}{\lambda_{79}-w}\right] - \ \lambda_{67}\, v_1 \left[\frac{e^{(\lambda_{79}-u)t}}{\lambda_{79}-u} - \frac{1}{\lambda_{79}-u}\right]$$

$$P_{57}(t) = \frac{\lambda_{67}\, v_1}{\lambda_{79} - w}\left[e^{-wt} - e^{-\lambda_{79} t}\right] - \frac{\lambda_{67}\, v_1}{\lambda_{79} - u}\left[e^{-ut} - e^{-\lambda_{79} t}\right] = \frac{\lambda_{67}\, v_1}{\lambda_{79} - w}\, e^{-wt} - \frac{\lambda_{67}\, v_1}{\lambda_{79} - u} e^{-ut} + \left(\frac{\lambda_{67}\, v_1}{\lambda_{79} - u} - \frac{\lambda_{67}\, v_1}{\lambda_{79} - w}\right) e^{-\lambda_{79} t}$$

$$let \quad \frac{\lambda_{67}\, v_1}{\lambda_{79} - w} = v_2 \;, \quad -\frac{\lambda_{67}\, v_1}{\lambda_{79} - u} = v_3 \;, \quad \left(\frac{\lambda_{67}\, v_1}{\lambda_{79} - u} - \frac{\lambda_{67}\, v_1}{\lambda_{79} - w}\right) = v_4 \;, so \quad P_{57}(t) = v_2\, e^{-wt} + v_3\, e^{-ut} + v_4\, e^{-\lambda_{79} t}$$

$$\frac{d\, P_{58}(t)}{dt} = \lambda_{58}\, P_{55}(t) - \lambda_{89}\, P_{58}(t) \,, \quad so \quad e^{\lambda_{89} t}\, \frac{d\, P_{58}(t)}{dt} + \lambda_{89}\, e^{\lambda_{89} t}\, P_{58}(t) = e^{\lambda_{89} t}\, \lambda_{58}\, e^{-wt}$$

$$\frac{d}{dt}\, e^{\lambda_{89} t}\, P_{58}(t) = \lambda_{58}\, e^{(\lambda_{89} - w)t} \,, \quad then \quad e^{\lambda_{89} t}\, P_{58}(t) = \lambda_{58}\left[\frac{e^{(\lambda_{89} - w)t}}{\lambda_{89} - w} - \frac{1}{\lambda_{89} - w}\right]$$

$$P_{58}(t) = \frac{\lambda_{58}}{\lambda_{89} - w}\left[e^{-wt} - e^{-\lambda_{89} t}\right] \,, \; let: \quad \frac{\lambda_{58}}{\lambda_{89} - w} = v_5 \;, \quad so \; P_{58}(t) = v_5\left(e^{-wt} - e^{-\lambda_{89} t}\right)$$

$$\frac{d\, P_{59}(t)}{dt} = \lambda_{59}\, P_{55}(t) + \lambda_{69}\, P_{56}(t) + \lambda_{79}\, P_{57}(t) + \lambda_{89}\, P_{58}(t)$$

$$\frac{d\, P_{59}(t)}{dt} = \lambda_{59} e^{-wt} + \lambda_{69}\, v_1\, \left[e^{-wt} - e^{-ut}\right] + \lambda_{79}\left[v_2\, e^{-wt} + v_3\, e^{-ut} + v_4\, e^{-\lambda_{79} t}\right] + \lambda_{89}\, v_5\left[e^{-wt} - e^{-\lambda_{89} t}\right]$$

$$let:\; (\lambda_{59} + \lambda_{69}\, v_1 + \lambda_{79} v_2 + \lambda_{89}\, v_5) = v_6 \;, \; (\lambda_{79}\, v_3 - \lambda_{69}\, v_1) = v_7 \,, \quad \lambda_{79}\, v_4 = v_8 \,, \quad -\lambda_{89}\, v_5 = v_9$$

$$\frac{d\, P_{59}(t)}{dt} = v_6 e^{-wt} + v_7\, e^{-ut} + v_8\, e^{-\lambda_{79} t} + v_9\, e^{-\lambda_{89} t}$$

$$P_{59}(t) = v_6\left[\frac{e^{-wt}}{-w} - \frac{1}{-w}\right] + v_7\left[\frac{e^{-ut}}{-u} - \frac{1}{-u}\right] + v_8\left[\frac{e^{-\lambda_{79} t}}{-\lambda_{79}} - \frac{1}{-\lambda_{79}}\right] + v_9\left[\frac{e^{-\lambda_{89} t}}{-\lambda_{89}} - \frac{1}{-\lambda_{89}}\right]$$

$$P_{59}(t) = \frac{-v_6}{w}\, e^{-wt} - \frac{v_7}{u}\, e^{-ut} - \frac{v_8}{\lambda_{79}}\, e^{-\lambda_{79} t} - \frac{v_9}{\lambda_{89}}\, e^{-\lambda_{89} t} + \left(\frac{v_6}{w} + \frac{v_7}{u} + \frac{v_8}{\lambda_{79}} + \frac{v_9}{\lambda_{58}}\right)$$

$$let: \; \frac{-v_6}{w} = v_{10} \,, \; \frac{-v_7}{u} = v_{11} \,, \; \frac{-v_8}{\lambda_{79}} = v_{12} \,, \; \frac{-v_9}{\lambda_{89}} = v_{13} \;, \; \left(\frac{v_6}{w} + \frac{v_7}{u} + \frac{v_8}{\lambda_{79}} + \frac{v_9}{\lambda_{58}}\right) = v_{14}$$

$$P_{59}(t) = v_{10}\, e^{-wt} + v_{11}\, e^{-ut} + v_{12}\, e^{-\lambda_{79} t} + v_{13}\, e^{-\lambda_{89} t} + v_{14}$$

$$\frac{d\, P_{66}(t)}{dt} = -(\lambda_{67} + \lambda_{69}) P_{66}(t) \;, let \quad (\lambda_{67} + \lambda_{69}) = u \;, \quad so \; \frac{d\, P_{66}(t)}{dt} = -u\, P_{66}(t) \quad and \; P_{66}(t) = e^{-ut}$$

$$\frac{d\, P_{67}(t)}{dt} = \lambda_{67}\, P_{66}(t) - \lambda_{79} P_{67}(t) \,, \quad so \quad e^{\lambda_{79} t}\, \frac{d\, P_{67}(t)}{dt} + \lambda_{79}\, e^{\lambda_{79} t}\, P_{67}(t) = e^{\lambda_{79} t}\, \lambda_{67}\, e^{-ut} = \lambda_{67}\, e^{(\lambda_{79} - u)t}$$

$$\frac{d}{dt}\, e^{\lambda_{79} t}\, P_{67}(t) = \lambda_{67}\, e^{(\lambda_{79} - u)t}$$

$$e^{\lambda_{79} t}\, P_{67}(t) = \lambda_{67}\left[\frac{e^{(\lambda_{79} - u)t}}{\lambda_{79} - u} - \frac{1}{\lambda_{79} - u}\right] = \frac{\lambda_{67}}{\lambda_{79} - u}\left[e^{(\lambda_{79} - u)t} - 1\right] \quad , so \; P_{67}(t) = \frac{\lambda_{67}}{\lambda_{79} - u}\left[e^{-ut} - e^{-\lambda_{79} t}\right]$$

$$Let \quad \frac{\lambda_{67}}{\lambda_{79} - u} = z_1 \quad , \quad \text{so} \quad P_{67}(t) = z_1\left(e^{-ut} - e^{-\lambda_{79} t}\right)$$

$$\frac{d\, P_{69}(t)}{dt} = \lambda_{69}\, P_{66}(t) + \lambda_{79} P_{67}(t)$$

$$\frac{d\, P_{69}(t)}{dt} = \lambda_{69}\, e^{-ut} + \lambda_{79} z_1\left(e^{-ut} - e^{-\lambda_{79} t}\right) = \lambda_{69}\, e^{-ut} + \lambda_{79} z_1 e^{-ut} - \lambda_{79} z_1 e^{-\lambda_{79} t} = (\lambda_{69} + \lambda_{79} z_1)\, e^{-ut} - \lambda_{79} z_1 e^{-\lambda_{79} t}$$

$$P_{69}(t) = (\lambda_{69} + \lambda_{79} z_1)\left[\frac{e^{-ut}}{-u} - \frac{1}{-u}\right] - \lambda_{79} z_1\left[\frac{e^{-\lambda_{79} t}}{-\lambda_{79}} - \frac{1}{-\lambda_{79}}\right] = \frac{(\lambda_{69} + \lambda_{79} z_1)}{-u}\left[e^{-ut} - 1\right] - \frac{\lambda_{79} z_1}{-\lambda_{79}}\left[e^{-\lambda_{79} t} - 1\right]$$

$$let: \; \frac{(\lambda_{69} + \lambda_{79} z_1)}{-u} = z_2 \quad , -\frac{\lambda_{79} z_1}{-\lambda_{79}} = z_3 \;, \; so \quad P_{69}(t) = z_2\left(e^{-ut} - 1\right) + z_3\left(e^{-\lambda_{79} t} - 1\right) = z_2 e^{-ut} + z_3 e^{-\lambda_{79} t} - z_2 - z_3$$

$$\frac{d\,P_{77}(t)}{dt} = -\lambda_{79}P_{77}(t) \quad , \; so \;\; P_{77}(t) = e^{-\lambda_{79}t}$$

$$\frac{d\,P_{79}(t)}{dt} = \lambda_{79}P_{77}(t) = \lambda_{79}\, e^{-\lambda_{79}t} \;\; so \;\; P_{79}(t) \; = \lambda_{79}\left[\frac{e^{-\lambda_{79}t}}{-\lambda_{79}} - \frac{1}{-\lambda_{79}}\right] = \frac{\lambda_{79}}{-\lambda_{79}}\left[e^{-\lambda_{79}t} \; - 1\right] = 1 - e^{-\lambda_{79}t}$$

$$\frac{d\,P_{88}(t)}{dt} = -\,\lambda_{89}P_{88}(t) \;\; , \; so \quad P_{88}(t) \; = e^{-\,\lambda_{89}t}$$

$$\frac{d\,P_{89}(t)}{dt} = \;\lambda_{89}P_{88}(t) \quad , \rightarrow \;\; P_{89}(t) \; = \; \lambda_{89}\left[\frac{e^{-\,\lambda_{89}t}}{-\,\lambda_{89}} - \frac{1}{-\,\lambda_{89}}\right] = \frac{\lambda_{89}}{-\,\lambda_{89}}\left[\,e^{-\,\lambda_{89}t} \; - 1\right] = 1 - \; e^{-\,\lambda_{89}t} \;\; and \;\; finaly \;\; P_{99}(t) = 1$$

## 5.2. Estimation Of The Q Rate transiiton Matrix:

The Q matrix

$$Q = \begin{bmatrix} -\gamma_1 & \lambda_{12} & 0 & 0 & 0 & 0 & 0 & \lambda_{18} & \lambda_{19} \\ \mu_{21} & -\gamma_2 & \lambda_{23} & 0 & 0 & 0 & 0 & \lambda_{28} & \lambda_{29} \\ 0 & \mu_{32} & -\gamma_3 & \lambda_{34} & 0 & 0 & 0 & \lambda_{38} & \lambda_{39} \\ 0 & 0 & \mu_{43} & -\gamma_4 & \lambda_{45} & 0 & 0 & \lambda_{48} & \lambda_{49} \\ 0 & 0 & 0 & 0 & -\gamma_5 & \lambda_{56} & 0 & \lambda_{58} & \lambda_{59} \\ 0 & 0 & 0 & 0 & 0 & -\gamma_6 & \lambda_{67} & 0 & \lambda_{69} \\ 0 & 0 & 0 & 0 & 0 & 0 & -\lambda_{79} & 0 & \lambda_{79} \\ 0 & 0 & 0 & 0 & 0 & 0 & 0 & -\lambda_{89} & \lambda_{89} \\ 0 & 0 & 0 & 0 & 0 & 0 & 0 & 0 & 0 \end{bmatrix}$$

$let$: $\quad (\lambda_{12} + \lambda_{18} + \lambda_{19}) = \gamma_1 \quad , \qquad -\gamma_1 - \rho = r$

$(\lambda_{23} + \lambda_{28} + \lambda_{29} + \mu_{21}) = \gamma_2 \quad , \qquad -\;\gamma_2 - \rho = x$

$(\lambda_{34} + \lambda_{38} + \lambda_{39} + \mu_{32}) = \gamma_3 \quad , \qquad -\;\gamma_3 - \rho = y$

$(\lambda_{45} + \lambda_{48} + \lambda_{49} + \mu_{43}) = \gamma_4 \quad , \qquad -\;\gamma_4 - \rho = z$

$(\lambda_{56} + \lambda_{58} + \lambda_{59}) = \gamma_5 \qquad , \qquad (\lambda_{67} + \lambda_{69}) = \gamma_6$

The eigen-values of this matrix are obtained by finding the zeros of the characteristic polynomial (polynomial of the 9 th degree) ,this is achieved by solving the following equation $|Q - \rho I| = 0$ ,

$the\ following\ is\ the\ determinant\ of\ this\ (Q - \rho I)$

$$[-\,\mu_{32}\,\lambda_{23}rz + (-\mu_{21}\,\lambda_{12} + rx)(yz - \mu_{43}\lambda_{34}\,)](-\,\gamma_5 - \rho)(-\,\gamma_6 - \rho)(-\,\lambda_{79} - \rho)(-\,\lambda_{89} - \rho)(-\rho) = 0$$

$$-[-\,\mu_{32}\,\lambda_{23}rz + (-\mu_{21}\,\lambda_{12} + rx)(yz - \mu_{43}\lambda_{34}\,)](\,\gamma_5 + \rho)(\,\gamma_6 + \rho)(\,\lambda_{79} + \rho)(\,\lambda_{89} + \rho)(\rho) = 0$$

As obviously seen this 9th degree polynomial has 9 eigen-values and the full model has 22 transition rates.

$\rho_i = \{\,0\,, zeros\ of\ \ both\ \ quadratics\ \ (\,\gamma_5 + \rho)(\,\gamma_6 + \rho)\ \ and\ \ (\,\lambda_{79} + \rho)(\ \lambda_{89} + \rho)$

$and\ zeros\ of\ the\ \ 4th\ degree\ polynomial\ \ [-\,\mu_{32}\,\lambda_{23}rz + (-\mu_{21}\,\lambda_{12} + rx)(yz - \mu_{43}\lambda_{34}\,)]\,\}$

$to\ obtain\ \ the\ \ zeros\ of\ the\ \ 4th\ degree\ polynomial\ \ [-\,\mu_{32}\,\lambda_{23}rz + (-\mu_{21}\,\lambda_{12} + rx)(yz - \mu_{43}\lambda_{34}\,)]$:

$rearrange\ the\ polynomial\ \ and\ substitute\ for\ (\,r, x, y, z)\ \ to\ have\ the\ expanded\ form\ of\ the\ polynomial\ \ which\ is\ the\ following$:

$f(\rho) = \rho^4 + (w_1 + w_3)\rho^3 + (w_2 + w_4 + \ w_1w_3 - \lambda_{23}\,\mu_{32}\,)\,\rho^2 + (w_2w_3 + w_1w_4 - \ w_5)\rho + \ (w_2w_4 - \ w_6)\,,$

$where\ \ all\ \ ws'\ \ are\ \ as\ previously\ \ defined$

$The\ quartic\ polynomial\ in\ the\ form\ of\ \ \ (a\,x^4 + b\,x^3 + c\,x^2 + dx + \ e\ \ )\ has\ four\ roots$ :

$$\rho_1 = -\frac{b}{4a} - S + \frac{1}{2}\sqrt{-4S^2 - 2p + \frac{q}{S}} \quad , \quad \rho_2 = -\frac{b}{4a} - S - \frac{1}{2}\sqrt{-4S^2 - 2p + \frac{q}{S}}$$

$$\rho_3 = -\frac{b}{4a} + S + \frac{1}{2}\sqrt{-4S^2 - 2p - \frac{q}{S}} \quad , \quad \rho_4 = -\frac{b}{4a} + S - \frac{1}{2}\sqrt{-4S^2 - 2p - \frac{q}{S}}$$

$$p = \frac{8ac - 3b^2}{8a^2} \quad , \quad q = \frac{b^3 - 4abc + 8a^2 d}{8a^3} \quad , \quad S = \frac{1}{2}\sqrt{\frac{-2p}{3} + \frac{2}{3a}\sqrt{\Delta_0}\cos\frac{\emptyset}{3}} \quad , \quad \emptyset = arccos\left(\frac{\Delta_1}{(2)\sqrt{\Delta_0^3}}\right) \quad , \quad \frac{d}{dx}\,[cos^{-1}u] = \frac{-1}{\sqrt{1-u^2}}\frac{du}{dx} \quad ,$$

$$\Delta_0 = c^2 - 3bd + 12ae \ , \quad \Delta_1 = 2c^3 - 9bcd + 27b^2 e + 27ad^2 - 72ace$$

The other roots:

The zeros for quadratic $(\gamma_5 + \rho)(\gamma_6 + \rho)$:

$$\rho_5 = \frac{-(\gamma_5 + \gamma_6) + \sqrt{(\gamma_5 + \gamma_6)^2 - 4\gamma_5\gamma_6}}{2} = \frac{-(\gamma_5 + \gamma_6) + \sqrt{.}}{2}$$

$$= \frac{-(\lambda_{56} + \lambda_{58} + \lambda_{59} + \lambda_{67} + \lambda_{69}) + (\lambda_{56}^2 + \lambda_{58}^2 + \lambda_{59}^2 + \lambda_{67}^2 + \lambda_{69}^2 + 2\lambda_{56}\lambda_{58} + 2\lambda_{56}\lambda_{59} + 2\lambda_{58}\lambda_{59} + 2\lambda_{67}\lambda_{69} - 2\lambda_{56}\lambda_{67} - 2\lambda_{56}\lambda_{69} - 2\lambda_{58}\lambda_{67} - 2\lambda_{58}\lambda_{69} - 2\lambda_{59}\lambda_{67} - 2\lambda_{59}\lambda_{69})^{.5}}{2}$$

$$\rho_6 = \frac{-(\gamma_5 + \gamma_6) - \sqrt{(\gamma_5 + \gamma_6)^2 - 4\gamma_5\gamma_6}}{2} = \frac{-(\gamma_5 + \gamma_6) - \sqrt{.}}{2}$$

$$= \frac{-(\lambda_{56} + \lambda_{58} + \lambda_{59} + \lambda_{67} + \lambda_{69}) - (\lambda_{56}^2 + \lambda_{58}^2 + \lambda_{59}^2 + \lambda_{67}^2 + \lambda_{69}^2 + 2\lambda_{56}\lambda_{58} + 2\lambda_{56}\lambda_{59} + 2\lambda_{58}\lambda_{59} + 2\lambda_{67}\lambda_{69} - 2\lambda_{56}\lambda_{67} - 2\lambda_{56}\lambda_{69} - 2\lambda_{58}\lambda_{67} - 2\lambda_{58}\lambda_{69} - 2\lambda_{59}\lambda_{67} - 2\lambda_{59}\lambda_{69})^{.5}}{2}$$

The zeros for quadratic $(\lambda_{79} + \rho)(\lambda_{89} + \rho)$:

$$\rho_7 = \frac{-(\lambda_{79} + \lambda_{89}) + \sqrt{(\lambda_{79} + \lambda_{89})^2 - 4\lambda_{79}\lambda_{89}}}{2} = \frac{-(\lambda_{79} + \lambda_{89}) + \sqrt{..}}{2} = \frac{-(\lambda_{79} + \lambda_{89}) + (\lambda_{79}^2 + \lambda_{89}^2 - 2\lambda_{79}\lambda_{89})^{.5}}{2}$$

$$\rho_8 = \frac{-(\lambda_{79} + \lambda_{89}) - \sqrt{(\lambda_{79} + \lambda_{89})^2 - 4\lambda_{79}\lambda_{89}}}{2} = \frac{-(\lambda_{79} + \lambda_{89}) - \sqrt{..}}{2} = \frac{-(\lambda_{79} + \lambda_{89}) - (\lambda_{79}^2 + \lambda_{89}^2 - 2\lambda_{79}\lambda_{89})^{.5}}{2}$$

Now differentiating each eigenvalue with respect to the rates forming this eigenvalue:

Starting with the last 4 roots or eigenvalues because they are simpler than the first 4 eigenvalues:

$$\frac{\partial}{\partial\lambda_{56}}\ \rho_5 = \frac{-1}{2} + \frac{1}{2}\frac{1}{2}(.)^{-.5}(2\lambda_{56} + 2\lambda_{58} + 2\lambda_{59} - 2\lambda_{67} - 2\lambda_{69}) = \frac{-1}{2} + \frac{1}{2}(.)^{-.5}(\lambda_{56} + \lambda_{58} + \lambda_{59} - \lambda_{67} - \lambda_{69})$$

$$\frac{\partial}{\partial\lambda_{58}}\ \rho_5 = \frac{-1}{2} + \frac{1}{2}\frac{1}{2}(.)^{-.5}(2\lambda_{58} + 2\lambda_{56} + 2\lambda_{59} - 2\lambda_{67} - 2\lambda_{69}) = \frac{-1}{2} + \frac{1}{2}(.)^{-.5}(\lambda_{58} + \lambda_{56} + \lambda_{59} - \lambda_{67} - \lambda_{69})$$

$$\frac{\partial}{\partial\lambda_{59}}\ \rho_5 = \frac{-1}{2} + \frac{1}{2}\frac{1}{2}(.)^{-.5}(2\lambda_{59} + 2\lambda_{56} + 2\lambda_{58} - 2\lambda_{67} - 2\lambda_{69}) = \frac{-1}{2} + \frac{1}{2}(.)^{-.5}(\lambda_{59} + \lambda_{56} + \lambda_{58} - \lambda_{67} - \lambda_{69})$$

$$\frac{\partial}{\partial\lambda_{67}}\ \rho_5 = \frac{-1}{2} + \frac{1}{2}\frac{1}{2}(.)^{-.5}(2\lambda_{67} + 2\lambda_{69} - 2\lambda_{56} - 2\lambda_{58} - 2\lambda_{59}) = \frac{-1}{2} + \frac{1}{2}(.)^{-.5}(\lambda_{67} + \lambda_{69} - \lambda_{56} - \lambda_{58} - \lambda_{59})$$

$$\frac{\partial}{\partial\lambda_{69}}\ \rho_5 = \frac{-1}{2} + \frac{1}{2}\frac{1}{2}(.)^{-.5}(2\lambda_{69} + 2\lambda_{67} - 2\lambda_{56} - 2\lambda_{58} - 2\lambda_{59}) = \frac{-1}{2} + \frac{1}{2}(.)^{-.5}(\lambda_{69} + \lambda_{67} - \lambda_{56} - \lambda_{58} - \lambda_{59})$$

$$\frac{\partial}{\partial\lambda_{56}}\ \rho_6 = \frac{-1}{2} - \frac{1}{2}\frac{1}{2}(.)^{-.5}(2\lambda_{56} + 2\lambda_{58} + 2\lambda_{59} - 2\lambda_{67} - 2\lambda_{69}) = \frac{-1}{2} - \frac{1}{2}(.)^{-.5}(\lambda_{56} + \lambda_{58} + \lambda_{59} - \lambda_{67} - \lambda_{69})$$

$$\frac{\partial}{\partial\lambda_{58}}\ \rho_6 = \frac{-1}{2} - \frac{1}{2}\frac{1}{2}(.)^{-.5}(2\lambda_{58} + 2\lambda_{56} + 2\lambda_{59} - 2\lambda_{67} - 2\lambda_{69}) = \frac{-1}{2} - \frac{1}{2}(.)^{-.5}(\lambda_{58} + \lambda_{56} + \lambda_{59} - \lambda_{67} - \lambda_{69})$$

$$\frac{\partial}{\partial\lambda_{59}}\ \rho_6 = \frac{-1}{2} - \frac{1}{2}\frac{1}{2}(.)^{-.5}(2\lambda_{59} + 2\lambda_{56} + 2\lambda_{58} - 2\lambda_{67} - 2\lambda_{69}) = \frac{-1}{2} - \frac{1}{2}(.)^{-.5}(\lambda_{59} + \lambda_{56} + \lambda_{58} - \lambda_{67} - \lambda_{69})$$

$$\frac{\partial}{\partial\lambda_{67}}\ \rho_6 = \frac{-1}{2} - \frac{1}{2}\frac{1}{2}(.)^{-.5}(2\lambda_{67} + 2\lambda_{69} - 2\lambda_{56} - 2\lambda_{58} - 2\lambda_{59}) = \frac{-1}{2} - \frac{1}{2}(.)^{-.5}(\lambda_{67} + \lambda_{69} - \lambda_{56} - \lambda_{58} - \lambda_{59})$$

$$\frac{\partial}{\partial\lambda_{69}}\ \rho_6 = \frac{-1}{2} - \frac{1}{2}\frac{1}{2}(.)^{-.5}(2\lambda_{69} + 2\lambda_{67} - 2\lambda_{56} - 2\lambda_{58} - 2\lambda_{59}) = \frac{-1}{2} - \frac{1}{2}(.)^{-.5}(\lambda_{69} + \lambda_{67} - \lambda_{56} - \lambda_{58} - \lambda_{59})$$

$$\frac{\partial}{\partial\lambda_{79}}\ \rho_7 = \frac{-1}{2} + \frac{1}{2}\frac{1}{2}(.)^{-.5}(2\lambda_{79} - 2\lambda_{89}) = \frac{-1}{2} + \frac{1}{2}(.)^{-.5}(\lambda_{79} - \lambda_{89})$$

$$\frac{\partial}{\partial\lambda_{89}}\ \rho_7 = \frac{-1}{2} + \frac{1}{2}\frac{1}{2}(.)^{-.5}(2\lambda_{89} - 2\lambda_{79}) = \frac{-1}{2} + \frac{1}{2}(.)^{-.5}(\lambda_{89} - \lambda_{79})$$

$$\frac{\partial}{\partial\lambda_{79}}\ \rho_8 = \frac{-1}{2} - \frac{1}{2}\frac{1}{2}(.)^{-.5}(2\lambda_{79} - 2\lambda_{89}) = \frac{-1}{2} - \frac{1}{2}(.)^{-.5}(\lambda_{79} - \lambda_{89})$$

$$\frac{\partial}{\partial\lambda_{89}}\ \rho_8 = \frac{-1}{2} - \frac{1}{2}\frac{1}{2}(.)^{-.5}(2\lambda_{89} - 2\lambda_{79}) = \frac{-1}{2} - \frac{1}{2}(.)^{-.5}(\lambda_{89} - \lambda_{79})$$

The first 4 roots or eigenvalues will be discussed as follows :

$$\rho_1' = -\frac{b'}{4} - S' + \frac{1}{2}\ \frac{1}{2}\left(-4S^2 - 2p + \frac{q}{S}\right)^{-.5}\left(-8SS' - 2p' + \frac{q'S - S'q}{S^2}\right)$$

$$\rho_2' = -\frac{b'}{4} - S' - \frac{1}{2}\ \frac{1}{2}\left(-4S^2 - 2p + \frac{q}{S}\right)^{-.5}\left(-8SS' - 2p' + \frac{q'S - S'q}{S^2}\right)$$

$$\rho_3' = -\frac{b'}{4} + S' + \frac{1}{2}\ \frac{1}{2}\left(-4S^2 - 2p - \frac{q}{S}\right)^{-.5}\left(-8SS' - 2p' - \frac{q'S - S'q}{S^2}\right),$$

$$\rho_4' = -\frac{b'}{4} + S' - \frac{1}{2}\ \frac{1}{2}\left(-4S^2 - 2p - \frac{q}{S}\right)^{-.5}\left(-8SS' - 2p' - \frac{q'S - S'q}{S^2}\right)$$

$\because a = 1$

$$p = \frac{8ac - 3b^2}{8a^2} = \frac{8c - 3b^2}{8} \qquad \rightarrow \qquad p' = \frac{1}{8}\ (8c' - 6bb')$$

$$q = \frac{b^3 - 4abc + 8a^2 d}{8a^3} = \frac{b^3 - 4bc + 8d}{8} \qquad \rightarrow \qquad q' = \frac{1}{8}\ (3\ b^2 b' - 4b'c - 4bc' + 8d')$$

$$S = \frac{1}{2}\sqrt{\frac{-2p}{3} + \frac{2}{3a}\sqrt{\Delta_0}\ cos\frac{\emptyset}{3}} = \frac{1}{2}\sqrt{\frac{-2p}{3} + \frac{2}{3}\sqrt{\Delta_0}\ cos\frac{\emptyset}{3}} \ \rightarrow$$

$$S' = \frac{1}{2}\ \frac{1}{2}\left(\frac{-2p}{3} + \frac{2}{3}\sqrt{\Delta_0}\ cos\frac{\emptyset}{3}\right)^{-.5}\left(\frac{-2p'}{3} + \frac{2}{3}\left(\sqrt{\Delta_0}\right)' cos\frac{\emptyset}{3} + \frac{2}{3}\sqrt{\Delta_0}\ \left(cos\frac{\emptyset}{3}\right)'\right)$$

$$\sqrt{\Delta_0} = \ \sqrt{c^2 - 3bd + 12e} \ \rightarrow\ \left(\sqrt{\Delta_0}\right)' = \frac{1}{2}(c^2 - 3bd + 12e)^{-.5}\ (2cc' - 3b'd - 3bd' + 12e')$$

$$\frac{2}{3}\left(\sqrt{\Delta_0}\right)' cos\frac{\emptyset}{3} = \frac{2}{3}\frac{1}{2}(c^2 - 3bd + 12e)^{-.5}\ (2cc' - 3b'd - 3bd' + 12e') cos\frac{\emptyset}{3} = \frac{1}{3}(c^2 - 3bd + 12e)^{-.5}\ (2cc' - 3b'd - 3bd' + 12e') cos\frac{\emptyset}{3}$$

$$\frac{2}{3}\sqrt{\Delta_0}\left[\left(cos\frac{\emptyset}{3}\right)'\right] = \frac{2}{3}\sqrt{\Delta_0}\left[\left(-sin\frac{\emptyset}{3}\right)\frac{1}{3}(\emptyset')\right] \ ,\emptyset = arccos\left(\frac{\Delta_1}{(2)\sqrt{\Delta_0^3}}\right) = cos^{-1}\left(\frac{\Delta_1}{(2)\sqrt{\Delta_0^3}}\right) \ \rightarrow\ hint\!:\ \frac{d}{dx}\ [cos^{-1}u] = \frac{-1}{\sqrt{1-u^2}}\frac{du}{dx}$$

$$\emptyset' = \frac{\partial}{\partial\theta_i}\ \left[cos^{-1}\left(\frac{\Delta_1}{(2)\sqrt{\Delta_0^3}}\right)\right] = \frac{-1}{\sqrt{1 - \left(\frac{\Delta_1}{(2)\sqrt{\Delta_0^3}}\right)^2}}\frac{\partial}{\partial\theta_i}\left(\frac{\Delta_1}{(2)\sqrt{\Delta_0^3}}\right), \frac{\partial}{\partial\theta_i}\left(\frac{\Delta_1}{(2)\sqrt{\Delta_0^3}}\right) = \frac{1}{2}\left(\frac{\Delta_1'\sqrt{\Delta_0^3} - \Delta_1\left(\sqrt{\Delta_0^3}\right)'}{\left(\sqrt{\Delta_0^3}\right)^2}\right) = \frac{1}{2}\left(\frac{\Delta_1'\sqrt{\Delta_0^3} - \Delta_1\left(\sqrt{\Delta_0^3}\right)'}{\Delta_0^3}\right)$$

$$\frac{2}{9}\sqrt{\Delta_0}\left[\left(-sin\frac{\emptyset}{3}\right)(\emptyset')\right] = \frac{2}{9}\sqrt{\Delta_0}\ \left(-sin\frac{\emptyset}{3}\right)\frac{-1}{\sqrt{1-\left(\frac{\Delta_1}{(2)\sqrt{\Delta_0^3}}\right)^2}}\frac{1}{2}\left(\frac{\Delta_1'\sqrt{\Delta_0^3}-\Delta_1\left(\sqrt{\Delta_0^3}\right)'}{\Delta_0^3}\right) = \frac{-1}{9}\frac{\sqrt{\Delta_0}\ \left(-sin\frac{\emptyset}{3}\right)}{\sqrt{1-\left(\frac{\Delta_1}{(2)\sqrt{\Delta_0^3}}\right)^2}}\left(\frac{\Delta_1'\sqrt{\Delta_0^3}-\Delta_1\left(\sqrt{\Delta_0^3}\right)'}{\Delta_0^3}\right)$$

$$\left({\Delta_0}^{3/2}\right)' = \frac{3}{2}(c^2-3bd+12e)^{.5}(2cc'-3b'd-3bd'+12e')$$

$$(\Delta_1)' = 6c^2c' - 9b'cd - 9bc'd - 9bcd' + 54bb'e + 27b^2e' + 54dd' - 72c'e - 72ce'$$

$$\frac{\partial}{\partial\lambda_{12}}b = 1\ ,\frac{\partial}{\partial\lambda_{12}}c = \{(\lambda_{23}+\lambda_{28}+\lambda_{29})\} + \{\lambda_{34}+\lambda_{38}+\lambda_{39}+\mu_{32}+\ \lambda_{45}+\lambda_{48}+\lambda_{49}+\mu_{43}\}$$

$$\frac{\partial}{\partial\lambda_{12}}d = \{(\lambda_{23}+\lambda_{28}+\lambda_{29})\}\{\lambda_{34}+\lambda_{38}+\lambda_{39}+\mu_{32}+\ \lambda_{45}+\lambda_{48}+\lambda_{49}+\mu_{43}\} + \{(\lambda_{34}+\lambda_{38}+\lambda_{39}+\mu_{32})(\ \lambda_{45}+\lambda_{48}+\lambda_{49}) + \mu_{43}(\lambda_{38}+\lambda_{39}+\mu_{32})\} - \ \lambda_{23}\,\mu_{32}$$

$$\frac{\partial}{\partial\lambda_{12}}e = (\lambda_{23}+\lambda_{28}+\lambda_{29})\{(\lambda_{34}+\lambda_{38}+\lambda_{39}+\mu_{32})(\ \lambda_{45}+\lambda_{48}+\lambda_{49}) + \mu_{43}(\lambda_{38}+\lambda_{39}+\mu_{32})\} - \{\lambda_{23}\,\mu_{32}\,(\lambda_{45}+\lambda_{48}+\lambda_{49}+\mu_{43})\}$$

$$\frac{\partial}{\partial\lambda_{18}}b = 1\ \ ,\quad \frac{\partial}{\partial\lambda_{18}}c = (\lambda_{23}+\lambda_{28}+\lambda_{29}) + \mu_{21} + \{\lambda_{34}+\lambda_{38}+\lambda_{39}+\mu_{32}+\ \lambda_{45}+\lambda_{48}+\lambda_{49}+\mu_{43}\}$$

$$\frac{\partial}{\partial\lambda_{18}}d = \{(\lambda_{23}+\lambda_{28}+\lambda_{29}) + \mu_{21}\}\{\lambda_{34}+\lambda_{38}+\lambda_{39}+\mu_{32}+\ \lambda_{45}+\lambda_{48}+\lambda_{49}+\mu_{43}\} + \{(\lambda_{34}+\lambda_{38}+\lambda_{39}+\mu_{32})(\ \lambda_{45}+\lambda_{48}+\lambda_{49}) + \mu_{43}(\lambda_{38}+\lambda_{39}+\mu_{32})\}$$

$$-\lambda_{23}\,\mu_{32}$$

$$\frac{\partial}{\partial\lambda_{18}}e = \{(\lambda_{23}+\lambda_{28}+\lambda_{29}) + \mu_{21}\}\{(\lambda_{34}+\lambda_{38}+\lambda_{39}+\mu_{32})(\ \lambda_{45}+\lambda_{48}+\lambda_{49}) + \mu_{43}(\lambda_{38}+\lambda_{39}+\mu_{32})\} - \ \{\lambda_{23}\,\mu_{32}\,(\lambda_{45}+\lambda_{48}+\lambda_{49}+\mu_{43})\}$$

$$\frac{\partial}{\partial\lambda_{19}}b = 1\ ,\quad \frac{\partial}{\partial\lambda_{19}}c = (\lambda_{23}+\lambda_{28}+\lambda_{29}) + \mu_{21} + \{\lambda_{34}+\lambda_{38}+\lambda_{39}+\mu_{32}+\ \lambda_{45}+\lambda_{48}+\lambda_{49}+\mu_{43}\}$$

$$\frac{\partial}{\partial\lambda_{19}}d = \{(\lambda_{23}+\lambda_{28}+\lambda_{29}) + \mu_{21}\}\{\lambda_{34}+\lambda_{38}+\lambda_{39}+\mu_{32}+\ \lambda_{45}+\lambda_{48}+\lambda_{49}+\mu_{43}\} + \{(\lambda_{34}+\lambda_{38}+\lambda_{39}+\mu_{32})(\ \lambda_{45}+\lambda_{48}+\lambda_{49}) + \mu_{43}(\lambda_{38}+\lambda_{39}+\mu_{32})\}$$

$$-\lambda_{23}\,\mu_{32}$$

$$\frac{\partial}{\partial\lambda_{19}}e = \{(\lambda_{23}+\lambda_{28}+\lambda_{29}) + \mu_{21}\}\{(\lambda_{34}+\lambda_{38}+\lambda_{39}+\mu_{32})(\ \lambda_{45}+\lambda_{48}+\lambda_{49}) + \mu_{43}(\lambda_{38}+\lambda_{39}+\mu_{32})\} - \ \{\lambda_{23}\,\mu_{32}\,(\lambda_{45}+\lambda_{48}+\lambda_{49}+\mu_{43})\}$$

$$\frac{\partial}{\partial\lambda_{23}}b = 1\ ,\quad \frac{\partial}{\partial\lambda_{23}}c = \{(\lambda_{12}+\lambda_{18}+\lambda_{19})\} + \{\lambda_{34}+\lambda_{38}+\lambda_{39}+\mu_{32}+\ \lambda_{45}+\lambda_{48}+\lambda_{49}+\mu_{43}\} - \ \mu_{32}$$

$$\frac{\partial}{\partial\lambda_{23}}d = (\lambda_{12}+\lambda_{18}+\lambda_{19})\{\lambda_{34}+\lambda_{38}+\lambda_{39}+\mu_{32}+\ \lambda_{45}+\lambda_{48}+\lambda_{49}+\mu_{43}\} + \{(\lambda_{34}+\lambda_{38}+\lambda_{39}+\mu_{32})(\ \lambda_{45}+\lambda_{48}+\lambda_{49}) + \mu_{43}(\lambda_{38}+\lambda_{39}+\mu_{32})\}$$

$$-\ \mu_{32}\,(\lambda_{12}+\lambda_{18}+\lambda_{19}+\lambda_{45}+\lambda_{48}+\lambda_{49}+\mu_{43})$$

$$\frac{\partial}{\partial\lambda_{23}}e = (\lambda_{12}+\lambda_{18}+\lambda_{19})\{(\lambda_{34}+\lambda_{38}+\lambda_{39}+\mu_{32})(\ \lambda_{45}+\lambda_{48}+\lambda_{49}) + \mu_{43}(\lambda_{38}+\lambda_{39}+\mu_{32})\} - \ \mu_{32}\,(\lambda_{12}+\lambda_{18}+\lambda_{19})(\lambda_{45}+\lambda_{48}+\lambda_{49}+\mu_{43})$$

$$\frac{\partial}{\partial\lambda_{28}}b = 1\ ,\quad \frac{\partial}{\partial\lambda_{28}}c = \{(\lambda_{12}+\lambda_{18}+\lambda_{19})\} + \{\lambda_{34}+\lambda_{38}+\lambda_{39}+\mu_{32}+\ \lambda_{45}+\lambda_{48}+\lambda_{49}+\mu_{43}\}$$

$$\frac{\partial}{\partial\lambda_{28}}d = (\lambda_{12}+\lambda_{18}+\lambda_{19})\{\lambda_{34}+\lambda_{38}+\lambda_{39}+\mu_{32}+\ \lambda_{45}+\lambda_{48}+\lambda_{49}+\mu_{43}\} + \{(\lambda_{34}+\lambda_{38}+\lambda_{39}+\mu_{32})(\ \lambda_{45}+\lambda_{48}+\lambda_{49}) + \mu_{43}(\lambda_{38}+\lambda_{39}+\mu_{32})\}$$

$$\frac{\partial}{\partial\lambda_{28}}e = (\lambda_{12}+\lambda_{18}+\lambda_{19})\{(\lambda_{34}+\lambda_{38}+\lambda_{39}+\mu_{32})(\ \lambda_{45}+\lambda_{48}+\lambda_{49}) + \mu_{43}(\lambda_{38}+\lambda_{39}+\mu_{32})\}$$

$$\frac{\partial}{\partial\lambda_{29}}b = 1\ \ ,\quad \frac{\partial}{\partial\lambda_{29}}c = \{(\lambda_{12}+\lambda_{18}+\lambda_{19})\} + \{\lambda_{34}+\lambda_{38}+\lambda_{39}+\mu_{32}+\ \lambda_{45}+\lambda_{48}+\lambda_{49}+\mu_{43}\}$$

$$\frac{\partial}{\partial\lambda_{29}}d = (\lambda_{12}+\lambda_{18}+\lambda_{19})\{\lambda_{34}+\lambda_{38}+\lambda_{39}+\mu_{32}+\ \lambda_{45}+\lambda_{48}+\lambda_{49}+\mu_{43}\} + \{(\lambda_{34}+\lambda_{38}+\lambda_{39}+\mu_{32})(\ \lambda_{45}+\lambda_{48}+\lambda_{49}) + \mu_{43}(\lambda_{38}+\lambda_{39}+\mu_{32})\}$$

$$\frac{\partial}{\partial\lambda_{29}}e = (\lambda_{12}+\lambda_{18}+\lambda_{19})\{(\lambda_{34}+\lambda_{38}+\lambda_{39}+\mu_{32})(\lambda_{45}+\lambda_{48}+\lambda_{49})+\mu_{43}(\lambda_{38}+\lambda_{39}+\mu_{32})\}$$

$$\frac{\partial}{\partial\mu_{21}}b = 1\,, \frac{\partial}{\partial\mu_{21}}c = (\lambda_{18}+\lambda_{19})+\{\lambda_{34}+\lambda_{38}+\lambda_{39}+\mu_{32}+\lambda_{45}+\lambda_{48}+\lambda_{49}+\mu_{43}\}$$

$$\frac{\partial}{\partial\mu_{21}}d = (\lambda_{18}+\lambda_{19})\{\lambda_{34}+\lambda_{38}+\lambda_{39}+\mu_{32}+\lambda_{45}+\lambda_{48}+\lambda_{49}+\mu_{43}\}+\{(\lambda_{34}+\lambda_{38}+\lambda_{39}+\mu_{32})(\lambda_{45}+\lambda_{48}+\lambda_{49})+\mu_{43}(\lambda_{38}+\lambda_{39}+\mu_{32})\}$$

$$\frac{\partial}{\partial\mu_{21}}e = (\lambda_{18}+\lambda_{19})\{(\lambda_{34}+\lambda_{38}+\lambda_{39}+\mu_{32})(\lambda_{45}+\lambda_{48}+\lambda_{49})+\mu_{43}(\lambda_{38}+\lambda_{39}+\mu_{32})\}$$

$$\frac{\partial}{\partial\lambda_{34}}b = 1\,, \quad \frac{\partial}{\partial\lambda_{34}}c = (\lambda_{45}+\lambda_{48}+\lambda_{49})+\{\lambda_{12}+\lambda_{18}+\lambda_{19}+\lambda_{23}+\lambda_{28}+\lambda_{29}+\mu_{21}\}$$

$$\frac{\partial}{\partial\lambda_{34}}d = \{(\lambda_{12}+\lambda_{18}+\lambda_{19})(\lambda_{23}+\lambda_{28}+\lambda_{29})+\mu_{21}(\lambda_{18}+\lambda_{19})\}+\{\lambda_{12}+\lambda_{18}+\lambda_{19}+\lambda_{23}+\lambda_{28}+\lambda_{29}+\mu_{21}\}(\lambda_{45}+\lambda_{48}+\lambda_{49})$$

$$\frac{\partial}{\partial\lambda_{34}}e = \{(\lambda_{12}+\lambda_{18}+\lambda_{19})(\lambda_{23}+\lambda_{28}+\lambda_{29})+\mu_{21}(\lambda_{18}+\lambda_{19})\}(\lambda_{45}+\lambda_{48}+\lambda_{49})$$

$$\frac{\partial}{\partial\lambda_{38}}b = 1\,, \quad \frac{\partial}{\partial\lambda_{38}}c = \{(\lambda_{45}+\lambda_{48}+\lambda_{49})+\mu_{43}\}+\{\lambda_{12}+\lambda_{18}+\lambda_{19}+\lambda_{23}+\lambda_{28}+\lambda_{29}+\mu_{21}\}$$

$$\frac{\partial}{\partial\lambda_{38}}d = \{(\lambda_{12}+\lambda_{18}+\lambda_{19})(\lambda_{23}+\lambda_{28}+\lambda_{29})+\mu_{21}(\lambda_{18}+\lambda_{19})\}+\{\lambda_{12}+\lambda_{18}+\lambda_{19}+\lambda_{23}+\lambda_{28}+\lambda_{29}+\mu_{21}\}\{(\lambda_{45}+\lambda_{48}+\lambda_{49})+\mu_{43}\}$$

$$\frac{\partial}{\partial\lambda_{38}}e = \{(\lambda_{12}+\lambda_{18}+\lambda_{19})(\lambda_{23}+\lambda_{28}+\lambda_{29})+\mu_{21}(\lambda_{18}+\lambda_{19})\}\{(\lambda_{45}+\lambda_{48}+\lambda_{49})+\mu_{43}\}$$

$$\frac{\partial}{\partial\lambda_{39}}b = 1\,, \quad \frac{\partial}{\partial\lambda_{39}}c = \{(\lambda_{45}+\lambda_{48}+\lambda_{49})+\mu_{43}\}+\{\lambda_{12}+\lambda_{18}+\lambda_{19}+\lambda_{23}+\lambda_{28}+\lambda_{29}+\mu_{21}\}$$

$$\frac{\partial}{\partial\lambda_{39}}d = \{(\lambda_{12}+\lambda_{18}+\lambda_{19})(\lambda_{23}+\lambda_{28}+\lambda_{29})+\mu_{21}(\lambda_{18}+\lambda_{19})\}+\{\lambda_{12}+\lambda_{18}+\lambda_{19}+\lambda_{23}+\lambda_{28}+\lambda_{29}+\mu_{21}\}\{(\lambda_{45}+\lambda_{48}+\lambda_{49})+\mu_{43}\}$$

$$\frac{\partial}{\partial\lambda_{39}}e = \{(\lambda_{12}+\lambda_{18}+\lambda_{19})(\lambda_{23}+\lambda_{28}+\lambda_{29})+\mu_{21}(\lambda_{18}+\lambda_{19})\}\{(\lambda_{45}+\lambda_{48}+\lambda_{49})+\mu_{43}\}$$

$$\frac{\partial}{\partial\mu_{32}}b = 1\,, \quad \frac{\partial}{\partial\mu_{32}}c = \{(\lambda_{45}+\lambda_{48}+\lambda_{49})+\mu_{43}\}+\{\lambda_{12}+\lambda_{18}+\lambda_{19}+\lambda_{23}+\lambda_{28}+\lambda_{29}+\mu_{21}\}-\lambda_{23}$$

$$\frac{\partial}{\partial\mu_{32}}d = \{(\lambda_{12}+\lambda_{18}+\lambda_{19})(\lambda_{23}+\lambda_{28}+\lambda_{29})+\mu_{21}(\lambda_{18}+\lambda_{19})\}+\{\lambda_{12}+\lambda_{18}+\lambda_{19}+\lambda_{23}+\lambda_{28}+\lambda_{29}+\mu_{21}\}\{(\lambda_{45}+\lambda_{48}+\lambda_{49})+\mu_{43}\}$$

$$-\lambda_{23}\,(\lambda_{12}+\lambda_{18}+\lambda_{19}+\lambda_{45}+\lambda_{48}+\lambda_{49}+\mu_{43})$$

$$\frac{\partial}{\partial\mu_{32}}e = \{(\lambda_{12}+\lambda_{18}+\lambda_{19})(\lambda_{23}+\lambda_{28}+\lambda_{29})+\mu_{21}(\lambda_{18}+\lambda_{19})\}\{(\lambda_{45}+\lambda_{48}+\lambda_{49})+\mu_{43}\}-\lambda_{23}\,(\lambda_{12}+\lambda_{18}+\lambda_{19})(\lambda_{45}+\lambda_{48}+\lambda_{49}+\mu_{43})$$

$$\frac{\partial}{\partial\lambda_{45}}b = 1\,, \quad \frac{\partial}{\partial\lambda_{45}}c = (\lambda_{34}+\lambda_{38}+\lambda_{39}+\mu_{32})+\{\lambda_{12}+\lambda_{18}+\lambda_{19}+\lambda_{23}+\lambda_{28}+\lambda_{29}+\mu_{21}\}$$

$$\frac{\partial}{\partial\lambda_{45}}d = \{(\lambda_{12}+\lambda_{18}+\lambda_{19})(\lambda_{23}+\lambda_{28}+\lambda_{29})+\mu_{21}(\lambda_{18}+\lambda_{19})\}+\{\lambda_{12}+\lambda_{18}+\lambda_{19}+\lambda_{23}+\lambda_{28}+\lambda_{29}+\mu_{21}\}(\lambda_{34}+\lambda_{38}+\lambda_{39}+\mu_{32})-\lambda_{23}\,\mu_{32}$$

$$\frac{\partial}{\partial\lambda_{45}}e = \{(\lambda_{12}+\lambda_{18}+\lambda_{19})(\lambda_{23}+\lambda_{28}+\lambda_{29})+\mu_{21}(\lambda_{18}+\lambda_{19})\}(\lambda_{34}+\lambda_{38}+\lambda_{39}+\mu_{32})-\{\lambda_{23}\,\mu_{32}\,(\lambda_{12}+\lambda_{18}+\lambda_{19})\}$$

$$\frac{\partial}{\partial\lambda_{48}}b = 1, \quad \frac{\partial}{\partial\lambda_{48}}c = (\lambda_{34}+\lambda_{38}+\lambda_{39}+\mu_{32})+\{\lambda_{12}+\lambda_{18}+\lambda_{19}+\lambda_{23}+\lambda_{28}+\lambda_{29}+\mu_{21}\}$$

$$\frac{\partial}{\partial\lambda_{48}}d = \{(\lambda_{12}+\lambda_{18}+\lambda_{19})(\lambda_{23}+\lambda_{28}+\lambda_{29})+\mu_{21}(\lambda_{18}+\lambda_{19})\}+\{\lambda_{12}+\lambda_{18}+\lambda_{19}+\lambda_{23}+\lambda_{28}+\lambda_{29}+\mu_{21}\}(\lambda_{34}+\lambda_{38}+\lambda_{39}+\mu_{32})-\lambda_{23}\,\mu_{32}$$

$$\frac{\partial}{\partial\lambda_{48}}e = \{(\lambda_{12}+\lambda_{18}+\lambda_{19})(\lambda_{23}+\lambda_{28}+\lambda_{29})+\mu_{21}(\lambda_{18}+\lambda_{19})\}(\lambda_{34}+\lambda_{38}+\lambda_{39}+\mu_{32})-\{\lambda_{23}\,\mu_{32}\,(\lambda_{12}+\lambda_{18}+\lambda_{19})\}$$

$$\frac{\partial}{\partial \lambda_{49}} b = 1\ , \qquad \frac{\partial}{\partial \lambda_{49}} c = (\lambda_{34} + \lambda_{38} + \lambda_{39} + \mu_{32}) + \{\lambda_{12} + \lambda_{18} + \lambda_{19} + \lambda_{23} + \lambda_{28} + \lambda_{29} + \mu_{21}\}$$

$$\frac{\partial}{\partial \lambda_{49}} d = \{(\lambda_{12} + \lambda_{18} + \lambda_{19})(\lambda_{23} + \lambda_{28} + \lambda_{29}) + \mu_{21}(\lambda_{18} + \lambda_{19})\} + \{\lambda_{12} + \lambda_{18} + \lambda_{19} + \lambda_{23} + \lambda_{28} + \lambda_{29} + \mu_{21}\}(\lambda_{34} + \lambda_{38} + \lambda_{39} + \mu_{32}) - \lambda_{23}\,\mu_{32}$$

$$\frac{\partial}{\partial \lambda_{49}} e = \{(\lambda_{12} + \lambda_{18} + \lambda_{19})(\lambda_{23} + \lambda_{28} + \lambda_{29}) + \mu_{21}(\lambda_{18} + \lambda_{19})\}(\lambda_{34} + \lambda_{38} + \lambda_{39} + \mu_{32}) - \{\lambda_{23}\,\mu_{32}\,(\lambda_{12} + \lambda_{18} + \lambda_{19})\}$$

$$\frac{\partial}{\partial \mu_{43}} b = 1\ , \quad \frac{\partial}{\partial \mu_{43}} c = (\lambda_{38} + \lambda_{39} + \mu_{32}) + \{\lambda_{12} + \lambda_{18} + \lambda_{19} + \lambda_{23} + \lambda_{28} + \lambda_{29} + \mu_{21}\}$$

$$\frac{\partial}{\partial \mu_{43}} d = \{(\lambda_{12} + \lambda_{18} + \lambda_{19})(\lambda_{23} + \lambda_{28} + \lambda_{29}) + \mu_{21}(\lambda_{18} + \lambda_{19})\} + \{\lambda_{12} + \lambda_{18} + \lambda_{19} + \lambda_{23} + \lambda_{28} + \lambda_{29} + \mu_{21}\}(\lambda_{38} + \lambda_{39} + \mu_{32}) - \lambda_{23}\,\mu_{32}$$

$$\frac{\partial}{\partial \mu_{43}} e = \{(\lambda_{12} + \lambda_{18} + \lambda_{19})(\lambda_{23} + \lambda_{28} + \lambda_{29}) + \mu_{21}(\lambda_{18} + \lambda_{19})\}(\lambda_{38} + \lambda_{39} + \mu_{32}) - \lambda_{23}\,\mu_{32}\,(\lambda_{12} + \lambda_{18} + \lambda_{19})$$

$\frac{\partial}{\partial \theta_k} P_{ij}(t) = te^{\Lambda t} d\Lambda$ , where $t = \Delta t$ and $\Lambda$ *are eigenvalues* $\rho_i$ , to construct the score vector

$$te^{\Lambda t} d\,\Lambda = te^{\rho_1 t}\begin{bmatrix} \frac{\partial}{\partial \lambda_{12}}\rho_1 \\ \frac{\partial}{\partial \lambda_{18}}\rho_1 \\ \frac{\partial}{\partial \lambda_{19}}\rho_1 \\ \frac{\partial}{\partial \lambda_{23}}\rho_1 \\ \frac{\partial}{\partial \lambda_{28}}\rho_1 \\ \frac{\partial}{\partial \lambda_{29}}\rho_1 \\ \frac{\partial}{\partial \mu_{21}}\rho_1 \\ \frac{\partial}{\partial \lambda_{34}}\rho_1 \\ \frac{\partial}{\partial \lambda_{38}}\rho_1 \\ \frac{\partial}{\partial \lambda_{39}}\rho_1 \\ \frac{\partial}{\partial \mu_{32}}\rho_1 \\ \frac{\partial}{\partial \lambda_{45}}\rho_1 \\ \frac{\partial}{\partial \lambda_{48}}\rho_1 \\ \frac{\partial}{\partial \lambda_{49}}\rho_1 \\ \frac{\partial}{\partial \mu_{43}}\rho_1 \\ \frac{\partial}{\partial \lambda_{56}}\rho_1 \\ \frac{\partial}{\partial \lambda_{58}}\rho_1 \\ \frac{\partial}{\partial \lambda_{59}}\rho_1 \\ \frac{\partial}{\partial \lambda_{67}}\rho_1 \\ \frac{\partial}{\partial \lambda_{69}}\rho_1 \\ \frac{\partial}{\partial \lambda_{79}}\rho_1 \\ \frac{\partial}{\partial \lambda_{89}}\rho_1 \end{bmatrix} + te^{\rho_2 t}\begin{bmatrix} \frac{\partial}{\partial \lambda_{12}}\rho_2 \\ \frac{\partial}{\partial \lambda_{18}}\rho_2 \\ \frac{\partial}{\partial \lambda_{19}}\rho_2 \\ \frac{\partial}{\partial \lambda_{23}}\rho_2 \\ \frac{\partial}{\partial \lambda_{28}}\rho_2 \\ \frac{\partial}{\partial \lambda_{29}}\rho_2 \\ \frac{\partial}{\partial \mu_{21}}\rho_2 \\ \frac{\partial}{\partial \lambda_{34}}\rho_2 \\ \frac{\partial}{\partial \lambda_{38}}\rho_2 \\ \frac{\partial}{\partial \lambda_{39}}\rho_2 \\ \frac{\partial}{\partial \mu_{32}}\rho_2 \\ \frac{\partial}{\partial \lambda_{45}}\rho_2 \\ \frac{\partial}{\partial \lambda_{48}}\rho_2 \\ \frac{\partial}{\partial \lambda_{49}}\rho_2 \\ \frac{\partial}{\partial \mu_{43}}\rho_2 \\ \frac{\partial}{\partial \lambda_{56}}\rho_2 \\ \frac{\partial}{\partial \lambda_{58}}\rho_2 \\ \frac{\partial}{\partial \lambda_{59}}\rho_2 \\ \frac{\partial}{\partial \lambda_{67}}\rho_2 \\ \frac{\partial}{\partial \lambda_{69}}\rho_2 \\ \frac{\partial}{\partial \lambda_{79}}\rho_2 \\ \frac{\partial}{\partial \lambda_{89}}\rho_2 \end{bmatrix} + te^{\rho_3 t}\begin{bmatrix} \frac{\partial}{\partial \lambda_{12}}\rho_3 \\ \frac{\partial}{\partial \lambda_{18}}\rho_3 \\ \frac{\partial}{\partial \lambda_{19}}\rho_3 \\ \frac{\partial}{\partial \lambda_{23}}\rho_3 \\ \frac{\partial}{\partial \lambda_{28}}\rho_3 \\ \frac{\partial}{\partial \lambda_{29}}\rho_3 \\ \frac{\partial}{\partial \mu_{21}}\rho_3 \\ \frac{\partial}{\partial \lambda_{34}}\rho_3 \\ \frac{\partial}{\partial \lambda_{38}}\rho_3 \\ \frac{\partial}{\partial \lambda_{39}}\rho_3 \\ \frac{\partial}{\partial \mu_{32}}\rho_3 \\ \frac{\partial}{\partial \lambda_{45}}\rho_3 \\ \frac{\partial}{\partial \lambda_{48}}\rho_3 \\ \frac{\partial}{\partial \lambda_{49}}\rho_3 \\ \frac{\partial}{\partial \mu_{43}}\rho_3 \\ \frac{\partial}{\partial \lambda_{56}}\rho_3 \\ \frac{\partial}{\partial \lambda_{58}}\rho_3 \\ \frac{\partial}{\partial \lambda_{59}}\rho_3 \\ \frac{\partial}{\partial \lambda_{67}}\rho_3 \\ \frac{\partial}{\partial \lambda_{69}}\rho_3 \\ \frac{\partial}{\partial \lambda_{79}}\rho_3 \\ \frac{\partial}{\partial \lambda_{89}}\rho_3 \end{bmatrix} + te^{\rho_4 t}\begin{bmatrix} \frac{\partial}{\partial \lambda_{12}}\rho_4 \\ \frac{\partial}{\partial \lambda_{18}}\rho_4 \\ \frac{\partial}{\partial \lambda_{19}}\rho_4 \\ \frac{\partial}{\partial \lambda_{23}}\rho_4 \\ \frac{\partial}{\partial \lambda_{28}}\rho_4 \\ \frac{\partial}{\partial \lambda_{29}}\rho_4 \\ \frac{\partial}{\partial \mu_{21}}\rho_4 \\ \frac{\partial}{\partial \lambda_{34}}\rho_4 \\ \frac{\partial}{\partial \lambda_{38}}\rho_4 \\ \frac{\partial}{\partial \lambda_{39}}\rho_4 \\ \frac{\partial}{\partial \mu_{32}}\rho_4 \\ \frac{\partial}{\partial \lambda_{45}}\rho_4 \\ \frac{\partial}{\partial \lambda_{48}}\rho_4 \\ \frac{\partial}{\partial \lambda_{49}}\rho_4 \\ \frac{\partial}{\partial \mu_{43}}\rho_4 \\ \frac{\partial}{\partial \lambda_{56}}\rho_4 \\ \frac{\partial}{\partial \lambda_{58}}\rho_4 \\ \frac{\partial}{\partial \lambda_{59}}\rho_4 \\ \frac{\partial}{\partial \lambda_{67}}\rho_4 \\ \frac{\partial}{\partial \lambda_{69}}\rho_4 \\ \frac{\partial}{\partial \lambda_{79}}\rho_4 \\ \frac{\partial}{\partial \lambda_{89}}\rho_4 \end{bmatrix} + te^{\rho_5 t}\begin{bmatrix} \frac{\partial}{\partial \lambda_{12}}\rho_5 \\ \frac{\partial}{\partial \lambda_{18}}\rho_5 \\ \frac{\partial}{\partial \lambda_{19}}\rho_5 \\ \frac{\partial}{\partial \lambda_{23}}\rho_5 \\ \frac{\partial}{\partial \lambda_{28}}\rho_5 \\ \frac{\partial}{\partial \lambda_{29}}\rho_5 \\ \frac{\partial}{\partial \mu_{21}}\rho_5 \\ \frac{\partial}{\partial \lambda_{34}}\rho_5 \\ \frac{\partial}{\partial \lambda_{38}}\rho_5 \\ \frac{\partial}{\partial \lambda_{39}}\rho_5 \\ \frac{\partial}{\partial \mu_{32}}\rho_5 \\ \frac{\partial}{\partial \lambda_{45}}\rho_5 \\ \frac{\partial}{\partial \lambda_{48}}\rho_5 \\ \frac{\partial}{\partial \lambda_{49}}\rho_5 \\ \frac{\partial}{\partial \mu_{43}}\rho_5 \\ \frac{\partial}{\partial \lambda_{56}}\rho_5 \\ \frac{\partial}{\partial \lambda_{58}}\rho_5 \\ \frac{\partial}{\partial \lambda_{59}}\rho_5 \\ \frac{\partial}{\partial \lambda_{67}}\rho_5 \\ \frac{\partial}{\partial \lambda_{69}}\rho_5 \\ \frac{\partial}{\partial \lambda_{79}}\rho_5 \\ \frac{\partial}{\partial \lambda_{89}}\rho_5 \end{bmatrix} + te^{\rho_6 t}\begin{bmatrix} \frac{\partial}{\partial \lambda_{12}}\rho_6 \\ \frac{\partial}{\partial \lambda_{18}}\rho_6 \\ \frac{\partial}{\partial \lambda_{19}}\rho_6 \\ \frac{\partial}{\partial \lambda_{23}}\rho_6 \\ \frac{\partial}{\partial \lambda_{28}}\rho_6 \\ \frac{\partial}{\partial \lambda_{29}}\rho_6 \\ \frac{\partial}{\partial \mu_{21}}\rho_6 \\ \frac{\partial}{\partial \lambda_{34}}\rho_6 \\ \frac{\partial}{\partial \lambda_{38}}\rho_6 \\ \frac{\partial}{\partial \lambda_{39}}\rho_6 \\ \frac{\partial}{\partial \mu_{32}}\rho_6 \\ \frac{\partial}{\partial \lambda_{45}}\rho_6 \\ \frac{\partial}{\partial \lambda_{48}}\rho_6 \\ \frac{\partial}{\partial \lambda_{49}}\rho_6 \\ \frac{\partial}{\partial \mu_{43}}\rho_6 \\ \frac{\partial}{\partial \lambda_{56}}\rho_6 \\ \frac{\partial}{\partial \lambda_{58}}\rho_6 \\ \frac{\partial}{\partial \lambda_{59}}\rho_6 \\ \frac{\partial}{\partial \lambda_{67}}\rho_6 \\ \frac{\partial}{\partial \lambda_{69}}\rho_6 \\ \frac{\partial}{\partial \lambda_{79}}\rho_6 \\ \frac{\partial}{\partial \lambda_{89}}\rho_6 \end{bmatrix} + te^{\rho_7 t}\begin{bmatrix} \frac{\partial}{\partial \lambda_{12}}\rho_7 \\ \frac{\partial}{\partial \lambda_{18}}\rho_7 \\ \frac{\partial}{\partial \lambda_{19}}\rho_7 \\ \frac{\partial}{\partial \lambda_{23}}\rho_7 \\ \frac{\partial}{\partial \lambda_{28}}\rho_7 \\ \frac{\partial}{\partial \lambda_{29}}\rho_7 \\ \frac{\partial}{\partial \mu_{21}}\rho_7 \\ \frac{\partial}{\partial \lambda_{34}}\rho_7 \\ \frac{\partial}{\partial \lambda_{38}}\rho_7 \\ \frac{\partial}{\partial \lambda_{39}}\rho_7 \\ \frac{\partial}{\partial \mu_{32}}\rho_7 \\ \frac{\partial}{\partial \lambda_{45}}\rho_7 \\ \frac{\partial}{\partial \lambda_{48}}\rho_7 \\ \frac{\partial}{\partial \lambda_{49}}\rho_7 \\ \frac{\partial}{\partial \mu_{43}}\rho_7 \\ \frac{\partial}{\partial \lambda_{56}}\rho_7 \\ \frac{\partial}{\partial \lambda_{58}}\rho_7 \\ \frac{\partial}{\partial \lambda_{59}}\rho_7 \\ \frac{\partial}{\partial \lambda_{67}}\rho_7 \\ \frac{\partial}{\partial \lambda_{69}}\rho_7 \\ \frac{\partial}{\partial \lambda_{79}}\rho_7 \\ \frac{\partial}{\partial \lambda_{89}}\rho_7 \end{bmatrix} + te^{\rho_8 t}\begin{bmatrix} \frac{\partial}{\partial \lambda_{12}}\rho_8 \\ \frac{\partial}{\partial \lambda_{18}}\rho_8 \\ \frac{\partial}{\partial \lambda_{19}}\rho_8 \\ \frac{\partial}{\partial \lambda_{23}}\rho_8 \\ \frac{\partial}{\partial \lambda_{28}}\rho_8 \\ \frac{\partial}{\partial \lambda_{29}}\rho_8 \\ \frac{\partial}{\partial \mu_{21}}\rho_8 \\ \frac{\partial}{\partial \lambda_{34}}\rho_8 \\ \frac{\partial}{\partial \lambda_{38}}\rho_8 \\ \frac{\partial}{\partial \lambda_{39}}\rho_8 \\ \frac{\partial}{\partial \mu_{32}}\rho_8 \\ \frac{\partial}{\partial \lambda_{45}}\rho_8 \\ \frac{\partial}{\partial \lambda_{48}}\rho_8 \\ \frac{\partial}{\partial \lambda_{49}}\rho_8 \\ \frac{\partial}{\partial \mu_{43}}\rho_8 \\ \frac{\partial}{\partial \lambda_{56}}\rho_8 \\ \frac{\partial}{\partial \lambda_{58}}\rho_8 \\ \frac{\partial}{\partial \lambda_{59}}\rho_8 \\ \frac{\partial}{\partial \lambda_{67}}\rho_8 \\ \frac{\partial}{\partial \lambda_{69}}\rho_8 \\ \frac{\partial}{\partial \lambda_{79}}\rho_8 \\ \frac{\partial}{\partial \lambda_{89}}\rho_8 \end{bmatrix}$$

$$te^{\Lambda t}d\,\Lambda = te^{\rho_1 t}\begin{bmatrix} \frac{\partial}{\partial \boldsymbol{\lambda_{12}}}\rho_1 \\ \frac{\partial}{\partial \boldsymbol{\lambda_{18}}}\rho_1 \\ \frac{\partial}{\partial \boldsymbol{\lambda_{19}}}\rho_1 \\ \frac{\partial}{\partial \boldsymbol{\lambda_{23}}}\rho_1 \\ \frac{\partial}{\partial \boldsymbol{\lambda_{28}}}\rho_1 \\ \frac{\partial}{\partial \boldsymbol{\lambda_{29}}}\rho_1 \\ \frac{\partial}{\partial \boldsymbol{\mu_{21}}}\rho_1 \\ \frac{\partial}{\partial \boldsymbol{\lambda_{34}}}\rho_1 \\ \frac{\partial}{\partial \boldsymbol{\lambda_{38}}}\rho_1 \\ \frac{\partial}{\partial \boldsymbol{\lambda_{39}}}\rho_1 \\ \frac{\partial}{\partial \boldsymbol{\mu_{32}}}\rho_1 \\ \frac{\partial}{\partial \boldsymbol{\lambda_{45}}}\rho_1 \\ \frac{\partial}{\partial \boldsymbol{\lambda_{48}}}\rho_1 \\ \frac{\partial}{\partial \boldsymbol{\lambda_{49}}}\rho_1 \\ \frac{\partial}{\partial \boldsymbol{\mu_{43}}}\rho_1 \\ 0 \\ 0 \\ 0 \\ 0 \\ 0 \\ 0 \\ 0 \end{bmatrix} + te^{\rho_2 t}\begin{bmatrix} \frac{\partial}{\partial \boldsymbol{\lambda_{12}}}\rho_2 \\ \frac{\partial}{\partial \boldsymbol{\lambda_{18}}}\rho_2 \\ \frac{\partial}{\partial \boldsymbol{\lambda_{19}}}\rho_2 \\ \frac{\partial}{\partial \boldsymbol{\lambda_{23}}}\rho_2 \\ \frac{\partial}{\partial \boldsymbol{\lambda_{28}}}\rho_2 \\ \frac{\partial}{\partial \boldsymbol{\lambda_{29}}}\rho_2 \\ \frac{\partial}{\partial \boldsymbol{\mu_{21}}}\rho_2 \\ \frac{\partial}{\partial \boldsymbol{\lambda_{34}}}\rho_2 \\ \frac{\partial}{\partial \boldsymbol{\lambda_{38}}}\rho_2 \\ \frac{\partial}{\partial \boldsymbol{\lambda_{39}}}\rho_2 \\ \frac{\partial}{\partial \boldsymbol{\mu_{32}}}\rho_2 \\ \frac{\partial}{\partial \boldsymbol{\lambda_{45}}}\rho_2 \\ \frac{\partial}{\partial \boldsymbol{\lambda_{48}}}\rho_2 \\ \frac{\partial}{\partial \boldsymbol{\lambda_{49}}}\rho_2 \\ \frac{\partial}{\partial \boldsymbol{\mu_{43}}}\rho_2 \\ 0 \\ 0 \\ 0 \\ 0 \\ 0 \\ 0 \\ 0 \end{bmatrix} + te^{\rho_3 t}\begin{bmatrix} \frac{\partial}{\partial \boldsymbol{\lambda_{12}}}\rho_3 \\ \frac{\partial}{\partial \boldsymbol{\lambda_{18}}}\rho_3 \\ \frac{\partial}{\partial \boldsymbol{\lambda_{19}}}\rho_3 \\ \frac{\partial}{\partial \boldsymbol{\lambda_{23}}}\rho_3 \\ \frac{\partial}{\partial \boldsymbol{\lambda_{28}}}\rho_3 \\ \frac{\partial}{\partial \boldsymbol{\lambda_{29}}}\rho_3 \\ \frac{\partial}{\partial \boldsymbol{\mu_{21}}}\rho_3 \\ \frac{\partial}{\partial \boldsymbol{\lambda_{34}}}\rho_3 \\ \frac{\partial}{\partial \boldsymbol{\lambda_{38}}}\rho_3 \\ \frac{\partial}{\partial \boldsymbol{\lambda_{39}}}\rho_3 \\ \frac{\partial}{\partial \boldsymbol{\mu_{32}}}\rho_3 \\ \frac{\partial}{\partial \boldsymbol{\lambda_{45}}}\rho_3 \\ \frac{\partial}{\partial \boldsymbol{\lambda_{48}}}\rho_3 \\ \frac{\partial}{\partial \boldsymbol{\lambda_{49}}}\rho_3 \\ \frac{\partial}{\partial \boldsymbol{\mu_{43}}}\rho_3 \\ 0 \\ 0 \\ 0 \\ 0 \\ 0 \\ 0 \\ 0 \end{bmatrix} + te^{\rho_4 t}\begin{bmatrix} \frac{\partial}{\partial \boldsymbol{\lambda_{12}}}\rho_4 \\ \frac{\partial}{\partial \boldsymbol{\lambda_{18}}}\rho_4 \\ \frac{\partial}{\partial \boldsymbol{\lambda_{19}}}\rho_4 \\ \frac{\partial}{\partial \boldsymbol{\lambda_{23}}}\rho_4 \\ \frac{\partial}{\partial \boldsymbol{\lambda_{28}}}\rho_4 \\ \frac{\partial}{\partial \boldsymbol{\lambda_{29}}}\rho_4 \\ \frac{\partial}{\partial \boldsymbol{\mu_{21}}}\rho_4 \\ \frac{\partial}{\partial \boldsymbol{\lambda_{34}}}\rho_4 \\ \frac{\partial}{\partial \boldsymbol{\lambda_{38}}}\rho_4 \\ \frac{\partial}{\partial \boldsymbol{\lambda_{39}}}\rho_4 \\ \frac{\partial}{\partial \boldsymbol{\mu_{32}}}\rho_4 \\ \frac{\partial}{\partial \boldsymbol{\lambda_{45}}}\rho_4 \\ \frac{\partial}{\partial \boldsymbol{\lambda_{48}}}\rho_4 \\ \frac{\partial}{\partial \boldsymbol{\lambda_{49}}}\rho_4 \\ \frac{\partial}{\partial \boldsymbol{\mu_{43}}}\rho_4 \\ 0 \\ 0 \\ 0 \\ 0 \\ 0 \\ 0 \\ 0 \end{bmatrix} + te^{\rho_5 t}\begin{bmatrix} 0 \\ 0 \\ 0 \\ 0 \\ 0 \\ 0 \\ 0 \\ 0 \\ 0 \\ 0 \\ 0 \\ 0 \\ 0 \\ 0 \\ 0 \\ \frac{\partial}{\partial \boldsymbol{\lambda_{56}}}\rho_5 \\ \frac{\partial}{\partial \boldsymbol{\lambda_{58}}}\rho_5 \\ \frac{\partial}{\partial \boldsymbol{\lambda_{59}}}\rho_5 \\ \frac{\partial}{\partial \boldsymbol{\lambda_{67}}}\rho_5 \\ \frac{\partial}{\partial \boldsymbol{\lambda_{69}}}\rho_5 \\ 0 \\ 0 \end{bmatrix} + te^{\rho_6 t}\begin{bmatrix} 0 \\ 0 \\ 0 \\ 0 \\ 0 \\ 0 \\ 0 \\ 0 \\ 0 \\ 0 \\ 0 \\ 0 \\ 0 \\ 0 \\ 0 \\ \frac{\partial}{\partial \boldsymbol{\lambda_{56}}}\rho_6 \\ \frac{\partial}{\partial \boldsymbol{\lambda_{58}}}\rho_6 \\ \frac{\partial}{\partial \boldsymbol{\lambda_{59}}}\rho_6 \\ \frac{\partial}{\partial \boldsymbol{\lambda_{67}}}\rho_6 \\ \frac{\partial}{\partial \boldsymbol{\lambda_{69}}}\rho_6 \\ 0 \\ 0 \end{bmatrix} + te^{\rho_7 t}\begin{bmatrix} 0 \\ 0 \\ 0 \\ 0 \\ 0 \\ 0 \\ 0 \\ 0 \\ 0 \\ 0 \\ 0 \\ 0 \\ 0 \\ 0 \\ 0 \\ 0 \\ 0 \\ 0 \\ 0 \\ 0 \\ \frac{\partial}{\partial \boldsymbol{\lambda_{79}}}\rho_7 \\ \frac{\partial}{\partial \boldsymbol{\lambda_{89}}}\rho_7 \end{bmatrix} + te^{\rho_8 t}\begin{bmatrix} 0 \\ 0 \\ 0 \\ 0 \\ 0 \\ 0 \\ 0 \\ 0 \\ 0 \\ 0 \\ 0 \\ 0 \\ 0 \\ 0 \\ 0 \\ 0 \\ 0 \\ 0 \\ 0 \\ 0 \\ \frac{\partial}{\partial \boldsymbol{\lambda_{79}}}\rho_8 \\ \frac{\partial}{\partial \boldsymbol{\lambda_{89}}}\rho_8 \end{bmatrix}$$

$$te^{\Lambda t}d\,\Lambda = \begin{bmatrix} te^{\rho_1 t}\frac{\partial}{\partial \boldsymbol{\lambda_{12}}}\rho_1 + te^{\rho_2 t}\frac{\partial}{\partial \boldsymbol{\lambda_{12}}}\rho_2 + te^{\rho_3 t}\frac{\partial}{\partial \boldsymbol{\lambda_{12}}}\rho_3 + te^{\rho_4 t}\frac{\partial}{\partial \boldsymbol{\lambda_{12}}}\rho_4 \\ te^{\rho_1 t}\frac{\partial}{\partial \boldsymbol{\lambda_{18}}}\rho_1 + te^{\rho_2 t}\frac{\partial}{\partial \boldsymbol{\lambda_{18}}}\rho_2 + te^{\rho_3 t}\frac{\partial}{\partial \boldsymbol{\lambda_{18}}}\rho_3 + te^{\rho_4 t}\frac{\partial}{\partial \boldsymbol{\lambda_{18}}}\rho_4 \\ te^{\rho_1 t}\frac{\partial}{\partial \boldsymbol{\lambda_{19}}}\rho_1 + te^{\rho_2 t}\frac{\partial}{\partial \boldsymbol{\lambda_{19}}}\rho_2 + te^{\rho_3 t}\frac{\partial}{\partial \boldsymbol{\lambda_{19}}}\rho_3 + te^{\rho_4 t}\frac{\partial}{\partial \boldsymbol{\lambda_{19}}}\rho_4 \\ te^{\rho_1 t}\frac{\partial}{\partial \boldsymbol{\lambda_{23}}}\rho_1 + te^{\rho_2 t}\frac{\partial}{\partial \boldsymbol{\lambda_{23}}}\rho_2 + te^{\rho_3 t}\frac{\partial}{\partial \boldsymbol{\lambda_{23}}}\rho_3 + te^{\rho_4 t}\frac{\partial}{\partial \boldsymbol{\lambda_{23}}}\rho_4 \\ te^{\rho_1 t}\frac{\partial}{\partial \boldsymbol{\lambda_{28}}}\rho_1 + te^{\rho_2 t}\frac{\partial}{\partial \boldsymbol{\lambda_{28}}}\rho_2 + te^{\rho_3 t}\frac{\partial}{\partial \boldsymbol{\lambda_{28}}}\rho_3 + te^{\rho_4 t}\frac{\partial}{\partial \boldsymbol{\lambda_{28}}}\rho_4 \\ te^{\rho_1 t}\frac{\partial}{\partial \boldsymbol{\lambda_{29}}}\rho_1 + te^{\rho_2 t}\frac{\partial}{\partial \boldsymbol{\lambda_{29}}}\rho_2 + te^{\rho_3 t}\frac{\partial}{\partial \boldsymbol{\lambda_{29}}}\rho_3 + te^{\rho_4 t}\frac{\partial}{\partial \boldsymbol{\lambda_{29}}}\rho_4 \\ te^{\rho_1 t}\frac{\partial}{\partial \boldsymbol{\mu_{21}}}\rho_1 + te^{\rho_2 t}\frac{\partial}{\partial \boldsymbol{\mu_{21}}}\rho_2 + te^{\rho_3 t}\frac{\partial}{\partial \boldsymbol{\mu_{21}}}\rho_3 + te^{\rho_4 t}\frac{\partial}{\partial \boldsymbol{\mu_{21}}}\rho_4 \\ te^{\rho_1 t}\frac{\partial}{\partial \boldsymbol{\lambda_{34}}}\rho_1 + te^{\rho_2 t}\frac{\partial}{\partial \boldsymbol{\lambda_{34}}}\rho_2 + te^{\rho_3 t}\frac{\partial}{\partial \boldsymbol{\lambda_{34}}}\rho_3 + te^{\rho_4 t}\frac{\partial}{\partial \boldsymbol{\lambda_{34}}}\rho_4 \\ te^{\rho_1 t}\frac{\partial}{\partial \boldsymbol{\lambda_{38}}}\rho_1 + te^{\rho_2 t}\frac{\partial}{\partial \boldsymbol{\lambda_{38}}}\rho_2 + te^{\rho_3 t}\frac{\partial}{\partial \boldsymbol{\lambda_{38}}}\rho_3 + te^{\rho_4 t}\frac{\partial}{\partial \boldsymbol{\lambda_{38}}}\rho_4 \\ te^{\rho_1 t}\frac{\partial}{\partial \boldsymbol{\lambda_{39}}}\rho_1 + te^{\rho_2 t}\frac{\partial}{\partial \boldsymbol{\lambda_{39}}}\rho_2 + te^{\rho_3 t}\frac{\partial}{\partial \boldsymbol{\lambda_{39}}}\rho_3 + te^{\rho_4 t}\frac{\partial}{\partial \boldsymbol{\lambda_{39}}}\rho_4 \\ te^{\rho_1 t}\frac{\partial}{\partial \boldsymbol{\mu_{32}}}\rho_1 + te^{\rho_2 t}\frac{\partial}{\partial \boldsymbol{\mu_{32}}}\rho_2 + te^{\rho_3 t}\frac{\partial}{\partial \boldsymbol{\mu_{32}}}\rho_3 + te^{\rho_4 t}\frac{\partial}{\partial \boldsymbol{\mu_{32}}}\rho_4 \\ te^{\rho_1 t}\frac{\partial}{\partial \boldsymbol{\lambda_{45}}}\rho_1 + te^{\rho_2 t}\frac{\partial}{\partial \boldsymbol{\lambda_{45}}}\rho_2 + te^{\rho_3 t}\frac{\partial}{\partial \boldsymbol{\lambda_{45}}}\rho_3 + te^{\rho_4 t}\frac{\partial}{\partial \boldsymbol{\lambda_{45}}}\rho_4 \\ te^{\rho_1 t}\frac{\partial}{\partial \boldsymbol{\lambda_{48}}}\rho_1 + te^{\rho_2 t}\frac{\partial}{\partial \boldsymbol{\lambda_{48}}}\rho_2 + te^{\rho_3 t}\frac{\partial}{\partial \boldsymbol{\lambda_{48}}}\rho_3 + te^{\rho_4 t}\frac{\partial}{\partial \boldsymbol{\lambda_{48}}}\rho_4 \\ te^{\rho_1 t}\frac{\partial}{\partial \boldsymbol{\lambda_{49}}}\rho_1 + te^{\rho_2 t}\frac{\partial}{\partial \boldsymbol{\lambda_{49}}}\rho_2 + te^{\rho_3 t}\frac{\partial}{\partial \boldsymbol{\lambda_{49}}}\rho_3 + te^{\rho_4 t}\frac{\partial}{\partial \boldsymbol{\lambda_{49}}}\rho_4 \\ te^{\rho_1 t}\frac{\partial}{\partial \boldsymbol{\mu_{43}}}\rho_1 + te^{\rho_2 t}\frac{\partial}{\partial \boldsymbol{\mu_{43}}}\rho_2 + te^{\rho_3 t}\frac{\partial}{\partial \boldsymbol{\mu_{43}}}\rho_3 + te^{\rho_4 t}\frac{\partial}{\partial \boldsymbol{\mu_{43}}}\rho_4 \\ te^{\rho_5 t}\frac{\partial}{\partial \boldsymbol{\lambda_{56}}}\rho_5 + te^{\rho_6 t}\frac{\partial}{\partial \boldsymbol{\lambda_{56}}}\rho_6 \\ te^{\rho_5 t}\frac{\partial}{\partial \boldsymbol{\lambda_{58}}}\rho_5 + te^{\rho_6 t}\frac{\partial}{\partial \boldsymbol{\lambda_{58}}}\rho_6 \\ te^{\rho_5 t}\frac{\partial}{\partial \boldsymbol{\lambda_{59}}}\rho_5 + te^{\rho_6 t}\frac{\partial}{\partial \boldsymbol{\lambda_{59}}}\rho_6 \\ te^{\rho_5 t}\frac{\partial}{\partial \boldsymbol{\lambda_{67}}}\rho_5 + te^{\rho_6 t}\frac{\partial}{\partial \boldsymbol{\lambda_{67}}}\rho_6 \\ te^{\rho_5 t}\frac{\partial}{\partial \boldsymbol{\lambda_{69}}}\rho_5 + te^{\rho_6 t}\frac{\partial}{\partial \boldsymbol{\lambda_{69}}}\rho_6 \\ te^{\rho_7 t}\frac{\partial}{\partial \boldsymbol{\lambda_{79}}}\rho_7 + te^{\rho_8 t}\frac{\partial}{\partial \boldsymbol{\lambda_{79}}}\rho_8 \\ te^{\rho_7 t}\frac{\partial}{\partial \boldsymbol{\lambda_{89}}}\rho_7 + te^{\rho_8 t}\frac{\partial}{\partial \boldsymbol{\lambda_{89}}}\rho_8 \end{bmatrix}$$

$the\ vectort\ (te^{\Lambda t}\ d\ \Lambda)\ is\ scaled\ by\ a\ factor = \left(\frac{n_{ij}(\Delta t)}{P_{ij}(\Delta t)}\right)\ according\ to\ counts\ in\ each\ cell\ ; the\ followings\ are\ the\ scalars:$

$$i.e\ \frac{n_{11}}{p_{11}} = n_{1+}, \frac{n_{12}}{p_{12}} = n_{1+}, \frac{n_{13}}{p_{13}} = n_{1+}, \frac{n_{14}}{p_{14}} = n_{1+}, \frac{n_{15}}{p_{15}} = n_{1+}, \frac{n_{16}}{p_{16}} = n_{1+}, \frac{n_{17}}{p_{17}} = n_{1+}, \frac{n_{18}}{p_{18}} = n_{1+}, \frac{n_{19}}{p_{19}} = n_{1+}$$

$$\frac{n_{21}}{p_{21}} = n_{2+}, \frac{n_{22}}{p_{22}} = n_{2+}, \frac{n_{23}}{p_{23}} = n_{2+}, \frac{n_{24}}{p_{24}} = n_{2+}, \frac{n_{25}}{p_{25}} = n_{2+}, \frac{n_{26}}{p_{26}} = n_{2+}, \frac{n_{27}}{p_{27}} = n_{2+}, \frac{n_{28}}{p_{28}} = n_{2+}, \frac{n_{29}}{p_{29}} = n_{2+}$$

$$\frac{n_{31}}{p_{31}} = n_{3+}, \frac{n_{32}}{p_{32}} = n_{3+}, \frac{n_{33}}{p_{33}} = n_{3+}, \frac{n_{34}}{p_{34}} = n_{3+}, \frac{n_{35}}{p_{35}} = n_{3+}, \frac{n_{36}}{p_{36}} = n_{3+}, \frac{n_{37}}{p_{37}} = n_{3+}, \frac{n_{38}}{p_{38}} = n_{3+}, \frac{n_{39}}{p_{39}} = n_{3+}$$

$$\frac{n_{41}}{p_{41}} = n_{4+}, \frac{n_{42}}{p_{42}} = n_{4+}, \frac{n_{43}}{p_{43}} = n_{4+}, \frac{n_{44}}{p_{44}} = n_{4+}, \frac{n_{45}}{p_{45}} = n_{4+}, \frac{n_{46}}{p_{46}} = n_{4+}, \frac{n_{47}}{p_{47}} = n_{4+}, \frac{n_{48}}{p_{48}} = n_{4+}, \frac{n_{49}}{p_{49}} = n_{4+}$$

$$\frac{n_{55}}{p_{55}} = n_{5+}, \frac{n_{56}}{p_{56}} = n_{5+}, \frac{n_{57}}{p_{57}} = n_{5+}, \frac{n_{58}}{p_{58}} = n_{5+}, \frac{n_{59}}{p_{59}} = n_{5+}$$

$$, \frac{n_{66}}{p_{66}} = n_{6+}, \frac{n_{67}}{p_{67}} = n_{6+}, \frac{n_{69}}{p_{69}} = n_{6+}$$

$$\frac{n_{77}}{p_{77}} = n_{7+}, \frac{n_{79}}{p_{79}} = n_{7+}, \quad \frac{n_{88}}{p_{88}} = n_{8+}, \quad \frac{n_{89}}{p_{89}} = n_{8+}$$

*then the scaled vectors are summed up to get the score function*

Hint: if the cell does not have counts, so the scalar corresponding to this cell is dropped.

$$S(\theta) = [9(n_{1+} + n_{2+} + n_{3+} + n_{4+}) + 5n_{5+} + 3n_{6+} + 2(n_{7+} + n_{8+})]te^{\Lambda t} d\ \Lambda$$

The scaled score function is cross product with itself i.e

$$M(\theta) = scaled\ S(\theta) \times [scaled\ S(\theta)]^T$$

*This score function is 22 by 1 vector and it is used in quasi − Newton Raphson method*

According to Kalbfliesch and Lawless (1985) the second derivative is assumed to be zero , the score function is crossed product and scaled for each pdf with the scalars (hint if the cell does not have counts, so the scalar of this cell is dropped ) :

$$i.e\ \frac{n_{1+}}{p_{11}}, \frac{n_{1+}}{p_{12}}, \dots, \frac{n_{1+}}{p_{19}}, \frac{n_{2+}}{p_{21}}, \frac{n_{2+}}{p_{22}}, \dots, \frac{n_{2+}}{p_{29}}, \frac{n_{3+}}{p_{31}}, \frac{n_{3+}}{p_{32}}, \dots, \frac{n_{3+}}{p_{39}}, \frac{n_{4+}}{p_{41}}, \frac{n_{4+}}{p_{42}}, \dots, \frac{n_{4+}}{p_{49}}, \frac{n_{5+}}{p_{55}}, \frac{n_{5+}}{p_{56}}, \dots, \frac{n_{5+}}{p_{59}}, \frac{n_{6+}}{p_{66}}, \frac{n_{6+}}{p_{67}}, \frac{n_{6+}}{p_{69}}, \frac{n_{7+}}{p_{77}}, \frac{n_{7+}}{p_{79}}, \frac{n_{8+}}{p_{88}}, \frac{n_{8+}}{p_{89}}$$

The scaled matrices are summed up to get the scaled hessian matrix $M(\theta_0)$

$$\frac{\partial^2 Log\ L}{\partial\theta_g \partial\theta_h} = \sum_{\Delta t \geq 1}^{3} \sum_{i,j}^{k=9} n_{ij} \left[\frac{\partial^2 P_{ij}(\Delta t)/\partial\theta_g \partial\theta_h}{P_{ij}(\Delta t)} - \frac{\partial P_{ij}(\Delta t)/\partial\theta_g \partial P_{ij}(\Delta t)/\partial\theta_h}{P_{ij}^2(\Delta t)}\right], k\ is\ index\ of\ states, \quad where\ P_{ij} = \frac{n_{ij}}{n_{i+}}$$

$$\frac{\partial^2 Log\ L}{\partial\theta_g \partial\theta_h} = \sum_{\Delta t \geq 1}^{3} \sum_{i,j}^{k=9} (P_{ij}\ n_{i+}) \left[\frac{\partial^2 P_{ij}(\Delta t)/\partial\theta_g \partial\theta_h}{P_{ij}(\Delta t)} - \frac{\partial P_{ij}(\Delta t)/\partial\theta_g \partial P_{ij}(\Delta t)/\partial\theta_h}{P_{ij}^2(\Delta t)}\right]$$

$$\frac{\partial^2 Log\ L}{\partial\theta_g \partial\theta_h} = \sum_{\Delta t \geq 1}^{3} \sum_{i,j}^{k=9} (P_{ij}\ n_{i+}) \left[\frac{0}{P_{ij}(\Delta t)} - \frac{\partial P_{ij}(\Delta t)/\partial\theta_g \partial P_{ij}(\Delta t)/\partial\theta_h}{P_{ij}^2(\Delta t)}\right] = -\sum_{\Delta t \geq 1}^{3} \sum_{i,j}^{k=9} n_{i+} \frac{\partial P_{ij}(\Delta t)/\partial\theta_g \partial P_{ij}(\Delta t)/\partial\theta_h}{P_{ij}(\Delta t)}$$

$Applying\quad Quasi - Newton\ Raphson\ method\ formula: \theta_1 = \theta_0 + M(\theta_0)^{-1} S(\theta_0)$

with initial theta according to Klotz and Sharples (1994); the initial $\theta$ is $P_{ij} = \frac{n_{ij}}{n_{i+}}\ in\ the\ interval\ \Delta\ \mathrm{t} = 1$

Substituting in Quasi-Newton method by the initial values, then the score and inverse of the hessian matrix are calculated to give the estimated rates.

$M(\theta) = [scaled\ S(\theta)][scaled\ S(\theta)]^T\ ,\ \ it\ is\ (22\ by\ 22)\ matrix$

$scaled\ M(\theta)\ \ will\ \ be\ inverted\ to\ be\ used\ \ in\ \ Quasi - Newton\ formula$

**5.3. Mean Sojourn Time**
These times are independent so covariance between them is zero

$$var(s_i) = \left[\left(q_{ii}(\hat{\theta})\right)^{-2}\right]^2 \sum_{h=1}^{22}\sum_{g=1}^{22} \frac{\partial q_{ii}}{\partial \theta_g}\frac{\partial q_{ii}}{\partial \theta_h}\ [\mathrm{M(\theta)}]^{-1}|_{\theta=\hat{\theta}}$$

$$var(s_i) = \left[\left(q_{ii}(\hat{\theta})\right)^{-2}\right]^2 \sum_{h=1}^{22}\sum_{g=1}^{22} \left[\frac{\partial q_{ii}}{\partial \theta_h}\right]^T\ [\mathrm{M(\theta)}]^{-1}|_{\theta=\hat{\theta}}\frac{\partial q_{ii}}{\partial \theta_g}$$

$\left[\frac{\partial q_{ii}}{\partial \theta_h}\right]$ $is\ a\ vector\ of\ size\ 22\ by\ 1\ and\ obtained\ by\ differentiating\ the\ diagonal\ elements\ of\ the\ Q\ matrix$

$with\ respect\ to\ each\ rate.$

$$var(s_i) = \left[\left(q_{ii}(\hat{\theta})\right)^{-2}\right]^2 \sum_{h=1}^{22}\sum_{g=1}^{22} \left[\frac{\partial q_{ii}}{\partial \theta_h}\right]^T\ [\mathrm{M(\theta)}]^{-1}|_{\theta=\hat{\theta}}\frac{\partial q_{ii}}{\partial \theta_g}$$

$$var(s_1) = \frac{1}{(\lambda_{12}+\lambda_{18}+\lambda_{19})^4}[-1\quad -1\quad \cdots\quad -1\quad -1][\mathrm{M(\theta)}]^{-1}|_{\theta=\hat{\theta}}\begin{bmatrix}-1\\-1\\ \vdots\\-1\\-1\end{bmatrix}$$

$The\ same\ procedure\ is\ applied\ to\ obtain\ the\ sojourn\ time\ for\ the\ other\ states\ \ and\ [\mathrm{M(\theta)}]^{-1}|_{\theta=\hat{\theta}}\ as\ calculated\ before$

## 5.4. State Probability Distribution:

To get the probability distribution of the states after a certain period of time the following equation should be solved:

$\pi = \ \pi(0)P_{ij}(t)$

$$[\pi_{01}\quad \pi_{02}\quad \pi_{03}\quad \pi_{04}\quad \pi_{05}\quad \pi_{06}\quad \pi_{07}\quad \pi_{08}\quad \pi_{09}] \times \begin{bmatrix} P_{11} & P_{12} & P_{13} & P_{14} & P_{15} & P_{16} & P_{17} & P_{18} & P_{19}\\ P_{21} & P_{22} & P_{23} & P_{24} & P_{25} & P_{26} & P_{27} & P_{28} & P_{29}\\ P_{31} & P_{32} & P_{33} & P_{34} & P_{35} & P_{36} & P_{37} & P_{38} & P_{39}\\ P_{41} & P_{42} & P_{43} & P_{44} & P_{45} & P_{46} & P_{47} & P_{48} & P_{49}\\ 0 & 0 & 0 & 0 & P_{55} & P_{56} & P_{57} & P_{58} & P_{59}\\ 0 & 0 & 0 & 0 & 0 & P_{66} & P_{67} & 0 & P_{69}\\ 0 & 0 & 0 & 0 & 0 & 0 & P_{77} & 0 & P_{79}\\ 0 & 0 & 0 & 0 & 0 & 0 & 0 & P_{88} & P_{89}\\ 0 & 0 & 0 & 0 & 0 & 0 & 0 & 0 & P_{99} \end{bmatrix}$$
$$= [\pi_1\quad \pi_2\quad \pi_3\quad \pi_4\quad \pi_5\quad \pi_6\quad \pi_7\quad \pi_8\quad \pi_9]$$

## 5.4.1. Asymptotic Covariance of the Stationary Distribution

To obtain stationary probability distribution when t goes to infinity or in other words when the process does not depend on time the following equation is solved for $\pi$ , $once\ the\ Q\ matrix\ is\ estimated$

$$\pi Q = 0\ \ with\ the\ following\ constraint\ \ \sum_{all\ z} \pi_z = 1$$

$$Q = \begin{bmatrix} -\gamma_1 & \lambda_{12} & 0 & 0 & 0 & 0 & 0 & \lambda_{18} & \lambda_{19} \\ \mu_{21} & -\gamma_2 & \lambda_{23} & 0 & 0 & 0 & 0 & \lambda_{28} & \lambda_{29} \\ 0 & \mu_{32} & -\gamma_3 & \lambda_{34} & 0 & 0 & 0 & \lambda_{38} & \lambda_{39} \\ 0 & 0 & \mu_{43} & -\gamma_4 & \lambda_{45} & 0 & 0 & \lambda_{48} & \lambda_{49} \\ 0 & 0 & 0 & 0 & -\gamma_5 & \lambda_{56} & 0 & \lambda_{58} & \lambda_{59} \\ 0 & 0 & 0 & 0 & 0 & -\gamma_6 & \lambda_{67} & 0 & \lambda_{69} \\ 0 & 0 & 0 & 0 & 0 & 0 & -\lambda_{79} & 0 & \lambda_{79} \\ 0 & 0 & 0 & 0 & 0 & 0 & 0 & -\lambda_{89} & \lambda_{89} \\ 0 & 0 & 0 & 0 & 0 & 0 & 0 & 0 & 0 \end{bmatrix}$$

$let$: $(\lambda_{12} + \lambda_{18} + \lambda_{19}) = \gamma_1$ , $(\lambda_{23} + \lambda_{28} + \lambda_{29} + \mu_{21}) = \gamma_2$ , $(\lambda_{34} + \lambda_{38} + \lambda_{39} + \mu_{32}) = \gamma_3$, $(\lambda_{45} + \lambda_{48} + \lambda_{49} + \mu_{43}) = \gamma_4$, $(\lambda_{56} + \lambda_{58} + \lambda_{59}) = \gamma_5$

$(\lambda_{67} + \lambda_{69}) = \gamma_6$

$$[\pi_1 \quad \pi_2 \quad \pi_3 \quad \pi_4 \quad \pi_5 \quad \pi_6 \quad \pi_7 \quad \pi_8 \quad \pi_9] \begin{bmatrix} -\gamma_1 & \lambda_{12} & 0 & 0 & 0 & 0 & 0 & \lambda_{18} & \lambda_{19} \\ \mu_{21} & -\gamma_2 & \lambda_{23} & 0 & 0 & 0 & 0 & \lambda_{28} & \lambda_{29} \\ 0 & \mu_{32} & -\gamma_3 & \lambda_{34} & 0 & 0 & 0 & \lambda_{38} & \lambda_{39} \\ 0 & 0 & \mu_{43} & -\gamma_4 & \lambda_{45} & 0 & 0 & \lambda_{48} & \lambda_{49} \\ 0 & 0 & 0 & 0 & -\gamma_5 & \lambda_{56} & 0 & \lambda_{58} & \lambda_{59} \\ 0 & 0 & 0 & 0 & 0 & -\gamma_6 & \lambda_{67} & 0 & \lambda_{69} \\ 0 & 0 & 0 & 0 & 0 & 0 & -\lambda_{79} & 0 & \lambda_{79} \\ 0 & 0 & 0 & 0 & 0 & 0 & 0 & -\lambda_{89} & \lambda_{89} \\ 0 & 0 & 0 & 0 & 0 & 0 & 0 & 0 & 0 \end{bmatrix} = [0 \quad 0 \quad 0 \quad 0 \quad 0 \quad 0 \quad 0 \quad 0 \quad 0]$$

That is to mean solve the following system of equations:

$-\pi_1\gamma_1 + \pi_2\mu_{21} = 0$ , $\pi_1\lambda_{12} - \pi_2\,\gamma_2 + \pi_3\,\mu_{32} = 0$ , $\pi_2\lambda_{23} - \pi_3\,\gamma_3 + \pi_4\,\mu_{43} = 0$ , $\pi_3\lambda_{34} - \pi_4\,\gamma_4 = 0$ , $\pi_4\lambda_{45} - \pi_5\,\gamma_5 = 0$

$\pi_5\lambda_{56} - \pi_6\,\gamma_6 = 0$ , $\pi_6\lambda_{67} - \pi_7\,\lambda_{79} = 0$, $\pi_1\lambda_{18} + \pi_2\lambda_{28} + \pi_3\lambda_{38} + \pi_4\lambda_{48} + \pi_5\lambda_{58} - \pi_8\lambda_{89} = 0$

$\pi_1\lambda_{19} + \pi_2\lambda_{29} + \pi_3\lambda_{39} + \pi_4\lambda_{49} + \pi_5\lambda_{59} + \pi_6\lambda_{69} + \pi_7\lambda_{79} + \pi_8\lambda_{89} = 0$

$subject\ to$ : $\pi_1 + \pi_2 + \pi_3 + \pi_4 + \pi_5 + \pi_6 + \pi_7 + \pi_8 + \pi_9 = 1$ ,

In matrix notation: $X\pi = y$

$$\begin{bmatrix} -\gamma_1 & \mu_{21} & 0 & 0 & 0 & 0 & 0 & 0 & 0 \\ \lambda_{12} & -\gamma_2 & \mu_{32} & 0 & 0 & 0 & 0 & 0 & 0 \\ 0 & \lambda_{23} & -\gamma_3 & \mu_{43} & 0 & 0 & 0 & 0 & 0 \\ 0 & 0 & \lambda_{34} & -\gamma_4 & 0 & 0 & 0 & 0 & 0 \\ 0 & 0 & 0 & \lambda_{45} & -\gamma_5 & 0 & 0 & 0 & 0 \\ 0 & 0 & 0 & 0 & \lambda_{56} & -\gamma_6 & 0 & 0 & 0 \\ 0 & 0 & 0 & 0 & 0 & \lambda_{67} & -\lambda_{79} & 0 & 0 \\ \lambda_{18} & \lambda_{28} & \lambda_{38} & \lambda_{48} & \lambda_{58} & 0 & 0 & -\lambda_{89} & 0 \\ \lambda_{19} & \lambda_{29} & \lambda_{39} & \lambda_{49} & \lambda_{59} & \lambda_{69} & \lambda_{79} & \lambda_{89} & 0 \\ 1 & 1 & 1 & 1 & 1 & 1 & 1 & 1 & 1 \end{bmatrix} \begin{bmatrix} \pi_1 \\ \pi_2 \\ \pi_3 \\ \pi_4 \\ \pi_5 \\ \pi_6 \\ \pi_7 \\ \pi_8 \\ \pi_9 \end{bmatrix} = \begin{bmatrix} 0 \\ 0 \\ 0 \\ 0 \\ 0 \\ 0 \\ 0 \\ 0 \\ 0 \\ 1 \end{bmatrix}$$

$$where \quad X = \begin{bmatrix} -\gamma_1 & \mu_{21} & 0 & 0 & 0 & 0 & 0 & 0 & 0 \\ \lambda_{12} & -\gamma_2 & \mu_{32} & 0 & 0 & 0 & 0 & 0 & 0 \\ 0 & \lambda_{23} & -\gamma_3 & \mu_{43} & 0 & 0 & 0 & 0 & 0 \\ 0 & 0 & \lambda_{34} & -\gamma_4 & 0 & 0 & 0 & 0 & 0 \\ 0 & 0 & 0 & \lambda_{45} & -\gamma_5 & 0 & 0 & 0 & 0 \\ 0 & 0 & 0 & 0 & \lambda_{56} & -\gamma_6 & 0 & 0 & 0 \\ 0 & 0 & 0 & 0 & 0 & \lambda_{67} & -\lambda_{79} & 0 & 0 \\ \lambda_{18} & \lambda_{28} & \lambda_{38} & \lambda_{48} & \lambda_{58} & 0 & 0 & -\lambda_{89} & 0 \\ \lambda_{19} & \lambda_{29} & \lambda_{39} & \lambda_{49} & \lambda_{59} & \lambda_{69} & \lambda_{79} & \lambda_{89} & 0 \\ 1 & 1 & 1 & 1 & 1 & 1 & 1 & 1 & 1 \end{bmatrix} \quad and \quad y = \begin{bmatrix} 0 \\ 0 \\ 0 \\ 0 \\ 0 \\ 0 \\ 0 \\ 0 \\ 0 \\ 1 \end{bmatrix}$$

$$X\pi = y \quad \rightarrow \quad \pi = (X^T X)^{-1} X^T y$$

To get the asymptotic covariance matrix of the state probability distribution, the derivative of the state probability distribution with respect to each parameter rate $\theta$ should be calculated as following:

$$F(\theta_h,\pi_i) = Q'\pi_i = 0$$

$$\frac{\partial}{\partial\theta}F(\theta_h,\pi_i) = \frac{\partial}{\partial\theta_h}(Q'\pi_i) = 0 \quad , \qquad with\ \ implicit\ differentiation \quad ,$$

$$\frac{\partial}{\partial\theta_h}F(\theta_h,\pi_i) = \frac{\partial}{\partial\theta_h}(Q'\pi_i) = [Q']\ \left[\frac{\partial}{\partial\theta_h}\pi_i\right] + \ \pi_i\left[\frac{\partial}{\partial\theta_h}Q'\right]^T \quad ,let's\ call \quad \pi_i\left[\frac{\partial}{\partial\theta_h}Q'\right]^T = C(\theta)\ \ is\ a\ matrix$$

$\left[\frac{\partial}{\partial\theta_h}\pi\right]$ *this is a matrix that gives all derivatives of* $\pi_1, \pi_2, \pi_3, \pi_4, \pi_5, \pi_6, \pi_7, \pi_8, \pi_9$ *with respect to each of the 22* $\theta$*'s*

$$\left[\frac{\partial}{\partial\theta_h}\pi\right] = -[Q']^{-1}C(\theta)\,, \quad \pi(\theta) = \begin{bmatrix}\pi_1\\ \pi_2\\ \pi_3\\ \pi_4\\ \pi_5\\ \pi_6\\ \pi_7\\ \pi_8\\ \pi_9\end{bmatrix}\ is\ a\ column\ vector\ ,and\ \ when\ t\ goes\ to\ infinity\ this\ vector = \ \pi(\theta) = \begin{bmatrix}0\\0\\0\\0\\0\\0\\0\\0\\1\end{bmatrix}$$

$$C(\theta) = \pi(\theta)\left[\frac{\partial}{\partial\theta_h}Q'\right]^T = \begin{bmatrix}\pi_1\\ \pi_2\\ \pi_3\\ \pi_4\\ \pi_5\\ \pi_6\\ \pi_7\\ \pi_8\\ \pi_9\end{bmatrix}[1 \quad 1 \quad \cdots \quad 1 \quad 1]\ \ ,where\ [1 \quad 1 \quad \cdots \quad 1 \quad 1]\ is\ a\ row\ vector\ of\ size\ \ (1\times 22)$$

$$\pi(\theta)\left[\frac{\partial}{\partial\theta_h}Q'\right]^T = C(\theta)\ \ is\ a\ matrix\ of\ \ size\ \ (9\ \times\ 22)$$

*$Q'$ is a singular matrix , and its inverse requires calculating the pseudoinverse using SVD*

$[Q']^{-1}$ *obtained by pseudoinverse is 9 by 9 matrix*

$$let: \qquad A(\theta) = \left[\frac{\partial}{\partial\theta_h}\pi_i\right] = -[Q']^{-1}C(\theta)\ \ \ is\ \ (9\ \times\ 22)\ matrix$$

$A(\theta)$

$$= \begin{bmatrix}
\frac{\partial\pi_1}{\partial\lambda_{12}} & \frac{\partial\pi_1}{\partial\lambda_{18}} & \frac{\partial\pi_1}{\partial\lambda_{19}} & \frac{\partial\pi_1}{\partial\lambda_{23}} & \frac{\partial\pi_1}{\partial\lambda_{28}} & \frac{\partial\pi_1}{\partial\lambda_{29}} & \frac{\partial\pi_1}{\partial\mu_{21}} & \frac{\partial\pi_1}{\partial\lambda_{34}} & \frac{\partial\pi_1}{\partial\lambda_{38}} & \frac{\partial\pi_1}{\partial\lambda_{39}} & \frac{\partial\pi_1}{\partial\mu_{32}} & \frac{\partial\pi_1}{\partial\lambda_{45}} & \frac{\partial\pi_1}{\partial\lambda_{48}} & \frac{\partial\pi_1}{\partial\lambda_{49}} & \frac{\partial\pi_1}{\partial\mu_{43}} & \frac{\partial\pi_1}{\partial\lambda_{56}} & \frac{\partial\pi_1}{\partial\lambda_{58}} & \frac{\partial\pi_1}{\partial\lambda_{59}} & \frac{\partial\pi_1}{\partial\lambda_{67}} & \frac{\partial\pi_1}{\partial\lambda_{69}} & \frac{\partial\pi_1}{\partial\lambda_{79}} & \frac{\partial\pi_1}{\partial\lambda_{89}} \\
\frac{\partial\pi_2}{\partial\lambda_{12}} & \frac{\partial\pi_2}{\partial\lambda_{18}} & \frac{\partial\pi_2}{\partial\lambda_{19}} & \frac{\partial\pi_2}{\partial\lambda_{23}} & \frac{\partial\pi_2}{\partial\lambda_{28}} & \frac{\partial\pi_2}{\partial\lambda_{29}} & \frac{\partial\pi_2}{\partial\mu_{21}} & \frac{\partial\pi_2}{\partial\lambda_{34}} & \frac{\partial\pi_2}{\partial\lambda_{38}} & \frac{\partial\pi_2}{\partial\lambda_{39}} & \frac{\partial\pi_2}{\partial\mu_{32}} & \frac{\partial\pi_2}{\partial\lambda_{45}} & \frac{\partial\pi_2}{\partial\lambda_{48}} & \frac{\partial\pi_2}{\partial\lambda_{49}} & \frac{\partial\pi_2}{\partial\mu_{43}} & \frac{\partial\pi_2}{\partial\lambda_{56}} & \frac{\partial\pi_2}{\partial\lambda_{58}} & \frac{\partial\pi_2}{\partial\lambda_{59}} & \frac{\partial\pi_2}{\partial\lambda_{67}} & \frac{\partial\pi_2}{\partial\lambda_{69}} & \frac{\partial\pi_2}{\partial\lambda_{79}} & \frac{\partial\pi_2}{\partial\lambda_{89}} \\
\frac{\partial\pi_3}{\partial\lambda_{12}} & \frac{\partial\pi_3}{\partial\lambda_{18}} & \frac{\partial\pi_3}{\partial\lambda_{19}} & \frac{\partial\pi_3}{\partial\lambda_{23}} & \frac{\partial\pi_3}{\partial\lambda_{28}} & \frac{\partial\pi_3}{\partial\lambda_{29}} & \frac{\partial\pi_3}{\partial\mu_{21}} & \frac{\partial\pi_3}{\partial\lambda_{34}} & \frac{\partial\pi_3}{\partial\lambda_{38}} & \frac{\partial\pi_3}{\partial\lambda_{39}} & \frac{\partial\pi_3}{\partial\mu_{32}} & \frac{\partial\pi_3}{\partial\lambda_{45}} & \frac{\partial\pi_3}{\partial\lambda_{48}} & \frac{\partial\pi_3}{\partial\lambda_{49}} & \frac{\partial\pi_3}{\partial\mu_{43}} & \frac{\partial\pi_3}{\partial\lambda_{56}} & \frac{\partial\pi_3}{\partial\lambda_{58}} & \frac{\partial\pi_3}{\partial\lambda_{59}} & \frac{\partial\pi_3}{\partial\lambda_{67}} & \frac{\partial\pi_3}{\partial\lambda_{69}} & \frac{\partial\pi_3}{\partial\lambda_{79}} & \frac{\partial\pi_3}{\partial\lambda_{89}} \\
\frac{\partial\pi_4}{\partial\lambda_{12}} & \frac{\partial\pi_4}{\partial\lambda_{18}} & \frac{\partial\pi_4}{\partial\lambda_{19}} & \frac{\partial\pi_4}{\partial\lambda_{23}} & \frac{\partial\pi_4}{\partial\lambda_{28}} & \frac{\partial\pi_4}{\partial\lambda_{29}} & \frac{\partial\pi_4}{\partial\mu_{21}} & \frac{\partial\pi_4}{\partial\lambda_{34}} & \frac{\partial\pi_4}{\partial\lambda_{38}} & \frac{\partial\pi_4}{\partial\lambda_{39}} & \frac{\partial\pi_4}{\partial\mu_{32}} & \frac{\partial\pi_4}{\partial\lambda_{45}} & \frac{\partial\pi_4}{\partial\lambda_{48}} & \frac{\partial\pi_4}{\partial\lambda_{49}} & \frac{\partial\pi_4}{\partial\mu_{43}} & \frac{\partial\pi_4}{\partial\lambda_{56}} & \frac{\partial\pi_4}{\partial\lambda_{58}} & \frac{\partial\pi_4}{\partial\lambda_{59}} & \frac{\partial\pi_4}{\partial\lambda_{67}} & \frac{\partial\pi_4}{\partial\lambda_{69}} & \frac{\partial\pi_4}{\partial\lambda_{79}} & \frac{\partial\pi_4}{\partial\lambda_{89}} \\
\frac{\partial\pi_5}{\partial\lambda_{12}} & \frac{\partial\pi_5}{\partial\lambda_{18}} & \frac{\partial\pi_5}{\partial\lambda_{19}} & \frac{\partial\pi_5}{\partial\lambda_{23}} & \frac{\partial\pi_5}{\partial\lambda_{28}} & \frac{\partial\pi_5}{\partial\lambda_{29}} & \frac{\partial\pi_5}{\partial\mu_{21}} & \frac{\partial\pi_5}{\partial\lambda_{34}} & \frac{\partial\pi_5}{\partial\lambda_{38}} & \frac{\partial\pi_5}{\partial\lambda_{39}} & \frac{\partial\pi_5}{\partial\mu_{32}} & \frac{\partial\pi_5}{\partial\lambda_{45}} & \frac{\partial\pi_5}{\partial\lambda_{48}} & \frac{\partial\pi_5}{\partial\lambda_{49}} & \frac{\partial\pi_5}{\partial\mu_{43}} & \frac{\partial\pi_5}{\partial\lambda_{56}} & \frac{\partial\pi_5}{\partial\lambda_{58}} & \frac{\partial\pi_5}{\partial\lambda_{59}} & \frac{\partial\pi_5}{\partial\lambda_{67}} & \frac{\partial\pi_5}{\partial\lambda_{69}} & \frac{\partial\pi_5}{\partial\lambda_{79}} & \frac{\partial\pi_5}{\partial\lambda_{89}} \\
\frac{\partial\pi_6}{\partial\lambda_{12}} & \frac{\partial\pi_6}{\partial\lambda_{18}} & \frac{\partial\pi_6}{\partial\lambda_{19}} & \frac{\partial\pi_6}{\partial\lambda_{23}} & \frac{\partial\pi_6}{\partial\lambda_{28}} & \frac{\partial\pi_6}{\partial\lambda_{29}} & \frac{\partial\pi_6}{\partial\mu_{21}} & \frac{\partial\pi_6}{\partial\lambda_{34}} & \frac{\partial\pi_6}{\partial\lambda_{38}} & \frac{\partial\pi_6}{\partial\lambda_{39}} & \frac{\partial\pi_6}{\partial\mu_{32}} & \frac{\partial\pi_6}{\partial\lambda_{45}} & \frac{\partial\pi_6}{\partial\lambda_{48}} & \frac{\partial\pi_6}{\partial\lambda_{49}} & \frac{\partial\pi_6}{\partial\mu_{43}} & \frac{\partial\pi_6}{\partial\lambda_{56}} & \frac{\partial\pi_6}{\partial\lambda_{58}} & \frac{\partial\pi_6}{\partial\lambda_{59}} & \frac{\partial\pi_6}{\partial\lambda_{67}} & \frac{\partial\pi_6}{\partial\lambda_{69}} & \frac{\partial\pi_6}{\partial\lambda_{79}} & \frac{\partial\pi_6}{\partial\lambda_{89}} \\
\frac{\partial\pi_7}{\partial\lambda_{12}} & \frac{\partial\pi_7}{\partial\lambda_{18}} & \frac{\partial\pi_7}{\partial\lambda_{19}} & \frac{\partial\pi_7}{\partial\lambda_{23}} & \frac{\partial\pi_7}{\partial\lambda_{28}} & \frac{\partial\pi_7}{\partial\lambda_{29}} & \frac{\partial\pi_7}{\partial\mu_{21}} & \frac{\partial\pi_7}{\partial\lambda_{34}} & \frac{\partial\pi_7}{\partial\lambda_{38}} & \frac{\partial\pi_7}{\partial\lambda_{39}} & \frac{\partial\pi_7}{\partial\mu_{32}} & \frac{\partial\pi_7}{\partial\lambda_{45}} & \frac{\partial\pi_7}{\partial\lambda_{48}} & \frac{\partial\pi_7}{\partial\lambda_{49}} & \frac{\partial\pi_7}{\partial\mu_{43}} & \frac{\partial\pi_7}{\partial\lambda_{56}} & \frac{\partial\pi_7}{\partial\lambda_{58}} & \frac{\partial\pi_7}{\partial\lambda_{59}} & \frac{\partial\pi_7}{\partial\lambda_{67}} & \frac{\partial\pi_7}{\partial\lambda_{69}} & \frac{\partial\pi_7}{\partial\lambda_{79}} & \frac{\partial\pi_7}{\partial\lambda_{89}} \\
\frac{\partial\pi_8}{\partial\lambda_{12}} & \frac{\partial\pi_8}{\partial\lambda_{18}} & \frac{\partial\pi_8}{\partial\lambda_{19}} & \frac{\partial\pi_8}{\partial\lambda_{23}} & \frac{\partial\pi_8}{\partial\lambda_{28}} & \frac{\partial\pi_8}{\partial\lambda_{29}} & \frac{\partial\pi_8}{\partial\mu_{21}} & \frac{\partial\pi_8}{\partial\lambda_{34}} & \frac{\partial\pi_8}{\partial\lambda_{38}} & \frac{\partial\pi_8}{\partial\lambda_{39}} & \frac{\partial\pi_8}{\partial\mu_{32}} & \frac{\partial\pi_8}{\partial\lambda_{45}} & \frac{\partial\pi_8}{\partial\lambda_{48}} & \frac{\partial\pi_8}{\partial\lambda_{49}} & \frac{\partial\pi_8}{\partial\mu_{43}} & \frac{\partial\pi_8}{\partial\lambda_{56}} & \frac{\partial\pi_8}{\partial\lambda_{58}} & \frac{\partial\pi_8}{\partial\lambda_{59}} & \frac{\partial\pi_8}{\partial\lambda_{67}} & \frac{\partial\pi_8}{\partial\lambda_{69}} & \frac{\partial\pi_8}{\partial\lambda_{79}} & \frac{\partial\pi_8}{\partial\lambda_{89}} \\
\frac{\partial\pi_9}{\partial\lambda_{12}} & \frac{\partial\pi_9}{\partial\lambda_{18}} & \frac{\partial\pi_9}{\partial\lambda_{19}} & \frac{\partial\pi_9}{\partial\lambda_{23}} & \frac{\partial\pi_9}{\partial\lambda_{28}} & \frac{\partial\pi_9}{\partial\lambda_{29}} & \frac{\partial\pi_9}{\partial\mu_{21}} & \frac{\partial\pi_9}{\partial\lambda_{34}} & \frac{\partial\pi_9}{\partial\lambda_{38}} & \frac{\partial\pi_9}{\partial\lambda_{39}} & \frac{\partial\pi_9}{\partial\mu_{32}} & \frac{\partial\pi_9}{\partial\lambda_{45}} & \frac{\partial\pi_9}{\partial\lambda_{48}} & \frac{\partial\pi_9}{\partial\lambda_{49}} & \frac{\partial\pi_9}{\partial\mu_{43}} & \frac{\partial\pi_9}{\partial\lambda_{56}} & \frac{\partial\pi_9}{\partial\lambda_{58}} & \frac{\partial\pi_9}{\partial\lambda_{59}} & \frac{\partial\pi_9}{\partial\lambda_{67}} & \frac{\partial\pi_9}{\partial\lambda_{69}} & \frac{\partial\pi_9}{\partial\lambda_{79}} & \frac{\partial\pi_9}{\partial\lambda_{89}}
\end{bmatrix}$$

Using multivariate delta method

$$var(\pi_i) = A(\theta)var(\theta)A(\theta)' \quad , \qquad where\ \ var(\theta) = [M(\theta)]^{-1}\ \ and \quad i = 1,2,\cdots,9.$$

### 5.5. Expected Number of Patients in Each State:

$$[m_1(0) \quad m_2(0) \quad m_3(0) \quad m_4(0) \quad m_5(0) \quad m_6(0) \quad m_7(0) \quad m_8(0) \quad m_9(0)] \begin{bmatrix} P_{11} & P_{12} & P_{13} & P_{14} & P_{15} & P_{16} & P_{17} & P_{18} & P_{19} \\ P_{21} & P_{22} & P_{23} & P_{24} & P_{25} & P_{26} & P_{27} & P_{28} & P_{29} \\ P_{31} & P_{32} & P_{33} & P_{34} & P_{35} & P_{36} & P_{37} & P_{38} & P_{39} \\ P_{41} & P_{42} & P_{43} & P_{44} & P_{45} & P_{46} & P_{47} & P_{48} & P_{49} \\ 0 & 0 & 0 & 0 & P_{55} & P_{56} & P_{57} & P_{58} & P_{59} \\ 0 & 0 & 0 & 0 & 0 & P_{66} & P_{67} & 0 & P_{69} \\ 0 & 0 & 0 & 0 & 0 & 0 & P_{77} & 0 & P_{79} \\ 0 & 0 & 0 & 0 & 0 & 0 & 0 & P_{88} & P_{89} \\ 0 & 0 & 0 & 0 & 0 & 0 & 0 & 0 & P_{99} \end{bmatrix}$$

$$= [m_1(t) \quad m_2(t) \quad m_3(t) \quad m_4(t) \quad m_5(t) \quad m_6(t) \quad m_7(t) \quad m_8(t) \quad m_9(t)]$$

### 5.6. Life Expectancy of a Patient Suffering from NAFLD in Various Stages:

The same procedure used in the previous chapter is used in this chapter.

Partitioning the differential equation into the following :

$$[\acute{P}(t) \quad \acute{P}_k(t)] = [P(t) \quad P_k(t)] \begin{bmatrix} B & A \\ 0 & 0 \end{bmatrix}$$

B is the transition rate matrix among the transient states and the column vector A is the transition rate from each transient state to the absorbing (death) state .

$$A = -B1^T \quad such\ that\ 1^T\ is\ a\ column\ vector\ of\ k-1\ \times 1\ with\ all\ its\ elements\ equal\ to\ one$$

$$Mean\ time\ to\ absorption\text{:}\ E(\tau_k) = (-1)\frac{df^*{}_k(s)}{ds}\bigg|_{s=0} = (-1)P(0)[sI-B]^{-2}A|_{s=0} = P(0)[-B]^{-1}\,1^T$$

$$B = \begin{bmatrix} -\gamma_1 & \lambda_{12} & 0 & 0 & 0 & 0 & 0 & \lambda_{18} \\ \mu_{21} & -\gamma_2 & \lambda_{23} & 0 & 0 & 0 & 0 & \lambda_{28} \\ 0 & \mu_{32} & -\gamma_3 & \lambda_{34} & 0 & 0 & 0 & \lambda_{38} \\ 0 & 0 & \mu_{43} & -\gamma_4 & \lambda_{45} & 0 & 0 & \lambda_{48} \\ 0 & 0 & 0 & 0 & -\gamma_5 & \lambda_{56} & 0 & \lambda_{58} \\ 0 & 0 & 0 & 0 & 0 & -\gamma_6 & \lambda_{67} & 0 \\ 0 & 0 & 0 & 0 & 0 & 0 & -\lambda_{79} & 0 \\ 0 & 0 & 0 & 0 & 0 & 0 & 0 & -\lambda_{89} \end{bmatrix}$$

Steps:

1. First: specify the Q matrix
2. Second: remove the last column and the last row from Q matrix to obtain B matrix
3. Third: get the inverse of the B matrix
4. Forth: multiply the inverse by -1
5. Lastly and fifth: Apply the following formula of mean time to absorption

### 5.7. Hypothetical Model:

To illustrate the above concepts and discussion, a hypothetical numerical example is introduced. It does not represent real data but it is for demonstrative purposes

A study was conducted over 15 years on 1050 patients with risk factors for developing NAFLD such as type 2 diabetes mellitus, obesity, and hypertension acting alone or together as a metabolic syndrome. The patients were decided to be followed up every year by a liver biopsy to identify the NAFLD cases, but the actual observations were recorded with different intervals. The following is the final estimated rate matrix and its variance , this is followed by elaborate discussion of the steps; the estimated final transition rate matrix "Q" is :

$$\hat{Q} = \begin{bmatrix} -.397 & .39 & 0 & 0 & 0 & 0 & 0 & 0 & .007 \\ .02 & -.281 & .25 & 0 & 0 & 0 & 0 & 0 & .011 \\ 0 & .05 & -.365 & .225 & 0 & 0 & 0 & .047 & .043 \\ 0 & 0 & .041 & -.538 & .281 & 0 & 0 & .109 & .107 \\ 0 & 0 & 0 & 0 & -.348 & .19 & 0 & .059 & .099 \\ 0 & 0 & 0 & 0 & 0 & -.934 & .767 & 0 & .167 \\ 0 & 0 & 0 & 0 & 0 & 0 & -.421 & 0 & .421 \\ 0 & 0 & 0 & 0 & 0 & 0 & 0 & -.745 & .745 \\ 0 & 0 & 0 & 0 & 0 & 0 & 0 & 0 & 0 \end{bmatrix}$$

$var(\hat{\theta}) = 1 \times 10^{-13} \begin{bmatrix} v_1 & v_2 \\ v_3 & v_4 \end{bmatrix}$ where

$$v_1 = \begin{bmatrix} .3292 & .0327 & .4414 & .0827 & .4004 & .0268 & .2540 & .1517 \\ .0327 & .0064 & .0428 & .0108 & .0385 & .0053 & .0268 & .0159 \\ .4414 & .0428 & .5922 & .1100 & .5373 & .0350 & .3400 & .2032 \\ .0827 & .0108 & .1100 & .0228 & .0995 & .0088 & .0650 & .0387 \\ .4004 & .0385 & .5373 & .0995 & .4876 & .0315 & .3082 & .1843 \\ .0268 & .0053 & .0350 & .0088 & .0315 & .0044 & .0219 & .0130 \\ .2540 & .0268 & .3400 & .0650 & .3082 & .0219 & .1967 & .1174 \\ .1517 & .0159 & .2032 & .0387 & .1843 & .0130 & .1174 & .0702 \end{bmatrix}$$

$v_2$ , $v_3$ $and$ $v_4$ are all zero matrices of size ( 8 by 14 ) , ( 14 by 8 ) and ( 14 by 14 ) respectively.

The estimated transition rate matrix Q is calculated utilizing procedure that is similar to the one used in the small model, as shown below in the following steps:

For each Δt , the observed transition counts in this interval are obtained, then applying the successive steps to get the estimated rate matrix:

**Step 1**: calculate the eigenvalues for the initial Q matrix obtained from the observed transition counts in this interval.

**Step 2**: calculate $te^{\Lambda t} d\,\Lambda$= t * e^(eigenvalue *t)* partial derivative of each eigenvalues with each rate or theta to get score function.

**Step 3:** rearrange the score function, then scale it .

**Step 4**: multiply the rearranged scaled scored function with the transposed rearranged scaled score function to get the hessian matrix (22 by 22 matrix).

**Step 5**: scale the above hessian matrix, then resultant matrix can be partitioned into 4 matrices.

**Step 6**: invert the scaled hessian matrix (only the upper left is invertible).

**Step 7**: multiply the inverted scaled hessian matrix by the scaled score function.

**Step 8**: apply quasi-newton formula.

| Observed counts of transitions during time interval Δt=1 | | | | | | | | | | |
|---|---|---|---|---|---|---|---|---|---|---|
| | **State 1** | **State 2** | **State3** | **State 4** | **State 5** | **State 6** | **State 7** | **State 8** | **State 9** | **total** |
| **State 1** | 784 | 573 | 74 | 21 | 20 | 0 | 0 | 0 | 10 | 1482 |
| **State 2** | 9 | 333 | 130 | 21 | 19 | 0 | 0 | 0 | 6 | 518 |
| **State 3** | 4 | 11 | 103 | 44 | 19 | 0 | 0 | 10 | 9 | 200 |
| **State 4** | 0 | 0 | 4 | 33 | 21 | 0 | 0 | 8 | 7 | 73 |
| **State 5** | 0 | 0 | 0 | 0 | 35 | 12 | 10 | 4 | 6 | 67 |
| **State 6** | 0 | 0 | 0 | 0 | 0 | 1 | 8 | 0 | 1 | 10 |
| **State 7** | 0 | 0 | 0 | 0 | 0 | 0 | 10 | 0 | 7 | 17 |
| **State 8** | 0 | 0 | 0 | 0 | 0 | 0 | 0 | 5 | 15 | 20 |
| **State 9** | 0 | 0 | 0 | 0 | 0 | 0 | 0 | 0 | 0 | 0 |
| | | | | | | | | | | 2387 |

| Initial Q matrix : | | | | | | | | | Step 1 : calculate eigenvalues for this Q matrix : | |
|---|---|---|---|---|---|---|---|---|---|---|
| -.397 | .39 | 0 | 0 | 0 | 0 | 0 | 0 | .007 | **Eigenvalue 1** | -.45792 |
| .02 | -.28 | .25 | 0 | 0 | 0 | 0 | 0 | .01 | **Eigenvalue 2** | -.58993 |
| 0 | .05 | -.36 | .22 | 0 | 0 | 0 | .05 | .04 | **Eigenvalue 3** | -.16836 |
| 0 | 0 | .05 | -.53 | .28 | 0 | 0 | .11 | .09 | **Eigenvalue 4** | -.35079 |
| 0 | 0 | 0 | 0 | -.33 | .18 | 0 | .06 | .09 | **Eigenvalue 5** | -.33 |
| 0 | 0 | 0 | 0 | 0 | -.9 | .8 | 0 | .1 | **Eigenvalue 6** | -.9 |
| 0 | 0 | 0 | 0 | 0 | 0 | -.41 | 0 | .41 | **Eigenvalue 7** | -.41 |
| 0 | 0 | 0 | 0 | 0 | 0 | 0 | -.75 | .75 | **Eigenvalue 8** | -.75 |
| 0 | 0 | 0 | 0 | 0 | 0 | 0 | 0 | 0 | **Eigenvalue 9** | 0 |

| Step 2 : calculate $te^{\Lambda t} d\,\Lambda$ | | Step 3 : rearrange the score function then scale it by a factor =14204 | | |
|---|---|---|---|---|
| **Lambda 12** | -.6608 | **Lambda 12** | -.6608 | -9385 |
| **Lambda 18** | -.6751 | **Mu 21** | -.4841 | -6876 |
| **Lambda 19** | -.6751 | **Lambda 23** | -.7267 | -10323 |
| **Lambda 23** | -.7267 | **Mu 32** | -.5233 | -7433 |
| **Lambda 28** | -.7632 | **Lambda 34** | -.6736 | -9568 |
| **Lambda 29** | -.7632 | **Mu 43** | -.4503 | -6396 |
| **Mu 21** | -.4841 | **Lambda 79** | -.6637 | -9426 |
| **Lambda 34** | -.6736 | **Lambda 89** | -.4724 | -6709 |
| **Lambda 38** | -.7058 | **Lambda 18** | -.6751 | -9589 |
| **Lambda 39** | -.7058 | **Lambda 19** | -.6751 | -9589 |
| **Mu 32** | -.5233 | **Lambda 28** | -.7632 | -10841 |
| **Lambda 45** | -.592 | **Lambda 29** | -.7632 | -10841 |
| **Lambda 48** | -.592 | **Lambda 38** | -.7058 | -10025 |
| **Lambda 49** | -.592 | **Lambda 39** | -.7058 | -10025 |
| **Mu 43** | -.4503 | **Lambda 45** | -.5920 | -8409 |
| **Lambda 56** | -.7189 | **Lambda 48** | -.5920 | -8409 |
| **Lambda 58** | -.7189 | **Lambda 49** | -.5920 | -8409 |
| **Lambda 59** | -.7189 | **Lambda 56** | -.7189 | -10212 |
| **Lambda 67** | -.4066 | **Lambda 58** | -.7189 | -10212 |
| **Lambda 69** | -.4066 | **Lambda 59** | -.7189 | -10212 |
| **Lambda 79** | -.6637 | **Lambda 67** | -.4066 | -5775 |
| **Lambda 89** | -.4724 | **Lambda 69** | -.4066 | -5775 |

**Step 4**: multiply the rearranged scaled scored function with the transposed rearranged scaled score function to get the hessian matrix (22 by 22 matrix)

**Step 5**: scale the above hessian matrix by a factor of 606523.3, the resultant matrix can be partitioned into 4 matrices, the upper left matrix of size (8 by 8) is the matrix needs to be inverted but it may be singular or close to be singular so the pseudo-inverse using singular value decomposition is applied. The lower right matrix is (14 by 14) has repeated rows. The matrices on the secondary diagonal, the upper right (8 by 14) and the lower left (14 by 8) have repeated rows so the only invertible matrix is the upper left.

| Step 6: invert the scaled hessian matrix ( only the upper left is invertible) | | | | | | | |
|---|---|---|---|---|---|---|---|
| $10^{-15}\times$ | | | | | | | |
| .4574 | .3351 | .503 | .3622 | .4663 | .3117 | .4594 | .327 |
| .3351 | .2455 | .3686 | .2654 | .3416 | .2284 | .3366 | .2396 |
| .503 | .3686 | .5533 | .3984 | .5128 | .3428 | .5052 | .3596 |
| .3622 | .2654 | .3984 | .2869 | .3693 | .2469 | .3638 | .2589 |
| .4663 | .3416 | .5128 | .3693 | .4753 | .3178 | .4683 | .3333 |
| .3117 | .2284 | .3428 | .2469 | .3178 | .2124 | .3131 | .2228 |
| .4594 | .3366 | .5052 | .3638 | .4683 | .3131 | .4614 | .3284 |
| .327 | .2396 | .3596 | .2589 | .3333 | .2228 | .3284 | .2337 |

| Step 7 : multiply inverted hessian matrix by scaled score function | Step 8: apply quasi-Newton formula to get the estimated rates in Δt=1 | |
|---|---|---|
| $-.2746 * 10^{-10}$ | **Lambda 12** | .39 |
| $-.2012 * 10^{-10}$ | **Mu 21** | .02 |
| $-.3020 * 10^{-10}$ | **Lambda 23** | .25 |
| $-.2175 * 10^{-10}$ | **Mu 32** | .05 |
| $-.2799 * 10^{-10}$ | **Lambda 34** | .22 |
| $-.1871 * 10^{-10}$ | **Mu 43** | .05 |
| $-.2758 * 10^{-10}$ | **Lambda 79** | .41 |
| $-.1963 * 10^{-10}$ | **Lambda 89** | .75 |
| 0 | **Lambda 18** | 0 |
| 0 | **Lambda 19** | .007 |
| 0 | **Lambda 28** | 0 |
| 0 | **Lambda 29** | .01 |
| 0 | **Lambda 38** | .05 |
| 0 | **Lambda 39** | .04 |
| 0 | **Lambda 45** | .28 |
| 0 | **Lambda 48** | .11 |
| 0 | **Lambda 49** | .09 |
| 0 | **Lambda 56** | .18 |
| 0 | **Lambda 58** | .06 |
| 0 | **Lambda 59** | .09 |
| 0 | **Lambda 67** | .8 |
| 0 | **Lambda 69** | .1 |

The same steps are performed for the transitions occurred in time interval =2：

| **Observed counts of transitions during time interval Δt=2** | | | | | | | | | | |
|---|---|---|---|---|---|---|---|---|---|---|
| | **State 1** | **State 2** | **State3** | **State 4** | **State 5** | **State 6** | **State 7** | **State 8** | **State 9** | **total** |
| **State 1** | 313 | 229 | 30 | 9 | 8 | 0 | 0 | 0 | 4 | 593 |
| **State 2** | 4 | 133 | 52 | 8 | 7 | 0 | 0 | 0 | 3 | 207 |
| **State 3** | 2 | 4 | 40 | 19 | 8 | 0 | 0 | 3 | 4 | 80 |
| **State 4** | 0 | 0 | 1 | 13 | 8 | 0 | 0 | 3 | 4 | 29 |
| **State 5** | 0 | 0 | 0 | 0 | 13 | 5 | 4 | 2 | 3 | 27 |
| **State 6** | 0 | 0 | 0 | 0 | 0 | 0 | 3 | 0 | 1 | 4 |
| **State 7** | 0 | 0 | 0 | 0 | 0 | 0 | 4 | 0 | 3 | 7 |
| **State 8** | 0 | 0 | 0 | 0 | 0 | 0 | 0 | 2 | 6 | 8 |
| **State 9** | 0 | 0 | 0 | 0 | 0 | 0 | 0 | 0 | 0 | 0 |
| | | | | | | | | | | 955 |

| **Initial Q matrix :** | | | | | | | | | **Step 1 : calculate eigenvalues for this Q matrix :** | |
|---|---|---|---|---|---|---|---|---|---|---|
| -.397 | .39 | 0 | 0 | 0 | 0 | 0 | 0 | .007 | **Eigenvalue 1** | -.46595 |
| .02 | -.28 | .25 | 0 | 0 | 0 | 0 | 0 | .01 | **Eigenvalue 2** | -.59304 |
| 0 | .05 | -.38 | .24 | 0 | 0 | 0 | .04 | .05 | **Eigenvalue 3** | -.17668 |
| 0 | 0 | .03 | -.55 | .28 | 0 | 0 | .1 | .14 | **Eigenvalue 4** | -.37133 |
| 0 | 0 | 0 | 0 | -.37 | .19 | 0 | .07 | .11 | **Eigenvalue 5** | -.37 |
| 0 | 0 | 0 | 0 | 0 | -1 | .75 | 0 | .25 | **Eigenvalue 6** | -1 |
| 0 | 0 | 0 | 0 | 0 | 0 | -.43 | 0 | .43 | **Eigenvalue 7** | -.43 |
| 0 | 0 | 0 | 0 | 0 | 0 | 0 | -.75 | .75 | **Eigenvalue 8** | -.75 |
| 0 | 0 | 0 | 0 | 0 | 0 | 0 | 0 | 0 | **Eigenvalue 9** | 0 |

| Step 2 : calculate $te^{\Lambda t}d\ \Lambda$ | | Step 3 : rearrange the score function then scale it by a factor =5678 | | |
|---|---|---|---|---|
| **Lambda 12** | -.8781 | **Lambda 12** | -.8781 | -4986.1 |
| **Lambda 18** | -.9194 | **Mu 21** | -.3807 | -2161.4 |
| **Lambda 19** | -.9194 | **Lambda 23** | -1.0806 | -6135.9 |
| **Lambda 23** | -1.0806 | **Mu 32** | -.4458 | -2531.3 |
| **Lambda 28** | -1.186 | **Lambda 34** | -.9245 | -5249.1 |
| **Lambda 29** | -1.186 | **Mu 43** | -.2905 | -1649.6 |
| **Mu 21** | -.3807 | **Lambda 79** | -.8463 | -4805.4 |
| **Lambda 34** | -.9245 | **Lambda 89** | -.4463 | -2533.9 |
| **Lambda 38** | -.9727 | **Lambda 18** | -.9194 | -5220.6 |
| **Lambda 39** | -.9727 | **Lambda 19** | -.9194 | -5220.6 |
| **Mu 32** | -.4458 | **Lambda 28** | -1.186 | -6734.2 |
| **Lambda 45** | -.6766 | **Lambda 29** | -1.186 | -6734.2 |
| **Lambda 48** | -.6766 | **Lambda 38** | -.9727 | -5523.2 |
| **Lambda 49** | -.6766 | **Lambda 39** | -.9727 | -5523.2 |
| **Mu 43** | -.2905 | **Lambda 45** | -.6766 | -3841.7 |
| **Lambda 56** | -.9542 | **Lambda 48** | -.6766 | -3841.7 |
| **Lambda 58** | -.9542 | **Lambda 49** | -.6766 | -3841.7 |
| **Lambda 59** | -.9542 | **Lambda 56** | -.9542 | -5418.1 |
| **Lambda 67** | -.2707 | **Lambda 58** | -.9542 | -5418.1 |
| **Lambda 69** | -.2707 | **Lambda 59** | -.9542 | -5418.1 |
| **Lambda 79** | -.8463 | **Lambda 67** | -.2707 | -1536.9 |
| **Lambda 89** | -.4463 | **Lambda 69** | -.2707 | -1536.9 |

**Step 4**: multiply the rearranged scaled scored function with the transposed rearranged scaled score function to get the hessian matrix (22 by 22 matrix)

**Step 5**: scale the above hessian matrix by a factor of 235355.8, the resultant matrix can be partitioned into 4 matrices, the upper left matrix of size (8 by 8) is the matrix needs to be inverted but it may be singular or close to be singular so the pseudo-inverse using singular value decomposition is applied. The lower right matrix is (14 by 14) has repeated rows. The matrices on the secondary diagonal, the upper right (8 by 14) and the lower left (14 by 8) have repeated rows so the only invertible matrix is the upper left.

| Step 6: invert the scaled hessian matrix ( only the upper left is invertible) | | | | | | | |
|---|---|---|---|---|---|---|---|
| $10^{-14}\times$ | | | | | | | |
| .5938 | .2574 | .7307 | .3015 | .6251 | .1964 | .5723 | .3018 |
| .2574 | .1116 | .3168 | .1307 | .271 | .0852 | .2481 | .1308 |
| .7307 | .3168 | .8992 | .371 | .7693 | .2417 | .7042 | .3713 |
| .3015 | .1307 | .371 | .153 | .3174 | .0997 | .2905 | .1532 |
| .6251 | .271 | .7693 | .3174 | .6581 | .2068 | .6025 | .3177 |
| .1964 | .0852 | .2417 | .0997 | .2068 | .065 | .1893 | .0998 |
| .5723 | .2481 | .7042 | .2905 | .6025 | .1893 | .5515 | .2908 |
| .3018 | .1308 | .3713 | .1532 | .3177 | .0998 | .2908 | .1533 |

| Step 7 : multiply inverted hessian matrix by scaled score function | Step 8: apply quasi-Newton formula to get the estimated rates in Δt=2 | |
|---|---|---|
| $-.1588 * 10^{-9}$ | **Lambda 12** | .39 |
| $-.0689 * 10^{-9}$ | **Mu 21** | .02 |
| $-.1955 * 10^{-9}$ | **Lambda 23** | .25 |
| $-.0806 * 10^{-9}$ | **Mu 32** | .05 |
| $-.1672 * 10^{-9}$ | **Lambda 34** | .24 |
| $-.0525 * 10^{-9}$ | **Mu 43** | .03 |
| $-.1531 * 10^{-9}$ | **Lambda 79** | .43 |
| $-.0807 * 10^{-9}$ | **Lambda 89** | .75 |
| 0 | **Lambda 18** | 0 |
| 0 | **Lambda 19** | .007 |
| 0 | **Lambda 28** | 0 |
| 0 | **Lambda 29** | .01 |
| 0 | **Lambda 38** | .04 |
| 0 | **Lambda 39** | .05 |
| 0 | **Lambda 45** | .28 |
| 0 | **Lambda 48** | .1 |
| 0 | **Lambda 49** | .14 |
| 0 | **Lambda 56** | .19 |
| 0 | **Lambda 58** | .07 |
| 0 | **Lambda 59** | .11 |
| 0 | **Lambda 67** | .75 |
| 0 | **Lambda 69** | .25 |

The same steps are performed for transitions occurred in time interval =3

| **Observed counts of transitions during time interval Δ t=3** | | | | | | | | | | |
|---|---|---|---|---|---|---|---|---|---|---|
| | **State 1** | **State 2** | **State3** | **State 4** | **State 5** | **State 6** | **State 7** | **State 8** | **State 9** | **total** |
| **State 1** | 78 | 57 | 8 | 2 | 2 | 0 | 0 | 0 | 1 | 148 |
| **State 2** | 1 | 32 | 13 | 2 | 2 | 0 | 0 | 0 | 1 | 51 |
| **State 3** | 0 | 0 | 9 | 4 | 2 | 0 | 0 | 2 | 1 | 18 |
| **State 4** | 0 | 0 | 0 | 3 | 2 | 0 | 0 | 1 | 1 | 7 |
| **State 5** | 0 | 0 | 0 | 0 | 3 | 2 | 1 | 0 | 1 | 7 |
| **State 6** | 0 | 0 | 0 | 0 | 0 | 0 | 1 | 0 | 1 | 2 |
| **State 7** | 0 | 0 | 0 | 0 | 0 | 0 | 1 | 0 | 1 | 2 |
| **State 8** | 0 | 0 | 0 | 0 | 0 | 0 | 0 | 1 | 2 | 3 |
| **State 9** | 0 | 0 | 0 | 0 | 0 | 0 | 0 | 0 | 0 | 0 |
| | | | | | | | | | | 238 |

| **Initial Q matrix :** | | | | | | | | | **Step 1 : calculate eigenvalues for this Q matrix :** | |
|---|---|---|---|---|---|---|---|---|---|---|
| -.397 | .39 | 0 | 0 | 0 | 0 | 0 | 0 | .007 | **Eigenvalue 1** | -.48277 |
| .02 | -.29 | .25 | 0 | 0 | 0 | 0 | 0 | .02 | **Eigenvalue 2** | -.57 |
| 0 | .05 | -.36 | .21 | 0 | 0 | 0 | .05 | .05 | **Eigenvalue 3** | -.18296 |
| 0 | 0 | 0 | -.57 | .29 | 0 | 0 | .14 | .14 | **Eigenvalue 4** | -.38128 |
| 0 | 0 | 0 | 0 | -.43 | .29 | 0 | 0 | .14 | **Eigenvalue 5** | -.43 |
| 0 | 0 | 0 | 0 | 0 | -1 | .5 | 0 | .5 | **Eigenvalue 6** | -1 |
| 0 | 0 | 0 | 0 | 0 | 0 | -.5 | 0 | .5 | **Eigenvalue 7** | -.5 |
| 0 | 0 | 0 | 0 | 0 | 0 | 0 | -.67 | .67 | **Eigenvalue 8** | -.67 |
| 0 | 0 | 0 | 0 | 0 | 0 | 0 | 0 | 0 | **Eigenvalue 9** | 0 |

| **Step 2 : calculate $te^{\Lambda t} d\,\Lambda$** | | **Step 3 : rearrange the score function then scale it by a factor =1354** | | |
|---|---|---|---|---|
| **Lambda 12** | -.8815 | **Lambda 12** | -.8815 | -1193.6 |
| **Lambda 18** | -.9480 | **Mu 21** | -.0671 | -90.9 |
| **Lambda 19** | -.9480 | **Lambda 23** | -1.189 | -1609.9 |
| **Lambda 23** | -1.189 | **Mu 32** | -.2052 | -277.8 |
| **Lambda 28** | -1.3642 | **Lambda 34** | -1.0813 | -1464 |
| **Lambda 29** | -1.3642 | **Mu 43** | -.0554 | -75 |
| **Mu 21** | -.0671 | **Lambda 79** | -.6694 | -906.4 |
| **Lambda 34** | -1.0813 | **Lambda 89** | -.4020 | -544.3 |
| **Lambda 38** | -1.0813 | **Lambda 18** | -.9480 | -1283.6 |
| **Lambda 39** | -1.0813 | **Lambda 19** | -.9480 | -1283.6 |
| **Mu 32** | -.2052 | **Lambda 28** | -1.3642 | -1847.1 |
| **Lambda 45** | -.5426 | **Lambda 29** | -1.3642 | -1847.1 |
| **Lambda 48** | -.5426 | **Lambda 38** | -1.0813 | -1464 |
| **Lambda 49** | -.5426 | **Lambda 39** | -1.0813 | -1464 |
| **Mu 43** | -.0554 | **Lambda 45** | -.5426 | -734.7 |
| **Lambda 56** | -.8258 | **Lambda 48** | -.5426 | -734.7 |
| **Lambda 58** | -.8258 | **Lambda 49** | -.5426 | -734.7 |
| **Lambda 59** | -.8258 | **Lambda 56** | -.8258 | -1118.1 |
| **Lambda 67** | -.1494 | **Lambda 58** | -.8258 | -1118.1 |
| **Lambda 69** | -.1494 | **Lambda 59** | -.8258 | -1118.1 |
| **Lambda 79** | -.6694 | **Lambda 67** | -.1494 | -202.2 |
| **Lambda 89** | -.4020 | **Lambda 69** | -.1494 | -202.2 |

**Step 4**: multiply the rearranged scaled scored function with the transposed rearranged scaled score function to get the hessian matrix (22 by 22 matrix)

**Step 5**: scale the above hessian matrix by a factor of 56367.63, the resultant matrix can be partitioned into 4 matrices, the upper left matrix of size (8 by 8) is the matrix needs to be inverted but it may be singular or close to be singular so the pseudo-inverse using singular value decomposition is applied. The lower right matrix is (14 by 14) has repeated rows. The matrices on the secondary diagonal, the upper right (8 by 14) and the lower left (14 by 8) have repeated rows so the only invertible matrix is the upper left.

| **Step 6: invert the scaled hessian matrix ( only the upper left is invertible)** | | | | | | | |
|---|---|---|---|---|---|---|---|
| $10^{-12}\times$ | | | | | | | |
| .4655 | .0354 | .6278 | .1084 | .571 | .0292 | .3535 | .2123 |
| .0354 | .0027 | .0478 | .0082 | .0435 | .0022 | .0269 | .0162 |
| .6278 | .0478 | .8468 | .1461 | .7701 | .0394 | .4768 | .2863 |
| .1084 | .0082 | .1461 | .0252 | .1329 | .0068 | .0823 | .0494 |
| .571 | .0435 | .7701 | .1329 | .7004 | .0359 | .4336 | .2604 |
| .0292 | .0022 | .0394 | .0068 | .0359 | .0018 | .0222 | .0133 |
| .3535 | .0269 | .4768 | .0823 | .4336 | .0222 | .2684 | .1612 |
| .2123 | .0162 | .2863 | .0494 | .2604 | .0133 | .1612 | .0968 |

| **Step 7 : multiply inverted hessian matrix by scaled score function** | **Step 8: apply quasi-Newton formula to get the estimated rates in Δt=3** | |
|---|---|---|
| $-.2874*10^{-8}$ | **Lambda 12** | .39 |
| $-.0219*10^{-8}$ | **Mu 21** | .02 |
| $-.3876*10^{-8}$ | **Lambda 23** | .25 |
| $-.0669*10^{-8}$ | **Mu 32** | .05 |
| $-.3525*10^{-8}$ | **Lambda 34** | .21 |
| $-.0180*10^{-8}$ | **Mu 43** | 0 |
| $-.2182*10^{-8}$ | **Lambda 79** | .5 |
| $-.1310*10^{-8}$ | **Lambda 89** | .67 |
| 0 | **Lambda 18** | 0 |
| 0 | **Lambda 19** | .007 |
| 0 | **Lambda 28** | 0 |
| 0 | **Lambda 29** | .02 |
| 0 | **Lambda 38** | .05 |
| 0 | **Lambda 39** | .05 |
| 0 | **Lambda 45** | .29 |
| 0 | **Lambda 48** | .14 |
| 0 | **Lambda 49** | .14 |
| 0 | **Lambda 56** | .29 |
| 0 | **Lambda 58** | 0 |
| 0 | **Lambda 59** | .14 |
| 0 | **Lambda 67** | .5 |
| 0 | **Lambda 69** | .5 |

Number of transitions observed in first interval corresponds to (2/3) of the total 3580 transitions while the number of transitions observed in the second interval corresponds to (4/15) of the total 3580 transitions and the number of transitions observed in the third interval corresponds to (1/15) of the total 3580 transitions.

Scaling the vector of theta or rates estimated in each interval by these correspondent weights in each interval and then summing up the weighted vectors, the final vector of rates or thetas is obtained which is:

| $\lambda_{12}$ | $\mu_{21}$ | $\lambda_{23}$ | $\mu_{32}$ | $\lambda_{34}$ | $\mu_{43}$ | $\lambda_{79}$ | $\lambda_{89}$ | $\lambda_{18}$ | $\lambda_{19}$ | $\lambda_{28}$ |
|---|---|---|---|---|---|---|---|---|---|---|
| 0.39 | 0.02 | 0.25 | 0.05 | 0.225 | 0.041 | 0.421 | 0.745 | 0 | .007 | 0 |

| $\lambda_{29}$ | $\lambda_{38}$ | $\lambda_{39}$ | $\lambda_{45}$ | $\lambda_{48}$ | $\lambda_{49}$ | $\lambda_{56}$ | $\lambda_{58}$ | $\lambda_{59}$ | $\lambda_{67}$ | $\lambda_{69}$ |
|---|---|---|---|---|---|---|---|---|---|---|
| 0.011 | 0.047 | 0.043 | 0.281 | 0.109 | 0.107 | 0.19 | 0.059 | 0.099 | 0.767 | 0.167 |

Doing the same procedure for the inverted scaled hessian matrix the final matrix which is the estimated variance of the rates:

| $10^{-13}\times$ | | | | | | | |
|---|---|---|---|---|---|---|---|
| .3293 | .0327 | .4414 | .0827 | .4004 | .0268 | .2540 | .1517 |
| .0327 | .0064 | .0428 | .0108 | .0385 | .0053 | .0268 | .0159 |
| .4414 | .0428 | .5922 | .1100 | .5373 | .0350 | .3400 | .2032 |
| .0827 | .0108 | .1100 | .0228 | .0995 | .0088 | .0650 | .0387 |
| .4004 | .0385 | .5373 | .0995 | .4876 | .0315 | .3082 | .1843 |
| .0268 | .0053 | .0350 | .0088 | .0315 | .0044 | .0219 | .0130 |
| .254 | .0268 | .3400 | .0650 | .3082 | .0219 | .1967 | .1174 |
| .1517 | .0159 | .2032 | .0387 | .1843 | .0130 | .1174 | .0702 |

**Mean sojourn time for each state:**

| For state 1 | For state 2 | For state 3 | For state 4 | For state 5 | For state 6 | For state 7 | For state 8 |
|---|---|---|---|---|---|---|---|
| 2.5189 years | 3.5587 years | 2.7397 years | 1.8587 years | 2.8736 years | 1.0707 years | 2.3753 years | 1.3423 years |

Mean time spent by the patient in state 1 is approximately 2 years and 6 months, in state 2 the mean sojourn time is approximately 3 years and 6 months , in state 3 it is approximately 2 years and 9 months , in state 4 it is approximately 1 years and 10 months , in state 5 it is approximately 2 years and 10 months, in state 6 it is approximately 1 years and 1 month , in state 7 it is approximately 2 years and 5 months and lastly in state 8 the mean sojourn time is approximately 1 years and 4 months.

**Variance of the sojourn time**

| For state 1 | For state 2 | For state 3 | For state 4 | For state 5 | For state 6 | For state 7 | For state 8 |
|---|---|---|---|---|---|---|---|
| $.0362\times10^{-9}$ | $.1443\times10^{-9}$ | $.0507\times10^{-9}$ | $.0107\times10^{-9}$ | $.0613\times10^{-9}$ | $.0012\times10^{-9}$ | $.0286\times10^{-9}$ | $.0029\times10^{-9}$ |

**The life expectancy of NAFLD patient or the mean time to absorption:**

| From state 1 | From state 2 | From state 3 | From state 4 | From state 5 | From state 6 | From state 7 | From state 8 |
|---|---|---|---|---|---|---|---|
| 13.6762 years | 11.3576 years | 7.6718 years | 5.1966 years | 4.7507 years | 3.0213 years | 2.3753 years | 1.3423 years |

Mean time for a patient in state 1 to absorption or death is approximately 13 years and 8 month, for a patient in state 2 it is approximately 11 years and 4 months, for a patient in state 3 it is approximately 7 years and 2 months, for a patient is state 4 it is approximately 5 years and 2 months ,for a patient is state 5 it is approximately 4 years and 9 months, for a patient is state 6 it is approximately 3 years, for a patient in state 7 it is approximately 2 years and 4 and a half months, for patient in state 8 it is approximately 1 year and 4 months.

Transition probability matrix at 1 year:

$$P(1)=\begin{bmatrix} .6751 & .279 & .0345 & .0024 & .0002 & .0000 & .0000 & .0006 & .0082 \\ .0143 & .7625 & .1819 & .0188 & .0019 & .0000 & .0000 & .0044 & .016 \\ .0004 & .0364 & .7017 & .1439 & .0206 & .0012 & .0002 & .0346 & .0604 \\ 0 & .0007 & .0262 & .5868 & .1800 & .0145 & .0039 & .0616 & .1215 \\ 0 & 0 & 0 & 0 & .7061 & .1015 & .0416 & .0344 & .1163 \\ 0 & 0 & 0 & 0 & 0 & .3930 & .3938 & 0 & .2132 \\ 0 & 0 & 0 & 0 & 0 & 0 & .6564 & 0 & .3436 \\ 0 & 0 & 0 & 0 & 0 & 0 & 0 & .4747 & .5253 \\ 0 & 0 & 0 & 0 & 0 & 0 & 0 & 0 & 1 \end{bmatrix}$$

If a cohort of 5000 NAFLD patients have initial distribution of $[.62 \quad .22 \quad .081 \quad .03 \quad .028 \quad .005 \quad .007 \quad .009 \quad 0]$ and initial counts of patients in each state are$[3100 \quad 1100 \quad 405 \quad 150 \quad 140 \quad 25 \quad 35 \quad 45 \quad 0]$then at 1 year the state probability distribution

is$[.4217 \quad .3437 \quad .1191 \quad .035 \quad .0274 \quad .0054 \quad .0079 \quad .0111 \quad .0287]$ and the expected counts of patients are $[2109 \quad 1718 \quad 595 \quad 175 \quad 137 \quad 27 \quad 39 \quad 56 \quad 144]$.

Transition probability matrix at 50 year:

$$P(50) = \begin{bmatrix} 0 & 0 & 0 & 0 & 0 & 0 & 0 & 0 & 1 \\ 0 & 0 & 0 & 0 & 0 & 0 & 0 & 0 & 1 \\ 0 & 0 & 0 & 0 & 0 & 0 & 0 & 0 & 1 \\ 0 & 0 & 0 & 0 & 0 & 0 & 0 & 0 & 1 \\ 0 & 0 & 0 & 0 & 0 & 0 & 0 & 0 & 1 \\ 0 & 0 & 0 & 0 & 0 & 0 & 0 & 0 & 1 \\ 0 & 0 & 0 & 0 & 0 & 0 & 0 & 0 & 1 \\ 0 & 0 & 0 & 0 & 0 & 0 & 0 & 0 & 1 \\ 0 & 0 & 0 & 0 & 0 & 0 & 0 & 0 & 1 \end{bmatrix}$$

For the above same cohort of 5000 NAFLD patients, at 50 year the state probability distribution is$[0 \quad 0 \quad 0 \quad 0 \quad 0 \quad 0 \quad 0 \quad 0 \quad 1]$, and the estimated variance of this distribution is

| $\mathbf{10^{-11}} \times$ | | | | | | | | |
|---|---|---|---|---|---|---|---|---|
| .01 | .0316 | .033 | .02 | .0257 | .0088 | .0239 | .0115 | 0 |
| .0316 | .0998 | .1043 | .0632 | .0813 | .0278 | .0757 | .0364 | 0 |
| .0330 | .1043 | .1090 | .0661 | .0850 | .0291 | .0791 | .0380 | 0 |
| .02 | .0632 | .0661 | .0400 | .0515 | .0176 | .0479 | .0230 | 0 |
| .0257 | .0813 | .0850 | .0515 | .0662 | .0226 | .0616 | .0296 | 0 |
| .0088 | .0278 | .0291 | .0176 | .0226 | .0077 | .0211 | .0101 | 0 |
| .0239 | .0757 | .0791 | .0479 | .0616 | .0211 | .0574 | .0276 | 0 |
| .0115 | .0364 | .0380 | .0230 | .0296 | .0101 | .0276 | .0133 | 0 |
| 0 | 0 | 0 | 0 | 0 | 0 | 0 | 0 | 0 |

steps for evaluation of the transition probabilities are demonstrated and as explained in the text :

**Step 1**: Laplace transform method is applied to solve the 4 differential equations in the first four rows along with Cramer rule using initial value , the determinant is a 4$^{th}$ degree polynomial which has 4 roots equal to the first four eigenvalues of the Q transition rate matrix. So once the Q rate matrix is estimated, the 4$^{th}$ degree polynomial can be solved for its roots$r_1 = -.4609, r_2 = -.5898, r_3 = -.1719 \;\; , r_4 = -.3584$ , $using \;\; Cramer \; rule \; P^*_{ij}(s) = \frac{D_{P^*_{ij}(s)}}{D}$ ,

**Step2**: the numerator for each probability is given by substitution for $D_{P^*_{ij}(s)}$ as illustrated in the discussion in text, this can be summarized in the following table:

| | Coefficient of $s^3$ | Coefficient of $s^2$ | Coefficient of s | Constant |
|---|---|---|---|---|
| $DP^*(s)_{11}$ | 1 | 1.184 | .428388 | .045862745 |
| $DP^*(s)_{12}$ | 0 | .39 | .35217 | .07298655 |
| $DP^*(s)_{13}$ | 0 | 0 | .0975 | .052455 |
| $DP^*(s)_{14}$ | 0 | 0 | 0 | .0219375 |
| $DP^*(s)_{21}$ | 0 | .02 | .01806 | .0037429 |
| $DP^*(s)_{22}$ | 1 | 1.3 | .545636 | .074296565 |
| $DP^*(s)_{23}$ | 0 | .25 | .23375 | .0533965 |
| $DP^*(s)_{24}$ | 0 | 0 | .05625 | .02233125 |
| $DP^*(s)_{31}$ | 0 | 0 | .001 | .000538 |
| $DP^*(s)_{32}$ | 0 | .05 | .04675 | .0106793 |
| $DP^*(s)_{33}$ | 1 | 1.216 | .468521 | .055821266 |
| $DP^*(s)_{34}$ | 0 | .225 | .15255 | .023345325 |
| $DP^*(s)_{41}$ | 0 | 0 | 0 | .000041 |
| $DP^*(s)_{42}$ | 0 | 0 | .00205 | .00081385 |
| $DP^*(s)_{43}$ | 0 | .041 | .027798 | .004254037 |
| $DP^*(s)_{44}$ | 1 | 1.043 | .338727 | .032908805 |

To get the inverse Laplace, partial fraction method is used and this needs the following calculations :

| **Step 3** : construct (K) matrix | | | |
|---|---|---|---|
| 1 | 1 | 1 | 1 |
| 1.12005634 | .991227769 | 1.409099675 | 1.222616213 |
| .37435313 | .306037387 | .648411308 | .452470172 |
| .03633368 | .028397028 | .097427266 | .046731407 |

| **Step 4** : invert (K) matrix | | | |
|---|---|---|---|
| -22.644304 | 55.6343565 | -120.696652 | 261.84686 |
| 16.4685712 | -22.92361 | 47.346434 | -80.27918609 |
| -.2255193 | 1.311918 | -7.631857 | 44.39699213 |
| 10.4012519 | -19.02266 | 80.9820756 | -225.9646 |

**Step 5**: calculate the coefficient of the first four $P_{ij}(t)$ in the first four rows; let them be called $A_{ij}$, $B_{ij}$, $C_{ij}$, $D_{ij}$ , it is calculated by multiplying the inverted K matrix by the $D_{P^*_{ij}(s)}$

| **Coefficient of $P_{11}$** | **Coefficient of $P_{12}$** | **Coefficient of $P_{13}$** | **Coefficient of $P_{14}$** |
|---|---|---|---|
| **$A_{11}$** = .5307928 | **$A_{12}$** = -1.69704167 | **$A_{13}$** = 1.9672538 | **$A_{14}$** = 5.74426567 |
| **$B_{11}$** = .00784714 | **$B_{12}$** = -.07551577 | **$B_{13}$** = .40523268 | **$B_{14}$** = -1.7611246 |
| **$C_{11}$** = .094564365 | **$C_{12}$** = 1.06432039 | **$C_{13}$** = 1.58473813 | **$C_{14}$** = .97395901 |
| **$D_{11}$** = .366808118 | **$D_{12}$** = .70823704 | **$D_{13}$** = -3.95722461 | **$D_{14}$** = -4.9571 |
| **Coefficient of $P_{21}$** | **Coefficient of $P_{22}$** | **Coefficient of $P_{23}$** | **Coefficient of $P_{24}$** |
| **$A_{21}$** = -.0870278 | **$A_{22}$** = .2782437 | **$A_{23}$** = -.322547186 | **$A_{24}$** = -.9418189 |
| **$B_{21}$** = -.0038726 | **$B_{22}$** = .0373265 | **$B_{23}$** = -.200301556 | **$B_{24}$** = .87050238 |
| **$C_{21}$** = .0545805 | **$C_{22}$** = .614303 | **$C_{23}$** = .914677019 | **$C_{24}$** = .56214835 |
| **$D_{21}$** = .0363198 | **$D_{22}$** = .0701267 | **$D_{23}$** = -.391828278 | **$D_{24}$** = -.4908319 |
| **Coefficient of $P_{31}$** | **Coefficient of $P_{32}$** | **Coefficient of $P_{33}$** | **Coefficient of $P_{34}$** |
| **$A_{31}$** = .020176962 | **$A_{32}$** = -.06450944 | **$A_{33}$** = .07478098 | **$A_{34}$** = .21835606 |
| **$B_{31}$** = .004156233 | **$B_{32}$** = -.04006031 | **$B_{33}$** = .21497162 | **$B_{34}$** = -.9342579 |
| **$C_{31}$** = .016253724 | **$C_{32}$** = .1829354 | **$C_{33}$** = .27238481 | **$C_{34}$** = .16740409 |
| **$D_{31}$** = -.04058692 | **$D_{32}$** = -.07836566 | **$D_{33}$** = .43786259 | **$D_{34}$** = .54849772 |
| **Coefficient of $P_{41}$** | **Coefficient of $P_{42}$** | **Coefficient of $P_{43}$** | **Coefficient of $P_{44}$** |
| **$A_{41}$** = .0107357 | **$A_{42}$** = -.0343241 | **$A_{43}$** = .0397889327 | **$A_{44}$** = .1161825 |
| **$B_{41}$** = -.0032914 | **$B_{42}$** = .031725 | **$B_{43}$** = -.17024246 | **$B_{44}$** = .73986714 |
| **$C_{41}$** = .0018203 | **$C_{42}$** = .02204872 | **$C_{43}$** = .030504745 | **$C_{44}$** = .01874781 |
| **$D_{41}$** = -.0092646 | **$D_{42}$** = -.0178881 | **$D_{43}$** = .099948473 | **$D_{44}$** = .12520254 |

**Step 6**: calculate the inverse Laplace for each PDF, which equals:

$A_{ij}e^{(r_1 t)} + B_{ij}e^{(r_2 t)} + C_{ij}e^{(r_3 t)} + D_{ij}e^{(r_4 t)}$ , $t = 1$

| | | | |
|---|---|---|---|
| **$P_{11}$** = .675065196 | **$P_{12}$** =.27897024 | **$P_{13}$** = .03452831 | **$P_{14}$** = .00246831 |
| **$P_{21}$** = .0143062 | **$P_{22}$** = .7624677 | **$P_{23}$** = .181919657 | **$P_{24}$** = .01902779 |
| **$P_{31}$** = .000354137 | **$P_{32}$** = .03638393 | **$P_{33}$** = .70170095 | **$P_{34}$** = .14396966 |
| **$P_{41}$** = 4.61E-06 | **$P_{42}$** = .0006935 | **$P_{43}$** = .026234471 | **$P_{44}$** = .58677589 |

**Step 7**: calculation of the last four probabilities in the first row:

| For $P_{15}$: | | | |
|---|---|---|---|
| $G_1=\lambda_{45}*A_{14}$ | 1.6141387 | $G_1/(w+r_1)=F_1$ | -14.29154 |
| $G_2=\lambda_{45}*B_{14}$ | -.494876 | $G_2/(w+r_2)=F_2$ | 2.04686 |
| $G_3=\lambda_{45}*C_{14}$ | .2736825 | $G_3/(w+r_3)=F_3$ | 1.5541339 |
| $G_4=\lambda_{45}*D_{14}$ | -1.392945 | $G_4/(w+r_4)=F_4$ | 134.14616 |
| | | $F_5=-(F_1+F_2+F_3+F_4)$ | -1234556 |

$$P_{15} = F_1 e^{(r_1 t)} + F_2 e^{(r_2 t)} + F_3 e^{(r_3 t)} + F_4 e^{(r_4 t)} + F_5 e^{(-wt)} = .00017491$$

| For $P_{16}$: | | | |
|---|---|---|---|
| $G_5=\lambda_{56}*F_1$ | -2.715392 | $G_5/(u+r_1)=F_6$ | -5.740103 |
| $G_6=\lambda_{56}*F_2$ | .3889051 | $G_6/(u+r_2)=F_7$ | 1.1297899 |
| $G_7=\lambda_{56}*F_3$ | .2952855 | $G_7/(u+r_3)=F_8$ | .387463 |
| $G_8=\lambda_{56}*F_4$ | 25.48777 | $G_8/(u+r_4)=F_9$ | 44.279103 |
| $G_9=\lambda_{56}*F_5$ | -23.45657 | $G_9/(u-w)=F_{10}$ | -40.02827 |
| | | $F_{11}=-(F_6+F_7+F_8+F_9+F_{10})$ | -.02798 |

$$P_{16} = F_6 e^{(r_1 t)} + F_7 e^{(r_2 t)} + F_8 e^{(r_3 t)} + F_9 e^{(r_4 t)} + F_{10} e^{(-wt)} + F_{11} e^{(-ut)} = 6.0813E-6$$

| For $P_{17}$: | | | |
|---|---|---|---|
| $G_{10}=\lambda_{67}*F_6$ | -4.402659 | $G_{10}/(\lambda_{79}+r_1)=F_{12}$ | 110.2217 |
| $G_{11}=\lambda_{67}*F_7$ | .86654887 | $G_{11}/(\lambda_{79}+r_2)=F_{13}$ | -5.134428 |
| $G_{12}=\lambda_{67}*F_8$ | .29718414 | $G_{12}/(\lambda_{79}+r_3)=F_{14}$ | 1.193033 |
| $G_{13}=\lambda_{67}*F_9$ | 33.9620721 | $G_{13}/(\lambda_{79}+r_4)=F_{15}$ | 542.38463 |
| $G_{14}=\lambda_{67}*F_{10}$ | -30.701685 | $G_{14}/(\lambda_{79}-w)=F_{16}$ | -420.57103 |
| $G_{15}=\lambda_{67}*F_{11}$ | -.0214606 | $G_{15}/(\lambda_{79}-u)=F_{17}$ | .04183347 |
| | | $F_{18}=-(F_{12}+F_{13}+F_{14}+F_{15}+F_{16}+F_{17})$ | -228.13579 |

$$P_{17} = F_{12} e^{(r_1 t)} + F_{13} e^{(r_2 t)} + F_{14} e^{(r_3 t)} + F_{15} e^{(r_4 t)} + F_{16} e^{(-wt)} + F_{17} e^{(-ut)} + F_{18} e^{(-\lambda_{79} t)} = 7.828E-7$$

| For $P_{18}$: | | | |
|---|---|---|---|
| $G_{16}=\lambda_{18}*A_{11}+\lambda_{28}*A_{12}+\lambda_{38}*A_{13}+\lambda_{48}*A_{14}+\lambda_{58}*F_1$ | -.1246149 | $G_{10}/(\lambda_{89}+r_1)=F_{19}$ | -.438697828 |
| $G_{17}=\lambda_{18}*B_{11}+\lambda_{28}*B_{12}+\lambda_{38}*B_{13}+\lambda_{48}*B_{14}+\lambda_{58}*F_2$ | -.0521514 | $G_{10}/(\lambda_{89}+r_2)=F_{20}$ | -.335966911 |
| $G_{18}=\lambda_{18}*C_{11}+\lambda_{28}*C_{12}+\lambda_{38}*C_{13}+\lambda_{48}*C_{14}+\lambda_{58}*F_3$ | .2723381 | $G_{10}/(\lambda_{89}+r_3)=F_{21}$ | .475202028 |
| $G_{19}=\lambda_{18}*D_{11}+\lambda_{28}*D_{12}+\lambda_{38}*D_{13}+\lambda_{48}*D_{14}+\lambda_{58}*F_4$ | 7.1883097 | $G_{10}/(\lambda_{89}+r_4)=F_{22}$ | 18.5928823 |
| $G_{20}=\lambda_{58}*F_5$ | -7.2838816 | $G_{10}/(\lambda_{89}-w)=F_{23}$ | -18.34730875 |
| | | $F_{24}=-(F_{19}+F_{20}+F_{21}+F_{22}+F_{23})$ | .0538892 |

$$P_{18} = F_{19}e^{(r_1 t)} + F_{20}e^{(r_2 t)} + F_{21}e^{(r_3 t)} + F_{22}e^{(r_4 t)} + F_{23}e^{(-wt)} + F_{24}e^{(-\lambda_{89} t)} = .00055542$$

| for P19: | | | |
|---|---|---|---|
| $G_{21}=\lambda_{19}*A_{11}+\lambda_{29}*A_{12}+\lambda_{39}*A_{13}+\lambda_{49}*A_{14}+\lambda_{59}*F_1+\lambda_{69}*F_6+\lambda_{79}*F_{12}+\lambda_{89}*F_{19}$ | 44.387339 | $G_{21}/r_1$=F25 | -96.296671 |
| $G_{22}=\lambda_{19}*B_{11}+\lambda_{29}*B_{12}+\lambda_{39}*B_{13}+\lambda_{49}*B_{14}+\lambda_{59}*F_2+\lambda_{69}*F_7+\lambda_{79}*F_{13}+\lambda_{89}*F_{20}$ | -2.1923658 | $G_{22}/r_2$=F26 | 3.71730926 |
| $G_{23}=\lambda_{19}*C_{11}+\lambda_{29}*C_{12}+\lambda_{39}*C_{13}+\lambda_{49}*C_{14}+\lambda_{59}*F_3+\lambda_{69}*F_8+\lambda_{79}*F_{14}+\lambda_{89}*F_{21}$ | 1.2595848 | $G_{23}/r_3$=F27 | -7.327414 |
| $G_{24}=\lambda_{19}*D_{11}+\lambda_{29}*D_{12}+\lambda_{39}*D_{13}+\lambda_{49}*D_{14}+\lambda_{59}*F_4+\lambda_{69}*F_9+\lambda_{79}*F_{15}+\lambda_{89}*F_{22}$ | 262.1805 | $G_{24}$/r4=F28 | -731.5635 |
| $G_{25}=\lambda_{59}*F_5+\lambda_{69}*$F10$+\lambda_{79}*F_{16}+\lambda_{89}*F_{23}$ | -209.63598 | $F_{29}=-G_{25}/w$ | 602.402231 |
| $G_{26}=\lambda_{69}*F_{11}+\lambda_{79}*F_{17}$ | 0.0129392 | $F_{30}=-G_{26}/u$ | -0.0138536 |
| $G_{27}=\lambda_{79}*F_{18}$ | -96.045167 | $F_{31}=-G_{27}/\lambda_{79}$ | 228.135788 |
| $G_{28}=\lambda_{89}*F_{24}$ | 0.0401474 | $F_{32}=-G_{28}/\lambda_{89}$ | -0.0538892 |
| | | $F_{33}=-F_{25}-F_{26}-F_{27}-F_{28}+F_{29}+F_{30}+F_{31}+F_{32}$ | 1 |
| | | | |

$$P_{19} = F_{25}e^{(r_1 t)} + F_{26}e^{(r_2 t)} + F_{27}e^{(r_3 t)} + F_{28}e^{(r_4 t)} + F_{29}e^{(-wt)} + F_{30}e^{(-ut)} + F_{31}e^{(-\lambda_{79} t)} + F_{32}e^{(-\lambda_{89} t)} + F_{33} = .008230754$$

The same substitution is used to calculate the last 5 probabilities in 2nd, 3rd, 4th rows as demonstrated in text . The same is true for the last 12 PDFs' in the subsequent 4 rows while $P_{99}=1$

The following tables show the results for the last PDFs' using the same example:

| $P_{55}$ | $v_1$ | $p_{56}$ | $v_2$ | $v_3$ | $v_4$ |
|---|---|---|---|---|---|
| 0.706098876 | 0.32423208 | 0.101523624 | 3.40665763 | 0.484768044 | -3.891425672 |

| $p_{57}$ | $v_5$ | $p_{58}$ | $v_6$ | $v_7$ | $v_8$ |
|---|---|---|---|---|---|
| 0.041647322 | 0.14861461 | 0.034384156 | 1.698067503 | 0.149940589 | -1.63829021 |

| $v_9$ | $v_{10}$ | $v_{11}$ | $v_{12}$ | $v_{13}$ | $v_{14}$ |
|---|---|---|---|---|---|
| -0.11071788 | -4.87950432 | -0.16053596 | 3.891425672 | 0.14861461 | 1 |

| $p_{59}$ | $p_{66}$ | $z_1$ | $p_{67}$ | $z_2$ | $z_3$ |
|---|---|---|---|---|---|
| 0.11634602 | 0.39297865 | -1.4951267 | 0.3938335 | 0.49512671 | -1.4951267 |

| $p_{69}$ | $p_{77}$ | $p_{79}$ | $p_{88}$ | $p_{89}$ |
|---|---|---|---|---|
| 0.21318785 | 0.6563901 | 0.3436099 | 0.4747343 | 0.5252657 |

The transition probability matrix at 1 year is

$$P_{ij}(t=1) = \begin{bmatrix} .6751 & .279 & .0345 & .0024 & .0002 & .0000 & .0000 & .0006 & .0082 \\ .0143 & .7625 & .1819 & .0188 & .0019 & .0000 & .0000 & .0044 & .016 \\ .0004 & .0364 & .7017 & .1439 & .0206 & .0012 & .0002 & .0346 & .0604 \\ 0 & .0007 & .0262 & .5868 & .1800 & .0145 & .0039 & .0616 & .1215 \\ 0 & 0 & 0 & 0 & .7061 & .1015 & .0416 & .0344 & .1163 \\ 0 & 0 & 0 & 0 & 0 & .3930 & .3938 & 0 & .2132 \\ 0 & 0 & 0 & 0 & 0 & 0 & .6564 & 0 & .3436 \\ 0 & 0 & 0 & 0 & 0 & 0 & 0 & .4747 & .5253 \\ 0 & 0 & 0 & 0 & 0 & 0 & 0 & 0 & 1 \end{bmatrix}$$

Of those patients (patients with susceptible risk factors) starting at stage S1 (NAFLD with no fibrosis), that is to mean only steatosis, about 30 % of them will move to S2 (NASH with no fibrosis), 3.5% will move to S3 (NASH with fibrosis whether F1, F2 or F3). They are less likely to develop liver cirrhosis whether compensated or decompensated (S4 and S5 respectively), as only; 0.24% of them will get S4 and 0.02% will get S5. Less than 1% of them, about 0.06% will develop HCC (S8=hepatocellular carcinoma) and 0.82% will die (S9). While the majority, about 67.51 % will remain stable in S1. These patients are not candidate for liver transplantation.

Once the patient starts to develop S2 (NASH with no fibrosis), 18.2% of those patients will progress to develop S3 and about 2% will develop S4 (compensated liver cirrhosis), while; they are less likely to get S5 as only 0.19% will get S5 (decompensated liver cirrhosis).They are more likely to develop S8 (0.44% v.s. 0.06% for those starting at S1) and 1.6% will die, however; only 1.4% will regress to S1. 76.25 % will remain stable is S2. These patients are also not candidate for liver transplantation.

Those patients at S3, about 14.39% will progress to S4 (compensated liver cirrhosis), while 2.1% will develop S5 (decompensated liver cirrhosis). Because they are not highly recommended for liver transplantation, only 0.12% of them will survive the first year after liver transplantation (S6) and 0.02% will survive longer than the first year after liver transplantation (S7), however about 3.5% of them will get HCC (S8) and 6.04% will die. Only 0.04 % of them will regress to S1, 3.64% will regress to S2, and 70.17 % will remain stable is S3.

Of the compensated liver cirrhosis patients (S4), about 18% will be decompensated (S5) and because they are putting on waiting list for liver transplantation, so; 1.5% of them will survive the first year after liver transplantation (S6) and 0.39% will survive longer than the first year after liver transplantation (S7), however about 6.16 % of them will get HCC (S8) and 12.15% will die, no one will regress to S1, 0.07% will regress to S2, 2.62% will regress to S3 and 58.68 % will remain stable is S4.

Once the patient develops decompensated liver cirrhosis (S5), they are highly candidate for liver transplantation, thus; 10.15% of them will survive the first year after liver transplantation (S6) and 4.16% will survive longer than the first year after liver transplantation (S7), however about 3.44% of them will get HCC (S8) and 11.63% will die, and 70.61 % will remain stable is S5, but no regression to previous stages (S1, S2, S3, S4).

Of the patients who had survived the first year, 39.38 % of them would be surviving after this year and 21.32% would die.

The patients who had survived for more than one year after liver transplantation, 34.36 % of them would die. And 52.53 % of patients with HCC (S8) will die.

The above hypothetical example is coded in matlab as illustrated in the appendix A and B. the code is also published in code ocean site with the following URL:
codeocean.com/capsule/7628018/tree/v1

# Chapter six: Incorporation of covariates in the CTMC:

In the present chapter, the deleterious effects of obesity, type 2diabetes and insulin resistance, systolic and diastolic hypertension on the rate of progression of fibrosis in non-alcoholic fatty liver disease (NAFLD) patients are illustrated using a new approach utilizing the Poisson regression to model the transition rate matrix. The observed counts in the transition counts matrix are used as response variables and the covariates are the risk factors for fatty liver. Then the estimated counts from running the Poisson regression are used to estimate the transition rates using the continuous time Markov chains (CTMC) followed by exponentiation of the estimated rate matrix to obtain the transition probability matrix at specific time points.

Singh et al. 2015 conducted a meta-analysis to evaluate the rate of fibrosis progression and thus searched multiple databases through a thoroughly systematic manner associated with author contact and found 11 cohort studies on NAFLD adult patients having at least one year apart paired liver biopsy specimens, from which they calculated a pooled-weighted annual fibrosis progression rate (number of stages changed between the 2 biopsy samples) with 95% confidence interval (CIs), and characterized the clinical risk factors accompanying this progression. They identified 411 patients with biopsy-proven NAFLD (150 with NAFL and 261 with NASH) included in those studies. Initially, the distribution of fibrosis for stages 0,1,2,3 and 4 was 35.8%, 32.5%, 16.7 %, 9.3% and 5.7% respectively, and over 2145.5 person-years of follow-up evaluation, 33.6% had fibrosis progression, 43.1% had stable fibrosis, and 22.3% had an improvement in fibrosis stage. The annual fibrosis progression rate in patients with NAFL who had stage 0 fibrosis at baseline was .07 stages (95% CI, 0.02-0.11 stages), compared with 0.14 stages in patients with NASH (95% CI, 0.07-0.21 stages). These findings correspond to 1 stage of progression over 14.3 years for patients with NAFL (95% CI, 9.1-50.0 y) and 7.1 years for patients with NASH (95% CI, 4.8-14.3 y).

Poisson regression is used to model the rates among states. The counts of each transition can be modeled as a function of some explanatory variables reflecting the characteristics of the patients. This can be accomplished by using Poisson regression model or log-linear model. The Poisson regression model specifies that each response $y_i$ is drawn from a Poisson population with parameter $\lambda_i$ , which is related to the regressors or the covariates. The primary equation of the model is

$$P(Y = y_i|x_i) = \frac{e^{-\lambda_i}\lambda_i^{y_i}}{y_i!}$$

$$the\ most\ common\ formulation\ for\ the\ \lambda_i\ is\ the\ log-linear\ model{:}\ \ ln\,\lambda_i = x_i'B,$$

$$And\ the\ expected\ number\ of\ events\ per\ person\ year\ is\ given\ by:\ E[y_i|x_i] = var[y_i|x_i] = \lambda_i = e^{x_i'B}$$

The observed counts in the transition counts matrix is used as response variables and the covariates are the risk factors for fatty liver. Then the estimated counts obtained from the Poisson regression model are used to estimate the rates using the CTMC, as the initially observed transition rates approximately equal the estimated transition rates among states, as illustrated in previous 2 chapters, followed by exponentiation of the estimated rate matrix. To expound this procedure a hypothetical example is used and it is in the form of a study conducted on 150 participants over 28 years to follow the progression of the NAFLD from F0 to F4.

A subset of the states that explicitly illustrates the phases of fibrosis process, which develops early in disease evolution cycle if the risk factors are not treated or eliminated, is modeled with CTMC to demonstrate: how covariates incorporated in a log-linear model can relate these predictors to transition rates among states, as illustrated in figure (6.1) (Younossi et al. 2020),(Singh et al. 2015). The presence of fibrosis is considered an ominous predictor for disease progression. This subset is a subset of states from the expanded model especially early phases or stages where reversibility of conditions in each stage can be achieved if properly treated and controlled so as to prevent reaching the irreversible damaged state which is liver cirrhosis or F4.

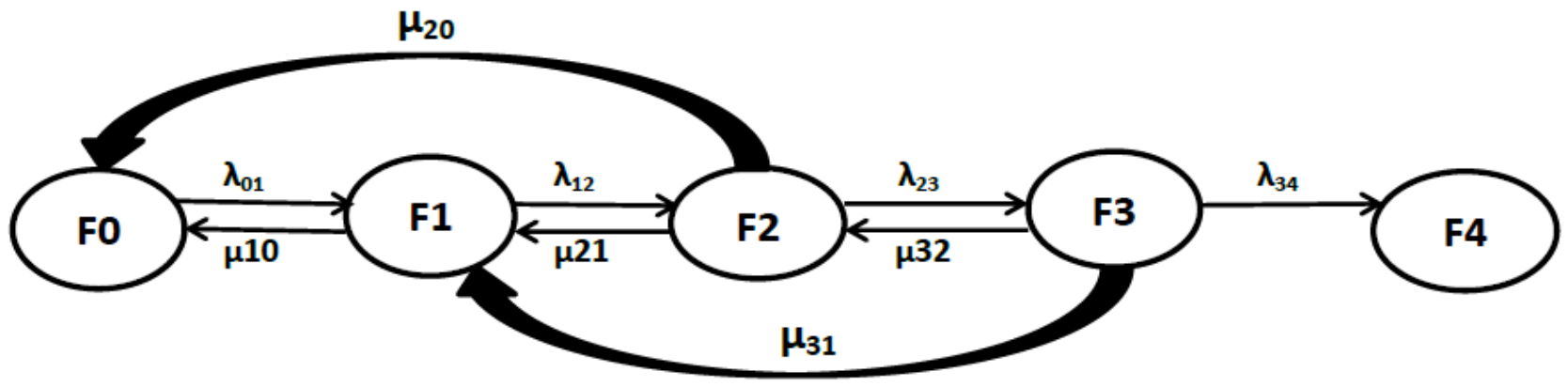

Figure (6.1) NAFLD with fibrosis stages.
*F0= no fibrosis (stage 0) whether hepatic steatosis is present or not . NASH-FB-1 =nonalcoholic steatohepatitis with mild fibrosis (stage 1). NASH -FB-2 = NASH with moderate fibrosis (stage 2). NASH -FB-3 = NASH with advanced or severe fibrosis (stage 3). CC= compensated cirrhosis (stage 4) which is the more severe or advanced form of fibrosis. This classification is based on the NAFLD activity score .*

### 6.1. Study Design

One hundred fifty participants were followed up every year for 28 years, and at each visit the characteristics of the participants were recorded like sex (0=female,1=male), age, BMI, LDL-chol, HOMA2_IR, systolic blood pressure and diastolic pressure as shown in the table (1) in appendix D. For each participant the recorded value in the table is the mean of the follow up measurements. Fitting the Poisson regression and the estimated counts for each transition were calculated using Stata 14. A summary statistics for the patients' characteristics is shown in table (6.1). The participants were categorized according to these demographic characteristics as shown in table (6.2), while in table (6.3) summary of the categorical groups according to the participants' characteristics like: age category BMI category, LDL-chol category, systolic and diastolic blood pressure category. There are high correlations between the continuous predictor variables as shown in table (6.4). In table (2), see appendix D, the transition counts accomplished by each participant in these 28 years are illustrated. Summary of transition counts among the states in these 28 years is clarified in table (6.5).The timeline for each participant is shown in table(3), see appendix D. The distribution of the transition counts among the states is Poisson as illustrated in the following successive figures using Statgraphics-19 software. For transition from 0 to 1, see figure (6.2). For transition from 1 to 2, see figure (6.3). For transition from 2 to 3, see figure (6.4). For transition from 3 to 4, see figure (6.5). For transition from 1 to 0, see figure (6.6). For transition from 2 to 1, see figure (6.7). For transition from 3 to 2, see figure (6.8). For transition from 2 to 0, see figure (6.9). For transition from 3 to 1, see figure (6.10). The observed transition counts are illustrated in table (6.6).

Table (6.1): statistical summary of the patients' characteristics:

| Variable | Observations | mean | Std. Dev. | Min | Max |
|---|---|---|---|---|---|
| **Gender:** | 150 | | | | |
| **Female=0** | 69 (0.46) | | | | |
| **Male=1** | 81(0.54) | | | | |
| **Age** | 150 | 40.2 | 4.93 | 27 | 53 |
| **LDL-chol** | 150 | 94.81 | 15.41 | 59.89 | 133.1 |
| **HOMA2-IR** | 150 | 2.28 | .71 | .49 | 4.36 |
| **BMI** | 150 | 28.28 | 2.991 | 20.3 | 35.16 |
| **Sys.Bl.Pr.** | 150 | 149.73 | 10.434 | 123.4 | 175.75 |
| **Dias.Bl.Pr.** | 150 | 94.25 | 11.39 | 70 | 124 |

Table (6.2): table summarizing the categorical groups of patients according to the previous characteristics:

| Variable | Group1 (desirable) | Group2 ( borderline) | Group3 (high) |
|---|---|---|---|
| **Age** | Age ≤ 35 | 35 < age ≤ 45 | Age > 45 |
| **LDL-chol** | LDL ≤ 70 | 70 < LDL < 100 | LDL ≥ 100 |
| **HOMA2-IR** | HOMA < 1.22 | 1.22 ≤ HOMA < 2.7 | HOMA ≥ 2.7 |
| **BMI** | BMI ≤ 25 | 25 < BMI < 30 | BMI ≥ 30 |
| **Systolic blood pressure** | Sys.Pr. ≤ 130 | 130 < Sys.Pr. < 160 | Sys.Pr. ≥ 160 |
| **Diastolic blood pressure** | Dias. Pr. ≤ 85 | 85 < Dias. Pr. < 100 | Dias. Pr. ≥ 100 |

Table (6.3): summary of categorical groups of the patients' characteristics regarding age category, BMI category, LDL-chol category, systolic and diastolic blood pressure category:

| category | Age | | | BMI | | | LDL-chol | | |
|---|---|---|---|---|---|---|---|---|---|
| | Frequency | Percent | Cum. | Frequency | Percent | Cum. | Frequency | Percent | Cum. |
| **1** | 22 | 14.67 | 14.67 | 22 | 14.67 | 14.67 | 5 | 3.33 | 3.33 |
| **2** | 109 | 72.67 | 87.33 | 83 | 55.33 | 70 | 93 | 62.00 | 65.33 |
| **3** | 19 | 12.67 | 100 | 45 | 30 | 100 | 52 | 34.67 | 100 |
| **total** | 150 | 100 | | 150 | 100 | | 150 | 100 | |
| **category** | **HOMA2-IR** | | | **Systolic Blood Pressure** | | | **Systolic Blood Pressure** | | |
| | **Frequency** | **Percent** | **Cum.** | **Frequency** | **Percent** | **Cum.** | **Frequency** | **Percent** | **Cum.** |
| **1** | 10 | 6.67 | 6.67 | 4 | 2.67 | 2.67 | 33 | 22 | 22 |
| **2** | 93 | 62.00 | 68.67 | 123 | 82.00 | 84.67 | 69 | 46 | 68 |
| **3** | 47 | 31.33 | 100 | 23 | 15.33 | 100 | 48 | 32 | 100 |
| **total** | 100 | 100 | | 150 | 100 | | 150 | 100 | |

Table (6.4): correlation between continuous predictor variables

| | age | LDL-chol | HOMA2-IR | BMI | Sys. Bl.Pr. | Dias. Bl.Pr. |
|---|---|---|---|---|---|---|
| **Age** | 1 | | | | | |
| **LDL-chol** | .9919 | 1 | | | | |
| **HOMA2-IR** | .9941 | .9947 | 1 | | | |
| **BMI** | .9938 | .9948 | .996 | 1 | | |
| **Sys. Bl.Pr.** | .9958 | .9953 | .9958 | .9962 | 1 | |
| **Dias. Bl.Pr.** | .9915 | .9951 | .9962 | .9945 | .9949 | 1 |

Table (6.5): summary transition counts between the states:

| Counts | Transition 0→1 | Transition 1→2 | Transition 2→3 | Transition 3→4 | Transition 1→0 | Transition 2→1 | Transition 3→2 | Transition 2→0 | Transition 3→1 |
|---|---|---|---|---|---|---|---|---|---|
| **0** | 63 | 96 | 121 | 128 | 121 | 127 | 130 | 138 | 139 |
| **1** | 58 | 43 | 23 | 22 | 24 | 17 | 17 | 11 | 9 |
| **2** | 25 | 9 | 4 | | 3 | 5 | 3 | 1 | 2 |
| **3** | 4 | 2 | 2 | | 2 | 1 | | | |

Table (6.6): Observed transitions counts of the patients over the 28 years

| | State 0 | State1 | State2 | State3 | State4 | total |
|---|---|---|---|---|---|---|
| **State0** | 1909 | 120 | 15 | 6 | 0 | 2050 |
| **State1** | 36 | 1116 | 67 | 28 | 0 | 1247 |
| **State2** | 13 | 30 | 703 | 37 | 0 | 783 |
| **State3** | 11 | 14 | 23 | 50 | 22 | 120 |
| **State4** | 0 | 0 | 0 | 0 | 0 | 0 |
| | | | | | | 4200 |

Initial observed rates are:

$$\lambda_{01} = \frac{120}{2050} = .059 \;, \lambda_{12} = \frac{67}{1247} = .0537 \;, \; \lambda_{23} = \frac{37}{783} = .047 \;, \lambda_{34} = \frac{22}{120} = .183$$

$$\mu_{10} = \frac{36}{1247} = .0288 \;, \mu_{21} = \frac{30}{783} = .0383 \,, \mu_{32} = \frac{23}{120} = .191 \,, \mu_{20} = \frac{13}{783} = .016 \,, \mu_{31} = \frac{14}{120} = .116$$

Using CTMC, the estimated rates approximately equal the initially observed rates, as illustrated in previous 2 chapters utilizing the simplest small model and the expanded model, where no covariates were included in the analysis.

The distribution of the transition counts is Poisson as illustrated in the following figures using the Statgraphics-19 software.

Figure 6.2: For transition from 0 to 1:

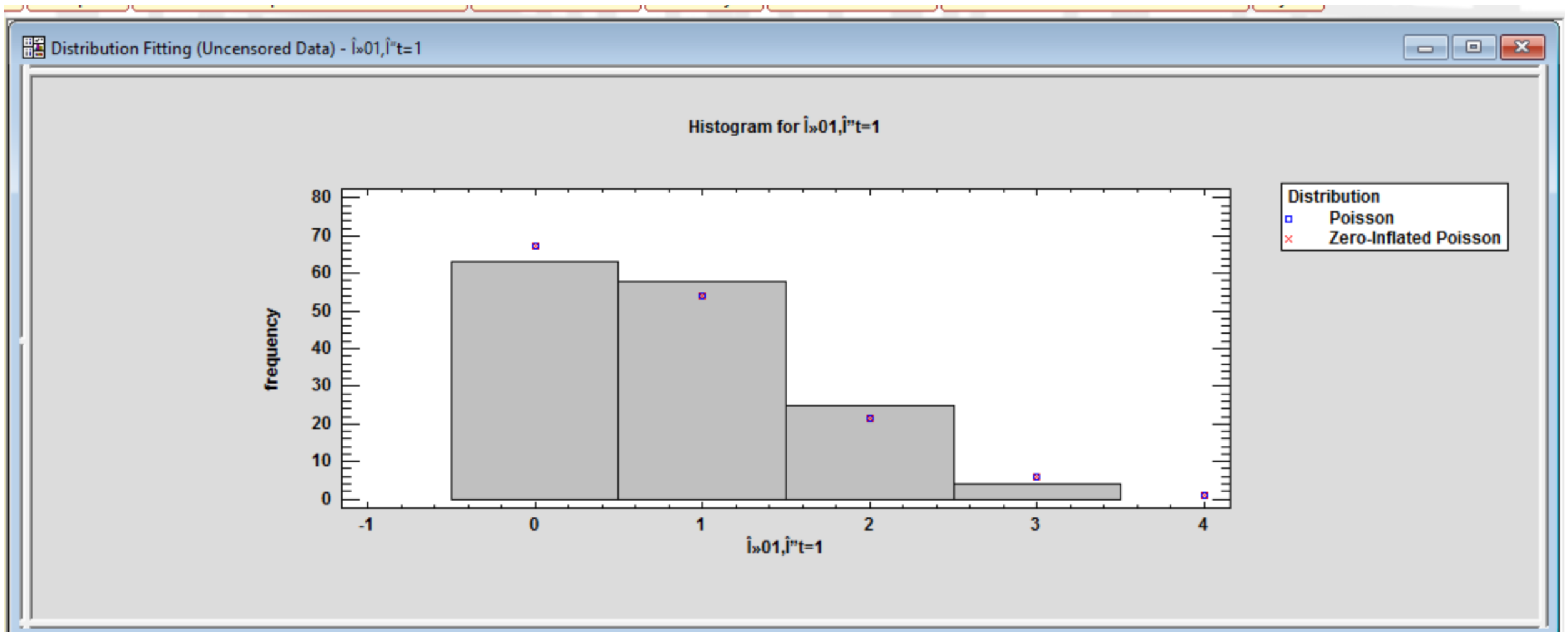

Figure 6.3: for transition from 1 to 2:

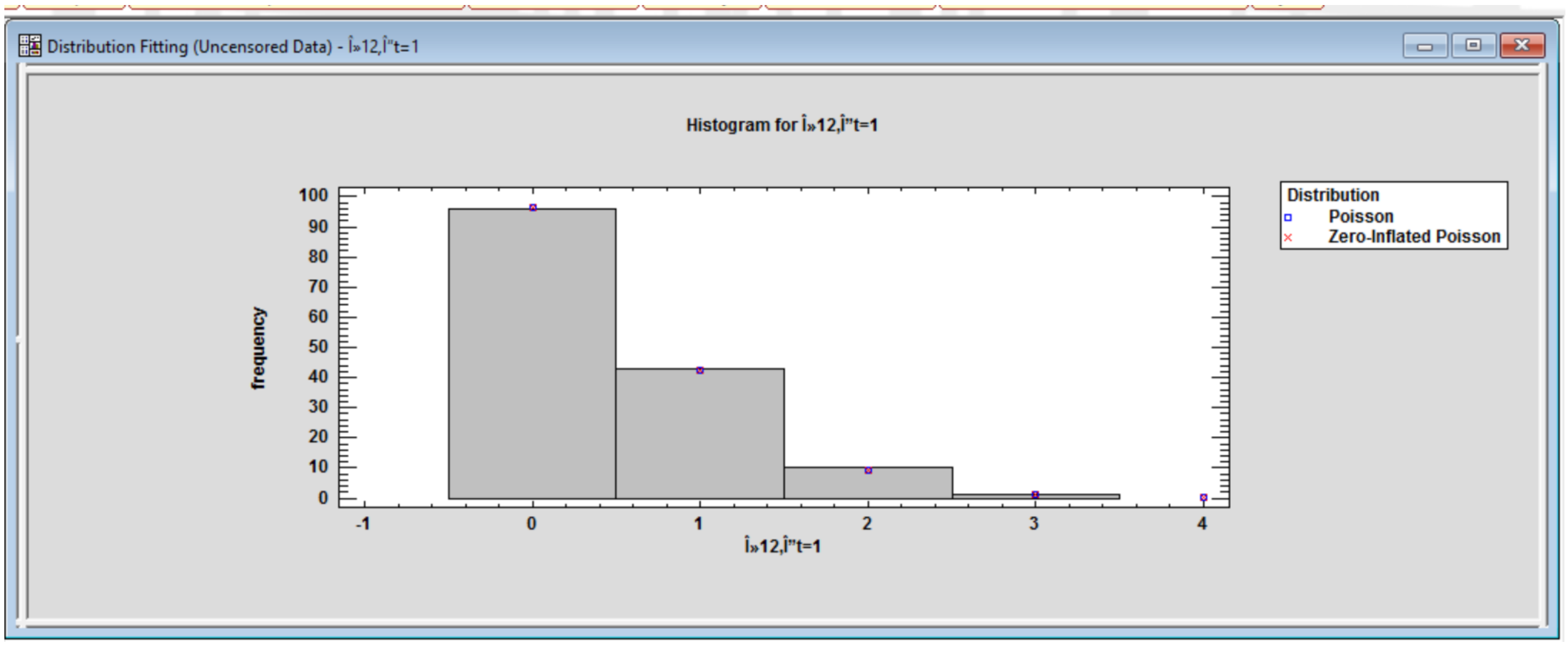

Figure 6.4: for transition from 2 to 3:

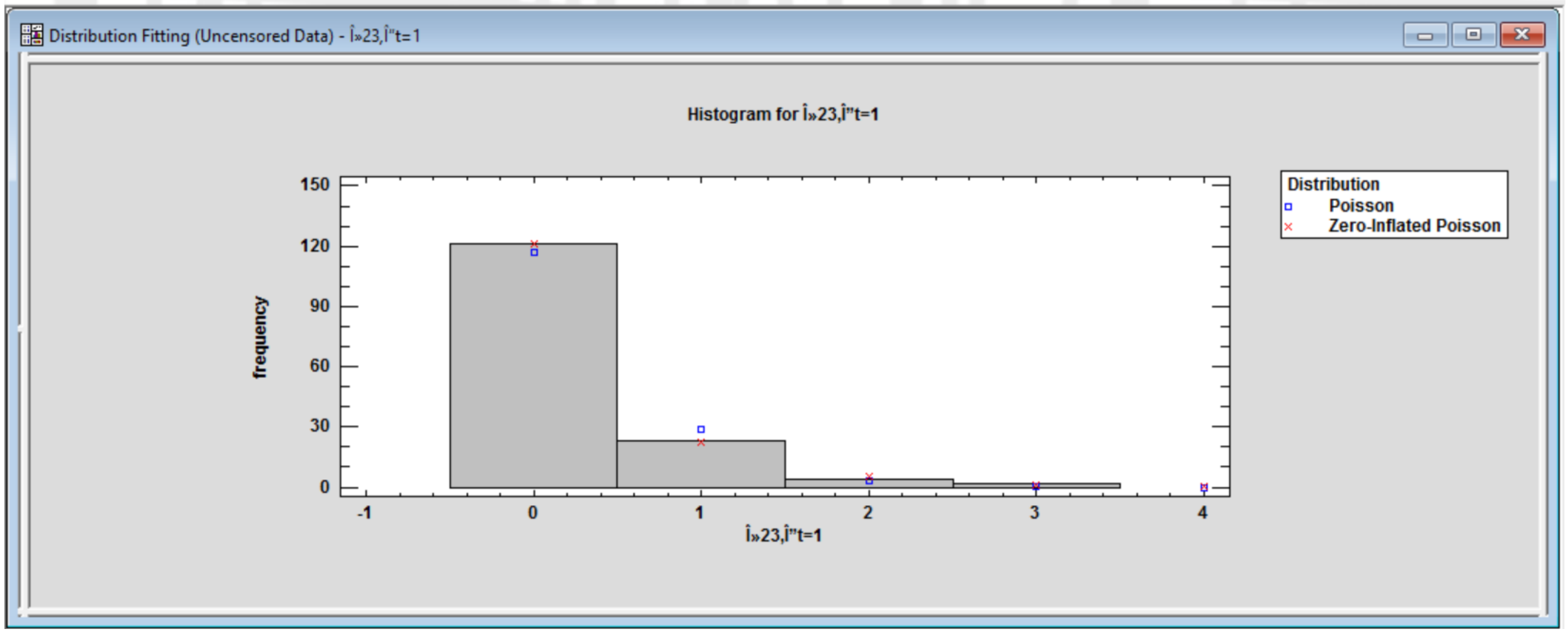

Figure 6.5: for transition from 3 to 4:

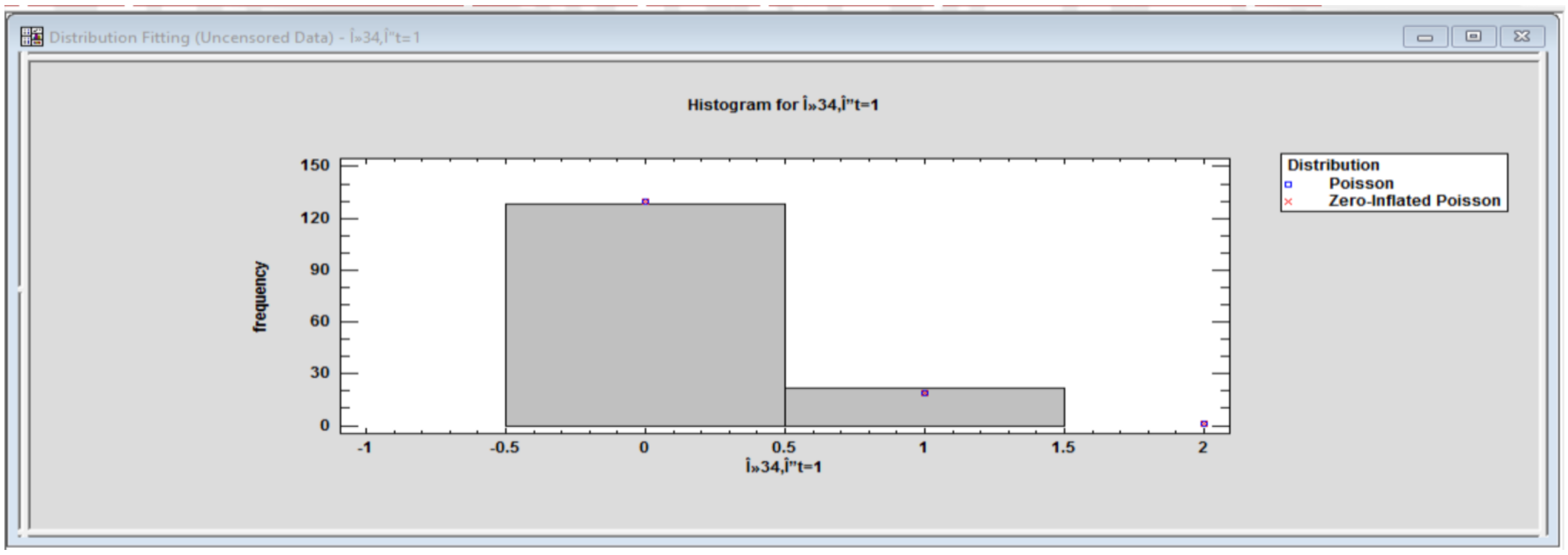

Figure 6.6 : for transition from 1 to 0 :

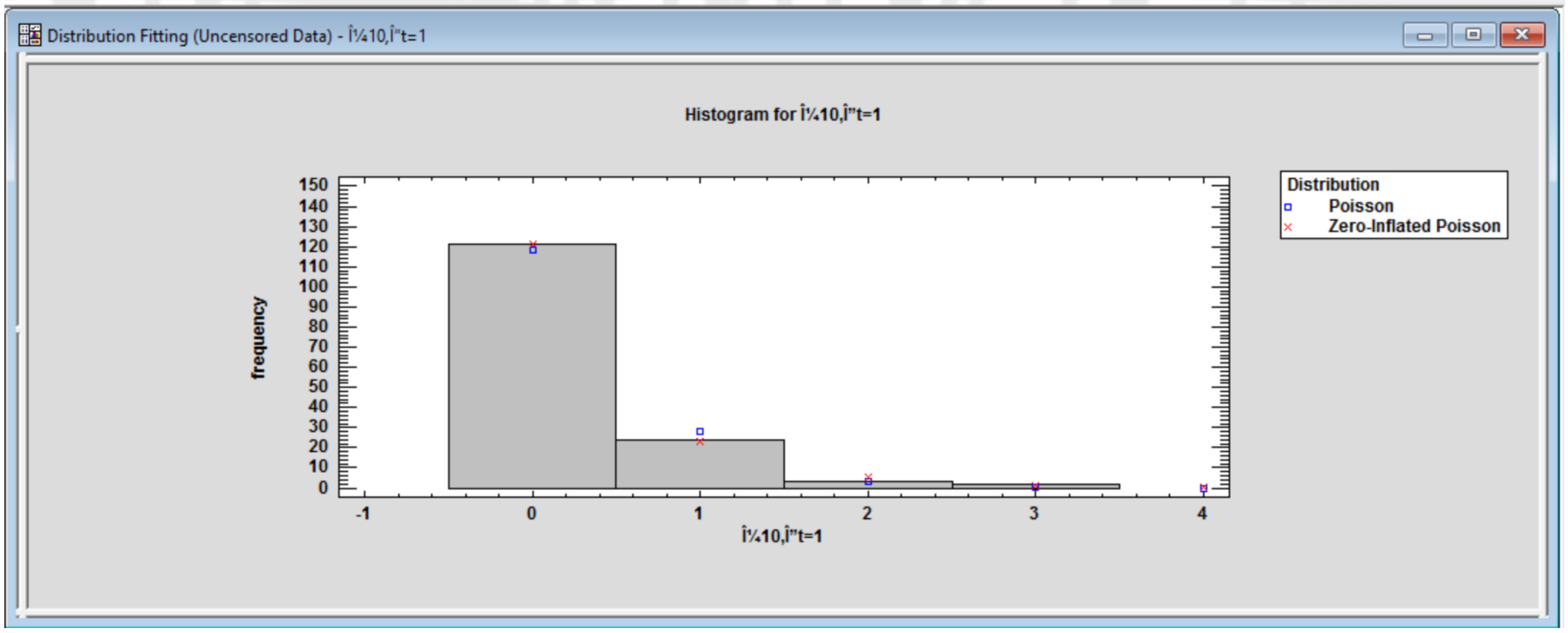

Figure 6.7: for transition from 2 to 1:

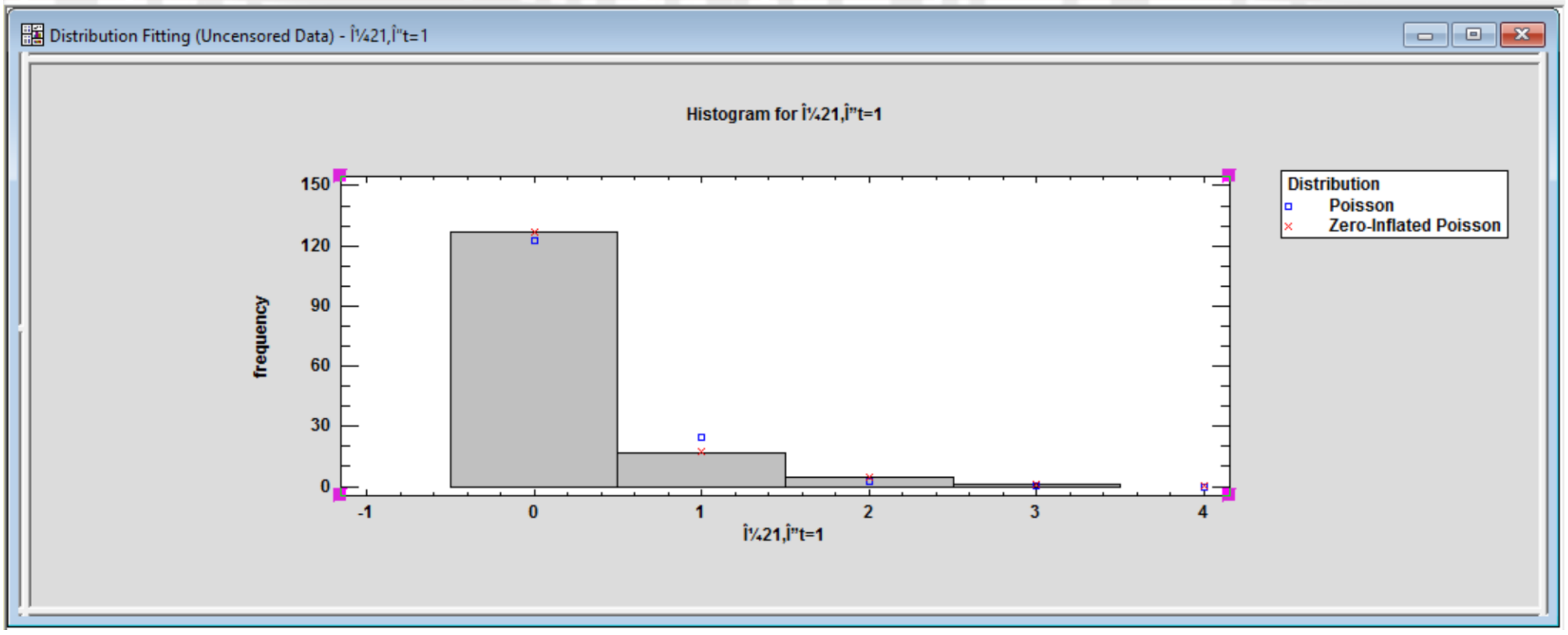

Figure 6.8: for transition from 3 to 2:

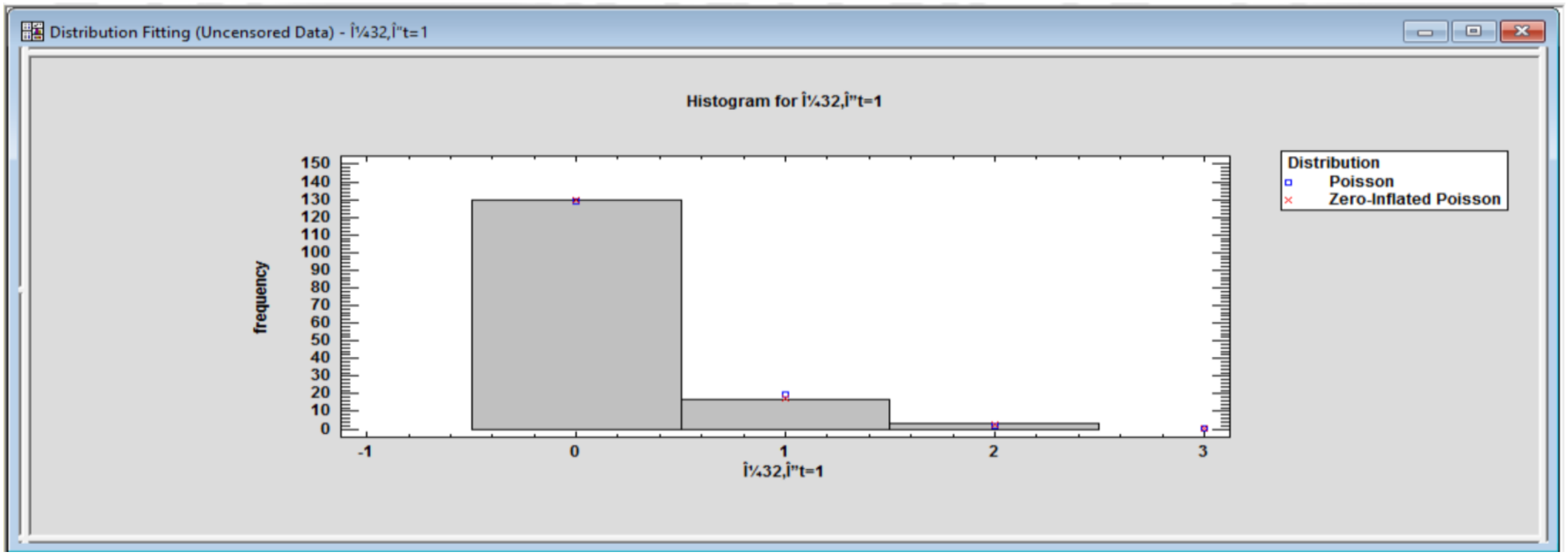

Figure 6.9: for transition from 2 to 0:

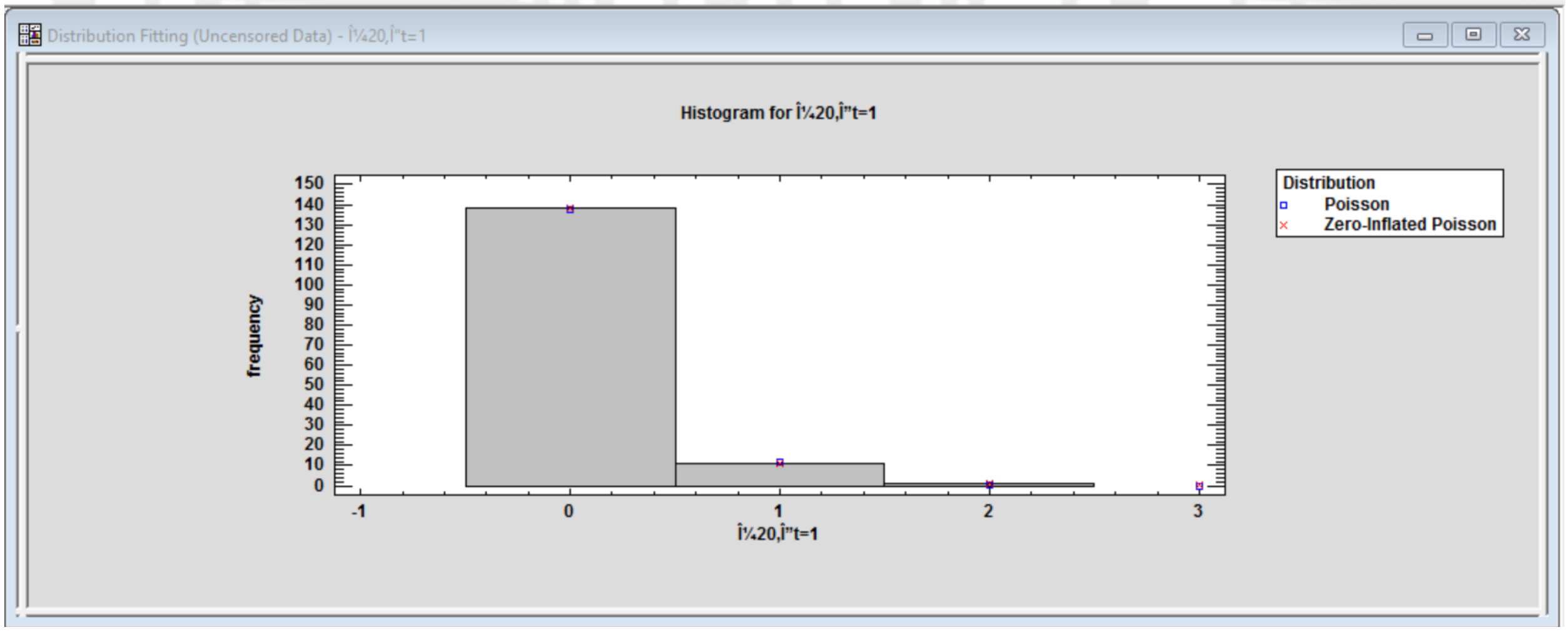

Figure 6.10 : for transition from 3 to 1:

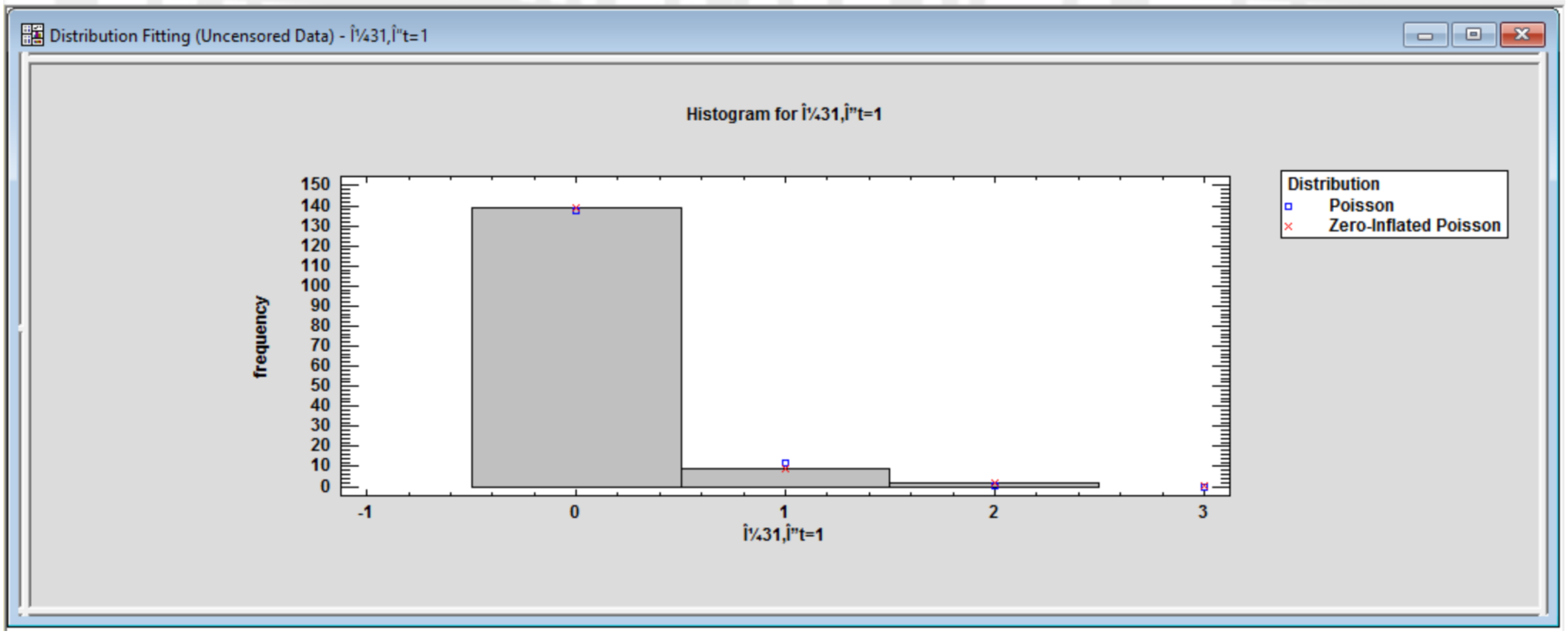

Lowess smoother illustrates that the relationships between each of the response rate and each variable is not strictly linear, but it is curvilinear relationship, with initial part of this relation being nearly horizontal and it starts to curve upwards at some predictor point located inside the second category of each predictor. The figures illustrating these relations are in the following successive figures from figure (6.11) to figure (6.19) for each response rate to the 7 variables. For example, relationship between number of transitions from state 0 to state 1 starts to bends up where each of the six predictors are located inside the second category; where age is approximately ≥37, BMI is approximately ≥ 26, LDL-chol is approximately ≥ 85 mg/dL, HOMA-IR is approximately ≥1.7, systolic blood pressure is approximately 142 mmHg, and diastolic blood pressure is approximately ≥ 85 mmHg. All these values are located in the second category. This can give good orientation to the functional form of the variables to be used in the regression model and avoid the misspecification resulting from mal-functional form of the predictors. In this work the restricted cubic splines are used for the predictors with 5 knots using Harrell approach which is the default procedure utilized by Stata 14 software. The locations of knots are illustrated in table (6.7) and correlations between the transformed variables are presented in table (6.8).

Table (6.7): location of knots for specified variables using Harrell approach (the default used in Stata 14)

| | Knot 1 | Knot 2 | Knot 3 | Knot 4 | Knot 5 |
|---|---|---|---|---|---|
| LDL-chol | 71.22 | 83.7 | 94.62 | 104.48 | 124.14 |
| HOMA2-IR | 1.09 | 1.8 | 2.26 | 2.75 | 3.48 |
| sysBloodPr. | 133.09 | 143.88 | 149.41 | 255.58 | 168.04 |
| Dias b lood Pr. | 74.45 | 87.44 | 94.07 | 101.11 | 114.49 |

Table (6.8): correlation between the transformed variables used in the Poisson regression models

| | LDLsp2 | HOMAsp1 | SYSsp2 | HOMAsp2 | DiasSP2 |
|---|---|---|---|---|---|
| LDLsp2 | 1 | | | | |
| HOMAsp1 | .8572 | 1 | | | |
| sysSP2 | .9959 | .8674 | 1 | .9908 | |
| HOMAsp2 | .9893 | | | 1 | |
| DiasSP2 | .9944 | | .9929 | .995 | 1 |

Figure 6.11: Transition 0 to 1:

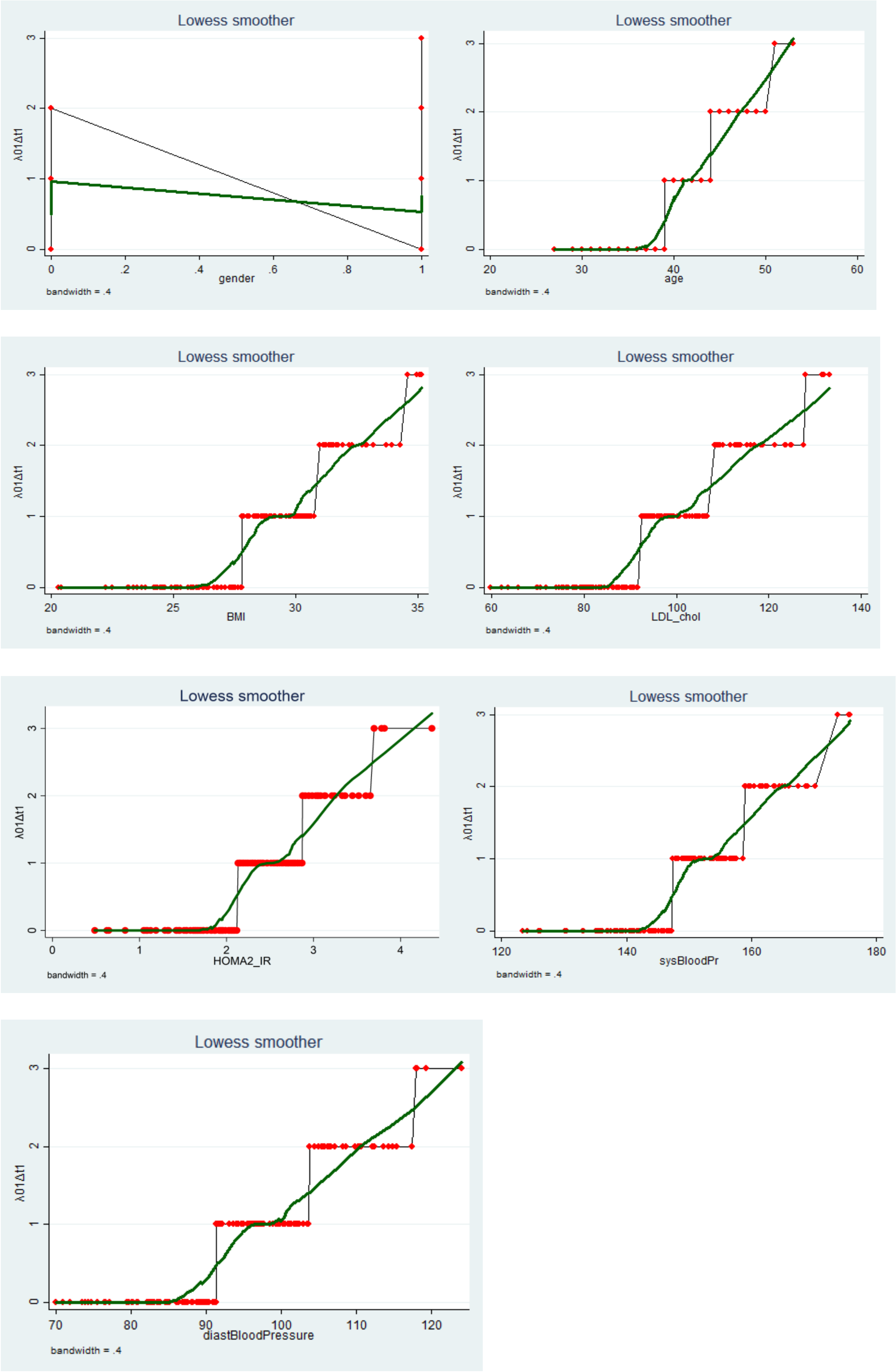

Figure 6.12: Transition 1 to 2:

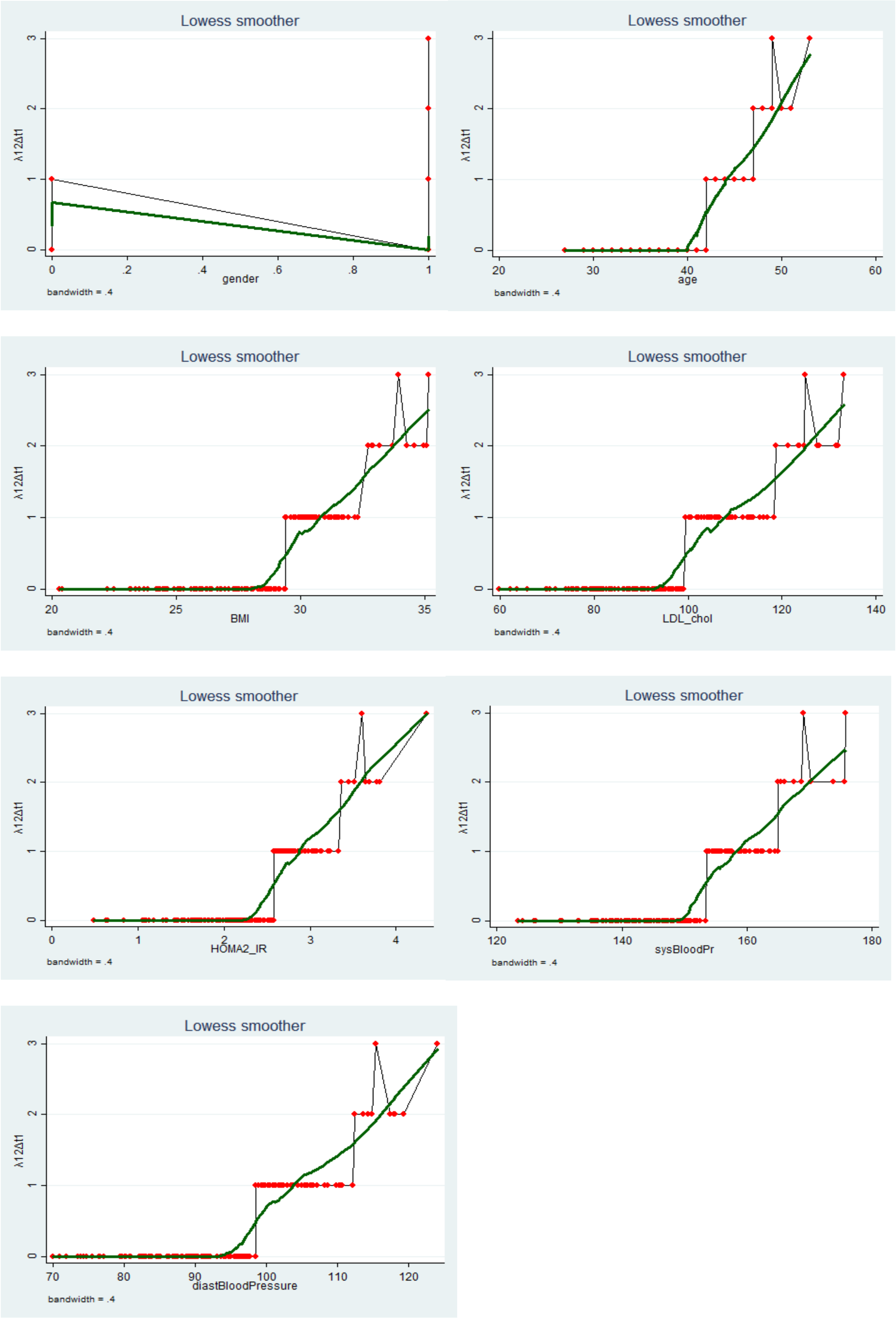

Figure 6.13: Transition 2 to 3:

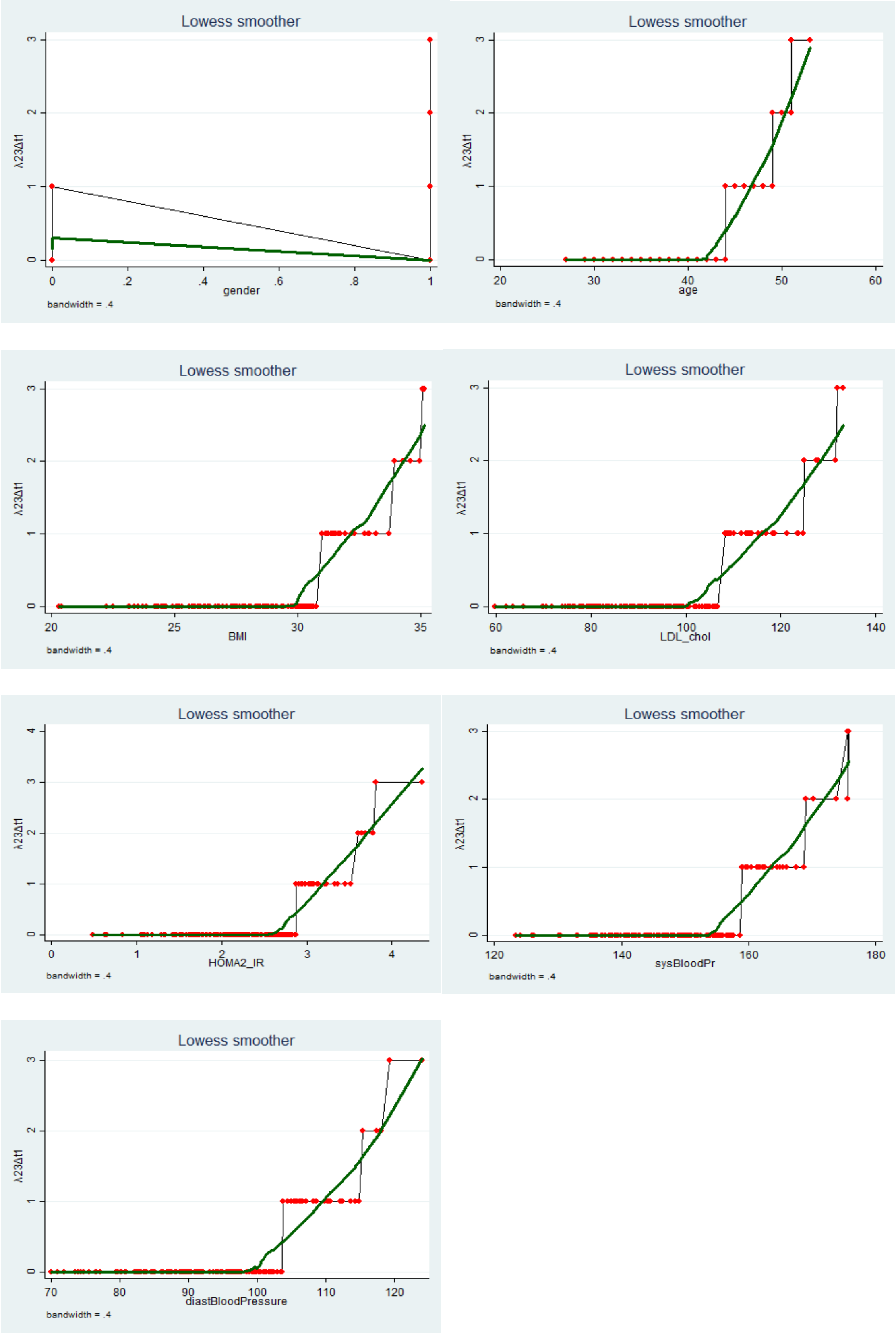

Figure 6.14: Transition 3 to 4:

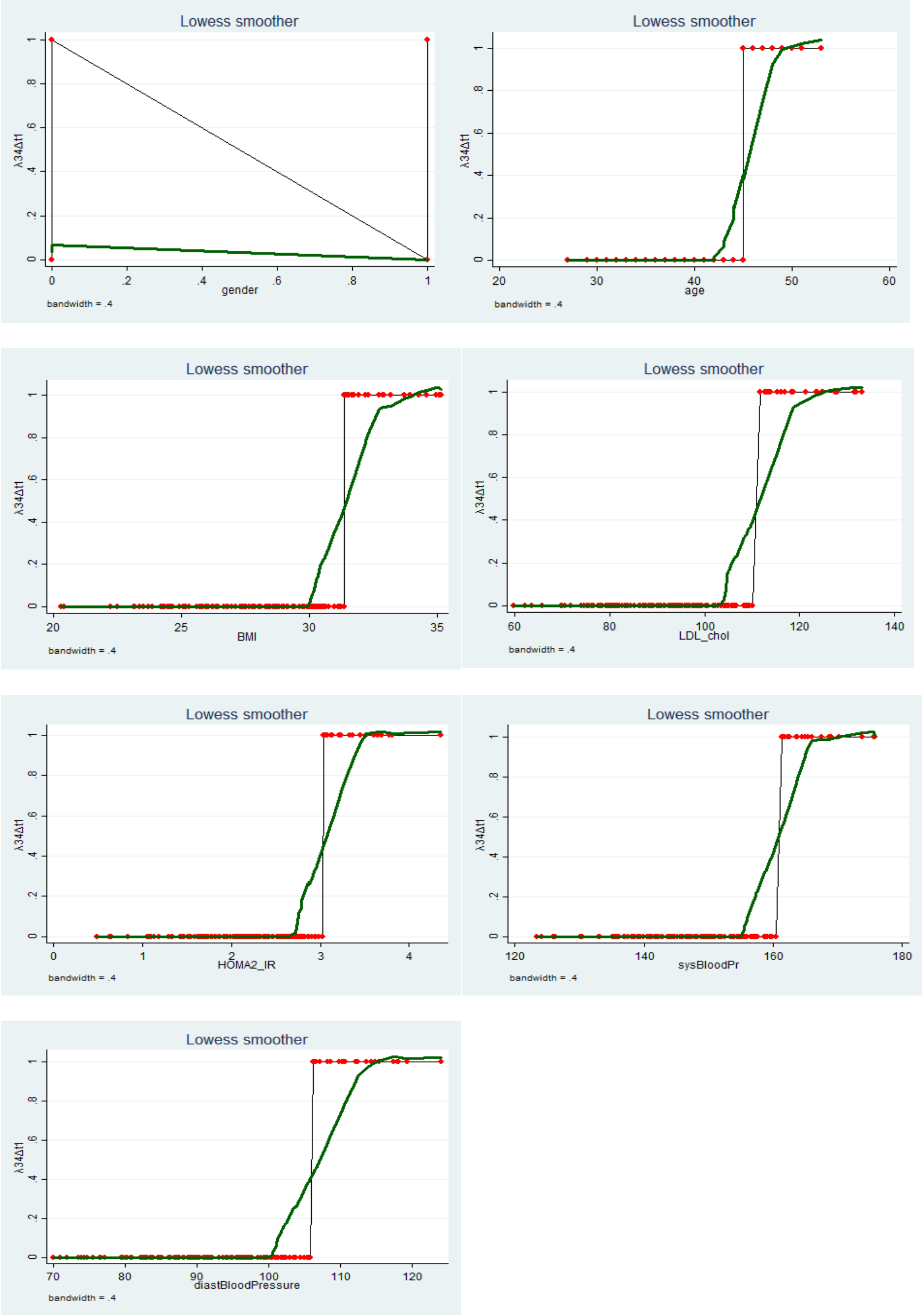

Figure 6.15: Transition 1 to 0:

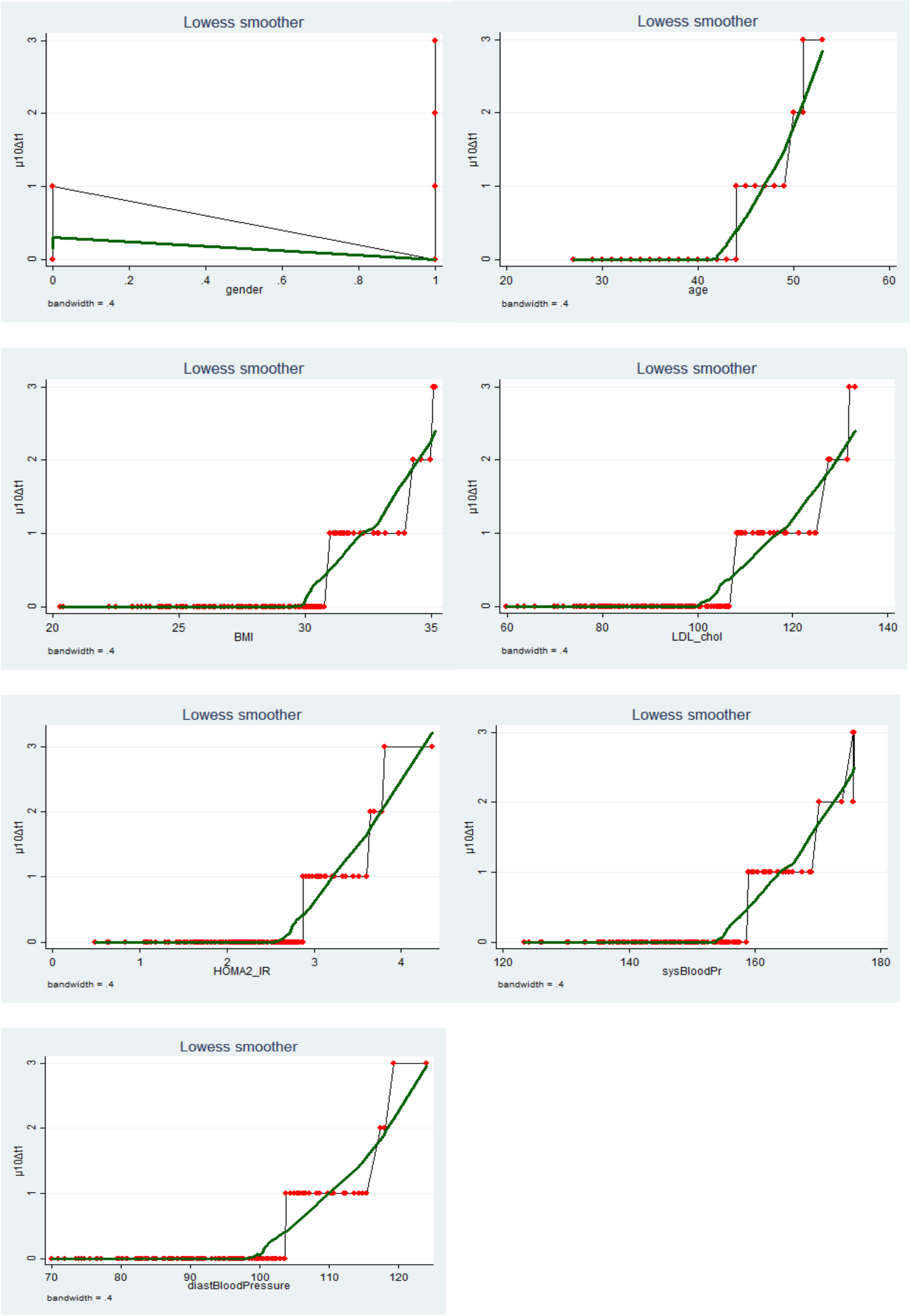

Figure 6.16: Transition 2 to 1:

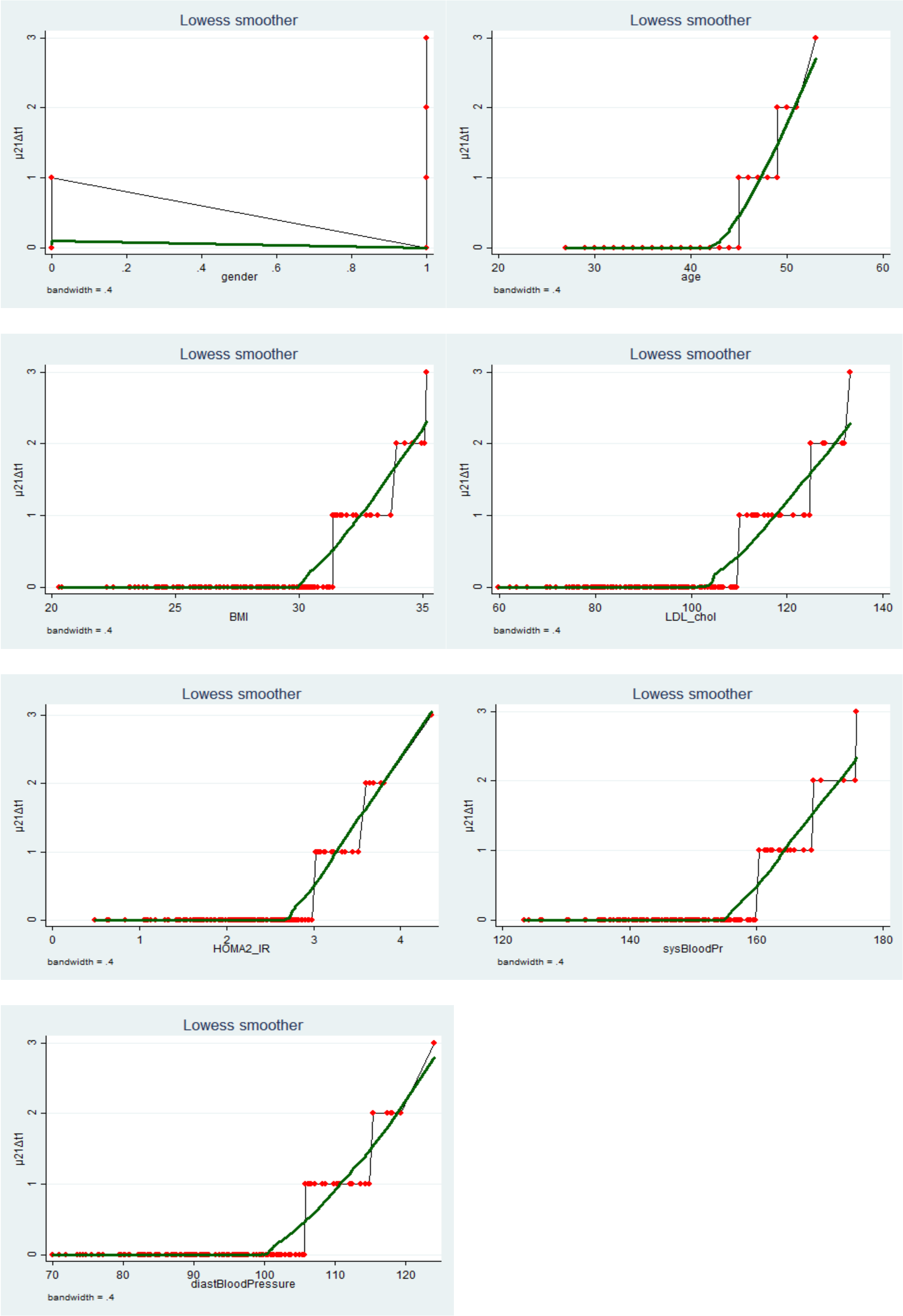

Figure 6.17: Transition 3 to 2 :

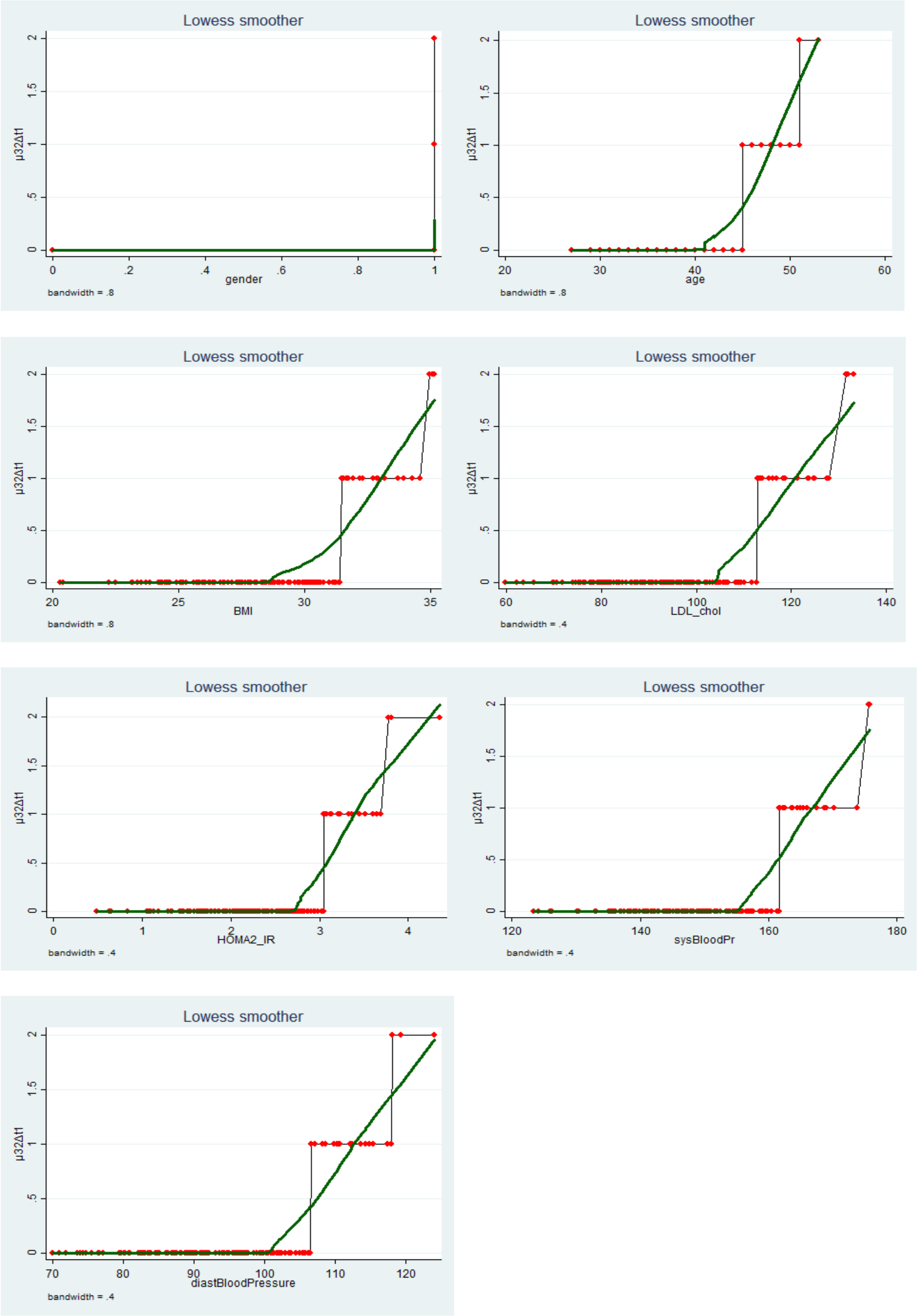

Figure 6.18: Transition 2 to 0 :

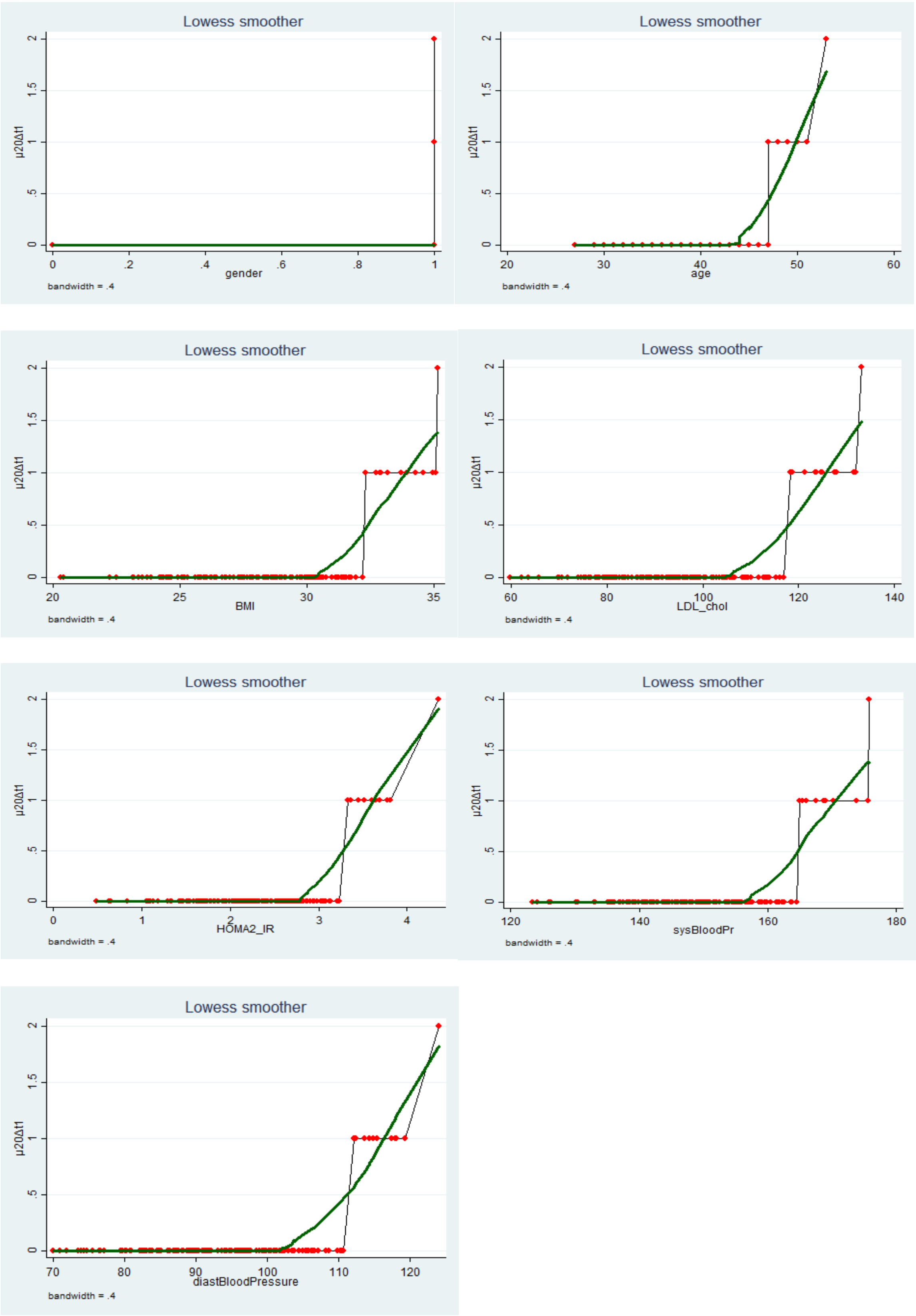

Figure 6.19: Transition 3 to 1:

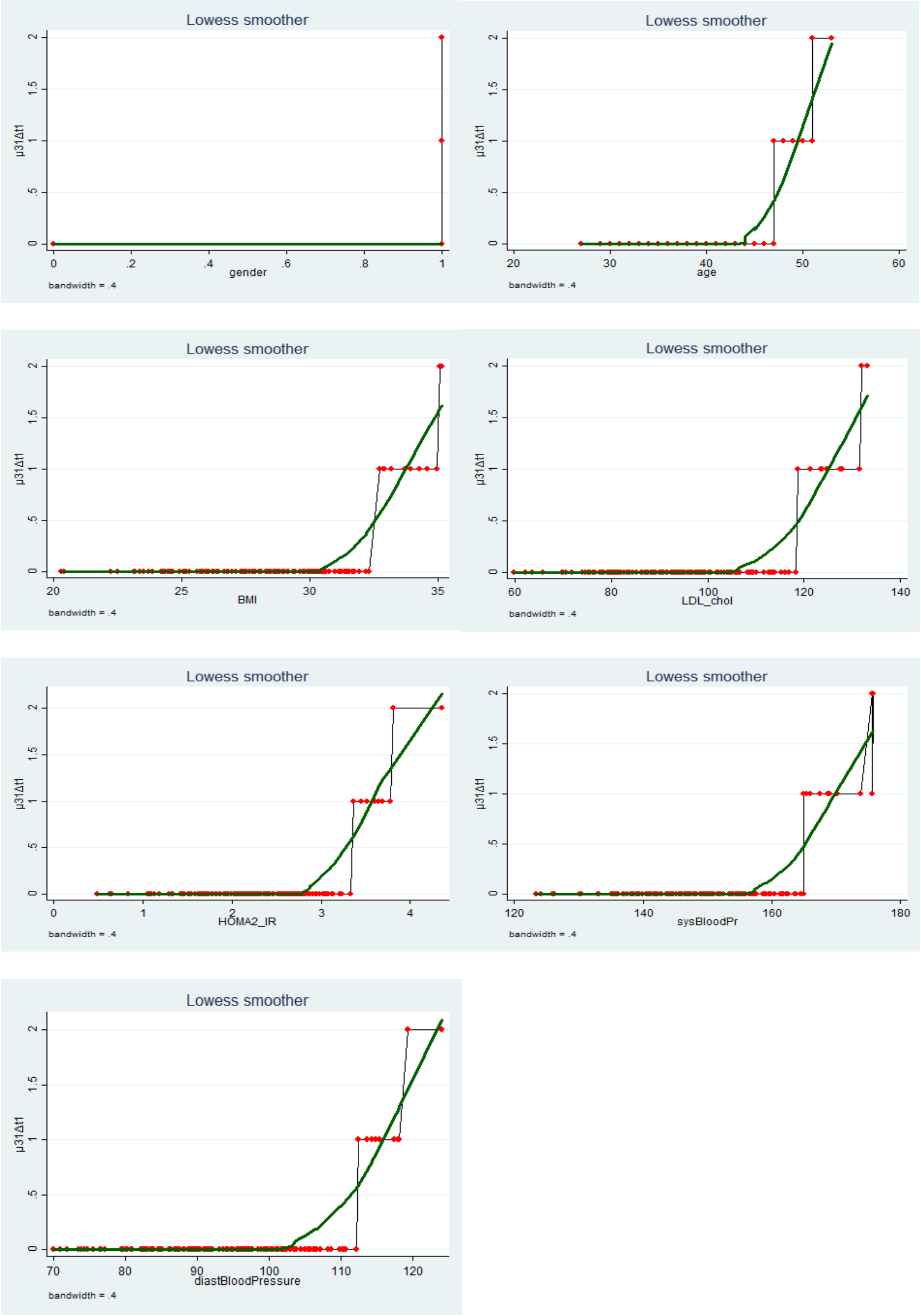

The Poisson regression was applied using the observed counts of the transition counts matrix as response variable, and the following results are obtained as discussed below in the next section.

### 6.2. Results and Discussion:

Fitting Poisson regression resulted in the following estimated counts as shown in table (6.9) and comparison of their distribution with the observed rates as shown in table (6.10):

Table 6.9: the estimated counts for each transition:

| **Counts** | **Transition 0→1** | **Transition 1→2** | **Transition 2→3** | **Transition 3→4** | **Transition 1→0** | **Transition 2→1** | **Transition 3→2** | **Transition 2→0** | **Transition 3→1** |
|---|---|---|---|---|---|---|---|---|---|
| **0** | 75 | 102 | 125 | 133 | 126 | 132 | 135 | 140 | 140 |
| **1** | 34 | 35 | 18 | 14 | 15 | 12 | 12 | 8 | 7 |
| **2** | 37 | 11 | 4 | 3 | 7 | 4 | 2 | 2 | 3 |
| **3** | 4 | 1 | 3 | 0 | 1 | 2 | 1 | 0 | 0 |
| **4** | 0 | 1 | 0 | 0 | 1 | 0 | 0 | 0 | 0 |
| **Total** | 150 | 150 | 150 | 150 | 150 | 150 | 150 | 150 | 150 |

Table 6.10: illustrates the comparisons between the distribution of the response rates and the estimated rates.

| | | Observed response count | Estimated mean response count |
|---|---|---|---|
| 0→1 | Mean | .8 | .8 |
| | Variance | .658 | .619 |
| | Std.dev. | .811 | .787 |
| 1→2 | Mean | .45 | .45 |
| | Variance | .45 | .45 |
| | Std.dev. | .671 | .671 |
| 2→3 | Mean | .25 | .25 |
| | Variance | .32 | .318 |
| | Std.dev. | .567 | .564 |
| 3→4 | Mean | .15 | .15 |
| | Variance | .126 | .126 |
| | Std.dev. | .355 | .355 |
| 1→0 | Mean | .24 | .24 |
| | Variance | .305 | .314 |
| | Std.dev. | .552 | .56 |
| 2→1 | Mean | .2 | .2 |
| | Variance | .268 | .284 |
| | Std.dev. | .518 | .533 |
| 3→2 | Mean | .15 | .15 |
| | Variance | .171 | .173 |
| | Std.dev. | .413 | .416 |
| 2→0 | Mean | .09 | .09 |
| | Variance | .093 | .101 |
| | Std.dev. | .305 | .318 |
| 3→1 | Mean | .09 | .09 |
| | Variance | .106 | .11 |
| | Std.dev. | .326 | .336 |

The results for each transition are demonstrated in the following tables from table (6.11) up to table (6.19) with a detailed discussion of the results.

Table 6.11: For transitions from 0 to 1:

| | Coefficient | Robust Std.dev. | z | p>\|z\| | 95%CI for B coeff. | IRR | Robust Std.dev. | 95% CI for IRR |
|---|---|---|---|---|---|---|---|---|
| **LDLsp2** | 0.523 | 0.243 | 2.149 | 0.032 | (0.046 , 1.000) | 1.687 | 0.411 | (1.047 , 2.718) |
| **HOMAsp1** | 4.096 | 0.328 | 12.470 | 0.000 | (3.452 , 4.740) | 60.097 | 19.739 | (31.569 , 114.403) |
| **sysSP2** | -0.628 | 0.347 | -1.809 | 0.070 | (-1.308 , 0.052) | 0.534 | 0.185 | (0.270 , 1.054) |
| **c.LDLsp2# c.HOMAsp1** | -0.179 | 0.070 | -2.540 | 0.011 | (-0.317 , -0.041) | 0.836 | 0.059 | (0.728 , 0.960) |
| **c.LDLsp2# c.sysSP2** | 0.003 | 0.000 | 8.144 | 0.000 | (0.002 , 0.003) | 1.003 | 0.000 | (1.002 , 1.003) |
| **c.sysSP2# c.HOMAsp1** | 0.151 | 0.098 | 1.547 | 0.122 | (-0.040 , 0.342) | 1.163 | 0.113 | (0.960 , 1.408) |
| **cons** | -9.510 | 0.725 | -13.122 | 0.000 | (-10.930 , -8.089) | 0.000 | 0.000 | (0.000 , 0.000) |
| **Log pseudolikelihood** = -110.43 | | | | | **Pseudo $R^2$** = .355 | | | |
| **Wald chi2(6)** = 535.34 | | | | | **Prob>chi2** =.000 | | | |
| **Deviance gof** = 27.55 | | | | | **Prob>chi2(143)** = 1 | | | |
| **Pearson gof** = 24.458 | | | | | **Prob>chi2(143)** = 1 | | | |
| **Log pseudolikelihood(full model )** = -110.43 | | | | | **Log pseudolikelihood(null model )** = -171.273 | | | |
| **DF** = 7 | | | **AIC** = 234.86 | | | **BIC** = 255.935 | | |

Using the above results to predict the $\ln \lambda_{01} = x_i' B$ then calculating $E[y_i|x_i] = \lambda_{01} = e^{x_i' B}$ where $\lambda_{01}$ denotes the transition counts from state 0 to state 1 and as the time for the follow up is equal to all patients. The expected counts for the transition from state 0 to state 1 are 120 transitions and the observed count is 120. The expected increase in log count for one-unit increase in transformed LDL cholesterol is (0.523), which is not highly statistically significant (P=0.032), and for one-unit increase in transformed HOMA is (4.096), which is highly statistically significant (P=0.000), as both are considered risk factors for NAFLD to progress from F0 to F1. The expected decrease in log count for one-unit increase in transformed systolic blood pressure is (0.628) which is not statistically significant (P=0.07). For every unit increase in transformed LDL, the incident rate ratio is increased (increase in transition counts) by 68.7%; while, for transformed HOMA, it is increased by 5909.7%, with 95% confidence that this increase is between 3056.9% and 11340.3%. The expected decrease in log count for one unit increase in interaction between the transformed LDL and transformed HOMA is (0.179) with high statistical significance (P=0.11), in other word, the rise in one predictor variable decreases the rising effect of the other on the response variable (expected log count) but not reverse it. And for every unit increase in this interaction, the incident rate ratio is decreased (i.e. decrease in transition counts) by 16.4%, with 95% confidence, that this decrease lies between 4% and 27.2%. While, the expected increase in log count for one unit increase in interaction between the transformed systolic blood pressure and transformed LDL is (0.003) with high statistical significance (P=0.000), and this increase in log count for one unit increase in interaction between transformed systolic blood pressure and transformed HOMA is (0.151) which is not statistically significant (P=0.122), in other word, the rise in one predictor variable increases the rising effect of the other on the response variable (expected log count).However, for every unit increase in the first interaction, the incident rate ratio is only increased (i.e. increase in transition counts) by 0.3%, with 95% confidence, that this increase lies between .2% and .31% , while for the second interaction; the IRR is increased by 16.3%. Poisson model fits the data as goodness of fit is not statistically significant (P=1).

Table 6.12 : For transitions from 1 to 2:

| | Coefficient | Robust Std.dev. | z | p>\|z\| | 95%CI for B coeff. | IRR | Robust Std.dev. | 95% CI for IRR |
|---|---|---|---|---|---|---|---|---|
| **LDLsp2** | 0.311 | 0.396 | 0.785 | 0.432 | (-0.465 , 1.086) | 1.364 | 0.540 | (0.628 , 2.962) |
| **HOMAsp1** | 5.486 | 0.571 | 9.599 | 0.000 | (4.366 , 6.606) | 241.179 | 137.821 | (78.690 , 739.192) |
| **sysSP2** | -0.314 | 0.545 | -0.577 | 0.564 | (-1.383 , 0.754) | 0.730 | 0.398 | (0.251 , 2.126) |
| **c.LDLsp2# c.HOMAsp1** | -0.105 | 0.116 | -0.902 | 0.367 | (-0.332 , 0.123) | 0.901 | 0.105 | (0.717 , 1.131) |
| **c.sysSP2# c.HOMAsp1** | 0.079 | 0.158 | 0.502 | 0.616 | (-0.231 , 0.389) | 1.083 | 0.171 | (0.794 , 1.476) |
| **cons** | -14.884 | 1.363 | -10.923 | 0.000 | (-17.555 , -12.213) | 0.000 | 0.000 | (0.000 , 0.000) |
| **Log pseudolikelihood** = -67.887 | | | | | **Pseudo $R^2$** = 0.481 | | | |
| **Wald chi2(6)** = 284.3 | | | | | **Prob>chi2** = 0.000 | | | |
| **Deviance gof** = 20.27 | | | | | **Prob>chi2(144)** = 1 | | | |
| **Pearson gof** = 18.12 | | | | | **Prob>chi2(144)** = 1 | | | |
| **Log pseudolikelihood( full model )** = -67.887 | | | | | **Log pseudolikelihood( null model )** = -130.82 | | | |
| **DF** = 6 | | | **AIC** = 147.77 | | | **BIC** = 165.84 | | |

Using the above results to predict the $\ln \lambda_{12} = x_i' B$ then calculating $E[y_i|x_i] = \lambda_{12} = e^{x_i' B}$ where $\lambda_{12}$ denotes the transition counts from state 1 to state 2 and as the time for the follow up is equal to all patients the expected counts for the transition from state 1 to state 2 are 64 transitions with while the observed count is 67 transitions . The expected increase in log count for one-unit increase in transformed LDL cholesterol is (0.311), which is not highly statistically significant (P=0.432), and for one-unit increase in transformed HOMA is (5.486), which is highly statistically significant (P=0.000), as both are considered risk factors for NAFLD to progress from F1 to F2. The expected decrease in log count for one-unit increase in transformed systolic blood pressure is (0.314), which is not statistically significant (P= 0.564). For every unit increase in transformed LDL the incident rate ratio is increased (increase in transition counts) by 36.4%; while, for transformed HOMA it is increased by 24017.9%, with 95% confidence that this increase is between 7769% and 73819.2%. The expected decrease in log count for one unit increase in interaction between the transformed LDL and transformed HOMA is (0.105) with no statistical significance (P=0.367), in other word, the rise in one predictor variable decreases the rising effect of the other on the response variable (expected log count) but not reverse it. While, the expected increase in log count for one unit increase in interaction between the transformed systolic blood pressure and transformed HOMA is (0.079) with no statistical significance (P=0.616). Poisson model fits the data as goodness of fit is not statistically significant (P=1).

Table 6.13: For transition from 2 to 3 :

| | Coefficient | Robust Std.dev. | z | p>\|z\| | 95%CI for B coeff. | IRR | Robust Std.dev. | 95% CI for IRR |
|---|---|---|---|---|---|---|---|---|
| **LDLsp2** | -1.480 | 0.685 | -2.159 | 0.031 | (-2.823 , -0.137) | 0.228 | 0.156 | (0.059 , 0.872) |
| **HOMAsp1** | 6.174 | 3.093 | 1.996 | 0.046 | (0.112 , 12.237) | 480.318 | 1485.815 | (1.118 , 2.06e+5) |
| **sysSP2** | 2.497 | 0.967 | 2.583 | 0.010 | (0.602 , 4.391) | 12.143 | 11.738 | (1.826 , 80.754) |
| **c.LDLsp2# c.HOMAsp1** | 0.390 | 0.192 | 2.032 | 0.042 | (0.014 , 0.766) | 1.477 | 0.283 | (1.014 , 2.151) |
| **c.LDLsp2# c.sysSP2** | -0.001 | 0.002 | -0.403 | 0.687 | (-0.005 , 0.004) | 0.999 | 0.002 | (0.995 , 1.004) |
| **c.sysSP2# c.HOMAsp1** | -0.655 | 0.274 | -2.392 | 0.017 | (-1.191 , -0.118) | 0.520 | 0.142 | (0.304 , 0.889) |
| **cons** | -20.866 | 7.293 | -2.861 | 0.004 | (-35.160 , -6.572) | 0.000 | 0.000 | (0.000 , 0.001) |
| **Log pseudolikelihood** = -37.87 | | | | | **Pseudo $R^2$** = 0.6020 | | | |
| **Wald chi2(6)** = 191.48 | | | | | **Prob>chi2** = 0.000 | | | |
| **Deviance gof** = 13.29 | | | | | **Prob>chi2(143)** = 1 | | | |
| **Pearson gof** = 12.42 | | | | | **Prob>chi2(143)** = 1 | | | |
| **Log pseudolikelihood(full model )** = -37.87 | | | | | **Log pseudolikelihood(null model )** = -95.15 | | | |
| **DF** = 7 | | | **AIC** = 89.73 | | | **BIC** = 110.81 | | |

Using the above results to predict the $\ln \lambda_{23} = x_i'B$ then calculating $E[y_i|x_i] = \lambda_{23} = e^{x_i'B}$
Where $\lambda_{23}$ denotes the transition counts from state 2 to state 3 and as the time for the follow up is equal to all patients the expected counts for the transition from state 2 to state 3 are 35 transitions with good fit of the model while the observed count is 37 transitions . The expected decrease in log count for one-unit increase in transformed LDL cholesterol is (1.48) which is not highly statistically significant (P=0.031). This effect is not explainable. The expected increase in log count for one-unit increase in transformed HOMA is (6.174) which is not highly statistically significant (P=0.046), and for one-unit increase in transformed systolic blood pressure is (2.497) which is statistically significant (P=0.01), as both are considered risk factors for NAFLD to progress from F2 to F3. For every unit increase in transformed systolic blood pressure, the incident rate ratio is increased (increase in transition counts) by 1114.3%, with 95% confidence that this increase is between 82.6% and 7975.4%. The expected increase in log count for one unit increase in interaction between the transformed LDL and transformed HOMA is (0.39) with no high statistical significance (P=0.042), in other word, the rise in one predictor variable increases the rising effect of the other on the response variable (expected log count). While, the expected decrease in log count for one unit increase in interaction between the transformed systolic blood pressure and transformed LDL is (0.001) with no statistical significance (P=0.687), and this decrease in log count for one unit increase in interaction between transformed systolic blood pressure and transformed HOMA is (0.655) which is highly statistically significant (P=0.017), in other word, the rise in one predictor variable decreases the rising effect of the other on the response variable (expected log count) but not reverse it. However, for every unit increase in the first interaction, the incident rate ratio is only decreased (i.e. decrease in transition counts) by 0.1%, and for the second interaction; the IRR is decreased by 48%, with 95% confidence that this decrease is between 11.1% and 69.6%. Poisson model fits the data as goodness of fit is not statistically significant.

Table 6.14: For transition from 3 to 4:

| | Coefficient | Robust Std.dev. | z | p>\|z\| | 95%CI for B coeff. | IRR | Robust Std.dev. | 95%CI for IRR |
|---|---|---|---|---|---|---|---|---|
| **LDLsp2** | 0.452 | 0.055 | 8.278 | 0.000 | (0.345 , 0.559) | 1.571 | 0.086 | (1.412 , 1.748) |
| **HOMAsp1** | 10.866 | 1.402 | 7.753 | 0.000 | (8.119 , 13.613) | 52375.984 | 73411.34 | (3357.9 , 8.17e+5) |
| **sysSP2** | 0.073 | 0.050 | 1.472 | 0.141 | (-0.024 , 0.171) | 1.076 | 0.054 | (0.976 , 1.187) |
| **c.LDLsp2# c.HOMAsp1** | -0.166 | 0.018 | -9.320 | 0.000 | (-0.201 , -0.131) | 0.847 | 0.015 | (0.818 , 0.877) |
| **cons** | -34.034 | 3.865 | -8.806 | 0.000 | (-41.608 , -26.459) | 0.000 | 0.000 | (0.000 , 0.000) |
| **Log pseudolikelihood** = -26.97 | | | | | **Pseudo $R^2$** = 0.58 | | | |
| **Wald chi2(6)** = 122.33 | | | | | **Prob>chi2** = 0.000 | | | |
| **Deviance gof** = 9.94 | | | | | **Prob>chi2(145)** = 1 | | | |
| **Pearson gof** = 8.96 | | | | | **Prob>chi2(145)** = 1 | | | |
| **Log pseudolikelihood(full model )** = -26.97 | | | | | **Log pseudolikelihood(null model )** = -64.23 | | | |
| **DF** = 5 | | | **AIC** = 63.94 | | | **BIC** = 78.99 | | |

Using the above results to predict the $\ln \lambda_{34} = x_i' B$ then calculating $E[y_i|x_i] = \lambda_{34} = e^{x_i' B}$ Where $\lambda_{34}$ denotes the transition counts from state 3 to state 4 and as the time for the follow up is equal to all patients the expected counts for the transition from state 3 to state 4 are 20 transitions with good fit of the model while the observed count is 22 transitions . The expected increase in log count for one-unit increase in transformed LDL cholesterol is (0.452) which is highly statistically significant (P=0.000), this expected increase in log count for one-unit increase in transformed HOMA is (10.866) which is also highly statistically significant (P=0.000), and it is for one-unit increase in transformed systolic blood pressure is (0.073) which is not statistically significant (P=0.141), as all are considered risk factors for NAFLD to progress from F3 to F4. For every unit increase in transformed LDL, the incident rate ratio is increased (increase in transition counts) by 57.1%, with 95% confidence that this increase is between 41.2% and 74.8%. Furthermore, for every unit increase in transformed HOMA, the incident rate ratio is increased (increase in transition counts) by 5237498.4%, with 95% confidence that this increase is between 335691.1% and 81.7e+6%. For every unit increase in this interaction, the incident rate ratio is decreased (i.e. decrease in transition counts) by 15.3%, with 95% confidence that this decrease is between 12.3% and 18.2%. Poisson model fits the data as goodness of fit is not statistically significant (P=1).

Table 6.15: For transition from 1 to 0 :

| | Coefficient | Robust Std.dev. | z | p>\|z\| | 95%CI for B coeff. | IRR | Robust Std.dev. | 95% CI for IRR |
|---|---|---|---|---|---|---|---|---|
| **LDLsp2** | -0.454 | 0.244 | -1.862 | 0.063 | (-0.932 , 0.024) | 0.635 | 0.155 | (0.394 , 1.024) |
| **HOMAsp2** | -4.489 | 2.962 | -1.515 | 0.130 | (-10.294 , 1.316) | 0.011 | 0.033 | (0.000 , 3.730) |
| **sysSP2** | 1.340 | 0.312 | 4.301 | 0.000 | (0.729 , 1.951) | 3.820 | 1.190 | (2.074 , 7.034) |
| **c.LDLsp2# c.HOMAsp2** | 0.290 | 0.096 | 3.029 | 0.002 | (0.102 , 0.478) | 1.337 | 0.128 | (1.108 , 1.612) |
| **c.LDLsp2# c.sysSP2** | -0.010 | 0.004 | -2.789 | 0.005 | (-0.017 , - 0.003) | 0.990 | 0.004 | (0.983 , 0.997) |
| **c.sysSP2# c.HOMAsp2** | -0.286 | 0.145 | -1.974 | 0.048 | (-0.571 , -0.002) | 0.751 | 0.109 | (0.565 , 0.998) |
| **cons** | -5.916 | 0.508 | -11.651 | 0.000 | (-6.912 , -4.921) | 0.003 | 0.001 | (0.001 , 0.007) |
| **Log pseudolikelihood** = -38.14 | | | | | **Pseudo $R^2$** = 0.59 | | | |
| **Wald chi2(6)** = 331.08 | | | | | **Prob>chi2** = 0.000 | | | |
| **Deviance gof** = 14.46 | | | | | **Prob>chi2(143)** = 1 | | | |
| **Pearson gof** = 13.55 | | | | | **Prob>chi2(143)** = 1 | | | |
| **Log pseudolikelihood(full model )** = -38.14 | | | | | **Log pseudolikelihood(null model )** = -93.04 | | | |
| **DF** = 7 | | | **AIC** = 90.29 | | | **BIC** = 111.36 | | |

Using the above results to predict the $\ln \mu_{10} = x_i' B$ then calculating $E[y_i|x_i] = \mu_{10} = e^{x_i' B}$
Where $\mu_{10}$ denotes the transition counts from state 1 to state 0 and as the time for the follow up is equal to all patients the expected counts for the transition from state 1 to state 0 are 36 transitions with good fit of the model while the observed count is 36 transitions . The expected decrease in log count for one-unit increase in transformed LDL cholesterol is (0.454), which is not highly statistically significant (P= 0.063), and the expected decrease in log count for one-unit increase in transformed HOMA is (4.489), which is not statistically significant (P= 0.13). As better management for both of these risk factors enhance the transition from F1 to F0. While the expected increase in log count for one-unit increase in transformed systolic blood pressure is (1.34), which is highly statistically significant (p=0.000). For every unit increase in transformed systolic blood pressure, the incident rate ratio is increased (increase in transition counts) by 282%, with 95% confidence that this increase is between 107.4% and 603.4%. This effect is not really explainable and further studies are needed to evaluate such effect as there may be some confounder substances that could induce such effect. The expected increase in log count for one unit increase in interaction between the transformed LDL and transformed HOMA is (0.29) with high statistical significance (p=0.002), in other word, the rise in one predictor variable increases the effect of the other on the response variable (expected log count). For every unit increase of this interaction, the incident rate ratio is increased (increase in transition counts) by 33.7%, with 95% confidence that this increase is between 10.8% and 61.2%. While, the expected decrease in log count for one unit increase in interaction between the transformed systolic blood pressure and transformed LDL is (0.01) with high statistical significance (p=0.005), and this decrease in log count for one unit increase in interaction between transformed systolic blood pressure and transformed HOMA is (0.286) which is not highly statistically significant (p=0.048), in other word, the rise in one predictor variable decreases the effect of the other on the response variable (expected log count). However, for every unit increase in the first interaction, the incident rate ratio is only decreased (i.e. decrease in transition counts) by 1%, with 95% confidence that this decrease is between 0.3% and 1.7%. Moreover, for the second interaction; the IRR is decreased by 24.9%. Poisson model fits the data as goodness of fit is not statistically significant (P=1).

Figure 6.16 : Transition from 2 to 1:

| | Coefficient | Robust Std.dev. | z | p>\|z\| | 95%CI for B coeff. | IRR | Robust Std.dev. | 95%CI for IRR |
|---|---|---|---|---|---|---|---|---|
| **LDLsp2** | -0.128 | 0.189 | -0.675 | 0.499 | (-0.499 , 0.243) | 0.880 | 0.167 | (0.607 , 1.275) |
| **HOMAsp2** | -3.288 | 2.812 | -1.169 | 0.242 | (-8.800 , 2.224) | 0.037 | 0.105 | (0.000 , 9.244) |
| **sysSP2** | 0.913 | 0.201 | 4.546 | 0.000 | (0.519 , 1.307) | 2.492 | 0.501 | (1.681 , 3.694) |
| **c.LDLsp2# c.HOMAsp2** | 0.152 | 0.066 | 2.288 | 0.022 | (0.022 , 0.282) | 1.164 | 0.077 | (1.022 , 1.326) |
| **c.LDLsp2# c.sysSP2** | -0.010 | 0.003 | -2.950 | 0.003 | (-0.017 , -0.003) | 0.990 | 0.003 | (0.983 , 0.997) |
| **c.sysSP2# c.HOMAsp2** | -0.114 | 0.114 | -1.001 | 0.317 | (-0.338 , 0.109) | 0.892 | 0.102 | (0.713 , 1.116) |
| **cons** | -7.666 | 0.617 | -12.426 | 0.000 | (-8.875 , -6.457) | 0.000 | 0.000 | (0.000 , 0.002) |
| **Log pseudolikelihood** = -29.96 | | | | | **Pseudo $R^2$** = 0.64 | | | |
| **Wald chi2(6)** = 304.94 | | | | | **Prob>chi2** = 0.000 | | | |
| **Deviance gof** = 9.86 | | | | | **Prob>chi2(143)** = 1 | | | |
| **Pearson gof** = 8.97 | | | | | **Prob>chi2(143)** = 1 | | | |
| **Log pseudolikelihood(full model )** = -29.96 | | | | | **Log pseudolikelihood(null model )** = -83.54 | | | |
| **DF** = 7 | | | **AIC** = 73.92 | | | **BIC** = 94.99 | | |

Using the above results to predict the $\ln \mu_{21} = x_i' B$ then calculating $E[y_i|x_i] = \mu_{21} = e^{x_i' B}$
Where $\mu_{21}$ denotes the transition counts from state 2 to state 1 and as the time for the follow up is equal to all patients the expected counts for the transition from state 2 to state 1 are 26 transitions with good fit of the model while the observed count is 30 transitions. The expected decrease in log count for one-unit increase in transformed LDL cholesterol is (0.128) which is not statistically significant(P=0.499), and the expected decrease in log count for one-unit increase in transformed HOMA is (3.288) which is not statistically significant (P=0.242). As better management of both of these risk factors, enhances the transition from F2 to F1. While the expected increase in log count for one-unit increase in transformed systolic blood pressure is (0.913) which is highly statistically significant (P=0.000). For every unit increase in transformed systolic blood pressure, the incident rate ratio is increased (increase in transition counts) by 149.2%, with 95% confidence that this increase is between 68.1% and 269.4%. This effect is not really explainable and further studies are needed to evaluate such effect as there may be some confounder substances that could induce such effect. The expected increase in log count for one unit increase in interaction between the transformed LDL and transformed HOMA is (0.152) with high statistical significance (P=0.022), in other word, the rise in one predictor variable increases the effect of the other on the response variable (expected log count). For every unit increase of this interaction, the incident rate ratio is increased (increase in transition counts) by 16.4%, with 95% confidence that this increase is between 2.2% and 32.6%. While, the expected decrease in log count for one unit increase in interaction between the transformed systolic blood pressure and transformed LDL is (0.01) with high statistical significance (P=0.003), and this decrease in log count for one unit increase in interaction between transformed systolic blood pressure and transformed HOMA is (0.114) which is not statistically significant (P=0.317), in other word, the rise in one predictor variable decreases the effect of the other on the response variable (expected log count) but not reverse it. However, for every unit increase in the first interaction, the incident rate ratio is only decreased (i.e. decrease in transition counts) by 1%, with 95% confidence that this decrease is between 0.3% and 1.7%. Moreover, for the second interaction; the IRR is decreased by 10.8%. Poisson model fits the data as goodness of fit is not statistically significant .

Table 6.17: Transition from 3 to 2 :

| | Coefficient | Robust Std.dev. | z | p>\|z\| | 95%CI for B coeff. | IRR | Robust Std.dev. | 95% CI for IRR |
|---|---|---|---|---|---|---|---|---|
| **LDLsp2** | .302 | .211 | 1.427 | .154 | (-.113 , 0.716) | 1.352 | .286 | (0.893 , 2.047) |
| **HOMAsp2** | -5.214 | 3.196 | -1.631 | .103 | (-11.478 , 1.05) | .005 | .017 | (0.000 , 2.859) |
| **sysSP2** | .422 | .288 | 1.467 | .142 | (-0.142 , .987) | 1.526 | .439 | (.868 , 2.683) |
| **c.LDLsp2# c.HOMAsp2** | .002 | .102 | .019 | .984 | (-.198 , .202) | 1.002 | .102 | (.821 , 1.223) |
| **c.LDLsp2# c.sysSP2** | -.012 | .004 | -2.749 | .006 | (-.02 ,- .003) | 0.998 | .004 | (0.98 , .997) |
| **c.sysSP2# c.HOMAsp2** | .132 | .149 | 0.888 | .375 | (-0.16 , .425) | 1.142 | .17 | (.852 , 1.529) |
| **cons** | -7.363 | .761 | -9.671 | .000 | (-8.855 , -5.871) | .001 | .000 | (.000 , .003) |
| **Log pseudolikelihood** = -26.37 | | | | | **Pseudo $R^2$** = .61 | | | |
| **Wald chi2(6)** = 175.47 | | | | | **Prob>chi2** =.000 | | | |
| **Deviance gof** = 10.89 | | | | | **Prob>chi2(143)** = 1 | | | |
| **Pearson gof** = 9.77 | | | | | **Prob>chi2(143)** = 1 | | | |
| **Log pseudolikelihood(full model )** = -26.37 | | | | | **Log pseudolikelihood(null model )** = -68.21 | | | |
| **DF** = 7 | | | **AIC** = 66.74 | | | **BIC** = 87.81 | | |

Using the above results to predict the $\ln \mu_{32} = x_i' B$ then calculating $E[y_i|x_i] = \mu_{32} = e^{x_i' B}$
Where $\mu_{32}$ denotes the transition counts from state 3 to state 2 and as the time for the follow up is equal to all patients the expected counts for the transition from state 3 to state 2 are 19 transitions with good fit of the model while the observed count is 23 transitions . The expected increase in log count for one-unit increase in transformed LDL cholesterol is (0.302) which is not statistically significant (P=0.154), and the expected increase in log count for one-unit increase in transformed systolic blood pressure is (0.422) which is not statistically significant (P=0.142). These effects are not really explainable and further studies are needed to evaluate such effects as there may be some confounder substances that could induce such effects. While the expected decrease in log count for one-unit increase in transformed HOMA is (5.214) which is not statistically significant (P=0.103). As better management of HOMA, enhances the transition from F3 to F2. The expected increase in log count for one unit increase in interaction between the transformed LDL and transformed HOMA is (0.002) with no high statistical significance (P=0.984), and this expected increase for one unit increase in interaction between transformed HOMA and transformed systolic blood pressure is (0.132), which is not statistically significant (P=0.375), in other word, the rise in one predictor variable increases the effect of the other on the response variable (expected log count). While, the expected decrease in log count for one unit increase in interaction between the transformed systolic blood pressure and transformed LDL is (.012) with high statistical significance (P=0.006), in other word, the rise in one predictor variable decreases the effect of the other on the response variable (expected log count) but not reverse it. For every unit increase in this interaction, the incident rate ratio is deceased (decrease in transition counts) by 1.2%, with 95% confidence that this decrease is between .3% and 2%. Poisson model fits the data as goodness of fit is not statistically significant.

Table 6.18 : Transition from 2 to 0:

| | Coefficient | Robust Std.dev. | z | p>\|z\| | 95%CI for B coeff. | IRR | Robust Std.dev. | 95% CI for IRR |
|---|---|---|---|---|---|---|---|---|
| **LDLsp2** | 0.076 | 0.079 | 0.965 | 0.335 | (-0.079 , 0.231) | 1.079 | 0.085 | (0.924 , 1.260) |
| **HOMAsp2** | -2.713 | 0.709 | -3.829 | 0.000 | (-4.102 , -1.324) | 0.066 | 0.047 | (0.017 , 0.266) |
| **sysSP2** | -0.123 | 0.047 | -2.593 | 0.010 | (-0.216 , -0.030) | 0.884 | 0.042 | (0.806 , 0.970) |
| **DiasSP2** | 0.358 | 0.110 | 3.266 | 0.001 | (0.143 , 0.573) | 1.430 | 0.157 | (1.154 , 1.773) |
| **cons** | -7.034 | 0.501 | -14.052 | 0.000 | (-8.015 , -6.053) | 0.001 | 0.000 | (0.000 , 0.002) |
| **Log pseudolikelihood** = -15.63 | | | | | **Pseudo $R^2$** = 0.656 | | | |
| **Wald chi2(6)** = 263.12 | | | | | **Prob>chi2** = 0.000 | | | |
| **Deviance gof** = 6.65 | | | | | **Prob>chi2(145)** = 1 | | | |
| **Pearson gof** = 7.36 | | | | | **Prob>chi2(145)** = 1 | | | |
| **Log pseudolikelihood(full model )** = -15.63 | | | | | **Log pseudolikelihood(null model )** = -45.49 | | | |
| **DF** = 5 | | | **AIC** = 41.26 | | | **BIC** = 56.31 | | |

Using the above results to predict the $\ln \mu_{20} = x_i' B$ then calculating $E[y_i|x_i] = \mu_{20} = e^{x_i' B}$ Where $\mu_{20}$ denotes the transition counts from state 2 to state 0 and as the time for the follow up is equal to all patients the expected counts for the transition from state 2 to state 0 are 12 transitions with good fit of the model while the observed count is 13 transitions . The expected increase in log count for one-unit increase in transformed LDL cholesterol is (0.076) which is not statistically significant (P=0.335), and the expected increase in log count for one-unit increase in transformed diastolic blood pressure is (0.358) which is highly statistically significant (P=0.001). These effects are not really explainable and further studies are needed to evaluate such effects as there may be some confounder substances that could induce such effects. For every unit increase of diastolic blood pressure, the incident rate ratio is increased (increase in transition counts) by 43%, with 95% confidence that this increase is between 15.4% and 77.3%. While the expected decrease in log count for one-unit increase in transformed HOMA is (2.713) which is highly statistically significant (P=0.000), and the expected decrease in log count for one-unit increase in transformed systolic blood pressure is (0.123) which is highly statistically significant (P=0.01). As better management of these risk factors, enhances the transition from F2 to F0. For every unit increase of transformed HOMA, the incident rate ratio is decreased (decrease in transition counts) by 93.4%, with 95% confidence that this decrease is between 73.4% and 98.3%. For every unit increase of transformed systolic blood pressure, the incident rate ratio is decreased (decrease in transition counts) by 11.6%, with 95% confidence that this decrease is between 3% and 19.4%. Poisson model fits the data as goodness of fit is not statistically significant (P=1).

Table 6.19 : Transition from 3 to 1:

| | Coefficient | Robust Std.dev. | z | p>\|z\| | 95%CI for B coeff. | IRR | Robust Std.dev. | 95%CI for B coeff. |
|---|---|---|---|---|---|---|---|---|
| **LDLsp2** | 0.145 | 0.070 | 2.079 | 0.038 | (0.008 , 0.282) | 1.156 | 0.081 | (1.008 , 1.326) |
| **HOMAsp2** | -2.476 | 0.660 | -3.754 | 0.000 | (-3.769 , -1.183) | 0.084 | 0.055 | (0.023 , 0.306) |
| **sysSP2** | -0.129 | 0.045 | -2.899 | 0.004 | (-0.216 , -0.042) | 0.879 | 0.039 | (0.805 , 0.959) |
| **DiasSP2** | 0.276 | 0.093 | 2.962 | 0.003 | (0.093 , 0.459) | 1.318 | 0.123 | (1.098 , 1.582) |
| **cons** | -7.584 | 0.688 | -11.017 | 0.000 | (-8.934 , -6.235) | 0.001 | 0.000 | (0.000 , 0.002) |
| **Log pseudolikelihood** = -14.18 | | | | | **Pseudo $R^2$** = 0.693 | | | |
| **Wald chi2(6)** = 202.29 | | | | | **Prob>chi2** = 0.000 | | | |
| **Deviance gof** = 5.14 | | | | | **Prob>chi2(145)** = 1 | | | |
| **Pearson gof** = 6.09 | | | | | **Prob>chi2(145)** = 1 | | | |
| **Log pseudolikelihood(full model )** = -14.18 | | | | | **Log pseudolikelihood(null model )** = -46.18 | | | |
| **DF** = 5 | | | **AIC** = 38.36 | | | **BIC** = 53.42 | | |

Using the above results to predict the $\ln \mu_{31} = x_i' B$ then calculating $E[y_i|x_i] = \mu_{31} = e^{x_i' B}$ Where $\mu_{31}$ denotes the transition counts from state 3 to state 1 and as the time for the follow up is equal to all patients the expected counts for the transition from state 3 to state 1 are 13 transitions with good fit of the model while the observed count is 14 transitions. The expected increase in log count for one-unit increase in transformed LDL cholesterol is (0.145) which is not highly statistically significant (P=.038), and the expected increase in log count for one-unit increase of transformed diastolic blood pressure is (0.276) which is highly statistically significant (P=0.003). These effects are not really explainable and further studies are needed to evaluate such effects as there may be some confounder substances that could induce such effects. For every unit increase in diastolic blood pressure, the incident rate ratio is increased (increase in transition counts) by 31.8%, with 95% confidence that this increase is between 9.8% and 58.2%. While the expected decrease in log count for one-unit increase in transformed HOMA is (2.476) which is highly statistically significant (P= 0.000), and the expected decrease in log count for one-unit increase in transformed systolic blood pressure is (0.129) which is highly statistically significant (P=0.004). As better management of these risk factors, enhances the transition from F3 to F1. For every unit increase of transformed HOMA, the incident rate ratio is decreased (decrease in transition counts) by 91.6%, with 95% confidence that this decrease is between 69.4% and 97.7%. Furthermore, for every unit increase in transformed systolic blood pressure, the incident rate ratio is decreased (decrease in transition counts) by 12.1%, with 95% confidence that this decrease is between 4.1% and 19.5%. Poisson model fits the data as goodness of fit is not statistically significant (P=1).

As the estimated rates approximately equal the observed rates obtained by CTMC especially when using the initial rates calculated as $\theta_0 = \frac{n_{ijr}}{n_{i+}}$ where the $n_{ijr}$ is the transition counts from state $i$ to state $j$ and the $n_{i+}$ is the total marginal transition counts out of this state $i$ , as verified in previous 2 chapters, and assuming that the marginal counts are the same, so the estimated Q transition rate matrix according to the estimated counts obtained by fitting Poisson regression is:

$$Q = \begin{bmatrix} -.059 & .059 & 0 & 0 & 0 \\ .029 & -.080 & .051 & 0 & 0 \\ .015 & .033 & -.093 & .045 & 0 \\ 0 & .108 & .158 & -.409 & .167 \\ 0 & 0 & 0 & 0 & 0 \end{bmatrix} \quad where$$

$$\lambda_{01} = \frac{120}{2050} = .059 \ , \lambda_{12} = \frac{64}{1247} = .051 \ , \ \lambda_{23} = \frac{35}{783} = .045 \ \ , \lambda_{34} = \frac{20}{120} = .167$$

$$\mu_{10} = \frac{36}{1247} = .029 \ , \mu_{21} = \frac{26}{783} = .033 \ , \mu_{32} = \frac{19}{120} = .158 \ , \mu_{20} = \frac{12}{783} = .015 \ , \mu_{31} = \frac{13}{120} = .108$$

Probability transition matrix is obtained from exponentiating this Q matrix after 1 year:

$$P(t=1)=\begin{bmatrix} .9435 & .0551 & .0014 & 0 & 0 \\ .0274 & .9247 & .0469 & .0009 & .0001 \\ .0144 & .0327 & .9149 & .0348 & .0032 \\ .0023 & .0863 & .1245 & .6512 & .1357 \\ 0 & 0 & 0 & 0 & 1 \end{bmatrix}$$

Of those patients starting at F0 ,only 5.51% will move to F1 in one year, this declines to 4.69% of patients starting at F1 moving to F2 ,while 3.48% of patients starting at F2 will move to F3 ; however, 13.57% of patients starting in F3 will move to F4, and this high percentage of patients moving towards advanced fibrosis may be due to the fact that advanced fibrosis is considered to be F3 and F4 and once the patient reaches F3 , his chance to progress to F4 is higher than being in any starting stage considered less advanced fibrosis including F0 to F2 ( by definition ), and this is obvious as shown by incidence rate ratio of this transition being the highest (81.7e+6) . It is shown that progression from F0 to F1 and from F1 to F2 is approximately equal, while transition from F2 to F3 is less and this may be to more aggressive intervention taken by the patients to hinder the progression of fibrosis by applying more intensive lifestyle modifications, but once the patient reaches stage F3 the progression to F4 is by far the most among the forward transitions. There are 2.74% of patients starting at F1 will move to F0 while this percentage decreases to 1.44% if starting at F2, and it is even less if starting at F3 (only .23 % of patients can achieve this task); hence it is more feasible to move from F1 to F0 than to move from F2 to F0 than to move from F3 to F0; that is to mean, the more advanced the stage of fibrosis the patient experiences, the less likely movement to F0 he affords to do. There is a paradox if the starting stage is F2 or F3 to F1. The movement to F1 is more obvious if the patient is in F3 ( 8.63% of patients move to F1) than if he is in F2 ( 3.27 % of patients move to F1); therefore, the more advanced fibrosis stage the patient recognizes , the more likely movement to F1 he can do, and may be this is due to the extensive lifestyle modification he performs to achieve less degree of fibrosis, but it remains a little bit difficult to reach F0 ( only .23 % of patient can move from F3 to F0). It is also noted that 2.74% of patients move from F1 to F0 , 3.27% of patients move from F2 to F1 while 12.45% of patients move from F3 to F2 ; in other words the more advanced the fibrosis stage is, the more likely the movement to the immediately previous stage is. Moreover if the starting stage is F3, then 13.57% of these patients move to F4, a little bit higher than moving to F2 (12.45% of the patients); whereas, movement to F1 and F0 declines (8.63% of the patients and .23% of the patients respectively, approximately movement to F0 is 2.66% that to F1). Of those patients starting in F2, 3.48% move to F3, a little bit more than moving to F1 (3.27 % of patients); nevertheless, movement to F0 is almost 44% that to F1 ( 1.44% of the patients move to F0).

Mean time spent by the patient in state 0 is approximately 17 years , that declines to 12 years and 6 months spent in state 1, which further declines to approximately 10 years and 9 months spent in state 2, and ultimately reaching 2 years and 3.7 months spent in state 3 . It is shown that, there is decrease in time spent in each stage as the disease process evolves over time. This huge rapid decline in time spent in state 3 is due to advanced fibrosis induced by dead hepatocytes, especially if no treatment is introduced as regards: lifestyle modification ,risk factors treatment, as well as anti-inflammatory and anti-fibrotic drugs, and if so, it is a matter of time to reach state 4, which is irreversible stage of damaged liver cells that will soon manifest with reduction in liver cell functions, and may be to hepatocellular carcinoma, and eventually death, if not managed with liver transplantation.

### 6.3. Conclusions:

Insulin resistance is a cornerstone for initializing all the fatty liver disease abnormalities. The findings revealed that insulin resistance expressed by MOMA-IR 2 has the most deleterious effects among other factors for increasing the rate of forward progression of patients from state 1 to state 2 as well as from state 2 to state 3 and from state 3 to state 4. The higher the level of HOMA-IR is, the more rapid the rate of progression is. As concluded from the hypothetical model that for every unit increase in the transformed HOMA, the incidence rate ratio for transition from state 0 to state 1 is increased by 5909.7% and this elevation is kept rising while moving forward from subsequent state to the immediately next state, that is to mean, for every unit increase in the transformed HOMA, the incidence rate ratio (IRR) for transition from state 1 to state 2 is increased by 24017.9%, while for the transition from state 2 to state 3, it is increased by 47931.8% , and for transition from state 3 to state 4 it is increased by 5237498.4%. This increment is almost always highly statistically significant. This is in comparison with transformed LDL, as for every unit increase in the transformed LDL, the IRR for transition from state 0 to state 1 is increased by 68.7%, while for the transition from state 1 to state 2, it is increased by 36.4% , and for transition from state 3 to state 4 it is increased by 57.1%. And it is only highly statistically significant for transition from state 3 to state 4. However the systolic blood pressure is almost highly statistically significant for the transition from state 2 to state 3 as obvious by for every unit increase in the transformed systolic pressure, the IRR for this transition to occur is increased by 1114.3%. Moreover, for every unit decrease in the transformed HOMA, the IRR for transition from state 1 to state 0 is increased by 1.1%, for transition from state 2 to state 1 it is increased by 3.7%, for transition from state 3 to state 2 it is increased by 0.5%, for transition from state 2 to state 0 it is increased by 6.6%, and for transition from state 3 to state 1 it is increased by 8.4%. This emphasizes that better control of insulin resistance helps the patient to reverse his condition. To sum up, the precipitating factors should be rigorously and extensively treated and controlled by life style modifications represented by dietary restriction of high calorie diet and sedentary life, thus the predisposed persons should consume healthy diets and regularly practicing physical exercises suitable for their medical conditions. The newly discovered drugs like anti-fibrotic drugs that treat the fibrotic changes in the liver are promising drugs and await further longitudinal studies, to reveal the most effective protocol, by which they are administered to the patients, for better control of the rate of progression of liver fibrosis. This control keeps the patient out of loss of liver functions, and subsequently away from end stage liver disease, which necessitates liver transplantation with all its accompanying post transplantation complications.

This hypothetical example is coded by stata-14 and is published in code ocean site with the following URL:
codeocean.com/capsule/4752445/tree/v1

# Chapter Seven: Conclusions and Recommendations

## 7.1. Conclusions:

Continuous time Markov chains are suitable mathematical and statistical tools to be used for analysis of disease evolution over time. CTMCs being a type of multistate models are utilized to study this evolution in NAFLD patients, with its main phenotypes NAFLD and NASH, as well as the associated presence of fibrosis and its stages. The prevalence of NAFLD is rapidly increasing worldwide, and parallels the epidemics of obesity and type 2 diabetes. Metabolic syndrome is a well-known risk factor.

In the present book, NAFLD is modeled using the simplest form for health, disease, and death model, with one state for susceptible individuals with risk factors, such as: type 2 diabetes, dyslipidemia and hypertension, the other state is the NAFLD phenotypes, and two competing states for death: one for liver-related mortality as a complication of NAFLD, and the other death state is death causes unrelated to liver disease. This is covered in chapter four.

In addition, NAFLD is modeled in more elaborative expanded form, which includes nine states: the first eight states are the states of disease progression as time elapses; while, the ninth state is the death state. This is outlined in chapter five.

The maximum likelihood estimation and the quasi-newton method are used to estimate the transition rate matrix, although they are well-known methods, they are used in a new approach that compensates for the missing values in the follow up period of patients during the longitudinal studies. Also this new approach yields an estimated transition rates that are approximately equal the observed rates. These are the advantages of this new approach as regards obtaining the transition rates matrix. The disadvantages are that the results cannot be generalized. Once the rate matrix is obtained, the probability rate matrix can be estimated by exponentiating this rate matrix but the results are unstable, while solving forward Kolmogorov differential equations gives more stable results.

Moreover, a subset of the states that explicitly illustrates the phases of fibrosis process, which develops early in disease evolution cycle if the risk factors are not treated or eliminated- is modeled with CTMC to demonstrate: how covariates incorporated in a log-linear model can relate these predictors to transition rates among states. It has been shown that the most detrimental risk factor for disease progression was insulin resistance, the more resistant to insulin the cells were, the higher the rate of transition to advanced liver fibrosis was. This is elucidated in chapter six.

As long as the estimated rates are approximately equal the observed rates, these observed rates are used in the Poisson regression to relate the risk factors of non-alcoholic fatty liver disease to the rates, and the estimated rates obtained from running this regression are used by exponentiating this rate matrix and thus obtaining the probability transition matrix and other indices like the mean sojourn time.

For each of the three models depicted above, hypothetical examples of factitious non-real data are used to emphasize the attributes need to be estimated:

- Transition rate matrix among the various states.
- Transition probability matrix among states.
- Mean sojourn time in each state.
- Life expectancy in each state; in other words, mean time to absorption (death state).
- Expected number of patients in each state.

❖ State probability distribution at specific time point in the future.

As explained in this book, using MLE method to obtain the rate matrix yields a singular variance-covariance matrix and thus the observed rates of transition among the states were equal to the estimated rates. This was achieved in both the simplest model and the expanded model. Therefore in this disease model, after collecting the data about the counts for each transition, calculating the observed rate for each transition in each time interval and then taking weight for each of these rates in each transition corresponding to the percent of the transition count in each interval from the total transition counts as shown in the discussion of the hypothetical model, the researcher can sum up these weighted rates to get final transition rate matrix. Exponentiating this rate matrix will yield the transition probability matrix. However, this method gives less stable result than the analytic solution of the differential equation as explained in the study. Using Poisson regression to relate the covariates with the observed counts of transition, then the estimated counts can be used in the CTMC to get the transition rate matrix and the probability matrix.

Such analysis may give better insights to physicians, especially when new drug classes will soon be released in the market. What drug classes are to be used first? How to monitor the disease throughout the journey of treatment? What investigations to be used in such monitoring? How to modify the drug treatment? What is the target that needs to be achieved and how to maintain this target? And what is more to be said; that in late stage of the disease, when patients suffer from decompensated liver cirrhosis; liver transplantation is the treatment of choice to such patients, which increases the economic burden of NAFLD as it has done during its treatment in early stages. Also, the load of what are the best economic noninvasive tests to be used in primary health care units for stratification and identification of high risk patients, whether to do genetic tests in health insurance setting, and when to refer for liver biopsy in secondary or special clinics, could be answered from such longitudinal studies conducted on susceptible individuals. Over and above, some of the recently investigated noninvasive scoring systems of fibrosis need further external validation so as to be generalized in ethnicities other than the one tested upon. There are some controversies of cutoff points of these scoring systems among countries, and among ethnicities within the same country. Although liver biopsy is considered the standard method for diagnosis of NAFLD and staging it; its limitations encourage the development of various noninvasive tests, which necessitate better correlation between the findings obtained from the biopsy and the results of these tests to minimize the misclassification errors, which hamper good diagnosis and prognosis of the patient. These tests should be easy, feasible, convenient and with high safety profile to be used repeatedly in patients for follow up in such longitudinal studies. Discussion of the above items, hand by hand, with some of the recent and applicable guidelines declared by European society for study of obesity, diabetes and liver disease - are expounded in chapter three.

Multistate model represented by CTMC is a valuable statistical methodology, for longitudinal studies in medical researches to better comprehend and understand the pathophysiology, or the mechanism of the NAFLD process, and the interactions between the different modifiers either the external, or the internal modifiers. The external modifiers reside in bad dietary habits with excessive fat and carbohydrate ingestion, as well as sedentary life; while, the internal modifiers are represented in genetic factors affecting the metabolism of the food stuff (fat and carbohydrates) and other cellular functions such as risk factors for fibro-genesis (formation of fibrous tissue); as fibrosis is a detrimental predictor factor for disease progression to liver cirrhosis and its complications. The importance of such understanding has a great impact to reveal the genes that must be tested if ever needed, for whom to do such a test, and should it be in the utilities or services offered by the medical insurance. Moreover, should the degradation byproducts resulting from extracellular matrix destruction be used in routine clinical practice to mirror the fibrosis stages? Some of these medical concepts are clarified with simple explanation in chapter three.

Furthermore, insulin resistance is a key stone for triggering all these abnormalities, the more sensitive the body cells to insulin, the less likely the complications of NALFD will develop. The effect of risk factors or covariates as a mainstay players, like: increased insulin resistance, hyperlipidemia with increased LDL-cholesterol, high systolic and diastolic blood pressure - are thoroughly explained in chapter six.

## 7.2. Recommendations and Future Works

In Egypt, there are scare data, or may be no available data, about the prevalence of NAFLD and its phenotypes. Guidelines for risk stratification and identification are also lacking. Thus, more longitudinal studies are needed to cover these issues.

Further longitudinal studies are required to reveal the effects of risk factors and their interactions on the rates of progression, among stages of fibrosis in NAFLD patients. For the non-obese, or the so called metabolically obese normal weight patients with NAFLD, more studies are needed to clarify the mechanisms behind their illness.

Patients with impaired absorption of essential fatty acids or choline may suffer NASH that may be rapidly progressive even to death before anybody can be aware of this process; convenient supplementations of these essential fatty acids or choline are a demand to these patients to enhance their metabolic profile. More studies are entailed to uncover their effects on amelioration of progression rates among stages of NASH.

Multistate models can also be used for analysis of competing risks to death in such patients, as the first and second most common causes of death in NAFLD patients are the cardiovascular diseases (CVD) and kidney diseases, while the liver-related mortality is the third common cause of death.

Some other statistical methodologies, like : semi Markov and hidden Markov chains can be used to model NAFLD, especially hidden Markov CTMC can be used to model misclassification errors encountered in studies conducted by time homogenous CTMC.

# Appendix A: Matlab code for rate estimation

Matlab code for the rate estimation in Δt=1

```
clear
clc
close all

Qq1=[-.397 .39 0 0 0 0 0 0 .007;
    .02  -.28 .25 0 0 0 0 0 .01;
    0 .05  -.36 .22 0 0 0 .05 .04 ;
    0 0 .05 -.53 .28 0 0 .11 .09;
    0 0 0 0 -.33 .18 0 .06 .09;
    0 0 0 0 0 -.9 .8 0 .1 ;
    0 0 0 0 0 0 -.41 0 .41;
    0 0 0 0 0 0 0 -.75 .75 ;
    0 0 0 0 0 0 0 0 0];
lambda12=Qq1(1,2);
lambda18=Qq1(1,8);
lambda19=Qq1(1,9);
mu21=Qq1(2,1);
lambda23=Qq1(2,3);
lambda28=Qq1(2,8);
lambda29=Qq1(2,9);
mu32=Qq1(3,2);
lambda34=Qq1(3,4);
lambda38=Qq1(3,8);
lambda39=Qq1(3,9);
mu43=Qq1(4,3);
lambda45=Qq1(4,5);
lambda48=Qq1(4,8);
lambda49=Qq1(4,9);
lambda56=Qq1(5,6);
lambda58=Qq1(5,8);
lambda59=Qq1(5,9);
lambda67=Qq1(6,7);
lambda69=Qq1(6,9);
lambda79=Qq1(7,9);
lambda89=Qq1(8,9);

% calculate eigenvalues of the Qq matrix
% eigenValuesVector= eig(Qq1);

% calculate 4 th polynomial
g1=-1*Qq1(1,1);
g2=-1*Qq1(2,2);
g3=-1*Qq1(3,3);
g4=-1*Qq1(4,4);

w1= g1+g2;
w2= g1*(lambda23+lambda28+lambda29)+mu21*(lambda18+lambda19);
w3=g3+g4;
w4=g3*(lambda45+lambda48+lambda49)+mu43*(lambda38+lambda39+mu32);
w5=lambda23*mu32*(g1+g4);
w6=lambda23*mu32*g1*g4;

a=1;
b=w1+w3;
c=w2+w4+w1*w3-lambda23*mu32;
d=w2*w3+w1*w4-w5;
e=w2*w4-w6;

p=(8*c-3*b^2)/8;
q=(b^3-4*b*c+8*d)/8;
delta0=c^2-3*b*d+12*e;
sqrdelta0=sqrt(delta0);
delta0cubed=delta0^3;
sqrdelta0cubed=sqrt(delta0cubed);
delta1=2*c^3-9*b*c*d+27*b^2*e+27*d^2-72*c*e;
theta=acos(delta1/(2*sqrt(delta0^3)));
```

```
m=cos(theta/3);
s=.5*sqrt(2/3*(-1*p+sqrdelta0*m));

% differentiate each b,c,d,e, with respect to each rate
%(diff [B], [C], [D], [E] w.r.t. lambda12)

diffBwrtlambda12=1;
diffCwrtlambda12=lambda23+lambda28+lambda29+lambda34+lambda38+lambda39+mu32+lambda45+lambda48+lam
bda49+mu43;
diffDwrtlambda12=(lambda23+lambda28+lambda29)*(lambda34+lambda38+lambda39+mu32+lambda45+lambda48+
lambda49+mu43)+((lambda34+lambda38+lambda39+mu32)*(lambda45+lambda48+lambda49)+mu43*(lambda38+lam
bda39+mu32))-lambda23*mu32;
diffEwrtlambda12=(lambda23+lambda28+lambda29)*((lambda34+lambda38+lambda39+mu32)*(lambda45+lambda
48+lambda49)+mu43*(lambda38+lambda39+mu32))-lambda23*mu32*(lambda45+lambda48+lambda49+mu43);

%(diff [B], [C], [D], [E] w.r.t. lambda18)

diffBwrtlambda18=1;
diffCwrtlambda18=lambda23+lambda28+lambda29+mu21+lambda34+lambda38+lambda39+mu32+lambda45+lambda4
8+lambda49+mu43;
diffDwrtlambda18=(lambda23+lambda28+lambda29+mu21)*(lambda34+lambda38+lambda39+mu32+lambda45+lamb
da48+lambda49+mu43)+((lambda34+lambda38+lambda39+mu32)*(lambda45+lambda48+lambda49)+mu43*(lambda3
8+lambda39+mu32))-lambda23*mu32;
diffEwrtlambda18=(lambda23+lambda28+lambda29+mu21)*((lambda34+lambda38+lambda39+mu32)*(lambda45+l
ambda48+lambda49)+mu43*(lambda38+lambda39+mu32))-lambda23*mu32*(lambda45+lambda48+lambda49+mu43);

%(diff [B], [C], [D], [E] w.r.t. lambda19)

diffBwrtlambda19=1;
diffCwrtlambda19=lambda23+lambda28+lambda29+mu21+lambda34+lambda38+lambda39+mu32+lambda45+lambda4
8+lambda49+mu43;
diffDwrtlambda19=(lambda23+lambda28+lambda29+mu21)*(lambda34+lambda38+lambda39+mu32+lambda45+lamb
da48+lambda49+mu43)+((lambda34+lambda38+lambda39+mu32)*(lambda45+lambda48+lambda49)+mu43*(lambda3
8+lambda39+mu32))-lambda23*mu32;
diffEwrtlambda19=(lambda23+lambda28+lambda29+mu21)*((lambda34+lambda38+lambda39+mu32)*(lambda45+l
ambda48+lambda49)+mu43*(lambda38+lambda39+mu32))-lambda23*mu32*(lambda45+lambda48+lambda49+mu43);

%(diff [B], [C], [D], [E] w.r.t. lambda23)

diffBwrtlambda23=1;
diffCwrtlambda23=lambda12+lambda18+lambda19+lambda34+lambda38+lambda39+mu32+lambda45+lambda48+lam
bda49+mu43-mu32;
diffDwrtlambda23=(lambda12+lambda18+lambda19)*(lambda34+lambda38+lambda39+mu32+lambda45+lambda48+
lambda49+mu43)+((lambda34+lambda38+lambda39+mu32)*(lambda45+lambda48+lambda49)+mu43*(lambda38+lam
bda39+mu32))-mu32*(lambda12+lambda18+lambda19+lambda45+lambda48+lambda49+mu43);
diffEwrtlambda23=(lambda12+lambda18+lambda19)*((lambda34+lambda38+lambda39+mu32)*(lambda45+lambda
48+lambda49)+mu43*(lambda38+lambda39+mu32))-
mu32*(lambda12+lambda18+lambda19)*(lambda45+lambda48+lambda49+mu43);

%(diff [B], [C], [D], [E] w.r.t. lambda28)

diffBwrtlambda28=1;
diffCwrtlambda28=lambda12+lambda18+lambda19+lambda34+lambda38+lambda39+mu32+lambda45+lambda48+lam
bda49+mu43;
diffDwrtlambda28=(lambda12+lambda18+lambda19)*(lambda34+lambda38+lambda39+mu32+lambda45+lambda48+
lambda49+mu43)+((lambda34+lambda38+lambda39+mu32)*(lambda45+lambda48+lambda49)+mu43*(lambda38+lam
bda39+mu32));
diffEwrtlambda28=(lambda12+lambda18+lambda19)*((lambda34+lambda38+lambda39+mu32)*(lambda45+lambda
48+lambda49)+mu43*(lambda38+lambda39+mu32));


%(diff [B], [C], [D], [E] w.r.t. lambda29)

diffBwrtlambda29=1;
diffCwrtlambda29=lambda12+lambda18+lambda19+lambda34+lambda38+lambda39+mu32+lambda45+lambda48+lam
bda49+mu43;
diffDwrtlambda29=(lambda12+lambda18+lambda19)*(lambda34+lambda38+lambda39+mu32+lambda45+lambda48+
lambda49+mu43)+((lambda34+lambda38+lambda39+mu32)*(lambda45+lambda48+lambda49)+mu43*(lambda38+lam
bda39+mu32));
diffEwrtlambda29=(lambda12+lambda18+lambda19)*((lambda34+lambda38+lambda39+mu32)*(lambda45+lambda
48+lambda49)+mu43*(lambda38+lambda39+mu32));
```

```
%(diff [B], [C], [D], [E] w.r.t. mu21)

diffBwrtmu21=1;
diffCwrtmu21=lambda18+lambda19+lambda34+lambda38+lambda39+mu32+lambda45+lambda48+lambda49+mu43;
diffDwrtmu21=(lambda18+lambda19)*(lambda34+lambda38+lambda39+mu32+lambda45+lambda48+lambda49+mu43
)+((lambda34+lambda38+lambda39+mu32)*(lambda45+lambda48+lambda49)+mu43*(lambda38+lambda39+mu32));
diffEwrtmu21=(lambda18+lambda19)*((lambda34+lambda38+lambda39+mu32)*(lambda45+lambda48+lambda49)+
mu43*(lambda38+lambda39+mu32));


%(diff [B], [C], [D], [E] w.r.t. lambda34)

diffBwrtlambda34=1;
diffCwrtlambda34=lambda12+lambda18+lambda19+lambda23+lambda28+lambda29+mu21+lambda45+lambda48+lam
bda49;
diffDwrtlambda34=((lambda12+lambda18+lambda19)*(lambda23+lambda28+lambda29)+mu21*(lambda18+lambda
19))+((lambda12+lambda18+lambda19+lambda23+lambda28+lambda29+mu21)*(lambda45+lambda48+lambda49));
diffEwrtlambda34=((lambda12+lambda18+lambda19)*(lambda23+lambda28+lambda29)+mu21*(lambda18+lambda
19))*(lambda45+lambda48+lambda49);

%(diff [B], [C], [D], [E] w.r.t. lambda38)

diffBwrtlambda38=1;
diffCwrtlambda38=lambda12+lambda18+lambda19+lambda23+lambda28+lambda29+mu21+lambda45+lambda48+lam
bda49+mu43;
diffDwrtlambda38=((lambda12+lambda18+lambda19)*(lambda23+lambda28+lambda29)+mu21*(lambda18+lambda
19))+((lambda12+lambda18+lambda19+lambda23+lambda28+lambda29+mu21)*(lambda45+lambda48+lambda49+mu
43));
diffEwrtlambda38=((lambda12+lambda18+lambda19)*(lambda23+lambda28+lambda29)+mu21*(lambda18+lambda
19))*(lambda45+lambda48+lambda49+mu43);

%(diff [B], [C], [D], [E] w.r.t. lambda39)

diffBwrtlambda39=1;
diffCwrtlambda39=lambda12+lambda18+lambda19+lambda23+lambda28+lambda29+mu21+lambda45+lambda48+lam
bda49+mu43;
diffDwrtlambda39=((lambda12+lambda18+lambda19)*(lambda23+lambda28+lambda29)+mu21*(lambda18+lambda
19))+((lambda12+lambda18+lambda19+lambda23+lambda28+lambda29+mu21)*(lambda45+lambda48+lambda49+mu
43));
diffEwrtlambda39=((lambda12+lambda18+lambda19)*(lambda23+lambda28+lambda29)+mu21*(lambda18+lambda
19))*(lambda45+lambda48+lambda49+mu43);

%(diff [B], [C], [D], [E] w.r.t. mu32)

diffBwrtmu32=1;
diffCwrtmu32=lambda12+lambda18+lambda19+lambda23+lambda28+lambda29+mu21+lambda45+lambda48+lambda4
9+mu43-lambda23;
diffDwrtmu32=((lambda12+lambda18+lambda19)*(lambda23+lambda28+lambda29)+mu21*(lambda18+lambda19))
+((lambda12+lambda18+lambda19+lambda23+lambda28+lambda29+mu21)*(lambda45+lambda48+lambda49+mu43))
-lambda23*(lambda12+lambda18+lambda19+lambda45+lambda48+lambda49+mu43);
diffEwrtmu32=((lambda12+lambda18+lambda19)*(lambda23+lambda28+lambda29)+mu21*(lambda18+lambda19))
*(lambda45+lambda48+lambda49+mu43)-
lambda23*(lambda12+lambda18+lambda19)*(lambda45+lambda48+lambda49+mu43);

%(diff [B], [C], [D], [E] w.r.t. lambda45)

diffBwrtlambda45=1;
diffCwrtlambda45=lambda12+lambda18+lambda19+lambda23+lambda28+lambda29+mu21+lambda34+lambda38+lam
bda39+mu32;
diffDwrtlambda45=((lambda12+lambda18+lambda19)*(lambda23+lambda28+lambda29)+mu21*(lambda18+lambda
19))+((lambda12+lambda18+lambda19+lambda23+lambda28+lambda29+mu21)*(lambda34+lambda38+lambda39+mu
32))-lambda23*mu32;
diffEwrtlambda45=((lambda12+lambda18+lambda19)*(lambda23+lambda28+lambda29)+mu21*(lambda18+lambda
19))*(lambda34+lambda38+lambda39+mu32)-lambda23*mu32*(lambda12+lambda18+lambda19);


%(diff [B], [C], [D], [E] w.r.t. lambda48)

diffBwrtlambda48=1;
```

```
diffCwrtlambda48=lambda12+lambda18+lambda19+lambda23+lambda28+lambda29+mu21+lambda34+lambda38+lam
bda39+mu32;
diffDwrtlambda48=((lambda12+lambda18+lambda19)*(lambda23+lambda28+lambda29)+mu21*(lambda18+lambda
19))+((lambda12+lambda18+lambda19+lambda23+lambda28+lambda29+mu21)*(lambda34+lambda38+lambda39+mu
32))-lambda23*mu32;
diffEwrtlambda48=((lambda12+lambda18+lambda19)*(lambda23+lambda28+lambda29)+mu21*(lambda18+lambda
19))*(lambda34+lambda38+lambda39+mu32)-lambda23*mu32*(lambda12+lambda18+lambda19);


%(diff [B], [C], [D], [E] w.r.t. lambda49)

diffBwrtlambda49=1;
diffCwrtlambda49=lambda12+lambda18+lambda19+lambda23+lambda28+lambda29+mu21+lambda34+lambda38+lam
bda39+mu32;
diffDwrtlambda49=((lambda12+lambda18+lambda19)*(lambda23+lambda28+lambda29)+mu21*(lambda18+lambda
19))+((lambda12+lambda18+lambda19+lambda23+lambda28+lambda29+mu21)*(lambda34+lambda38+lambda39+mu
32))-lambda23*mu32;
diffEwrtlambda49=((lambda12+lambda18+lambda19)*(lambda23+lambda28+lambda29)+mu21*(lambda18+lambda
19))*(lambda34+lambda38+lambda39+mu32)-lambda23*mu32*(lambda12+lambda18+lambda19);

%(diff [B], [C], [D], [E] w.r.t. mu43)

diffBwrtmu43=1;
diffCwrtmu43=lambda12+lambda18+lambda19+lambda23+lambda28+lambda29+mu21+lambda38+lambda39+mu32;
diffDwrtmu43=((lambda12+lambda18+lambda19)*(lambda23+lambda28+lambda29)+mu21*(lambda18+lambda19))
+((lambda12+lambda18+lambda19+lambda23+lambda28+lambda29+mu21)*(lambda38+lambda39+mu32))-
lambda23*mu32;
diffEwrtmu43=((lambda12+lambda18+lambda19)*(lambda23+lambda28+lambda29)+mu21*(lambda18+lambda19))
*(lambda38+lambda39+mu32)-lambda23*mu32*(lambda12+lambda18+lambda19);

% diff p w.r.t. lambda 12
diffPwrtlambda12=(8*diffCwrtlambda12-6*b*diffBwrtlambda12)/8;

% diff p w.r.t. lambda 18
diffPwrtlambda18=(8*diffCwrtlambda18-6*b*diffBwrtlambda18)/8;

% diff p w.r.t. lambda 19
diffPwrtlambda19=(8*diffCwrtlambda19-6*b*diffBwrtlambda19)/8;

% diff p w.r.t. lambda 23
diffPwrtlambda23=(8*diffCwrtlambda23-6*b*diffBwrtlambda23)/8;

% diff p w.r.t. lambda 28
diffPwrtlambda28=(8*diffCwrtlambda28-6*b*diffBwrtlambda28)/8;

% diff p w.r.t. lambda 29
diffPwrtlambda29=(8*diffCwrtlambda29-6*b*diffBwrtlambda29)/8;

% diff p w.r.t. mu 21
diffPwrtmu21=(8*diffCwrtmu21-6*b*diffBwrtmu21)/8;

% diff p w.r.t. lambda 34
diffPwrtlambda34=(8*diffCwrtlambda34-6*b*diffBwrtlambda34)/8;

% diff p w.r.t. lambda 38
diffPwrtlambda38=(8*diffCwrtlambda38-6*b*diffBwrtlambda38)/8;

% diff p w.r.t. lambda 39
diffPwrtlambda39=(8*diffCwrtlambda39-6*b*diffBwrtlambda39)/8;

% diff p w.r.t. mu 32
diffPwrtmu32=(8*diffCwrtmu32-6*b*diffBwrtmu32)/8;

% diff p w.r.t. lambda 45
diffPwrtlambda45=(8*diffCwrtlambda45-6*b*diffBwrtlambda45)/8;

% diff p w.r.t. lambda 48
diffPwrtlambda48=(8*diffCwrtlambda48-6*b*diffBwrtlambda48)/8;

% diff p w.r.t. lambda 49
diffPwrtlambda49=(8*diffCwrtlambda49-6*b*diffBwrtlambda49)/8;
```

```
% diff p w.r.t. mu 43
diffPwrtmu43=(8*diffCwrtmu43-6*b*diffBwrtmu43)/8;

% diff q w.r.t. lambda 12
diffQwrtlambda12=(3*b^2-4*diffBwrtlambda12*c-4*diffCwrtlambda12*b+8*diffDwrtlambda12)/8;

% diff q w.r.t. lambda 18
diffQwrtlambda18=(3*b^2-4*diffBwrtlambda18*c-4*diffCwrtlambda18*b+8*diffDwrtlambda18)/8;

% diff q w.r.t. lambda 19
diffQwrtlambda19=(3*b^2-4*diffBwrtlambda19*c-4*diffCwrtlambda19*b+8*diffDwrtlambda19)/8;

% diff q w.r.t. lambda 23
diffQwrtlambda23=(3*b^2-4*diffBwrtlambda23*c-4*diffCwrtlambda23*b+8*diffDwrtlambda23)/8;

% diff q w.r.t. lambda 28
diffQwrtlambda28=(3*b^2-4*diffBwrtlambda28*c-4*diffCwrtlambda28*b+8*diffDwrtlambda28)/8;

% diff q w.r.t. lambda 29
diffQwrtlambda29=(3*b^2-4*diffBwrtlambda29*c-4*diffCwrtlambda29*b+8*diffDwrtlambda29)/8 ;

% diff q w.r.t. mu 21
diffQwrtmu21=(3*b^2-4*diffBwrtmu21*c-4*diffCwrtmu21*b+8*diffDwrtmu21)/8;

% diff q w.r.t. lambda 34
diffQwrtlambda34=(3*b^2-4*diffBwrtlambda34*c-4*diffCwrtlambda34*b+8*diffDwrtlambda34)/8;

% diff q w.r.t. lambda 38
diffQwrtlambda38=(3*b^2-4*diffBwrtlambda38*c-4*diffCwrtlambda38*b+8*diffDwrtlambda38)/8;

% diff q w.r.t. lambda 39
diffQwrtlambda39=(3*b^2-4*diffBwrtlambda39*c-4*diffCwrtlambda39*b+8*diffDwrtlambda39)/8;

% diff q w.r.t. mu 32
diffQwrtmu32=(3*b^2-4*diffBwrtmu32*c-4*diffCwrtmu32*b+8*diffDwrtmu32)/8;

% diff q w.r.t. lambda 45
diffQwrtlambda45=(3*b^2-4*diffBwrtlambda45*c-4*diffCwrtlambda45*b+8*diffDwrtlambda45)/8;

% diff q w.r.t. lambda 48
diffQwrtlambda48=(3*b^2-4*diffBwrtlambda48*c-4*diffCwrtlambda48*b+8*diffDwrtlambda48)/8;

% diff q w.r.t. lambda 49
diffQwrtlambda49=(3*b^2-4*diffBwrtlambda49*c-4*diffCwrtlambda49*b+8*diffDwrtlambda49)/8;

% diff q w.r.t. mu 43
diffQwrtmu43=(3*b^2-4*diffBwrtmu43*c-4*diffCwrtmu43*b+8*diffDwrtmu43)/8;


% multiply diff p by  -2/3
lambda12FirstTermSDiff=-1*(2/3)*diffPwrtlambda12;
lambda18FirstTermSDiff=-1*(2/3)*diffPwrtlambda18;
lambda19FirstTermSDiff=-1*(2/3)*diffPwrtlambda19;
lambda23FirstTermSDiff=-1*(2/3)*diffPwrtlambda23;
lambda28FirstTermSDiff=-1*(2/3)*diffPwrtlambda28;
lambda29FirstTermSDiff=-1*(2/3)*diffPwrtlambda29;
mu21FirstTermSDiff=-1*(2/3)*diffPwrtmu21;
lambda34FirstTermSDiff=-1*(2/3)*diffPwrtlambda34;
lambda38FirstTermSDiff=-1*(2/3)*diffPwrtlambda38;
lambda39FirstTermSDiff=-1*(2/3)*diffPwrtlambda39;
mu32FirstTermSDiff=-1*(2/3)*diffPwrtmu32;
lambda45FirstTermSDiff=-1*(2/3)*diffPwrtlambda45;
lambda48FirstTermSDiff=-1*(2/3)*diffPwrtlambda48;
lambda49FirstTermSDiff=-1*(2/3)*diffPwrtlambda49;
mu43FirstTermSDiff=-1*(2/3)*diffPwrtmu43;

% calculate Diff of squre root of delta0:
% step 1 = reciprocal of sqrdelta0
%step 2 = diff delta0 w.r.t. each rate
```

```
reciprocalOfsqrdelta0=1/sqrdelta0;

% diff sqrt delta0 w.r.t. lambda 12
diffDelta0wrtlambda12=2*c*diffCwrtlambda12-3*diffBwrtlambda12*d-
3*diffDwrtlambda12*b+12*diffEwrtlambda12;
diffSqrDelta0wrtlambda12=.5*reciprocalOfsqrdelta0*diffDelta0wrtlambda12;

% diff sqrt delta0 w.r.t. lambda 18
diffDelta0wrtlambda18=2*c*diffCwrtlambda18-3*diffBwrtlambda18*d-
3*diffDwrtlambda18*b+12*diffEwrtlambda18;
diffSqrDelta0wrtlambda18=.5*reciprocalOfsqrdelta0*diffDelta0wrtlambda18;

% diff sqrt delta0 w.r.t. lambda 19
diffDelta0wrtlambda19=2*c*diffCwrtlambda19-3*diffBwrtlambda19*d-
3*diffDwrtlambda19*b+12*diffEwrtlambda19;
diffSqrDelta0wrtlambda19=.5*reciprocalOfsqrdelta0*diffDelta0wrtlambda19;

% diff sqrt delta0 w.r.t. lambda 23
diffDelta0wrtlambda23=2*c*diffCwrtlambda23-3*diffBwrtlambda23*d-
3*diffDwrtlambda23*b+12*diffEwrtlambda23;
diffSqrDelta0wrtlambda23=.5*reciprocalOfsqrdelta0*diffDelta0wrtlambda23;

% diff sqrt  delta0 w.r.t. lambda 28
diffDelta0wrtlambda28=2*c*diffCwrtlambda28-3*diffBwrtlambda28*d-
3*diffDwrtlambda28*b+12*diffEwrtlambda28;
diffSqrDelta0wrtlambda28=.5*reciprocalOfsqrdelta0*diffDelta0wrtlambda28;

% diff sqrt delta0 w.r.t. lambda 29
diffDelta0wrtlambda29=2*c*diffCwrtlambda29-3*diffBwrtlambda29*d-
3*diffDwrtlambda29*b+12*diffEwrtlambda29;
diffSqrDelta0wrtlambda29=.5*reciprocalOfsqrdelta0*diffDelta0wrtlambda29;

% diff sqrt  delta0 w.r.t. mu 21
diffDelta0wrtmu21=2*c*diffCwrtmu21-3*diffBwrtmu21*d-3*diffDwrtmu21*b+12*diffEwrtmu21;
diffSqrDelta0wrtmu21=.5*reciprocalOfsqrdelta0*diffDelta0wrtmu21;

% diff sqrt delta0 w.r.t. lambda 34
diffDelta0wrtlambda34=2*c*diffCwrtlambda34-3*diffBwrtlambda34*d-
3*diffDwrtlambda34*b+12*diffEwrtlambda34;
diffSqrDelta0wrtlambda34=.5*reciprocalOfsqrdelta0*diffDelta0wrtlambda34;

% diff sqrt delta0 w.r.t. lambda 38
diffDelta0wrtlambda38=2*c*diffCwrtlambda38-3*diffBwrtlambda38*d-
3*diffDwrtlambda38*b+12*diffEwrtlambda38;
diffSqrDelta0wrtlambda38=.5*reciprocalOfsqrdelta0*diffDelta0wrtlambda38;

% diff sqrt delta0 w.r.t. lambda 39
diffDelta0wrtlambda39=2*c*diffCwrtlambda39-3*diffBwrtlambda39*d-
3*diffDwrtlambda39*b+12*diffEwrtlambda39;
diffSqrDelta0wrtlambda39=.5*reciprocalOfsqrdelta0*diffDelta0wrtlambda39;

% diff sqrt delta0 w.r.t. mu 32
diffDelta0wrtmu32=2*c*diffCwrtmu32-3*diffBwrtmu32*d-3*diffDwrtmu32*b+12*diffEwrtmu32;
diffSqrDelta0wrtmu32=.5*reciprocalOfsqrdelta0*diffDelta0wrtmu32;

% diff sqrt delta0 w.r.t. lambda 45
diffDelta0wrtlambda45=2*c*diffCwrtlambda45-3*diffBwrtlambda45*d-
3*diffDwrtlambda45*b+12*diffEwrtlambda45;
diffSqrDelta0wrtlambda45=.5*reciprocalOfsqrdelta0*diffDelta0wrtlambda45;

% diff sqrt delta0 w.r.t. lambda 48
diffDelta0wrtlambda48=2*c*diffCwrtlambda49-3*diffBwrtlambda48*d-
3*diffDwrtlambda48*b+12*diffEwrtlambda48;
diffSqrDelta0wrtlambda48=.5*reciprocalOfsqrdelta0*diffDelta0wrtlambda48;

% diff sqrt delta0 w.r.t. lambda 49
diffDelta0wrtlambda49=2*c*diffCwrtlambda49-3*diffBwrtlambda49*d-
3*diffDwrtlambda49*b+12*diffEwrtlambda49;
diffSqrDelta0wrtlambda49=.5*reciprocalOfsqrdelta0*diffDelta0wrtlambda49;

% diff sqrt delta0 w.r.t. mu 43
```

```
diffDelta0wrtmu43=2*c*diffCwrtmu43-3*diffBwrtmu43*d-3*diffDwrtmu43*b+12*diffEwrtmu43;
diffSqrDelta0wrtmu43=.5*reciprocalOfsqrdelta0*diffDelta0wrtmu43;

% calculate the second term in diff s

lambda12SecondTermSDiff=(2/3)*diffSqrDelta0wrtlambda12*m;
lambda18SecondTermSDiff=(2/3)*diffSqrDelta0wrtlambda18*m ;
lambda19SecondTermSDiff=(2/3)*diffSqrDelta0wrtlambda19*m;
lambda23SecondTermSDiff=(2/3)*diffSqrDelta0wrtlambda23*m;
lambda28SecondTermSDiff=(2/3)*diffSqrDelta0wrtlambda28*m;
lambda29SecondTermSDiff=(2/3)*diffSqrDelta0wrtlambda29*m;
mu21SecondTermSDiff=(2/3)*diffSqrDelta0wrtmu21*m;
lambda34SecondTermSDiff=(2/3)*diffSqrDelta0wrtlambda34*m;
lambda38SecondTermSDiff=(2/3)*diffSqrDelta0wrtlambda38*m;
lambda39SecondTermSDiff=(2/3)*diffSqrDelta0wrtlambda39*m;
mu32SecondTermSDiff=(2/3)*diffSqrDelta0wrtmu32*m;
lambda45SecondTermSDiff=(2/3)*diffSqrDelta0wrtlambda45*m;
lambda48SecondTermSDiff=(2/3)*diffSqrDelta0wrtlambda48*m;
lambda49SecondTermSDiff=(2/3)*diffSqrDelta0wrtlambda49*m;
mu43SecondTermSDiff=(2/3)*diffSqrDelta0wrtmu43*m;

% calculate third term in diff s :
sintheta=-sin(theta/3);
sqrt(delta0);
num1=sintheta*sqrt(delta0);
den1=sqrt(1-(delta1/(2*sqrt(delta0^3)))^2);
k=(-1/9)*(num1/den1);

% calculate num2/den2:
% where  num2= diffdelta1*sqrt of delta0cubed - delta1*diffsqrtdelta0cubed
% where  den2=delta0cubed:
% calculate diff of delta1:
DiffOfdelta1wrtlambda12=6*c^2*diffCwrtlambda12-9*diffBwrtlambda12*c*d-9*diffCwrtlambda12*b*d-
9*diffDwrtlambda12*b*c+54*b*e*diffBwrtlambda12+27*b^2*diffEwrtlambda12+54*d*diffDwrtlambda12-
72*diffCwrtlambda12*e-72*c*diffEwrtlambda12;
DiffOfdelta1wrtlambda18=6*c^2*diffCwrtlambda18-9*diffBwrtlambda18*c*d-9*diffCwrtlambda18*b*d-
9*diffDwrtlambda18*b*c+54*b*e*diffBwrtlambda18+27*b^2*diffEwrtlambda18+54*d*diffDwrtlambda18-
72*diffCwrtlambda18*e-72*c*diffEwrtlambda18;
DiffOfdelta1wrtlambda19=6*c^2*diffCwrtlambda19-9*diffBwrtlambda19*c*d-9*diffCwrtlambda19*b*d-
9*diffDwrtlambda19*b*c+54*b*e*diffBwrtlambda19+27*b^2*diffEwrtlambda19+54*d*diffDwrtlambda19-
72*diffCwrtlambda19*e-72*c*diffEwrtlambda19;
DiffOfdelta1wrtlambda23=6*c^2*diffCwrtlambda23-9*diffBwrtlambda23*c*d-9*diffCwrtlambda23*b*d-
9*diffDwrtlambda23*b*c+54*b*e*diffBwrtlambda23+27*b^2*diffEwrtlambda23+54*d*diffDwrtlambda23-
72*diffCwrtlambda23*e-72*c*diffEwrtlambda23;
DiffOfdelta1wrtlambda28=6*c^2*diffCwrtlambda28-9*diffBwrtlambda28*c*d-9*diffCwrtlambda28*b*d-
9*diffDwrtlambda28*b*c+54*b*e*diffBwrtlambda28+27*b^2*diffEwrtlambda28+54*d*diffDwrtlambda28-
72*diffCwrtlambda28*e-72*c*diffEwrtlambda28;
DiffOfdelta1wrtlambda29=6*c^2*diffCwrtlambda29-9*diffBwrtlambda29*c*d-9*diffCwrtlambda29*b*d-
9*diffDwrtlambda29*b*c+54*b*e*diffBwrtlambda29+27*b^2*diffEwrtlambda29+54*d*diffDwrtlambda29-
72*diffCwrtlambda29*e-72*c*diffEwrtlambda29;
DiffOfdelta1wrtmu21=6*c^2*diffCwrtmu21-9*diffBwrtmu21*c*d-9*diffCwrtmu21*b*d-
9*diffDwrtmu21*b*c+54*b*e*diffBwrtmu21+27*b^2*diffEwrtmu21+54*d*diffDwrtmu21-72*diffCwrtmu21*e-
72*c*diffEwrtmu21;
DiffOfdelta1wrtlambda34=6*c^2*diffCwrtlambda34-9*diffBwrtlambda34*c*d-9*diffCwrtlambda34*b*d-
9*diffDwrtlambda34*b*c+54*b*e*diffBwrtlambda34+27*b^2*diffEwrtlambda34+54*d*diffDwrtlambda34-
72*diffCwrtlambda34*e-72*c*diffEwrtlambda34;
DiffOfdelta1wrtlambda38=6*c^2*diffCwrtlambda38-9*diffBwrtlambda38*c*d-9*diffCwrtlambda38*b*d-
9*diffDwrtlambda38*b*c+54*b*e*diffBwrtlambda38+27*b^2*diffEwrtlambda38+54*d*diffDwrtlambda38-
72*diffCwrtlambda38*e-72*c*diffEwrtlambda38;
DiffOfdelta1wrtlambda39=6*c^2*diffCwrtlambda39-9*diffBwrtlambda39*c*d-9*diffCwrtlambda39*b*d-
9*diffDwrtlambda39*b*c+54*b*e*diffBwrtlambda39+27*b^2*diffEwrtlambda39+54*d*diffDwrtlambda39-
72*diffCwrtlambda39*e-72*c*diffEwrtlambda39;
DiffOfdelta1wrtmu32=6*c^2*diffCwrtmu32-9*diffBwrtmu32*c*d-9*diffCwrtmu32*b*d-
9*diffDwrtmu32*b*c+54*b*e*diffBwrtmu32+27*b^2*diffEwrtmu32+54*d*diffDwrtmu32-72*diffCwrtmu32*e-
72*c*diffEwrtmu32;
DiffOfdelta1wrtlambda45=6*c^2*diffCwrtlambda45-9*diffBwrtlambda45*c*d-9*diffCwrtlambda45*b*d-
9*diffDwrtlambda45*b*c+54*b*e*diffBwrtlambda45+27*b^2*diffEwrtlambda45+54*d*diffDwrtlambda45-
72*diffCwrtlambda45*e-72*c*diffEwrtlambda45;
DiffOfdelta1wrtlambda48=6*c^2*diffCwrtlambda48-9*diffBwrtlambda48*c*d-9*diffCwrtlambda48*b*d-
9*diffDwrtlambda48*b*c+54*b*e*diffBwrtlambda48+27*b^2*diffEwrtlambda48+54*d*diffDwrtlambda48-
72*diffCwrtlambda48*e-72*c*diffEwrtlambda48;
```

```
DiffOfdelta1wrtlambda49=6*c^2*diffCwrtlambda49-9*diffBwrtlambda49*c*d-9*diffCwrtlambda49*b*d-
9*diffDwrtlambda49*b*c+54*b*e*diffBwrtlambda49+27*b^2*diffEwrtlambda49+54*d*diffDwrtlambda49-
72*diffCwrtlambda49*e-72*c*diffEwrtlambda49;
DiffOfdelta1wrtmu43=6*c^2*diffCwrtmu43-9*diffBwrtmu43*c*d-9*diffCwrtmu43*b*d-
9*diffDwrtmu43*b*c+54*b*e*diffBwrtmu43+27*b^2*diffEwrtmu43+54*d*diffDwrtmu43-72*diffCwrtmu43*e-
72*c*diffEwrtmu43;

% calculate diff of sqrt of delta0cubed :
sqrdelta0=sqrt(delta0);
DiffsqrtOfdelta0cubedwrtlambda12=(3/2)*sqrdelta0*(2*c*diffCwrtlambda12-3*diffBwrtlambda12*d-
3*diffDwrtlambda12*b+12*diffEwrtlambda12);
DiffsqrtOfdelta0cubedwrtlambda18=(3/2)*sqrdelta0*(2*c*diffCwrtlambda18-3*diffBwrtlambda18*d-
3*diffDwrtlambda18*b+12*diffEwrtlambda18);
DiffsqrtOfdelta0cubedwrtlambda19=(3/2)*sqrdelta0*(2*c*diffCwrtlambda19-3*diffBwrtlambda19*d-
3*diffDwrtlambda19*b+12*diffEwrtlambda19);
DiffsqrtOfdelta0cubedwrtlambda23=(3/2)*sqrdelta0*(2*c*diffCwrtlambda23-3*diffBwrtlambda23*d-
3*diffDwrtlambda23*b+12*diffEwrtlambda23);
DiffsqrtOfdelta0cubedwrtlambda28=(3/2)*sqrdelta0*(2*c*diffCwrtlambda28-3*diffBwrtlambda28*d-
3*diffDwrtlambda28*b+12*diffEwrtlambda28);
DiffsqrtOfdelta0cubedwrtlambda29=(3/2)*sqrdelta0*(2*c*diffCwrtlambda29-3*diffBwrtlambda29*d-
3*diffDwrtlambda29*b+12*diffEwrtlambda29);
DiffsqrtOfdelta0cubedwrtmu21=(3/2)*sqrdelta0*(2*c*diffCwrtmu21-3*diffBwrtmu21*d-
3*diffDwrtmu21*b+12*diffEwrtmu21);
DiffsqrtOfdelta0cubedwrtlambda34=(3/2)*sqrdelta0*(2*c*diffCwrtlambda34-3*diffBwrtlambda34*d-
3*diffDwrtlambda34*b+12*diffEwrtlambda34);
DiffsqrtOfdelta0cubedwrtlambda38=(3/2)*sqrdelta0*(2*c*diffCwrtlambda38-3*diffBwrtlambda38*d-
3*diffDwrtlambda38*b+12*diffEwrtlambda38);
DiffsqrtOfdelta0cubedwrtlambda39=(3/2)*sqrdelta0*(2*c*diffCwrtlambda39-3*diffBwrtlambda39*d-
3*diffDwrtlambda39*b+12*diffEwrtlambda39);
DiffsqrtOfdelta0cubedwrtmu32=(3/2)*sqrdelta0*(2*c*diffCwrtmu32-3*diffBwrtmu32*d-
3*diffDwrtmu32*b+12*diffEwrtmu32);
DiffsqrtOfdelta0cubedwrtlambda45=(3/2)*sqrdelta0*(2*c*diffCwrtlambda45-3*diffBwrtlambda45*d-
3*diffDwrtlambda45*b+12*diffEwrtlambda45);
DiffsqrtOfdelta0cubedwrtlambda48=(3/2)*sqrdelta0*(2*c*diffCwrtlambda48-3*diffBwrtlambda48*d-
3*diffDwrtlambda48*b+12*diffEwrtlambda48);
DiffsqrtOfdelta0cubedwrtlambda49=(3/2)*sqrdelta0*(2*c*diffCwrtlambda49-3*diffBwrtlambda49*d-
3*diffDwrtlambda49*b+12*diffEwrtlambda49);
DiffsqrtOfdelta0cubedwrtmu43=(3/2)*sqrdelta0*(2*c*diffCwrtmu43-3*diffBwrtmu43*d-
3*diffDwrtmu43*b+12*diffEwrtmu43);

%calculate num2/den2 wrt each rate :
delta0cubed=delta0^3;
sqrdelta0cubed=sqrt(delta0cubed);
delta1=2*c^3-9*b*c*d+27*b^2*e+27*d^2-72*c*e;
k;
lambda12LastTermSDiff=k*(DiffOfdelta1wrtlambda12*sqrdelta0cubed-
delta1*DiffsqrtOfdelta0cubedwrtlambda12)/delta0cubed;
lambda18LastTermSDiff=k*(DiffOfdelta1wrtlambda18*sqrdelta0cubed-
delta1*DiffsqrtOfdelta0cubedwrtlambda18)/delta0cubed;
lambda19LastTermSDiff=k*(DiffOfdelta1wrtlambda19*sqrdelta0cubed-
delta1*DiffsqrtOfdelta0cubedwrtlambda19)/delta0cubed;
lambda23LastTermSDiff=k*(DiffOfdelta1wrtlambda23*sqrdelta0cubed-
delta1*DiffsqrtOfdelta0cubedwrtlambda23)/delta0cubed;
lambda28LastTermSDiff=k*(DiffOfdelta1wrtlambda28*sqrdelta0cubed-
delta1*DiffsqrtOfdelta0cubedwrtlambda28)/delta0cubed;
lambda29LastTermSDiff=k*(DiffOfdelta1wrtlambda29*sqrdelta0cubed-
delta1*DiffsqrtOfdelta0cubedwrtlambda29)/delta0cubed;
mu21LastTermSDiff=k*(DiffOfdelta1wrtmu21*sqrdelta0cubed-
delta1*DiffsqrtOfdelta0cubedwrtmu21)/delta0cubed;
lambda34LastTermSDiff=k*(DiffOfdelta1wrtlambda34*sqrdelta0cubed-
delta1*DiffsqrtOfdelta0cubedwrtlambda34)/delta0cubed;
lambda38LastTermSDiff=k*(DiffOfdelta1wrtlambda38*sqrdelta0cubed-
delta1*DiffsqrtOfdelta0cubedwrtlambda39)/delta0cubed;
lambda39LastTermSDiff=k*(DiffOfdelta1wrtlambda39*sqrdelta0cubed-
delta1*DiffsqrtOfdelta0cubedwrtlambda39)/delta0cubed;
mu32LastTermSDiff=k*(DiffOfdelta1wrtmu32*sqrdelta0cubed-
delta1*DiffsqrtOfdelta0cubedwrtmu32)/delta0cubed;
lambda45LastTermSDiff=k*(DiffOfdelta1wrtlambda45*sqrdelta0cubed-
delta1*DiffsqrtOfdelta0cubedwrtlambda45)/delta0cubed;
lambda48LastTermSDiff=k*(DiffOfdelta1wrtlambda48*sqrdelta0cubed-
delta1*DiffsqrtOfdelta0cubedwrtlambda48)/delta0cubed;
```

```
lambda49LastTermSDiff=k*(DiffOfdelta1wrtlambda49*sqrdelta0cubed-
delta1*DiffsqrtOfdelta0cubedwrtlambda49)/delta0cubed;
mu43LastTermSDiff=k*(DiffOfdelta1wrtmu43*sqrdelta0cubed-
delta1*DiffsqrtOfdelta0cubedwrtmu43)/delta0cubed;

% calculate diff of s w.r.t. each rate :
n=sqrt(2/3*(-1*p+sqrdelta0*m));
h=1/n;
diffSwrtlambda12=.25*h*(lambda12FirstTermSDiff+lambda12SecondTermSDiff+lambda12LastTermSDiff);
diffSwrtlambda18=.25*h*(lambda18FirstTermSDiff+lambda18SecondTermSDiff+lambda18LastTermSDiff);
diffSwrtlambda19=.25*h*(lambda19FirstTermSDiff+lambda19SecondTermSDiff+lambda19LastTermSDiff);
diffSwrtlambda23=.25*h*(lambda23FirstTermSDiff+lambda23SecondTermSDiff+lambda23LastTermSDiff);
diffSwrtlambda28=.25*h*(lambda28FirstTermSDiff+lambda28SecondTermSDiff+lambda28LastTermSDiff);
diffSwrtlambda29=.25*h*(lambda29FirstTermSDiff+lambda29SecondTermSDiff+lambda29LastTermSDiff);
diffSwrtmu21=.25*h*(mu21FirstTermSDiff+mu21SecondTermSDiff+mu21LastTermSDiff);
diffSwrtlambda34=.25*h*(lambda34FirstTermSDiff+lambda34SecondTermSDiff+lambda34LastTermSDiff);
diffSwrtlambda38=.25*h*(lambda38FirstTermSDiff+lambda38SecondTermSDiff+lambda38LastTermSDiff);
diffSwrtlambda39=.25*h*(lambda39FirstTermSDiff+lambda39SecondTermSDiff+lambda39LastTermSDiff);
diffSwrtmu32=.25*h*(mu32FirstTermSDiff+mu32SecondTermSDiff+mu32LastTermSDiff);
diffSwrtlambda45=.25*h*(lambda45FirstTermSDiff+lambda45SecondTermSDiff+lambda45LastTermSDiff);
diffSwrtlambda48=.25*h*(lambda48FirstTermSDiff+lambda48SecondTermSDiff+lambda48LastTermSDiff);
diffSwrtlambda49=.25*h*(lambda49FirstTermSDiff+lambda49SecondTermSDiff+lambda49LastTermSDiff);
diffSwrtmu43=.25*h*(mu43FirstTermSDiff+mu43SecondTermSDiff+mu43LastTermSDiff);

% differentiate roots( first 4 roots):

p=(8*c-3*b^2)/8;
q=(b^3-4*b*c+8*d)/8;
delta0=c^2-3*b*d+12*e;
sqrdelta0=sqrt(delta0);
delta0cubed=delta0^3;
sqrdelta0cubed=sqrt(delta0cubed);
delta1=2*c^3-9*b*c*d+27*b^2*e+27*d^2-72*c*e;
theta=acos(delta1/(2*sqrt(delta0^3)));
m=cos(theta/3);
s=.5*sqrt(2/3*(-1*p+sqrdelta0*m));

% for root 1,2
% diff root1= -b/4 - s'+ .25 (-4s^2-2p+q/s)^-.5 *(-8ss'-2p'+ (q'*s-s'*q)/s^2)
% diff root2= -b/4 - s'- .25 (-4s^2-2p+q/s)^-.5 *(-8ss'-2p'+ (q'*s-s'*q)/s^2)
% diff root3= -b/4 + s'+ .25 (-4s^2-2p-q/s)^-.5 *(-8ss'-2p'- (q'*s-s'*q)/s^2)
% diff root4= -b/4 + s'- .25 (-4s^2-2p-q/s)^-.5 *(-8ss'-2p'- (q'*s-s'*q)/s^2)

% root 1
f=1/sqrt(-4*s^2-2*p+(q/s));
fNeg=1/sqrt(-4*s^2-2*p-(q/s));

diffroot1wrtlambda12=-(diffBwrtlambda12/4)-diffSwrtlambda12+.25*f*(-8*s*diffSwrtlambda12-
2*diffPwrtlambda12+((diffQwrtlambda12*s-diffSwrtlambda12*q)/s^2));
diffroot1wrtlambda18=-(diffBwrtlambda18/4)-diffSwrtlambda18+.25*f*(-8*s*diffSwrtlambda18-
2*diffPwrtlambda18+((diffQwrtlambda18*s-diffSwrtlambda18*q)/s^2));
diffroot1wrtlambda19=-(diffBwrtlambda19/4)-diffSwrtlambda19+.25*f*(-8*s*diffSwrtlambda19-
2*diffPwrtlambda19+((diffQwrtlambda19*s-diffSwrtlambda19*q)/s^2));
diffroot1wrtlambda23=-(diffBwrtlambda23/4)-diffSwrtlambda23+.25*f*(-8*s*diffSwrtlambda23-
2*diffPwrtlambda23+((diffQwrtlambda23*s-diffSwrtlambda23*q)/s^2));
diffroot1wrtlambda28=-(diffBwrtlambda28/4)-diffSwrtlambda28+.25*f*(-8*s*diffSwrtlambda28-
2*diffPwrtlambda28+((diffQwrtlambda28*s-diffSwrtlambda28*q)/s^2));
diffroot1wrtlambda29=-(diffBwrtlambda29/4)-diffSwrtlambda29+.25*f*(-8*s*diffSwrtlambda29-
2*diffPwrtlambda29+((diffQwrtlambda29*s-diffSwrtlambda29*q)/s^2));
diffroot1wrtmu21=-(diffBwrtmu21/4)-diffSwrtmu21+.25*f*(-8*s*diffSwrtmu21-
2*diffPwrtmu21+((diffQwrtmu21*s-diffSwrtmu21*q)/s^2));
diffroot1wrtlambda34=-(diffBwrtlambda34/4)-diffSwrtlambda34+.25*f*(-8*s*diffSwrtlambda34-
2*diffPwrtlambda34+((diffQwrtlambda34*s-diffSwrtlambda34*q)/s^2));
diffroot1wrtlambda38=-(diffBwrtlambda38/4)-diffSwrtlambda38+.25*f*(-8*s*diffSwrtlambda38-
2*diffPwrtlambda38+((diffQwrtlambda38*s-diffSwrtlambda38*q)/s^2));
diffroot1wrtlambda39=-(diffBwrtlambda39/4)-diffSwrtlambda39+.25*f*(-8*s*diffSwrtlambda39-
2*diffPwrtlambda39+((diffQwrtlambda39*s-diffSwrtlambda39*q)/s^2));
diffroot1wrtmu32=-(diffBwrtmu32/4)-diffSwrtmu32+.25*f*(-8*s*diffSwrtmu32-
2*diffPwrtmu32+((diffQwrtmu32*s-diffSwrtmu32*q)/s^2));
diffroot1wrtlambda45=-(diffBwrtlambda45/4)-diffSwrtlambda45+.25*f*(-8*s*diffSwrtlambda45-
2*diffPwrtlambda45+((diffQwrtlambda45*s-diffSwrtlambda45*q)/s^2));
```

```
diffroot1wrtlambda48=-(diffBwrtlambda48/4)-diffSwrtlambda48+.25*f*(-8*s*diffSwrtlambda48-
2*diffPwrtlambda48+((diffQwrtlambda48*s-diffSwrtlambda48*q)/s^2));
diffroot1wrtlambda49=-(diffBwrtlambda49/4)-diffSwrtlambda49+.25*f*(-8*s*diffSwrtlambda49-
2*diffPwrtlambda49+((diffQwrtlambda49*s-diffSwrtlambda49*q)/s^2));
diffroot1wrtmu43=-(diffBwrtmu43/4)-diffSwrtmu43+.25*f*(-8*s*diffSwrtmu43-
2*diffPwrtmu43+((diffQwrtmu43*s-diffSwrtmu43*q)/s^2));

% root2
diffroot2wrtlambda12=-(diffBwrtlambda12/4)-diffSwrtlambda12-.25*f*(-8*s*diffSwrtlambda12-
2*diffPwrtlambda12+((diffQwrtlambda12*s-diffSwrtlambda12*q)/s^2));
diffroot2wrtlambda18=-(diffBwrtlambda18/4)-diffSwrtlambda18-.25*f*(-8*s*diffSwrtlambda18-
2*diffPwrtlambda18+((diffQwrtlambda18*s-diffSwrtlambda18*q)/s^2));
diffroot2wrtlambda19=-(diffBwrtlambda19/4)-diffSwrtlambda19-.25*f*(-8*s*diffSwrtlambda19-
2*diffPwrtlambda19+((diffQwrtlambda19*s-diffSwrtlambda19*q)/s^2));
diffroot2wrtlambda23=-(diffBwrtlambda23/4)-diffSwrtlambda23-.25*f*(-8*s*diffSwrtlambda23-
2*diffPwrtlambda23+((diffQwrtlambda23*s-diffSwrtlambda23*q)/s^2));
diffroot2wrtlambda28=-(diffBwrtlambda28/4)-diffSwrtlambda28-.25*f*(-8*s*diffSwrtlambda28-
2*diffPwrtlambda28+((diffQwrtlambda28*s-diffSwrtlambda28*q)/s^2));
diffroot2wrtlambda29=-(diffBwrtlambda29/4)-diffSwrtlambda29-.25*f*(-8*s*diffSwrtlambda29-
2*diffPwrtlambda29+((diffQwrtlambda29*s-diffSwrtlambda29*q)/s^2));
diffroot2wrtmu21=-(diffBwrtmu21/4)-diffSwrtmu21-.25*f*(-8*s*diffSwrtmu21-
2*diffPwrtmu21+((diffQwrtmu21*s-diffSwrtmu21*q)/s^2));
diffroot2wrtlambda34=-(diffBwrtlambda34/4)-diffSwrtlambda34-.25*f*(-8*s*diffSwrtlambda34-
2*diffPwrtlambda34+((diffQwrtlambda34*s-diffSwrtlambda34*q)/s^2));
diffroot2wrtlambda38=-(diffBwrtlambda38/4)-diffSwrtlambda38-.25*f*(-8*s*diffSwrtlambda38-
2*diffPwrtlambda38+((diffQwrtlambda38*s-diffSwrtlambda38*q)/s^2));
diffroot2wrtlambda39=-(diffBwrtlambda39/4)-diffSwrtlambda39-.25*f*(-8*s*diffSwrtlambda39-
2*diffPwrtlambda39+((diffQwrtlambda39*s-diffSwrtlambda39*q)/s^2));
diffroot2wrtmu32=-(diffBwrtmu32/4)-diffSwrtmu32-.25*f*(-8*s*diffSwrtmu32-
2*diffPwrtmu32+((diffQwrtmu32*s-diffSwrtmu32*q)/s^2));
diffroot2wrtlambda45=-(diffBwrtlambda45/4)-diffSwrtlambda45-.25*f*(-8*s*diffSwrtlambda45-
2*diffPwrtlambda45+((diffQwrtlambda45*s-diffSwrtlambda45*q)/s^2));
diffroot2wrtlambda48=-(diffBwrtlambda48/4)-diffSwrtlambda48-.25*f*(-8*s*diffSwrtlambda48-
2*diffPwrtlambda48+((diffQwrtlambda48*s-diffSwrtlambda48*q)/s^2));
diffroot2wrtlambda49=-(diffBwrtlambda49/4)-diffSwrtlambda49-.25*f*(-8*s*diffSwrtlambda49-
2*diffPwrtlambda49+((diffQwrtlambda49*s-diffSwrtlambda49*q)/s^2));
diffroot2wrtmu43=-(diffBwrtmu43/4)-diffSwrtmu43-.25*f*(-8*s*diffSwrtmu43-
2*diffPwrtmu43+((diffQwrtmu43*s-diffSwrtmu43*q)/s^2));

% root 3 :

diffroot3wrtlambda12=-(diffBwrtlambda12/4)+diffSwrtlambda12+.25*fNeg*(-8*s*diffSwrtlambda12-
2*diffPwrtlambda12-((diffQwrtlambda12*s-diffSwrtlambda12*q)/s^2));
diffroot3wrtlambda18=-(diffBwrtlambda18/4)+diffSwrtlambda18+.25*fNeg*(-8*s*diffSwrtlambda18-
2*diffPwrtlambda18-((diffQwrtlambda18*s-diffSwrtlambda18*q)/s^2));
diffroot3wrtlambda19=-(diffBwrtlambda19/4)+diffSwrtlambda19+.25*fNeg*(-8*s*diffSwrtlambda19-
2*diffPwrtlambda19-((diffQwrtlambda19*s-diffSwrtlambda19*q)/s^2));
diffroot3wrtlambda23=-(diffBwrtlambda23/4)+diffSwrtlambda23+.25*fNeg*(-8*s*diffSwrtlambda23-
2*diffPwrtlambda23-((diffQwrtlambda23*s-diffSwrtlambda23*q)/s^2));
diffroot3wrtlambda28=-(diffBwrtlambda28/4)+diffSwrtlambda28+.25*fNeg*(-8*s*diffSwrtlambda28-
2*diffPwrtlambda28-((diffQwrtlambda28*s-diffSwrtlambda28*q)/s^2));
diffroot3wrtlambda29=-(diffBwrtlambda29/4)+diffSwrtlambda29+.25*fNeg*(-8*s*diffSwrtlambda29-
2*diffPwrtlambda29-((diffQwrtlambda29*s-diffSwrtlambda29*q)/s^2));
diffroot3wrtmu21=-(diffBwrtmu21/4)+diffSwrtmu21+.25*fNeg*(-8*s*diffSwrtmu21-2*diffPwrtmu21-
((diffQwrtmu21*s-diffSwrtmu21*q)/s^2));
diffroot3wrtlambda34=-(diffBwrtlambda34/4)+diffSwrtlambda34+.25*fNeg*(-8*s*diffSwrtlambda34-
2*diffPwrtlambda34-((diffQwrtlambda34*s-diffSwrtlambda34*q)/s^2));
diffroot3wrtlambda38=-(diffBwrtlambda38/4)+diffSwrtlambda38+.25*fNeg*(-8*s*diffSwrtlambda38-
2*diffPwrtlambda38-((diffQwrtlambda38*s-diffSwrtlambda38*q)/s^2));
diffroot3wrtlambda39=-(diffBwrtlambda39/4)+diffSwrtlambda39+.25*fNeg*(-8*s*diffSwrtlambda39-
2*diffPwrtlambda39-((diffQwrtlambda39*s-diffSwrtlambda39*q)/s^2));
diffroot3wrtmu32=-(diffBwrtmu32/4)+diffSwrtmu32+.25*fNeg*(-8*s*diffSwrtmu32-2*diffPwrtmu32-
((diffQwrtmu32*s-diffSwrtmu32*q)/s^2));
diffroot3wrtlambda45=-(diffBwrtlambda45/4)+diffSwrtlambda45+.25*fNeg*(-8*s*diffSwrtlambda45-
2*diffPwrtlambda45-((diffQwrtlambda45*s-diffSwrtlambda45*q)/s^2));
diffroot3wrtlambda48=-(diffBwrtlambda48/4)+diffSwrtlambda48+.25*fNeg*(-8*s*diffSwrtlambda48-
2*diffPwrtlambda48-((diffQwrtlambda48*s-diffSwrtlambda48*q)/s^2));
diffroot3wrtlambda49=-(diffBwrtlambda49/4)+diffSwrtlambda49+.25*fNeg*(-8*s*diffSwrtlambda49-
2*diffPwrtlambda49-((diffQwrtlambda49*s-diffSwrtlambda49*q)/s^2));
diffroot3wrtmu43=-(diffBwrtmu43/4)+diffSwrtmu43+.25*fNeg*(-8*s*diffSwrtmu43-2*diffPwrtmu43-
((diffQwrtmu43*s-diffSwrtmu43*q)/s^2));
```

```
% root 4 :

diffroot4wrtlambda12=-(diffBwrtlambda12/4)+diffSwrtlambda12-.25*fNeg*(-8*s*diffSwrtlambda12-
2*diffPwrtlambda12-((diffQwrtlambda12*s-diffSwrtlambda12*q)/s^2));
diffroot4wrtlambda18=-(diffBwrtlambda18/4)+diffSwrtlambda18-.25*fNeg*(-8*s*diffSwrtlambda18-
2*diffPwrtlambda18-((diffQwrtlambda18*s-diffSwrtlambda18*q)/s^2));
diffroot4wrtlambda19=-(diffBwrtlambda19/4)+diffSwrtlambda19-.25*fNeg*(-8*s*diffSwrtlambda19-
2*diffPwrtlambda19-((diffQwrtlambda19*s-diffSwrtlambda19*q)/s^2));
diffroot4wrtlambda23=-(diffBwrtlambda23/4)+diffSwrtlambda23-.25*fNeg*(-8*s*diffSwrtlambda23-
2*diffPwrtlambda23-((diffQwrtlambda23*s-diffSwrtlambda23*q)/s^2));
diffroot4wrtlambda28=-(diffBwrtlambda28/4)+diffSwrtlambda28-.25*fNeg*(-8*s*diffSwrtlambda28-
2*diffPwrtlambda28-((diffQwrtlambda28*s-diffSwrtlambda28*q)/s^2));
diffroot4wrtlambda29=-(diffBwrtlambda29/4)+diffSwrtlambda29-.25*fNeg*(-8*s*diffSwrtlambda29-
2*diffPwrtlambda29-((diffQwrtlambda29*s-diffSwrtlambda29*q)/s^2));
diffroot4wrtmu21=-(diffBwrtmu21/4)+diffSwrtmu21-.25*fNeg*(-8*s*diffSwrtmu21-2*diffPwrtmu21-
((diffQwrtmu21*s-diffSwrtmu21*q)/s^2));
diffroot4wrtlambda34=-(diffBwrtlambda34/4)+diffSwrtlambda34-.25*fNeg*(-8*s*diffSwrtlambda34-
2*diffPwrtlambda34-((diffQwrtlambda34*s-diffSwrtlambda34*q)/s^2));
diffroot4wrtlambda38=-(diffBwrtlambda38/4)+diffSwrtlambda38-.25*fNeg*(-8*s*diffSwrtlambda38-
2*diffPwrtlambda38-((diffQwrtlambda38*s-diffSwrtlambda38*q)/s^2));
diffroot4wrtlambda39=-(diffBwrtlambda39/4)+diffSwrtlambda39-.25*fNeg*(-8*s*diffSwrtlambda39-
2*diffPwrtlambda39-((diffQwrtlambda39*s-diffSwrtlambda39*q)/s^2));
diffroot4wrtmu32=-(diffBwrtmu32/4)+diffSwrtmu32-.25*fNeg*(-8*s*diffSwrtmu32-2*diffPwrtmu32-
((diffQwrtmu32*s-diffSwrtmu32*q)/s^2));
diffroot4wrtlambda45=-(diffBwrtlambda45/4)+diffSwrtlambda45-.25*fNeg*(-8*s*diffSwrtlambda45-
2*diffPwrtlambda45-((diffQwrtlambda45*s-diffSwrtlambda45*q)/s^2));
diffroot4wrtlambda48=-(diffBwrtlambda48/4)+diffSwrtlambda48-.25*fNeg*(-8*s*diffSwrtlambda48-
2*diffPwrtlambda48-((diffQwrtlambda48*s-diffSwrtlambda48*q)/s^2));
diffroot4wrtlambda49=-(diffBwrtlambda49/4)+diffSwrtlambda49-.25*fNeg*(-8*s*diffSwrtlambda49-
2*diffPwrtlambda49-((diffQwrtlambda49*s-diffSwrtlambda49*q)/s^2));
diffroot4wrtmu43=-(diffBwrtmu43/4)+diffSwrtmu43-.25*fNeg*(-8*s*diffSwrtmu43-2*diffPwrtmu43-
((diffQwrtmu43*s-diffSwrtmu43*q)/s^2));

% calculate t*exp(rooti*t)*diffOfRoot  with respect to each root let's call
% rootiagain
t=1;
eigenValuesVector= eig(Qq1);
r1=-(b/4)-s+.5*sqrt(-4*s^2-2*p+(q/s));
r2=-(b/4)-s-.5*sqrt(-4*s^2-2*p+(q/s));
r3=-(b/4)+s+.5*sqrt(-4*s^2-2*p-(q/s));
r4=-(b/4)+s-.5*sqrt(-4*s^2-2*p-(q/s));

% root1again
root1againwrtlambda12=t*exp(r1*t)*diffroot1wrtlambda12;
root1againwrtlambda18=t*exp(r1*t)*diffroot1wrtlambda18;
root1againwrtlambda19=t*exp(r1*t)*diffroot1wrtlambda19;
root1againwrtlambda23=t*exp(r1*t)*diffroot1wrtlambda23;
root1againwrtlambda28=t*exp(r1*t)*diffroot1wrtlambda28;
root1againwrtlambda29=t*exp(r1*t)*diffroot1wrtlambda29;
root1againwrtmu21=t*exp(r1*t)*diffroot1wrtmu21;
root1againwrtlambda34=t*exp(r1*t)*diffroot1wrtlambda34;
root1againwrtlambda38=t*exp(r1*t)*diffroot1wrtlambda38;
root1againwrtlambda39=t*exp(r1*t)*diffroot1wrtlambda39;
root1againwrtmu32=t*exp(r1*t)*diffroot1wrtmu32;
root1againwrtlambda45=t*exp(r1*t)*diffroot1wrtlambda45;
root1againwrtlambda48=t*exp(r1*t)*diffroot1wrtlambda48;
root1againwrtlambda49=t*exp(r1*t)*diffroot1wrtlambda49;
root1againwrtmu43=t*exp(r1*t)*diffroot1wrtmu43;

% root2again
root2againwrtlambda12=t*exp(r2*t)*diffroot2wrtlambda12;
root2againwrtlambda18=t*exp(r2*t)*diffroot2wrtlambda18;
root2againwrtlambda19=t*exp(r2*t)*diffroot2wrtlambda19;
root2againwrtlambda23=t*exp(r2*t)*diffroot2wrtlambda23;
root2againwrtlambda28=t*exp(r2*t)*diffroot2wrtlambda28;
root2againwrtlambda29=t*exp(r2*t)*diffroot2wrtlambda29;
root2againwrtmu21=t*exp(r2*t)*diffroot2wrtmu21;
root2againwrtlambda34=t*exp(r2*t)*diffroot2wrtlambda34;
root2againwrtlambda38=t*exp(r2*t)*diffroot2wrtlambda38;
root2againwrtlambda39=t*exp(r2*t)*diffroot2wrtlambda39;
```

```
root2againwrtmu32=t*exp(r2*t)*diffroot2wrtmu32;
root2againwrtlambda45=t*exp(r2*t)*diffroot2wrtlambda45;
root2againwrtlambda48=t*exp(r2*t)*diffroot2wrtlambda48;
root2againwrtlambda49=t*exp(r2*t)*diffroot2wrtlambda49;
root2againwrtmu43=t*exp(r2*t)*diffroot2wrtmu43;

% root3again
root3againwrtlambda12=t*exp(r3*t)*diffroot3wrtlambda12;
root3againwrtlambda18=t*exp(r3*t)*diffroot3wrtlambda18;
root3againwrtlambda19=t*exp(r3*t)*diffroot3wrtlambda19;
root3againwrtlambda23=t*exp(r3*t)*diffroot3wrtlambda23;
root3againwrtlambda28=t*exp(r3*t)*diffroot3wrtlambda28;
root3againwrtlambda29=t*exp(r3*t)*diffroot3wrtlambda29;
root3againwrtmu21=t*exp(r3*t)*diffroot3wrtmu21;
root3againwrtlambda34=t*exp(r3*t)*diffroot3wrtlambda34;
root3againwrtlambda38=t*exp(r3*t)*diffroot3wrtlambda38;
root3againwrtlambda39=t*exp(r3*t)*diffroot3wrtlambda39;
root3againwrtmu32=t*exp(r3*t)*diffroot3wrtmu32;
root3againwrtlambda45=t*exp(r3*t)*diffroot3wrtlambda45;
root3againwrtlambda48=t*exp(r3*t)*diffroot3wrtlambda48;
root3againwrtlambda49=t*exp(r3*t)*diffroot3wrtlambda49;
root3againwrtmu43=t*exp(r3*t)*diffroot3wrtmu43;

% root4again
root4againwrtlambda12=t*exp(r4*t)*diffroot4wrtlambda12;
root4againwrtlambda18=t*exp(r4*t)*diffroot4wrtlambda18;
root4againwrtlambda19=t*exp(r4*t)*diffroot4wrtlambda19;
root4againwrtlambda23=t*exp(r4*t)*diffroot4wrtlambda23;
root4againwrtlambda28=t*exp(r4*t)*diffroot4wrtlambda28;
root4againwrtlambda29=t*exp(r4*t)*diffroot4wrtlambda29;
root4againwrtmu21=t*exp(r4*t)*diffroot4wrtmu21;
root4againwrtlambda34=t*exp(r4*t)*diffroot4wrtlambda34;
root4againwrtlambda38=t*exp(r4*t)*diffroot4wrtlambda38;
root4againwrtlambda39=t*exp(r4*t)*diffroot4wrtlambda39;
root4againwrtmu32=t*exp(r4*t)*diffroot4wrtmu32;
root4againwrtlambda45=t*exp(r4*t)*diffroot4wrtlambda45;
root4againwrtlambda48=t*exp(r4*t)*diffroot4wrtlambda48;
root4againwrtlambda49=t*exp(r4*t)*diffroot4wrtlambda49;
root4againwrtmu43=t*exp(r4*t)*diffroot4wrtmu43;

% sum the  4 row entries of the score function :
sumSCORE1=root1againwrtlambda12+root2againwrtlambda12+root3againwrtlambda12+root4againwrtlambda12
;
sumSCORE2=root1againwrtlambda18+root2againwrtlambda18+root3againwrtlambda18+root4againwrtlambda18
;
sumSCORE3=root1againwrtlambda19+root2againwrtlambda19+root3againwrtlambda19+root4againwrtlambda19
;
sumSCORE4=root1againwrtlambda23+root2againwrtlambda23+root3againwrtlambda23+root4againwrtlambda23
;
sumSCORE5=root1againwrtlambda28+root2againwrtlambda28+root3againwrtlambda28+root4againwrtlambda28
;
sumSCORE6=root1againwrtlambda29+root2againwrtlambda29+root3againwrtlambda29+root4againwrtlambda29
;
sumSCORE7=root1againwrtmu21+root2againwrtmu21+root3againwrtmu21+root4againwrtmu21;
sumSCORE8=root1againwrtlambda34+root2againwrtlambda34+root3againwrtlambda34+root4againwrtlambda34
;
sumSCORE9=root1againwrtlambda38+root2againwrtlambda38+root3againwrtlambda38+root4againwrtlambda38
;
sumSCORE10=root1againwrtlambda39+root2againwrtlambda39+root3againwrtlambda39+root4againwrtlambda3
9;
sumSCORE11=root1againwrtmu32+root2againwrtmu32+root3againwrtmu32+root4againwrtmu32;
sumSCORE12=root1againwrtlambda45+root2againwrtlambda45+root3againwrtlambda45+root4againwrtlambda4
5;
sumSCORE13=root1againwrtlambda48+root2againwrtlambda48+root3againwrtlambda48+root4againwrtlambda4
8;
sumSCORE14=root1againwrtlambda49+root2againwrtlambda49+root3againwrtlambda49+root4againwrtlambda4
9;
sumSCORE15=root1againwrtmu43+root2againwrtmu43+root3againwrtmu43+root4againwrtmu43;

% calculate the root5,root6,root7,root8
g5=-1*Qq1(5,5);
```

```
g6=-1*Qq1(6,6);
g7=-1*Qq1(7,7);
g8=-1*Qq1(8,8);

sqroot5=sqrt((g5+g6)^2-4*g5*g6);
sqroot6=-1*sqrt((g5+g6)^2-4*g5*g6);
sqroot7=sqrt((g7+g8)^2-4*g7*g8);
sqroot8=-1*sqrt((g7+g8)^2-4*g7*g8);

r5=.5*(-1*(g5+g6)+sqroot5);
r6=.5*(-1*(g5+g6)+sqroot6);
r7=.5*(-1*(g7+g8)+sqroot7);
r8=.5*(-1*(g7+g8)+sqroot8);

diffroot5wrtlambda56=-.5+.5*1/sqroot5*(g5-g6);
diffroot5wrtlambda58=-.5+.5*1/sqroot5*(g5-g6);
diffroot5wrtlambda59=-.5+.5*1/sqroot5*(g5-g6);
diffroot5wrtlambda67=-.5+.5*1/sqroot5*(g6-g5);
diffroot5wrtlambda69=-.5+.5*1/sqroot5*(g6-g5);

diffroot6wrtlambda56=-.5-.5*1/sqroot5*(g5-g6);
diffroot6wrtlambda58=-.5-.5*1/sqroot5*(g5-g6);
diffroot6wrtlambda59=-.5-.5*1/sqroot5*(g5-g6);
diffroot6wrtlambda67=-.5-.5*1/sqroot5*(g6-g5);
diffroot6wrtlambda69=-.5-.5*1/sqroot5*(g6-g5);

% calculate 16th to 20th enteries in the score function

root5againwrtlambda56=t*exp(r5*t)*diffroot5wrtlambda56;
root5againwrtlambda58=t*exp(r5*t)*diffroot5wrtlambda58;
root5againwrtlambda59=t*exp(r5*t)*diffroot5wrtlambda59;
root5againwrtlambda67=t*exp(r5*t)*diffroot5wrtlambda67;
root5againwrtlambda69=t*exp(r5*t)*diffroot5wrtlambda69;

root6againwrtlambda56=t*exp(r6*t)*diffroot6wrtlambda56;
root6againwrtlambda58=t*exp(r6*t)*diffroot6wrtlambda58;
root6againwrtlambda59=t*exp(r6*t)*diffroot6wrtlambda59;
root6againwrtlambda67=t*exp(r6*t)*diffroot6wrtlambda67;
root6againwrtlambda69=t*exp(r6*t)*diffroot6wrtlambda69;

sumSCORE16=root5againwrtlambda56+root6againwrtlambda56;
sumSCORE17=root5againwrtlambda58+root6againwrtlambda58;
sumSCORE18=root5againwrtlambda59+root6againwrtlambda59;
sumSCORE19=root5againwrtlambda67+root6againwrtlambda67;
sumSCORE20=root5againwrtlambda69+root6againwrtlambda69;


% differentiate root 7 , root 8

diffroot7wrtlambda79=-.5+.5*1/sqroot7*(g7-g8);
diffroot7wrtlambda89=-.5+.5*1/sqroot7*(g8-g7);
diffroot8wrtlambda79=-.5-.5*1/sqroot7*(g7-g8);
diffroot8wrtlambda89=-.5-.5*1/sqroot7*(g8-g7);


% calculate 21th to 22th enteries in the score function

root7againwrtlambda79=t*exp(r7*t)*diffroot7wrtlambda79;
root7againwrtlambda89=t*exp(r7*t)*diffroot7wrtlambda89;
root8againwrtlambda79=t*exp(r8*t)*diffroot8wrtlambda79;
root8againwrtlambda89=t*exp(r8*t)*diffroot8wrtlambda89;


sumSCORE21=root7againwrtlambda79+root8againwrtlambda79;
sumSCORE22=root7againwrtlambda89+root8againwrtlambda89;

% construct score function or score vector(column vector )
scoreFunction =[sumSCORE1;sumSCORE2;sumSCORE3;sumSCORE4;
                sumSCORE5;sumSCORE6;sumSCORE7;sumSCORE8;
                sumSCORE9;sumSCORE10;sumSCORE11;sumSCORE12;
                sumSCORE13;sumSCORE14;sumSCORE15;sumSCORE16;
```

```
                sumSCORE17;sumSCORE18;sumSCORE19;sumSCORE20;
                sumSCORE21;sumSCORE22 ];
reArrangedScoreFunction =[ sumSCORE1;sumSCORE7;sumSCORE4;sumSCORE11;
                           sumSCORE8;sumSCORE15;sumSCORE21;sumSCORE22;
                           sumSCORE2;sumSCORE3;sumSCORE5;sumSCORE6;
                           sumSCORE9;sumSCORE10;sumSCORE12;sumSCORE13;
                           sumSCORE14;sumSCORE16;sumSCORE17;sumSCORE18;
                           sumSCORE19;sumSCORE20 ];
% scale the score function by 14204
VectorScalar=14204;
scaledScoreFunction=14204*reArrangedScoreFunction;

% construct the M matrix (hessian matrix), scale this by 606523.3 then invert
% the result
MatrixScalar=606523.3;
M_matrix=scaledScoreFunction*scaledScoreFunction' ;
scaledM_matrix=MatrixScalar*M_matrix;

invertible_Mmatrix=scaledM_matrix(1:8,1:8);
% invertible_Mmatrix is singular so calculate pseudoinverse
invertM_matrix=pinv(invertible_Mmatrix)

% construct inverted scaled hessian matrix :
M1=invertM_matrix;
M2=zeros(8,14);
M3=zeros(14,8);
M4=zeros(14,14);

invertedScaledHessianMatrix1=[ M1 M2 ;M3 M4 ];

% apply quasiNewton formula :
initialRateVector=[ lambda12;mu21;lambda23;mu32;lambda34;mu43;lambda79;lambda89;lambda18
                   lambda19;lambda28;lambda29;lambda38;lambda39;lambda45;lambda48;lambda49;
                   lambda56;lambda58;lambda59;lambda67;lambda69 ];
invertedScaledHessianMatrix1*scaledScoreFunction

desiredVector1= initialRateVector + (invertedScaledHessianMatrix1*scaledScoreFunction);
```

The same code is used for Δt=2, Δt=3 with specific scalar for score function and hessian matrix as demonstrated in the text.

# Appendix B: Matlab code for probability matrix estimation

Matlab code for probability matrix estimation

```
clear
clc
close all
Qq = [-.397  .39 0 0 0 0 0 0 .007;
    .02   -.281   .25 0 0 0 0 0 .011;
    0 .05  -.365  .225 0 0 0 .047 .043;
    0 0 .041  -.538 .281 0 0  .109  .107;
    0 0 0 0 -.348 .19 0 .059 .099;
    0 0 0 0 0 -.934 .767 0 .167;
    0 0 0 0 0 0 -.421 0 .421;
    0 0 0 0 0 0 0 -.745 .745;
    0 0 0 0 0 0 0 0 0];
eig(Qq);
lambda12=Qq(1,2);
lambda18=Qq(1,8);
lambda19=Qq(1,9);
mu21=Qq(2,1);
lambda23=Qq(2,3);
lambda28=Qq(2,8);
lambda29=Qq(2,9);
mu32=Qq(3,2);
lambda34=Qq(3,4);
lambda38=Qq(3,8);
lambda39=Qq(3,9);
mu43=Qq(4,3);
lambda45=Qq(4,5);
```

```
lambda48=Qq(4,8);
lambda49=Qq(4,9);
lambda56=Qq(5,6);
lambda58=Qq(5,8);
lambda59=Qq(5,9);
lambda67=Qq(6,7);
lambda69=Qq(6,9);
lambda79=Qq(7,9);
lambda89=Qq(8,9);

% calculate eigenvalues of the Qq matrix
eigenValuesVector= eig(Qq);

%  calculate the roots for the 4 th first polynomial in this manner
% the first 4  roots are the first 4  eiqenvalues
% bur using the formula will get the roots in this order

g1=-1*Qq(1,1);
g2=-1*Qq(2,2);
g3=-1*Qq(3,3);
g4=-1*Qq(4,4);

w1= g1+g2;
w2= g1*(lambda23+lambda28+lambda29)+mu21*(lambda18+lambda19);
w3=g3+g4;
w4=g3*(lambda45+lambda48+lambda49)+mu43*(lambda38+lambda39+mu32);
w5=lambda23*mu32*(g1+g4);
w6=lambda23*mu32*g1*g4;

a=1;
b=w1+w3;
c=w2+w4+w1*w3-lambda23*mu32;
d=w2*w3+w1*w4-w5;
e=w2*w4-w6;

p=(8*c-3*b^2)/8;
q=(b^3-4*b*c+8*d)/8;
delta0=c^2-3*b*d+12*e;
sqrdelta0=sqrt(delta0);
delta0cubed=delta0^3;
sqrdelta0cubed=sqrt(delta0cubed);
delta1=2*c^3-9*b*c*d+27*b^2*e+27*d^2-72*c*e;
theta=acos(delta1/(2*sqrt(delta0^3)));
m=cos(theta/3);
s=.5*sqrt(2/3*(-1*p+sqrdelta0*m));


r1=-(b/4)-s+.5*sqrt(-4*s^2-2*p+(q/s))
r2=-(b/4)-s-.5*sqrt(-4*s^2-2*p+(q/s))
r3=-(b/4)+s+.5*sqrt(-4*s^2-2*p-(q/s))
r4=-(b/4)+s-.5*sqrt(-4*s^2-2*p-(q/s))



% calculate probabilities when time  t=1 for the first elements in first
% row
t=1;
% m = [ exp(-.4609*t) exp(-.5898*t) exp(-.1719*t) exp(-.3584*t) ];

g1=-1*Qq(1,1);
g2=-1*Qq(2,2);
g3=-1*Qq(3,3);
g4=-1*Qq(4,4);
%calculate DPij*(s)-12 equations
DP11sss= 1;
DP11ss=g2+g3+g4;
DP11s=g2*g3 + g2*g4+ g3*g4 -lambda34*mu43 -lambda23*mu32;
DP11constant=g2*g3*g4 -lambda23*mu32*g4 -lambda34*mu43*g2;
DP11col=[DP11sss ;DP11ss ;DP11s ;DP11constant ];
% DP12
DP12sss= 0;
```

```
DP12ss=lambda12;
DP12s=lambda12*(g3+g4);
DP12constant=lambda12*g3*g4-lambda12*lambda34*mu43;
DP12col=[DP12sss ;DP12ss ;DP12s ;DP12constant ]
% DP13
DP13sss= 0;
DP13ss=0;
DP13s=lambda12*lambda23;
DP13constant=lambda12*lambda23*g4;
DP13col=[DP13sss ;DP13ss ;DP13s ;DP13constant ]
% DP14
DP14sss= 0;
DP14ss=0;
DP14s=0;
DP14constant=lambda12*lambda23*lambda34;
DP14col=[DP14sss ;DP14ss ;DP14s ;DP14constant ]
% DP21
DP21sss= 0;
DP21ss=mu21;
DP21s=mu21*(g3+g4);
DP21constant=mu21*g3*g4-mu21*lambda34*mu43;
DP21col=[DP21sss ;DP21ss ;DP21s ;DP21constant ];
% DP22
DP22sss= 1;
DP22ss=g1+g3+g4;
DP22s=g1*g3+g1*g4+g3*g4-lambda34*mu43;
DP22constant=g1*g3*g4-lambda34*mu43*g1;
DP22col=[DP22sss ;DP22ss ;DP22s ;DP22constant ];
% DP23
DP23sss= 0;
DP23ss=lambda23;
DP23s=lambda23*(g4+g1);
DP23constant=lambda23*g4*g1;
DP23col=[DP23sss ;DP23ss ;DP23s ;DP23constant ];
% DP24
DP24sss= 0;
DP24ss=0;
DP24s=lambda23*lambda34;
DP24constant=lambda23*lambda34*g1;
DP24col=[DP24sss ;DP24ss ;DP24s ;DP24constant ];
% DP31
DP31sss= 0;
DP31ss=0;
DP31s=mu21*mu32;
DP31constant=mu21*mu32*g4;
DP31col=[DP31sss ;DP31ss ;DP31s ;DP31constant ];
% DP32
DP32sss= 0;
DP32ss=mu32;
DP32s=mu32*(g1+g4);
DP32constant=mu32*g1*g4;
DP32col=[DP32sss ;DP32ss ;DP32s ;DP32constant ];
% DP33
DP33sss= 1;
DP33ss=g1+g2+g4;
DP33s=g1*g2+g1*g4+g2*g4-mu21*lambda12;
DP33constant=g1*g2*g4-mu21*lambda12*g4;
DP33col=[DP33sss ;DP33ss ;DP33s ;DP33constant ];
% DP34
DP34sss= 0;
DP34ss=lambda34;
DP34s=lambda34*(g1+g2);
DP34constant=lambda34*g1*g2-lambda12*lambda34*mu21;
DP34col=[DP34sss ;DP34ss ;DP34s ;DP34constant ];
% DP41
DP41sss=0;
DP41ss=0;
DP41s=0;
DP41constant=mu21*mu32*mu43;
DP41col=[DP41sss ;DP41ss ;DP41s ;DP41constant ];
% DP42
```

```
DP42sss=0;
DP42ss=0;
DP42s= mu32*mu43;
DP42constant= mu32*mu43*g1;
DP42col=[DP42sss ;DP42ss ;DP42s ;DP42constant ];
% DP43
DP43sss=0;
DP43ss=mu43;
DP43s= mu43*(g1+g2);
DP43constant= g1*g2*mu43-lambda12*mu21*mu43;
DP43col=[DP43sss ;DP43ss ;DP43s ;DP43constant ];
% DP44
DP44sss=1;
DP44ss=g1+g2+g3;
DP44s= g1*g2+g2*g3+g1*g3-lambda23*mu32-mu21*lambda12;
DP44constant= g1*g2*g3-lambda23*mu32*g1-mu21*lambda12*g3;
DP44col=[DP44sss ;DP44ss ;DP44s ;DP44constant ];
% construct K matrix
%  eigvalues= [  -.4609  -.5898  -.1719  -.3584  ];
%  r1=eigvalues(1,1);
%  r2=eigvalues(1,2);
%  r3=eigvalues(1,3);
%  r4=eigvalues(1,4);

krow1=[1 1 1 1 ];
%krow2
krow2Element1=-1*(r2+r3+r4);
krow2Element2=-1*(r1+r3+r4);
krow2Element3=-1*(r1+r2+r4);
krow2Element4=-1*(r1+r2+r3);
krow2=[krow2Element1 krow2Element2 krow2Element3 krow2Element4];
%krow3
krow3Element1=(r2*r3+r3*r4+r2*r4);
krow3Element2=(r1*r3+r3*r4+r1*r4);
krow3Element3=(r1*r4+r1*r2+r2*r4);
krow3Element4=(r1*r3+r2*r1+r2*r3);
krow3=[krow3Element1 krow3Element2 krow3Element3 krow3Element4];
%krow4
krow4Element1=-1*(r2*r3*r4);
krow4Element2=-1*(r1*r3*r4);
krow4Element3=-1*(r1*r2*r4);
krow4Element4=-1*(r1*r2*r3);
krow4=[krow4Element1 krow4Element2 krow4Element3 krow4Element4];
K=[krow1;krow2;krow3;krow4];
Kin=inv(K);
%calculate  Aij,Bij,Cij,Dij for each P11,P12,P13,P14
coeffP11= Kin*DP11col
coeffP12= Kin*DP12col
coeffP13=Kin*DP13col
coeffP14=Kin*DP14col
%calculate  Aij,Bij,Cij,Dij for each P21,P22,P23,P24
coeffP21=Kin*DP21col;
coeffP22=Kin*DP22col;
coeffP23=Kin*DP23col;
coeffP24=Kin*DP24col;
%calculate  Aij,Bij,Cij,Dij for each P31,P32,P33,P34
coeffP31=Kin*DP31col;
coeffP32=Kin*DP32col;
coeffP33=Kin*DP33col;
coeffP34=Kin*DP34col;
%calculate  Aij,Bij,Cij,Dij for each P41,P42,P43,P44
coeffP41=Kin*DP41col;
coeffP42=Kin*DP42col;
coeffP43=Kin*DP43col;
coeffP44=Kin*DP44col;
m = [ exp(r1*t) exp(r2*t) exp(r3*t) exp(r4*t) ];
%calculate first 4  probabilities for the first row :
P11=m*coeffP11
P12=m*coeffP12;
P13=m*coeffP13;
P14=m*coeffP14;
```

```
%calculate first 4  probabilities for the second  row :
P21=m*coeffP21;
P22=m*coeffP22;
P23=m*coeffP23;
P24=m*coeffP24;
%calculate first 4  probabilities for the third row :
P31=m*coeffP31;
P32=m*coeffP32;
P33=m*coeffP33;
P34=m*coeffP34;
%calculate first 4  probabilities for the fourth row :
P41=m*coeffP41;
P42=m*coeffP42;
P43=m*coeffP43;
P44=m*coeffP44;

%calculate P15:

fifthGs=lambda45.*coeffP14;
G1=fifthGs(1,1);
G2=fifthGs(2,1);
G3=fifthGs(3,1);
G4=fifthGs(4,1);


w=lambda56+lambda58+lambda59;
F1=G1/(w+r1);
F2=G2/(w+r2);
F3=G3/(w+r3);
F4=G4/(w+r4);
F5=-1*(F1+F2+F3+F4);

m15 = [  exp(r1*t) exp(r2*t) exp(r3*t) exp(r4*t)  exp(-w*t) ];

F1To5vector=[F1; F2; F3; F4; F5];
P15=m15*F1To5vector;
%calculate P25:

fifthHs=lambda45.*coeffP24;
H1=fifthHs(1,1);
H2=fifthHs(2,1);
H3=fifthHs(3,1);
H4=fifthHs(4,1);


w=lambda56+lambda58+lambda59;
K1=H1/(w+r1);
K2=H2/(w+r2);
K3=H3/(w+r3);
K4=H4/(w+r4);
K5=-1*(K1+K2+K3+K4);

m25 = [ exp(r1*t) exp(r2*t) exp(r3*t) exp(r4*t) exp(-w*t) ];

K1To5vector=[K1; K2; K3; K4; K5];
P25=m25*K1To5vector;
% calculate P35:

fifthLs=lambda45.*coeffP34;
L1=fifthLs(1,1);
L2=fifthLs(2,1);
L3=fifthLs(3,1);
L4=fifthLs(4,1);


w=lambda56+lambda58+lambda59;
M1=L1/(w+r1);
M2=L2/(w+r2);
M3=L3/(w+r3);
M4=L4/(w+r4);
M5=-1*(M1+M2+M3+M4);
```

```
m35 = [ exp(r1*t) exp(r2*t) exp(r3*t) exp(r4*t) exp(-w*t) ];

M1To5vector=[M1; M2; M3; M4; M5];
P35=m35*M1To5vector;
%calculate P45 :

fifthYs=lambda45.*coeffP44;
Y1=fifthYs(1,1);
Y2=fifthYs(2,1);
Y3=fifthYs(3,1);
Y4=fifthYs(4,1);


w=lambda56+lambda58+lambda59;
R1=Y1/(w+r1);
R2=Y2/(w+r2);
R3=Y3/(w+r3);
R4=Y4/(w+r4);
R5=-1*(R1+R2+R3+R4);

m45 = [ exp(r1*t) exp(r2*t) exp(r3*t) exp(r4*t) exp(-w*t) ];

R1To5vector=[R1; R2; R3; R4; R5];
P45=m45*R1To5vector;
%calculate P16

u=lambda67+lambda69;

F1To5vector=[F1; F2; F3; F4; F5];
sithGs=lambda56.*F1To5vector;
G5=sithGs(1,1);
G6=sithGs(2,1);
G7=sithGs(3,1);
G8=sithGs(4,1);
G9=sithGs(5,1);

F6=G5/(u+r1);
F7=G6/(u+r2);
F8=G7/(u+r3);
F9=G8/(u+r4);
F10=G9/(u-w);
F11=-1*(F6+F7+F8+F9+F10);

m16 = [ exp(r1*t) exp(r2*t) exp(r3*t) exp(r4*t) exp(-w*t) exp(-u*t)];
F6To11vector=[F6;F7;F8;F9;F10;F11];
P16=m16*F6To11vector;

%calculate P26:
K1To5vector=[K1; K2; K3; K4; K5];
sithHs=lambda56.*K1To5vector;
H5=sithHs(1,1);
H6=sithHs(2,1);
H7=sithHs(3,1);
H8=sithHs(4,1);
H9=sithHs(5,1);

K6=H5/(u+r1);
K7=H6/(u+r2);
K8=H7/(u+r3);
K9=H8/(u+r4);
K10=H9/(u-w);
K11=-1*(K6+K7+K8+K9+K10);

m26 = [ exp(r1*t) exp(r2*t) exp(r3*t) exp(r4*t) exp(-w*t) exp(-u*t)];
K6To11vector=[K6;K7;K8;K9;K10;K11];
P26=m26*K6To11vector;

%calculate P36:
M1To5vector=[M1; M2; M3; M4; M5];
sithLs=lambda56.*M1To5vector;
```

```
L5=sithLs(1,1);
L6=sithLs(2,1);
L7=sithLs(3,1);
L8=sithLs(4,1);
L9=sithLs(5,1);

M6=L5/(u+r1);
M7=L6/(u+r2);
M8=L7/(u+r3);
M9=L8/(u+r4);
M10=L9/(u-w);
M11=-1*(M6+M7+M8+M9+M10);

m36 = [ exp(r1*t) exp(r2*t) exp(r3*t) exp(r4*t) exp(-w*t) exp(-u*t)];
M6To11vector=[M6;M7;M8;M9;M10;M11];
P36=m36*M6To11vector;

%calculate P46:
R1To5vector=[R1; R2; R3; R4; R5];
sithYs=lambda56.*R1To5vector;
Y5=sithYs(1,1);
Y6=sithYs(2,1);
Y7=sithYs(3,1);
Y8=sithYs(4,1);
Y9=sithYs(5,1);

R6=Y5/(u+r1);
R7=Y6/(u+r2);
R8=Y7/(u+r3);
R9=Y8/(u+r4);
R10=Y9/(u-w);
R11=-1*(R6+R7+R8+R9+R10);

m46 = [ exp(r1*t) exp(r2*t) exp(r3*t) exp(r4*t) exp(-w*t) exp(-u*t)];
R6To11vector=[R6;R7;R8;R9;R10;R11];
P46=m46*R6To11vector;

%calculate P17:
F6To11vector=[F6;F7;F8;F9;F10;F11];

seventhGs=lambda67.*F6To11vector;
G10=seventhGs(1,1);
G11=seventhGs(2,1);
G12=seventhGs(3,1);
G13=seventhGs(4,1);
G14=seventhGs(5,1);
G15=seventhGs(6,1);

F12=G10/(lambda79+r1);
F13=G11/(lambda79+r2);
F14=G12/(lambda79+r3);
F15=G13/(lambda79+r4);
F16=G14/(lambda79-w);
F17=G15/(lambda79-u);
F18=-1*(F12+F13+F14+F15+F16+F17);

m17 = [ exp(r1*t) exp(r2*t) exp(r3*t) exp(r4*t) exp(-w*t) exp(-u*t) exp(-lambda79*t)];
F12To18vector=[F12;F13;F14;F15;F16;F17;F18];
P17=m17*F12To18vector;

%calculate P27:
K6To11vector=[K6;K7;K8;K9;K10;K11];

seventhHs=lambda67.*K6To11vector;
H10=seventhHs(1,1);
H11=seventhHs(2,1);
H12=seventhHs(3,1);
H13=seventhHs(4,1);
H14=seventhHs(5,1);
H15=seventhHs(6,1);
```

```
K12=H10/(lambda79+r1);
K13=H11/(lambda79+r2);
K14=H12/(lambda79+r3);
K15=H13/(lambda79+r4);
K16=H14/(lambda79-w);
K17=H15/(lambda79-u);
K18=-1*(K12+K13+K14+K15+K16+K17);

m27 = [ exp(r1*t) exp(r2*t) exp(r3*t) exp(r4*t) exp(-w*t) exp(-u*t) exp(-lambda79*t)];
K12To18vector=[K12;K13;K14;K15;K16;K17;K18];
P27=m27*K12To18vector;

%calculate P37:
M6To11vector=[M6;M7;M8;M9;M10;M11];

seventhLs=lambda67.*M6To11vector;
L10=seventhLs(1,1);
L11=seventhLs(2,1);
L12=seventhLs(3,1);
L13=seventhLs(4,1);
L14=seventhLs(5,1);
L15=seventhLs(6,1);

M12=L10/(lambda79+r1);
M13=L11/(lambda79+r2);
M14=L12/(lambda79+r3);
M15=L13/(lambda79+r4);
M16=L14/(lambda79-w);
M17=L15/(lambda79-u);
M18=-1*(M12+M13+M14+M15+M16+M17);

m37 = [ exp(r1*t) exp(r2*t) exp(r3*t) exp(r4*t) exp(-w*t) exp(-u*t) exp(-lambda79*t)];
M12To18vector=[M12;M13;M14;M15;M16;M17;M18];
P37=m37*M12To18vector

%calculate P47:
R6To11vector=[R6;R7;R8;R9;R10;R11];

seventhYs=lambda67.*R6To11vector;
Y10=seventhYs(1,1);
Y11=seventhYs(2,1);
Y12=seventhYs(3,1);
Y13=seventhYs(4,1);
Y14=seventhYs(5,1);
Y15=seventhYs(6,1);

R12=Y10/(lambda79+r1);
R13=Y11/(lambda79+r2);
R14=Y12/(lambda79+r3);
R15=Y13/(lambda79+r4);
R16=Y14/(lambda79-w);
R17=Y15/(lambda79-u);
R18=-1*(R12+R13+R14+R15+R16+R17);

m47 = [ exp(r1*t) exp(r2*t) exp(r3*t) exp(r4*t) exp(-w*t) exp(-u*t) exp(-lambda79*t)];
R12To18vector=[R12;R13;R14;R15;R16;R17;R18];
P47=m47*R12To18vector;

%calculate P18:
% define  Aij,Bij,Cij,Dij for the first 4 probabilities in first 4 rows :
A11=coeffP11(1,1);
B11=coeffP11(2,1);
C11=coeffP11(3,1);
D11=coeffP11(4,1);

A12=coeffP12(1,1);
B12=coeffP12(2,1);
C12=coeffP12(3,1);
D12=coeffP12(4,1);

A13=coeffP13(1,1);
```

```
B13=coeffP13(2,1);
C13=coeffP13(3,1);
D13=coeffP13(4,1);

A14=coeffP14(1,1);
B14=coeffP14(2,1);
C14=coeffP14(3,1);
D14=coeffP14(4,1);

A21=coeffP21(1,1);
B21=coeffP21(2,1);
C21=coeffP21(3,1);
D21=coeffP21(4,1);

A22=coeffP22(1,1);
B22=coeffP22(2,1);
C22=coeffP22(3,1);
D22=coeffP22(4,1);

A23=coeffP23(1,1);
B23=coeffP23(2,1);
C23=coeffP23(3,1);
D23=coeffP23(4,1);

A24=coeffP24(1,1);
B24=coeffP24(2,1);
C24=coeffP24(3,1);
D24=coeffP24(4,1);

A31=coeffP31(1,1);
B31=coeffP31(2,1);
C31=coeffP31(3,1);
D31=coeffP31(4,1);

A32=coeffP32(1,1);
B32=coeffP32(2,1);
C32=coeffP32(3,1);
D32=coeffP32(4,1);

A33=coeffP33(1,1);
B33=coeffP33(2,1);
C33=coeffP33(3,1);
D33=coeffP33(4,1);

A34=coeffP34(1,1);
B34=coeffP34(2,1);
C34=coeffP34(3,1);
D34=coeffP34(4,1);

A41=coeffP41(1,1);
B41=coeffP41(2,1);
C41=coeffP41(3,1);
D41=coeffP41(4,1);

A42=coeffP42(1,1);
B42=coeffP42(2,1);
C42=coeffP42(3,1);
D42=coeffP42(4,1);

A43=coeffP43(1,1);
B43=coeffP43(2,1);
C43=coeffP43(3,1);
D43=coeffP43(4,1);

A44=coeffP44(1,1);
B44=coeffP44(2,1);
C44=coeffP44(3,1);
D44=coeffP44(4,1);

% F1To5vector=[F1; F2; F3; F4; F5]
```

```
lambda8vector=[lambda18; lambda28; lambda38; lambda48 ;lambda58];

g16=[A11 A12 A13 A14 F1];
g17=[B11 B12 B13 B14 F2];
g18=[C11 C12 C13 C14 F3];
g19=[D11 D12 D13 D14 F4];

G16=g16*lambda8vector;
G17=g17*lambda8vector;
G18=g18*lambda8vector;
G19=g19*lambda8vector;
G20=lambda58*F5;

F19=G16/(lambda89+r1);
F20=G17/(lambda89+r2);
F21=G18/(lambda89+r3);
F22=G19/(lambda89+r4);
F23=G20/(lambda89-w);
F24=-1*(F19+F20+F21+F22+F23);

m18 = [ exp(r1*t) exp(r2*t) exp(r3*t) exp(r4*t) exp(-w*t) exp(-lambda89*t)];
F19To24vector=[F19;F20;F21;F22;F23;F24];
P18=m18*F19To24vector;

% calculate P28 :
% K1To5vector=[K1; K2; K3; K4; K5]

h16=[A21 A22 A23 A24 K1];
h17=[B21 B22 B23 B24 K2];
h18=[C21 C22 C23 C24 K3];
h19=[D21 D22 D23 D24 K4];

H16=h16*lambda8vector;
H17=h17*lambda8vector;
H18=h18*lambda8vector;
H19=h19*lambda8vector;
H20=lambda58*K5;

K19=H16/(lambda89+r1);
K20=H17/(lambda89+r2);
K21=H18/(lambda89+r3);
K22=H19/(lambda89+r4);
K23=H20/(lambda89-w);
K24=-1*(K19+K20+K21+K22+K23);

m28 = [ exp(r1*t) exp(r2*t) exp(r3*t) exp(r4*t) exp(-w*t) exp(-lambda89*t)];
K19To24vector=[K19;K20;K21;K22;K23;K24];
P28=m28*K19To24vector;

% calculate P38 :
% M1To5vector=[M1; M2; M3; M4; M5]

l16=[A31 A32 A33 A34 M1];
l17=[B31 B32 B33 B34 M2];
l18=[C31 C32 C33 C34 M3];
l19=[D31 D32 D33 D34 M4];

L16=l16*lambda8vector;
L17=l17*lambda8vector;
L18=l18*lambda8vector;
L19=l19*lambda8vector;
L20=lambda58*M5;

M19=L16/(lambda89+r1);
M20=L17/(lambda89+r2);
M21=L18/(lambda89+r3);
M22=L19/(lambda89+r4);
M23=L20/(lambda89-w);
M24=-1*(M19+M20+M21+M22+M23);

m38 = [ exp(r1*t) exp(r2*t) exp(r3*t) exp(r4*t) exp(-w*t) exp(-lambda89*t)];
```

```
M19To24vector=[M19;M20;M21;M22;M23;M24];
P38=m38*M19To24vector;

% calculate P48 :
% R1To5vector=[R1; R2; R3; R4; R5]

y16=[A41 A42 A43 A44 R1];
y17=[B41 B42 B43 B44 R2];
y18=[C41 C42 C43 C44 R3];
y19=[D41 D42 D43 D44 R4];

Y16=y16*lambda8vector;
Y17=y17*lambda8vector;
Y18=y18*lambda8vector;
Y19=y19*lambda8vector;
Y20=lambda58*R5;

R19=Y16/(lambda89+r1);
R20=Y17/(lambda89+r2);
R21=Y18/(lambda89+r3);
R22=Y19/(lambda89+r4);
R23=Y20/(lambda89-w);
R24=-1*(R19+R20+R21+R22+R23);

m48 = [ exp(r1*t) exp(r2*t) exp(r3*t) exp(r4*t) exp(-w*t) exp(-lambda89*t)];
R19To24vector=[R19;R20;R21;R22;R23;R24];
P48=m48*R19To24vector;

%calculate P19:


lambda9vector=[lambda19; lambda29; lambda39; lambda49 ;lambda59 ;lambda69; lambda79; lambda89 ];

g21=[A11 A12 A13 A14 F1 F6 F12 F19];
g22=[B11 B12 B13 B14 F2 F7 F13 F20];
g23=[C11 C12 C13 C14 F3 F8 F14 F21];
g24=[D11 D12 D13 D14 F4 F9 F15 F22];
g25=[F5 F10 F16 F23];
g26=[F11 F17 ];
g27=F18;
g28=F24;

G21=g21*lambda9vector;
G22=g22*lambda9vector;
G23=g23*lambda9vector;
G24=g24*lambda9vector;
G25=g25*[lambda59; lambda69; lambda79; lambda89];
G26=g26*[lambda69; lambda79];
G27=g27*lambda79;
G28=g28*lambda89;

F25=G21/r1;
F26=G22/r2;
F27=G23/r3;
F28=G24/r4;
F29=-1*(G25/w);
F30=-1*(G26/u);
F31=-1*(G27/lambda79);
F32=-1*(G28/lambda89);
F33=-1*(F25+F26+F27+F28)-1*(F29+F30+F31+F32);

m19 = [ exp(r1*t) exp(r2*t) exp(r3*t) exp(r4*t) exp(-w*t) exp(-u*t) exp(-lambda79*t) exp(-
lambda89*t) exp(0)];
F25To33vector=[F25;F26;F27;F28;F29;F30;F31;F32;F33 ];
P19=m19*F25To33vector;

% calculate P29:

h21=[A21 A22 A23 A24 K1 K6 K12 K19];
h22=[B21 B22 B23 B24 K2 K7 K13 K20];
h23=[C21 C22 C23 C24 K3 K8 K14 K21];
```

```
h24=[D21 D22 D23 D24 K4 K9 K15 K22];
h25=[K5 K10 K16 K23];
h26=[K11 K17 ];
h27=K18;
h28=K24;

H21=h21*lambda9vector;
H22=h22*lambda9vector;
H23=h23*lambda9vector;
H24=h24*lambda9vector;
H25=h25*[lambda59; lambda69; lambda79; lambda89];
H26=h26*[lambda69; lambda79];
H27=h27*lambda79;
H28=h28*lambda89;

K25=H21/r1;
K26=H22/r2;
K27=H23/r3;
K28=H24/r4;
K29=-1*(H25/w);
K30=-1*(H26/u);
K31=-1*(H27/lambda79);
K32=-1*(H28/lambda89);
K33=-1*(K25+K26+K27+K28)-1*(K29+K30+K31+K32);

m29 = [ exp(r1*t) exp(r2*t) exp(r3*t) exp(r4*t) exp(-w*t) exp(-u*t) exp(-lambda79*t) exp(-
lambda89*t) exp(0)];
K25To33vector=[K25;K26;K27;K28;K29;K30;K31;K32;K33 ];
P29=m29*K25To33vector;

%calculate P39
l21=[A31 A32 A33 A34 M1 M6 M12 M19];
l22=[B31 B32 B33 B34 M2 M7 M13 M20];
l23=[C31 C32 C33 C34 M3 M8 M14 M21];
l24=[D31 D32 D33 D34 M4 M9 M15 M22];
l25=[M5 M10 M16 M23];
l26=[M11 M17 ];
l27=M18;
l28=M24;

L21=l21*lambda9vector;
L22=l22*lambda9vector;
L23=l23*lambda9vector;
L24=l24*lambda9vector;
L25=l25*[lambda59; lambda69; lambda79; lambda89];
L26=l26*[lambda69; lambda79];
L27=l27*lambda79;
L28=l28*lambda89;

M25=L21/r1;
M26=L22/r2;
M27=L23/r3;
M28=L24/r4;
M29=-1*(L25/w);
M30=-1*(L26/u);
M31=-1*(L27/lambda79);
M32=-1*(L28/lambda89);
M33=-1*(M25+M26+M27+M28)-1*(M29+M30+M31+M32);

m39 = [ exp(r1*t) exp(r2*t) exp(r3*t) exp(r4*t) exp(-w*t) exp(-u*t) exp(-lambda79*t) exp(-
lambda89*t) exp(0)];
M25To33vector=[M25;M26;M27;M28;M29;M30;M31;M32;M33 ];
P39=m39*M25To33vector;

%calculate P49:
y21=[A41 A42 A43 A44 R1 R6 R12 R19];
y22=[B41 B42 B43 B44 R2 R7 R13 R20];
y23=[C41 C42 C43 C44 R3 R8 R14 R21];
y24=[D41 D42 D43 D44 R4 R9 R15 R22];
y25=[R5 R10 R16 R23];
y26=[R11 R17 ];
```

```
y27=R18;
y28=R24;

Y21=y21*lambda9vector;
Y22=y22*lambda9vector;
Y23=y23*lambda9vector;
Y24=y24*lambda9vector;
Y25=y25*[lambda59; lambda69; lambda79; lambda89];
Y26=y26*[lambda69; lambda79];
Y27=y27*lambda79;
Y28=y28*lambda89;

R25=Y21/r1;
R26=Y22/r2;
R27=Y23/r3;
R28=Y24/r4;
R29=-1*(Y25/w);
R30=-1*(Y26/u);
R31=-1*(Y27/lambda79);
R32=-1*(Y28/lambda89);
R33=-1*(R25+R26+R27+R28)-1*(R29+R30+R31+R32);

m49 = [ exp(r1*t) exp(r2*t) exp(r3*t) exp(r4*t) exp(-w*t) exp(-u*t) exp(-lambda79*t) exp(-
lambda89*t) exp(0)];
R25To33vector=[R25;R26;R27;R28;R29;R30;R31;R32;R33 ];
P49=m49*R25To33vector;

Sum1=P11+P12+P13+P14+P15+P16+P17+P18+P19
Sum2=P21+P22+P23+P24+P25+P26+P27+P28+P29
Sum3=P31+P32+P33+P34+P35+P36+P37+P38+P39
Sum4=P41+P42+P43+P44+P45+P46+P47+P48+P49

% calculate probabilities of the 5th row:
% calculate P55

P55=exp(-w*t);

% calculate P56:
V1=lambda56/(u-w);
P56=V1*(exp(-w*t)-exp(-u*t));

% calculate P57
V2=(lambda67*V1)/(lambda79-w);
V3=-1*(lambda67*V1)/(lambda79-u);
V4=-1*(V2+V3);
m57=[exp(-w*t) exp(-u*t) exp(-lambda79*t)];
V2ToV4vector=[V2;V3;V4];
P57=m57*V2ToV4vector;

% calculate P58
V5=lambda58/(lambda89-w);
P58=V5*(exp(-w*t)-exp(-lambda89*t));

% calculate P59
V6=lambda59+lambda69*V1+lambda79*V2+lambda89*V5;
V7=lambda79*V3-lambda69*V1;
V8=lambda79*V4;
V9=-lambda89*V5;
V10=-V6/w;
V11=-V7/u;
V12=-V8/lambda79;
V13=-V9/lambda89;
V14=-1*(V10+V11+V12+V13);
m59=[exp(-w*t) exp(-u*t) exp(-lambda79*t) exp(-lambda89*t) exp(0) ];
V10ToV14vector=[V10; V11; V12; V13; V14];
P59=m59*V10ToV14vector

Sum5=P55+P56+P57+P58+P59;

% calculate probability of 6th row
% calculate P66
```

```
P66=exp(-u*t);

% calculate P67:
Z1=lambda67/(lambda79-u);
P67=Z1*(exp(-u*t)-exp(-lambda79*t));

% calculate P69 :
Z2=-1*(lambda69+lambda79*Z1)/u;
Z3=Z1;
m69=[exp(-u*t) exp(-lambda79*t) exp(0) exp(0)];
Z2Toz3vector=[Z2;Z3;-Z2;-Z3];
P69=m69*Z2Toz3vector;

Sum6=P66+P67+P69;

%calculate 7 th row
% calculate P77
P77=exp(-lambda79*t);
P79=1-P77;

%calculate 8 th row
P88=exp(-lambda89*t);
P89=1-P88;

Sum5=P55+P56+P57+P58+P59
Sum6=P66+P67+P69
Sum7=P77+P79
Sum8=P88+P89

% calculate P's in each row
P1vector=[P11 P12 P13 P14 P15 P16 P17 P18 P19];
P2vector=[P21 P22 P23 P24 P25 P26 P27 P28 P29];
P3vector=[P31 P32 P33 P34 P35 P36 P37 P38 P39];
P4vector=[P41 P42 P43 P44 P45 P46 P47 P48 P49];
P5vector=[0    0   0   0  P55 P56 P57 P58 P59];
P6vector=[0    0   0   0  0   P66 P67  0  P69];
P7vector=[0    0   0   0  0   0   P77  0  P79];
P8vector=[0    0   0   0  0   0    0  P88 P89];
P9vector=[0    0   0   0  0   0    0   0   1 ];
Pij=[P1vector;P2vector;P3vector;P4vector;P5vector;P6vector;P7vector;P8vector;P9vector]
initialNumOfPatients=[3100 1100 405 150 140 25 35 45 0 ];
initialProbOfPatient= [ .62 .22 .081 .03 .028 .005 .007 .009 0 ];
expectedNumOfPatients= initialNumOfPatients*Pij
oneYearProbOfPatient = initialProbOfPatient*Pij
```

# Appendix C: Goodness of Fit

## C.1. Small model:

To calculate goodness of fit for multistate model used in the small model, it is like the procedure used in contingency table, and it is calculated in each interval then sum up:

Step 1 : $H_0 =$ *future state does not depend on the current state.*

$H_1 =$ *future state does depend on the current state*

Step 2: calculate the $P_{ij}(\Delta t = 1) = \begin{bmatrix} .7338 & .2139 & .0247 & .0277 \\ .0206 & .7411 & .1793 & .059 \\ 0 & 0 & 1 & 0 \\ 0 & 0 & 0 & 1 \end{bmatrix}$ by exponentiation of the estimated Q matrix step 3 :calculate the expected counts in this interval by multiplying each row in the probability matrix with the corresponding total marginal counts in the observed transition counts matrix in the same interval to get the expected counts.

| | State 1 | State 2 | State3 | State4 | total |
|---|---|---|---|---|---|
| State1 | 403.59 | 117.645 | 13.585 | 15.235 | 550.055 |
| State2 | 5.15 | 185.275 | 44.825 | 14.75 | 250 |
| State3 | 0 | 0 | 0 | 0 | 0 |
| State4 | 0 | 0 | 0 | 0 | 0 |

Step 4: apply $\sum_{i=1}^{4}\sum_{j=1}^{4}\frac{(O_{ij}-E_{ij})^2}{E_{ij}} = 104.247 \sim \chi^2_{(4-1)(4-1)(.05)}$

The same steps are used for the observed transition counts in the $\Delta t = 2\ and\ \Delta t = 3$ with the following results:

$$P_{ij}(\Delta t = 2) = \begin{bmatrix} .5428 & .3154 & .0811 & .0607 \\ .0304 & .5537 & .3126 & .1033 \\ 0 & 0 & 1 & 0 \\ 0 & 0 & 0 & 1 \end{bmatrix}$$

The expected counts:

| | State 1 | State 2 | State3 | State4 | total |
|---|---|---|---|---|---|
| State1 | 60.2508 | 35.0094 | 9.0021 | 6.7377 | 111 |
| State2 | 1.1856 | 21.5943 | 12.1914 | 4.0287 | 39 |
| State3 | 0 | 0 | 0 | 0 | 0 |
| State4 | 0 | 0 | 0 | 0 | 0 |

$$\sum_{i=1}^{4}\sum_{j=1}^{4}\frac{(O_{ij}-E_{ij})^2}{E_{ij}} = 8.022 \sim \chi^2_{(4-1)(4-1)(.05)}$$

The same steps are used for the observed transition counts in $\Delta t = 3$ with the following results:

$$P_{ij}(\Delta t = 3) = \begin{bmatrix} .4048 & .3499 & .151 & .0943 \\ .0337 & .4168 & .4126 & .1368 \\ 0 & 0 & 1 & 0 \\ 0 & 0 & 0 & 1 \end{bmatrix}$$

The expected counts:

| | State 1 | State 2 | State3 | State4 | total |
|---|---|---|---|---|---|
| State1 | 15.7872 | 13.6461 | 5.889 | 3.6777 | 39 |
| State2 | .3707 | 4.5848 | 4.5386 | 1.5048 | 11 |
| State3 | 0 | 0 | 0 | 0 | 0 |
| State4 | 0 | 0 | 0 | 0 | 0 |

$$\sum_{i=1}^{4}\sum_{j=1}^{4}\frac{(O_{ij}-E_{ij})^2}{E_{ij}} = 6.588 \sim \chi^2_{(4-1)(4-1)(.05)}$$

Step 5: sum up the above results to get:

$$\sum_{i=1}^{4}\sum_{j=1}^{4}\sum_{l=1}^{t=3}\frac{(O_{ijl}-E_{ijl})^2}{E_{ijl}} = 118.857 \sim \chi^2_{(df=27)(.05)}$$

So from the above results the null hypothesis is rejected while the alternative hypothesis is accepted and the model fits the data that is to mean the future state depends on the current state with the estimated transition rate and probability matrices as obtained.

**C.2. Big model:**

Step 1: $H_0 =$ *future state does not depend on the current state*

$H_1 =$ *future state does depend on the current state*

Step 2: calculate the $P_{ij}(\Delta t = 1)$

$$P_{ij}(\Delta t = 1) = \begin{bmatrix} .6751 & .279 & .0345 & .0024 & .0002 & .0000 & .0000 & .0006 & .0082 \\ .0143 & .7625 & .1819 & .0188 & .0018 & .0001 & .0000 & .0044 & .016 \\ .0004 & .0364 & .7017 & .1409 & .0206 & .0012 & .0002 & .0346 & .0604 \\ 0 & .0007 & .0257 & .561 & .1771 & .0145 & .0039 & .0616 & .1215 \\ 0 & 0 & 0 & 0 & .7061 & .1015 & .0416 & .0344 & .1163 \\ 0 & 0 & 0 & 0 & 0 & .3930 & .3938 & 0 & .2132 \\ 0 & 0 & 0 & 0 & 0 & 0 & .6564 & 0 & .3436 \\ 0 & 0 & 0 & 0 & 0 & 0 & 0 & .4747 & .5253 \\ 0 & 0 & 0 & 0 & 0 & 0 & 0 & 0 & 1 \end{bmatrix}$$

Step 3: calculate the expected counts in this interval by multiplying each row in the probability matrix with the corresponding total marginal counts in the observed transition counts matrix in the same interval to get the expected counts

| | State1 | State2 | State3 | State4 | State5 | State6 | State7 | State8 | State9 | total |
|---|---|---|---|---|---|---|---|---|---|---|
| State1 | 1000.5 | 413.5 | 51.13 | 3.6 | .3 | 0 | 0 | .8892 | 12.15 | 1482 |
| State2 | 7.4074 | 394.97 | 94.22 | 9.738 | .9324 | .0518 | 0 | 2.2792 | 8.288 | 517.89 |
| State3 | .08 | 7.28 | 140.34 | 28.18 | 4.12 | .24 | .04 | 6.92 | 12.08 | 199.28 |
| State4 | 0 | .0511 | 1.8761 | 40.953 | 12.9283 | 1.0585 | .2847 | 4.4968 | 8.8695 | 70.518 |
| State5 | 0 | 0 | 0 | 0 | 47.3087 | 6.8005 | 2.7872 | 2.3048 | 7.7921 | 66.9933 |
| State6 | 0 | 0 | 0 | 0 | 0 | 3.93 | 3.938 | 0 | 2.132 | 10 |
| State7 | 0 | 0 | 0 | 0 | 0 | 0 | 11.1588 | 0 | 5.8412 | 17 |
| State8 | 0 | 0 | 0 | 0 | 0 | 0 | 0 | 9.494 | 10.506 | 20 |
| State9 | 0 | 0 | 0 | 0 | 0 | 0 | 0 | 0 | 0 | 0 |

Step 4: apply $\sum_{i=1}^{9}\sum_{j=1}^{9}\frac{(O_{ij}-E_{ij})^2}{E_{ij}} = 2226.362 \sim \chi^2_{(9-1)(9-1)(.05)}$

The same steps are used for the observed transition counts in the $\Delta t = 2$ *and* $\Delta t = 3$ with the following results:

$$P_{ij}(\Delta t = 2) = \begin{bmatrix} .4597 & .4023 & .0983 & .0131 & .0019 & .0001 & .0000 & .0032 & .0209 \\ .0206 & .5920 & .2673 & .0505 & .0097 & .0007 & .0002 & .0129 & .0441 \\ .001 & .0535 & .5026 & .1786 & .054 & .0054 & .0022 & .0503 & .1414 \\ 0 & .0018 & .0325 & .3183 & .2249 & .0318 & .0178 & .0708 & .2486 \\ 0 & 0 & 0 & 0 & .4986 & .1116 & .0967 & .0406 & .2525 \\ 0 & 0 & 0 & 0 & 0 & .1544 & .4133 & 0 & .4323 \\ 0 & 0 & 0 & 0 & 0 & 0 & .4308 & 0 & .5692 \\ 0 & 0 & 0 & 0 & 0 & 0 & 0 & .2254 & .7746 \\ 0 & 0 & 0 & 0 & 0 & 0 & 0 & 0 & 1 \end{bmatrix}$$

The expected counts

| | State1 | State2 | State3 | State4 | State5 | State6 | State7 | State8 | State9 | total |
|---|---|---|---|---|---|---|---|---|---|---|
| State1 | 272.6 | 238.56 | 58.292 | 7.7683 | 1.1267 | .0593 | 0 | 1.8976 | 12.394 | 592.704 |
| State2 | 4.2642 | 122.54 | 55.33 | 10.454 | 2.0079 | .1449 | .0414 | 2.6703 | 9.129 | 206.59 |
| State3 | .08 | 4.28 | 40.208 | 14.288 | 4.32 | .432 | .176 | 4.024 | 11.312 | 79.12 |
| State4 | 0 | .0522 | .9425 | 9.231 | 6.5221 | .922 | .5162 | 2.0532 | 7.209 | 27.45 |
| State5 | 0 | 0 | 0 | 0 | 13.462 | 3.013 | 2.6109 | 1.0962 | 6.818 | 27 |
| State6 | 0 | 0 | 0 | 0 | 0 | .6176 | 1.6532 | 0 | 1.729 | 4 |
| State7 | 0 | 0 | 0 | 0 | 0 | 0 | 3.0156 | 0 | 3.984 | 7 |
| State8 | 0 | 0 | 0 | 0 | 0 | 0 | 0 | 1.8032 | 6.197 | 8 |
| State9 | 0 | 0 | 0 | 0 | 0 | 0 | 0 | 0 | 0 | 0 |

$\sum_{i=1}^{9}\sum_{j=1}^{9}\frac{(O_{ij}-E_{ij})^2}{E_{ij}} = 160.115 \sim \chi^2_{(9-1)(9-1)(.05)}$

The same steps are used for the observed transition counts in $\Delta t = 3$ with the following results:

$$P_{ij}(\Delta t=3)=\begin{bmatrix} .3161 & .4386 & .1584 & .0299 & .0065 & .0006 & .0002 & .0078 & .0406 \\ .0225 & .4669 & .2973 & .0771 & .0223 & .0024 & .0011 & .0214 & .0842 \\ .0016 & .0595 & .367 & .172 & .0802 & .0108 & .0066 & .0544 & .2289 \\ .0001 & .0028 & .0313 & .1832 & .2159 & .0400 & .0348 & .0621 & .3655 \\ 0 & 0 & 0 & 0 & .352 & .0945 & .1282 & .0364 & .3889 \\ 0 & 0 & 0 & 0 & 0 & .0607 & .3321 & 0 & .6072 \\ 0 & 0 & 0 & 0 & 0 & 0 & .2828 & 0 & .7172 \\ 0 & 0 & 0 & 0 & 0 & 0 & 0 & .107 & .893 \\ 0 & 0 & 0 & 0 & 0 & 0 & 0 & 0 & 1 \end{bmatrix}$$

The expected counts

| | State1 | State2 | State3 | State4 | State5 | State6 | State7 | State8 | State9 | total |
|---|---|---|---|---|---|---|---|---|---|---|
| State1 | 46.783 | 64.913 | 23.443 | 4.4252 | .962 | .0888 | .0296 | 1.1544 | 6.0088 | 147.81 |
| State2 | 1.148 | 23.812 | 15.162 | 3.9321 | 1.1373 | .1224 | .0561 | 1.3914 | 4.2942 | 50.76 |
| State3 | .0288 | 1.071 | 6.606 | 3.096 | 1.4436 | .1944 | .1188 | .9792 | 4.1202 | 17.66 |
| State4 | .0007 | .0196 | .2191 | 1.2824 | 1.5113 | .28 | .2436 | .4347 | 2.5585 | 6.55 |
| State5 | 0 | 0 | 0 | 0 | 2.464 | .6615 | .8974 | .2548 | 2.7223 | 7 |
| State6 | 0 | 0 | 0 | 0 | 0 | .1214 | .6642 | 0 | 1.2144 | 2 |
| State7 | 0 | 0 | 0 | 0 | 0 | 0 | .5656 | 0 | 1.4344 | 2 |
| State8 | 0 | 0 | 0 | 0 | 0 | 0 | 0 | .321 | 2.679 | 3 |
| State9 | 0 | 0 | 0 | 0 | 0 | 0 | 0 | 0 | 0 | 0 |

$\sum_{i=1}^{9}\sum_{j=1}^{9}\frac{\left(O_{ij}-E_{ij}\right)^2}{E_{ij}}=69.778\sim\chi^2_{(9-1)(9-1)(.05)}$

Step 5: sum up the above results to get:

$$\sum_{i=1}^{9}\sum_{j=1}^{9}\sum_{l=1}^{t=3}\frac{\left(O_{ijl}-E_{ijl}\right)^2}{E_{ijl}}=2456.255\sim\chi^2_{(df=192)(.05)}$$

So from the above results the null hypothesis is rejected while the alternative hypothesis is accepted and the model fits the data that is to mean the future state depends on the current state with the estimated transition rate and probability matrices as obtained.

**C.3. Model with Covariates:**

Step 1 : $H_0 = \textit{future state does not depend on the current state}$

$H_1 = \textit{future state does depend on the current state.}$

Step 2: calculate the $p_{ij}(\Delta t=1)=\begin{bmatrix} .9435 & .0551 & .0014 & 0 & 0 \\ .0274 & .9247 & .0469 & .0009 & .0001 \\ .0144 & .0327 & .9149 & .0348 & .0032 \\ .0023 & .0863 & .1245 & .6512 & .1357 \\ 0 & 0 & 0 & 0 & 1 \end{bmatrix}$

Using exponentiation of the estimated Q matrix

Step3: calculate the expected counts in this interval by multiplying each row in the probability matrix with the corresponding total marginal counts in the observed transition counts matrix in the same interval to get the expected counts

| | State 0 | State 1 | State 2 | State 3 | State 4 | Total |
|---|---|---|---|---|---|---|
| State 0 | 1934.175 | 112.955 | 2.87 | 0 | 0 | 2050 |
| State 1 | 34.1678 | 1153.101 | 58.4843 | 1.1223 | 0.1247 | 1247 |
| State 2 | 11.2752 | 25.6041 | 716.3667 | 27.2484 | 2.5056 | 783 |
| State 3 | .276 | 10.356 | 14.94 | 78.144 | 16.284 | 120 |
| State 4 | 0 | 0 | 0 | 0 | 0 | 0 |

Step 4: apply $\sum_{i=1}^{5}\sum_{j=1}^{5}\frac{\left(O_{ij}-E_{ij}\right)^2}{E_{ij}}=1140.097\sim\chi^2_{(5-1)(5-1)(.05)}$ . So from the above results the null hypothesis is rejected while the alternative hypothesis is accepted and the model fits the data that is to mean the future state depends on the current state with the estimated transition rates and probability matrices as obtained.

**Appendix D: selected tables as referred in chapter 6:** Table (1): patients' characteristics.

| ID | Sex | age | Age Cat. | LDL chol | LDL Cat | Homa 2-IR | Homa Cat | BMI | BMI cat | Sys pre | Sys cat | Dias pre | Dias cat |
|---|---|---|---|---|---|---|---|---|---|---|---|---|---|
| **1** | 0 | 27 | 1 | 59.89 | 1 | 0.49 | 1 | 20.30 | 1 | 123.40 | 1 | 70.00 | 1 |
| **2** | 0 | 27 | 1 | 62.18 | 1 | 0.63 | 1 | 20.39 | 1 | 124.18 | 1 | 70.96 | 1 |
| **3** | 0 | 29 | 1 | 63.65 | 1 | 0.65 | 1 | 22.21 | 1 | 125.90 | 1 | 71.78 | 1 |
| **4** | 0 | 29 | 1 | 65.81 | 1 | 0.84 | 1 | 22.50 | 1 | 126.18 | 1 | 71.91 | 1 |
| **5** | 0 | 30 | 1 | 69.79 | 1 | 1.05 | 1 | 23.14 | 1 | 130.08 | 2 | 73.52 | 1 |
| **6** | 0 | 31 | 1 | 70.11 | 2 | 1.06 | 1 | 23.18 | 1 | 130.43 | 2 | 73.89 | 1 |
| **7** | 0 | 31 | 1 | 70.60 | 2 | 1.08 | 1 | 23.35 | 1 | 132.98 | 2 | 74.24 | 1 |
| **8** | 0 | 32 | 1 | 71.74 | 2 | 1.09 | 1 | 23.36 | 1 | 133.19 | 2 | 74.61 | 1 |
| **9** | 1 | 33 | 1 | 74.00 | 2 | 1.13 | 1 | 23.52 | 1 | 135.13 | 2 | 74.70 | 1 |
| **10** | 0 | 34 | 1 | 74.63 | 2 | 1.19 | 1 | 23.70 | 1 | 135.27 | 2 | 75.54 | 1 |
| **11** | 0 | 34 | 1 | 74.72 | 2 | 1.29 | 2 | 23.85 | 1 | 135.50 | 2 | 76.49 | 1 |
| **12** | 0 | 34 | 1 | 75.17 | 2 | 1.32 | 2 | 24.18 | 1 | 135.66 | 2 | 76.52 | 1 |
| **13** | 0 | 34 | 1 | 75.18 | 2 | 1.34 | 2 | 24.23 | 1 | 136.00 | 2 | 77.16 | 1 |
| **14** | 0 | 34 | 1 | 75.23 | 2 | 1.42 | 2 | 24.29 | 1 | 136.10 | 2 | 79.35 | 1 |
| **15** | 0 | 34 | 1 | 75.62 | 2 | 1.43 | 2 | 24.36 | 1 | 136.90 | 2 | 79.74 | 1 |
| **16** | 0 | 35 | 1 | 76.18 | 2 | 1.45 | 2 | 24.44 | 1 | 137.08 | 2 | 80.02 | 1 |
| **17** | 0 | 35 | 1 | 76.24 | 2 | 1.45 | 2 | 24.54 | 1 | 137.09 | 2 | 80.16 | 1 |
| **18** | 0 | 35 | 1 | 76.30 | 2 | 1.46 | 2 | 24.61 | 1 | 137.56 | 2 | 80.75 | 1 |
| **19** | 0 | 35 | 1 | 76.70 | 2 | 1.46 | 2 | 24.62 | 1 | 138.07 | 2 | 80.85 | 1 |
| **20** | 0 | 35 | 1 | 77.59 | 2 | 1.51 | 2 | 24.64 | 1 | 138.32 | 2 | 82.06 | 1 |
| **21** | 0 | 35 | 1 | 77.71 | 2 | 1.54 | 2 | 24.91 | 1 | 138.59 | 2 | 82.10 | 1 |
| **22** | 0 | 35 | 1 | 77.95 | 2 | 1.54 | 2 | 24.92 | 1 | 139.01 | 2 | 82.32 | 1 |
| **23** | 0 | 36 | 2 | 78.06 | 2 | 1.59 | 2 | 25.06 | 2 | 139.18 | 2 | 82.49 | 1 |
| **24** | 0 | 36 | 2 | 78.26 | 2 | 1.62 | 2 | 25.10 | 2 | 139.42 | 2 | 82.77 | 1 |
| **25** | 0 | 36 | 2 | 78.44 | 2 | 1.62 | 2 | 25.12 | 2 | 139.89 | 2 | 83.11 | 1 |
| **26** | 0 | 36 | 2 | 79.18 | 2 | 1.63 | 2 | 25.13 | 2 | 139.91 | 2 | 83.43 | 1 |
| **27** | 0 | 36 | 2 | 79.34 | 2 | 1.64 | 2 | 25.18 | 2 | 140.56 | 2 | 83.54 | 1 |
| **28** | 0 | 36 | 2 | 79.90 | 2 | 1.64 | 2 | 25.29 | 2 | 140.70 | 2 | 83.81 | 1 |
| **29** | 0 | 36 | 2 | 80.23 | 2 | 1.65 | 2 | 25.61 | 2 | 140.72 | 2 | 83.84 | 1 |
| **30** | 0 | 36 | 2 | 80.54 | 2 | 1.65 | 2 | 25.70 | 2 | 141.08 | 2 | 84.51 | 1 |
| **31** | 0 | 36 | 2 | 80.79 | 2 | 1.67 | 2 | 25.71 | 2 | 141.14 | 2 | 84.66 | 1 |
| **32** | 0 | 36 | 2 | 80.95 | 2 | 1.67 | 2 | 25.73 | 2 | 141.30 | 2 | 84.77 | 1 |
| **33** | 0 | 36 | 2 | 81.40 | 2 | 1.68 | 2 | 25.80 | 2 | 141.74 | 2 | 84.94 | 1 |
| **34** | 0 | 36 | 2 | 81.52 | 2 | 1.69 | 2 | 25.83 | 2 | 142.02 | 2 | 85.05 | 2 |
| **35** | 0 | 36 | 2 | 81.63 | 2 | 1.69 | 2 | 25.96 | 2 | 142.14 | 2 | 85.08 | 2 |
| **36** | 0 | 36 | 2 | 81.64 | 2 | 1.71 | 2 | 25.98 | 2 | 142.27 | 2 | 85.97 | 2 |
| **37** | 0 | 37 | 2 | 81.8 | 2 | 1.73 | 2 | 26.03 | 2 | 142.75 | 2 | 86.08 | 2 |
| **38** | 0 | 37 | 2 | 82.10 | 2 | 1.73 | 2 | 26.05 | 2 | 142.88 | 2 | 86.34 | 2 |
| **39** | 0 | 37 | 2 | 82.73 | 2 | 1.74 | 2 | 26.20 | 2 | 143.64 | 2 | 86.78 | 2 |
| **40** | 0 | 37 | 2 | 83.19 | 2 | 1.78 | 2 | 26.25 | 2 | 143.65 | 2 | 87.11 | 2 |
| **41** | 0 | 37 | 2 | 83.51 | 2 | 1.78 | 2 | 26.35 | 2 | 143.84 | 2 | 87.38 | 2 |
| **42** | 0 | 37 | 2 | 83.88 | 2 | 1.82 | 2 | 26.36 | 2 | 143.90 | 2 | 87.50 | 2 |
| **43** | 0 | 37 | 2 | 83.91 | 2 | 1.83 | 2 | 26.44 | 2 | 144.10 | 2 | 88.10 | 2 |
| **44** | 0 | 37 | 2 | 84.40 | 2 | 1.86 | 2 | 26.65 | 2 | 144.15 | 2 | 88.17 | 2 |
| **45** | 0 | 37 | 2 | 85.18 | 2 | 1.87 | 2 | 26.67 | 2 | 144.35 | 2 | 88.77 | 2 |
| **46** | 0 | 38 | 2 | 86.24 | 2 | 1.87 | 2 | 26.78 | 2 | 144.51 | 2 | 88.78 | 2 |
| **47** | 0 | 38 | 2 | 86.42 | 2 | 1.92 | 2 | 26.81 | 2 | 144.52 | 2 | 89.01 | 2 |
| **48** | 0 | 38 | 2 | 86.97 | 2 | 1.92 | 2 | 26.96 | 2 | 144.74 | 2 | 89.32 | 2 |
| **49** | 0 | 38 | 2 | 87.03 | 2 | 1.93 | 2 | 27.11 | 2 | 145.11 | 2 | 89.34 | 2 |
| **50** | 0 | 38 | 2 | 87.21 | 2 | 1.94 | 2 | 27.11 | 2 | 145.17 | 2 | 89.38 | 2 |
| **51** | 1 | 38 | 2 | 87.60 | 2 | 1.94 | 2 | 27.13 | 2 | 145.63 | 2 | 89.50 | 2 |
| **52** | 1 | 38 | 2 | 87.65 | 2 | 1.95 | 2 | 27.31 | 2 | 145.78 | 2 | 89.53 | 2 |

| ID | Sex | age | Age Cat. | LDL-chol | LDL Cat. | HOMA2-IR | Homa Cat. | BMI | BMI Cat. | Sys. Pre. | Sys. Cat. | Dias. Pre. | Dias Cat. |
|---|---|---|---|---|---|---|---|---|---|---|---|---|---|
| **53** | 1 | 38 | 2 | 88.22 | 2 | 1.95 | 2 | 27.31 | 2 | 145.84 | 2 | 89.56 | 2 |
| **54** | 1 | 39 | 2 | 88.34 | 2 | 1.96 | 2 | 27.34 | 2 | 145.85 | 2 | 89.82 | 2 |
| **55** | 1 | 39 | 2 | 88.74 | 2 | 1.96 | 2 | 27.36 | 2 | 145.95 | 2 | 89.95 | 2 |
| **56** | 1 | 39 | 2 | 89.00 | 2 | 1.99 | 2 | 27.42 | 2 | 146.00 | 2 | 90.18 | 2 |
| **57** | 1 | 39 | 2 | 89.31 | 2 | 2.03 | 2 | 27.43 | 2 | 146.44 | 2 | 90.34 | 2 |
| **58** | 1 | 39 | 2 | 90.06 | 2 | 2.03 | 2 | 27.45 | 2 | 146.60 | 2 | 90.87 | 2 |
| **59** | 1 | 39 | 2 | 90.42 | 2 | 2.05 | 2 | 27.51 | 2 | 146.73 | 2 | 90.92 | 2 |
| **60** | 1 | 39 | 2 | 90.67 | 2 | 2.07 | 2 | 27.54 | 2 | 146.74 | 2 | 91.06 | 2 |
| **61** | 1 | 39 | 2 | 90.92 | 2 | 2.08 | 2 | 27.64 | 2 | 146.88 | 2 | 91.07 | 2 |
| **62** | 1 | 39 | 2 | 91.45 | 2 | 2.10 | 2 | 27.76 | 2 | 146.91 | 2 | 91.12 | 2 |
| **63** | 1 | 39 | 2 | 91.74 | 2 | 2.12 | 2 | 27.78 | 2 | 147.31 | 2 | 91.30 | 2 |
| **64** | 1 | 39 | 2 | 92.51 | 2 | 2.13 | 2 | 27.79 | 2 | 147.39 | 2 | 91.32 | 2 |
| **65** | 1 | 39 | 2 | 92.73 | 2 | 2.15 | 2 | 27.80 | 2 | 147.39 | 2 | 91.45 | 2 |
| **66** | 1 | 40 | 2 | 92.90 | 2 | 2.15 | 2 | 27.85 | 2 | 147.77 | 2 | 91.50 | 2 |
| **67** | 1 | 40 | 2 | 92.90 | 2 | 2.16 | 2 | 27.87 | 2 | 148.11 | 2 | 91.60 | 2 |
| **68** | 1 | 40 | 2 | 93.15 | 2 | 2.17 | 2 | 28.02 | 2 | 148.21 | 2 | 91.61 | 2 |
| **69** | 1 | 40 | 2 | 93.18 | 2 | 2.19 | 2 | 28.02 | 2 | 148.23 | 2 | 91.67 | 2 |
| **70** | 1 | 40 | 2 | 93.54 | 2 | 2.19 | 2 | 28.11 | 2 | 148.27 | 2 | 91.70 | 2 |
| **71** | 1 | 40 | 2 | 93.57 | 2 | 2.20 | 2 | 28.28 | 2 | 148.37 | 2 | 92.01 | 2 |
| **72** | 1 | 40 | 2 | 93.97 | 2 | 2.21 | 2 | 28.32 | 2 | 148.85 | 2 | 92.04 | 2 |
| **73** | 1 | 40 | 2 | 94.15 | 2 | 2.23 | 2 | 28.35 | 2 | 149.01 | 2 | 92.99 | 2 |
| **74** | 1 | 40 | 2 | 94.27 | 2 | 2.24 | 2 | 28.36 | 2 | 149.08 | 2 | 93.62 | 2 |
| **75** | 1 | 40 | 2 | 94.57 | 2 | 2.25 | 2 | 28.38 | 2 | 149.38 | 2 | 93.99 | 2 |
| **76** | 1 | 40 | 2 | 94.68 | 2 | 2.28 | 2 | 28.42 | 2 | 149.44 | 2 | 94.15 | 2 |
| **77** | 1 | 40 | 2 | 94.73 | 2 | 2.29 | 2 | 28.45 | 2 | 149.69 | 2 | 94.28 | 2 |
| **78** | 1 | 40 | 2 | 94.81 | 2 | 2.30 | 2 | 28.45 | 2 | 149.96 | 2 | 94.65 | 2 |
| **79** | 1 | 40 | 2 | 95.22 | 2 | 2.31 | 2 | 28.49 | 2 | 149.98 | 2 | 94.68 | 2 |
| **80** | 1 | 41 | 2 | 95.32 | 2 | 2.33 | 2 | 28.61 | 2 | 150.00 | 2 | 94.95 | 2 |
| **81** | 1 | 41 | 2 | 95.81 | 2 | 2.33 | 2 | 28.61 | 2 | 150.03 | 2 | 95.50 | 2 |
| **82** | 1 | 41 | 2 | 95.88 | 2 | 2.36 | 2 | 28.62 | 2 | 150.47 | 2 | 95.68 | 2 |
| **83** | 1 | 41 | 2 | 96.00 | 2 | 2.36 | 2 | 28.62 | 2 | 150.63 | 2 | 95.72 | 2 |
| **84** | 1 | 41 | 2 | 96.11 | 2 | 2.37 | 2 | 28.67 | 2 | 150.76 | 2 | 95.92 | 2 |
| **85** | 1 | 41 | 2 | 96.40 | 2 | 2.39 | 2 | 28.71 | 2 | 150.84 | 2 | 96.07 | 2 |
| **86** | 1 | 41 | 2 | 96.77 | 2 | 2.39 | 2 | 28.76 | 2 | 150.86 | 2 | 96.34 | 2 |
| **87** | 1 | 41 | 2 | 96.81 | 2 | 2.40 | 2 | 28.81 | 2 | 150.91 | 2 | 96.48 | 2 |
| **88** | 1 | 41 | 2 | 96.90 | 2 | 2.41 | 2 | 28.92 | 2 | 150.92 | 2 | 96.52 | 2 |
| **89** | 1 | 41 | 2 | 96.91 | 2 | 2.42 | 2 | 28.99 | 2 | 151.06 | 2 | 96.61 | 2 |
| **90** | 1 | 41 | 2 | 97.55 | 2 | 2.43 | 2 | 29.10 | 2 | 151.58 | 2 | 96.78 | 2 |
| **91** | 1 | 41 | 2 | 97.58 | 2 | 2.46 | 2 | 29.11 | 2 | 151.69 | 2 | 97.16 | 2 |
| **92** | 1 | 41 | 2 | 97.82 | 2 | 2.47 | 2 | 29.14 | 2 | 152.01 | 2 | 97.39 | 2 |
| **93** | 1 | 41 | 2 | 98.19 | 2 | 2.51 | 2 | 29.17 | 2 | 152.49 | 2 | 97.51 | 2 |
| **94** | 1 | 41 | 2 | 98.65 | 2 | 2.54 | 2 | 29.18 | 2 | 152.54 | 2 | 97.58 | 2 |
| **95** | 1 | 42 | 2 | 98.68 | 2 | 2.56 | 2 | 29.34 | 2 | 153.38 | 2 | 98.40 | 2 |
| **96** | 1 | 42 | 2 | 99.00 | 2 | 2.58 | 2 | 29.37 | 2 | 153.46 | 2 | 98.42 | 2 |
| **97** | 1 | 42 | 2 | 99.47 | 2 | 2.58 | 2 | 29.41 | 2 | 153.50 | 2 | 98.49 | 2 |
| **98** | 1 | 42 | 2 | 99.57 | 2 | 2.59 | 2 | 29.41 | 2 | 153.63 | 2 | 98.93 | 2 |
| **99** | 1 | 42 | 2 | 100.28 | 3 | 2.61 | 2 | 29.60 | 2 | 153.91 | 2 | 99.20 | 2 |
| **100** | 1 | 42 | 2 | 100.57 | 3 | 2.64 | 2 | 29.72 | 2 | 153.94 | 2 | 99.39 | 2 |
| **101** | 1 | 42 | 2 | 100.60 | 3 | 2.65 | 2 | 29.80 | 2 | 154.46 | 2 | 99.72 | 2 |
| **102** | 1 | 42 | 2 | 101.85 | 3 | 2.68 | 2 | 29.94 | 2 | 154.56 | 2 | 99.74 | 2 |

| ID | Sex | age | Age Cat. | LDL chol | LDL Cat. | Homa 2-IR | Homa Cat. | BMI | BMI Cat. | Sys Pre. | Sys Cat. | Dias Pre. | Dias Cat. |
|---|---|---|---|---|---|---|---|---|---|---|---|---|---|
| **103** | 1 | 42 | 2 | 101.95 | 3 | 2.69 | 2 | 29.97 | 2 | 154.82 | 2 | 100.16 | 3 |
| **104** | 1 | 42 | 2 | 102.25 | 3 | 2.70 | 3 | 29.97 | 2 | 154.84 | 2 | 100.45 | 3 |
| **105** | 1 | 43 | 2 | 102.91 | 3 | 2.71 | 3 | 29.97 | 2 | 155.02 | 2 | 100.54 | 3 |
| **106** | 1 | 43 | 2 | 103.48 | 3 | 2.72 | 3 | 30.00 | 3 | 155.03 | 2 | 100.95 | 3 |
| **107** | 1 | 43 | 2 | 104.09 | 3 | 2.72 | 3 | 30.03 | 3 | 155.41 | 2 | 100.98 | 3 |
| **108** | 1 | 43 | 2 | 104.37 | 3 | 2.74 | 3 | 30.11 | 3 | 155.50 | 2 | 101.08 | 3 |
| **109** | 1 | 43 | 2 | 104.39 | 3 | 2.75 | 3 | 30.15 | 3 | 155.57 | 2 | 101.08 | 3 |
| **110** | 1 | 43 | 2 | 104.58 | 3 | 2.76 | 3 | 30.18 | 3 | 155.59 | 2 | 101.14 | 3 |
| **111** | 0 | 43 | 2 | 104.66 | 3 | 2.77 | 3 | 30.28 | 3 | 155.63 | 2 | 101.34 | 3 |
| **112** | 0 | 43 | 2 | 104.72 | 3 | 2.77 | 3 | 30.34 | 3 | 155.96 | 2 | 101.66 | 3 |
| **113** | 0 | 44 | 2 | 104.85 | 3 | 2.78 | 3 | 30.35 | 3 | 155.97 | 2 | 101.66 | 3 |
| **114** | 0 | 44 | 2 | 104.92 | 3 | 2.79 | 3 | 30.36 | 3 | 156.53 | 2 | 101.77 | 3 |
| **115** | 0 | 44 | 2 | 105.04 | 3 | 2.79 | 3 | 30.42 | 3 | 156.62 | 2 | 101.84 | 3 |
| **116** | 0 | 44 | 2 | 105.77 | 3 | 2.81 | 3 | 30.46 | 3 | 156.97 | 2 | 101.87 | 3 |
| **117** | 0 | 44 | 2 | 106.05 | 3 | 2.82 | 3 | 30.52 | 3 | 157.23 | 2 | 102.14 | 3 |
| **118** | 0 | 44 | 2 | 106.32 | 3 | 2.84 | 3 | 30.54 | 3 | 157.36 | 2 | 102.32 | 3 |
| **119** | 0 | 44 | 2 | 106.36 | 3 | 2.86 | 3 | 30.55 | 3 | 157.53 | 2 | 102.93 | 3 |
| **120** | 0 | 44 | 2 | 106.50 | 3 | 2.87 | 3 | 30.63 | 3 | 157.67 | 2 | 103.46 | 3 |
| **121** | 0 | 44 | 2 | 106.60 | 3 | 2.87 | 3 | 30.74 | 3 | 158.72 | 2 | 103.61 | 3 |
| **122** | 0 | 44 | 2 | 108.32 | 3 | 2.87 | 3 | 31.00 | 3 | 158.92 | 2 | 103.73 | 3 |
| **123** | 0 | 45 | 2 | 108.59 | 3 | 2.88 | 3 | 31.12 | 3 | 159.14 | 2 | 103.75 | 3 |
| **124** | 0 | 45 | 2 | 108.82 | 3 | 2.91 | 3 | 31.18 | 3 | 159.54 | 2 | 104.41 | 3 |
| **125** | 0 | 45 | 2 | 108.92 | 3 | 2.94 | 3 | 31.19 | 3 | 159.60 | 2 | 104.98 | 3 |
| **126** | 0 | 45 | 2 | 109.36 | 3 | 2.95 | 3 | 31.24 | 3 | 159.73 | 2 | 105.40 | 3 |
| **127** | 0 | 45 | 2 | 109.44 | 3 | 2.99 | 3 | 31.34 | 3 | 159.83 | 2 | 105.58 | 3 |
| **128** | 0 | 45 | 2 | 110.00 | 3 | 3.03 | 3 | 31.35 | 3 | 160.49 | 3 | 105.80 | 3 |
| **129** | 0 | 45 | 2 | 111.71 | 3 | 3.04 | 3 | 31.35 | 3 | 161.40 | 3 | 106.17 | 3 |
| **130** | 0 | 45 | 2 | 112.73 | 3 | 3.05 | 3 | 31.39 | 3 | 161.58 | 3 | 106.43 | 3 |
| **131** | 1 | 45 | 2 | 112.96 | 3 | 3.05 | 3 | 31.47 | 3 | 161.62 | 3 | 106.51 | 3 |
| **132** | 1 | 46 | 3 | 113.07 | 3 | 3.06 | 3 | 31.51 | 3 | 161.78 | 3 | 107.06 | 3 |
| **133** | 1 | 46 | 3 | 113.08 | 3 | 3.08 | 3 | 31.56 | 3 | 162.26 | 3 | 108.16 | 3 |
| **134** | 1 | 47 | 3 | 113.50 | 3 | 3.08 | 3 | 31.67 | 3 | 162.37 | 3 | 108.61 | 3 |
| **135** | 1 | 47 | 3 | 113.97 | 3 | 3.13 | 3 | 31.75 | 3 | 162.53 | 3 | 109.79 | 3 |
| **136** | 1 | 47 | 3 | 115.21 | 3 | 3.13 | 3 | 31.94 | 3 | 163.57 | 3 | 110.16 | 3 |
| **137** | 1 | 47 | 3 | 116.16 | 3 | 3.21 | 3 | 32.18 | 3 | 163.78 | 3 | 110.53 | 3 |
| **138** | 1 | 47 | 3 | 116.90 | 3 | 3.24 | 3 | 32.20 | 3 | 164.47 | 3 | 110.68 | 3 |
| **139** | 1 | 47 | 3 | 118.37 | 3 | 3.33 | 3 | 32.31 | 3 | 164.93 | 3 | 112.09 | 3 |
| **140** | 1 | 47 | 3 | 118.75 | 3 | 3.36 | 3 | 32.71 | 3 | 164.93 | 3 | 112.44 | 3 |
| **141** | 1 | 48 | 3 | 121.35 | 3 | 3.36 | 3 | 32.87 | 3 | 165.41 | 3 | 113.56 | 3 |
| **142** | 1 | 48 | 3 | 123.60 | 3 | 3.36 | 3 | 32.91 | 3 | 165.96 | 3 | 114.21 | 3 |
| **143** | 1 | 49 | 3 | 123.70 | 3 | 3.45 | 3 | 33.17 | 3 | 167.51 | 3 | 114.27 | 3 |
| **144** | 1 | 49 | 3 | 124.69 | 3 | 3.52 | 3 | 33.73 | 3 | 168.68 | 3 | 114.77 | 3 |
| **145** | 1 | 49 | 3 | 124.96 | 3 | 3.60 | 3 | 33.93 | 3 | 169.07 | 3 | 115.36 | 3 |
| **146** | 1 | 50 | 3 | 127.56 | 3 | 3.65 | 3 | 34.29 | 3 | 170.23 | 3 | 117.39 | 3 |
| **147** | 1 | 51 | 3 | 127.83 | 3 | 3.70 | 3 | 34.59 | 3 | 173.73 | 3 | 117.88 | 3 |
| **148** | 1 | 51 | 3 | 131.57 | 3 | 3.78 | 3 | 34.95 | 3 | 175.61 | 3 | 118.11 | 3 |
| **149** | 1 | 51 | 3 | 131.89 | 3 | 3.82 | 3 | 35.09 | 3 | 175.59 | 3 | 119.27 | 3 |
| **150** | 1 | 53 | 3 | 133.13 | 3 | 4.36 | 3 | 35.16 | 3 | 175.75 | 3 | 124.04 | 3 |

The following table are the actual data used in Stata 14 software, not the above rounded values. The above values are only used for the space to demonstrate all the variables and corresponding category each participant belong to.

| patient_ID | LDL_chol | HOMA2_IR | BMI | sysBloodPr | diastBloodPressure |
|---|---|---|---|---|---|
| 1 | 59.88626253 | 0.486660227 | 20.30083374 | 123.4009178 | 69.99634935 |
| 2 | 62.17804017 | 0.625517898 | 20.38539827 | 124.1760587 | 70.95836483 |
| 3 | 63.65385754 | 0.648810155 | 22.20504654 | 125.8975997 | 71.77720889 |
| 4 | 65.80681914 | 0.835501959 | 22.50371191 | 126.1787116 | 71.90626924 |
| 5 | 69.79260909 | 1.048423757 | 23.13687005 | 130.0785608 | 73.51626359 |
| 6 | 70.10500285 | 1.063198485 | 23.18066959 | 130.4312855 | 73.89087266 |
| 7 | 70.5951389 | 1.082373227 | 23.34615748 | 132.9771952 | 74.24442658 |
| 8 | 71.73608315 | 1.09242144 | 23.3553593 | 133.1941031 | 74.60984969 |
| 9 | 73.99519667 | 1.129448668 | 23.51913306 | 135.1296929 | 74.69569464 |
| 10 | 74.63244092 | 1.189798489 | 23.7000037 | 135.2716671 | 75.54314352 |
| 11 | 74.71785909 | 1.29368181 | 23.84731358 | 135.4981258 | 76.48721248 |
| 12 | 75.17096758 | 1.324724984 | 24.18061722 | 135.6569022 | 76.52476289 |
| 13 | 75.18235196 | 1.335093535 | 24.23270373 | 136.0038509 | 77.16186785 |
| 14 | 75.23138391 | 1.422111938 | 24.28506674 | 136.0953597 | 79.34849803 |
| 15 | 75.61753709 | 1.432224008 | 24.36230869 | 136.9038556 | 79.73530992 |
| 16 | 76.18401022 | 1.453228249 | 24.44359033 | 137.0809407 | 80.01683653 |
| 17 | 76.24358806 | 1.453815821 | 24.53815568 | 137.0869254 | 80.15864331 |
| 18 | 76.29881294 | 1.456305613 | 24.61359837 | 137.5649291 | 80.74565456 |
| 19 | 76.69659462 | 1.464666603 | 24.61722818 | 138.0735912 | 80.85114089 |
| 20 | 77.58989329 | 1.51356061 | 24.64186842 | 138.3241151 | 82.05654214 |
| 21 | 77.70521508 | 1.536136161 | 24.9131702 | 138.5853047 | 82.1038819 |
| 22 | 77.94998599 | 1.544577706 | 24.92131773 | 139.008886 | 82.31728002 |
| 23 | 78.05838748 | 1.586554018 | 25.06314382 | 139.181514 | 82.48948116 |
| 24 | 78.25702966 | 1.617951836 | 25.09951711 | 139.4240568 | 82.76779198 |
| 25 | 78.4403701 | 1.619881 | 25.11813044 | 139.889396 | 83.11299612 |
| 26 | 79.17897107 | 1.626478279 | 25.13316489 | 139.9072013 | 83.42940556 |
| 27 | 79.34469678 | 1.639802471 | 25.17546941 | 140.5566385 | 83.53598548 |
| 28 | 79.90002937 | 1.640401777 | 25.28627577 | 140.7048667 | 83.80597309 |
| 29 | 80.23286431 | 1.650270486 | 25.60867263 | 140.7248329 | 83.83555089 |
| 30 | 80.53759625 | 1.651985736 | 25.70466017 | 141.0786766 | 84.5109185 |
| 31 | 80.78692327 | 1.665993018 | 25.71210314 | 141.1392888 | 84.6618672 |
| 32 | 80.95179537 | 1.666095395 | 25.73388846 | 141.3048354 | 84.77066523 |
| 33 | 81.39997551 | 1.676923007 | 25.80137142 | 141.7431129 | 84.94485471 |
| 34 | 81.5230473 | 1.694165523 | 25.83013284 | 142.0180379 | 85.04737784 |
| 35 | 81.6333478 | 1.694587966 | 25.95726452 | 142.1419579 | 85.08232589 |
| 36 | 81.63910707 | 1.7096777 | 25.9830848 | 142.2669347 | 85.9701009 |
| 37 | 81.79551762 | 1.728720127 | 26.03356639 | 142.7497928 | 86.07727811 |
| 38 | 82.09730138 | 1.732354693 | 26.05377139 | 142.8777895 | 86.33759816 |
| 39 | 82.72673203 | 1.738136697 | 26.19980058 | 143.6355794 | 86.78184428 |
| 40 | 83.19282441 | 1.775434606 | 26.25343572 | 143.6518685 | 87.11256427 |
| 41 | 83.51015721 | 1.780812997 | 26.35494787 | 143.8440352 | 87.37727918 |
| 42 | 83.87801001 | 1.821641325 | 26.3591208 | 143.904684 | 87.49567067 |
| 43 | 83.90994145 | 1.831844486 | 26.44070213 | 144.0988788 | 88.0956395 |
| 44 | 84.40043839 | 1.862877807 | 26.65017692 | 144.1541475 | 88.16578459 |
| 45 | 85.17501733 | 1.86912154 | 26.66541993 | 144.3474132 | 88.76568544 |
| 46 | 86.24150134 | 1.872825229 | 26.78361722 | 144.509008 | 88.7825505 |
| 47 | 86.41587725 | 1.916601042 | 26.812301 | 144.5242085 | 89.0125243 |
| 48 | 86.96738711 | 1.92222757 | 26.96295086 | 144.7370245 | 89.31527986 |
| 49 | 87.02828486 | 1.932907728 | 27.11405208 | 145.1060697 | 89.33794082 |
| 50 | 87.21342521 | 1.939328867 | 27.11442109 | 145.1690792 | 89.38452095 |
| 51 | 87.59738179 | 1.941058513 | 27.12546756 | 145.6324734 | 89.5019489 |
| 52 | 87.65253761 | 1.951682018 | 27.30688699 | 145.7802001 | 89.5308856 |
| 53 | 88.21762356 | 1.952190865 | 27.30755139 | 145.8370062 | 89.55774174 |
| 54 | 88.34186837 | 1.958971905 | 27.34284662 | 145.8501257 | 89.81576935 |
| 55 | 88.7409777 | 1.962141305 | 27.36056248 | 145.9489973 | 89.95193185 |
| 56 | 88.99684398 | 1.989914501 | 27.42257654 | 145.9993347 | 90.17512035 |
| 57 | 89.31278846 | 2.02697825 | 27.43065895 | 146.4363352 | 90.33968778 |
| 58 | 90.05869996 | 2.033050053 | 27.45034168 | 146.6040765 | 90.86615116 |
| 59 | 90.42098152 | 2.052561065 | 27.50839688 | 146.7290227 | 90.91646661 |
| 60 | 90.66844195 | 2.065170296 | 27.54237172 | 146.7365686 | 91.06059405 |
| 61 | 90.91962854 | 2.082371342 | 27.63616749 | 146.8836997 | 91.07330339 |
| 62 | 91.45092179 | 2.098034701 | 27.75959177 | 146.9104233 | 91.12413523 |
| 63 | 91.73729771 | 2.1191684 | 27.78172501 | 147.3086787 | 91.29977663 |
| 64 | 92.51302825 | 2.132822463 | 27.79279514 | 147.3877559 | 91.32199752 |
| 65 | 92.73033242 | 2.147590744 | 27.7992254 | 147.3889239 | 91.44767437 |
| 66 | 92.90181516 | 2.15186423 | 27.84995994 | 147.7746667 | 91.49574549 |
| 67 | 92.90450229 | 2.159333468 | 27.86907689 | 148.1125105 | 91.60054843 |
| 68 | 93.15072278 | 2.170694946 | 28.01690012 | 148.2112376 | 91.61239313 |
| 69 | 93.18080864 | 2.187319216 | 28.01932658 | 148.2342406 | 91.66833284 |
| 70 | 93.53782288 | 2.190678916 | 28.11007119 | 148.2676521 | 91.69940127 |
| 71 | 93.5737695 | 2.196355685 | 28.28004598 | 148.3681924 | 92.00989498 |
| 72 | 93.96977742 | 2.211907172 | 28.31709791 | 148.8537684 | 92.04176836 |
| 73 | 94.14544773 | 2.226152678 | 28.34828762 | 149.0057149 | 92.99337843 |
| 74 | 94.26851067 | 2.240807892 | 28.36225579 | 149.0833154 | 93.62273135 |
| 75 | 94.57093547 | 2.250136361 | 28.37706409 | 149.3803847 | 93.98582303 |
| 76 | 94.68147986 | 2.276625971 | 28.42003727 | 149.4410218 | 94.14513547 |
| 77 | 94.73162406 | 2.291658478 | 28.44672231 | 149.6870214 | 94.28270759 |
| 78 | 94.8091201 | 2.300868336 | 28.45466819 | 149.9554549 | 94.64978228 |
| 79 | 95.22476536 | 2.307544036 | 28.4938252 | 149.9772856 | 94.67519469 |
| 80 | 95.31849196 | 2.326027215 | 28.6073615 | 150.0036963 | 94.95138569 |
| 81 | 95.80827063 | 2.332456523 | 28.60853748 | 150.0317356 | 95.504466 |
| 82 | 95.87960534 | 2.35945621 | 28.62039083 | 150.4667957 | 95.68060593 |
| 83 | 96.00168055 | 2.359833761 | 28.62136969 | 150.6259156 | 95.72396957 |
| 84 | 96.10649581 | 2.373084069 | 28.67169603 | 150.7608258 | 95.92103971 |
| 85 | 96.40483791 | 2.387931316 | 28.70849209 | 150.8378482 | 96.06573604 |
| 86 | 96.76889879 | 2.390069935 | 28.76101083 | 150.8610255 | 96.34116959 |
| 87 | 96.80646738 | 2.399277902 | 28.80966656 | 150.905197 | 96.47836375 |
| 88 | 96.90122271 | 2.411984544 | 28.92018631 | 150.9217811 | 96.51952908 |
| 89 | 96.90967787 | 2.4218036 | 28.9943167 | 151.0591804 | 96.60639678 |
| 90 | 97.54518517 | 2.425332347 | 29.10458361 | 151.5783406 | 96.78398633 |
| 91 | 97.57532315 | 2.464533455 | 29.11393665 | 151.6895192 | 97.16179582 |
| 92 | 97.82240849 | 2.472015524 | 29.14182311 | 152.0147919 | 97.38518017 |
| 93 | 98.19383152 | 2.512852519 | 29.17354849 | 152.4897552 | 97.50577044 |
| 94 | 98.65168935 | 2.538669397 | 29.17974603 | 152.5383571 | 97.583515 |
| 95 | 98.67628649 | 2.557198677 | 29.33804648 | 153.3760363 | 98.40263236 |
| 96 | 99.000552 | 2.579092668 | 29.37220989 | 153.4606462 | 98.42450822 |
| 97 | 99.47036072 | 2.583321929 | 29.40508095 | 153.4953371 | 98.4857294 |
| 98 | 99.56796044 | 2.589431917 | 29.40925322 | 153.6290617 | 98.93001208 |
| 99 | 100.2809094 | 2.610692002 | 29.60369794 | 153.9099293 | 99.20391047 |
| 100 | 100.5717195 | 2.636885226 | 29.71572889 | 153.9424972 | 99.38911282 |
| 101 | 100.6047924 | 2.651751393 | 29.79977925 | 154.4560393 | 99.71821712 |
| 102 | 101.8494775 | 2.678843061 | 29.93617461 | 154.558702 | 99.7448965 |
| 103 | 101.9492914 | 2.69069223 | 29.96661684 | 154.8245324 | 100.1596441 |
| 104 | 102.2515504 | 2.698054556 | 29.96835067 | 154.837513 | 100.4489565 |
| 105 | 102.9114573 | 2.708479035 | 29.96957974 | 155.019825 | 100.5387061 |
| 106 | 103.4806938 | 2.71543582 | 30.00466141 | 155.0302859 | 100.9490535 |
| 107 | 104.0881578 | 2.720640069 | 30.03225126 | 155.4068362 | 100.9756056 |
| 108 | 104.3713252 | 2.736640518 | 30.10960076 | 155.4976388 | 101.0822111 |
| 109 | 104.3914483 | 2.748266639 | 30.14865929 | 155.5674547 | 101.0830179 |
| 110 | 104.5798686 | 2.758139944 | 30.18347007 | 155.5914364 | 101.1394993 |
| 111 | 104.6607348 | 2.768605796 | 30.27695636 | 155.6261913 | 101.3435673 |
| 112 | 104.7238725 | 2.772175859 | 30.33694066 | 155.9660494 | 101.6553383 |
| 113 | 104.8524726 | 2.784769065 | 30.34647086 | 155.972805 | 101.6564651 |
| 114 | 104.9203266 | 2.788978027 | 30.35791172 | 156.5262548 | 101.7654722 |
| 115 | 105.0374154 | 2.792846389 | 30.42338805 | 156.623788 | 101.8369263 |
| 116 | 105.7708525 | 2.813543425 | 30.45991065 | 156.9746708 | 101.8745766 |
| 117 | 106.0526701 | 2.8216319 | 30.52013788 | 157.2333107 | 102.1438818 |
| 118 | 106.3224205 | 2.836471693 | 30.54321084 | 157.3606742 | 102.3191044 |
| 119 | 106.3588698 | 2.86156774 | 30.54610609 | 157.5349887 | 102.9303476 |
| 120 | 106.4972799 | 2.870049811 | 30.62901275 | 157.6709272 | 103.4613788 |
| 121 | 106.604525 | 2.870113146 | 30.7433647 | 158.7207682 | 103.6144469 |
| 122 | 108.3227542 | 2.874031357 | 30.99738238 | 158.9222397 | 103.7258609 |
| 123 | 108.5904841 | 2.87941575 | 31.12302385 | 159.1374886 | 103.7531646 |
| 124 | 108.8231986 | 2.906358239 | 31.18244663 | 159.5381187 | 104.4083546 |
| 125 | 108.9212838 | 2.944579876 | 31.19197902 | 159.599158 | 104.9786418 |
| 126 | 109.3646854 | 2.946595176 | 31.24239694 | 159.7268598 | 105.3992209 |
| 127 | 109.443203 | 2.985029147 | 31.34270287 | 159.830625 | 105.5849072 |
| 128 | 110.0035342 | 3.027057171 | 31.34638621 | 160.4869579 | 105.8003815 |
| 129 | 111.7142213 | 3.0415236 | 31.35100678 | 161.3990233 | 106.1653939 |
| 130 | 112.727007 | 3.045917418 | 31.39241181 | 161.5790007 | 106.4294163 |
| 131 | 112.9599582 | 3.054332948 | 31.4744049 | 161.620505 | 106.5062134 |
| 132 | 113.0742184 | 3.059351383 | 31.51252628 | 161.7783022 | 107.0606273 |
| 133 | 113.0811469 | 3.080270062 | 31.56477963 | 162.2555993 | 108.1639294 |
| 134 | 113.5011891 | 3.082093216 | 31.6663631 | 162.3732048 | 108.6064678 |
| 135 | 113.9678215 | 3.12902167 | 31.74508879 | 162.5259587 | 109.7874429 |
| 136 | 115.2098298 | 3.133993 | 31.93917448 | 163.5672102 | 110.1637194 |
| 137 | 116.1608298 | 3.212666172 | 32.18283095 | 163.7750487 | 110.5316899 |
| 138 | 116.8971631 | 3.239904042 | 32.19956838 | 164.4663885 | 110.6818642 |
| 139 | 118.3710727 | 3.331810851 | 32.31424866 | 164.9260434 | 112.0894894 |
| 140 | 118.750906 | 3.361332552 | 32.70754103 | 164.9274447 | 112.4385468 |
| 141 | 121.3539927 | 3.361957484 | 32.86909366 | 165.4101566 | 113.5642057 |
| 142 | 123.6046089 | 3.363641715 | 32.90902767 | 165.9629702 | 114.2073523 |
| 143 | 123.6953321 | 3.44841049 | 33.17015687 | 167.5063644 | 114.2730905 |
| 144 | 124.6863766 | 3.520206957 | 33.72790263 | 168.6830254 | 114.7742868 |
| 145 | 124.9633003 | 3.600924971 | 33.9273159 | 169.066132 | 115.3553829 |
| 146 | 127.5631494 | 3.652359416 | 34.29045674 | 170.229292 | 117.3937688 |
| 147 | 127.8271221 | 3.697602312 | 34.58909385 | 173.7274296 | 117.8790236 |
| 148 | 131.5713536 | 3.782653123 | 34.95279071 | 175.6065684 | 118.1096726 |
| 149 | 131.8938225 | 3.815486059 | 35.0914694 | 175.5867168 | 119.268837 |
| 150 | 133.1308799 | 4.361195597 | 35.16374084 | 175.7543407 | 124.037896 |

Table (2): transition counts for each patient.

| ID | 0→1 | 1→2 | 2→3 | 3→4 | 1→0 | 2→1 | 3→2 | 2→0 | 3→1 |
|---|---|---|---|---|---|---|---|---|---|
| 1 | 0 | 0 | 0 | 0 | 0 | 0 | 0 | 0 | 0 |
| 2 | 0 | 0 | 0 | 0 | 0 | 0 | 0 | 0 | 0 |
| 3 | 0 | 0 | 0 | 0 | 0 | 0 | 0 | 0 | 0 |
| 4 | 0 | 0 | 0 | 0 | 0 | 0 | 0 | 0 | 0 |
| 5 | 0 | 0 | 0 | 0 | 0 | 0 | 0 | 0 | 0 |
| 6 | 0 | 0 | 0 | 0 | 0 | 0 | 0 | 0 | 0 |
| 7 | 0 | 0 | 0 | 0 | 0 | 0 | 0 | 0 | 0 |
| 8 | 0 | 0 | 0 | 0 | 0 | 0 | 0 | 0 | 0 |
| 9 | 0 | 0 | 0 | 0 | 0 | 0 | 0 | 0 | 0 |
| 10 | 0 | 0 | 0 | 0 | 0 | 0 | 0 | 0 | 0 |
| 11 | 0 | 0 | 0 | 0 | 0 | 0 | 0 | 0 | 0 |
| 12 | 0 | 0 | 0 | 0 | 0 | 0 | 0 | 0 | 0 |
| 13 | 0 | 0 | 0 | 0 | 0 | 0 | 0 | 0 | 0 |
| 14 | 0 | 0 | 0 | 0 | 0 | 0 | 0 | 0 | 0 |
| 15 | 0 | 0 | 0 | 0 | 0 | 0 | 0 | 0 | 0 |
| 16 | 0 | 0 | 0 | 0 | 0 | 0 | 0 | 0 | 0 |
| 17 | 0 | 0 | 0 | 0 | 0 | 0 | 0 | 0 | 0 |
| 18 | 0 | 0 | 0 | 0 | 0 | 0 | 0 | 0 | 0 |
| 19 | 0 | 0 | 0 | 0 | 0 | 0 | 0 | 0 | 0 |
| 20 | 0 | 0 | 0 | 0 | 0 | 0 | 0 | 0 | 0 |
| 21 | 0 | 0 | 0 | 0 | 0 | 0 | 0 | 0 | 0 |
| 22 | 0 | 0 | 0 | 0 | 0 | 0 | 0 | 0 | 0 |
| 23 | 0 | 0 | 0 | 0 | 0 | 0 | 0 | 0 | 0 |
| 24 | 0 | 0 | 0 | 0 | 0 | 0 | 0 | 0 | 0 |
| 25 | 0 | 0 | 0 | 0 | 0 | 0 | 0 | 0 | 0 |
| 26 | 0 | 0 | 0 | 0 | 0 | 0 | 0 | 0 | 0 |
| 27 | 0 | 0 | 0 | 0 | 0 | 0 | 0 | 0 | 0 |
| 28 | 0 | 0 | 0 | 0 | 0 | 0 | 0 | 0 | 0 |
| 29 | 0 | 0 | 0 | 0 | 0 | 0 | 0 | 0 | 0 |
| 30 | 0 | 0 | 0 | 0 | 0 | 0 | 0 | 0 | 0 |
| 31 | 0 | 0 | 0 | 0 | 0 | 0 | 0 | 0 | 0 |
| 32 | 0 | 0 | 0 | 0 | 0 | 0 | 0 | 0 | 0 |
| 33 | 0 | 0 | 0 | 0 | 0 | 0 | 0 | 0 | 0 |
| 34 | 0 | 0 | 0 | 0 | 0 | 0 | 0 | 0 | 0 |
| 35 | 0 | 0 | 0 | 0 | 0 | 0 | 0 | 0 | 0 |
| 36 | 0 | 0 | 0 | 0 | 0 | 0 | 0 | 0 | 0 |
| 37 | 0 | 0 | 0 | 0 | 0 | 0 | 0 | 0 | 0 |
| 38 | 0 | 0 | 0 | 0 | 0 | 0 | 0 | 0 | 0 |
| 39 | 0 | 0 | 0 | 0 | 0 | 0 | 0 | 0 | 0 |
| 40 | 0 | 0 | 0 | 0 | 0 | 0 | 0 | 0 | 0 |
| 41 | 0 | 0 | 0 | 0 | 0 | 0 | 0 | 0 | 0 |
| 42 | 0 | 0 | 0 | 0 | 0 | 0 | 0 | 0 | 0 |
| 43 | 0 | 0 | 0 | 0 | 0 | 0 | 0 | 0 | 0 |
| 44 | 0 | 0 | 0 | 0 | 0 | 0 | 0 | 0 | 0 |
| 45 | 0 | 0 | 0 | 0 | 0 | 0 | 0 | 0 | 0 |
| 46 | 0 | 0 | 0 | 0 | 0 | 0 | 0 | 0 | 0 |
| 47 | 0 | 0 | 0 | 0 | 0 | 0 | 0 | 0 | 0 |
| 48 | 0 | 0 | 0 | 0 | 0 | 0 | 0 | 0 | 0 |
| 49 | 0 | 0 | 0 | 0 | 0 | 0 | 0 | 0 | 0 |
| 50 | 0 | 0 | 0 | 0 | 0 | 0 | 0 | 0 | 0 |
| 51 | 0 | 0 | 0 | 0 | 0 | 0 | 0 | 0 | 0 |
| 52 | 0 | 0 | 0 | 0 | 0 | 0 | 0 | 0 | 0 |
| 53 | 0 | 0 | 0 | 0 | 0 | 0 | 0 | 0 | 0 |

| ID | 0→1 | 1→2 | 2→3 | 3→4 | 1→0 | 2→1 | 3→2 | 2→0 | 3→1 |
|---|---|---|---|---|---|---|---|---|---|
| **54** | 0 | 0 | 0 | 0 | 0 | 0 | 0 | 0 | 0 |
| **55** | 0 | 0 | 0 | 0 | 0 | 0 | 0 | 0 | 0 |
| **56** | 0 | 0 | 0 | 0 | 0 | 0 | 0 | 0 | 0 |
| **57** | 0 | 0 | 0 | 0 | 0 | 0 | 0 | 0 | 0 |
| **58** | 0 | 0 | 0 | 0 | 0 | 0 | 0 | 0 | 0 |
| **59** | 0 | 0 | 0 | 0 | 0 | 0 | 0 | 0 | 0 |
| **60** | 0 | 0 | 0 | 0 | 0 | 0 | 0 | 0 | 0 |
| **61** | 0 | 0 | 0 | 0 | 0 | 0 | 0 | 0 | 0 |
| **62** | 0 | 0 | 0 | 0 | 0 | 0 | 0 | 0 | 0 |
| **63** | 0 | 0 | 0 | 0 | 0 | 0 | 0 | 0 | 0 |
| **64** | 1 | 0 | 0 | 0 | 0 | 0 | 0 | 0 | 0 |
| **65** | 1 | 0 | 0 | 0 | 0 | 0 | 0 | 0 | 0 |
| **66** | 1 | 0 | 0 | 0 | 0 | 0 | 0 | 0 | 0 |
| **67** | 1 | 0 | 0 | 0 | 0 | 0 | 0 | 0 | 0 |
| **68** | 1 | 0 | 0 | 0 | 0 | 0 | 0 | 0 | 0 |
| **69** | 1 | 0 | 0 | 0 | 0 | 0 | 0 | 0 | 0 |
| **70** | 1 | 0 | 0 | 0 | 0 | 0 | 0 | 0 | 0 |
| **71** | 1 | 0 | 0 | 0 | 0 | 0 | 0 | 0 | 0 |
| **72** | 1 | 0 | 0 | 0 | 0 | 0 | 0 | 0 | 0 |
| **73** | 1 | 0 | 0 | 0 | 0 | 0 | 0 | 0 | 0 |
| **74** | 1 | 0 | 0 | 0 | 0 | 0 | 0 | 0 | 0 |
| **75** | 1 | 0 | 0 | 0 | 0 | 0 | 0 | 0 | 0 |
| **76** | 1 | 0 | 0 | 0 | 0 | 0 | 0 | 0 | 0 |
| **77** | 1 | 0 | 0 | 0 | 0 | 0 | 0 | 0 | 0 |
| **78** | 1 | 0 | 0 | 0 | 0 | 0 | 0 | 0 | 0 |
| **79** | 1 | 0 | 0 | 0 | 0 | 0 | 0 | 0 | 0 |
| **80** | 1 | 0 | 0 | 0 | 0 | 0 | 0 | 0 | 0 |
| **81** | 1 | 0 | 0 | 0 | 0 | 0 | 0 | 0 | 0 |
| **82** | 1 | 0 | 0 | 0 | 0 | 0 | 0 | 0 | 0 |
| **83** | 1 | 0 | 0 | 0 | 0 | 0 | 0 | 0 | 0 |
| **84** | 1 | 0 | 0 | 0 | 0 | 0 | 0 | 0 | 0 |
| **85** | 1 | 0 | 0 | 0 | 0 | 0 | 0 | 0 | 0 |
| **86** | 1 | 0 | 0 | 0 | 0 | 0 | 0 | 0 | 0 |
| **87** | 1 | 0 | 0 | 0 | 0 | 0 | 0 | 0 | 0 |
| **88** | 1 | 0 | 0 | 0 | 0 | 0 | 0 | 0 | 0 |
| **89** | 1 | 0 | 0 | 0 | 0 | 0 | 0 | 0 | 0 |
| **90** | 1 | 0 | 0 | 0 | 0 | 0 | 0 | 0 | 0 |
| **91** | 1 | 0 | 0 | 0 | 0 | 0 | 0 | 0 | 0 |
| **92** | 1 | 0 | 0 | 0 | 0 | 0 | 0 | 0 | 0 |
| **93** | 1 | 0 | 0 | 0 | 0 | 0 | 0 | 0 | 0 |
| **94** | 1 | 0 | 0 | 0 | 0 | 0 | 0 | 0 | 0 |
| **95** | 1 | 0 | 0 | 0 | 0 | 0 | 0 | 0 | 0 |
| **96** | 1 | 0 | 0 | 0 | 0 | 0 | 0 | 0 | 0 |
| **97** | 1 | 1 | 0 | 0 | 0 | 0 | 0 | 0 | 0 |
| **98** | 1 | 1 | 0 | 0 | 0 | 0 | 0 | 0 | 0 |
| **99** | 1 | 1 | 0 | 0 | 0 | 0 | 0 | 0 | 0 |
| **100** | 1 | 1 | 0 | 0 | 0 | 0 | 0 | 0 | 0 |
| **101** | 1 | 1 | 0 | 0 | 0 | 0 | 0 | 0 | 0 |
| **102** | 1 | 1 | 0 | 0 | 0 | 0 | 0 | 0 | 0 |
| **103** | 1 | 1 | 0 | 0 | 0 | 0 | 0 | 0 | 0 |
| **104** | 1 | 1 | 0 | 0 | 0 | 0 | 0 | 0 | 0 |
| **105** | 1 | 1 | 0 | 0 | 0 | 0 | 0 | 0 | 0 |
| **106** | 1 | 1 | 0 | 0 | 0 | 0 | 0 | 0 | 0 |
| **107** | 1 | 1 | 0 | 0 | 0 | 0 | 0 | 0 | 0 |
| **108** | 1 | 1 | 0 | 0 | 0 | 0 | 0 | 0 | 0 |

| ID | 0→1 | 1→2 | 2→3 | 3→4 | 1→0 | 2→1 | 3→2 | 2→0 | 3→1 |
|---|---|---|---|---|---|---|---|---|---|
| **109** | 1 | 1 | 0 | 0 | 0 | 0 | 0 | 0 | 0 |
| **110** | 1 | 1 | 0 | 0 | 0 | 0 | 0 | 0 | 0 |
| **111** | 1 | 1 | 0 | 0 | 0 | 0 | 0 | 0 | 0 |
| **112** | 1 | 1 | 0 | 0 | 0 | 0 | 0 | 0 | 0 |
| **113** | 1 | 1 | 0 | 0 | 0 | 0 | 0 | 0 | 0 |
| **114** | 1 | 1 | 0 | 0 | 0 | 0 | 0 | 0 | 0 |
| **115** | 1 | 1 | 0 | 0 | 0 | 0 | 0 | 0 | 0 |
| **116** | 1 | 1 | 0 | 0 | 0 | 0 | 0 | 0 | 0 |
| **117** | 1 | 1 | 0 | 0 | 0 | 0 | 0 | 0 | 0 |
| **118** | 1 | 1 | 0 | 0 | 0 | 0 | 0 | 0 | 0 |
| **119** | 1 | 1 | 0 | 0 | 0 | 0 | 0 | 0 | 0 |
| **120** | 1 | 1 | 0 | 0 | 0 | 0 | 0 | 0 | 0 |
| **121** | 1 | 1 | 0 | 0 | 0 | 0 | 0 | 0 | 0 |
| **122** | 2 | 1 | 1 | 0 | 1 | 0 | 0 | 0 | 0 |
| **123** | 2 | 1 | 1 | 0 | 1 | 0 | 0 | 0 | 0 |
| **124** | 2 | 1 | 1 | 0 | 1 | 0 | 0 | 0 | 0 |
| **125** | 2 | 1 | 1 | 0 | 1 | 0 | 0 | 0 | 0 |
| **126** | 2 | 1 | 1 | 0 | 1 | 0 | 0 | 0 | 0 |
| **127** | 2 | 1 | 1 | 0 | 1 | 0 | 0 | 0 | 0 |
| **128** | 2 | 1 | 1 | 0 | 1 | 1 | 0 | 0 | 0 |
| **129** | 2 | 1 | 1 | 1 | 1 | 1 | 0 | 0 | 0 |
| **130** | 2 | 1 | 1 | 1 | 1 | 1 | 0 | 0 | 0 |
| **131** | 2 | 1 | 1 | 1 | 1 | 1 | 1 | 0 | 0 |
| **132** | 2 | 1 | 1 | 1 | 1 | 1 | 1 | 0 | 0 |
| **133** | 2 | 1 | 1 | 1 | 1 | 1 | 1 | 0 | 0 |
| **134** | 2 | 1 | 1 | 1 | 1 | 1 | 1 | 0 | 0 |
| **135** | 2 | 1 | 1 | 1 | 1 | 1 | 1 | 0 | 0 |
| **136** | 2 | 1 | 1 | 1 | 1 | 1 | 1 | 0 | 0 |
| **137** | 2 | 1 | 1 | 1 | 1 | 1 | 1 | 0 | 0 |
| **138** | 2 | 1 | 1 | 1 | 1 | 1 | 1 | 0 | 0 |
| **139** | 2 | 1 | 1 | 1 | 1 | 1 | 1 | 1 | 0 |
| **140** | 2 | 2 | 1 | 1 | 1 | 1 | 1 | 1 | 1 |
| **141** | 2 | 2 | 1 | 1 | 1 | 1 | 1 | 1 | 1 |
| **142** | 2 | 2 | 1 | 1 | 1 | 1 | 1 | 1 | 1 |
| **143** | 2 | 2 | 1 | 1 | 1 | 1 | 1 | 1 | 1 |
| **144** | 2 | 2 | 1 | 1 | 1 | 1 | 1 | 1 | 1 |
| **145** | 2 | 3 | 2 | 1 | 1 | 2 | 1 | 1 | 1 |
| **146** | 2 | 2 | 2 | 1 | 2 | 2 | 1 | 1 | 1 |
| **147** | 3 | 2 | 2 | 1 | 2 | 2 | 1 | 1 | 1 |
| **148** | 3 | 2 | 2 | 1 | 2 | 2 | 2 | 1 | 1 |
| **149** | 3 | 2 | 3 | 1 | 3 | 2 | 2 | 1 | 2 |
| **150** | 3 | 3 | 3 | 1 | 3 | 3 | 2 | 2 | 2 |

Table(3): time line for each patient. First column is the patient's ID.

| | | | | | | | | | | | | | | | | | | | | | | | | | | | | | |
|---|---|---|---|---|---|---|---|---|---|---|---|---|---|---|---|---|---|---|---|---|---|---|---|---|---|---|---|---|---|
| 1 | 0 | 0 | 0 | 0 | 0 | 0 | 0 | 0 | 0 | 0 | 0 | 0 | 0 | 0 | 0 | 0 | 0 | 0 | 0 | 0 | 0 | 0 | 0 | 0 | 0 | 0 | 0 | 0 | 0 |
| 2 | 0 | 0 | 0 | 0 | 0 | 0 | 0 | 0 | 0 | 0 | 0 | 0 | 0 | 0 | 0 | 0 | 0 | 0 | 0 | 0 | 0 | 0 | 0 | 0 | 0 | 0 | 0 | 0 | 0 |
| 3 | 0 | 0 | 0 | 0 | 0 | 0 | 0 | 0 | 0 | 0 | 0 | 0 | 0 | 0 | 0 | 0 | 0 | 0 | 0 | 0 | 0 | 0 | 0 | 0 | 0 | 0 | 0 | 0 | 0 |
| 4 | 0 | 0 | 0 | 0 | 0 | 0 | 0 | 0 | 0 | 0 | 0 | 0 | 0 | 0 | 0 | 0 | 0 | 0 | 0 | 0 | 0 | 0 | 0 | 0 | 0 | 0 | 0 | 0 | 0 |
| 5 | 0 | 0 | 0 | 0 | 0 | 0 | 0 | 0 | 0 | 0 | 0 | 0 | 0 | 0 | 0 | 0 | 0 | 0 | 0 | 0 | 0 | 0 | 0 | 0 | 0 | 0 | 0 | 0 | 0 |
| 6 | 0 | 0 | 0 | 0 | 0 | 0 | 0 | 0 | 0 | 0 | 0 | 0 | 0 | 0 | 0 | 0 | 0 | 0 | 0 | 0 | 0 | 0 | 0 | 0 | 0 | 0 | 0 | 0 | 0 |
| 7 | 0 | 0 | 0 | 0 | 0 | 0 | 0 | 0 | 0 | 0 | 0 | 0 | 0 | 0 | 0 | 0 | 0 | 0 | 0 | 0 | 0 | 0 | 0 | 0 | 0 | 0 | 0 | 0 | 0 |
| 8 | 0 | 0 | 0 | 0 | 0 | 0 | 0 | 0 | 0 | 0 | 0 | 0 | 0 | 0 | 0 | 0 | 0 | 0 | 0 | 0 | 0 | 0 | 0 | 0 | 0 | 0 | 0 | 0 | 0 |
| 9 | 0 | 0 | 0 | 0 | 0 | 0 | 0 | 0 | 0 | 0 | 0 | 0 | 0 | 0 | 0 | 0 | 0 | 0 | 0 | 0 | 0 | 0 | 0 | 0 | 0 | 0 | 0 | 0 | 0 |
| 10 | 0 | 0 | 0 | 0 | 0 | 0 | 0 | 0 | 0 | 0 | 0 | 0 | 0 | 0 | 0 | 0 | 0 | 0 | 0 | 0 | 0 | 0 | 0 | 0 | 0 | 0 | 0 | 0 | 0 |
| 11 | 0 | 0 | 0 | 0 | 0 | 0 | 0 | 0 | 0 | 0 | 0 | 0 | 0 | 0 | 0 | 0 | 0 | 0 | 0 | 0 | 0 | 0 | 0 | 0 | 0 | 0 | 0 | 0 | 0 |
| 12 | 0 | 0 | 0 | 0 | 0 | 0 | 0 | 0 | 0 | 0 | 0 | 0 | 0 | 0 | 0 | 0 | 0 | 0 | 0 | 0 | 0 | 0 | 0 | 0 | 0 | 0 | 0 | 0 | 0 |
| 13 | 0 | 0 | 0 | 0 | 0 | 0 | 0 | 0 | 0 | 0 | 0 | 0 | 0 | 0 | 0 | 0 | 0 | 0 | 0 | 0 | 0 | 0 | 0 | 0 | 0 | 0 | 0 | 0 | 0 |
| 14 | 0 | 0 | 0 | 0 | 0 | 0 | 0 | 0 | 0 | 0 | 0 | 0 | 0 | 0 | 0 | 0 | 0 | 0 | 0 | 0 | 0 | 0 | 0 | 0 | 0 | 0 | 0 | 0 | 0 |
| 15 | 0 | 0 | 0 | 0 | 0 | 0 | 0 | 0 | 0 | 0 | 0 | 0 | 0 | 0 | 0 | 0 | 0 | 0 | 0 | 0 | 0 | 0 | 0 | 0 | 0 | 0 | 0 | 0 | 0 |
| 16 | 0 | 0 | 0 | 0 | 0 | 0 | 0 | 0 | 0 | 0 | 0 | 0 | 0 | 0 | 0 | 0 | 0 | 0 | 0 | 0 | 0 | 0 | 0 | 0 | 0 | 0 | 0 | 0 | 0 |
| 17 | 0 | 0 | 0 | 0 | 0 | 0 | 0 | 0 | 0 | 0 | 0 | 0 | 0 | 0 | 0 | 0 | 0 | 0 | 0 | 0 | 0 | 0 | 0 | 0 | 0 | 0 | 0 | 0 | 0 |
| 18 | 0 | 0 | 0 | 0 | 0 | 0 | 0 | 0 | 0 | 0 | 0 | 0 | 0 | 0 | 0 | 0 | 0 | 0 | 0 | 0 | 0 | 0 | 0 | 0 | 0 | 0 | 0 | 0 | 0 |
| 19 | 0 | 0 | 0 | 0 | 0 | 0 | 0 | 0 | 0 | 0 | 0 | 0 | 0 | 0 | 0 | 0 | 0 | 0 | 0 | 0 | 0 | 0 | 0 | 0 | 0 | 0 | 0 | 0 | 0 |
| 20 | 0 | 0 | 0 | 0 | 0 | 0 | 0 | 0 | 0 | 0 | 0 | 0 | 0 | 0 | 0 | 0 | 0 | 0 | 0 | 0 | 0 | 0 | 0 | 0 | 0 | 0 | 0 | 0 | 0 |
| 21 | 0 | 0 | 0 | 0 | 0 | 0 | 0 | 0 | 0 | 0 | 0 | 0 | 0 | 0 | 0 | 0 | 0 | 0 | 0 | 0 | 0 | 0 | 0 | 0 | 0 | 0 | 0 | 0 | 0 |
| 22 | 0 | 0 | 0 | 0 | 0 | 0 | 0 | 0 | 0 | 0 | 0 | 0 | 0 | 0 | 0 | 0 | 0 | 0 | 0 | 0 | 0 | 0 | 0 | 0 | 0 | 0 | 0 | 0 | 0 |
| 23 | 0 | 0 | 0 | 0 | 0 | 0 | 0 | 0 | 0 | 0 | 0 | 0 | 0 | 0 | 0 | 0 | 0 | 0 | 0 | 0 | 0 | 0 | 0 | 0 | 0 | 0 | 0 | 0 | 0 |
| 24 | 0 | 0 | 0 | 0 | 0 | 0 | 0 | 0 | 0 | 0 | 0 | 0 | 0 | 0 | 0 | 0 | 0 | 0 | 0 | 0 | 0 | 0 | 0 | 0 | 0 | 0 | 0 | 0 | 0 |
| 25 | 0 | 0 | 0 | 0 | 0 | 0 | 0 | 0 | 0 | 0 | 0 | 0 | 0 | 0 | 0 | 0 | 0 | 0 | 0 | 0 | 0 | 0 | 0 | 0 | 0 | 0 | 0 | 0 | 0 |
| 26 | 0 | 0 | 0 | 0 | 0 | 0 | 0 | 0 | 0 | 0 | 0 | 0 | 0 | 0 | 0 | 0 | 0 | 0 | 0 | 0 | 0 | 0 | 0 | 0 | 0 | 0 | 0 | 0 | 0 |
| 27 | 0 | 0 | 0 | 0 | 0 | 0 | 0 | 0 | 0 | 0 | 0 | 0 | 0 | 0 | 0 | 0 | 0 | 0 | 0 | 0 | 0 | 0 | 0 | 0 | 0 | 0 | 0 | 0 | 0 |
| 28 | 0 | 0 | 0 | 0 | 0 | 0 | 0 | 0 | 0 | 0 | 0 | 0 | 0 | 0 | 0 | 0 | 0 | 0 | 0 | 0 | 0 | 0 | 0 | 0 | 0 | 0 | 0 | 0 | 0 |
| 29 | 0 | 0 | 0 | 0 | 0 | 0 | 0 | 0 | 0 | 0 | 0 | 0 | 0 | 0 | 0 | 0 | 0 | 0 | 0 | 0 | 0 | 0 | 0 | 0 | 0 | 0 | 0 | 0 | 0 |
| 30 | 0 | 0 | 0 | 0 | 0 | 0 | 0 | 0 | 0 | 0 | 0 | 0 | 0 | 0 | 0 | 0 | 0 | 0 | 0 | 0 | 0 | 0 | 0 | 0 | 0 | 0 | 0 | 0 | 0 |
| 31 | 0 | 0 | 0 | 0 | 0 | 0 | 0 | 0 | 0 | 0 | 0 | 0 | 0 | 0 | 0 | 0 | 0 | 0 | 0 | 0 | 0 | 0 | 0 | 0 | 0 | 0 | 0 | 0 | 0 |
| 32 | 0 | 0 | 0 | 0 | 0 | 0 | 0 | 0 | 0 | 0 | 0 | 0 | 0 | 0 | 0 | 0 | 0 | 0 | 0 | 0 | 0 | 0 | 0 | 0 | 0 | 0 | 0 | 0 | 0 |
| 33 | 0 | 0 | 0 | 0 | 0 | 0 | 0 | 0 | 0 | 0 | 0 | 0 | 0 | 0 | 0 | 0 | 0 | 0 | 0 | 0 | 0 | 0 | 0 | 0 | 0 | 0 | 0 | 0 | 0 |
| 34 | 0 | 0 | 0 | 0 | 0 | 0 | 0 | 0 | 0 | 0 | 0 | 0 | 0 | 0 | 0 | 0 | 0 | 0 | 0 | 0 | 0 | 0 | 0 | 0 | 0 | 0 | 0 | 0 | 0 |
| 35 | 0 | 0 | 0 | 0 | 0 | 0 | 0 | 0 | 0 | 0 | 0 | 0 | 0 | 0 | 0 | 0 | 0 | 0 | 0 | 0 | 0 | 0 | 0 | 0 | 0 | 0 | 0 | 0 | 0 |
| 36 | 0 | 0 | 0 | 0 | 0 | 0 | 0 | 0 | 0 | 0 | 0 | 0 | 0 | 0 | 0 | 0 | 0 | 0 | 0 | 0 | 0 | 0 | 0 | 0 | 0 | 0 | 0 | 0 | 0 |
| 37 | 0 | 0 | 0 | 0 | 0 | 0 | 0 | 0 | 0 | 0 | 0 | 0 | 0 | 0 | 0 | 0 | 0 | 0 | 0 | 0 | 0 | 0 | 0 | 0 | 0 | 0 | 0 | 0 | 0 |
| 38 | 0 | 0 | 0 | 0 | 0 | 0 | 0 | 0 | 0 | 0 | 0 | 0 | 0 | 0 | 0 | 0 | 0 | 0 | 0 | 0 | 0 | 0 | 0 | 0 | 0 | 0 | 0 | 0 | 0 |
| 39 | 0 | 0 | 0 | 0 | 0 | 0 | 0 | 0 | 0 | 0 | 0 | 0 | 0 | 0 | 0 | 0 | 0 | 0 | 0 | 0 | 0 | 0 | 0 | 0 | 0 | 0 | 0 | 0 | 0 |
| 40 | 0 | 0 | 0 | 0 | 0 | 0 | 0 | 0 | 0 | 0 | 0 | 0 | 0 | 0 | 0 | 0 | 0 | 0 | 0 | 0 | 0 | 0 | 0 | 0 | 0 | 0 | 0 | 0 | 0 |
| 41 | 0 | 0 | 0 | 0 | 0 | 0 | 0 | 0 | 0 | 0 | 0 | 0 | 0 | 0 | 0 | 0 | 0 | 0 | 0 | 0 | 0 | 0 | 0 | 0 | 0 | 0 | 0 | 0 | 0 |
| 42 | 0 | 0 | 0 | 0 | 0 | 0 | 0 | 0 | 0 | 0 | 0 | 0 | 0 | 0 | 0 | 0 | 0 | 0 | 0 | 0 | 0 | 0 | 0 | 0 | 0 | 0 | 0 | 0 | 0 |
| 43 | 0 | 0 | 0 | 0 | 0 | 0 | 0 | 0 | 0 | 0 | 0 | 0 | 0 | 0 | 0 | 0 | 0 | 0 | 0 | 0 | 0 | 0 | 0 | 0 | 0 | 0 | 0 | 0 | 0 |
| 44 | 0 | 0 | 0 | 0 | 0 | 0 | 0 | 0 | 0 | 0 | 0 | 0 | 0 | 0 | 0 | 0 | 0 | 0 | 0 | 0 | 0 | 0 | 0 | 0 | 0 | 0 | 0 | 0 | 0 |
| 45 | 0 | 0 | 0 | 0 | 0 | 0 | 0 | 0 | 0 | 0 | 0 | 0 | 0 | 0 | 0 | 0 | 0 | 0 | 0 | 0 | 0 | 0 | 0 | 0 | 0 | 0 | 0 | 0 | 0 |
| 46 | 0 | 0 | 0 | 0 | 0 | 0 | 0 | 0 | 0 | 0 | 0 | 0 | 0 | 0 | 0 | 0 | 0 | 0 | 0 | 0 | 0 | 0 | 0 | 0 | 0 | 0 | 0 | 0 | 0 |
| 47 | 0 | 0 | 0 | 0 | 0 | 0 | 0 | 0 | 0 | 0 | 0 | 0 | 0 | 0 | 0 | 0 | 0 | 0 | 0 | 0 | 0 | 0 | 0 | 0 | 0 | 0 | 0 | 0 | 0 |
| 48 | 0 | 0 | 0 | 0 | 0 | 0 | 0 | 0 | 0 | 0 | 0 | 0 | 0 | 0 | 0 | 0 | 0 | 0 | 0 | 0 | 0 | 0 | 0 | 0 | 0 | 0 | 0 | 0 | 0 |
| 49 | 0 | 0 | 0 | 0 | 0 | 0 | 0 | 0 | 0 | 0 | 0 | 0 | 0 | 0 | 0 | 0 | 0 | 0 | 0 | 0 | 0 | 0 | 0 | 0 | 0 | 0 | 0 | 0 | 0 |
| 50 | 0 | 0 | 0 | 0 | 0 | 0 | 0 | 0 | 0 | 0 | 0 | 0 | 0 | 0 | 0 | 0 | 0 | 0 | 0 | 0 | 0 | 0 | 0 | 0 | 0 | 0 | 0 | 0 | 0 |
| 51 | 0 | 0 | 0 | 0 | 0 | 0 | 0 | 0 | 0 | 0 | 0 | 0 | 0 | 0 | 0 | 0 | 0 | 0 | 0 | 0 | 0 | 0 | 0 | 0 | 0 | 0 | 0 | 0 | 0 |
| 52 | 0 | 0 | 0 | 0 | 0 | 0 | 0 | 0 | 0 | 0 | 0 | 0 | 0 | 0 | 0 | 0 | 0 | 0 | 0 | 0 | 0 | 0 | 0 | 0 | 0 | 0 | 0 | 0 | 0 |
| 53 | 0 | 0 | 0 | 0 | 0 | 0 | 0 | 0 | 0 | 0 | 0 | 0 | 0 | 0 | 0 | 0 | 0 | 0 | 0 | 0 | 0 | 0 | 0 | 0 | 0 | 0 | 0 | 0 | 0 |
| 54 | 0 | 0 | 0 | 0 | 0 | 0 | 0 | 0 | 0 | 0 | 0 | 0 | 0 | 0 | 0 | 0 | 0 | 0 | 0 | 0 | 0 | 0 | 0 | 0 | 0 | 0 | 0 | 0 | 0 |
| 55 | 0 | 0 | 0 | 0 | 0 | 0 | 0 | 0 | 0 | 0 | 0 | 0 | 0 | 0 | 0 | 0 | 0 | 0 | 0 | 0 | 0 | 0 | 0 | 0 | 0 | 0 | 0 | 0 | 0 |
| 56 | 0 | 0 | 0 | 0 | 0 | 0 | 0 | 0 | 0 | 0 | 0 | 0 | 0 | 0 | 0 | 0 | 0 | 0 | 0 | 0 | 0 | 0 | 0 | 0 | 0 | 0 | 0 | 0 | 0 |
| 57 | 0 | 0 | 0 | 0 | 0 | 0 | 0 | 0 | 0 | 0 | 0 | 0 | 0 | 0 | 0 | 0 | 0 | 0 | 0 | 0 | 0 | 0 | 0 | 0 | 0 | 0 | 0 | 0 | 0 |
| 58 | 0 | 0 | 0 | 0 | 0 | 0 | 0 | 0 | 0 | 0 | 0 | 0 | 0 | 0 | 0 | 0 | 0 | 0 | 0 | 0 | 0 | 0 | 0 | 0 | 0 | 0 | 0 | 0 | 0 |
| 59 | 0 | 0 | 0 | 0 | 0 | 0 | 0 | 0 | 0 | 0 | 0 | 0 | 0 | 0 | 0 | 0 | 0 | 0 | 0 | 0 | 0 | 0 | 0 | 0 | 0 | 0 | 0 | 0 | 0 |
| 60 | 0 | 0 | 0 | 0 | 0 | 0 | 0 | 0 | 0 | 0 | 0 | 0 | 0 | 0 | 0 | 0 | 0 | 0 | 0 | 0 | 0 | 0 | 0 | 0 | 0 | 0 | 0 | 0 | 0 |
| 61 | 0 | 0 | 0 | 0 | 0 | 0 | 0 | 0 | 0 | 0 | 0 | 0 | 0 | 0 | 0 | 0 | 0 | 0 | 0 | 0 | 0 | 0 | 0 | 0 | 0 | 0 | 0 | 0 | 0 |
| 62 | 0 | 0 | 0 | 0 | 0 | 0 | 0 | 0 | 0 | 0 | 0 | 0 | 0 | 0 | 0 | 0 | 0 | 0 | 0 | 0 | 0 | 0 | 0 | 0 | 0 | 0 | 0 | 0 | 0 |
| 63 | 0 | 0 | 0 | 0 | 0 | 0 | 0 | 0 | 0 | 0 | 0 | 0 | 0 | 0 | 0 | 0 | 0 | 0 | 0 | 0 | 0 | 0 | 0 | 0 | 0 | 0 | 0 | 0 | 0 |
| 64 | 0 | 1 | 1 | 1 | 1 | 1 | 1 | 1 | 1 | 1 | 1 | 1 | 1 | 1 | 1 | 1 | 1 | 1 | 1 | 1 | 1 | 1 | 1 | 1 | 1 | 1 | 1 | 1 | 1 |
| 65 | 0 | 1 | 1 | 1 | 1 | 1 | 1 | 1 | 1 | 1 | 1 | 1 | 1 | 1 | 1 | 1 | 1 | 1 | 1 | 1 | 1 | 1 | 1 | 1 | 1 | 1 | 1 | 1 | 1 |
| 66 | 0 | 1 | 1 | 1 | 1 | 1 | 1 | 1 | 1 | 1 | 1 | 1 | 1 | 1 | 1 | 1 | 1 | 1 | 1 | 1 | 1 | 1 | 1 | 1 | 1 | 1 | 1 | 1 | 1 |
| 67 | 0 | 1 | 1 | 1 | 1 | 1 | 1 | 1 | 1 | 1 | 1 | 1 | 1 | 1 | 1 | 1 | 1 | 1 | 1 | 1 | 1 | 1 | 1 | 1 | 1 | 1 | 1 | 1 | 1 |
| 68 | 0 | 1 | 1 | 1 | 1 | 1 | 1 | 1 | 1 | 1 | 1 | 1 | 1 | 1 | 1 | 1 | 1 | 1 | 1 | 1 | 1 | 1 | 1 | 1 | 1 | 1 | 1 | 1 | 1 |
| 69 | 0 | 1 | 1 | 1 | 1 | 1 | 1 | 1 | 1 | 1 | 1 | 1 | 1 | 1 | 1 | 1 | 1 | 1 | 1 | 1 | 1 | 1 | 1 | 1 | 1 | 1 | 1 | 1 | 1 |
| 70 | 0 | 1 | 1 | 1 | 1 | 1 | 1 | 1 | 1 | 1 | 1 | 1 | 1 | 1 | 1 | 1 | 1 | 1 | 1 | 1 | 1 | 1 | 1 | 1 | 1 | 1 | 1 | 1 | 1 |
| 71 | 0 | 1 | 1 | 1 | 1 | 1 | 1 | 1 | 1 | 1 | 1 | 1 | 1 | 1 | 1 | 1 | 1 | 1 | 1 | 1 | 1 | 1 | 1 | 1 | 1 | 1 | 1 | 1 | 1 |
| 72 | 0 | 1 | 1 | 1 | 1 | 1 | 1 | 1 | 1 | 1 | 1 | 1 | 1 | 1 | 1 | 1 | 1 | 1 | 1 | 1 | 1 | 1 | 1 | 1 | 1 | 1 | 1 | 1 | 1 |
| 73 | 0 | 1 | 1 | 1 | 1 | 1 | 1 | 1 | 1 | 1 | 1 | 1 | 1 | 1 | 1 | 1 | 1 | 1 | 1 | 1 | 1 | 1 | 1 | 1 | 1 | 1 | 1 | 1 | 1 |
| 74 | 0 | 1 | 1 | 1 | 1 | 1 | 1 | 1 | 1 | 1 | 1 | 1 | 1 | 1 | 1 | 1 | 1 | 1 | 1 | 1 | 1 | 1 | 1 | 1 | 1 | 1 | 1 | 1 | 1 |
| 75 | 0 | 1 | 1 | 1 | 1 | 1 | 1 | 1 | 1 | 1 | 1 | 1 | 1 | 1 | 1 | 1 | 1 | 1 | 1 | 1 | 1 | 1 | 1 | 1 | 1 | 1 | 1 | 1 | 1 |
| 76 | 0 | 1 | 1 | 1 | 1 | 1 | 1 | 1 | 1 | 1 | 1 | 1 | 1 | 1 | 1 | 1 | 1 | 1 | 1 | 1 | 1 | 1 | 1 | 1 | 1 | 1 | 1 | 1 | 1 |
| 77 | 0 | 1 | 1 | 1 | 1 | 1 | 1 | 1 | 1 | 1 | 1 | 1 | 1 | 1 | 1 | 1 | 1 | 1 | 1 | 1 | 1 | 1 | 1 | 1 | 1 | 1 | 1 | 1 | 1 |
| 78 | 0 | 1 | 1 | 1 | 1 | 1 | 1 | 1 | 1 | 1 | 1 | 1 | 1 | 1 | 1 | 1 | 1 | 1 | 1 | 1 | 1 | 1 | 1 | 1 | 1 | 1 | 1 | 1 | 1 |
| 79 | 0 | 1 | 1 | 1 | 1 | 1 | 1 | 1 | 1 | 1 | 1 | 1 | 1 | 1 | 1 | 1 | 1 | 1 | 1 | 1 | 1 | 1 | 1 | 1 | 1 | 1 | 1 | 1 | 1 |
| 80 | 0 | 1 | 1 | 1 | 1 | 1 | 1 | 1 | 1 | 1 | 1 | 1 | 1 | 1 | 1 | 1 | 1 | 1 | 1 | 1 | 1 | 1 | 1 | 1 | 1 | 1 | 1 | 1 | 1 |
| 81 | 0 | 1 | 1 | 1 | 1 | 1 | 1 | 1 | 1 | 1 | 1 | 1 | 1 | 1 | 1 | 1 | 1 | 1 | 1 | 1 | 1 | 1 | 1 | 1 | 1 | 1 | 1 | 1 | 1 |
| 82 | 0 | 1 | 1 | 1 | 1 | 1 | 1 | 1 | 1 | 1 | 1 | 1 | 1 | 1 | 1 | 1 | 1 | 1 | 1 | 1 | 1 | 1 | 1 | 1 | 1 | 1 | 1 | 1 | 1 |
| 83 | 0 | 1 | 1 | 1 | 1 | 1 | 1 | 1 | 1 | 1 | 1 | 1 | 1 | 1 | 1 | 1 | 1 | 1 | 1 | 1 | 1 | 1 | 1 | 1 | 1 | 1 | 1 | 1 | 1 |
| 84 | 0 | 1 | 1 | 1 | 1 | 1 | 1 | 1 | 1 | 1 | 1 | 1 | 1 | 1 | 1 | 1 | 1 | 1 | 1 | 1 | 1 | 1 | 1 | 1 | 1 | 1 | 1 | 1 | 1 |
| 85 | 0 | 1 | 1 | 1 | 1 | 1 | 1 | 1 | 1 | 1 | 1 | 1 | 1 | 1 | 1 | 1 | 1 | 1 | 1 | 1 | 1 | 1 | 1 | 1 | 1 | 1 | 1 | 1 | 1 |
| 86 | 0 | 1 | 1 | 1 | 1 | 1 | 1 | 1 | 1 | 1 | 1 | 1 | 1 | 1 | 1 | 1 | 1 | 1 | 1 | 1 | 1 | 1 | 1 | 1 | 1 | 1 | 1 | 1 | 1 |
| 87 | 0 | 1 | 1 | 1 | 1 | 1 | 1 | 1 | 1 | 1 | 1 | 1 | 1 | 1 | 1 | 1 | 1 | 1 | 1 | 1 | 1 | 1 | 1 | 1 | 1 | 1 | 1 | 1 | 1 |
| 88 | 0 | 1 | 1 | 1 | 1 | 1 | 1 | 1 | 1 | 1 | 1 | 1 | 1 | 1 | 1 | 1 | 1 | 1 | 1 | 1 | 1 | 1 | 1 | 1 | 1 | 1 | 1 | 1 | 1 |
| 89 | 0 | 1 | 1 | 1 | 1 | 1 | 1 | 1 | 1 | 1 | 1 | 1 | 1 | 1 | 1 | 1 | 1 | 1 | 1 | 1 | 1 | 1 | 1 | 1 | 1 | 1 | 1 | 1 | 1 |
| 90 | 0 | 1 | 1 | 1 | 1 | 1 | 1 | 1 | 1 | 1 | 1 | 1 | 1 | 1 | 1 | 1 | 1 | 1 | 1 | 1 | 1 | 1 | 1 | 1 | 1 | 1 | 1 | 1 | 1 |
| 91 | 0 | 1 | 1 | 1 | 1 | 1 | 1 | 1 | 1 | 1 | 1 | 1 | 1 | 1 | 1 | 1 | 1 | 1 | 1 | 1 | 1 | 1 | 1 | 1 | 1 | 1 | 1 | 1 | 1 |
| 92 | 0 | 1 | 1 | 1 | 1 | 1 | 1 | 1 | 1 | 1 | 1 | 1 | 1 | 1 | 1 | 1 | 1 | 1 | 1 | 1 | 1 | 1 | 1 | 1 | 1 | 1 | 1 | 1 | 1 |
| 93 | 0 | 1 | 1 | 1 | 1 | 1 | 1 | 1 | 1 | 1 | 1 | 1 | 1 | 1 | 1 | 1 | 1 | 1 | 1 | 1 | 1 | 1 | 1 | 1 | 1 | 1 | 1 | 1 | 1 |
| 94 | 0 | 1 | 1 | 1 | 1 | 1 | 1 | 1 | 1 | 1 | 1 | 1 | 1 | 1 | 1 | 1 | 1 | 1 | 1 | 1 | 1 | 1 | 1 | 1 | 1 | 1 | 1 | 1 | 1 |
| 95 | 0 | 1 | 1 | 1 | 1 | 1 | 1 | 1 | 1 | 1 | 1 | 1 | 1 | 1 | 1 | 1 | 1 | 1 | 1 | 1 | 1 | 1 | 1 | 1 | 1 | 1 | 1 | 1 | 1 |
| 96 | 0 | 1 | 1 | 1 | 1 | 1 | 1 | 1 | 1 | 1 | 1 | 1 | 1 | 1 | 1 | 1 | 1 | 1 | 1 | 1 | 1 | 1 | 1 | 1 | 1 | 1 | 1 | 1 | 1 |
| 97 | 0 | 1 | 2 | 2 | 2 | 2 | 2 | 2 | 2 | 2 | 2 | 2 | 2 | 2 | 2 | 2 | 2 | 2 | 2 | 2 | 2 | 2 | 2 | 2 | 2 | 2 | 2 | 2 | 2 |
| 98 | 0 | 1 | 2 | 2 | 2 | 2 | 2 | 2 | 2 | 2 | 2 | 2 | 2 | 2 | 2 | 2 | 2 | 2 | 2 | 2 | 2 | 2 | 2 | 2 | 2 | 2 | 2 | 2 | 2 |
| 99 | 0 | 1 | 2 | 2 | 2 | 2 | 2 | 2 | 2 | 2 | 2 | 2 | 2 | 2 | 2 | 2 | 2 | 2 | 2 | 2 | 2 | 2 | 2 | 2 | 2 | 2 | 2 | 2 | 2 |
| 100 | 0 | 1 | 2 | 2 | 2 | 2 | 2 | 2 | 2 | 2 | 2 | 2 | 2 | 2 | 2 | 2 | 2 | 2 | 2 | 2 | 2 | 2 | 2 | 2 | 2 | 2 | 2 | 2 | 2 |
| 101 | 0 | 1 | 2 | 2 | 2 | 2 | 2 | 2 | 2 | 2 | 2 | 2 | 2 | 2 | 2 | 2 | 2 | 2 | 2 | 2 | 2 | 2 | 2 | 2 | 2 | 2 | 2 | 2 | 2 |
| 102 | 0 | 1 | 2 | 2 | 2 | 2 | 2 | 2 | 2 | 2 | 2 | 2 | 2 | 2 | 2 | 2 | 2 | 2 | 2 | 2 | 2 | 2 | 2 | 2 | 2 | 2 | 2 | 2 | 2 |
| 103 | 0 | 1 | 2 | 2 | 2 | 2 | 2 | 2 | 2 | 2 | 2 | 2 | 2 | 2 | 2 | 2 | 2 | 2 | 2 | 2 | 2 | 2 | 2 | 2 | 2 | 2 | 2 | 2 | 2 |
| 104 | 0 | 1 | 2 | 2 | 2 | 2 | 2 | 2 | 2 | 2 | 2 | 2 | 2 | 2 | 2 | 2 | 2 | 2 | 2 | 2 | 2 | 2 | 2 | 2 | 2 | 2 | 2 | 2 | 2 |
| 105 | 0 | 1 | 2 | 2 | 2 | 2 | 2 | 2 | 2 | 2 | 2 | 2 | 2 | 2 | 2 | 2 | 2 | 2 | 2 | 2 | 2 | 2 | 2 | 2 | 2 | 2 | 2 | 2 | 2 |
| 106 | 0 | 1 | 2 | 2 | 2 | 2 | 2 | 2 | 2 | 2 | 2 | 2 | 2 | 2 | 2 | 2 | 2 | 2 | 2 | 2 | 2 | 2 | 2 | 2 | 2 | 2 | 2 | 2 | 2 |
| 107 | 0 | 1 | 2 | 2 | 2 | 2 | 2 | 2 | 2 | 2 | 2 | 2 | 2 | 2 | 2 | 2 | 2 | 2 | 2 | 2 | 2 | 2 | 2 | 2 | 2 | 2 | 2 | 2 | 2 |
| 108 | 0 | 1 | 2 | 2 | 2 | 2 | 2 | 2 | 2 | 2 | 2 | 2 | 2 | 2 | 2 | 2 | 2 | 2 | 2 | 2 | 2 | 2 | 2 | 2 | 2 | 2 | 2 | 2 | 2 |
| 109 | 0 | 1 | 2 | 2 | 2 | 2 | 2 | 2 | 2 | 2 | 2 | 2 | 2 | 2 | 2 | 2 | 2 | 2 | 2 | 2 | 2 | 2 | 2 | 2 | 2 | 2 | 2 | 2 | 2 |
| 110 | 0 | 1 | 2 | 2 | 2 | 2 | 2 | 2 | 2 | 2 | 2 | 2 | 2 | 2 | 2 | 2 | 2 | 2 | 2 | 2 | 2 | 2 | 2 | 2 | 2 | 2 | 2 | 2 | 2 |
| 111 | 0 | 1 | 2 | 2 | 2 | 2 | 2 | 2 | 2 | 2 | 2 | 2 | 2 | 2 | 2 | 2 | 2 | 2 | 2 | 2 | 2 | 2 | 2 | 2 | 2 | 2 | 2 | 2 | 2 |
| 112 | 0 | 1 | 2 | 2 | 2 | 2 | 2 | 2 | 2 | 2 | 2 | 2 | 2 | 2 | 2 | 2 | 2 | 2 | 2 | 2 | 2 | 2 | 2 | 2 | 2 | 2 | 2 | 2 | 2 |
| 113 | 0 | 1 | 2 | 2 | 2 | 2 | 2 | 2 | 2 | 2 | 2 | 2 | 2 | 2 | 2 | 2 | 2 | 2 | 2 | 2 | 2 | 2 | 2 | 2 | 2 | 2 | 2 | 2 | 2 |
| 114 | 0 | 1 | 2 | 2 | 2 | 2 | 2 | 2 | 2 | 2 | 2 | 2 | 2 | 2 | 2 | 2 | 2 | 2 | 2 | 2 | 2 | 2 | 2 | 2 | 2 | 2 | 2 | 2 | 2 |
| 115 | 0 | 1 | 2 | 2 | 2 | 2 | 2 | 2 | 2 | 2 | 2 | 2 | 2 | 2 | 2 | 2 | 2 | 2 | 2 | 2 | 2 | 2 | 2 | 2 | 2 | 2 | 2 | 2 | 2 |
| 116 | 0 | 1 | 2 | 2 | 2 | 2 | 2 | 2 | 2 | 2 | 2 | 2 | 2 | 2 | 2 | 2 | 2 | 2 | 2 | 2 | 2 | 2 | 2 | 2 | 2 | 2 | 2 | 2 | 2 |
| 117 | 0 | 1 | 2 | 2 | 2 | 2 | 2 | 2 | 2 | 2 | 2 | 2 | 2 | 2 | 2 | 2 | 2 | 2 | 2 | 2 | 2 | 2 | 2 | 2 | 2 | 2 | 2 | 2 | 2 |
| 118 | 0 | 1 | 2 | 2 | 2 | 2 | 2 | 2 | 2 | 2 | 2 | 2 | 2 | 2 | 2 | 2 | 2 | 2 | 2 | 2 | 2 | 2 | 2 | 2 | 2 | 2 | 2 | 2 | 2 |
| 119 | 0 | 1 | 2 | 2 | 2 | 2 | 2 | 2 | 2 | 2 | 2 | 2 | 2 | 2 | 2 | 2 | 2 | 2 | 2 | 2 | 2 | 2 | 2 | 2 | 2 | 2 | 2 | 2 | 2 |
| 120 | 0 | 1 | 2 | 2 | 2 | 2 | 2 | 2 | 2 | 2 | 2 | 2 | 2 | 2 | 2 | 2 | 2 | 2 | 2 | 2 | 2 | 2 | 2 | 2 | 2 | 2 | 2 | 2 | 2 |
| 121 | 0 | 1 | 2 | 2 | 2 | 2 | 2 | 2 | 2 | 2 | 2 | 2 | 2 | 2 | 2 | 2 | 2 | 2 | 2 | 2 | 2 | 2 | 2 | 2 | 2 | 2 | 2 | 2 | 2 |
| 122 | 0 | 1 | 1 | 1 | 1 | 1 | 2 | 3 | 0 | 0 | 0 | 0 | 0 | 1 | 0 | 0 | 0 | 0 | 0 | 0 | 0 | 0 | 0 | 0 | 0 | 0 | 0 | 0 | 0 |
| 123 | 0 | 1 | 1 | 1 | 1 | 1 | 2 | 3 | 0 | 0 | 0 | 0 | 0 | 1 | 0 | 0 | 0 | 0 | 0 | 0 | 0 | 0 | 0 | 0 | 0 | 0 | 0 | 0 | 0 |
| 124 | 0 | 1 | 1 | 1 | 1 | 1 | 2 | 3 | 0 | 0 | 0 | 0 | 0 | 1 | 0 | 0 | 0 | 0 | 0 | 0 | 0 | 0 | 0 | 0 | 0 | 0 | 0 | 0 | 0 |
| 125 | 0 | 1 | 1 | 1 | 1 | 1 | 2 | 3 | 0 | 0 | 0 | 0 | 0 | 1 | 0 | 0 | 0 | 0 | 0 | 0 | 0 | 0 | 0 | 0 | 0 | 0 | 0 | 0 | 0 |
| 126 | 0 | 1 | 1 | 1 | 1 | 1 | 2 | 3 | 0 | 0 | 0 | 0 | 0 | 1 | 0 | 0 | 0 | 0 | 0 | 0 | 0 | 0 | 0 | 0 | 0 | 0 | 0 | 0 | 0 |
| 127 | 0 | 1 | 1 | 1 | 1 | 1 | 2 | 3 | 0 | 0 | 0 | 0 | 0 | 1 | 0 | 0 | 0 | 0 | 0 | 0 | 0 | 0 | 0 | 0 | 0 | 0 | 0 | 0 | 0 |
| 128 | 0 | 1 | 1 | 1 | 1 | 1 | 2 | 3 | 0 | 2 | 2 | 2 | 2 | 1 | 0 | 0 | 0 | 1 | 1 | 1 | 1 | 1 | 1 | 1 | 1 | 1 | 3 | 3 | 3 |
| 129 | 0 | 1 | 1 | 1 | 1 | 1 | 2 | 3 | 0 | 2 | 2 | 2 | 2 | 1 | 0 | 0 | 0 | 1 | 1 | 1 | 1 | 1 | 1 | 1 | 1 | 1 | 3 | 3 | 4 |
| 130 | 0 | 1 | 1 | 1 | 1 | 1 | 2 | 3 | 0 | 2 | 2 | 2 | 2 | 1 | 0 | 0 | 0 | 1 | 1 | 1 | 1 | 1 | 1 | 1 | 1 | 1 | 3 | 3 | 4 |
| 131 | 0 | 1 | 1 | 1 | 1 | 1 | 2 | 3 | 2 | 2 | 2 | 2 | 2 | 1 | 0 | 0 | 0 | 1 | 1 | 1 | 1 | 1 | 1 | 1 | 1 | 1 | 3 | 3 | 4 |
| 132 | 0 | 1 | 1 | 1 | 1 | 1 | 2 | 3 | 2 | 2 | 2 | 2 | 2 | 1 | 0 | 0 | 0 | 1 | 1 | 1 | 1 | 1 | 1 | 1 | 1 | 1 | 3 | 3 | 4 |
| 133 | 0 | 1 | 1 | 1 | 1 | 1 | 2 | 3 | 2 | 2 | 2 | 2 | 2 | 1 | 0 | 0 | 0 | 1 | 1 | 1 | 1 | 1 | 1 | 1 | 1 | 1 | 3 | 3 | 4 |
| 134 | 0 | 1 | 1 | 1 | 1 | 1 | 2 | 3 | 2 | 2 | 2 | 2 | 2 | 1 | 0 | 0 | 0 | 1 | 1 | 1 | 1 | 1 | 1 | 1 | 1 | 1 | 3 | 3 | 4 |
| 135 | 0 | 1 | 1 | 1 | 1 | 1 | 2 | 3 | 2 | 2 | 2 | 2 | 2 | 1 | 0 | 0 | 0 | 1 | 1 | 1 | 1 | 1 | 1 | 1 | 1 | 1 | 3 | 3 | 4 |
| 136 | 0 | 1 | 1 | 1 | 1 | 1 | 2 | 3 | 2 | 2 | 2 | 2 | 2 | 1 | 0 | 0 | 0 | 1 | 1 | 1 | 1 | 1 | 1 | 1 | 1 | 1 | 3 | 3 | 4 |
| 137 | 0 | 1 | 1 | 1 | 1 | 1 | 2 | 3 | 2 | 2 | 2 | 2 | 2 | 1 | 0 | 0 | 0 | 1 | 1 | 1 | 1 | 1 | 1 | 1 | 1 | 1 | 3 | 3 | 4 |
| 138 | 0 | 1 | 1 | 1 | 1 | 1 | 2 | 3 | 2 | 2 | 2 | 2 | 2 | 1 | 0 | 0 | 0 | 1 | 1 | 1 | 1 | 1 | 1 | 1 | 1 | 1 | 3 | 3 | 4 |
| 139 | 0 | 1 | 1 | 1 | 3 | 3 | 3 | 3 | 2 | 3 | 3 | 1 | 2 | 1 | 0 | 2 | 0 | 1 | 1 | 1 | 1 | 1 | 1 | 1 | 1 | 1 | 3 | 3 | 4 |
| 140 | 0 | 1 | 1 | 1 | 3 | 3 | 3 | 3 | 2 | 3 | 3 | 1 | 2 | 1 | 0 | 0 | 0 | 1 | 1 | 1 | 1 | 1 | 1 | 1 | 2 | 0 | 3 | 3 | 4 |
| 141 | 0 | 1 | 1 | 1 | 3 | 3 | 3 | 3 | 2 | 3 | 3 | 1 | 2 | 1 | 0 | 0 | 0 | 1 | 1 | 1 | 1 | 1 | 1 | 1 | 2 | 0 | 3 | 3 | 4 |
| 142 | 0 | 1 | 1 | 1 | 3 | 3 | 3 | 3 | 2 | 3 | 3 | 1 | 2 | 1 | 0 | 0 | 0 | 1 | 1 | 1 | 1 | 1 | 1 | 1 | 2 | 0 | 3 | 3 | 4 |
| 143 | 0 | 1 | 1 | 1 | 3 | 3 | 3 | 3 | 2 | 3 | 3 | 1 | 2 | 1 | 0 | 0 | 0 | 1 | 1 | 1 | 1 | 1 | 1 | 1 | 2 | 0 | 3 | 3 | 4 |
| 144 | 0 | 1 | 1 | 1 | 3 | 3 | 3 | 3 | 2 | 3 | 3 | 1 | 2 | 1 | 0 | 0 | 0 | 1 | 1 | 1 | 1 | 1 | 1 | 1 | 2 | 0 | 3 | 3 | 4 |
| 145 | 0 | 1 | 0 | 1 | 1 | 3 | 3 | 3 | 2 | 3 | 3 | 1 | 2 | 1 | 2 | 2 | 2 | 2 | 2 | 2 | 2 | 2 | 2 | 1 | 2 | 0 | 2 | 3 | 4 |
| 146 | 0 | 1 | 0 | 1 | 0 | 0 | 0 | 3 | 2 | 3 | 3 | 1 | 2 | 1 | 3 | 0 | 2 | 1 | 1 | 1 | 1 | 1 | 1 | 1 | 2 | 0 | 2 | 3 | 4 |
| 147 | 0 | 1 | 0 | 1 | 0 | 2 | 1 | 3 | 2 | 3 | 3 | 1 | 2 | 1 | 3 | 0 | 0 | 1 | 1 | 1 | 1 | 1 | 1 | 1 | 2 | 0 | 2 | 3 | 4 |
| 148 | 0 | 1 | 0 | 1 | 0 | 2 | 1 | 3 | 2 | 3 | 3 | 1 | 2 | 1 | 3 | 2 | 0 | 1 | 1 | 1 | 1 | 1 | 1 | 1 | 2 | 2 | 2 | 3 | 4 |
| 149 | 0 | 1 | 0 | 1 | 0 | 2 | 1 | 3 | 2 | 3 | 3 | 1 | 2 | 1 | 3 | 2 | 0 | 1 | 0 | 0 | 0 | 2 | 3 | 1 | 2 | 2 | 2 | 3 | 4 |
| 150 | 0 | 1 | 0 | 1 | 0 | 2 | 1 | 3 | 2 | 3 | 3 | 1 | 2 | 1 | 3 | 2 | 0 | 1 | 0 | 2 | 0 | 2 | 3 | 1 | 2 | 1 | 2 | 3 | 4 |

The first column in the above table is the ID, the next column (year index) is t=0 and the last column is t=28

## References:

Akshintala, Divya, Radhika Chugh, Farah Amer, and Kenneth Cusi. 2021. "Nonalcoholic Fatty Liver Disease : The Overlooked Complication of Type 2 Diabetes."

Akyuz, Umit, Atakan Yesil, and Yusuf Yilmaz. 2014. "Characterization of Lean Patients with Nonalcoholic Fatty Liver Disease: Potential Role of High Hemoglobin Levels." *Scandinavian Journal of Gastroenterology* 50(3):341–46. doi: 10.3109/00365521.2014.983160.

Alam, Shahinul, Utpal Das Gupta, Mahbubul Alam, Jahangir Kabir, Ziaur Rahman Chowdhury, and A. K. M. Khorshe. Alam. 2014. "Clinical, Anthropometric, Biochemical, and Histological Characteristics of Nonobese Nonalcoholic Fatty Liver Disease Patients of Bangladesh." *Indian Journal of Gastroenterology* 33(5):452–57. doi: 10.1007/s12664-014-0488-5.

Alkhouri, Naim, Eric Lawitz, and Mazen Noureddin. 2019. "Looking Into the Crystal Ball: Predicting the Future Challenges of Fibrotic NASH Treatment." *Hepatology Communications* 3(5):605–13. doi: 10.1002/hep4.1342.

Alkhouri, Naim, Fred Poordad, and Eric Lawitz. 2018. "Management of Nonalcoholic Fatty Liver Disease: Lessons Learned from Type 2 Diabetes." *Hepatology Communications* 2(7):778–85.

Allen, Linda J. S. 2010. *An Introduction to Stochastic Processes with Applications to Biology*. CRC Press.

Angulo, Paul, Jason M. Hui, Giulio Marchesini, Ellisabetta Bugianesi, Jacob George, Geoffrey C. Farrell, Felicity Enders, Sushma Saksena, Alastair D. Burt, John P. Bida, Keith Lindor, Schuyler O. Sanderson, Marco Lenzi, Leon A. Adams, James Kench, Terry M. Therneau, and Christopher P. Day. 2007. "The NAFLD Fibrosis Score: A Noninvasive System That Identifies Liver Fibrosis in Patients with NAFLD." *Hepatology* 45(4):846–54. doi: 10.1002/hep.21496.

Anwar, Noura, and M. Riad Mahmoud. 2014. "A Stochastic Model for the Progression of Chronic Kidney Disease." *Journal of Engineering Research and Applications [Internet]* 4(11):8–19.

Baffy, Gyorgy. 2005. "Uncoupling Protein-2 and Non-Alcoholic Fatty Liver Disease." *Frontiers in Bioscience*.

Bagheri Lankarani, Kamran, Fariborz Ghaffarpasand, Mojtaba Mahmoodi, Mehrzad Lotfi, Nima Zamiri, Sayed Taghi Heydari, Mohammad K. Kazem Fallahzadeh, Najmeh Maharlouei, Meisam Babaeinejad, Soheila Mehravar, and Bita Geramizadeh. 2013. "Non Alcoholic Fatty Liver Disease in Southern Iran: A Population Based Study." *Hepatitis Monthly* 13(5):3–9. doi: 10.5812/hepatmon.9248.

Bailey, N. T. J. (1975). *The mathematical theory of infectious diseases and its applications* (Issue 2nd ediition). Charles Griffin & Company Ltd 5a Crendon Street, High Wycombe, Bucks HP13 6LE.

Barendregt, J. J., Van Oortmarssen, G. J., Vos, T., & Murray, C. J. L. (2003). A generic model for the assessment of disease epidemiology: the computational basis of DisMod II. *Population Health Metrics*, *1*(1), 1–8.

Bartolomeo, Nicola, Paolo Trerotoli, and Gabriella Serio. 2011. "Progression of Liver Cirrhosis to HCC: An Application of Hidden Markov Model." *BMC Medical Research Methodology* 11(1):1–8.

Bedogni, Giorgio, Stefano Bellentani, Lucia Miglioli, Flora Masutti, Marilena Passalacqua, Anna Castiglione, and Claudio Tiribelli. 2006. "The Fatty Liver Index: A Simple and Accurate Predictor of Hepatic Steatosis in the General Population." *BMC Gastroenterology* 6(1):1–7.

Bedossa, Pierre, Christine Poitou, Nicolas Veyrie, Jean-Luc Bouillot, Arnaud Basdevant, Valerie Paradis, Joan Tordjman, and Karine Clement. 2012. "Histopathological Algorithm and Scoring System for Evaluation of Liver Lesions in Morbidly Obese Patients." *Hepatology* 56(5):1751–59.

Bellentani, Stefano, Gioconda Saccoccio, Flora Masutti, Lory S. Crocè, Giovanni Brandi, Franco Sasso, Giovanni Cristanini, and Claudio Tiribelli. 2000. "Prevalence of and Risk Factors for Hepatic Steatosis in Northern Italy." *Annals of Internal Medicine* 132(2):112–17. doi: 10.7326/0003-4819-132-2-200001180-00004.

Blank, Valentin, David Petroff, Sebastian Beer, Albrecht Böhlig, Maria Heni, Thomas Berg, Yvonne Bausback, Arne Dietrich, Anke Tönjes, Marcus Hollenbach, Matthias Blüher, Volker Keim, Johannes Wiegand, and Thomas Karlas. 2020. "Current NAFLD Guidelines for Risk Stratification in Diabetic Patients Have Poor Diagnostic Discrimination." *Scientific Reports* 10(1):1–11. doi: 10.1038/s41598-020-75227-x.

Boyer, Thomas D., and Keith D. Lindor. 2016. *Zakim and Boyer's Hepatology: A Textbook of Liver Disease e-Book*. Elsevier Health Sciences.

Brand, Martin D., and Telma C. Esteves. 2005. "Physiological Functions of the Mitochondrial Uncoupling Proteins UCP2 and UCP3." *Cell Metabolism* 2(2):85–93.

Browning, Jeffrey D., Lidia S. Szczepaniak, Robert Dobbins, Pamela Nuremberg, Jay D. Horton, Jonathan C. Cohen, Scott M. Grundy, and Helen H. Hobbs. 2004. "Prevalence of Hepatic Steatosis in an Urban Population in the United States: Impact of Ethnicity." *Hepatology* 40(6):1387–95. doi: 10.1002/hep.20466.

Cassandras, Christos G., and Stephane Lafortune. 2009. *Introduction to Discrete Event Systems*. Springer Science & Business Media.

Castañeda, Liliana Blanco, Viswanathan Arunachalam, and Selvamuthu Dharmaraja. 2012. *Introduction to Probability and Stochastic Processes with Applications*. Wiley Online Library.

Castera, Laurent, Mireen Friedrich-Rust, and Rohit Loomba. 2019. "Noninvasive Assessment of Liver Disease in Patients With Nonalcoholic Fatty Liver Disease." *Gastroenterology* 156(5):1264-1281.e4. doi: 10.1053/j.gastro.2018.12.036.

Chalasani, Naga, Laura Wilson, David E. Kleiner, Oscar W. Cummings, Elizabeth M. Brunt, Aynur Ünalp, and NASH Clinical Research Network. 2008. "Relationship of Steatosis Grade and Zonal Location to Histological Features of Steatohepatitis in Adult Patients with Non-Alcoholic Fatty Liver Disease." *Journal of Hepatology* 48(5):829–34.

Chalasani, Naga, Zobair Younossi, Joel E. Lavine, Michael Charlton, Kenneth Cusi, Mary Rinella, Stephen A. Harrison, Elizabeth M. Brunt, and Arun J. Sanyal. 2018. "The Diagnosis and Management of Nonalcoholic Fatty Liver Disease: Practice Guidance from the American Association for the Study of Liver Diseases." *Hepatology* 67(1):328–57. doi: 10.1002/hep.29367.

Chen, Chien-Hua, Min-Ho Huang, Jee-Chun Yang, Chiu-Kue Nien, Chi-Chieh Yang, Yung-Hsiang Yeh, and Sen-Kou Yueh. 2006. "Prevalence and Risk Factors of Nonalcoholic Fatty Liver Disease in an Adult Population of Taiwan: Metabolic Significance of Nonalcoholic Fatty Liver Disease in Nonobese Adults." *Journal of Clinical Gastroenterology* 40(8):745–52.

Chiang, Chin Long. 1968. "Introduction to Stochastic Processes in Biostatistics."

Conus, Florence, David B. Allison, Remi Rabasa-Lhoret, Maxime St.-Onge, David H. St.-Pierre, Andréanne Tremblay-Lebeau, and Eric T. Poehlman. 2004. "Metabolic and Behavioral Characteristics of Metabolically Obese but Normal-Weight Women." *Journal of Clinical Endocrinology and Metabolism* 89(10):5013–20. doi: 10.1210/jc.2004-0265.

Conus, Florence, Rémi Rabasa-Lhoret, and François Péronnet. 2007. "Characteristics of Metabolically Obese Normal-Weight (MONW) Subjects." *Applied Physiology, Nutrition and Metabolism* 32(1):4–12. doi:

10.1139/H06-092.

Cruz, Anna Christina Dela, Elisabetta Bugianesi, Jacob George, Christopher P. Day, Hammad Liaquat, Phunchai Charatcharoenwitthaya, Peter R. Mills, Sanne Dam-Larsen, Einar S. Bjornsson, Svanhildur Haflidadottir, Leon A. Adams, Flemming Bendtsen, and Paul Angulo. 2014. "379 Characteristics and Long-Term Prognosis of Lean Patients With Nonalcoholic Fatty Liver Disease." *Gastroenterology* 146(5):S-909. doi: 10.1016/s0016-5085(14)63307-2.

Das, Kausik, Kshaunish Das, Partha S. Mukherjee, Alip Ghosh, Sumantra Ghosh, Asit R. Mridha, Tapan Dhibar, Bhaskar Bhattacharya, Dilip Bhattacharya, Byomkesh Manna, Gopal K. Dhali, Amal Santra, and Abhijit Chowdhury. 2010. "Nonobese Population in a Developing Country Has a High Prevalence of Nonalcoholic Fatty Liver and Significant Liver Disease." *Hepatology* 51(5):1593–1602. doi: 10.1002/hep.23567.

Dasarathy, Srinivasan, Jaividhya Dasarathy, Amer Khiyami, Rajesh Joseph, Rocio Lopez, and Arthur J. McCullough. 2009. "Validity of Real Time Ultrasound in the Diagnosis of Hepatic Steatosis: A Prospective Study." *Journal of Hepatology* 51(6):1061–67. doi: 10.1016/j.jhep.2009.09.001.

Dassanayake, Anuradha S., Anuradhani Kasturiratne, Shaman Rajindrajith, Udaya Kalubowila, Sureka Chakrawarthi, Arjuna P. De Silva, Miyuki Makaya, Tetsuya Mizoue, Norihiro Kato, A. Rajitha Wickremasinghe, and H. Janaka De Silva. 2009. "Prevalence and Risk Factors for Non-Alcoholic Fatty Liver Disease among Adults in an Urban Sri Lankan Population." *Journal of Gastroenterology and Hepatology (Australia)* 24(7):1284–88. doi: 10.1111/j.1440-1746.2009.05831.x.

De, Arka, and Ajay Duseja. 2020. "Natural History of Simple Steatosis or Nonalcoholic Fatty Liver." *Journal of Clinical and Experimental Hepatology* 10(3):255–62. doi: 10.1016/j.jceh.2019.09.005.

Dongiovanni, Paola, Quentin Anstee, and Luca Valenti. 2013. "Genetic Predisposition in NAFLD and NASH: Impact on Severity of Liver Disease and Response to Treatment." *Current Pharmaceutical Design* 19(29):5219–38. doi: 10.2174/13816128113199990381.

Dvorak, Roman V., Walter F. DeNino, Philip A. Ades, and Eric T. Poehlman. 1999. "Phenotypic Characteristics Associated with Insulin Resistance in Metabolically Obese but Normal-Weight Young Women." *Diabetes* 48(11):2210–14. doi: 10.2337/diabetes.48.11.2210.

Ehrlich, Avner, Daniel Duche, Gladys Ouedraogo, and Yaakov Nahmias. 2019. "Challenges and Opportunities in the Design of Liver-on-Chip Microdevices." *Annual Review of Biomedical Engineering* 21:219–39. doi: 10.1146/annurev-bioeng-060418-052305.

Estes, Chris, Homie Razavi, Rohit Loomba, Zobair Younossi, and Arun J. Sanyal. 2018. "Modeling the Epidemic of Nonalcoholic Fatty Liver Disease Demonstrates an Exponential Increase in Burden of Disease." *Hepatology* 67(1):123–33.

Fackrell, Mark. 2009. "Modelling Healthcare Systems with Phase-Type Distributions." *Health Care Management Science* 12(1):11.

Fan, Jian Gao, Jun Zhu, Xing Jian Li, Lan Chen, Lui Li, Fei Dai, Feng Li, and Shi Yao Chen. 2005. "Prevalence of and Risk Factors for Fatty Liver in a General Population of Shanghai, China." *Journal of Hepatology* 43(3):508–14. doi: 10.1016/j.jhep.2005.02.042.

Foster, Temitope, Frank A. Anania, Dong Li, Ronit Katz, and Matthew Budoff. 2013. "The Prevalence and Clinical Correlates of Nonalcoholic Fatty Liver Disease (NAFLD) in African Americans: The Multiethnic Study of Atherosclerosis (MESA)." *Digestive Diseases and Sciences* 58(8):2392–98. doi: 10.1007/s10620-013-2652-7.

Foucher, Yohann, Eve Mathieu, Philippe Saint-Pierre, Jean-François Durand, and Jean-Pierre Daurès. 2005. "A Semi-Markov Model Based on Generalized Weibull Distribution with an Illustration for HIV Disease."

*Biometrical Journal: Journal of Mathematical Methods in Biosciences* 47(6):825–33.

Fromenty, B., M. A. Robin, A. Igoudjil, A. Mansouri, and D. Pessayre. 2004. "The Ins and Outs of Mitochondrial Dysfunction in NASH." *Diabetes & Metabolism* 30(2):121–38.

Fu, Chen Chung, Ming Chen Chen, Yin Ming Li, Tso Tsai Liu, and Li Yu Wang. 2009. "The Risk Factors for Ultrasound-Diagnosed Non-Alcoholic Fatty Liver Disease among Adolescents." *Annals of the Academy of Medicine Singapore* 38(1):15–21.

Grover, Gurprit, Divya Seth, Vajala Ravi, and Prafulla Kumar Swain. 2014. "A Multistate Markov Model for the Progression of Liver Cirrhosis in the Presence of Various Prognostic Factors." *Chilean Journal of Statistics* 5:15–27.

Guirguis, Erenie, Yasmin Grace, Anthony Bolson, Matthew J. DellaVecchia, and Melissa Ruble. 2020. "Emerging Therapies for the Treatment of Non-Alcoholic Steatohepatitis: A Systematic Review." *Pharmacotherapy: The Journal of Human Pharmacology and Drug Therapy*.

Hae, Jin Kim, Jin Kim Hyeong, Eun Lee Kwang, Jung Kim Dae, Kyung Kim Soo, Woo Ahn Chul, Sung Kil Lim, Rae Kim Kyung, Chul Lee Hyun, Bum Huh Kap, and Soo Cha Bong. 2004. "Metabolic Significance of Nonalcoholic Fatty Liver Disease in Nonobese, Nondiabetic Adults." *Archives of Internal Medicine* 164(19):2169–75. doi: 10.1001/archinte.164.19.2169.

Ibe, Oliver. 2013. *Markov Processes for Stochastic Modeling*. Newnes.

Imbert-Bismut, Françoise, Vlad Ratziu, Laurence Pieroni, Frederic Charlotte, Yves Benhamou, and Thierry Poynard. 2001. "Biochemical Markers of Liver Fibrosis in Patients with Hepatitis C Virus Infection: A Prospective Study." *The Lancet* 357(9262):1069–75.

Jackson, Christopher H. 2011. "Multi-State Models for Panel Data: The Msm Package for R." *Journal of Statistical Software* 38(8):1–29.

Jackson, Christopher H., and Linda D. Sharples. 2002. "Hidden Markov Models for the Onset and Progression of Bronchiolitis Obliterans Syndrome in Lung Transplant Recipients." *Statistics in Medicine* 21(1):113–28.

Kalbfleisch, J. D., and Jerald Franklin Lawless. 1985. "The Analysis of Panel Data under a Markov Assumption." *Journal of the American Statistical Association* 80(392):863–71.

Katsuki, Akira, Yasuhiro Sumida, Hideki Urakawa, Esteban C. Gabazza, Shuichi Murashima, Noriko Maruyama, Kohei Morioka, Kaname Nakatani, Yutaka Yano, and Yukihiko Adachi. 2003. "Increased Visceral Fat and Serum Levels of Triglyceride Are Associated with Insulin Resistance in Japanese Metabolically Obese, Normal Weight Subjects with Normal Glucose Tolerance." *Diabetes Care* 26(8):2341–44. doi: 10.2337/diacare.26.8.2341.

Kim, Donghee, and W. Ray Kim. 2017. "Nonobese Fatty Liver Disease." *Clinical Gastroenterology and Hepatology* 15(4):474–85. doi: 10.1016/j.cgh.2016.08.028.

Kleiner, David E., Elizabeth M. Brunt, Mark Van Natta, Cynthia Behling, Melissa J. Contos, Oscar W. Cummings, Linda D. Ferrell, Yao-Chang Liu, Michael S. Torbenson, and Aynur Unalp-Arida. 2005. "Design and Validation of a Histological Scoring System for Nonalcoholic Fatty Liver Disease." *Hepatology* 41(6):1313–21.

Klotz, Jerome H., and Linda D. Sharples. 1994. "Estimation for a Markov Heart Transplant Model." *Journal of the Royal Statistical Society: Series D (The Statistician)* 43(3):431–38.

Konerman, Monica A., Jacob C. Jones, and Stephen A. Harrison. 2018. "Pharmacotherapy for NASH: Current and Emerging." *Journal of Hepatology* 68(2):362–75. doi: 10.1016/j.jhep.2017.10.015.

Kotronen, Anna, Markku Peltonen, Antti Hakkarainen, Ksenia Sevastianova, Robert Bergholm, Lina M. Johansson, Nina Lundbom, Aila Rissanen, Martin Ridderstråle, and Leif Groop. 2009. "Prediction of Non-Alcoholic Fatty Liver Disease and Liver Fat Using Metabolic and Genetic Factors." *Gastroenterology* 137(3):865–72.

Kruijshaar, M. E., Barendregt, J. J., & Hoeymans, N. (2002). The use of models in the estimation of disease epidemiology. *Bulletin of the World Health Organization*, *80*, 622–628.

Kwon, Young Min, Seung Won Oh, Seung Sik Hwang, Cheolmin Lee, Hygbrtae Kwon, and Goh Eun Chung. 2012. "Association of Nonalcoholic Fatty Liver Disease with Components of Metabolic Syndrome According to Body Mass Index in Korean Adults." *American Journal of Gastroenterology* 107(12):1852–58. doi: 10.1038/ajg.2012.314.

Leung, Jonathan Chung Fai, Thomson Chi Wang Loong, Jeremy Lok Wei, Grace Lai Hung Wong, Anthony Wing Hung Chan, Paul Cheung Lung Choi, Sally She Ting Shu, Angel Mei Ling Chim, Henry Lik Yuen Chan, and Vincent Wai Sun Wong. 2017. "Histological Severity and Clinical Outcomes of Nonalcoholic Fatty Liver Disease in Nonobese Patients." *Hepatology* 65(1):54–64. doi: 10.1002/hep.28697.

Lichtinghagen, Ralf, Daniel Pietsch, Heike Bantel, Michael P. Manns, Korbinian Brand, and Matthias J. Bahr. 2013. "The Enhanced Liver Fibrosis (ELF) Score: Normal Values, Influence Factors and Proposed Cut-off Values." *Journal of Hepatology* 59(2):236–42.

Longini Jr, Ira M., W. Scott Clark, Robert H. Byers, John W. Ward, William W. Darrow, George F. Lemp, and Herbert W. Hethcote. 1989. "Statistical Analysis of the Stages of HIV Infection Using a Markov Model." *Statistics in Medicine* 8(7):831–43.

Loomba, Rohit, and Arun J. Sanyal. 2013. "The Global NAFLD Epidemic." *Nature Reviews Gastroenterology & Hepatology* 10(11):686–90.

Machado, Mariana V., and Helena Cortez-Pinto. 2013. "Non-Invasive Diagnosis of Non-Alcoholic Fatty Liver Disease. A Critical Appraisal." *Journal of Hepatology* 58(5):1007–19. doi: 10.1016/j.jhep.2012.11.021.

Manco, Melania, Anna Alisi, Jose Manuel Fernandez Real, Francesco Equitani, Rita Devito, Luca Valenti, and Valerio Nobili. 2011. "Early Interplay of Intra-Hepatic Iron and Insulin Resistance in Children with Non-Alcoholic Fatty Liver Disease." *Journal of Hepatology* 55(3):647–53. doi: 10.1016/j.jhep.2010.12.007.

Marchesini, Giulio, Elisabetta Bugianesi, Gabriele Forlani, Fernanda Cerrelli, Marco Lenzi, Rita Manini, Stefania Natale, Ester Vanni, Nicola Villanova, Nazario Melchionda, and Mario Rizzetto. 2003. "Nonalcoholic Fatty Liver, Steatohepatitis, and the Metabolic Syndrome." *Hepatology* 37(4):917–23. doi: 10.1053/jhep.2003.50161.

Marchesini, Giulio, Christopher P. Day, Jean Francois Dufour, Ali Canbay, Valerio Nobili, Vlad Ratziu, Herbert Tilg, Michael Roden, Amalia Gastaldelli, Hannele Yki-Jarvinen, Fritz Schick, Roberto Vettor, Gema Fruhbeck, and Lisbeth Mathus-Vliegen. 2016. "EASL-EASD-EASO Clinical Practice Guidelines for the Management of Non-Alcoholic Fatty Liver Disease." *Journal of Hepatology* 64(6):1388–1402. doi: 10.1016/j.jhep.2015.11.004.

Marshall, Guillermo, and Richard H. Jones. 1995. "Multi-state Models and Diabetic Retinopathy." *Statistics in Medicine* 14(18):1975–83.

McPherson, Stuart, Tim Hardy, Elsbeth Henderson, Alastair D. Burt, Christopher P. Day, and Quentin M. Anstee. 2015. "Evidence of NAFLD Progression from Steatosis to Fibrosing-Steatohepatitis Using Paired Biopsies: Implications for Prognosis and Clinical Management." *Journal of Hepatology* 62(5):1148–55.

Meigs, James B., Peter W. F. Wilson, Caroline S. Fox, Ramachandran S. Vasan, David M. Nathan, Lisa M. Sullivan, and Ralph B. D'Agostino. 2006. "Body Mass Index, Metabolic Syndrome, and Risk of Type 2 Diabetes or Cardiovascular Disease." *Journal of Clinical Endocrinology and Metabolism* 91(8):2906–12. doi:

10.1210/jc.2006-0594.

Mofrad, Pouneh, Melissa J. Contos, Mahmadul Haque, Carol Sargeant, Robert A. Fisher, Velimir A. Luketic, Richard K. Sterling, Mitchell L. Shiffman, Richard T. Stravitz, and Arun J. Sanyal. 2003. “Clinical and Histologic Spectrum of Nonalcoholic Fatty Liver Disease Associated with Normal ALT Values.” *Hepatology* 37(6):1286–92. doi: 10.1053/jhep.2003.50229.

Molero-Conejo, Emperatriz, Luz Marina Morales, Virginia Fernández, Xiomara Raleigh, Maria Esther Gómez, Maritza Semprún-Fereira, Gilberto Campos, and Elena Ryder. 2003. “Lean Adolescents with Increased Risk for Metabolic Syndrome.” *Archivos Latinoamericanos de Nutricion* 53(1):39–46.

Neuschwander-Tetri, Brent A. 2017. “Non-Alcoholic Fatty Liver Disease.” *BMC Medicine* 15(1):1–6. doi: 10.1186/s12916-017-0806-8.

Newsome, Philip N., Magali Sasso, Jonathan J. Deeks, Angelo Paredes, Jérôme Boursier, Wah Kheong Chan, Yusuf Yilmaz, Sébastien Czernichow, Ming Hua Zheng, Vincent Wai Sun Wong, Michael Allison, Emmanuel Tsochatzis, Quentin M. Anstee, David A. Sheridan, Peter J. Eddowes, Indra N. Guha, Jeremy F. Cobbold, Valérie Paradis, Pierre Bedossa, Véronique Miette, Céline Fournier-Poizat, Laurent Sandrin, and Stephen A. Harrison. 2020. “FibroScan-AST (FAST) Score for the Non-Invasive Identification of Patients with Non-Alcoholic Steatohepatitis with Significant Activity and Fibrosis: A Prospective Derivation and Global Validation Study.” *The Lancet Gastroenterology and Hepatology* 5(4):362–73. doi: 10.1016/S2468-1253(19)30383-8.

Omagari, Katsuhisa, Yoshiko Kadokawa, Jun-ichi Masuda, and Ichiei Egawa. 2002. “NON-ALCOHOLIC FATTY LIVER IN JAPAN Fatty Liver in Non-Alcoholic Non-Overweight Japanese Adults :” *J Gastroenterol Hepatol.* 17(10):1–7.

Park, Joong-Won, Gyu Jeong, Sang Jin Kim, Mi Kyung Kim, and Sill Moo Park. 2007. “Predictors Reflecting the Pathological Severity of Non-alcoholic Fatty Liver Disease: Comprehensive Study of Clinical and Immunohistochemical Findings in Younger Asian Patients.” *Journal of Gastroenterology and Hepatology* 22(4):491–97.

Park, Seung H., Woo K. Jeon, Sang H. Kim, Hong J. Kim, Dong I. Park, Yong K. Cho, In K. Sung, Chong I. Sohn, Dong K. Keum, and Byung I. Kim. 2006. “Prevalence and Risk Factors of Non-Alcoholic Fatty Liver Disease among Korean Adults.” *Journal of Gastroenterology and Hepatology (Australia)* 21(1 PART1):138–43. doi: 10.1111/j.1440-1746.2005.04086.x.

Paul, Jayanta. 2020. “Recent Advances in Non-Invasive Diagnosis and Medical Management of Non-Alcoholic Fatty Liver Disease in Adult.” *Egyptian Liver Journal* 10(1). doi: 10.1186/s43066-020-00043-x.

Pérez-Ocón, Rafael, Juan Eloy Ruiz-Castro, and M. Luz Gámiz-Pérez. 2001. “A Piecewise Markov Process for Analysing Survival from Breast Cancer in Different Risk Groups.” *Statistics in Medicine* 20(1):109–22.

Petta, S., C. Muratore, and A. Craxì. 2009. “Non-Alcoholic Fatty Liver Disease Pathogenesis: The Present and the Future.” *Digestive and Liver Disease* 41(9):615–25. doi: 10.1016/j.dld.2009.01.004.

Portillo-Sanchez, Paola, Fernando Bril, Maryann Maximos, Romina Lomonaco, Diane Biernacki, Beverly Orsak, Sreevidya Subbarayan, Amy Webb, Joan Hecht, and Kenneth Cusi. 2015. “High Prevalence of Nonalcoholic Fatty Liver Disease in Patients with Type 2 Diabetes Mellitus and Normal Plasma Aminotransferase Levels.” *Journal of Clinical Endocrinology and Metabolism* 100(6):2231–38. doi: 10.1210/jc.2015-1966.

Prati, Daniele, Emanuela Taioli, Alberto Zanella, Emanuela Della Torre, Sonia Butelli, Emanuela Del Vecchio, Luciana Vianello, Francesco Zanuso, Fulvio Mozzi, Silvano Milani, Dario Conte, Massimo Colombo, and Girolamo Sirchia. 2002. “Updated Definitions of Healthy Ranges for Serum Alanine Aminotransferase Levels.” *Annals of Internal Medicine* 137(1):1–9. doi: 10.7326/0003-4819-137-1-200207020-00006.

Ribeiro, Paulo S., Helena Cortez-Pinto, Susana Solá, Rui E. Castro, Rita M. Ramalho, Amélia Baptista, Miguel C. Moura, Maria E. Camilo, and Cecília M. P. Rodrigues. 2004. “Hepatocyte Apoptosis, Expression of Death Receptors, and Activation of NF-κ B in the Liver of Nonalcoholic and Alcoholic Steatohepatitis Patients.” *Official Journal of the American College of Gastroenterology| ACG* 99(9):1708–17.

Ruderman, B., and H. Schneider. 1981. “The ‘Metabolically-Obese,’ Individual1 3.” *The American Journal OfClinical Nutrition* 34(April):1617–21.

Ruderman, Neil, D. Chisholm, X. Pi-Sunyer, and S. Schneider. 1998. “The Metabolically Obese, Normal-Weight Individual Revisited.” *Diabetes* 47(5):699–713. doi: 10.2337/diabetes.47.5.699.

Saint-Pierre, P., C. Combescure, J. P. Daures, and P. Godard. 2003. “The Analysis of Asthma Control under a Markov Assumption with Use of Covariates.” *Statistics in Medicine* 22(24):3755–70.

Santoro, Nicola, Clarence K. Zhang, Hongyu Zhao, Andrew J. Pakstis, Grace Kim, Romy Kursawe, Daniel J. Dykas, Allen E. Bale, Cosimo Giannini, and Bridget Pierpont. 2012. “Variant in the Glucokinase Regulatory Protein (GCKR) Gene Is Associated with Fatty Liver in Obese Children and Adolescents.” *Hepatology* 55(3):781–89.

Sberna, A. L., B. Bouillet, A. Rouland, M. C. Brindisi, A. Nguyen, T. Mouillot, L. Duvillard, D. Denimal, R. Loffroy, B. Vergès, P. Hillon, and J. M. Petit. 2018. “European Association for the Study of the Liver (EASL), European Association for the Study of Diabetes (EASD) and European Association for the Study of Obesity (EASO) Clinical Practice Recommendations for the Management of Non-Alcoholic Fatty Liver Diseas.” *Diabetic Medicine* 35(3):368–75. doi: 10.1111/dme.13565.

Shah, Amy G., Alison Lydecker, Karen Murray, Brent N. Tetri, Melissa J. Contos, Arun J. Sanyal, and Nash Clinical Research Network. 2009. “Comparison of Noninvasive Markers of Fibrosis in Patients with Nonalcoholic Fatty Liver Disease.” *Clinical Gastroenterology and Hepatology* 7(10):1104–12.

Sharples, Linda D. 1993. “Use of the Gibbs Sampler to Estimate Transition Rates between Grades of Coronary Disease Following Cardiac Transplantation.” *Statistics in Medicine* 12(12):1155–69.

Shortle, John F., James M. Thompson, Donald Gross, and Carl M. Harris. 2018. *Fundamentals of Queueing Theory*. Vol. 399. John Wiley & Sons.

Singh, Siddharth, Alina M. Allen, Zhen Wang, Larry J. Prokop, Mohammad H. Murad, and Rohit Loomba. 2015. “Fibrosis Progression in Nonalcoholic Fatty Liver vs Nonalcoholic Steatohepatitis: A Systematic Review and Meta-Analysis of Paired-Biopsy Studies.” *Clinical Gastroenterology and Hepatology* 13(4):643–54.

Sinn, Dong Hyun, Geum Youn Gwak, Ha Na Park, Jee Eun Kim, Yang Won Min, Kwang Min Kim, Yu Jin Kim, Moon Seok Choi, Joon Hyeok Lee, Kwang Cheol Koh, Seung Woon Paik, and Byung Chul Yoo. 2012. “Ultrasonographically Detected Non-Alcoholic Fatty Liver Disease Is an Independent Predictor for Identifying Patients with Insulin Resistance in Non-Obese, Non-Diabetic Middle-Aged Asian Adults.” *American Journal of Gastroenterology* 107(4):561–67. doi: 10.1038/ajg.2011.400.

St-Onge, Marie-Pierre, Ian Janssen, and Steven B. Heymsfield. 2004. “Metabolic Syndrome in Normal-Weight Americans: New Definition of the Metabolically Obese, Normal-Weight Individual.” *Diabetes Care* 27(9):2222–28.

Stewart, William J. 2009. *Probability, Markov Chains, Queues, and Simulation*. Princeton university press.

Tapper, Elliot B., Katherine Krajewski, Michelle Lai, Tracy Challies, Robert Kane, Nezam Afdhal, and Daryl Lau. 2014. “Simple Non-Invasive Biomarkers of Advanced Fibrosis in the Evaluation of Non-Alcoholic Fatty Liver Disease.” *Gastroenterology Report* 2(4):276–80. doi: 10.1093/gastro/gou034.

Vos, Bertrand, Christophe Moreno, Nathalie Nagy, Françoise Féry, Miriam Cnop, Pierre Vereerstraeten, Jacques

Devière, and Michael Adler. 2011. “Lean Non-Alcoholic Fatty Liver Disease (Lean-NAFLD): A Major Cause of Cryptogenic Liver Disease.” *Acta Gastro-Enterologica Belgica* 74(3):389–94.

Wei, Jeremy Lok, Jonathan Chung Fai Leung, Thomson Chi Wang Loong, Grace Lai Hung Wong, David Ka Wai Yeung, Ruth Suk Mei Chan, Henry Lik Yuen Chan, Angel Mei Ling Chim, Jean Woo, Winnie Chiu Wing Chu, and Vincent Wai Sun Wong. 2015. “Prevalence and Severity of Nonalcoholic Fatty Liver Disease in Non-Obese Patients: A Population Study Using Proton-Magnetic Resonance Spectroscopy.” *American Journal of Gastroenterology* 110(9):1306–14. doi: 10.1038/ajg.2015.235.

Wong, Vincent Wai Sun, Leon A. Adams, Victor de Lédinghen, Grace Lai Hung Wong, and Silvia Sookoian. 2018. “Noninvasive Biomarkers in NAFLD and NASH — Current Progress and Future Promise.” *Nature Reviews Gastroenterology and Hepatology* 15(8):461–78. doi: 10.1038/s41575-018-0014-9.

Wong, Vincent Wai Sun, Julien Vergniol, Grace Lai Hung Wong, Juliette Foucher, Henry Lik Yuen Chan, Brigitte Le Bail, Paul Cheung Lung Choi, Mathurin Kowo, Anthony Wing Hung Chan, Wassil Merrouche, Joseph Jao Yiu Sung, and Victor De Ĺedinghen. 2010. “Diagnosis of Fibrosis and Cirrhosis Using Liver Stiffness Measurement in Nonalcoholic Fatty Liver Disease.” *Hepatology* 51(2):454–62. doi: 10.1002/hep.23312.

Xu, Chengfu, Chaohui Yu, Han Ma, Lei Xu, Min Miao, and Youming Li. 2013. “Prevalence and Risk Factors for the Development of Nonalcoholic Fatty Liver Disease in a Nonobese Chinese Population : The Zhejiang Zhenhai Study.” (October 2012):1–6. doi: 10.1038/ajg.2013.104.

Yang, ShiQi, Hong Zhu, Yunbo Li, HuiZhi Lin, Kathleen Gabrielson, Michael A. Trush, and Anna Mae Diehl. 2000. “Mitochondrial Adaptations to Obesity-Related Oxidant Stress.” *Archives of Biochemistry and Biophysics* 378(2):259–68.

Yoneda, Masato, Kento Imajo, Hirokazu Takahashi, Yuji Ogawa, Yuichiro Eguchi, Yoshio Sumida, Masashi Yoneda, Miwa Kawanaka, Satoru Saito, Katsutoshi Tokushige, and Atsushi Nakajima. 2018. “Clinical Strategy of Diagnosing and Following Patients with Nonalcoholic Fatty Liver Disease Based on Invasive and Noninvasive Methods.” *Journal of Gastroenterology* 53(2):181–96. doi: 10.1007/s00535-017-1414-2.

Younes, Ramy, and Elisabetta Bugianesi. 2019. “NASH in Lean Individuals.” *Semin Liver Dis* 39(1):86–95.

Younossi, Zobair, Quentin M. Anstee, Milena Marietti, Timothy Hardy, Linda Henry, Mohammed Eslam, Jacob George, and Elisabetta Bugianesi. 2017. “Global Burden of NAFLD and NASH: Trends, Predictions, Risk Factors and Prevention.” *Nature Publishing Group* 14(1):11–20. doi: 10.1038/nrgastro.2017.109.

Younossi, Zobair M., Deirdre Blissett, Robert Blissett, Linda Henry, Maria Stepanova, Youssef Younossi, Andrei Racila, Sharon Hunt, and Rachel Beckerman. 2016. “The Economic and Clinical Burden of Nonalcoholic Fatty Liver Disease in the United States and Europe.” *Hepatology* 64(5):1577–86.

Younossi, Zobair M., Radhika P. Tampi, Andrei Racila, Ying Qiu, Leah Burns, Issah Younossi, and Fatema Nader. 2020. “Economic and Clinical Burden of Nonalcoholic Steatohepatitis in Patients with Type 2 Diabetes in the US.” *Diabetes Care* 43(2):283–89.

# الملخص العربى

إن هذا الكتاب المقدم من الطبيبة إيمان محمد عطية عبدالخالق الحاصلة على بكالوريوس الطب و الجراحة العامة دفعة 1996 من جامعة القاهرة ، و على درجة الماجستير أمراض باطنة عامة دفعة 2001 من جامعة القاهرة ، و المسجلة على درجة الماجستير بكلية الدراسات العليا للبحوث الإحصائية ؛ قسم الإحصاء الرياضى سنة 2019 فى جامعة القاهرة ، تتناول فيه إستخدام النماذج العشوائية لتحليل تطور الأمراض المزمنة .

**الفصل الأول**: يتناول مقدمة بسيطة عن أنواع النماذج العشوائية المستخدمة فى دراسة الأمراض، و تعريف بسيط عن فصول الكتاب .

**الفصل الثانى** : يلقى الضوء على الأساسيات و التعريفات الرياضية و الإحصائية لسلاسل ماركوف المحددة و المتصلة زمنيا، و الإشتقاق الرياضى لبعض هذه القوانين الرياضية و الإحصائية .

**الفصل الثالث** : يشرح طبيا بشكل مبسط المرض المراد دراسته بالتحديد فى هذه الرسالة وهو ترسيب الدهون بالكبد ، تعريفه و أسبابه، وتطور مراحله المختلفة عبر الزمن، وكيفية تشخيصه و معالجته، و الدراسات الدوائية التى مازالت قيد البحث حتى لحظة الإنتهاء من هذا الكتاب .

**الفصل الرابع** : يتناول الشرح العام المجرد لنموذج ماركوف، ألا وهو( الصحة-المرض-الوفاة) ، و كيفية الحصول على المؤشرات الإحصائية التى تفيد صانع القرار الطبى فى تحديد الموارد الطبية والإقتصادية اللازمة لتشخيص، وعلاج الأفراد الأكثر عرضة للإصابة بهذا المرض. و يتذيل هذا الفصل مثال إفتراضى للتوضيح .

**الفصل الخامس** : يعالج النموذج الأكثر تعقيدا من النموذج العام المجرد الذى تمت دراسته فى الفصل الرابع، حيث أن هذا الفصل يتناول بإسهاب المراحل التسعة المختلفة لمرض ترسيبات الدهون، و كيفية إنتقال المريض بين هذه المراحل المختلفة للمرض، و يمتاز هذا التفصيل للمراحل التسعة بإظهار قدرة المريض على التنقل تقدما و إيابا بين المراحل الأربعة لهذا المرض إنتهاء بالوصول إلى المرحلة المتقدمة من تليف الكبد و الفقدان الكامل لوظائف الكبد، و التى تستدعى عمل زراعة للكبد بما فيها من المضاعفات الناتجة عن مثل هذه الجراحة الكبرى . و فى هذا الفصل إستخدمت الباحثة دالة الترجيح أو الإحتمال الأعظم و معادلة نيوتين التقريبية بمنهجية جديدة ، وعلى الرغم أنهما من الطرق الإحصائية المعروفة لتقدير معدلات الإنتقال بين المراحل المرضية ،إلا أنها إستخدمتهما بمنهجية جديدة تعالج بها القيم المفقودة التى يمكن أن تتواجد فى ملفات متابعة المرضى خلال الدراسات الطولية ، كما أنها إستخدمتها بطريقة تفضى إلى أن معدلات الإنتقال الملاحظة بين المراحل المختلفة، تكاد تقارب تقديرات معدلات الإنتقال بين المراحل المرضية الناتجة عن إستخدام دالة الترجيح الأعظم و معادلة نيوتن التقريبية بهذه المنهجية الجديدة. و بمجرد الحصول على معدلات الإنتقال المقدرة فإنه يمكن الحصول على الدوال الإحتمالية للتنقلات بين هذه المراحل المرضية فى أى توقيت زمنى، ويتأتى هذا بطريقتين عرضتهما الباحثة: فبالطريقة الأولى يتم حل المعادلات التفاضلية بإستخدام قانون لابلاس ثم التعويض بهذه المقدرات فى الدوال الإحتمالية الناتجة عن حل هذه المعادلات التفاضلية، أما الطريقة الثانية فهى أن ترفع مصفوفة معدلات الإنتقال المقدرة الى الأس e .و ينتهى هذا الفصل بمثال إفتراضى لكيفية الحصول على هذه المؤشرات الإحصائية لمثل هذا النموذج الأكثر تعقيدا عن سابقه.

**الفصل السادس**: فيه تستخدم الباحثة كلا من نموذج إنحدار بواسون لربط العوامل المؤثرة على حدوث هذا المرض و معدلات التنقل بين مراحله المختلفة، لتفسير كيف تؤثر هذه العوامل الخطرة المسببة لهذا المرض من سكرى و إرتفاع نسبة الكولسترول و الدهون فى الدم و إرتفاع ضغط الدم على تطور المرض، و الإنتقال بالمريض من مرحلة إلى أخرى. حيث تعد مقاومة الجسم للإنسولين من أهم العوامل المؤثرة و الخطيرة و التى تسبب الإنتقال إلى المراحل المتقدمة و المتأخرة من المرض، و إلى تطور تكوين النسيج الليفى فى الكبد الذى يعد مؤشرا خطيرا لفشل الكبد فى أداء وظائفه. و ينتهى هذا الفصل أيضا بمثال إفتراضى لهذا النموذج .

**الفصل السابع** : تتناول فيه الباحثة النتائج و التوصيات.

و فى الملاحق المكملة لهذا العمل تتناول الباحثة كيفية إستخدام برنامج ماتلاب (matlab) فى تكويد و عمل كود لإجراء هذه العمليات الرياضية لحساب مثل هذه المؤشرات الإحصائية . كما أنه فى فى هذه الملاحق توجد النظريات الفرضية لهذه النماذج الثلاثة المستخدمة فى الفصول الثلاثة السابقة.

فى السطور القليلة القادمة تعريف و تلخيص بمحتوى الكتاب:

تعد سلاسل ماركوف المتصلة زمنيا و المتجانسة و المحددة أداة إحصائية ورياضية مناسبة وجيدة ، لتحليل و معرفة التطور الزمني للمرض عبر مراحله المختلفة . تُسْتَخْدم هذه السلاسل : لمعرفة و تحليل المراحل المختلفة لترسبات الدهون الكبدية، ولمعرفة كيفية تطور هذا المرض بنوعيه الإثنين : الأول ترسيب الدهون مع عدم حدوث إلتهاب كبدي ، و الثاني ترسيب الدهون مع حدوث إلتهاب كبدي ، و لدراسة المراحل المختلفة لتكوين النسيج الليفى المصاحب لهذه الترسبات الدهنية.

أصبح شيوع ترسبات الدهون الكبدية ظاهرة سريعة التوسع و الانتشار عبر العالم ، بالتوازي مع انتشار السمنة ، و سكر الدم من النوع الثاني ، اللذين أصبحا أقرب إلى كونهما جائحة . متلازمة الأيض هي الأخرى أحد الأسباب الخطيرة للإصابة بترسبات الدهون الكبدية.

فى هذا الكتاب ، تم دراسة هذه الترسبات إحصائيا ، بإستخدام نموذج عشوائي فى أبسط أشكاله ، لتمثيل مراحل (الصحة-المرض-الوفاة) . في هذا النموذج : تمثل المرحلة الأولى الأشخاص الأكثر عرضة للإصابة بالمرض ، مثل : مرضى السكري من النوع الثاني، و مرضى ارتفاع الدهون و الكولسترول بالدم، و مرضى ارتفاع ضغط الدم ، أما المرحلة الثانية فهى تمثل الإصابة بالترسبات الدهنية الكبدية بجميع أشكال هذا المرض و مضاعفاته . و عن آخر مرحلتين فى هذا النموذج : أحدهما مرحلة الوفاة نتيجة مضاعفات الإصابة بالترسبات الدهنية الكبدية ، و الأخرى تمثل مرحلة الوفاة نتيجة أي سبب آخر غير إصابة الكبد بمضاعفات ترسب هذه الدهون. وهذا النموذج يتناوله الفصل الرابع بالشرح و التفصيل .

بالإضافة لهذا النموذج البسيط ، يوجد النموذج الأكثر إسهابا و إستفاضة و الأوفر عمقا و تشريحا و تفصيلا لمراحل المرض المختلفة ؛ حيث يحتوى النموذج على ثمانية مراحل تفصيلية لتطور هذا المرض عبر الزمن ، و المرحلة التاسعة هى مرحلة الوفاة . و هذا الشرح التفصيلي يتناوله الفصل الخامس.

علاوة على ذلك ، يمكن دراسة مجموعة من المراحل المقتطعة أو المقتطفة من النموذج التفصيلى التوسعي ، و هى تركز على المراحل الأولى للمرض ، من حيث الإصابة بترسيب الدهون ، و بدء تكوين النسيج الليفى ، مع تطور مراحله المختلفة القابلة للتحسن فى حالة إزالة العوامل الخطيرة المسببة لتكوينه ، حتى الوصول الى تكوين نسيج ليفى ، لا يمكن التحسن منه ، و الذى يُفضي الى التليف الكبدي ؛ الذى هو بداية مرحلة فقدان الكبد الكامل لوظائفه الحيوية الشديدة الأهمية ؛ و هذه نتيجة حتمية لبقاء هذه العوامل، وتحفيزها لتكوين نسيج ليفى ، فى عضو لا يجب أن يتكون فيه مثل هذا النسيج الليفى ، دون إجراءات علاجية قوية وجدية. و تغطية هذه التأثيرات السلبية والتفاعلات المتداخلة بين المسببات المؤذية لهذا العضو الحيوى ، و المفضية الى هذا التدهور التدريجى المستمر فى وظائف الكبد ؛ فإن الفصل السادس يتناولها بالشرح والإسهاب فى العرض المفصل لكيفية استخدام نموذج انحدار بواسون لتقدير معدلات الإنتقال بين المراحل المختلفة لتكوين النسيج الليفى داخل الكبد.

و هذه النماذج الثلاثة المُسْتَخْدَمَة فى الفصول السابقة الذكر تَسْتَخدم بيانات غير حقيقية، و إنما هى لشرح ولتوضيح المفاهيم و المؤشرات الإحصائية التى يمكن الحصول عليها ، باستخدام سلاسل ماركوف المتصلة زمنيا والمتجانسة و المحددة ، و هذه المؤشرات هى كالآتى:

- معدلات الإنتقال المقدرة بين المراحل المختلفة المكونة للمرض.
- إحتمالات الإنتقال بين المراحل المختلفة المكونة للمرض.
- متوسط زمن بقاء المريض فى كل مرحلة.
- متوسط زمن إنتقال المريض من أى مرحلة وصولا إلى مرحلة الوفاة ؛ بكلمات أُخر، متوسط زمن بقاء المريض على قيد الحياة فى حالة تواجده فى أى مرحلة من مراحل المرض.
- متوسط العدد المتوقع للمرضى المتواجدين فى كل مرحلة.
- التوزيع الإحتمالى لمراحل المرض بعد مرور فترة معينة من الزمن .

إن مثل هذا التحليل الإحصائى لمراحل المرض يعطى رؤية أشمل، وأوضح، وأفضل لمتخذى القرار، و واضعى الخطط و السياسات الطبية الإقتصادية، وأيضا للأطباء فى كيفية الإكتشاف المبكر لمراحل المرض المختلفة ، خاصة لدى الأشخاص الأكثر عرضة للإصابة به ؛ لتجنب الفشل الكبدى ، خاصة مع قرب ظهور أدوية جديدة لمعالجة تَكْوين النسيج الليفى ، قبل أن يَسْتَشْري فى الكبد، ويصل بالمريض إلى مرحلة الفشل الكبدى الغير قابل للتحسن والشفاء. حيث أنه يقع على عاتق الطبيب قرارات : بأى العلاجات الدوائية يجب عليه أن يبدأ بها أولاً ؟ و ما هو الهدف المرجو الوصول إليه و تحقيقه؟ و كيفية الإبقاء على هذا الهدف؟ و ما هى الأدوية التى ممكن إضافتها أو استبدالها إذا لم يتم التوصل إلى الهدف المَرْجُوْ؟ وكيف يتابع نتيجة هذه الأدوية؛ بمعنى آخر، ما هى الفحوصات الطبية الواجب على الطبيب إجراؤها للمريض لمتابعة إستجابته للعلاج المقدم إليه؟ كل هذا يزيد من الأعباء الإقتصادية التى تقع على

عاتق المريض، و على عاتق المنظومة الصحية فى البلاد، وما تتكلفه ميزانية الدولة من أجل قطاع الصحة . و من ناحية أخرى، فى حالة إصابة المريض بالفشل الكبدى، و إحتياجه إلى إجراء عملية زراعة كبد ؛ فإن كل هذا يزيد من الأعباء الإقتصادية على ميزانية الدولة، و لهذا فإن هذه الدراسات الطولية مفيدة وناجعة، لمعرفة ما هى الفحوصات الطبية الإقتصادية ذات النتائج عالية الجودة التى يمكن إجراؤها للمريض فى وحدات الرعاية الصحية الأولية؟ و متى يجب تحويل المريض لعمل عينة كبدية فى المراكز التخصصية ؛ و هذا لما يصاحب إجراء العينة الكبدية من مضاعفات، و تحديات، و صعوبة إجرائها لكل مرضى الترسبات الدهنية الكبدية؟كما أن بعض الفحوصات الطبية هى نماذج رياضية تم تجربتها على قوميات معينة بذاتها ، و لم يتم اختبارها على قوميات أخرى، و هذا يستلزم دراسات إضافية ؛ لمعرفة فاعلية، و مشروعية هذه الفحوصات غير النافذة فى جسم المريض . بالإضافة أن تحديد المستويات الفاصلة و المسببة للإصابة بالمرض من عدمه غير ثابتة و متغيرة من دولة إلى أخرى ، و تختلف داخل الدولة الواحدة بناءاً على القوميات المختلفة فى هذه الدولة ؛ مما يحتاج إلى دراسة أكثر عمقا لتحديد هذه المستويات الفاصلة لهذه الفحوصات . علاوة على ذلك، فإن الحاجة إلى هذه الفحوصات غير النافذة و النتائج المترتبة عنها تستلزم معرفة الإرتباط بينها و بين نتائج عينة الكبد، فى كل مرحلة من مراحل المرض؛ لِيُمْكِن الإستعاضة عن عينة الكبد بالفحوصات غير النافذة، و تجنب المخاطر المحدقة و المصاحبة لإجراء عينة الكبد، و إقتصارها على النتائج الغير قطعية و الغير واضحة لهذه الفحوصات غير النافذة. و هذا كله يتطلب المزيد من الدراسات، خاصة الطولية منها لمعرفة مثل هذا الإرتباط ؛ وهذا كله يصب فى مصلحة التشخيص الصحيح ، الأقرب إلى مجافاة الأخطاء ، التى تتسبب فى تأخر التشخيص و بالتالى العلاج والشفاء ، و التى تتسبب أيضا فى تدهور حالة المريض الصحية . و يجب لمثل هذه الفحوصات أن تتميز ببعض الخصائص : كونها يمكن تنفيذها بسهولة على أرض الواقع، و عملية، و مناسبة، و معقولة، و آمنة، و ملائمة، و ذات تكلفة إقتصادية يمكن للمريض أ ن يتحملها ؛ و هذا كله لِيُمْكِن استخدامها فى المتابعة المتكررة للمريض. إضافة إلى كل ما سبق، فإن بعض البروتوكولات للجمعيات و الهيئات و المنظمات العالمية المهنية المتخصِّصَة فى دراسة أمراض الكبد، و الجهاز الهضمى، و أيضا فى دراسة السمنة ، و مرض السكرى مثل الجمعية الأوروبية لدراسة السمنة و السكرى وأمراض الكبد - تتواجد فى الفصل الثالث.

سلاسل ماركوف المتصلة زمنيا و المتجانسة و المحددة هى أحد الأدوات الإحصائية ذات القيمة العالية و النافعة فى الأبحاث الطبية، لفهم و إدراك الفسيولوجيا المرضية؛ أي كيفية حدوث المرض و تأثيره على عمل العضو المصاب . و هنا فى هذا الكتاب أخص بالذكر ضرورة فهم وإدراك التطور المرحلى و الزمني للترسبات الدهنية الكبدية، و ما تحدثه من تأثيرات فى الكبد، و أيضا ضرورة الفهم والإدراك الشامل للتفاعلات التى تحدث بين المتغيرات الداخلية و الخارجية المسببة لهذا المرض . حيث أنه من هذه المتغيرات الخارجية هو نمط الحياة الذى يتميز بالإسترخاء، و قلة ممارسة الرياضة، و تناول كميات كبيرة من الدهون و السكريات فى الطعام، بينما المتغيرات الداخلية تتمثل فى العوامل الوراثية و الخلفية الجينية التى تؤثر على عمليات الأيض : المتمثلة فى تكسير وأكسدة و تخزين ما يتناوله الإنسان، و غيره من العمليات الخلوية الأخرى مثل القدرة على تكوين النسيج الليفى الذى يُنْتِج عند تكوينه بعض المواد المفرزة فى الدم، و التى تدل على أنه قَد بَدأت بالفعل عملية تكوين هذا النسيج . و يُعد هذا النسيج الليفى عاملا مُتلفاً و ضاراً و مُؤْذياً للكبد، وعاملاً لتدهور الوظائف الحيوية للكبد، والمضاعفات الناتجة عن ذلك. و إذاً تَكْمُن أهمية هذا الفهم فى الإجابة على التساؤل عن أهمية و ضرورة إجراء تحليل وراثى للمرضى الأكثر عرضة للإصابة بتكوين هذا النسيج الليفى ؛ نتيجة الترسبات الدهنية الكبدية، و هل يجب أن تتضمنه منظومة التأمين الصحى؟ وهل يجب أيضا إدراج تحليل المواد التى تَنْتُج و تُفْرَز عند تكوين هذا النسيج ؟ و حيث أنَّ كِلا التحليلين بنوعيه باهظ و مكلف؛ فهل توجد جدوى إقتصادية من عمل مثل هذه الفحوصات للمريض على المدى الطويل و الأجل البعيد؟ فإن كل هذا يستلزم القيام بعمل دراسات طولية للإجابة عن مثل هذه التساؤلات. و هذا الشرح المبسط للفسيولوجيا المرضية يَحْتويه أيضا الفصل الثالث.

و حيث أن عدم قيام الإنسولين بوظائفه الخلوية على الوجه الأكمل ، هو أحد اللاعبين الرئيسيين فى تطور المرض عبر مراحله الزمنية المختلفة، و التى تكون قابلة للتحسن فى المراحل الأولى، بينما هى على النقيض غير قابلة للتحسن ، مع وصول تكوين النسيج الليفى إلى مستوياته الضارة ( مرحلة تليف الكبد و فقدان وظائفه الحيوية). فإن عمل الإنسولين هذا يتمثل فى المتغير المسمى (هوما 2 )، و الذى هو قياس لمستوى الإنسولين و السكر بالدم ، من خلال خوارزمية تماثل عملهما الفسيولوجى فى الجسم . وقد تبين أنه هو الفاعل و المؤثر الواضح فى تطور المرض عبر مراحله الزمنية المختلفة، و تفاعل هذا المؤثر مع المؤثرات الأخرى يحدد معدلات الإنتقال بين المراحل المختلفة للمرض، حيث أنه يوجد أيضا ارتباط بينه، و بين المؤثرات الأخرى ، مثل : ارتفاع نسبة الدهون و الكولسترول بالدم ، و ارتفاع ضغط الدم .و هنا يلوح فى الأفق تسؤال، عن إمكانية إستخدام قياسات الأحماض الدهنية الضارة فى الدم، كوسيلة لمتابعة تطور هذا المرض ومدى إستجابة المرضى للعلاج. وهذا موضح فى الفصل السادس.

التوصيات :

البيانات القادمة من مصر قليلة و ضئيلة عن انتشار مرض الترسبات الدهنية الكبدية بنوعيه ، و المضاعفات الناتجة عنه، بالإضافة إلى قلة البروتوكولات المستخدمة للتعرف على المرضى المصابين به، و تقسيمهم إلى طبقات تشخيصية ، لتقديم العلاج

المناسب لكل طبقة اعتمادا على هذا التقسيم التشخيصى . بالتالى فإن إجراء دراسات طولية تستخدم سلاسل ماركوف المتصلة زمنيا و متجانسة و محددة ، لبناء نموذج عشوائى يمثل مراحل المرض المختلفة و تطورها عبر الزمن تمثيلا جيدا بناءا على : الخصائص الديموجرافية للمرضى، و على المتغيرات و العوامل ذات الخطورة العالية، و المسببة للمرض ، سوف يساعد على تغطية هذه النقاط و القضايا الشائكة.

علاوة على ذلك، فإن بعض المرضى لا يعاني من السمنة بتعريفاتها المتعارف عليها، و لكن تركيبهم الأيضي يعاني بعض الخلل فى وظائف الأيض الحيوية الخلوية ؛ و بالتالى فإن مثل هذه الدراسات الطولية تساعد في فهم و إدراك طبيعة ما يعانيه هؤلاء الأشخاص، و كيفية متابعتهم عبر حياتهم المستقبلية.

أيضا المرضى الذين يعانون من خلل فى إمتصاص الطعام عبر الجهاز الهضمي ؛ فإنهم عرضة للإصابة بنقص بعض العوامل المهمة فى الجسم مثل : الأحماض الدهنية الأساسية، و الكولين الضرورى لتحريك الدهون من الكبد وعدم ترسبها فيه ، و بالتالى حماية الكبد من التليف و فقدان وظائفه الحيوية ؛ و هؤلاء بالضرورة فى حاجة إلى تعويض هذا النقص ، بزيادة تناول هذه المواد بالكيفية التى تكون فيها هذه المواد قابلة للإمتصاص . و إستخدام مثل هذه الدراسات يوفر القدرة على متابعة تطورات وظائف الكبد إستجابة لمثل هذه العلاجات .

كما أن النماذج متعددة المراحل يمكن استخدامها لتحليل مدى تأثير العوامل، التي تتنازع مثل هؤلاء المرضى، للوصول بهم إلى مرحلة الوفاة. حيث أن الأسباب الأكثر شيوعا للوفاة: هى تصلب الشرايين و أمراض القلب و الكلى، وهي بذلك تحتل المركز الأول و الثاني، و تليها فى المركز الثالث أسباب الوفاة الناجمة عن تليف الكبد، و سرطان الكبد، والمضاعفات المرتبطة بزراعة الكبد .

كما أنه يمكن إستخدام سلاسل شبه ماركوف، و نموذج ماركوف الخفي، و نماذج ماركوف الكامنة، لدراسة تطور الأمراض المزمنة عبر الزمن جنبا إلى جنب لسلاسل ماركوف المتصلة زمنيا و المتجانسة و المحددة.

# تحليل تطور الأمراض المزمنة باستخدام النماذج العشوائية

اعداد

**طبيبة / إيمان محمد عطية عبد الخالق**
**ماجستير أمراض باطنة عامة – جامعة القاهرة**
**دبلومة إحصاء- جامعة القاهرة**

**2021**